\documentclass[12pt, openany, titlepage]{report} 

\usepackage[
    letterpaper,
    left=1in,
    right=1in,
    top=1in,
    bottom=1in
    ]{geometry}
\usepackage{lmodern}
\usepackage{graphicx,caption}
\usepackage{booktabs}
\usepackage{moresize}
\usepackage{mathtools}
\usepackage{comment}
\usepackage{authblk}
\usepackage{titletoc}
\usepackage{titlesec}
\usepackage[dvipsnames]{xcolor}
\usepackage[hyphens]{url}
\usepackage[breaklinks=true,colorlinks=true, linkcolor=black, urlcolor=blue]{hyperref}
\usepackage{siunitx}
\usepackage{longtable}
\usepackage{xfrac}
\usepackage{caption}
\usepackage{amsmath}
\usepackage{amssymb}
\usepackage{gensymb}
\usepackage{multirow}
\usepackage{array}
\usepackage{textcomp}
\usepackage{xcolor}
\usepackage{placeins}
\usepackage[style=authoryear, backend=biber]{biblatex}
\definecolor{docblue}{RGB}{0,120,230}

\titleformat{\chapter}
  {\normalfont\sffamily\LARGE\bfseries\color{docblue}}{\thechapter}{1em}{}
\titleformat{\section}
  {\normalfont\sffamily\Large\bfseries\color{docblue}}
  {\thesection}{1em}{}
\titleformat{\subsection}
  {\normalfont\sffamily\bfseries\color{docblue}}
  {\thesubsection}{1em}{}
\titleformat{\subsubsection}
  {\normalfont\sffamily\bfseries}
  {\thesubsubsection}{1em}{}

\title{\vspace{-4.0cm}\includegraphics[width=6cm]{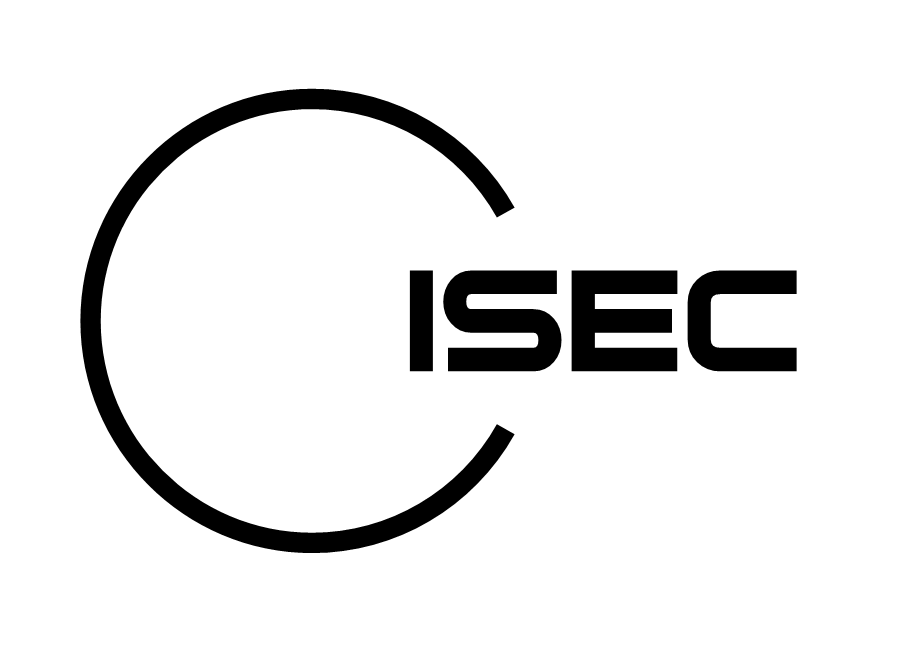}\\[1cm]Powering the Space Elevator}
\usepackage[titletoc]{appendix}

\begin{document}

 \textcolor{docblue}{\section*{}}

\thispagestyle{empty}

\vspace{-4.0 cm}

\textbf{\small International Space Elevator Consortium  \hspace{1cm}       ISEC Study Report \#2026-1}

\vspace{2.5 cm}

\begin{center}
\textbf{\HUGE Powering the Space Elevator}
\\

\vspace{1.0cm}

\includegraphics[width=\linewidth]{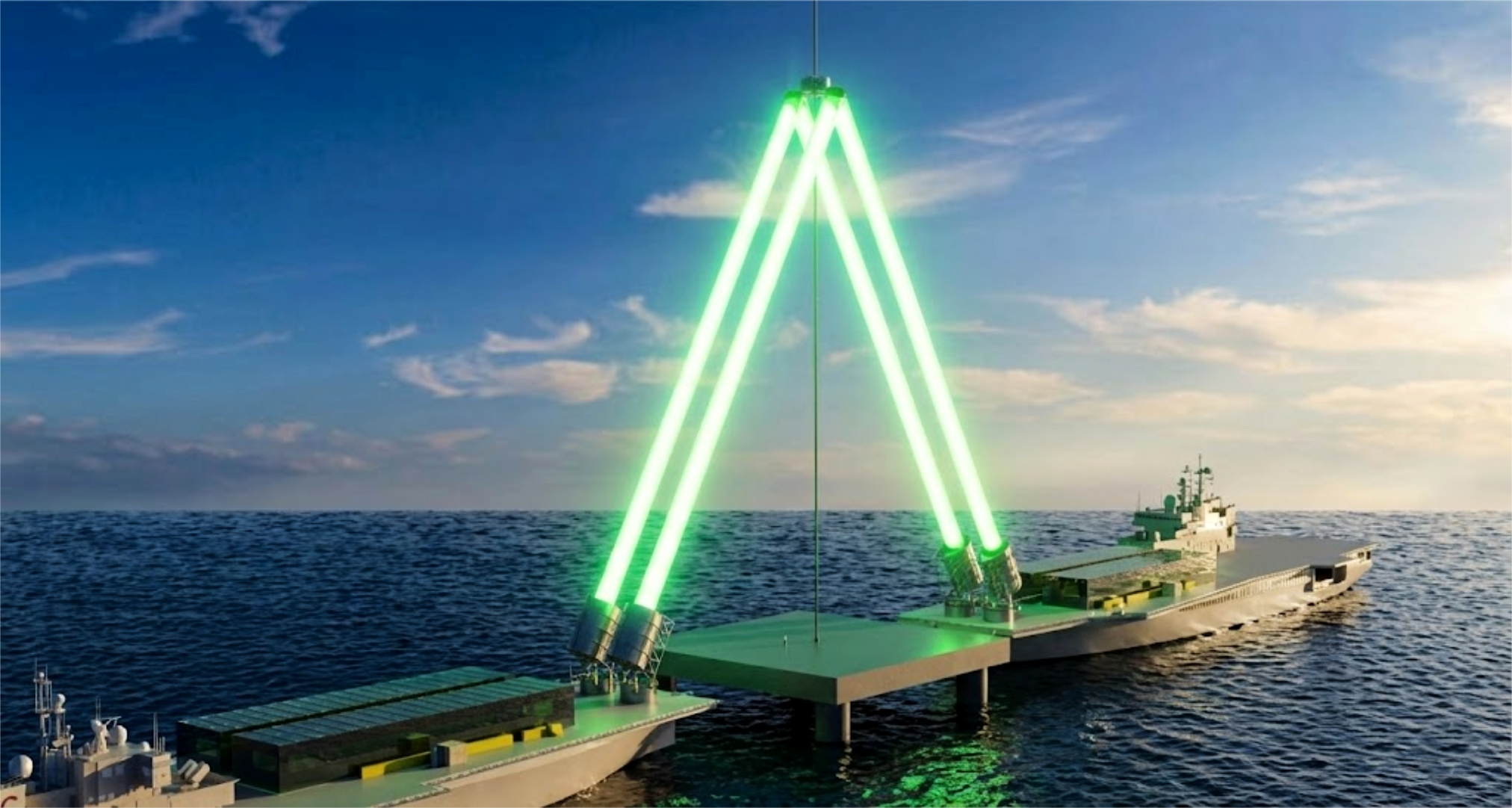}
\end{center}

\vspace{3.5cm}

  \begin{flushright}
  \begin{minipage}{8cm}
  \includegraphics[width=8cm]{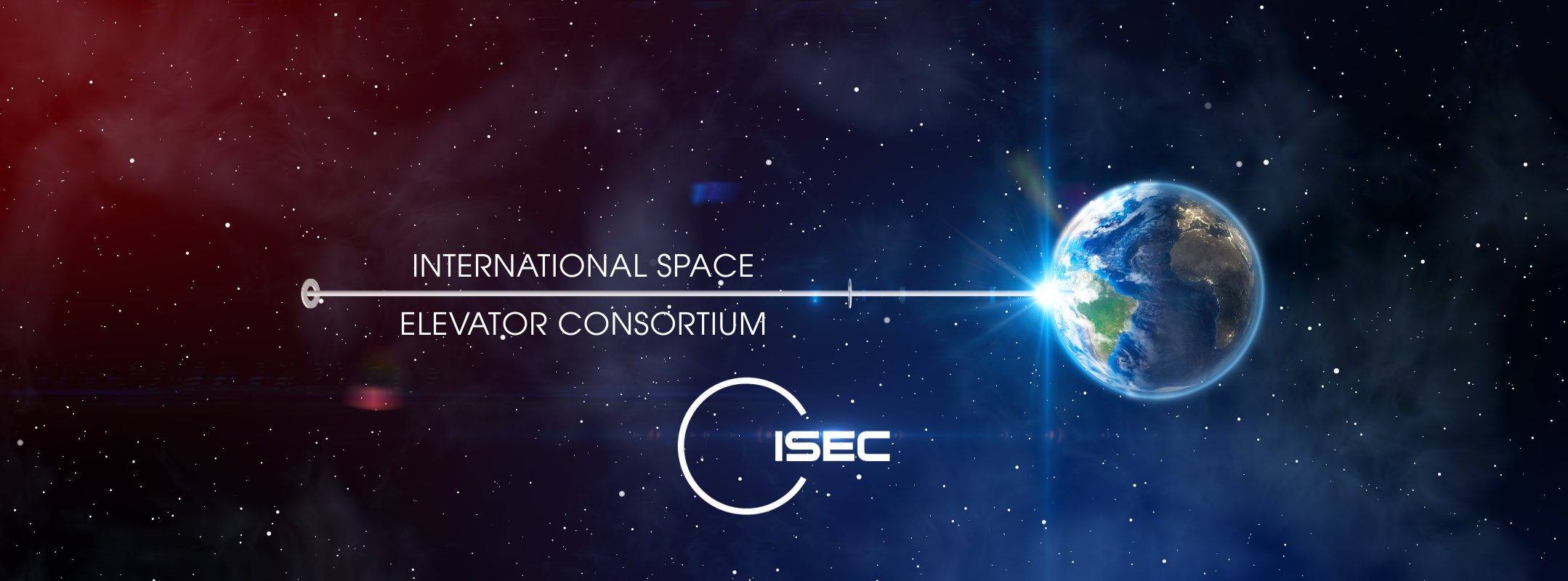}
  \end{minipage}
   \end{flushright}

\vspace{-5cm}
\noindent\textbf{\large Larry Bartoszek}\par
\noindent\textbf{\large A.J. Burke}\par
\noindent\textbf{\large Shreyansh Chauhan}\par
\noindent\textbf{\large Avery Davis}\par
\noindent\textbf{\large Hassan El Chab}\par
\noindent\textbf{\large Blaise Gassend}\par
\noindent\textbf{\large John Knapman}\par
\noindent\textbf{\large Martin Lades}\par
\noindent\textbf{\large George Neil}\par
\noindent\textbf{\large Adrian Nixon}\par
\noindent\textbf{\large Dennis H. Wright}\par

\textcolor{docblue}{\section*{}}

\thispagestyle{empty}

\vspace{-3.0 cm}

\begin{center}
\includegraphics[width=8cm]{ISECBW.png}\\

\vspace{2.0 cm}

\textbf{\LARGE Powering the Space Elevator}

\vspace{2.0 cm}

\textbf{\large International Space Elevator Consortium}\\
\textbf{\large 2026} \\

\vspace{2.0 cm}

\textbf{Authors:} \\
\textbf{Larry Bartoszek}\\
\textbf{A.J. Burke}\\
\textbf{Shreyansh Chauhan}\\
\textbf{Avery Davis}\\
\textbf{Hassan El Chab}\\
\textbf{Blaise Gassend}\\
\textbf{John Knapman}\\
\textbf{Martin Lades}\\
\textbf{George Neil}\\
\textbf{Adrian Nixon}\\
\textbf{Dennis H. Wright}\\

\vspace{1.0 cm}

\end{center}

\thispagestyle{empty}
\newpage
\textcolor{docblue}{\section*{}}

\vspace{-3.0 cm}

\begin{center}
{\Large Powering the Space Elevator}
\end{center}

\vspace{2.0 cm}

\noindent Copyright \textcopyright  \hspace{0.1 cm} 2026 by:

\noindent Larry Bartoszek\hfill\\
A.J. Burke\\
Shreyansh Chauhan\\
Avery Davis\\
Hassan El Chab\\
Blaise Gassend\\
John Knapman\\
Martin Lades\\
George Neil\\
Adrian Nixon\\
Dennis H. Wright

\vspace{1.0 cm}

\noindent All rights reserved, including the rights to reproduce this manuscript or portions thereof in any form.
\bigbreak
\noindent Published by lulu.com
\bigbreak
\noindent 978-1-105-42511-0
\bigbreak
\noindent Printed in the United States of America
\thispagestyle{empty}
\chapter*{Preface}
\addcontentsline{toc}{chapter}{\text{Preface}}

One of the missions of the International Space Elevator Consortium (ISEC) is to identify areas of study critical to the development of Earth-anchored space elevators. A typical study consists of an in-depth examination of the state of the art technologies needed for a given component, as well as proposals or conceptual designs that could meet the need. Previous studies indicated the physical conditions that must exist to allow the climber to ascend the space elevator tether; they also indicated a gap in our knowledge, both in concept and in detail, of how the climber could be powered. With the help of experts in the fields of power transmission and conversion, solar power, electric motors, mechanical engineering, strong materials and tether dynamics, the climber power study was able to fill some of the gaps and recommend power options. The conceptual designs presented here demonstrate that a climber could be powered with today's technology, but the rapid advances expected in this area all but guarantee changes in design and power recommendations. Hence, the research areas necessary for future development were identified. It is hoped that the results of this study will convince the reader that there are viable options for powering the climber and that they will inspire deeper research into this area.  
\bigbreak
 \noindent Signed:

    \includegraphics[width=0.4\linewidth]{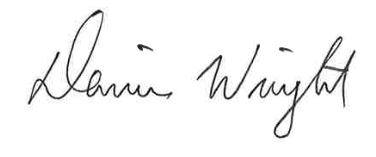}
 \\
 ISEC President
 \\
 31 January 2026

\setcounter{page}{1}
\chapter*{Executive Summary}
\addcontentsline{toc}{chapter}{\text{Executive Summary}}

This report presents the results of research into several different methods to power the climbers on the space elevator.  The climber in the report is the reference conceptual design from [\cite{wright2023climber}].  The first part of the report states the requirements common to any climber power system and establishes the power requirements for multiple climbers on the tether simultaneously.  Chapter 2 also identifies all of the components on a climber that use electrical power.

The power delivery methods shown are laser power beaming, microwave power beaming, solar-powered climbers, and climbers powered by an electrically conducting tether that is energized.

The short wavelength of the laser-powered climbers allows for the smallest, lightest receivers on the climbers.  Laser systems do not have the efficiency of microwave-powered systems, but the longer microwave wavelength leads to a rapidly diverging beam that forces the receivers on the climber to be very large for microwave systems.  Microwave power beaming may also be limited to a maximum power density in the ionosphere of 230 \si{W/m^2} to avoid communications disruptions.  Further work is needed to determine the appropriate power density limit.

Both laser and microwave power beaming systems require large and expensive infrastructure on the ground to power the transmission systems.  Laser transmitters on the ground are small enough that they can be made mobile on ships at sea.  Microwave transmitters on the ground are too large to mount on ships and must be in fixed positions.

The solar-powered climber does not need extensive ground power infrastructure, but its challenges come from the fact that sunlight is much lower in power (by up to a factor of 1,000) than the light intensities of power beaming, it is not available at night (for climbers at low altitudes), and the source moves around throughout the day.  The low intensity of sunlight forces the receiver to be larger than any of the power beaming receivers, and as such it is a large target to orbital debris in LEO.

The tether-powered climber requires an additional level of manufacturing complexity on the tether above the already daunting process of making a tether strong enough out of single-crystal graphene or hexagonal boron nitride.  The conductive tether requires interleaved layers of conducting graphene and insulating boron nitride to create a tether with sufficiently conductive "wires" to deliver power to the climber at any altitude.  This tether manufacturing process is currently completely unknown.

The difference between ascending and descending climbers is described, highlighting the radically different power infrastructures required by each.  No current technology exists to fully recover the energy of a descending climber.

\renewcommand{\contentsname}{\textcolor{docblue}{\textsf{Table of Contents}}}
\tableofcontents

\chapter{Introduction: The Problem of Powering Climbers on the Space Elevator}\label{chapt:intro}

A number of technologies to power space elevator climbers have been proposed over the years. Some of the more unusual and esoteric ones are discussed in Chapter \ref{chapt:OtherMethods}.  This study will primarily focus on climbers as electric vehicles that use electric motors to roll straight up the elevator tether and out into space.  This type of pinched-wheel climber needs to squeeze the tether with their wheels with tons of force to get enough traction to roll upwards on the tether.  The ISEC study [\cite{wright2023climber}] shows a reference conceptual design for a climber that is the basis of the climbers in this study.

A premise of this study is that we are designing for the first space elevator.  It is not reasonable to assume power from Solar Power Satellites if the space elevator will be built to deliver the mass of the SPS system to GEO.  The designs presented here attempt to be for the first generation space elevator, and only deviate from that where limitations in the power system design can be overcome by assuming features of a later iteration.

It takes megawatts of power to boost the climber away from the surface of the Earth.  The basic question is, where does all this power come from?  The first commercial climbers will only mass about 20 tonnes and more than half of that will be payload.  As will be shown, it requires mass in the traction drive just for the motors, axles, wheels, bearings, and all the structure.  There isn't room in the mass budget of the climber for batteries big enough to provide that much power.  Or, for a nuclear reactor.  A key differentiator between rockets and the space elevator is that climbers on the elevator do not carry fuel.

So the question becomes, if you can't store or release enough energy on the climber, where does it come from?

This study contains a compendium of methods we have studied over several years that are possible means of getting power to the climbers.  Broadly, the methods break down into two categories: transmitting power wirelessly, whether the source is the sun or an artificial source such as a laser or microwave transmitter, and using the tether as an electrical power cable.  (See Chapter \ref{chapt:OtherMethods} for alternatives.)  Some methods have more technical challenges than others, requiring new materials, technologies, or techniques that require research and development.  Some are just a matter of spending the money required to build them at the scale we need, but the technology is known.  As with any spacecraft, one of the biggest challenges is getting necessary functions (such as high voltage conversion) at a mass that fits within the budget of the climber.  This may ultimately derail some of the concepts shown here.

ISEC refers to the space elevator as the "green road to space".  Clearly, the greenest power architecture would be entirely solar-based.  An unfortunate fact of sunlight is that it is a relatively diffuse power source.  One can see by comparing the designs presented here where the challenges of each design come from.  And even if a power system is driven by a nuclear reactor as a power source, it is still greener than hundreds to thousands of rocket launches in a year.

We leave it to the reader to compare the various ideas shown and choose between them as if you were the CEO of the Space Elevator Company.  Someone will have to make these choices before a space elevator can become a reality.

\section{Constants and Nomenclature}
The current topic spans several disciplines and many orders of magnitude in scale.  This
often leads to conflicts in terminology and reserved symbols for various constants.   One 
particular case is the use of SI and English units.  Here, SI units are used wherever possible.
In some cases, though, where particular components are referred to, or software defaults
are assumed, English units will appear.  In these cases the corresponding conversion to 
SI will be included.

Other frequently used terms and constants are defined in the following tables.
\par
\bigskip
\noindent \textbf{Physical Constants}
\par
\begin{tabular}{l l}
c = 299,792,458 m/s & : speed of light in vacuum \\
$\epsilon_0 = 8.854 \times 10^{-12}$ F/m & : permittivity of free space \\
$h = 6.626 \times 10^{-34}$ \si{Joule \cdot sec} & : Planck's constant \\
$\mu_0 = 1.2566 \times 10^{-6}$ \si{N/A^2} & : permeability of free space \\ 
$\mu_E = 3.9860 \times 10^5$ km$^3$/s$^2$ & : gravitational parameter of Earth \\ &  \hspace{0.1cm} (Earth's mass $\times$ universal gravitational constant G) \\
$G_{SC} = 1,361 $ W/m$^2$ & : solar constant \\
$g = 9.80665 \; $ m/s$^2$ & : standard gravity \\
$\omega_E = 7.29212 \times 10^{-5}$ rad/s & : angular velocity of Earth's rotation \\
$R_E = 6,378$ km & : equatorial radius of Earth \\
$R_G = 42,164$ km & : radius of geosynchronous orbit \\
$\sigma_{SB} = 5.6704 \times 10^{-8}$ W/m$^2$/K$^4$ & : Stefan-Boltzmann constant 
\end{tabular}
\par
\bigskip
\noindent \textbf{Units}
\par

\begin{tabular}{l l}
AU & : Astronomical unit, distance between the Sun and Earth \\
\si{K} & : Kelvin \\
kg & : kilogram \\
km & : kilometers \\
kph & : kilometers per hour \\
ksi & : kilo-pounds per square inch = 6.895 MPa \\ 
N, kN, MN & : Newton, kiloNewton, megaNewton \\
Pa, MPa, GPa, TPa & : Pascal, megaPascal, gigaPascal, teraPascal \\
S/m & : Siemens per meter, electrical conductivity \\
t or tonne & : metric ton = 1000 kg \\
ton & : US ton =  0.907185 metric ton \\
V, kV, MV & : volt, kilovolt, megavolt \\
W, kW, MW & : watt, kilowatt, megawatt \\
\end{tabular}

\newpage
\noindent \textbf{Acronyms}

\begin{longtable}[l]{l l}
\endfirsthead
\multicolumn{2}{l}{\textbf{Continuation of Acronyms}}\\
\endhead
AESA & : Active Electronically Scanned Array \\ 
AO & : Adaptive Optics \\
BAF & : Beam Aspect Factor \\
BPP & : Beam Parameter Product \\
CFRC & : carbon fiber reinforced composite \\
CLMR & : abbreviation for Climber \\
CNT & : Carbon NanoTube \\
CW & : continuous wave \\
ER & : energy recovery \\
ETL & : electron-transporting layer \\
ETA & : Estimated Time of Arrival \\
FEA & : finite element analysis \\
FEL & : Free Electron Laser \\
FFAG & : Fixed Field Alternating Gradient (magnets) \\
GEO & : geostationary orbit \\
GSL & : graphene super-laminate \\
hBN & : hexagonal boron nitride \\
HEMM & : High Efficiency Megawatt Motor \\
HEU & : highly enriched uranium \\
HTL & : hole-transporting layer \\
HTS & : high temperature superconductor \\
IOC & : Initial Operational Capability \\
IR & : infrared light \\
LEO & : Low Earth Orbit \\
LEU & : low enriched uranium \\
LFR & : lead fast reactor \\
LMR & : liquid metal cooled reactor \\
LPC & : Laser Power Converter \\
LWR & : light water reactor \\
MEO & : Medium Earth Orbit \\
MINIX & : Microwave Ionosphere Nonlinear Interaction Experiment \\
MPT & : Microwave Power Transmission \\
MSR & : Molten Salt Reactor \\
NRE & : non-recurring engineering (costs) \\
OFHC & : Oxygen Free High Conductivity (refers to copper) \\
PB & : power beaming \\
PD & : Power Density \\
PPC & : Photonic Power Converter \\
PV & : photovoltaic \\
PWR & : pressurized water reactor \\
RCD & : reference conceptual design \\
ROM & : rough order of magnitude \\
RF & : radio frequency \\
SCG & : single crystal graphene \\
SFR & : sodium fast reactor \\
SMR & : Small Modular Reactor \\
SPG & : space power grid \\
SPS & : Solar Power Satellite \\
SRF & : Superconducting radio frequency \\
UV & : ultraviolet light \\
VEHSA & : Vertical Epitaxial Heterostructure Architecture \\
WPT & : Wireless Power Transmission \\
\end{longtable}
\bigskip

\noindent \textbf{Variable and Parameter Names}

\begin{longtable}[l]{l l}
\endfirsthead
\multicolumn{2}{l}{\textbf{Continuation of Variable and Parameter Names}}\\
\endhead
A & : area of aperture, cross-sectional area of a solenoid coil, area of PV array \\   
& \hspace{0.1cm} experiencing drag in the atmosphere, cross-sectional area of the tether, \\
& \hspace{0.1cm} or cross-sectional area of conductors in the solar receiver \\
$A_{beam}(z)$ & : area of the beam at the distance z \\
$A_{RX}$ & : area of receiver \\
$A_{TX}$ & : area of transmitter \\
$A_w$ & : a constant used in $w(s,t,\varphi)$ \\
A(z) & : amplitude of the wave equation normalized to beam power \\
b & : depth of focus \\
B & : magnetic field \\
$B_w$ & : Used as a constant in $w(s,t,\varphi)$ \\
$c_\parallel$ & : velocity of wave propagation longitudinally on the tether \\
$c_\perp$ & : velocity of wave propagation perpendicular to the tether \\
C & : capacitance, farads \\
$C_D$ & : drag coefficient \\
$C_l$ & : longitudinal compressive force \\
$C_{max}$ & : maximum longitudinal compressive force \\
d & : separation between electrodes in a capacitor \\
D & : diameter of aperture, subscript r for receiver, t for transmitter \\
E & : Young's modulus or stored energy in batteries, etc, or electric field \\ 
$E_t$ & : total power required to get from the surface of the Earth to GEO \\
f & : focal length of $D_t$ (transmitter diameter), safety factor on stress in the tether, \\
& \hspace{0.1cm} or the frequency of waves on the tether \\
$f_p$ & : power loss fraction \\
$F$ & : force from a tether along the center of the solar receiver sphere that acts \\
& \hspace{0.1cm} on the top and bottom of the rods \\
$F_b$ & : force pushing up on a rod in the solar receiver from the bottom \\
$F_c$ & : compressive force in a rod in the solar receiver \\
$F_t$ & : force pushing up on a rod in the solar receiver at the top (negative \\
& \hspace{0.1cm} if downward) \\
$F_y$ & : lateral force on a rod in the solar climber receiver \\
$F_z$ & : vertical force on a rod in the solar receiver \\
$F_Z$ & : drag force from moving the tether in tension \\
G & : \'Etendue, defined as the product of a light source's cross-sectional area and \\
& \hspace{0.1cm} the solid angle it subtends.  \\
$h_a$ & : characteristic height of the atmosphere \\
$h_h$ & : height of a higher climber \\
$h_l$ & : height of a lower climber \\
I & : electrical current, amps, or area moment of inertia of the cross section \\
& \hspace{0.1cm} of a beam \\
$I_0$ & : peak intensity in the middle of a circular PV array \\
$I_a$ & : average discharge current of a fuel cell \\
I(r,z) & : intensity or power density of laser light on a receiver as a function of r and z \\
$I_{SC}$ & : Short-Circuit Current of a photocell \\
$J$ & : rotational moment of inertia \\
$j$ & : heat energy flux \\
$K_{rms}$ & : dimensionless root-mean-square (RMS) deflection parameter or undulator \\
& \hspace{0.1cm} parameter for free electron lasers\\
$l$ & : length of a solenoid coil or length of conductor in the solar receiver \\
$l_c$ & : length of a climber \\
L & : characteristic length used to calculate Reynolds number, length of tether \\
& \hspace{0.1cm} to be spooled, or inductance \\
$M$ & : torque or bending moment \\
$m_c$ & : mass of climber \\
n & : number of rods in the solar climber receiver or number of cells in a fuel cell\\
N & : number of conductor turns in a solenoid \\
P & : power of the beam \\
$\mathcal{P}$ & : power put into the tether by inducing oscillations \\
$P_c(r)$ & : shaft power at the wheels of the climber \\
$P_{cl}(z)$ & : power received at the climber array at altitude z \\
$P_{eq}$ & : equivalent pressure from drag force \\
$P_{in}$ & : wall power input to the power beaming system  \\
$P_o$ & : power at the transmitter \\
$P_{out}$ & : shaft power at the wheels of the climber \\
$P_r$ & : received power \\
$P_{tG}$ & : power in a Gaussian beam at the transmitter \\
q(z) & : complex beam parameter \\
$r$ & : distance from the center of the Earth, or radial distance on a circular PV \\
& \hspace{0.1cm} array, or radius of a rod in solar receiver \\
$r_{rec}$ & : radius of either a PV or microwave receiver \\
R & : distance between source aperture and receive aperture, radius of the solar \\
& \hspace{0.1cm} receiver, resistance in an electrical circuit, or radiance\\
$R_c$ & : combined electrical resistance of conductors in the solar receiver\\
Re & : Reynolds number \\
$R_{PV}$ & : radius of the PV array \\
R(z) & : radius of curvature of the wavefront \\
s & : position along the length of a rod in the solar receiver \\
S & : Poynting vector \\
t & : thickness of the photocell or time of day in hours \\
$t_d$ & : discharge time of a fuel cell \\
T & : temperature \\
$T(s,t,\varphi)$ & : force per unit distance exerted on a rod by the tension in the membrane \\
& \hspace{0.1cm} in the solar receiver\\
$T^\prime$ & : integral of $T(s,t,\varphi)$ with respect to $s$ \\
$T_c$ & : constant component of simplified tension in the solar membrane per \\
& \hspace{0.1cm} unit length along a rod \\
$T_\theta$ & : component of tension in the solar membrane per unit length along a rod \\
$T_\phi$ & : component of tension in the solar membrane per unit length across a rod \\
$T_{max}$ & : maximum tension in the solar membrane per unit length along a rod \\
$\Delta T$ & : differential tether tension \\
u & : energy density (energy per unit volume) \\
$v$ & : speed of the tether where a wave has arrived \\
$v_0$ & : speed of vibrations of the tether at the bottom of the atmosphere \\
$v_1$ & : speed of vibrations of the tether at the top of the atmosphere \\
$v_c$ & : velocity of the climber \\
$v_h$ & : speed of higher climber \\
$v_l$ & : speed of a lower climber \\
$v_{max}$ & : maximum climber velocity \\
V & : voltage, speed of the climber, or the volume of tether to be spooled \\
$V(s,t,\varphi)$ & : function for the part of the power-converter weight that one rod supports \\
& \hspace{0.1cm} in the solar receiver \\
$V_d$ & : discharge voltage of a fuel cell \\
$V_{OC}$ & : Open-Circuit Voltage of a photovoltaic cell \\
$w_c$ & : weight of the power converter in the solar receiver \\
$w(s,t,\varphi)$ & : weight per unit length on each rod of the solar receiver \\
$w_o$ & : minimum waist of the power beam \\
$w_M$ & : weight of the membrane of the solar climber receiver per unit area \\
w(z) & : waist (the radius) of the beam at the distance z \\
$W_C$ & : weight per rod per meter of the power converter at the middle circle \\
x & : horizontal distance between the beam transmitter and the base of the tether \\
y & : height at which the climber starts receiving beamed power \\
z & : altitude from the surface of the Earth, or the distance between \\
& \hspace{0.1cm} a power-beaming transmitter and receiver \\
$z_R$ & : Rayleigh range \\ 
Z & : mechanical impedance of the tether \\
$\alpha$ & : absorptivity of a radiating body, or tilt angle of a disk in flow \\
$\alpha_c(r)$ & : acceleration of gravity at the distance r from the center of the Earth \\
$\epsilon_{motor}$ & :  electric motor efficiency \\
$\varepsilon$ & : permittivity or dielectric constant of a material, emissivity of a radiating \\
& \hspace{0.1cm} body, or power efficiency of MPT \\
$\varepsilon_{1-6}$ & : efficiency terms in the power beaming transfer function \\
$\varepsilon_r$ & : relative permittivity or dielectric constant of a material \\
$\eta$ & : efficiency of photovoltaic technology \\
$\theta$ & : gradient angle along a rod in the solar receiver or half angle of \\
& \hspace{0.1cm} beam divergence \\
$\Theta$ & : total angular spread, subscripts r for receiver, t for transmitter \\
$\kappa$ & : electrical conductivity \\
$\lambda$ & : wavelength \\
$\mu$ & : coefficient of friction \\
$\mu_r$ & : relative permeability of ferrous material \\
$\nu$ & : Poisson's ratio or the kinematic viscosity of air \\
$\Pi(x)$ & : rectangle function\\
$\rho$ & : mass density \\
$\rho_a$ & : mass density of aluminum or the density of the atmosphere at the surface \\
& \hspace{0.1cm} of the Earth \\
$\rho_c$ & : electrical resistivity \\
$\rho_M$ & : mass density of membrane in solar receiver \\
$\rho_r$ & : mass density of CFRP in solar receiver support rods \\
$\sigma$ & : compressive or tensile stress, or maximum working stress of the tether \\
$\tau$ & : shear stress, power-link parameter, or membrane thickness of solar receiver \\
$\varphi$ & : angle of a rod with respect to the afternoon side of the equator in the solar \\
& \hspace{0.1cm} spherical receiver \\
$\Psi(z)$ & : Gouy phase \\
$\omega$ & : angular velocity \\
\end{longtable}

A number of the variables in the preceding table have multiple meanings and uses throughout this report.  The meaning of a given variable should be obvious from the context in which it is used.
\chapter{Power Requirements for Ascending Climbers}\label{chapt:req}

\section{Calculating the power required to lift a climber to GEO}

Calculating the power required to lift the climber is actually very simple.  $P=F\times v$, where $P$ is the power, $F$ is the weight of the climber on the tether, and $v$ is the speed of the climber.  For simplicity, we assume that the speed of the climber is a constant 200 km/h because the acceleration and deceleration phases are short portions of the trip.  (We'll look at this in more detail later.)  The climber's weight changes as a function of its altitude above Earth as gravity falls off with $1/r^2$.

200 km/h has been set as a nominal cruising speed for the climber because it allows the climber to make the trip from the surface of the Earth to GEO in about a week.  We hope to be able to increase the speed of the climber above 200 km/h at higher altitudes to decrease the trip time, but more work is needed on the rotating components to see if they bear up to the increased stress.  (The fastest wheeled electric vehicle on Earth is the Buckeye Bullet 3, built by engineering students at Ohio State University.  It attained a top speed of 549.43 km/h.  Needless to say, the Bullet does not have to go straight up, continuously, and perfectly reliably for weeks without any service stations along the way.)
 
The first thing to do is calculate the weight of the climber at every altitude between the surface of the Earth and GEO.  The weight is the mass times the apparent gravitational acceleration, ($mg$), so we start by calculating the value of $g$ ($=a_c(r)$) at every altitude.

\begin{equation}\label{eqn1}
a_c(r)=\frac{\mu_E}{r^2}-r\omega_E^2
\end{equation}
where 

$a_c(r)$ is the acceleration of gravity at the distance $r$ from the center of the Earth

$\mu_E = 3.9860 \times 10^5$ km$^3$/s$^2$ is the gravitational parameter of Earth (Earth's mass $\times$ universal gravitational constant)

$\omega_E = 7.29212 \times 10^{-5}$ rad/s is the angular velocity of Earth's rotation

\begin{figure}
\begin{centering}
\includegraphics[width=\textwidth]{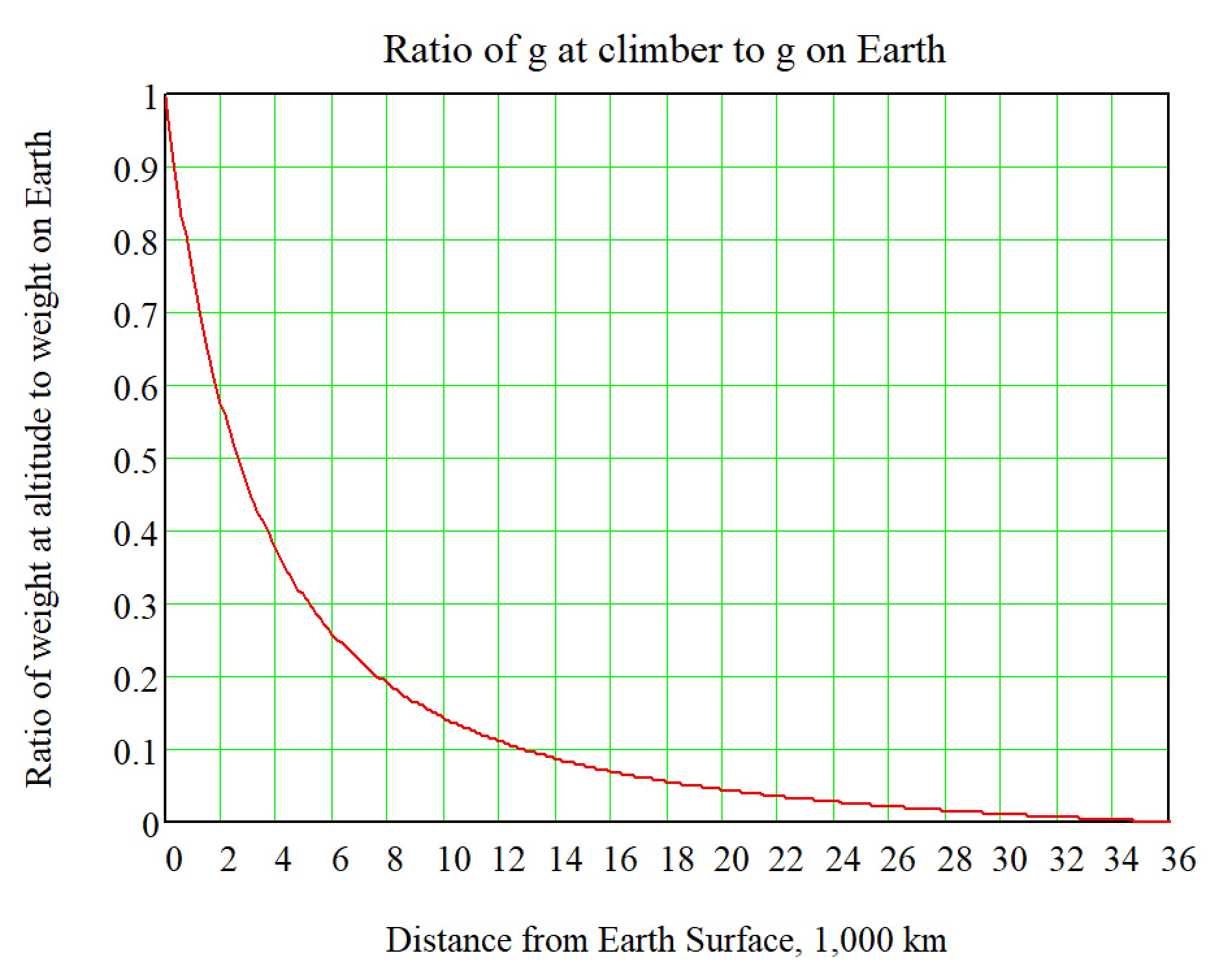}    
\caption{\label{fig:galt} This graph shows the ratio of the apparent gravitational acceleration felt by a climber at any altitude to the acceleration of gravity at the surface of the Earth.  The apparent gravity is not just the force of gravity caused by the mass of Earth, but it also accounts for the reduction in apparent weight of the climber the higher it goes caused by the rotation of the Earth.  The graph stops at GEO because above GEO it does not take external power to climb toward the Apex.  The rotation of the Earth overcomes the force of gravity and the climber experiences an apparent gravity where down is in the direction of the Apex.}
\end{centering}
\end{figure}

\vspace{5mm}
Figure \ref{fig:galt} shows a graph of the ratio of the weight of the climber at altitude compared to the weight of the climber at the surface of the Earth.  The weight of the climber is the force that the power is lifting at constant speed.
\newpage
\begin{equation}
    P_c(r) = m_c \cdot a_c(r) \cdot v_c
    \label{eq:eqn2}
\end{equation}

$P_c$ is power in watts at the wheels of the climber

$m_c$ is the mass of the climber, 20 tonnes (20,000 kg)

$a_c(r)$ is the acceleration of gravity at height r from the center of the Earth

$v_c$ is the velocity of the climber, 200 km/h \par
\bigskip
The mass of the climber is assumed to be 20 tonnes because that has historically been considered the minimum mass of a commercially viable climber.  Once the tether can lift a 20 tonne climber, Initial Operational Capability (IOC) has been reached.  [\cite{Edwards}]

\begin{figure}
\begin{centering}
\includegraphics[width=\textwidth]{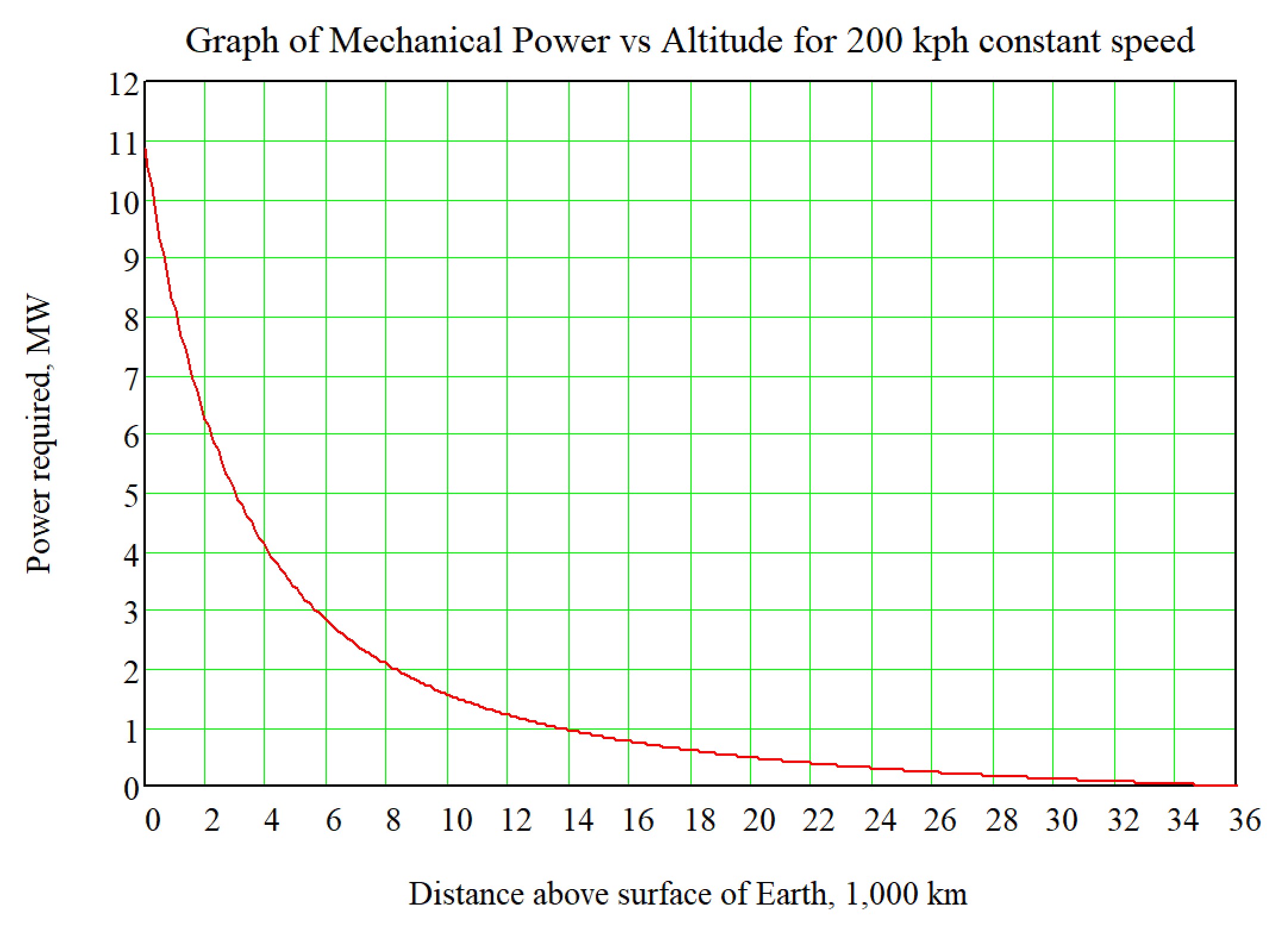}    
\caption{\label{fig:power} This graph shows the power required at the wheels of the climber (shaft power) to lift a 20 tonne climber at 200 km/h from the surface of the Earth to GEO.}
\end{centering}
\end{figure}

Equation \ref{eq:eqn2} gives the power required to lift the climber at 200 km/h as a function of altitude, and the graph is shown in Figure \ref{fig:power}. $P_c$ is the power at the wheels (sometimes called shaft power) because this is the power required to lift the climber.  It is not the power required to be put into the power transfer system (commonly referred to as wall-plug power). $P_c$ is the power left after all of the inefficiencies in the power system are accounted for and have reduced the input wall-plug power to $P_c$.  

To calculate the total energy required to be put into the climber (neglecting inefficiencies), we integrate the power equation over the range of altitude.

\begin{equation}
    E_t = \int_{R_E}^{R_G} m_c \cdot a_c(r) \, dr
    \label{eq:energy}
\end{equation}
where $E_t$ is the total power required to get from the surface of the Earth, $r=R_E$, to the altitude $R_G$ (GEO in this case).

We can see from Figure \ref{fig:power} that if we want to drive the climber at 200 km/h up at the surface of the Earth, we need to put almost 11 MW into the wheels.  There are reasons not to drive the climber this fast in the atmosphere, and we will see later how the initial climb out of the atmosphere changes the speed profile of the climber.

Equation \ref{eq:energy} gives the total energy required to put into the climber to get to GEO as $9.69 \times 10^{11}$ Joules, almost a terajoule.  If we take this number and divide it by the 20,000 kg mass of the climber, we find that we need to put 48.45 MJ into every kilogram that goes to GEO on the space elevator.

Another thing that Figure \ref{fig:power} shows us is that the power required to make the climber roll up the tether at 200 km/h drops off very fast the higher the climber gets, dropping to zero at GEO.  This has major implications for any power beaming system.

\newpage
\section{The subsystems of the climber that require power}
\begin{figure}
\begin{centering}
\includegraphics[width=\textwidth]{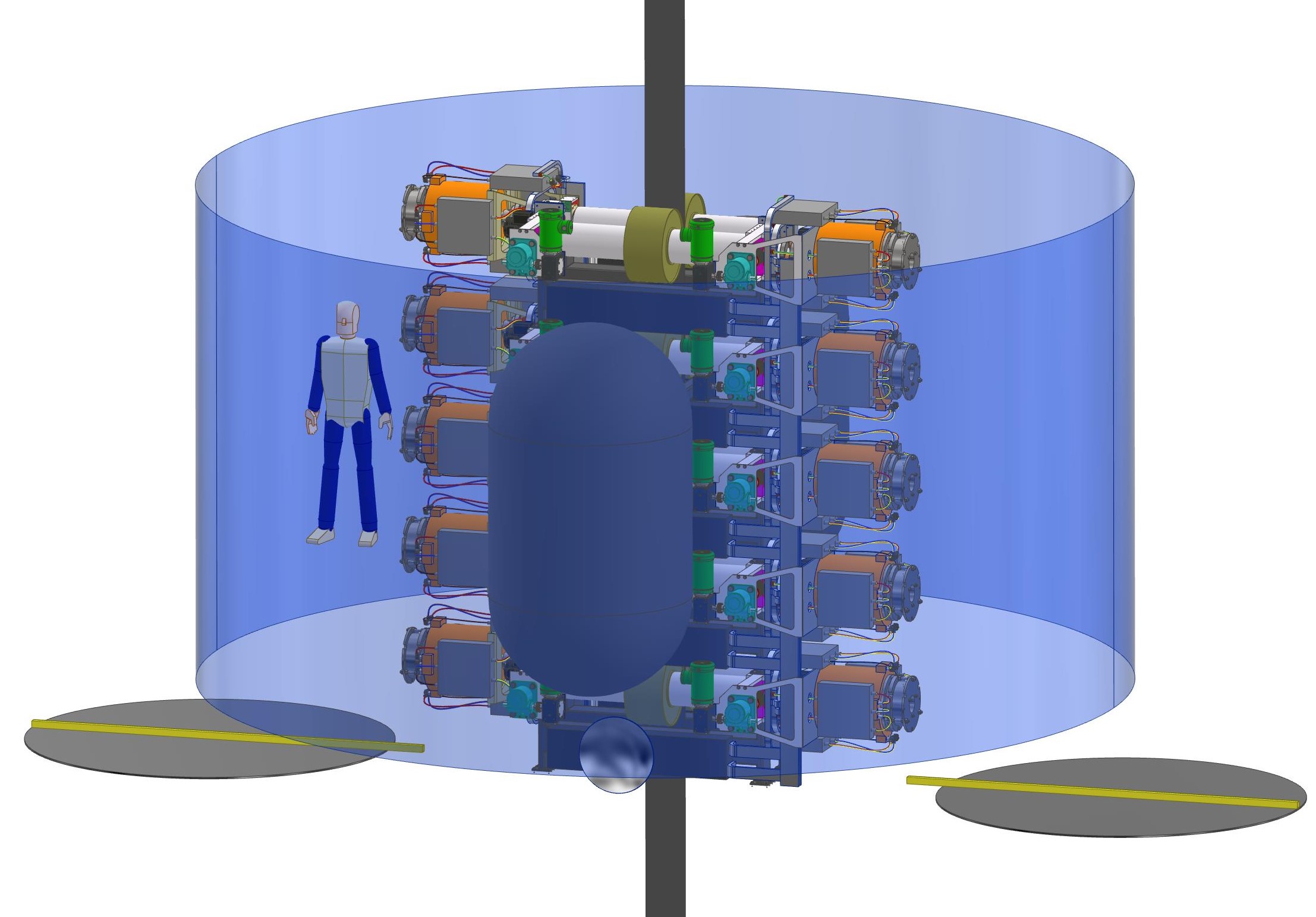}    
\caption{\label{fig:20tonne} Rendering of the Reference Conceptual Design Climber from the 2023 ISEC Study Report.  The PV arrays shown were placeholders only.  The lozenge-shaped tanks are intended to be LOX payload tanks for delivery to a GEO LOX and fuel depot.}
\end{centering}
\end{figure}

The 2023 ISEC Study Report on the Climber-Tether Interface, [\cite{wright2023climber}], contains a reference conceptual design (RCD) for a 20 tonne climber on the space elevator.  Figure~\ref{fig:20tonne} shows the main features of the climber.  The design concept is for five wheel pairs, with each wheel driven by its own motor. To achieve traction on the tether, the wheel pairs must be pressed against each other with a wheel compression mechanism.  The coefficient of friction between the wheels and tether is currently unknown, but the climber was designed assuming the coefficient of friction was at least 0.1. 

\begin{figure}
\begin{centering}
\includegraphics[width=\textwidth]{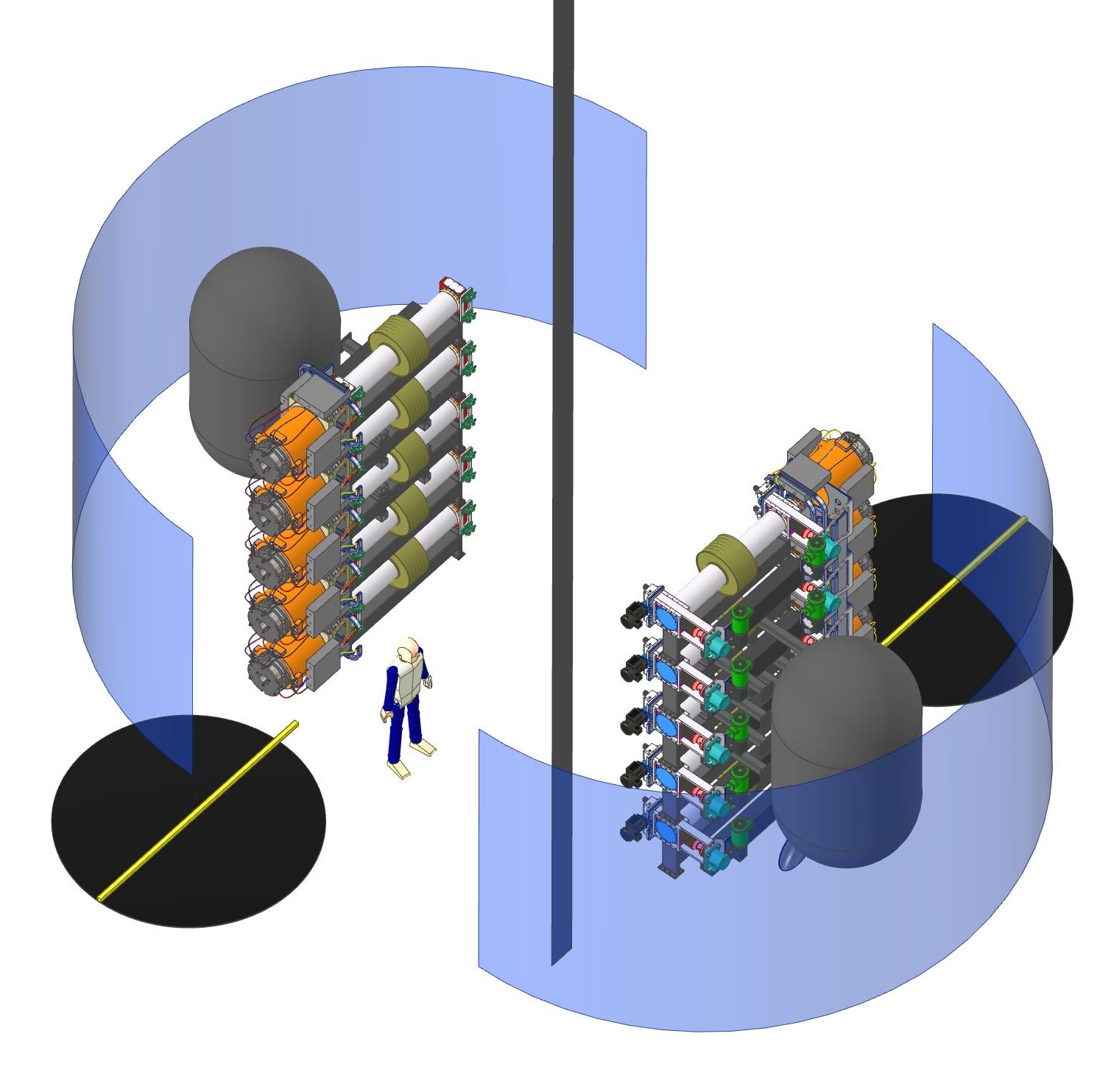}    
\caption{\label{fig:splitclimber} The two halves of the climber shown separated prior to assembly on the tether and the engagement of the quick disconnect mechanisms. (See Section \ref{sec:QDmech}).}
\end{centering}
\end{figure}

\begin{figure}
\begin{centering}
\includegraphics[width=.8\textwidth]{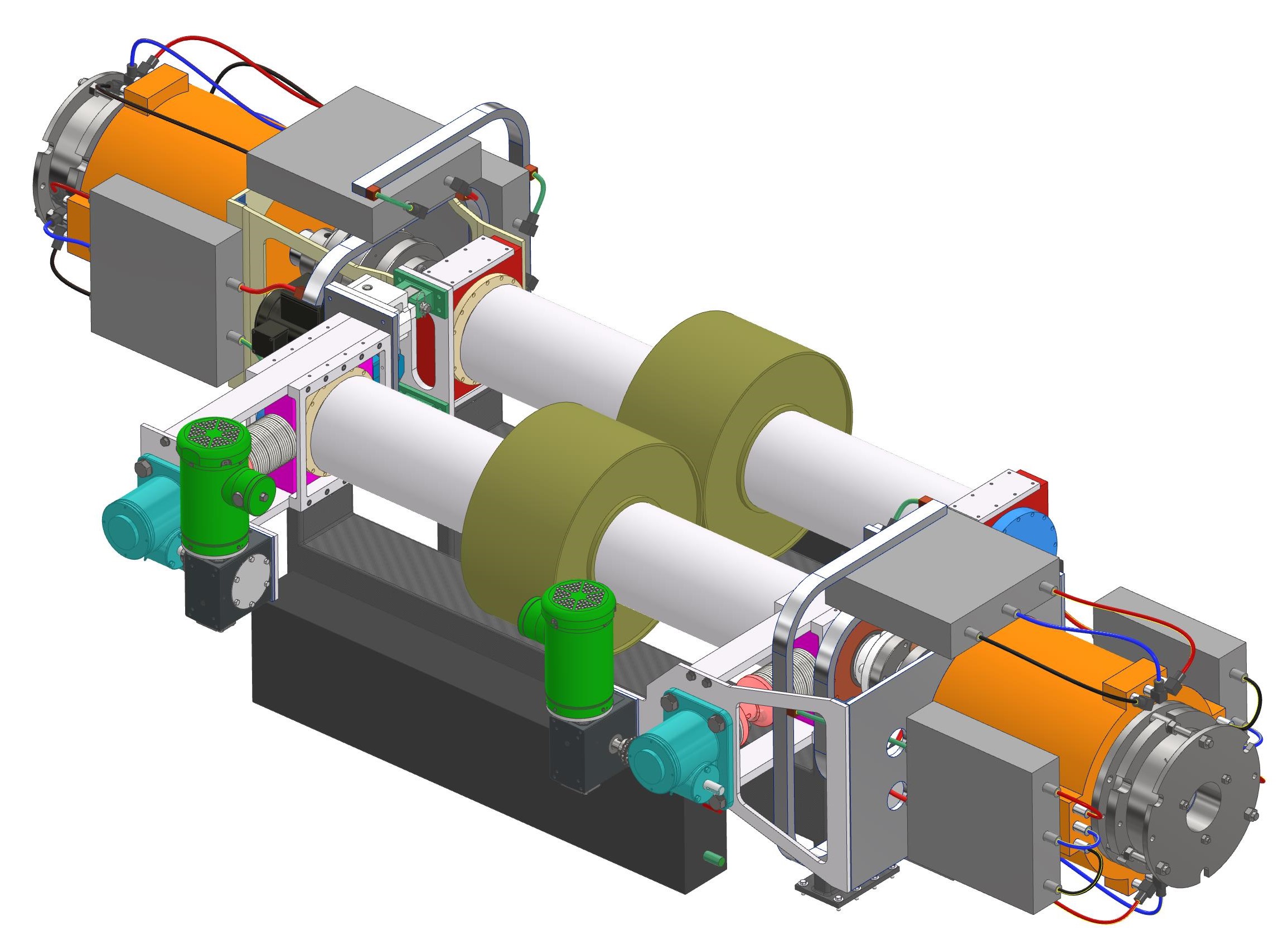}    
\caption{\label{fig:wheelpair} Close-up of a single wheel pair showing the primary drive motors (orange) and other subsystems.}
\end{centering}
\end{figure}

\begin{figure}
\begin{centering}
\includegraphics[width=.8\textwidth]{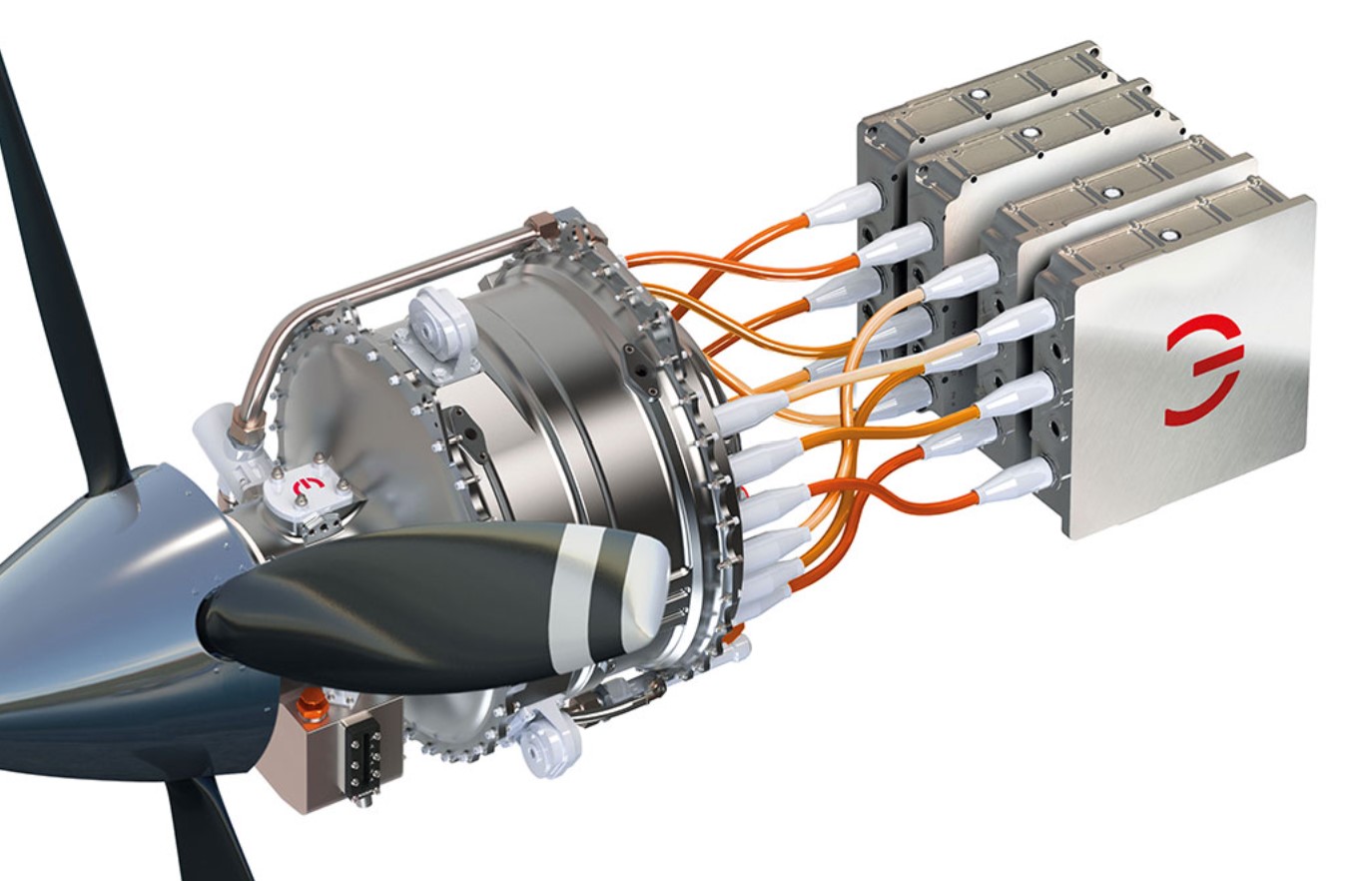}    
\caption{\label{fig:magnimotor} Marketing picture of the magni650 motor ca. 2020.  The CAD model of the Reference Conceptual Design was built around this product. [\cite{magniX}]}
\end{centering}
\end{figure}

\subsection{Primary motors to drive the wheels}

The volume and mass of the primary motors in the RCD were originally taken from an electric aircraft motor designed at the magniX company.  [\cite{magniX}]  Figure \ref{fig:magnimotor} shows a marketing image from the website of how the motor and its drive electronics were configured in an electric aircraft.  The magni650 motor was determined during the 2023 study to be insufficient in torque to raise the climber, even though its specifications were unique at the time.  The solution to continue the design was to assume an enhanced motor with twice the torque and power capability at the same mass as the magni650.  Since 2020, there have been several other companies developing megawatt-scale motors with the torque required to lift a five wheel-pair climber. 

The original specifications for the magni650 motor was that it had a continuous torque rating of 2,814 N-m, available from 0 to 1,900 RPM.  The mass of the motor was 200 kg.  Its rated power was 560 kW.  The ratings assumed for the enhanced motors of the five wheel-pair climber are 5,628 N-m and 1.12 MW per motor at full torque and speed.

This means that ten motors on the climber would consume 11.2 MW running at full torque and speed.  It will be seen later that the motors are not running at full power when they are outputting full torque because the climber is moving slower at the beginning of the trip.

\subsubsection{New room temperature conductor motor technologies for the wheeled climber}

Since the reference conceptual climber design was introduced in 2023 [\cite{wright2023climber}], several new aviation-class motors have entered the market. One-megawatt (1 MW) electric motors designed for aviation are powerful and specialized; most are either being developed or just starting production, primarily serving the needs of emerging electric and hybrid-electric aircraft. Typically, these motors are supplied directly to aerospace programs by manufacturers and are not yet widely available as off-the-shelf products.

Manufacturers developing or offering motors in this category include:

\begin{itemize}
    \item Collins Aerospace: The company has created a 1 MW permanent magnet motor for RTX's flight demonstrator program. Ground tests are underway, and flight testing is scheduled. More information can be found in their press releases. [\cite{Collins}]
    \item GE Aerospace: In partnership with NASA, GE is advancing a megawatt-class hybrid electric propulsion system through the Electrified Powertrain Flight Demonstration (EPFD) program. [\cite{GEEPFD}]
    \item Honeywell: Their compact 1 MW turbogenerator, which can also function as an electric motor without modification, is crafted for hybrid-electric aircraft. It serves as a key component in projects such as the Flying Whales LCA60T airship. Further details are available on the Honeywell Aerospace website. [\cite{Honeywell}]
    \item Safran: The ENGINeUS™ XL motor from Safran delivers 750 kW per unit and can potentially reach 1 MW. Intended for all-electric nine-passenger planes through to regional shuttles, their ePower range is showcased online. [\cite{Safran}]
    \item Wright Electric: This U.S. startup has successfully produced one megawatt of shaft power in bench tests for its experimental commercial aircraft motor-generator. These motors are reportedly capable of delivering 2.5 MW of power. [\cite{Wrightelectric}] With a power density of 16 kW/kg, a 2.5 MW motor would only mass 156.3 kg, less than the nominal mass of the 1.1 MW motors in the climber model that mass 200 kg.  A motor like this could significantly increase the payload capacity of the climber.
\end{itemize}

In addition, research bodies like MIT and NASA are actively involved in the design and testing of $>1$ MW motors to help advance foundational technologies for future electrified aircraft. [\cite{NASAHEMM}]

\subsubsection{Maximizing the torque and minimizing the motor mass}

An updated survey of the commercially available electric motors since the 2023 study indicates that light-weight electric motors used in the electric aircraft and eVTOL industries continue to be the best candidates to drive the SE climber.

To drive the wheel/axle combination, the motor must produce a torque
proportional to the volume of the motor rotor. The mass of the motor increases with
the torque it must produce. Thus, the torque-to-mass ratio is important, as is the
mass-to-power ratio.  Reference [\cite{shreyanshmot}] covers work done by an ISEC intern in 2025, Shreyansh Chauhan, on strategies to increase the torque of permanent magnet motors.  He showed the possibility of increasing the torque of a motor like the magni650 by 155\%.

Operating aviation-class electric motors in a vacuum introduces constraints not present in atmospheric designs. Without convective heat transfer to air, all waste heat must be removed through liquid cooling loops coupled to radiators sized specifically for radiative heat rejection, increasing total system mass. Bearings and lubricants must be redesigned or replaced with vacuum-qualified or magnetic alternatives to prevent outgassing and wear. Likewise, insulation systems (winding varnish, potting materials, and cable jackets) require vacuum-stable formulations capable of tolerating thermal cycling and radiation exposure. These modifications do not prevent the use of existing motor architectures, but they must be explicitly accounted for in the climber’s mass and reliability budget.

Climber traction motors operate on a much longer and steadier duty cycle than typical electric aircraft. Multi-hour continuous operation at high power makes thermal equilibrium the dominant design driver. Continuous losses (copper, core, switching) must be managed without reliance on short-burst peak ratings, and cooling systems must be sized for sustained heat rejection rather than transient peaks. This typically results in noticeable derating when compared to published aircraft power-density figures. Cooling loop capacity, radiator area, and pump redundancy therefore become integral to motor selection, and their mass contributions must be tied directly to the motor’s continuous-power profile rather than its peak specifications.

Given the mission duration and absence of in-flight servicing, a distributed motor architecture offers significant advantages. Using multiple moderate-power motors instead of a single high-power traction unit minimizes single-point failures and improves survivability in the event of inverter, motor, or wheel-pair faults. Distributed drives also spread thermal loads, simplify power routing, and allow the climber to maintain partial capability in degraded-performance scenarios. This redundancy philosophy aligns naturally with the multi-wheel configuration of the climber and enhances overall robustness with a relatively modest mass penalty.

\subsubsection{Superconducting motors}

The 1.4-megawatt (MW) High-Efficiency Megawatt Motor (HEMM) is being designed to meet the needs of electrified aircraft propulsion. HEMM has a target performance of 16 kW/kg, 99\% efficiency, and will have three times lower losses and weight than current aircraft motors and generators. [\cite{NASAHEMM}]

\begin{figure}
\begin{centering}
\includegraphics[width=\textwidth]{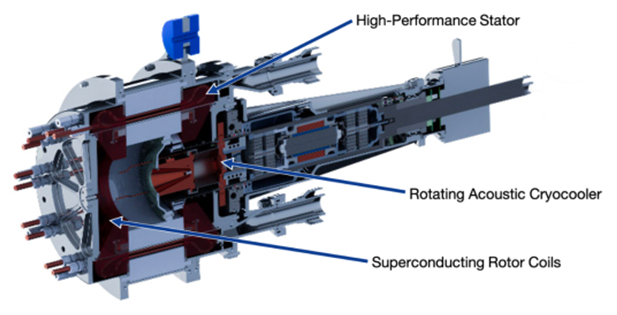}    
\caption{\label{fig:NASAHEMM} NASA High-Efficiency Megawatt Motor (HEMM) is rated at 1.4 megawatt}
\end{centering}
\end{figure}

The HEMM technology can be integrated with a typical aircraft cooling system and can be applied to a range of aircraft types that require megawatt-level electrical power. New technologies must be developed to ensure the motor can produce more power per mass while efficiently maintaining the required cryogenic temperatures.  Three components being built and tested are a rotating acoustic cryogenic cooler, superconducting rotor coils, and a high-performance stator.

\subsubsection{Primary motor conclusions}

The landscape of motors has changed dramatically in the years since the Reference Conceptual Design in [\cite{wright2023climber}] was published.  That design incorporated an aspirational motor that had twice the performance of the MagniX motor, but was unavailable on the market at the time. Since then, a number of companies have developed multi-megawatt motors for electric aircraft.  What was hoped for in the 2023 study has come to pass in terms of power and torque.  These motors are all liquid-cooled, but they were not developed to operate in a vacuum.  That is the next hurdle in the development of motors for the space elevator climber.

\subsection{Brake actuation}

Figure \ref{fig:drivetrain} shows how the original magni650 motor and its drive electronics were configured for the climber.  The metallic structure on the backside of the motor is an extrapolated version of a commercially available Ogura electromagnetic brake [\cite{ogura}] used to clamp the climber on the tether once it has come to a stop. It is not known at this time how much power this brake would consume.  It is possible that after initial actuation, the brake power could be disengaged allowing the climber to stop on the tether with a purely mechanical clamping force.  It is not expected that the brakes will need a substantial amount of power.  The batteries on the climber were designed to have enough power for emergency stop situations.

\subsection{The wheel compression mechanism motors}
\label{sec:wheelcompress}

Figure \ref{fig:wheelpair} shows two motors in green that drive the wheel compression mechanism.  The first thing to be noted about the RCD is that all of the motors were selected from off-the-shelf components and, as such, they are not space-worthy.  The slots in the end of the green motors are part of the motor's internal air cooling system.  For climbers, these motors need to be replaced with liquid-cooled units that can work in vacuum.  The power of each of these motors is estimated to be 3 horsepower (2.24 kW) to drive the 20 ton screw jacks in the mechanism.  

The wheel compression motors allow an adjustable force to be applied to each side of the axles.  As the climber becomes lighter higher up the tether, the motors can reduce the applied force on the axles in unison and maintain just the necessary traction at every altitude.  This reduces the stress on the rotating components and increases the fatigue life of the drivetrain.

A feature that is imagined but that has never been simulated or tested is that the wheel compression mechanisms might be used to steer the climber on the tether.  The concept is that if one side applies more force to the axle than the other, then the wheel will take on a slightly conical shape.  This should cause the wheel to track sideways along the tether.  This mechanism is intended to prevent the climber wheels from drifting too far to one side or the other on the tether.  Exactly how this differential force would be programmed into a climber to work with five wheel pairs is not currently known.

\subsection{The quick disconnect mechanism motors}
\label{sec:QDmech}

\begin{figure}
\begin{centering}
\includegraphics[width=\textwidth]{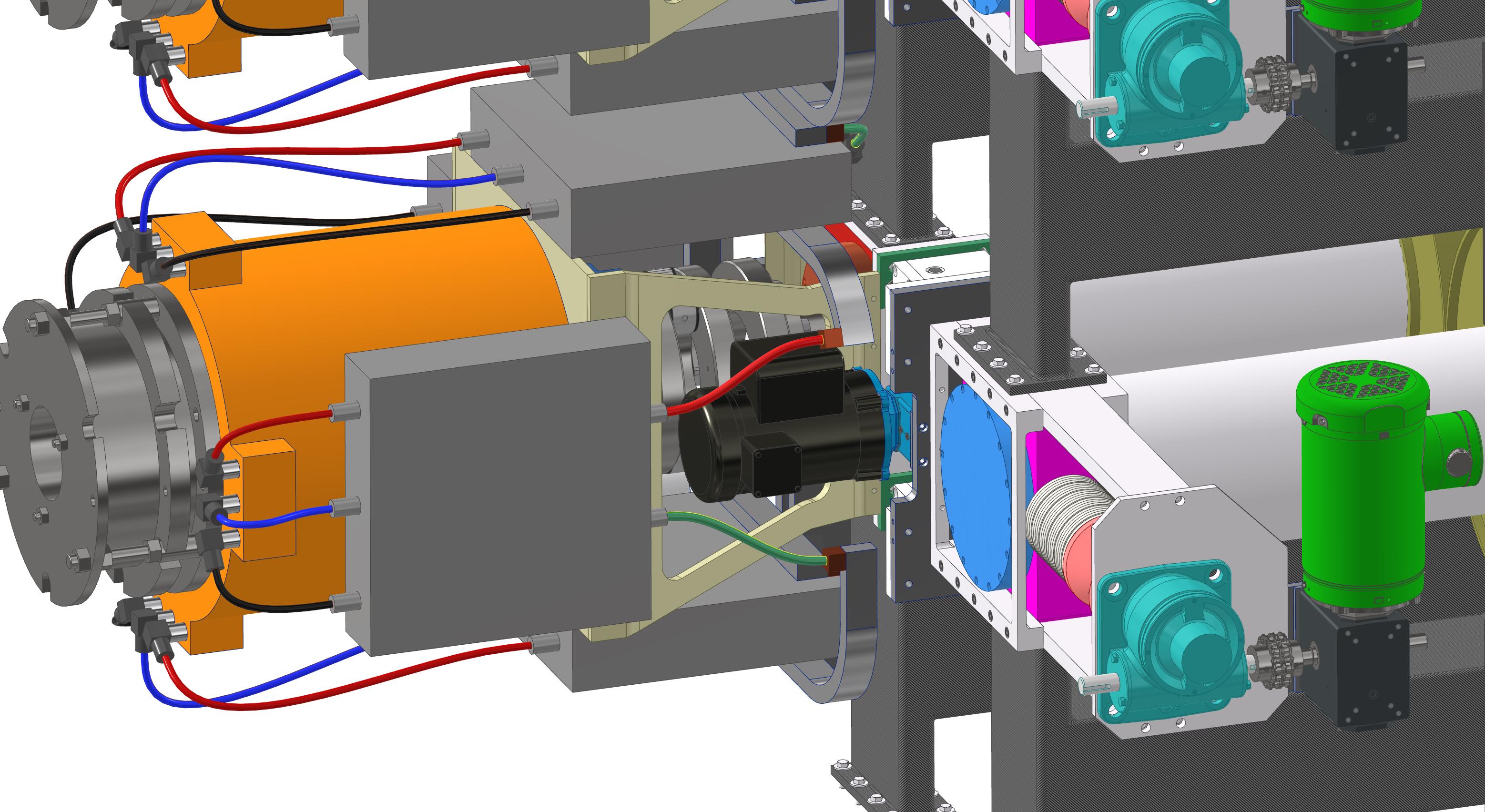}    
\caption{\label{fig:drivetrain} Close-up of one side of a climber showing the orange primary motor that drives the wheel on the fixed side of the climber and the smaller black motor that drives the quick disconnect mechanism on one side of the climber.  The design of the quick disconnect mechanism is based on a remotely actuated turnbuckle that can resist the force of the wheel compression mechanism (Section \ref{sec:wheelcompress}) and also release the two halves of the climber from the tether. The primary motor brake is the metallic-colored structure visible at the left end of the orange primary motor.}
\end{centering}
\end{figure}

\begin{figure}
\begin{centering}
\includegraphics[width=\textwidth]{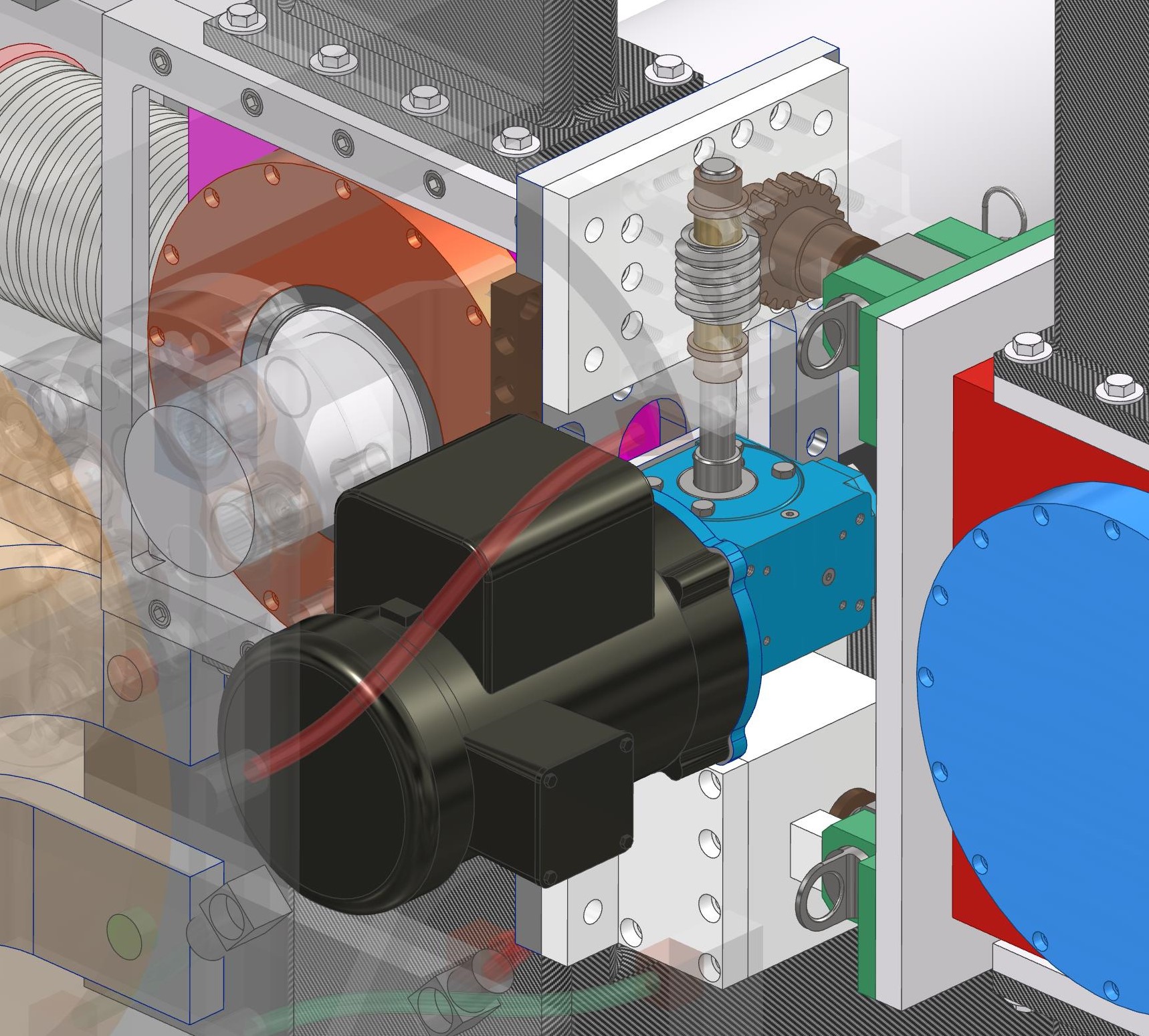}    
\caption{\label{fig:QDmech} Zoomed in view of the quick disconnect mechanism with foreground items rendered transparent.  The upper gearbox housing of the mechanism is also rendered transparent to show the worm and wheel gears that pull the two climber halves together around the tether. The center of the wheel gear is threaded to push or pull a turnbuckle shaft and eye (gray) that is pinned to the green clevis on the fixed half of the climber.}
\end{centering}
\end{figure}

Figures \ref{fig:drivetrain} and \ref{fig:QDmech} show a black motor that drives the quick disconnect mechanism that holds the two halves of the climber together.  One of these motors is mounted on each side of the climber, for every wheel pair.  This mechanism is based on a turnbuckle clamping scheme originally designed to clamp kiloamp-carrying conductors in a physics experiment.  The mechanism allows for easy assembly of the two halves of the climber on Earth, as shown in Figure \ref{fig:splitclimber}, and provides a remotely operated way to separate the two halves in space.  

The mechanism is intended to be actuated on Earth to tighten the connection between the two climber halves, but it can then be powered off because the turnbuckle clamp is self-locking.  It would only be needed in an emergency that forces the climber to be ejected from the tether, or at GEO to disassemble the climber from the tether.  This turnbuckle mechanism counteracts the compression between the wheels that would otherwise drive the two halves of the climber apart.

This is the mechanism discussed in Section \ref{sec:passingproblem} that causes a climber to surround the tether and prevent another climber from passing it on the tether.

The power of these motors is estimated to be on the order of .25 horsepower (.19 kW).  Their infrequent use means that they should be operable from the onboard battery power.

\FloatBarrier

\subsection{The coolant circulation system pump}

As has been noted, one of the subsystems that is not shown in the RCD and has no conceptual design associated with it yet is the fluid recirculation system and pump that cools the primary motors, the electronics, the gearboxes, and all of the bearings on the climber.  No estimates have been made of the power required by this subsystem, but it is not expected to be substantial compared to the power required by the primary motors.

\subsection{The refrigeration system compressor pump}

Figure \ref{fig:20tonne} and Figure \ref{fig:splitclimber} both show a transparent blue shell surrounding the climber.  This shell is a placeholder for a space radiator that is large enough to radiate the waste heat away from the climber.  The shell shown has an area sufficient to dissipate 160 kW at a temperature of \SI{200}{\degree C}.  That temperature was chosen in the RCD as the maximum temperature that the fluids cooling the active systems could tolerate.  The power capacity was based on the historical 4 MW power limit multiplied by the 4\% inefficiency of the motors.  (MagniX motors claimed 96\% efficiency at rated torque and speed.)

We know now that a 160 kW power rating for the thermal radiator is inadequate.  It did not take into account the power inefficiency of the motor drive electronics, or the waste heat in the PV arrays (for the laser-powered climber).  Keeping the PV arrays between \SI{80}{\degree C} and \SI{100}{\degree C} is an as-yet-unsolved engineering problem, but it motivates the refrigeration subsystem described below.

The mechanical details of the space radiator have never been worked out, but radiating waste heat away in space is a significant engineering challenge.  The only way to make the space radiator smaller, or radiate away more heat, is to raise its temperature.  To protect the electronics, motors, and bearings from higher temperatures, a refrigeration loop needs to be added.  This loop would involve a phase-changing liquid that would allow the space radiator to run hotter, and would output cooled fluid for the drivetrain and PV arrays.

The power consumed by this system is not expected to be significant compared to the power required by the primary drive motors at low altitudes.  Of more significance is the mass of such a system and its effect on the payload of the climber.  Every kilogram added to the drivetrain on a 20 tonne climber is a kilogram taken away from payload.

\subsection{Communications and control power}

Figure \ref{fig:20tonne} shows a small dish antenna at the bottom middle of the climber.  This is a placeholder for the climber control system.  The RCD assumed that the climber would be controlled by communicating through the Starlink satellite network.  No control electronics is shown in the RCD.  The power required of the communications system is assumed to be negligible.  Of more concern is the mass of a radiation-hard communications system, for the same reason mentioned above about payload trade-off.

\subsection{Onboard batteries}
\label{sec:onboardbatt}
Figure \ref{fig:wheelpair} shows a black rectangular solid just below the wheel compression mechanisms and parallel to the dark gray CFRC structure that supports the wheel bearings.  This is a battery and every wheel pair has two of them, one for each primary motor.  The volume of the batteries was originally estimated by assuming the need to store 50 kWh of energy at the same energy density as Tesla EV batteries.  Later, the volume was reduced by a factor of three because lithium sulfur batteries can store three times the energy of lithium-ion batteries. [\cite{xue2017gravimetric}], [\cite{Faradaybatt}].  Lithium sulfur batteries have a theoretical specific energy of 2,500 Wh/kg and an energy density of 2,800 Wh/l. However, in practice, they have only attained 300 Wh/l so far.  The biggest trade-off with Li-S batteries is cycle life.  The latest batteries can achieve more than 1,000 charging cycles.  Li-S has not achieved commercialization yet and has a number of technical difficulties in manufacturing that have caused some companies to abandon the technology.

It is a matter of engineering judgment and risk analysis to determine the amount of power the batteries should hold.  The primary motors operate at 1.12 MW of power at maximum output (near the surface of the Earth), so the batteries shown could operate the primary motors for 2.7 minutes if external power were cut off from the climber.  In the event of an external power failure, the batteries would power the electromagnetic brakes, the wheel compression mechanism motors, and the quick disconnect motors if needed.

\subsection{Attitude control system}

The tether has a very small torsional stiffness and may be prone to twisting.  Wind and impacts with orbital debris, especially impacts on the climber, can cause twisting of the tether.  The only way to mitigate this is to put an attitude control system, such as a gyroscope or reaction wheel, on the climber that can untwist the tether as needed.  This is another subsystem on the climber that has not yet been developed. At the 2025 Annual ISEC Conference, Blaise Gassend showed how the tether could buckle from torsional twisting [\cite{tethtwist}]. Having a climber roll over a buckled section of tether could damage the tether, hence the need for a look-ahead system that can detect the state of the ribbon and cause the climber to spin around to correct any twists.  Another reason for this system is that Coriolis forces cause the climber to be pushed to the West, putting a bend or kink in the tether right at the climber.  The tether should stay in a North-South orientation at the climber so that the kink in the tether bends it about its weak axis and not its strong axis.  It is not yet known what the mass of such a system would be or how much power it would require.  Its power requirement is expected to be small compared to the primary drive motors.

\section{The power requirements for multiple climbers on the tether at the same time}

\begin{figure}
\begin{centering}
\includegraphics[width=.8\textwidth]{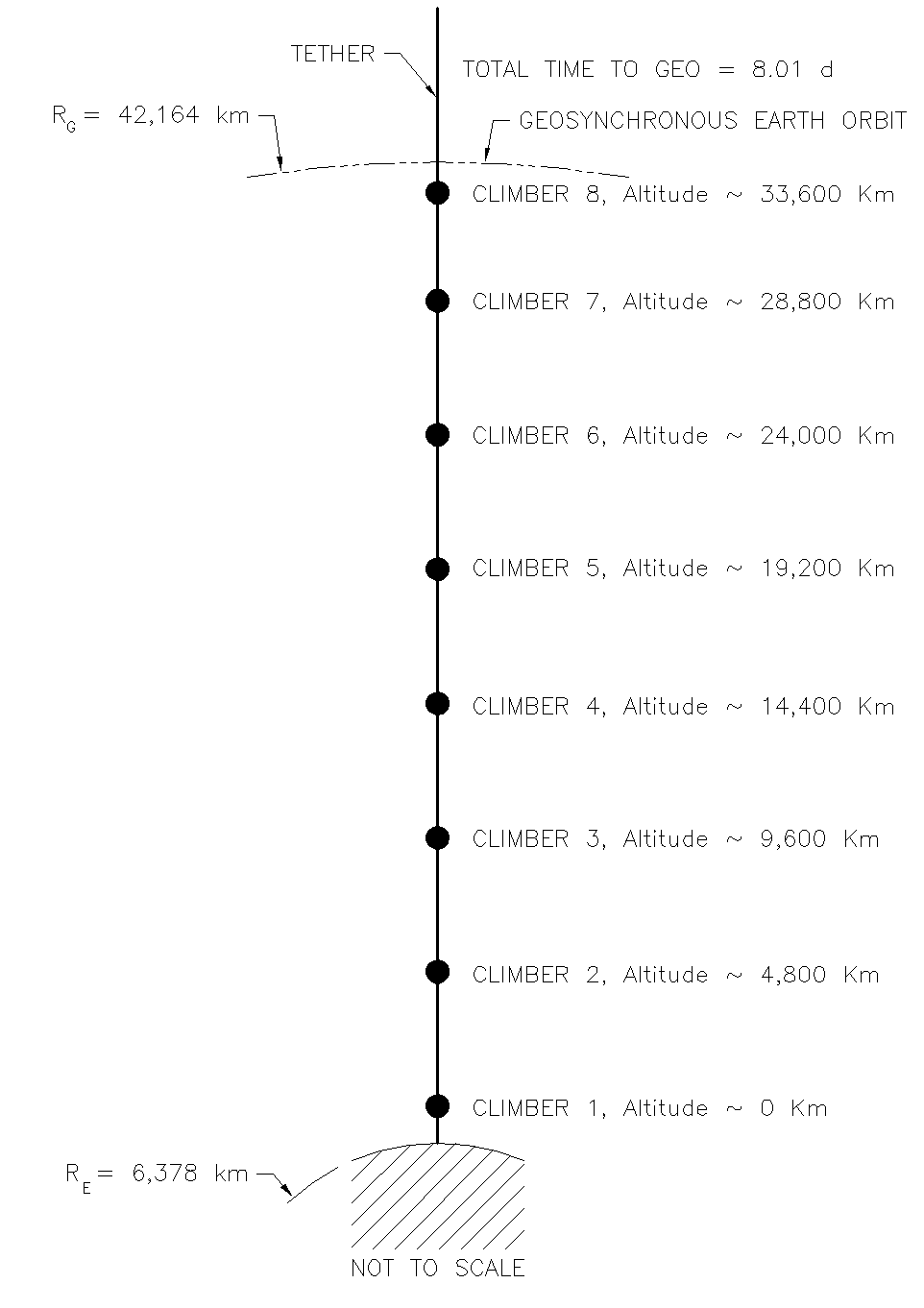}    
\caption{\label{fig:multiclimber} Schematic of eight climbers arrayed on the tether showing their relative altitudes after each day of climbing at 200 km/h.  Distances are approximate because at a constant 200 km/h from the ground, the trip would only take 7.5 days.  The acceleration at the beginning and deceleration at the end extends the time to about 8 days as calculated in [\cite{wright2023climber}].}
\end{centering}
\end{figure}

The goal of a mature space elevator infrastructure is to be able to launch at least one twenty tonne climber every day so that there is a constant stream of material being brought to GEO.  A simple-minded model of multiple climbers on the tether places eight climbers equally spaced between the ground and GEO, as shown in Figure \ref{fig:multiclimber}.  References [\cite{Robinson2022}], or [\cite{robinson2023space}], give a much more nuanced view of climber spacing and the optimal number of climbers on the tether, but the mass and design of the climbers in those papers are parametric and allowed to vary.  We are starting with a simplistic model using only 20 tonne climbers to bound the ranges of power transmitters.  Given the altitudes for each climber and assuming that most of them are moving at 200 km/h (except \#1 which cannot climb at 200 km/h until a higher altitude), we can refer to the graph in Figure \ref{fig:power} and assign the power required at the wheels for each of the climbers.

These numbers are a rough starting point to determine the range of power that the power delivery system has to provide for each climber.  A stretch goal is to have climbers roll up faster than 200 km/h when they are far enough away from Earth that there is sufficient power to do so, but it is currently unknown at which altitude the extra climber speed can be applied (or what the true maximum speed is based on the climbers' mechanisms.)  Climber \#1 starts with the highest power requirement, but by the end of its first day of climbing, it is now climber \#2 and only needs 3.5 MW of power.  A power beaming system can be tailored to trade off power transmitters from one climber to another each day.  This would allow each climber to be serviced by a power-beaming system optimized for its current altitude, with reduced power requirements but increasing distance requirements.

These numbers apply to any climber power system.  We will use these numbers later in Chapter \ref{chapt:PowerBeaming} to determine the number of laser transmitters and free electron lasers needed to be built to satisfy these power requirements.

\begin{table}
    \begin{centering}
    \begin{tabular}{ccc}
    \toprule
        Climber number & Altitude, km & Power Required \\
        \midrule
        1 & 0-100 & 10.9 MW\\
        2 & 4,800 & 3.5 MW\\
        3 & 9,600 & 1.6 MW\\
        4 & 14,400 & 904 kW\\
        5 & 19,200 & 526 kW \\
        6 & 24,000 & 301 kW \\
        7 & 28,800 & 150 kW \\
        8 & 33,600 & 41 kW \\
        \bottomrule
    \end{tabular}
    \caption{Table of the power required at the wheels of eight 20 tonne climbers on the tether as shown in Figure \ref{fig:multiclimber} if they are all moving at 200 km/h.  These numbers are to bound the ranges of powers needed at different altitudes for the power beaming facility.  The reason the beginning altitude for climber 1 is between 0 and 100 km is because the climber power source may be turned on after the climber has been lifted above the atmosphere as shown in Section \ref{sec:reelout}.  This alleviates most of the atmospheric drag issues that affect the climber.}
    \label{tab:sevenclimb}
    \end{centering}
\end{table}

\textbf{}
\chapter{Power Beaming to Ascending Climbers}\label{chapt:PowerBeaming}
\section{Power beaming framework}
\subsection{Introduction}

Powering the space elevator climber has many challenges. The general scenario requires sources on the ground or in space with a compact receiver on the climber that converts the transferred energy into electricity to power the infrastructure and the motors of the climber.

This chapter is dedicated to power beaming (PB), a method of wireless power transfer (WPT) that uses the electromagnetic spectrum to transport energy over relatively large distances. The ratio of the link distance to the overall dimension of the transmit- and receive-structures can be defined as a beam aspect factor (BAF). For power beaming the BAF is significantly larger than one. Therefore the link length is even larger compared to the wavelengths. See also~[\cite{jaffe2024power}, 1.4, "Defining Power Beaming"], for some extended definitions.  

\begin{figure}[hb]
\begin{centering}
\includegraphics[width=\textwidth]{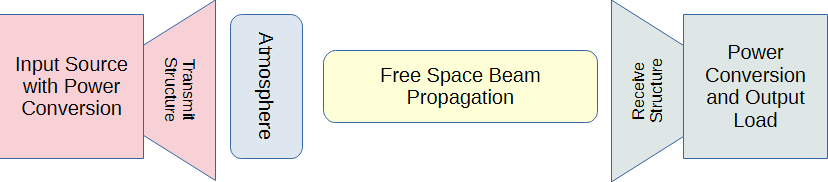}    
\caption{\label{fig:powerbeamingscheme} Schematic of power beaming components with source on Earth and receiver on the climber.}
\end{centering}
\end{figure}

A ground-based power beaming setup for the space elevator may then be schematically described as in Figure~\ref{fig:powerbeamingscheme}.\textbf{}
 
\FloatBarrier

Other modalities of power transfer are investigated elsewhere in this report.  These include transmission over the tether as an electrical conductor (Chapter~\ref{chapt:ACDC}), including hybrid electromagnetic modes (Section~\ref{sec:external-microwaves}) or through mechanical oscillations (Section~\ref{sec:mechanical-vibrations}).

This introduction to power beaming covers free-space propagation and, for ground-based transmitters, atmospheric effects on transmission (see Fig.~\ref{fig:powerbeamingscheme}, "Atmosphere and Free-Space Beam Propagation") in an idealized framework that is applicable across the entire PB frequency window considered. Dedicated sections describe how behavior at the power conversion points and the transmitter and receiver ends, depends on modality, whether lasers or microwaves. 

First, a common framework for power beaming is sketched using Gaussian beams and the geometric relationship between a round transmitter and receiver array for a power beam. This is equally applicable to all EM-frequencies considered for beam propagation through free space. The usual approximations are made for Gaussian beams and for the aperture width used to cut off the Gaussian beam and its power. Propagation through the atmosphere is complex and will only be covered as a
high-level scenario in the second part of this section, treating the atmosphere as a blur filter. Optional components addressing atmospheric propagation, such as methods that compensate for beam-quality degradation, are only briefly mentioned. These could be variants of active optics, such as using an active target on the climber, a guide star in the upper atmosphere to derive real-time corrections of the beams wave-shape at the source, or even high-gain phase conjugated mirrors,  to deliver maximal power at the receiver target on the climber. 

Without loss of generality, it is assumed that the path of the climber through the atmosphere can be covered by a reel-in-reel-out method of the tether, which does not require beamed power. (See Section~\ref{sec:reelout}) This helps to avoid large angles between the PB-transmitters and the receiver target on the climber when close to the ground. This implies that most of the power beam path is in vacuum and covered by free space propagation of the electromagnetic waves. The minimum distance, after reel-in-reel-out, between a source on the ground and the receiver target is set to be at least 100~km. The medium of space close to Earth is very thin and its optical influence or that of sizable bodies or fragments encountered in the beam path can be safely ignored for now.

Diffraction sets limits on the useful frequencies at the long-wavelength end of the spectrum due to the high power requirements and the relatively compact size of a mobile receiver structure on the space elevator climber. Our chosen limit for this was 1~m or 300~MHz for the longest microwave wavelength considered. Atmospheric scattering and opacity set the short wavelength limit. There are also limits to useful conversion at a high-power source or onboard a climber. Transmission in the atmosphere rapidly declines going from 400 to 300~nm, and high absorption by O$_3$, O$_2$, and N$_2$ molecules allows little solar radiation to reach the surface below 300~nm. Of course, this is a highly complex topic. See, for example, the HITRAN2024 molecular spectroscopic database~[\cite{2026JQSRT.35309807G}]. We set our lower wavelength limit at 400~nm. The combined upper and lower wavelength limits define our useful EM frequency window for power beaming to a space elevator from 1~m (300~MHz) to 400~nm (750~THz). See Figures~\ref{fig:window} and~\ref{fig:microwavewindow}.

\subsection{Beam propagation in free space}

Power transmission in free space (vacuum) is often described by the ideal case of a diffraction-limited Gaussian beam. This applies to lasers and microwaves alike within the frequency window outlined above.  Starting from the Helmholtz equation, choosing cylindrical coordinates $\mathrm{(r, \phi, z)}$ and making the paraxial approximations $\mathrm{tan(x)}\approx \mathrm{sin(x)}\approx x, ~\mathrm{cos(x)}\approx 1$~[\cite{ParaxialApprox}], and a slowly varying envelope~[\cite{SlowlyVaryingEnvelopeApprox}] we eventually arrive at an expression for the electric field of the Gaussian beam (see~[\cite{GaussianBeam}]):

\begin{equation}
E(r,z) = E_0 \frac{w_0}{w(z)} \exp\!\left(-\frac{r^2}{w(z)^2}\right)
\exp\!\left[-i\left(kz - \Psi(z) + \frac{k r^2}{2R(z)}\right)\right]
\label{eq:gaussianbeam}
\end{equation}

\noindent where
\begin{equation}
w(z) = w_0 \sqrt{1+\left(\frac{z}{z_R}\right)^2}, \qquad
R(z) = z\left[1+\left(\frac{z_R}{z}\right)^2\right], \qquad
\Psi(z) = \arctan\!\left(\frac{z}{z_R}\right)
\label{eq:gaussianbeamRadiusCurvatureGouy}
\end{equation}
and
\begin{equation}
z_R = \frac{\pi w_0^2}{\lambda}, \qquad k = \frac{2\pi}{\lambda}.
\label{eq:gaussianBeamRayleighK}
\end{equation}

\noindent $\mathrm{w_0}$ is the beam waist, $\mathrm{R(z)}$ is the radius of curvature of the wavefront, and $\mathrm{\Psi(z)}$ is the Gouy phase.

Here are a few examples using the Gaussian beam formalism. We use the definition of beam radius $w$, where the power has fallen off to $1/e^2$ of the on-axis maximum value. This implies that with a full power of the beam P the on-axis maximum irradiance is $2 P/{\pi w}$.

This sets a power cutoff for a Gaussian beam for a round aperture of radius w at around 86\% and almost 14\% being lost outside of it. However, this can be addressed by simple factors, scaling the beam radius down or the aperture up relative to each other. A receiver radius of $1.22 w$ captures 95\%, one of $2 w$ captures 99.97\%. We will use the simple w in our examples.

We want to transfer as much power as possible to a small receiver on the climber with 5~m diameter at the GEO distance of around 36000~km and lose as little power as possible of the whole beam by too large a beam size. So what does it take to have a beam of radius 2.5~m hit the climber?

For lasers, using the formulas from above, the $z_R = \pi {w_0}^2 / \lambda$ for $\lambda=500~\si{nm}$ and a transmitter aperture of 5~m diameter, is 
$z_R~\approx~39000~\si{km}$; for a smaller $w_0=2.292~\si{m}$ $z_R\approx 33000~\si{km}$; For the laser case and a highly collimated beam, the desired beaming distance for reasonable apertures is around the Rayleigh range $z_R$, influenced mostly by the Fresnel diffraction regime. 

For a microwave frequency of 10~GHz, $\lambda=3~\si{cm}$. What transmitter aperture is required to focus the beam on the 5~m receiver at GEO distance with an ideal Gaussian beam? How large is the required aperture? 
\begin{equation}
\begin{aligned}
\theta_{req}&=radius_{target}/distance=\frac{2.5~\si{m}}{36000~\si{km}}=6.9444\times10^{-8}~\si{rad}\approx 0.0694~\si{\mu rad} \\
D_T&= \frac{2 \lambda}{\pi~\theta_{req}}=\frac{2\cdot3~\si{cm}}{\pi~6.9444\times10^{-8}~\si{rad}} \approx 275~\si{km}, \mathrm{see~also~Eq.}~\ref{eq:Thetatheta}\\
z_R&=\frac{\pi {w_0}^2}{\lambda}= 1.98\times10^9 ~\si{km} = 13.240~AU
\label{eq:microwaveExample10GzAperture}
\end{aligned}
\end{equation}
This is quite large, certainly 5~m are insufficient, but microwave transmitters can produce waves from many transmitter elements and form a beam from a very distributed aperture. An advantage of the large aperture in the microwave case is that there is less danger of exposure to the beam due to the lower energy density near the ground. If the transmitter is significantly smaller, a significant amount of energy misses the receiver at larger distances.

\subsubsection{Beam quality parameter $M^2$}
The ideal diffraction-limited Gaussian beam is a useful reference. It maintains its Gaussian shape on propagation, with only the width and curvature changing. The Fourier transform of a Gaussian is another Gaussian, so Gaussian source profiles produce Gaussian far-field diffraction patterns (under linear, paraxial approximations).  
However, the real world is less than ideal. For lasers, a standardized Beam Quality Parameter $M^2$ (ISO Standard 11146, also~[\cite{Msquared}]), offering a first step to characterize deviation from the ideal Gaussian beam in a single number, is defined as
\begin{equation}
M^2 = \frac{\pi w_0 \theta_{real}}{\lambda} = \frac{\theta_{real}}{\theta_{Gaussian}} = \frac{w_{0~real}}{w_{0~Gaussian}} 
\label{eq:M2}
\end{equation}
with $w_0$ the beam waist radius where the intensity is $1/e^2$ of the maximum on axis, $\theta$ the half-angle far field divergence in radians, and $\lambda$ the wavelength. 

For an ideal TEM00 Gaussian, $M^2 = 1$. Real beams have an $M^2\ge 1$, providing a factor by which the focusable spot area and divergence exceeds the ideal Gaussian of the same wavelength and waist. The focusable spot radius scales as $w_{0~real}=M^2 \cdot w_{0~Gaussian}$. Beam propagation, radius vs. z  follows the same Gaussian formulas with $\lambda$ replaced by $\lambda\cdot M^2$.  

$M^2$ is measured according to ISO~11146 by recording the beam radius at multiple positions around the waist, fitting to the generalized Gaussian-propagation formula, and extracting $M^2$.

Of course the reality of wavefront shaping far exceeds the $M^2$ characterization but also of this description. The use of diverse forms of phased arrays (e.g., the AESA radar tracking multiple targets) opens a wide field of possible beam profiles and apertures, trying to image power to receivers. In the optical regime, miniaturization and nantennas may offer similar flexibility as microwaves, with challenges in the Space Elevator case expected beyond simple miniaturization due to the extremely high power demands.
 
\subsection{Power transfer with idealized Gaussian beams} 
So how do we characterize the efficiency of a power-beaming link? 

Optics uses the beam parameter product (BPP) $BPP=w_0 \theta= M^2 \lambda/\pi$ for a Gaussian beam. We can see $M^2=BPP_{measured}/{(\lambda/\pi)}$. The often used \'Etendue, in this case geometrical for the ideal Gaussian beam, is $G=\pi^2 {BPP}^2$. Other phase-space descriptions, such as the space-bandwidth product or the Wigner distribution, have their own advantages, but the two above are used here for the simple ideal case.

Radiance R is defined as Watts per $\si{m^2}$ per steradian, the power arriving at a surface per unit projected area and per unit solid angle at a given direction and wavelength. For our ideal case we can define Radiance R as $R = {Power P}/{Etendue}$ for a given wavelength with an assumed refraction index of $n=1$ or $R=P/(\pi^2 {BPP}^2)$ in terms of the BPP. Irradiance~I is the power of the integrated incident radiance over an area.  This is discussed in Section \ref{sec:PVarray} for the laser power beaming case.

Diving into the origins of diffraction theory between Huygens (spherical), Young (cylindrical/edge), and Rayleigh (plane) and expanding on EM designs could help for a better understanding of power beaming. Different languages for the same topic, just as between microwaves and lasers.

\subsubsection{Goubau relationship}
Looking at the efficiency of power transfer between source and receiver, we need to combine the radiance exiting the transmitter with the fraction hitting the receiver. This can be roughly captured by addressing only geometrical factors through the Goubau relationship as introduced in the microwave section, Section \ref{sec:microwave}. 
The power efficiency $\varepsilon$ as defined under~\ref{eq:GEpsilon} is just the fraction of a Gaussian inside the aperture defined by the power link transmission parameter $\tau$ with the fraction of the Gaussian contained within a round aperture being $1-e^{ -2~r^2/w^2}$ with w the beam (Gaussian) radius and r the radius of the aperture.

$\tau$ for a beaming distance d and wavelength $\lambda$, apertures (optical or microwave) $At, Ar$ as defined in~\ref{eq:GEpsilon} is
\begin{equation}
\tau = \frac{\sqrt{A_{t} A_{t}}}{\lambda d}
\label{eq:transmissionParameter}
\end{equation}

From the examples before, we can see that the Rayleigh range of the highly collimated power beaming link for the SE is huge. Trying to focus on a tiny receiver array of 5~m leads to a required link divergence of a fraction of a microradian and the link remaining in the Fresnel range of diffraction for a Gaussian beam. Assuming a waist size with radius of 2.5~m with a $\lambda$ of 500~nm, $z_R=\pi w0^2/\lambda=39300~km$ and for a microwave example with 5~km waist radius and $\lambda$ of 3~cm, $z_R=2.6\times10^6~km$.

Calculating the fraction of a Gaussian beam hitting a receiver with radius $r_r$ with the transmitter defined as the exit aperture $w_0$ we get for the ideal diffraction-limited Gaussian beam (power $P_{tG}$) with the $w_d$ defined under~\ref{eq:gaussianbeamRadiusCurvatureGouy}:
\begin{equation}
P_r=P_{tG} (1-e^{-2~\frac{r_{rec}^2}{w(d)^2}}) = P_{tG} (1-e^{-2~\frac{r_{rec}^2}{ (w_0 \sqrt{1+(d/z_R)^2})^2 }}) = P_{tG} (1-e^{-2~(\frac{r_{rec}}{w_0})^2~ \frac{1}{(1+(d/z_R)^2}}))
\label{eq:powerbeamGauss}
\end{equation}
For a power beam with the same transmit and receive apertures at the Rayleigh distance, as put forward before, the receiving aperture truncates the received beam power at $1-1/e^{2}$ or 86\% and loses the remaining 14\%. This formula~\ref{eq:powerbeamGauss} applies within the Rayleigh range not only in the far field.

\subsection{What about tophat beams?}
The target receiver on the climber should be as compact and lightweight as possible. An even illumination, a tophat beam, with a rectangular power profile, on that receiver appears to be desirable.

Various beam profiles are a possibility, but the natural mode of a beam propagation remains Gaussian. A tophat profile includes components that evolve differently, e.g., a different Guoy phase for the different Gaussian components, so a tophat profile will not remain so at different distances. The desired tophat on the climber needs to be produced by managing the transmitter's wavefront as a function of distance.

\subsection{Influence of the atmosphere}
\subsubsection{Atmospheric transmission windows, absorption, and scattering}

The atmosphere of Earth is relatively transparent to three bands of electromagnetic energy of use in wireless power transmission (WPT).  

\begin{figure}[h]
\begin{centering}
\includegraphics[width=\textwidth]{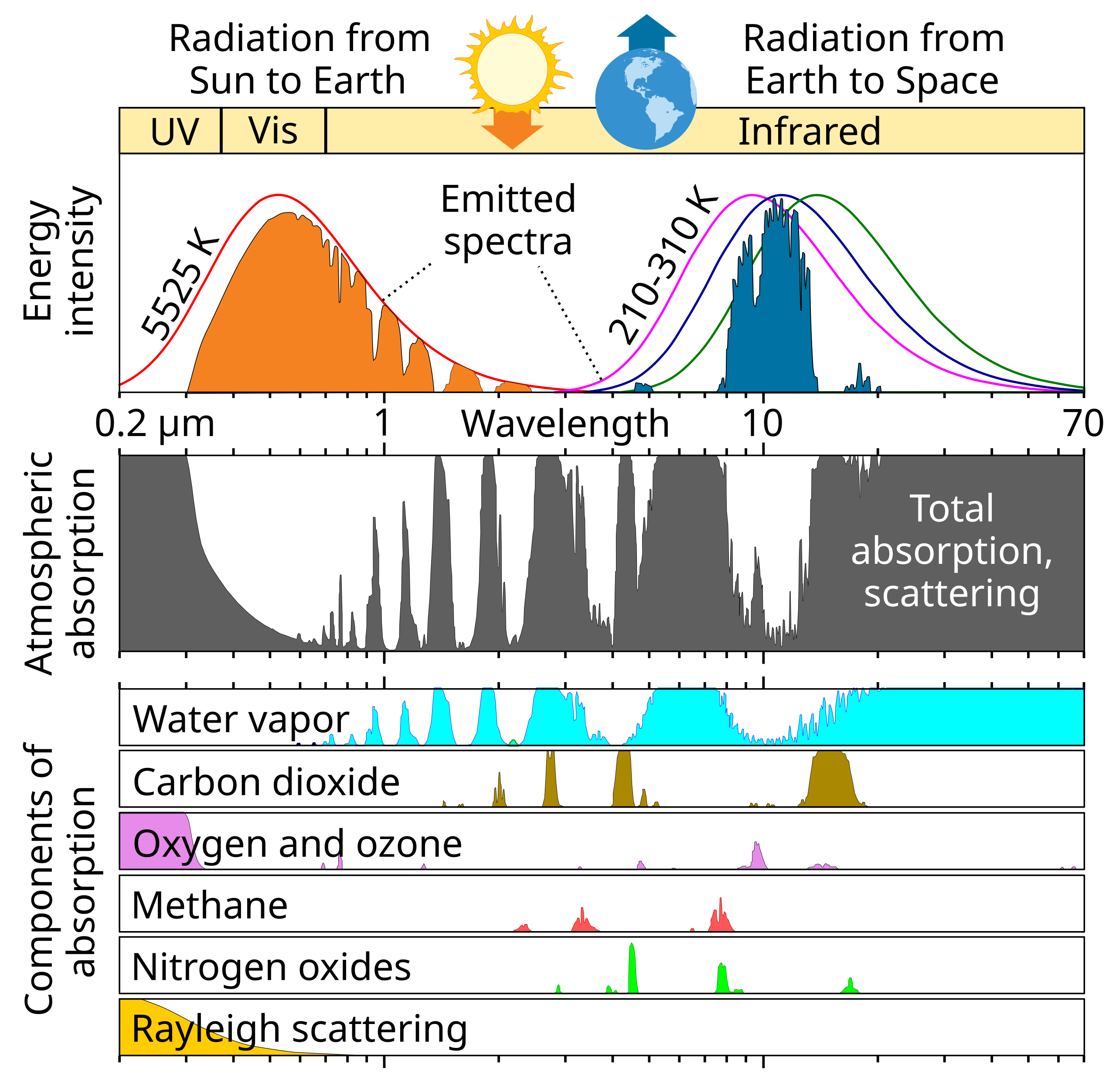}    
\caption{\label{fig:window} The windows for light transmission through the atmosphere. [\cite{window}]}
\end{centering}
\end{figure}

Figure \ref{fig:window} shows that there is a window through the atmosphere in the visible light wavelengths, peaked between 0.5 and 0.6 microns.  This window is how the visible portion of sunlight gets to Earth.  The next window is in the infrared, peaked at about 10 microns in wavelength.  This is the range of wavelengths that Earth radiates heat back into space at night.

\begin{figure}
\begin{centering}
\includegraphics[width=\textwidth]{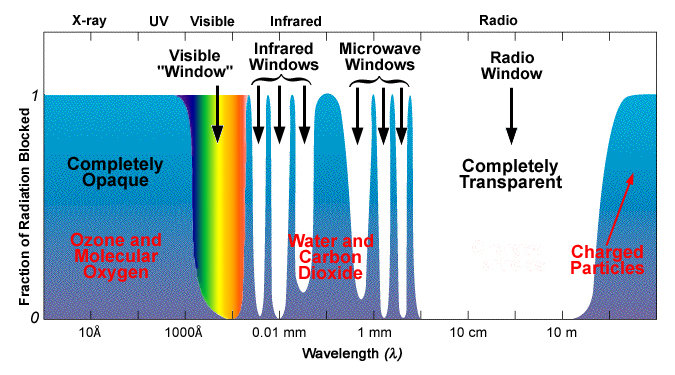}    
\caption{\label{fig:microwavewindow} The transmission curve for radio to X-ray wavelengths through the atmosphere. [\cite{pham2018satellite}]}
\end{centering}
\end{figure}

Figure \ref{fig:microwavewindow} shows the transmission curve for the range of frequencies from X-ray to radio waves.  The third window is the select frequencies in the microwave band that are not absorbed by water and carbon dioxide. As will be seen below in the section on microwave power transmission (Section \ref{sec:microwave}), 10 GHz frequency (3 cm wavelength) is one of the frequencies chosen for the microwave power beaming conceptual design.  It will become clear in the section on the divergence of beams in a vacuum why longer wavelengths in the radio bands were not considered, even though the atmosphere is transparent to them.

\FloatBarrier

\subsubsection{Atmospheric turbulence}

Atmospheric turbulence is a chaotic variation of air density, temperature, and pressure, that change the refractive index of the air a laser beam is passing through. Non-uniform heating of the atmosphere by the sun and mixing of air causes turbulent eddies of different sizes. These eddies act as lenses that can bend laser beams away from their intended target. These fluctuations distort, scatter, and attenuate laser beams, causing three primary detrimental effects: beam wander (misdirection), broadening (spreading), and scintillation (intensity dancing). Most of the impact of these effects on beams happens in the lowest 30 km of the atmosphere.  Defining the main detrimental effects:

\begin{enumerate}
    \item Beam Wander (Deflection): Large-scale fluctuations cause the entire beam's centroid to dance, making it difficult to keep the beam focused on a target, especially on ground-to-space links.
    \item Beam Spreading (Broadening): Small-scale turbulent eddies scatter the beam, causing it to broaden faster than the natural diffraction limit. This decreases intensity and reduces energy density at the target.
    \item Scintillation (Intensity Fluctuations): This refers to rapid, random temporal and spatial fluctuations in the intensity (brightness) of the beam, causing fading or high-peak power fluctuations.
    \item Phase Aberrations: Turbulent layers induce random phase shifts across the beam’s wavefront, degrading the wavefront quality and reducing the coherence length.
    \item Power Loss and Coupling: Turbulence causes power to couple from the desired transmitted mode into other, unwanted modes, reducing the total energy delivered, specifically for Gaussian beams.
\end{enumerate}

Adaptive Optics (AO) can mitigate some of these effects by counteracting wavefront distortions in real time. Without such a correction, the atmosphere, acting as a blur filter for a 500~nm laser beam, can make $M^2$ reach double or triple digits for a telescope-like transmitter with a 5~m diameter. 
This would severely degrade a power-beaming link. While it is understood that Adaptive Optics will be incorporated in the laser beam expander described in Section \ref{sec:beamexpand}, the details of how Adaptive Optics will be incorporated are beyond the scope of this report. 

Microwave beams are affected by atmospheric turbulence in the same ways that laser beams are, but to a lesser extent. Microwaves typically pass through the atmosphere with less disruption than lasers, making them more robust to poor weather, such as fog and rain. Techniques such as using diffractive optical elements or beam homogenizers can reduce the impact of turbulence, enabling stable, uniform power delivery.  The use of beam homogenizers in front of receiving panels (rectennas) can help diffuse high-intensity spots (speckles), stabilizing the power output of the rectenna.

\subsubsection{Some weather phenomena}

Laser beams are heavily affected by clouds.  The water droplets in clouds scatter 500~nm light very efficiently. While fast-pulsing terawatt-scale lasers have been used to punch holes in clouds to allow a communications beam to pass through, the hole is typically only a few centimeters wide; not wide enough to pass the large-diameter laser beams described in Section \ref{sec:beamexpand}.

This is one of the biggest motivations for mounting laser platforms on mobile ships so that they can move out from under cloud cover. As will be seen in Section~\ref{sec:FELdesign} and Section \ref{sec:beamexpand}, laser transmission systems are small enough that it is possible to mount them on ships.  The same cannot be said for microwave transmitters on the ground.

Microwave beams are generally not heavily affected by clouds, allowing them to penetrate through cloud cover, dust, smoke, and light rain. This makes them ideal for satellite imaging and radar. However, at higher frequencies (tens of GHz) or in heavy, thick, and precipitating clouds, microwaves can experience scattering, attenuation, or absorption.

\subsection{The goals for this chapter}
The goal of this chapter is to present several conceptual designs for wireless power beaming to the climbers.  The designs are entirely determined by the physics of the chosen electromagnetic wavelengths.  Power transfer efficiency is only one of the design parameters, but it determines the sizes of the transmitters and receivers.  A goal of any transmission technology is to develop a transfer function that allows the system efficiency to be calculated.  Only by knowing the system efficiency can the required wall-plug power be determined. 

We have all seen pictures of satellites and the International Space Station showing photovoltaic (PV) arrays that collect sunlight and convert it to electricity.  The primary difference between these receivers and those on the climbers of the space elevator is that the receivers on satellites and the ISS are in microgravity.  Those vehicles are in orbit around the Earth and so their PV array support structures do not need to be very strong.  They are not subjected to Earth's full gravity.

Climbers on the space elevator are not in orbit.  When they are taking off from ground level, every part of the climber is subjected to the full 1 G of Earth's gravity.  The receiving arrays on the climber, regardless of the wavelength they are designed to work with, must have a support structure similar to that of any receiver on Earth.  This makes climber design very different from satellite design.  This is discussed in more detail in Section \ref{sec:PVin1G}.
\section{Laser power beaming to the climbers}

In this section we will look at delivering power to the climbers by laser beam.  The questions to be answered are:

\begin{enumerate}
    \item What wavelength of light is the best to use when considering transmission from the ground to GEO?
    \item What laser technology produces that wavelength of light in the power needed?
    \item What receiver technology has the best efficiency and is most practical on the climber for the laser light received?
\end{enumerate}

\subsection{Determining the wavelength of the laser by study of the atmospheric windows}

Looking at the curves shown in Figure \ref{fig:window}, it is clear that light in the range of 0.5 to 0.7 microns in wavelength suffers the least absorption passing through the atmosphere compared to neighboring wavelengths.  Shorter wavelengths are more subject to absorption by oxygen, ozone, and Rayleigh scattering.  Longer wavelengths are more heavily absorbed by water vapor in the atmosphere.  Choosing light whose wavelength is around 0.5 microns, green light is a reasonable choice from the perspective of atmospheric absorption.  It will be shown later why choosing light in the shorter wavelength of this band is preferable to longer wavelengths.

\subsection{Estimating the total power lost by atmospheric absorption}

\begin{figure}[hb]
\begin{centering}
\includegraphics[width=.87\textwidth]{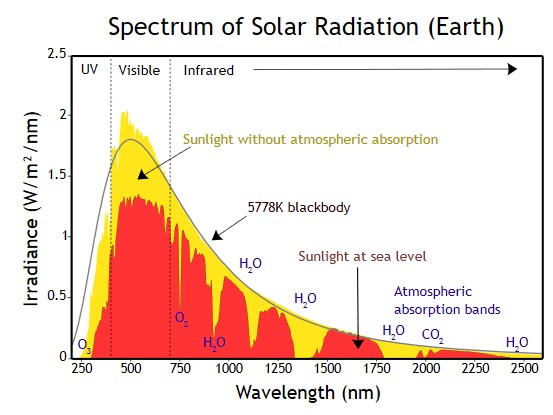}    
\caption{\label{fig:solarspec} The difference in power per square meter between sunlight just outside the atmosphere and sunlight at the surface of the Earth as a function of the wavelength of the light. [\cite{sunlight}]}
\end{centering}
\end{figure}

It is clear from Figure \ref{fig:solarspec} that the amount of sunlight available just outside the atmosphere of Earth is higher than that received at sea level.  The irradiance numbers shown are for rays of sunlight shining straight down vertically.  This graph provides a rough method of estimating the amount of power lost from the laser beam to absorption in the atmosphere.  We will assume that the wavelength of light we are interested in is 0.5 microns, or 500 nm.

From the graph, the solar spectral irradiance outside the atmosphere at 0.5 microns, (500 nm) wavelength, is approximately 1.9 \si{W/m^2/nm}.  The solar irradiance at sea level is approximately 1.3 \si{W/m^2/nm}.  Calculating the percent lost between the top of the atmosphere and sea level gives (1.9-1.3)/1.9 = .316, approximately a 31.6\% drop in power at the 500 nm wavelength of sunlight from the top of the atmosphere to the bottom at sea level.  It is reasonable to assume that 500 nm light starting out at sea level would lose about 31.6\% of its energy on the way out of the atmosphere.  This means that there is a 68.4\% transmittance of 0.5 micron laser light pointed straight out of the atmosphere.  This will become one of the terms in the transfer function of system efficiency (see Section \ref{sec:transfer}).

These numbers are for fair weather with no clouds.  This rough calculation sidesteps the complications of atmospheric modeling.  Understanding the mechanisms of light lost in the atmosphere (turbulence, absorption, and scattering) is important, especially in the design of adaptive optics systems to correct the beam for atmospheric variations.  That study is far beyond the scope of this document.  The intent of this calculation was to come up with an engineering estimate that would be good enough to bound the size of the power source needed on Earth or in space.

Clouds change the picture significantly.  Green laser light can penetrate thin clouds, but the water droplets in clouds scatter 500 nm light much more than microwave radiation (in the appropriate wavelength). Back-of-the-envelope calculations indicate that it would take a lot of energy (comparable to the amount of energy being beamed to the climbers,) to "burn" large enough holes through clouds to allow the power beam through.

Cloud cover is a strong reason to have the laser transmitters on mobile platforms so that they can move out from under cloud layers.

\FloatBarrier

\subsection{Diffraction-limited beams and the divergence of lasers in vacuum}
\label{sec:diff}

The photovoltaic arrays on the climber represent targets of a given aperture that the laser beams from the source must hit without a significant amount of light spilling over the sides of the arrays.  Unfortunately, a fundamental process in physics caused by the wave nature of light is that light diverges the further it gets from the source.  Effectively, the beam spot gets bigger the further away from the source the receiver is.  Climbers need to receive power up to 36,000 km up in space.  The beam must be of high quality so that it does not diverge much larger than the PV arrays.  The diffraction limit is the basic limit on the resolution of an optical system.  The variables to manipulate the divergence of the beam are the wavelength of the light and the size of the aperture.  Smaller wavelengths and larger apertures minimize the divergence of a beam.  No focusing of the beam is accounted for in this analysis.  Focusing may become part of the final system with a more detailed design.

Rearranging the terms of equation \ref{eq:gaussianbeamRadiusCurvatureGouy} to eliminate $z_R$, we get the following equation:

\begin{equation}
    w(z) = w_{0} \sqrt{1 + \left( \frac{\lambda z}{\pi w_{0}^{2}} \right)^{2}}
    \label{eq:beamradius}
\end{equation}

Where w(z) is the waist (the radius) of the beam at the distance z from the minimum waist, $w_0$. $\lambda$ is the wavelength of the light.  Figure \ref{fig:Fig4gaussian} shows the relationships between the variables.

The next step in the analysis is to determine the size of the laser beam needed to transmit up to GEO and what the loss as a function of distance is.  A simplifying assumption in this analysis is that the power loss is determined by the ratio of the size of the PV array to the size of the beam at a given distance.  Also for simplicity, the power distribution of the beam is assumed to be uniform, referred to as a flat top beam.  Real beams are somewhere in between flat top distributions and Gaussian distributions of power.  If the beam is larger than the array, then the power in the annulus between the beam diameter and the PV array diameter is lost.  The equation to determine this power loss is:

\begin{equation}
    P_{\mathrm{cl}}(z) = P_{o} \cdot \pi \cdot \frac{\left(R_{\mathrm{PV}}\right)^{2}}{A_{\mathrm{beam}}(z)}
    \label{eq:powerloss}
\end{equation}

Where $P_{cl}(z)$ is the power received at the climber array.  $P_o$ is the power transmitted from the source.  The graphs that follow assume that $P_o$ is 4 MW and ignore the power lost to the atmosphere.  4 MW is an arbitrary number chosen to study the effect of varying beam waist and wavelength. Actual power required is calculated later. $R_{PV}$ is the radius of the PV array, which is set equal to the beam waist at the transmitter ($R_{PV} = \omega_o$.) $A_{beam}(z)$ is the area of the beam at the distance z from the transmitter.  $A_{beam} = \pi \cdot \omega^2$.

\begin{figure}
\begin{centering}
\includegraphics[width=\textwidth]{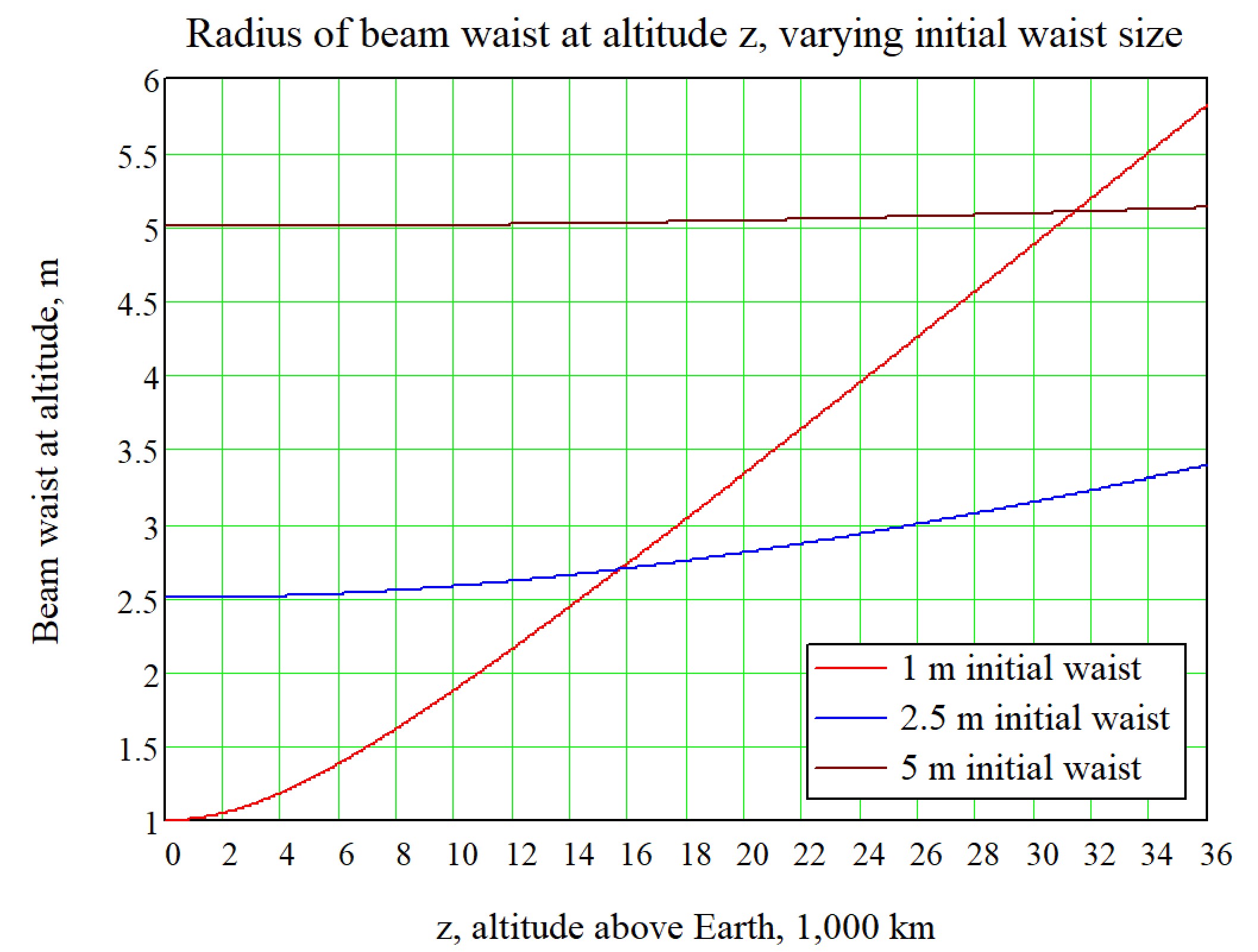}    
\caption{\label{fig:3waistvsalt} The radius of the beam waist at altitude z for three different starting waist beams (1 m, 2.5 m, 5 m,) at ground level, z=0, $\lambda = 0.5 \si{\mu m}$}
\end{centering}
\end{figure}

\begin{figure}[t]
\begin{centering}
\includegraphics[width=\textwidth]{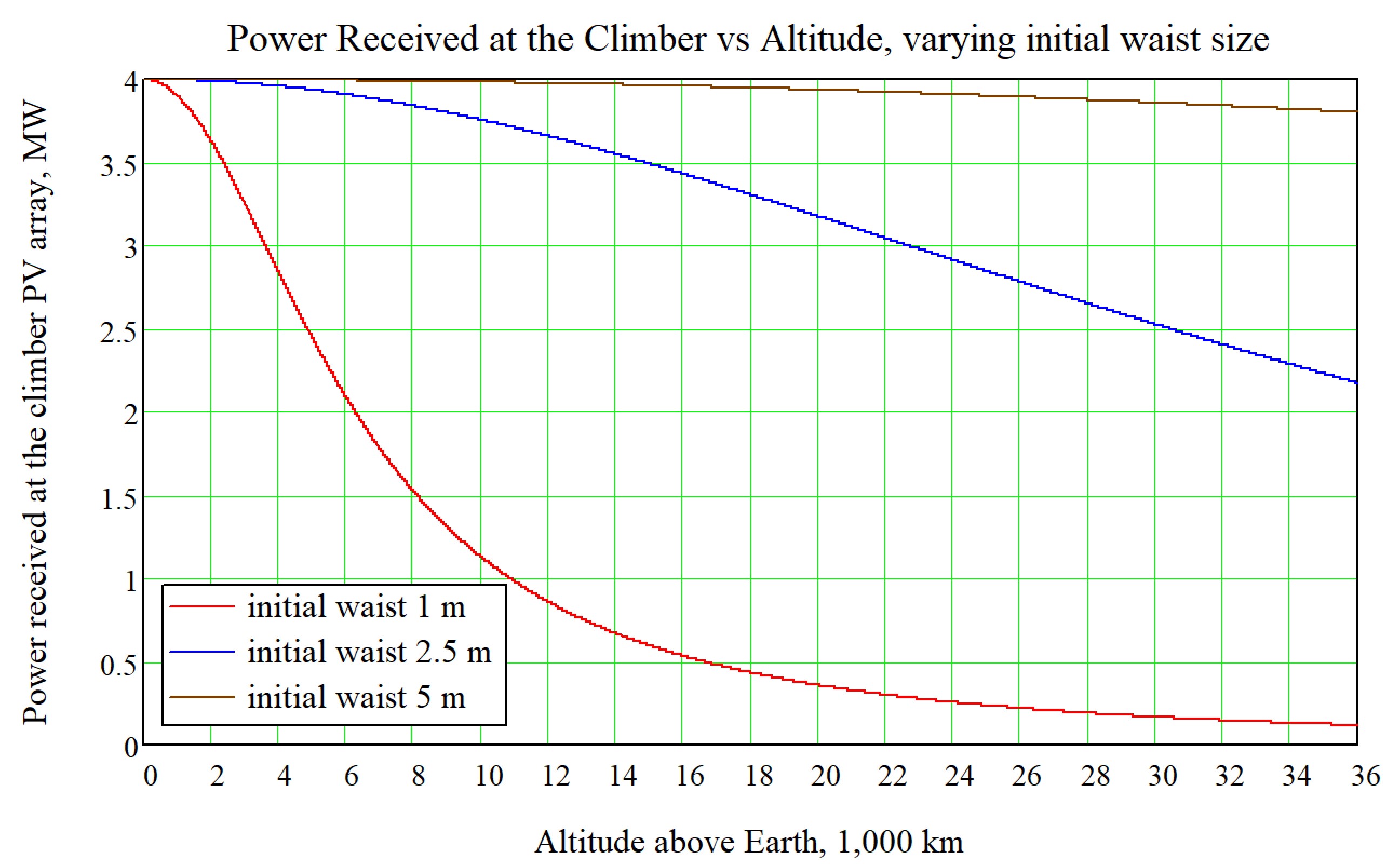}    
\caption{\label{fig:3waistpower} The power received at the PV array on the climber for three different waist beams (1 m, 2.5 m, 5 m,) and matching radius PV arrays, as a function of altitude. $\lambda = 0.5 \si{\mu m}$}
\end{centering}
\end{figure}

Figure \ref{fig:3waistvsalt} shows that a 1 m beam radius diverges to almost 5.8 m in radius at the altitude of GEO.  Figure \ref{fig:3waistpower} shows that the divergence from the 1 m beam waist causes the power received at the climber PV array at GEO to be down to 118.2 kW, less than 3\% of the starting power.

Figure \ref{fig:3waistvsalt} also shows that a 2.5 m beam radius diverges to almost 3.4 m in radius at the altitude of GEO.  Figure \ref{fig:3waistpower} shows that the power received at the climber PV array at GEO is reduced to 2.17 MW, 54.3\% of the starting power for the 2.5 m beam waist.

Figure \ref{fig:3waistvsalt} shows that a 5 m beam radius diverges to 5.13 m in radius at the altitude of GEO.  Figure \ref{fig:3waistpower} shows that the power received at the climber PV array at GEO is reduced to 3.8 MW, 95\% of the starting power for the 5 m beam waist.

From these graphs, we can see the effect of varying the beam waist.  A larger beam waist leads to a more tightly confined beam that delivers more power to the climber.  Power at the climber declines as a function of altitude because of the divergence, but we also know that the power required at the climber declines as a function of altitude as gravity weakens.  An optimum size transmitting and receiving aperture must be found that provides enough power to the climber but is the least cost solution to the problem.  A general rule of thumb to follow is that the smallest aperture that satisfies the climber requirements will also be the cheapest to build.

\begin{figure}
\begin{centering}
\includegraphics[width=\textwidth]{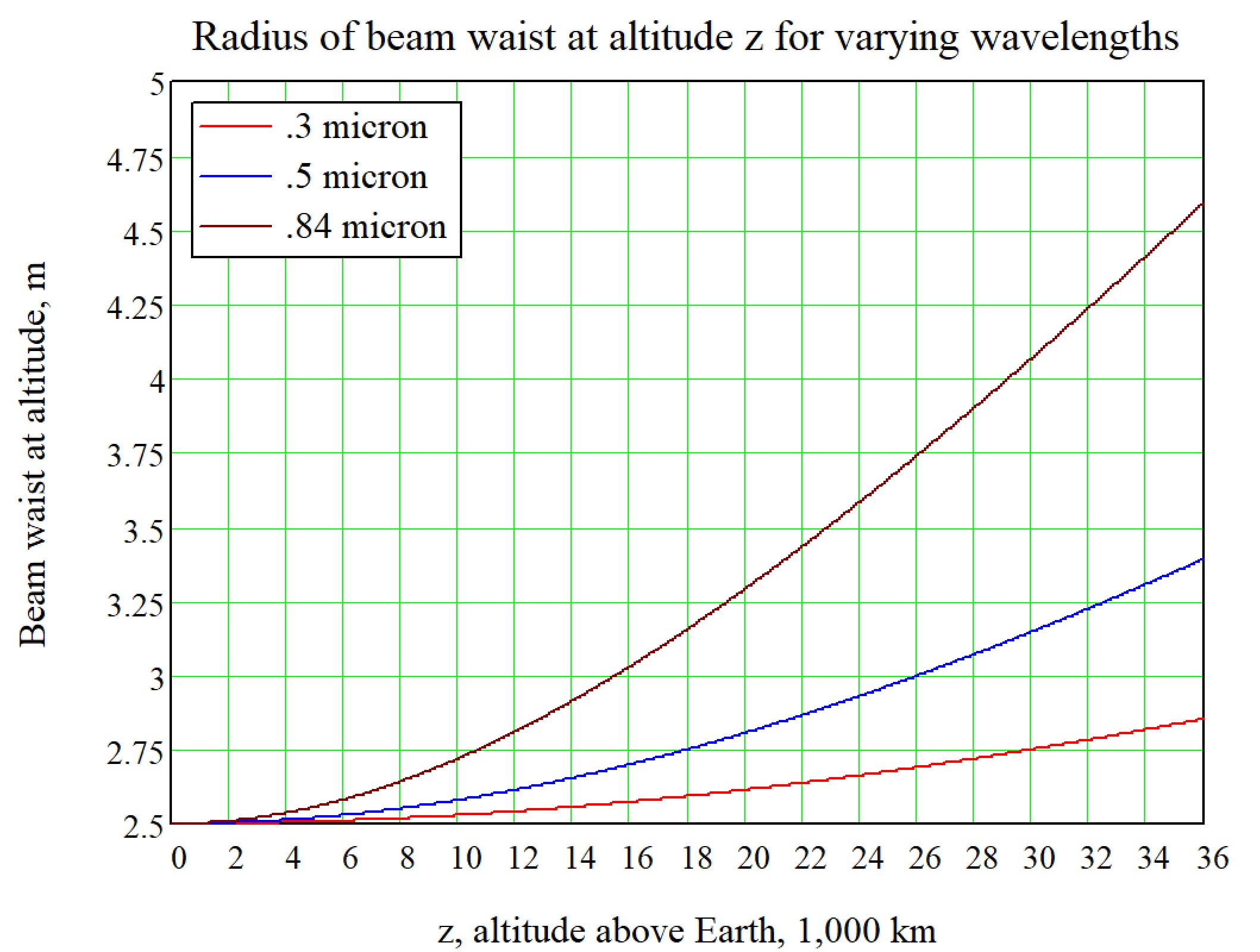}    
\caption{\label{fig:25mwv3usalt} The radius of the beam waist at altitude z for a 2.5 m starting waist beam at ground level, z=0, $\lambda$ = 0.3 $\mu \si{m}$, 0.5 $\mu \si{m}$, and 0.84 $\mu \si{m}$.}
\end{centering}
\end{figure}

\begin{figure}
\begin{centering}
\includegraphics[width=\textwidth]{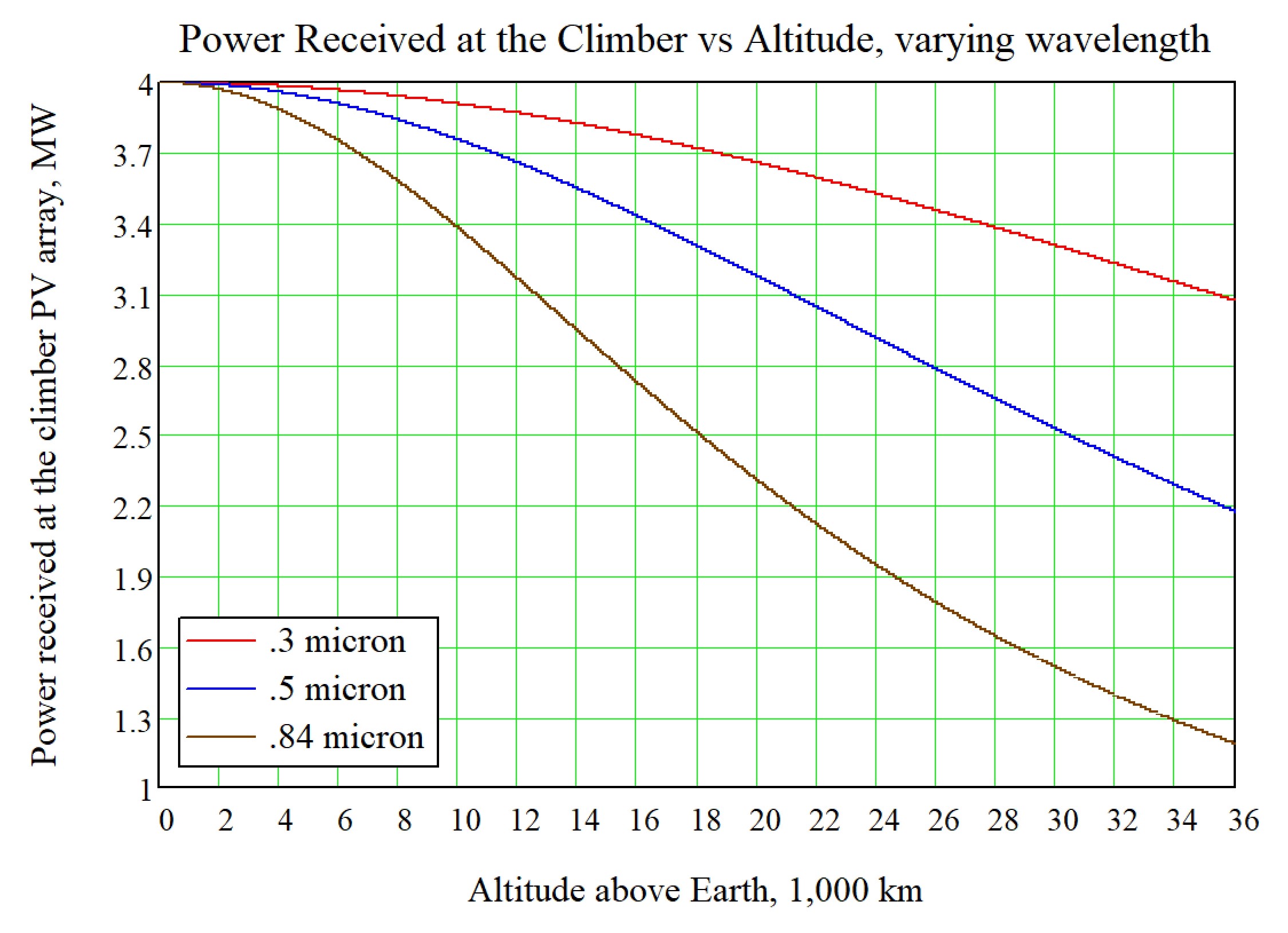}
\caption{\label{fig:25mpower3u} The power received at the PV array on the climber for a 2.5 m waist beam and 2.5 m radius PV arrays, as a function of altitude. $\lambda$ = 0.3 $\mu \si{m}$, 0.5 $\mu \si{m}$, and 0.84 $\mu \si{m}$.  (This calculation ignores any atmospheric scattering or absorption at the wavelength.)}
\end{centering}
\end{figure}

Another interesting study is to see the effect of varying the wavelength on the power received for a fixed aperture.  For these graphs, a 2.5 m beam waist and PV radius is assumed and the wavelength is varied from .3 microns to .84 microns.  .3 microns is in the blue part of the spectrum, and .84 microns is in the near infrared.
\newpage
Figure \ref{fig:25mwv3usalt} compares the waist of the beam holding the initial waist constant at 2.5 m, but varying the wavelength.  The figure shows that a 2.5 m beam waist of .3 $\si{\mu m}$ light diverges to only 2.85 m in radius at the altitude of GEO.  Figure \ref{fig:25mpower3u} compares the power received at the PV array as a function of altitude, but varying the wavelength.  This figure shows that the power received at the climber PV array at GEO is reduced to 3.07 MW, 76.8\% of the starting power.  These numbers make UV light look like the wavelength to deliver more power than 0.5 micron green light.  However, looking at Figure \ref{fig:window}, we see that .3 micron light is more heavily absorbed by oxygen and is Rayleigh scattered.  .3 micron light could be a viable contender for power beaming outside the atmosphere, but not when the beam starts at ground level.

Figure \ref{fig:25mwv3usalt} also shows that a 2.5 m beam radius of .84 $\si{\mu m}$ light diverges to almost 4.6 m in radius at the altitude of GEO.  Figure \ref{fig:25mpower3u} shows that the power received at the climber PV array at GEO is reduced to 1.18 MW, 29.5\% of the starting power.  Again, looking at Figure \ref{fig:window}, we see that .84 micron light is absorbed by water vapor. 0.84 micron light was checked because it is in the middle of the range of sensitivity of Gallium Arsenide (GaAs) PV cells.  It is not as good a candidate to deliver power at high altitude as 0.5 micron green light.

\subsection{Combining the beam power losses and comparing beam power to climber requirements}
\label{sec:powerlossreq}

The data presented in this chapter allows us to calculate how much the beam power is degraded by atmospheric losses and diffraction limit losses.  The only way to choose the smallest aperture transmitter that satisfies all the requirements is to compare the shaft power at the wheels to the required power to climb at 200 km/h at any altitude.  If the shaft power is higher than the required power, then the climber can ascend at 200 km/h or faster. If the shaft power is less than the required power, then the climber must climb at a slower rate.  Based on the speed calculations done in [\cite{wright2023climber}], another requirement besides a trip time to GEO on the order of a week is that the power must be sufficient to allow a climber to ascend at 200 km/h by an altitude of 4,000 km.  If this condition is not satisfied, trip times will be substantially longer than one week.

\begin{figure}
\begin{centering}
\includegraphics[width=\textwidth]{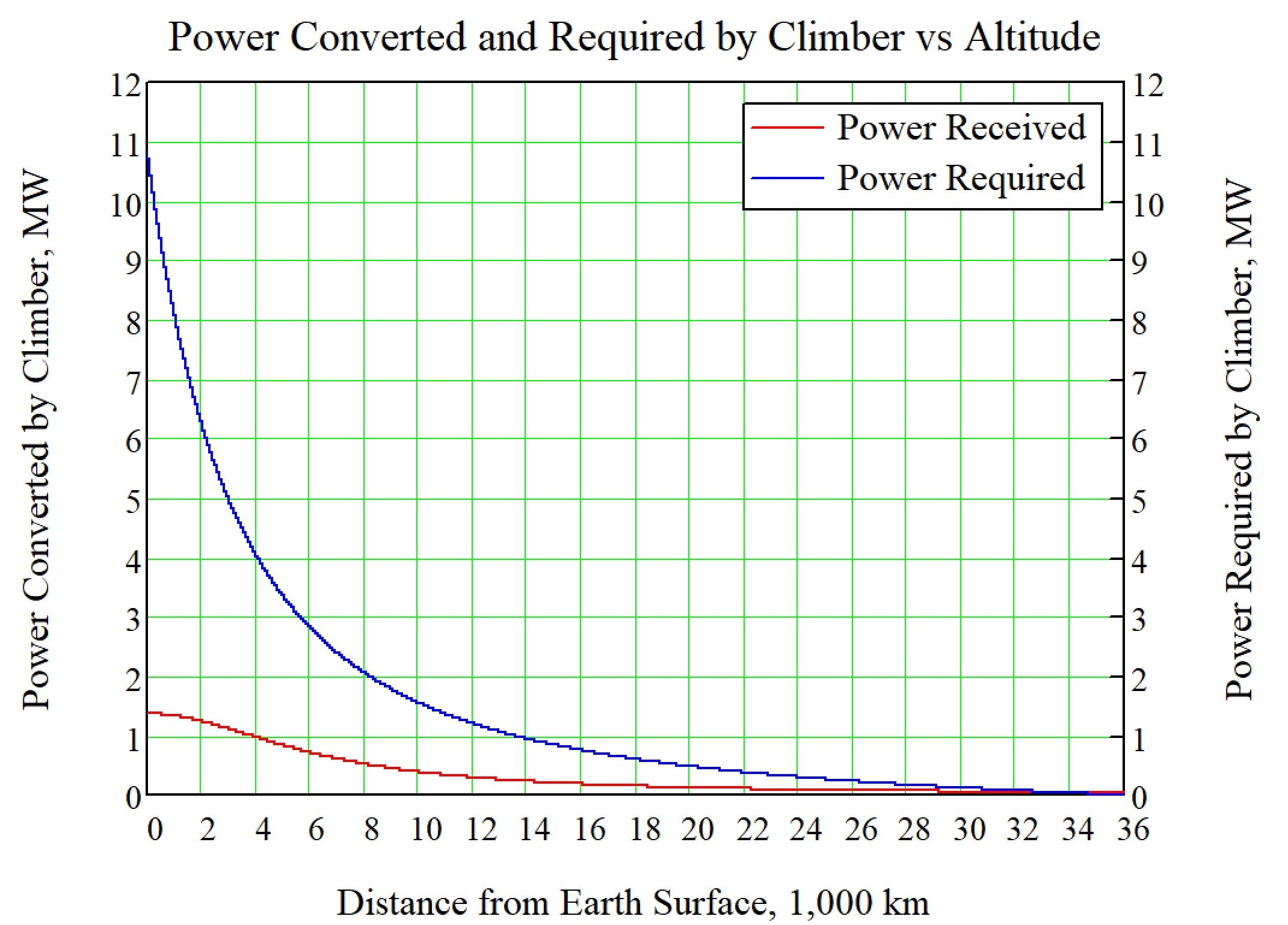}
\caption{\label{fig:1FEL1m} The blue curve is the power required by the climber to climb at 200 km/h. The red curve is the shaft power delivered by one 5 MW laser, 1 m waist, $\lambda$ = 0.5 $\mu m$.  The 5 MW laser at the ground is 3.42 MW above the atmosphere after the 31.6\% reduction caused by absorption in the atmosphere.  This arrangement never delivers enough power for the climber to climb at 200 km/h.}
\end{centering}
\end{figure}

There are more inefficiencies to consider in between the laser light received at the climber and the shaft power at the wheel.  Probably the largest additional inefficiency is the conversion efficiency of the photovoltaic arrays themselves.  After the sections on the technology of the laser beam and the beam expander, there is a discussion of the technology of the PV array on the climber.  For now, we can estimate the conversion efficiency to be 40\%.  40\% of the amount of laser power that is intercepted by the PV arrays becomes electricity to the primary motors of the climber.  This 40\% is the reason that the red curves in the graphs below start at lower values than the power of the lasers just above the atmosphere.  For example, in Figure \ref{fig:1FEL1m}, the red curve starts at 1.37 MW.  This is 40\% of 3.42 MW, the power of the 5 MW laser above the atmosphere.  

New work on VEHSA PV arrays (shown in Section \ref{sec:PVarray}) may allow the efficiency number to go up in future, but as shown in Figure \ref{fig:laserconv}, not enough work has been done for multijunction cells sensitive to 500 nm light yet.

Another assumption to be made here is that there is a maximum output from any laser technology chosen.  The physics and thermodynamics of lasers put stringent limits on their rate of heat transfer and cooling.  The assumption here is that the maximum power from any single laser will be 5 MW on the ground.  If more than 5 MW of ground power is needed, then more than one laser will be needed.  See the end of the section on the Free Electron Laser, \ref{sec:FELplimit}, for a more detailed discussion of why the conceptual design of the FEL was limited to 5 MW.  

So to determine the aperture and number of lasers needed, let's look at graphs that show how the beam power declines with altitude when several inefficiencies are combined.  Let's start with one 5 MW laser transmitting a beam with a waist at z=0 of 1 m in radius.  A laser starting at 5 MW at ground level is immediately reduced by 31.6\% above the atmosphere from absorption.  This means that 5 MW on the ground is 3.42 MW about 100 km up the tether.

Figure \ref{fig:1FEL1m} shows that with a beam waist of 1 m, the climber never gets enough power to climb at 200 km/h.  The red curve is always below the blue curve.

\begin{figure}
\begin{centering}
\includegraphics[width=\textwidth]{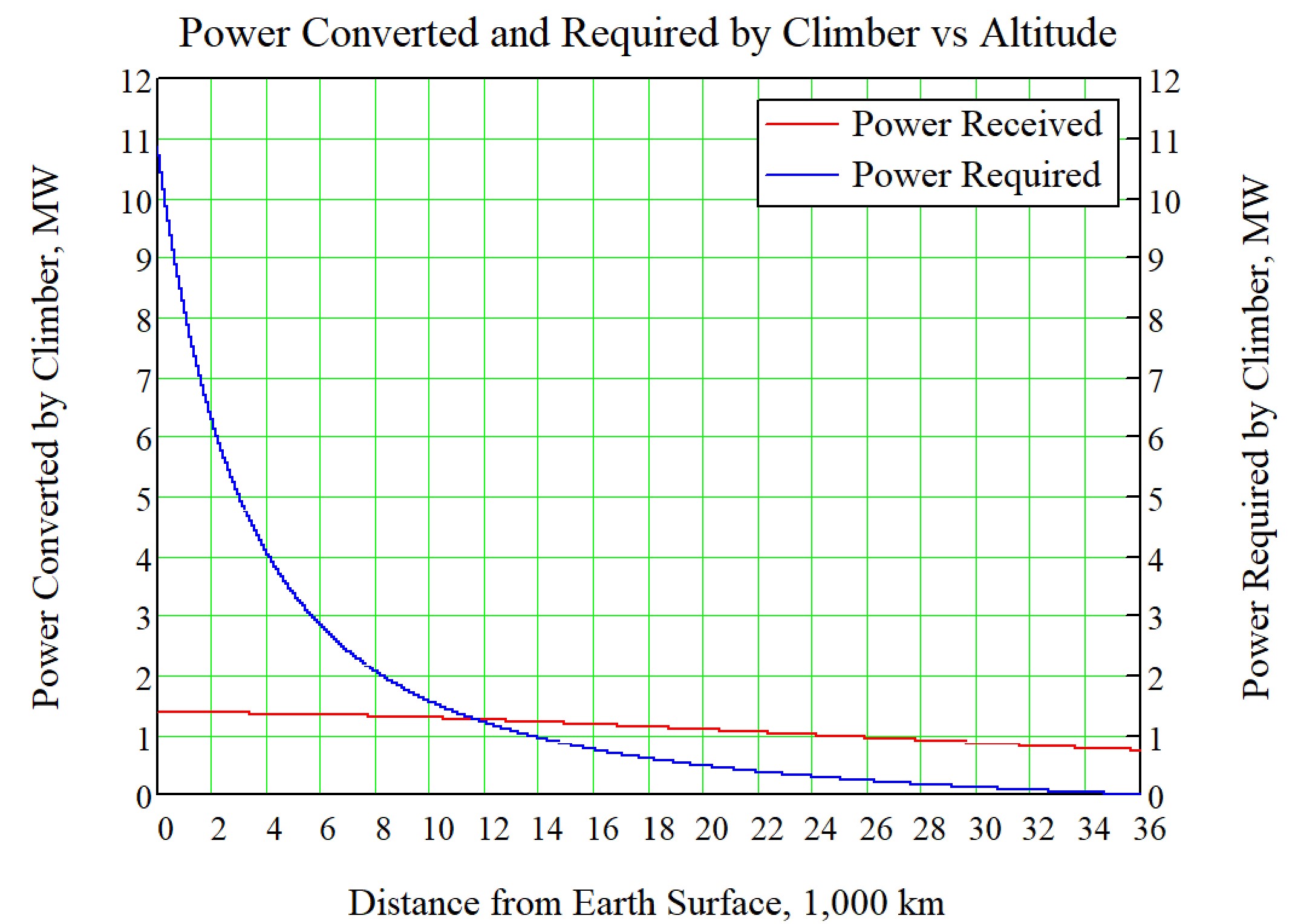}
\caption{\label{fig:1FEL25m} The blue curve is the power required by the climber to climb at 200 km/h. The red curve is the shaft power delivered by one 5 MW laser, 2.5 m waist, $\lambda$ = 0.5 $\si{\mu m}$.  The 5 MW laser at the ground is 3.42 MW above the atmosphere after the 31.6\% reduction caused by absorption in the atmosphere.  This arrangement only delivers enough power for the climber to climb at 200 km/h above 12,000 km.}
\end{centering}
\end{figure}

Figure \ref{fig:1FEL25m} shows that with a beam waist of 2.5 m, the climber only gets enough power to climb at 200 km/h above an altitude of 12,000 km.  This does not satisfy the requirement of attaining 200 km/h by 4,000 km up.

\begin{figure}
\begin{centering}
\includegraphics[width=\textwidth]{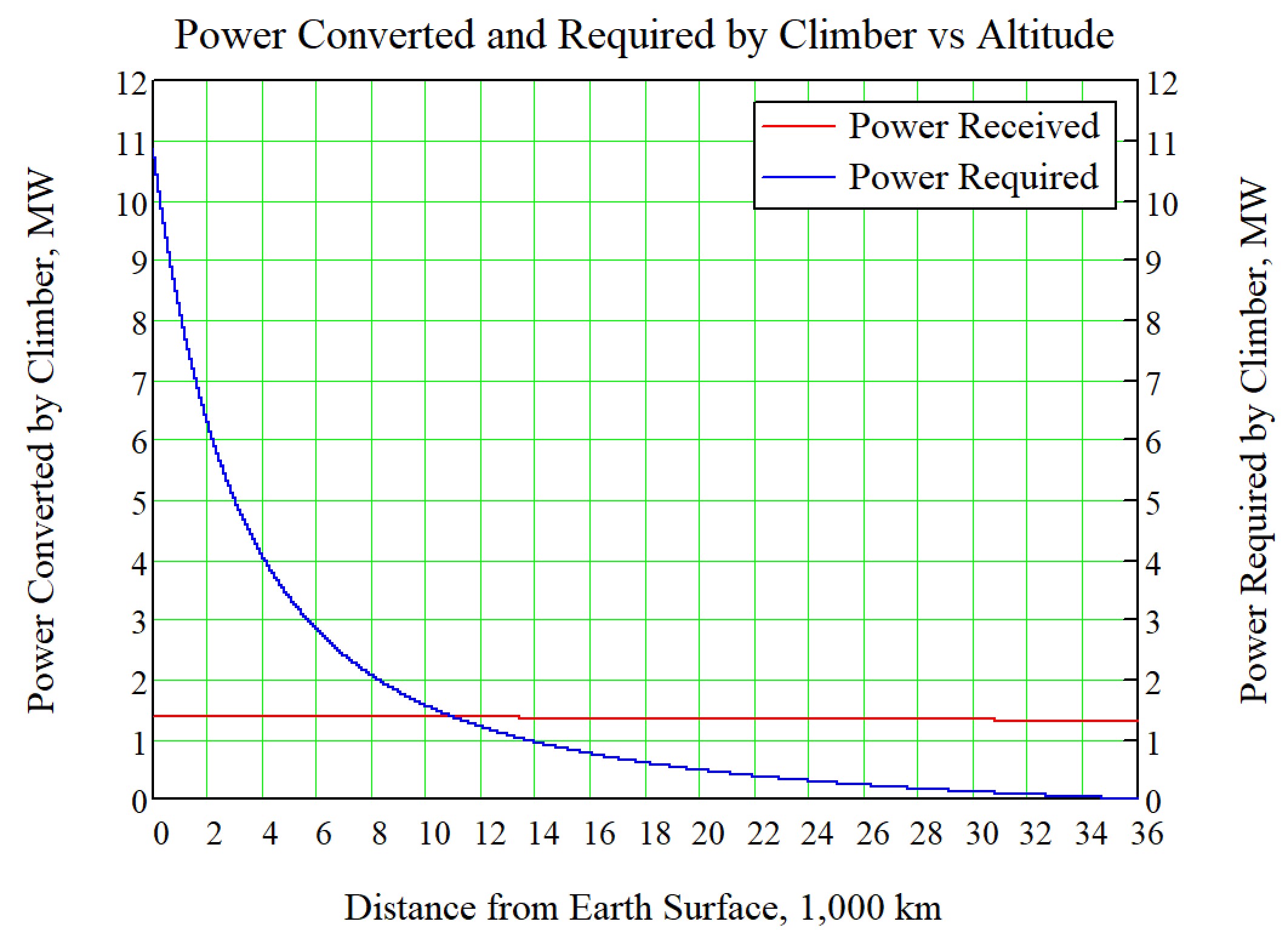}
\caption{\label{fig:1FEL5m} The blue curve is the power required by the climber to climb at 200 km/h. The red curve is the shaft power delivered by one 5 MW laser, 5 m waist, $\lambda$ = 0.5 $\si{\mu m}$.  The 5 MW laser at the ground is 3.42 MW above the atmosphere after the 31.6\% reduction caused by absorption in the atmosphere.  This arrangement only delivers enough power for the climber to climb at 200 km/h above 11,000 km.}
\end{centering}
\end{figure}

Figure \ref{fig:1FEL5m} shows that with a beam waist of 5 m, the climber only gets enough power to climb at 200 km/h above an altitude of 11,000 km.  This does not satisfy the requirement of attaining 200 km/h by 4,000 km up, and it is not much improvement for a laser transmitter that has a 10 m beam diameter.  This is the point of diminishing returns for laser aperture.  The next thing to do to get the climber going faster sooner is add another 5 MW laser.

\begin{figure}
\begin{centering}
\includegraphics[width=\textwidth]{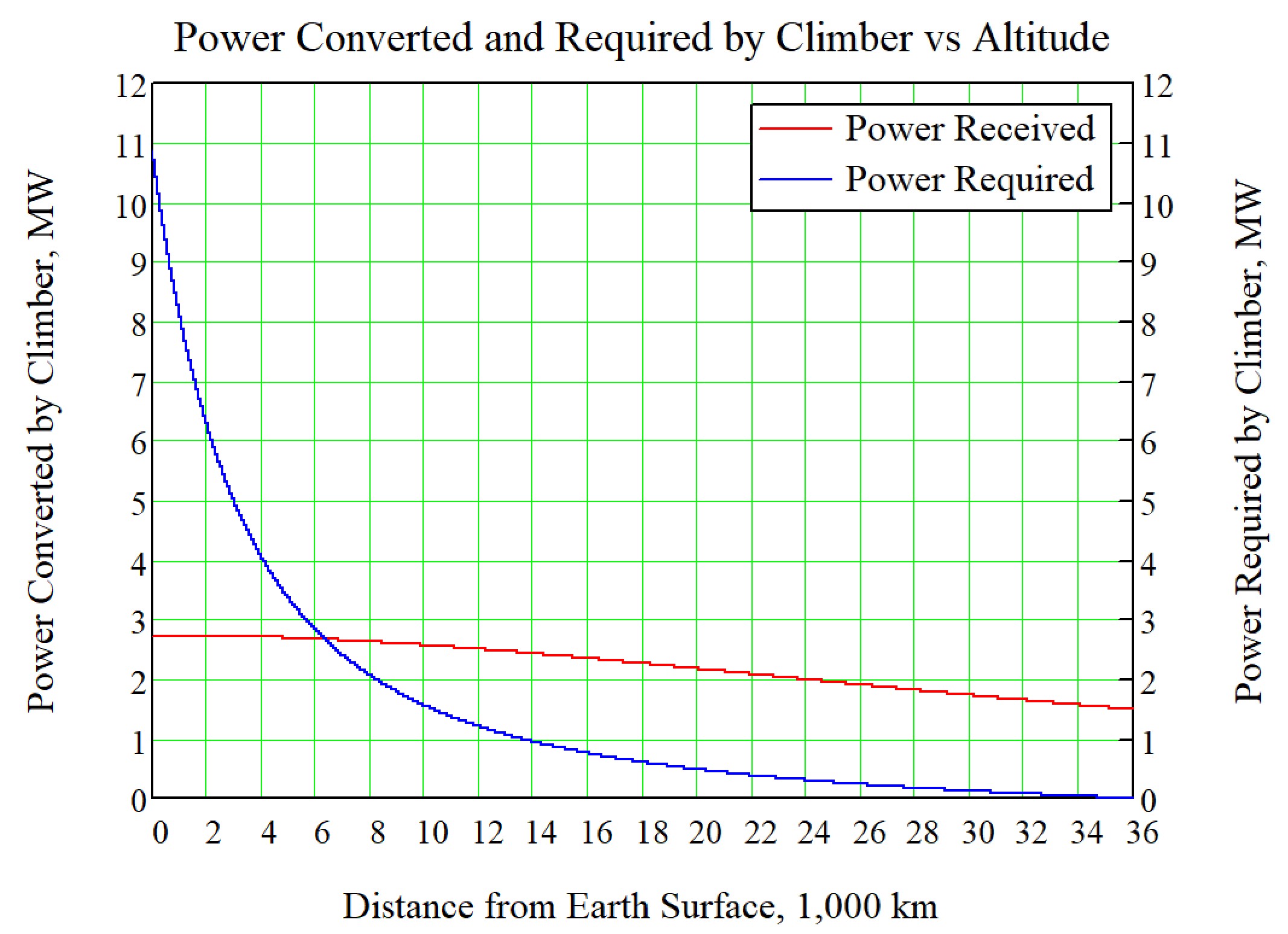}
\caption{\label{fig:2FEL25m} The blue curve is the power required by the climber to climb at 200 km/h. The red curve is the shaft power delivered by two 5 MW lasers, 2.5 m waist, $\lambda$ = 0.5~$\si{\mu m}$.  The 10 MW of laser power at the ground is 6.84 MW above the atmosphere after the 31.6\% reduction caused by absorption in the atmosphere.  This arrangement only delivers enough power for the climber to climb at 200 km/h above 6,500 km.  There is starting to be substantial marginal power at higher altitudes.}
\end{centering}
\end{figure}

Figure \ref{fig:2FEL25m} shows that with two 5 MW lasers and a beam waist of 2.5 m, the climber only gets enough power to climb at 200 km/h above an altitude of 6,500 km.  This does not satisfy the 4,000 km criterion, but it is starting to show substantial extra power at higher altitudes.  The blue curve can be modified to increase at higher altitudes by raising the speed of the climber, or the beam can be throttled back at higher altitudes for continued 200 km/h climbing.

\begin{figure}
\begin{centering}
\includegraphics[width=\textwidth]{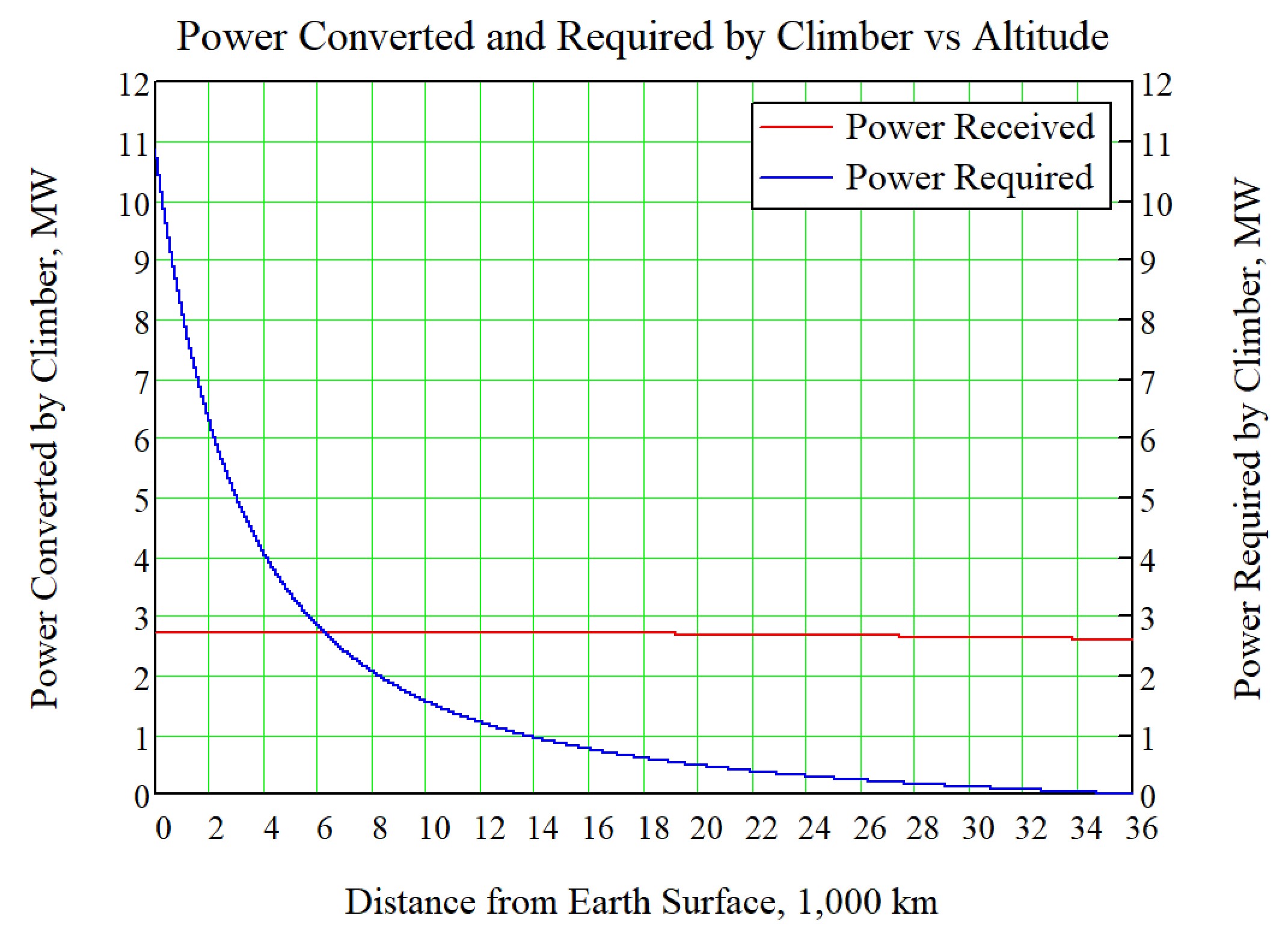}
\caption{\label{fig:2FEL5m} The blue curve is the power required by the climber to climb at 200 km/h. The red curve is the shaft power delivered by two 5 MW lasers, 5 m waist, $\lambda$ = 0.5~$\si{\mu m}$.  The 10 MW of laser power at the ground is 6.84 MW above the atmosphere after the 31.6\% reduction caused by absorption in the atmosphere.  This arrangement does not change the fact that the climber can only go 200 km/h above 6,500 km, but now there is a lot of extra power at higher altitudes.}
\end{centering}
\end{figure}

Figure \ref{fig:2FEL5m} shows that with two 5 MW lasers and a beam waist of 5 m, the climber still only gets enough power to climb at 200 km/h above an altitude of 6,500 km.  This does not satisfy the 4,000 km criterion, but there is now a substantial amount of extra power at higher altitudes.  The blue curve can only be modified so far to increase at higher altitudes by raising the speed of the climber (perhaps to 300 or 350 km/h at the most.)  The beam must be throttled back at higher altitudes to avoid wasting power into space.  Increasing the beam transmitter to deliver a 5 m waist beam is not the correct direction to go.  Let's see the effect of three 5 MW lasers with a 2.5 m waist beam.

\begin{figure}
\begin{centering}
\includegraphics[width=\textwidth]{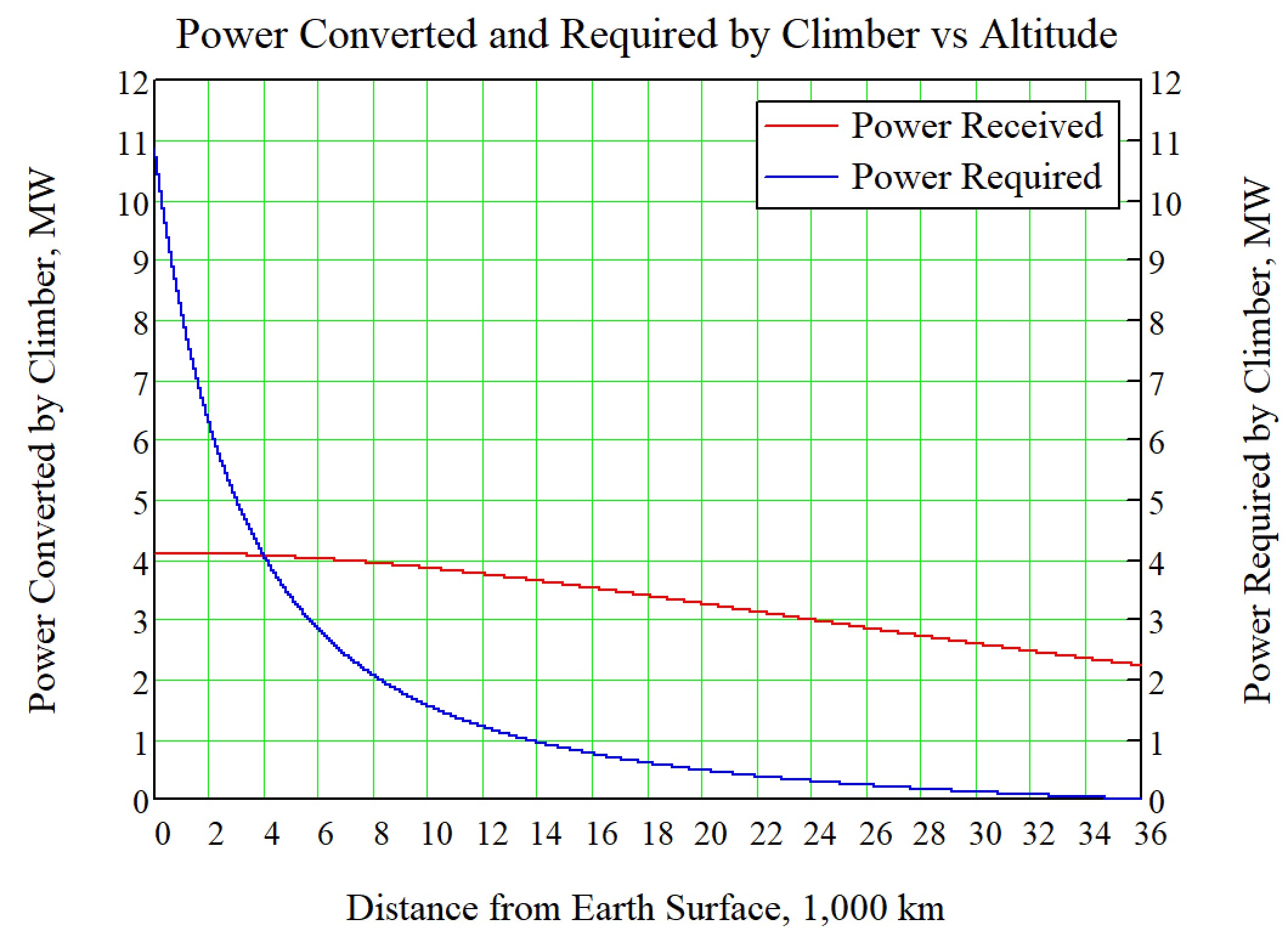} 
\caption{\label{fig:3FEL25m} The blue curve is the power required by the climber to climb at 200 km/h. The red curve is the shaft power delivered by three 5 MW lasers, 2.5 m waist, $\lambda$ = 0.5~$\si{\mu m}$.  The 15 MW of laser power at the ground is 10.26 MW above the atmosphere after the 31.6\% reduction caused by absorption in the atmosphere.  This arrangement is the first combination that allows the climber to climb at 200 km/h at 4,000 km and up. There is a lot of extra power at higher altitudes, so the solution is to power those higher climbers with scaled-back laser hardware.}
\end{centering}
\end{figure}

Figure \ref{fig:3FEL25m} shows that with three 5 MW lasers and a beam waist of 2.5 m, the climber finally gets enough power to climb at 200 km/h above an altitude of 4,000 km.  This satisfies the 4,000 km criterion, but there is now a substantial amount of extra power at higher altitudes.  If the speed of the climber were doubled above 9,000 km, the power required would not exceed the power available. The solution is to power the higher climbers with scaled-back laser hardware that is not delivering 15 MW at the ground.

\begin{figure}
\begin{centering}
\includegraphics[width=\textwidth]{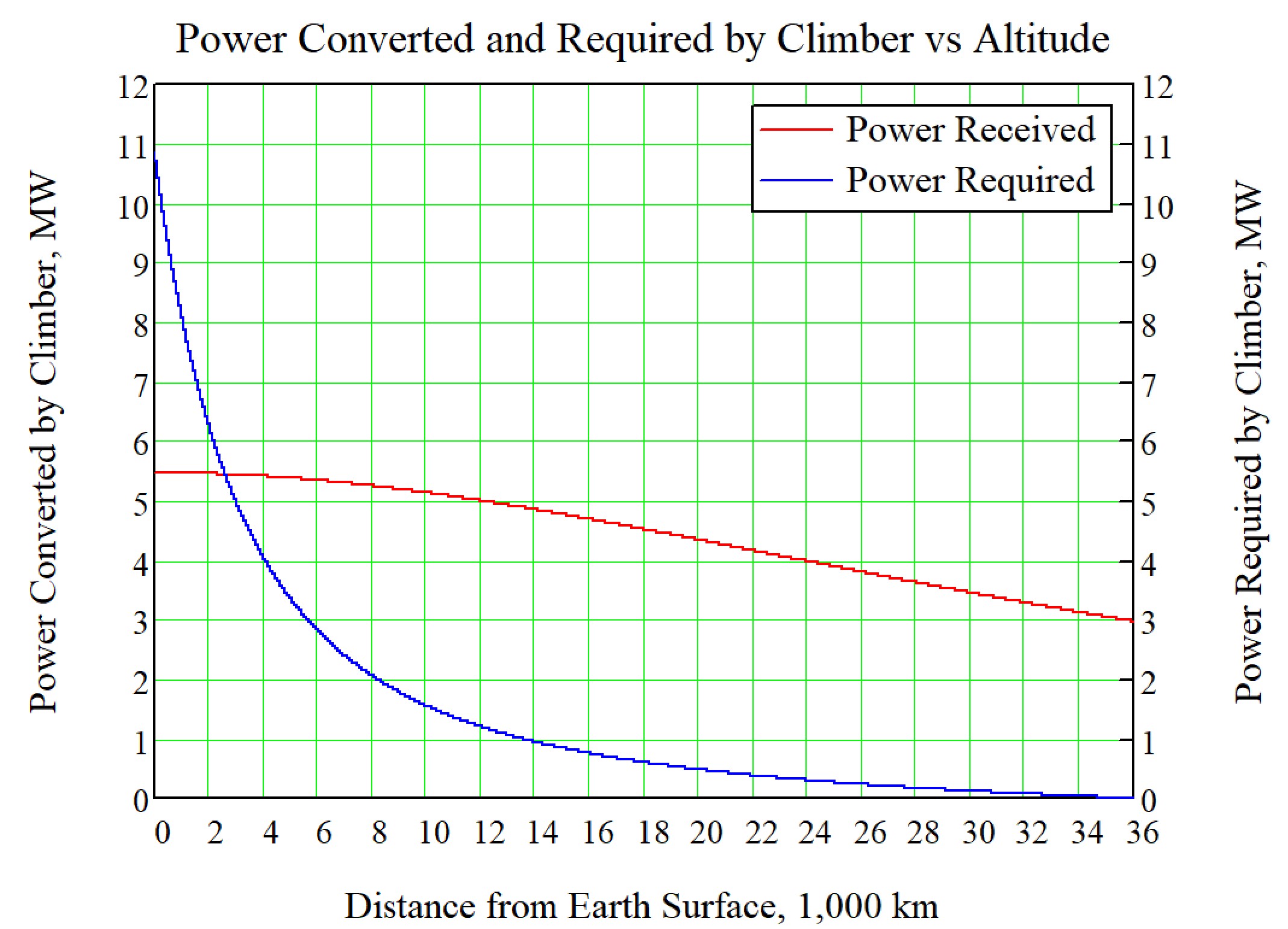} 
\caption{\label{fig:4FEL25m} The blue curve is the power required by the climber to climb at 200 km/h. The red curve is the shaft power delivered by four 5 MW lasers, 2.5 m waist, $\lambda$ = 0.5~$\si{\mu m}$.  The 20 MW of laser power at the ground is 13.68 MW above the atmosphere after the 31.6\% reduction caused by absorption in the atmosphere.  This arrangement allows the climber to climb at 200 km/h just below 3,000 km and up. Again, there is a lot of extra power at higher altitudes, so the solution is to power those higher climbers with scaled-back laser hardware.}
\end{centering}
\end{figure}

Figure \ref{fig:4FEL25m} shows that with four 5 MW lasers and a beam waist of 2.5 m, the climber gets enough power to climb at 200 km/h just below an altitude of 3,000 km.  Having this much laser power available at the beginning of a climb could reduce the total climb time even if the speed is not increased above 200 km/h.

\newpage
\subsubsection{Conclusions about laser power and beam size}

The preceding graphs comparing the required power to climb at 200 km/h and the available power from different combinations of lasers and beam waists leads to the conclusion that a 1 m beam waist is too small to deliver power to the climber efficiently because it diverges too quickly.

By similar reasoning, a single 5 MW laser beam, even put through a 5 m waist transmitter, does not deliver enough power to the climber to get it climbing at 200 km/h at the required 4,000 km.  Increasing the beam waist does not lower the altitude of sufficient power to climb at 200 km/h, it only increases the power margin at higher altitudes.

It is only with a minimum of three 5 MW lasers and a 2.5 m beam waist that the climber is able to climb at 200 km/h at 4,000 km up the tether.  Four lasers allow the climber to reach 200 km/h at an altitude just below 3,000 km.

It becomes an economic analysis of the trade-off between the number of 5 MW lasers and reduced transit time to GEO.  This optics analysis drives the conceptual design of the free electron laser (see section \ref{sec:FELdesign}) and 2.5 m beam waist expander (section \ref{sec:beamexpand}).

\subsection{Candidate laser technologies and the most likely laser technology}

The candidates in the following sections are all capable of producing light at or near 500 nm in wavelength.  [\cite{greenlasers}]  Basic requirements for the laser technology selected are that it be capable of CW running at 5 MW output (or less) for days or weeks at high efficiency and reliability.

All but one of the lasers described below involve inverting populations of atoms in solid, liquid, or gaseous media to produce the lasing effect.  To date, the highest power lasers are military lasers in the range of 500 kW.  These lasers are not CW, not high enough beam quality to reach climbers near GEO, and typically emit in wavelengths that are only capable of reaching targets tens of kilometers away.  Militaries have different goals for high-power lasers than the high-power lasers needed for power beaming to climbers.  All but one of the lasers described below have limits to their maximum power based on their ability to extract heat from the laser media.

\subsubsection{Argon ion lasers}

The maximum power of argon ion lasers is about 20 W, but their efficiency is very low at 0.1\%.  They are based on amplification in an argon plasma made by electrical discharge. [\cite{argonlasers}] Their low efficiency disqualifies them as a serious contender for power beaming.

\subsubsection{Copper vapor lasers}

Copper vapor lasers (CVL) can emit relatively high powers at 510.6 nm. They are based on a pulsed discharge in copper vapor and emit nanosecond pulses. [\cite{greenlasers}]  The average power of copper lasers is between 25 W to 2 kW. [\cite{copperlaser}] Pulsed lasers cause damage to photovoltaic cells, so even if it could be scaled up, CVLs are not a good choice for power beaming.

\subsubsection{Erbium-doped upconversion lasers}

Upconversion lasers have laser wavelengths that are shorter than their pumped wavelength.  Erbium-doped crystals such as Er:YLiF$_4$ and Er:LiLuF$_4$ generate tens of watts of power in green light at room temperature.  Upconversion reduces the efficiency of the laser.  Their low power and efficiency makes them unsuitable for power beaming.  

\subsubsection{Frequency doubled lasers}

There are various types of frequency-doubled solid-state lasers, where the actual laser emits in the 1 $\micro$m spectral region and a frequency doubler (based on a nonlinear crystal material such as LiNbO$_3$, LBO or KTP) converts this radiation into green light. The frequency doubling process can take place either outside the laser or within the resonator. Most common is an output wavelength of 532 nm, obtained by doubling 1064-nm radiation from a neodymium-doped 1064-nm Nd:YAG laser or Nd:YVO$_4$ laser, and 515 nm from doubling the output of a 1030-nm Yb:YAG laser.  [\cite{greenlasers}] Frequency doubling drastically reduces the efficiency of the laser system, making it unsuitable for power beaming.

\subsubsection{Dye lasers}

Liquid dye lasers can emit green light and can dissipate heat faster than solid state lasers, but dye lasers face problems like rapid photodegradation (bleaching), leading to short lifetimes, and handling hazardous, often carcinogenic, toxic dyes and flammable solvents. They also suffer from thermal effects, causing refractive index changes and output instability, plus require bulky pumping systems and constant dye circulation, making them complex and less commercially viable than solid-state alternatives.  The drawbacks to dye lasers make them unsuitable for power beaming.

\subsubsection{Free Electron Lasers}

A fundamentally different kind of laser is the Free Electron Laser (FEL). FELs start with high power beams of electrons circulating in a particle accelerator.  By changing the direction of travel of the electrons over very short distances and time scales with strong magnets called "wigglers" the electrons give off synchrotron light as a laser beam. Section \ref{sec:FELdesign} describes a new conceptual design for a 5 MW FEL capable of delivering power to climbers on the space elevator at any altitude up to GEO.
    
\subsubsection{Combining multiple small lasers for larger beam power}

The state of the art in power for military lasers is approximately 500 kW, sufficient to knock down a drone, but insufficient to knock down a missile.  Laser researchers have been trying to solve the problem of combining multiple small lasers to produce a megawatt-scale laser for years, but the problem is difficult and still not completely solved.  One class of solutions to the problem is referred to as \textit{coherent beam combining}, and is discussed in reference [\cite{brignon2013coherent}].  None of the architectures proposed in that book are suitable for multimegawatt beams.

\begin{figure}
\begin{centering}
\includegraphics[width=\textwidth]{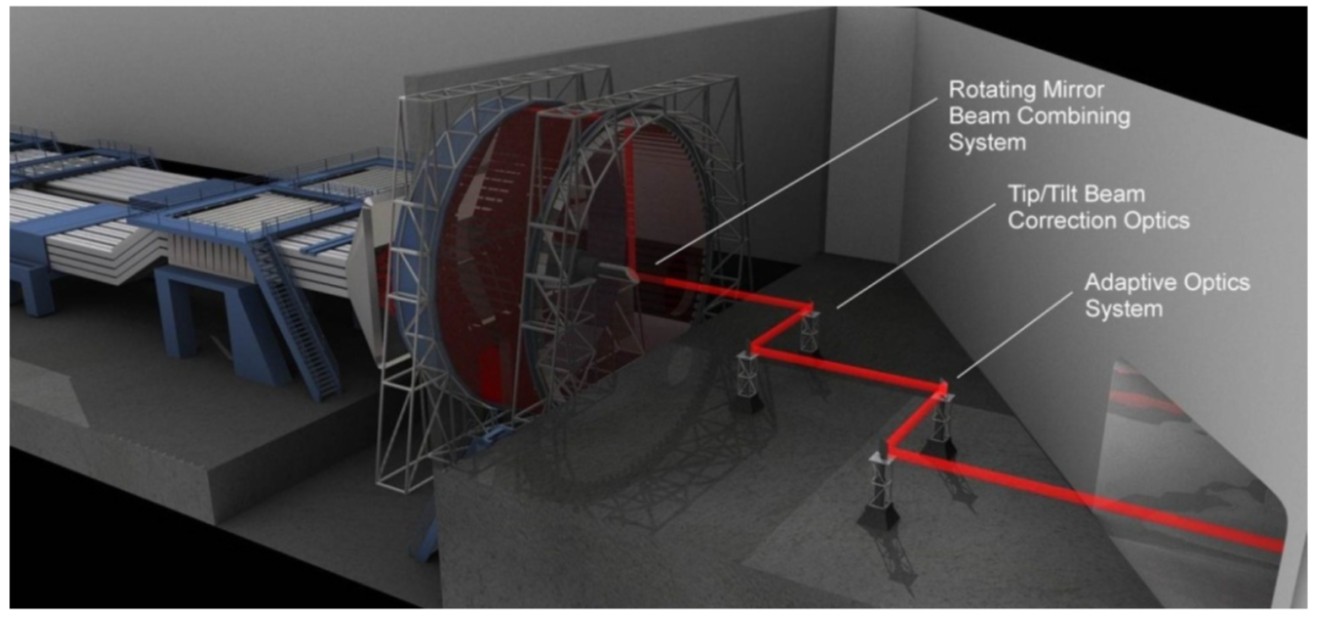} 
\caption{\label{fig:BEPlaser} Figure from Ref [\cite{BEP}]. "The boost laser consists of a continuous or a pulsed 140-MW (Figure 2.9) output laser constructed of up to 2900 (continuous or pulsed), 120-kW commercial-off-the-shelf (COTS) modules. Two options exist for the “main” laser: (1) a ground system of up to 7250 COTS lasers (20 kW each) or (2) a space-based relay station that utilizes the boost laser facility."}
\end{centering}
\end{figure}

NASA put out a study in 2012 on Beamed Energy Propulsion using very high-power lasers to drive spaceships from Earth to LEO and from LEO to GEO [\cite{BEP}]. The technologies reviewed in the study included thermal propulsion using beamed power and electric (plasma) propulsion using beamed power.  As seen in Figure \ref{fig:BEPlaser}, the power levels of the beams to orbit were far higher (28X) than those discussed in this report.  The figure shows a concept of how to combine the outputs of thousands of 120 kW commercial lasers.  The output of the "Gatling gun" of lasers was put through a 3 m telescope.  The beam combining was done temporally using the pulsed output of the lasers, not spatially.  Spatial beam combining is the more difficult technology discussed in reference [\cite{brignon2013coherent}].

The inherent problem of spatial beam combining is that each of the small lasers starts with a small beam. To hit the receivers on the climber, they must all go through beam expansion to be expanded up to the required 2.5 m beam waist.  The small laser beams can't present as 42 separate beams with slightly different trajectories going through a single beam expander (as might be the case with a Gatling-gun-type solution).  The angle differences would become magnified and the output beam would diverge too quickly.
The most simple-minded solution to make a 5 MW beam from 42 120 kW beams is to make 42 7.5 m diameter offset Cassegrain beam expanders.  This is clearly not a cost-effective solution compared to a single 5 MW laser and a single 7.5 m beam expander.  It remains to be seen whether coherent beam combining can create a 5 MW beam from many small lasers, which can then be delivered through a single beam expander.  As with every engineering project, the cost trade-off between a single 5 MW laser and multiple smaller-scale COTS lasers determines the final technology selection.

\clearpage
\newpage
\subsection{A conceptual design of a Free Electron Laser for powering ascending climbers} 
\label{sec:FELdesign}
\subsubsection{Abstract}

Free Electron Lasers (FELs) represent a possible radiation source for power beaming applications due to their inherent capability to produce any desired wavelength and the potential for efficient, high average power production. The paragraphs below represent a scoping study of what such an FEL system might look like. Rather than a conceptual design, the system presented should be viewed as a starting point for a serious design with present technology status guiding the system choices.

\subsubsection{Introduction}

FELs have been proposed as high-power laser candidates almost from inception in the mid-1970s. They formed the basis of major Strategic Defense Initiative efforts through the late 1980s. Despite this fact, the highest power generated by an FEL remains at 14 kW [\cite{neil2000sustained}], perhaps more from a lack of the sponsor’s appetite for the required investment in driving power than any fundamental technology limit. After all, circulating electron beams presently operate with average power exceeding 1.5 GW [\cite{BNLSLS}], so even a modest conversion of electron beam power to light could result in massive laser output.
For the purpose of this note, we will assume a desired output power in the 5-10 MW range with a wavelength of 500 nm in the visible for high photovoltaic conversion efficiency. The FEL physics guides the electron beam specifications while we rely on experience in the field for system designs to support production of the required electron beam parameters.

\subsubsection{System level design}

The overall accelerator system configuration is chosen as a superconducting RF (SRF) accelerator utilizing same-cell energy recovery.  The SRF choice is driven by the need for CW operation to achieve high average power.   The technology is mature with commercial vendors able to supply components. Figure \ref{fig:JLAB_SRF} is a photo of an SRF cavity from the free electron laser built at the Thomas Jefferson National Laboratory.  Same-cell energy recovery (ER) offers high potential system electrical efficiency since the waste beam from the FEL is sent back through the accelerating cavities 180 degrees out of phase.  The beam is decelerated instead of accelerated and its power is now available to accelerate subsequent electrons.  With SRF technology the loss in such conversion is inconsequential.  In addition to the efficiency advantage, ER reduces the required (expensive) RF power by more than an order of magnitude.  It also eliminates radioactivation by the dump beam and activation of ground water since the beam can be terminated below 10 MeV, the neutron photoproduction threshold.

\begin{figure}
\begin{centering}
\includegraphics[width=\textwidth]{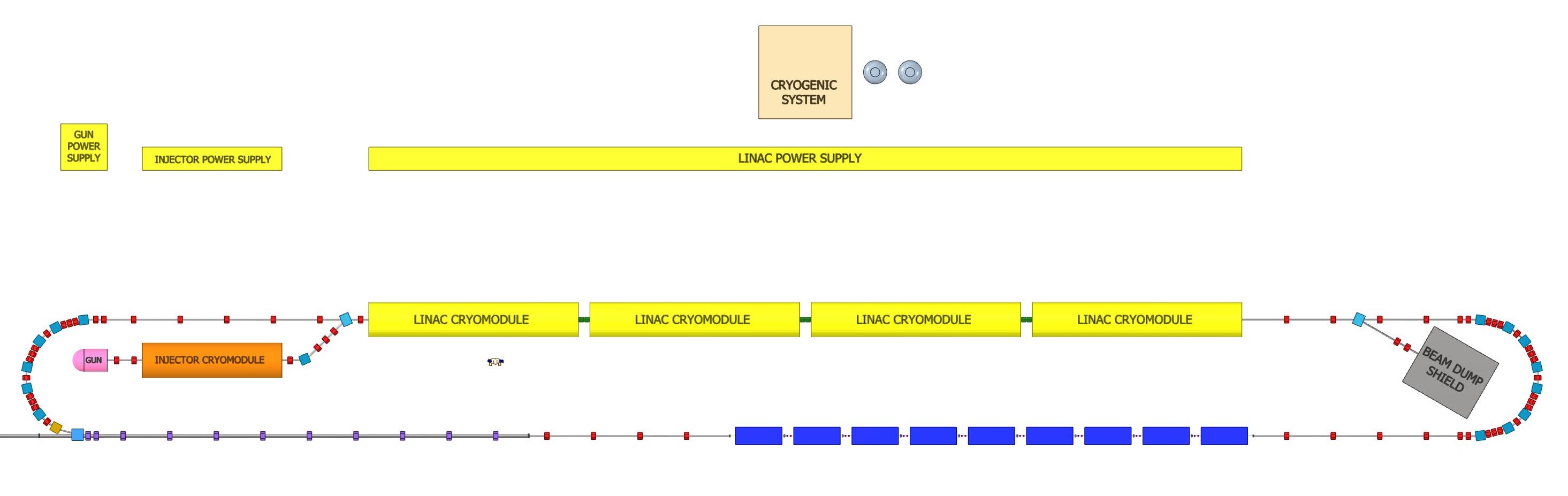}    
\caption{\label{fig:FEL-01} Top view of the CAD model of the concept electron accelerator for the Free Electron Laser.  The length of the straight sections is 60 m.}
\end{centering}
\end{figure}

Figure~\ref{fig:FEL-01} shows a top view of the conceptual design for the FEL.  The dark blue boxes in the lower straight section are wigglers that produce the laser light.  The yellow rectangles in the upper beam line are superconducting RF cavities (SRF) arranged as a linear accelerator (linac) that accelerates the beam going into the wigglers up to the required energy, and decelerates the beam exiting the wigglers to recover its energy.  The light blue boxes at every point where the beam changes direction are dipole magnets.  Red boxes are quadrupole magnets.

5 MW of laser light exits the accelerator at the lower left of the picture in a vacuum tube with a cooled sapphire window at the end.

We choose 5\%-10\% as a nominal FEL beam conversion efficiency to photons in the FEL.  This is above most FEL systems but not by a large factor [\cite{zen2020high}]. Because of ER the wall-plug efficiency will be substantially higher; we give an estimate later in this note.    If the conversion efficiency is 5\% and we want 5 MW that means we need 100 MW of average electron beam power.   The electron beam energy must be sufficient to produce 500 nm light with wigglers of practical dimensions, perhaps 200 to 500 MeV. We choose 300 MeV as a nominal operating point.  (Higher energies have higher beam power but lower FEL gain.)  Since we need 100 MW of beam power that means we need to produce 0.33 A of average current.  This average current is produced as a train of micropulses at up to the fundamental frequency of the accelerator’s RF cavities.  SRF cavity designs exist in the 500 MHz to 1500 MHz range with the lower frequency cavities being bulkier but having more stored energy to handle high average currents.  A 750 MHz cavity design seems reasonable which results in 440 pC charge per bunch at 750 MHz.

The overall system consists of an injector which initiates the relativistic electron beam, the linac to accelerate the electrons to the desired energy, a beam transport system to steer, condition, and control the electrons both accelerating and decelerating, the FEL itself, the electron beam dump to dispose of the end beam, and an optical transport system to condition and transport the optical beam to the projection mirror.  The sections below discuss each of these systems in turn.

\subsubsection{Injector}

The injector must produce the train of 440 pC bunches at 750 MHz at an energy of around 10 MeV, sufficiently relativistic that space charge forces are no longer significant in the beam transport.  The injector consists of a gun, a set of accelerating cavities, and transport magnetics.  Presently there are two gun technologies worth consideration, both are photocathode sources.  One possibility is a low frequency room temperature RF cavity photogun.  Present systems operate below 200 MHz which is incompatible with our 750 MHz pulse train.  The most mature approach is the DC photogun, having shown 75 mA average current at Cornell [\cite{loehl2010high}].  Guns typically operate with a DC supply in the 300 to 500 kV range and use a modelocked green or UV laser on a photocathode wafer to produce the 4-10 ps electron micropulses.  Typical quantum efficiencies are around 2\% so to produce 0.3 A around 50 W of drive laser power would be needed at 350 nm. 

Once extracted from the gun, the electron bunches are then accelerated in a series of SRF cavities, again at 750 MHz, though these cavities do not participate in the energy recovery process so the full accelerating power would need to be provided by RF supplies. We envision 10 single cell (or 5 double cell) SRF cavities in the injector, each accelerating the beam by 1 MeV or 330 kW/cavity of RF.  Total RF power required is 3.3 MW plus 10\% overhead for control.   Using solid state power the DC to RF conversion could be about 55\% efficient for approximately 5.5 MW power draw.

This specification keeps the RF power supplies and fundamental power couplers in each cavity at a large but manageable level.   It is conceivable to use two power couplers per cavity if desired to reduce the load on each coupler.  The 10 each, 20 cm long (plus entrance and exit couplers), single cell cavities could be mounted in a single or a pair of cryogenic dewars with transport magnets placed at necessary intervals until merging into the recirculating accelerator transport line of the main accelerator.

\begin{figure}
\begin{centering}
\includegraphics[width=\textwidth]{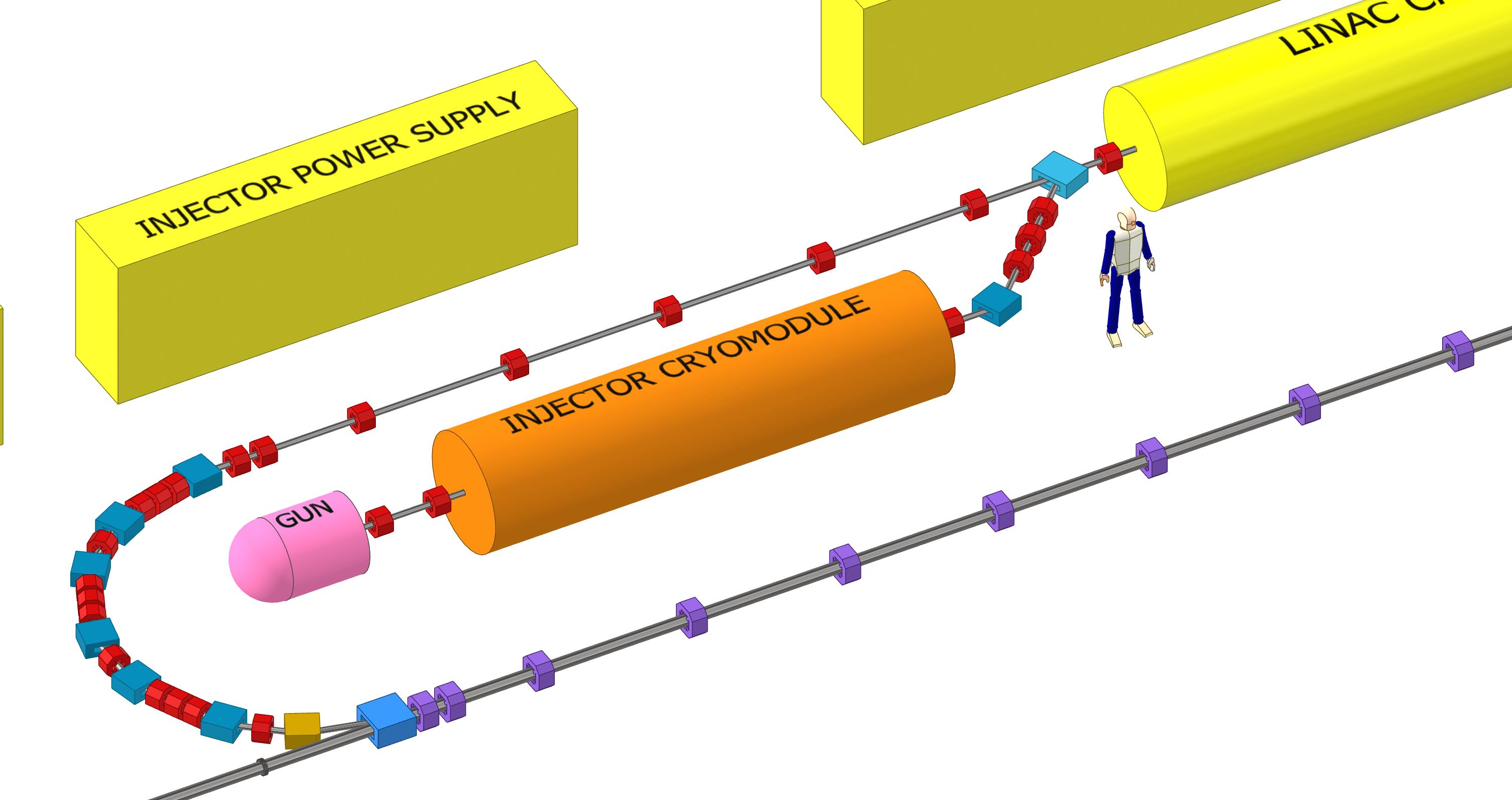}    
\caption{\label{fig:FEL-02} Closeup of the gun, injector cryomodule and beam steering and focusing elements that inject electrons into the linac.}
\end{centering}
\end{figure}
\newpage
\subsubsection{Beam transport} 

The accelerator could be laid out as a single linear length with an arc to transport beam around to the FEL followed by another bend, back through the accelerator (decelerator) then into the dump. Conceivably, it may be more advantageous to split the accelerator into two parallel parts with an arc in between, another higher energy arc to the FEL, an arc to the second half of the accelerator for initial deceleration and a final arc to the first half of the accelerator and subsequent dump.  Although this configuration requires four magnetic arcs rather than two, it provides the beam transport designer with desirable flexibility in the manipulation of the electron phase space.  The high-energy arc radius would be less than 5 m even utilizing conventional water-cooled magnets.

The electron bunches are kept long (ps) until just before injection into the FEL where the bunch must be short with high peak current for FEL gain and efficiency.  The bunch length is kept long during acceleration to minimize coherent synchrotron radiation instabilities and beam degradation in the bends which would render the electron beam quality insufficient to provide efficient lasing. The bunch is manipulated by accelerating the beam off the RF crest and providing an energy slew to the microbunch.  Then, through an energy-dependent-path-length transport, the bunch is compressed as it enters the FEL.  Several options exist for the design of the arc transport.  Resolution of the optimum would require beam transport modeling to validate the ability to transport such a high charge beam while maintaining beam quality.  Candidate arc approaches include the reverse bend, which MIT Bates used at Jefferson Lab, and the FFAG utilized at Cornell.  Other options may also be considered.

The electron bunch acquires a sizable (15\%-20\%: 45 to 60 MeV) energy spread as well as a 5\% (15 MeV) lower centroid energy in the FEL.   If nothing were done, this 60 MeV energy spread would be 6 times the beam energy when dumped at 10 MeV, an intolerable situation.  So the inverse of what was done during acceleration is performed.  The beam is decelerated off-trough and the higher energy electrons are decelerated more than the lower energy electrons.  The energy spread is compressed as the deceleration takes place.  The control required is exacting, but once set up properly, it is robust and tolerant of beam variations.  Nevertheless, the magnetic transport does have to handle comparatively large energy spreads without intercepting ANY of the beam.   Less than 1E-8 of the electron beam can be allowed to touch a solid pipe wall at any point during any of the transport process or the deposited energy would melt through the beam pipe letting the system up to air.  This was the normal operational level of containment in \cite{neil2000sustained}.  Magnet powers would not be significant in the overall system, perhaps 400 kW even using conventional copper-wound iron yoke magnets.

\subsubsection{Accelerator}

The accelerator must bring the beam up (and down) from 10 MeV to 300 MeV in a stable manner.  The SRF accelerator cavities could run at nominally 15 MV/m as assembled collectively in a dewar termed a cryomodule.  If the SRF cavities had 5 cells each at 750 MHz and there were 6 cavities per module with a filling factor of 0.7 then 4 cryomodules each 9 m long would form the accelerator.  In the linear sections, a 2 m spacing between them would allow for diagnostics and transport focusing.  Although energy recovery would supply most of the power needed for the beam, perhaps 40 kW/cavity would be needed to maintain stability. The lasing power of the FEL must come from the linac since the beam power from the injector and that to the dump roughly cancel. For 5 MW out of 24 cavities, 208 kW are required per cavity.  This results in beam and control power of 6000 kW RF and 11,000 kW of AC power draw. 

The injector and accelerator cavities, if made of pure niobium as is standard today, would operate at 2K supplied from a helium refrigeration plant.  Technologies are in development for coating cavities with niobium nitride or Nb3Sn which may permit operation at 4K vastly simplifying the refrigeration plant requirements and more than doubling electrical efficiency.   With today’s pure Nb cavities, total cavity loss and parasitic power to the cryogen might be 600 W leading to a 0.6 MW electrical draw for the power plant at nominally 1000W electric per 2KW deposited.

\begin{figure}
\begin{centering}
\includegraphics[width=\textwidth]{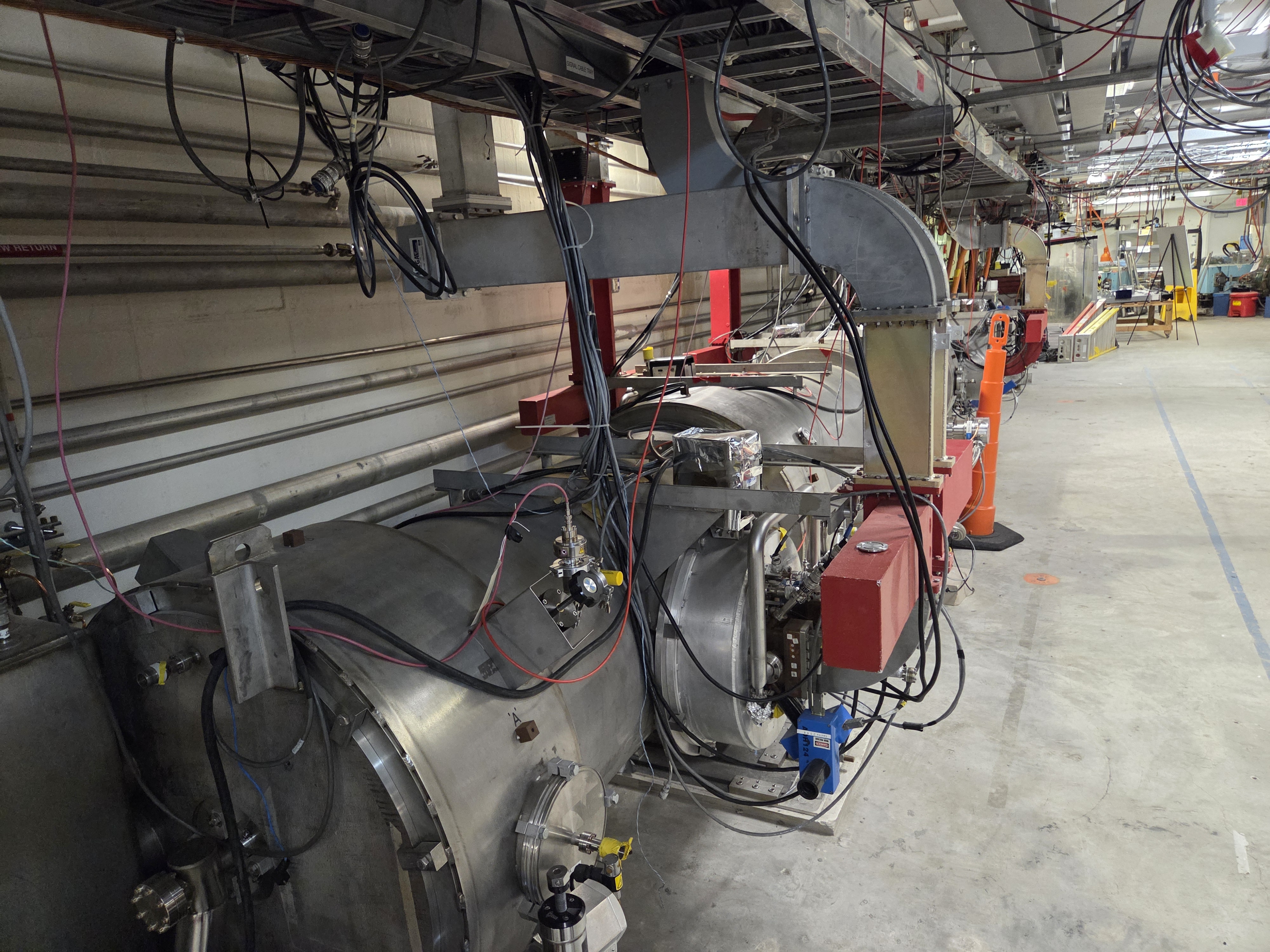}    
\caption{\label{fig:JLAB_SRF} Photo of an SRF cavity in the Jefferson Laboratory FEL.}
\end{centering}
\end{figure}
\FloatBarrier

\subsubsection{FEL}

Having discussed ebeam production, we now convert its power into photons via the FEL.   Several choices present themselves as to the actual FEL configuration:  Self-Amplified Spontaneous Emission (SASE) [\cite{fawley1997output}] where the beam noise is amplified to saturation in a long wiggler, regenerative amplifier FEL (RAFEL) [\cite{nguyen2000regenerative}] where a small portion of the output beam is used to seed the amplification process, and (if a wavelength is chosen where existing lasers operate) a power amplifier (PA).   While the RAFEL and PA might offer advantages of shorter wiggler and improved stability we will analyze the SASE system as a limiting case for requirements.  It is also the approach with a large experience base from X-ray FEL facilities. A pure high-power FEL oscillator approach is rejected because of extreme mirror thermal loading.

In all systems, the FEL would consist of a wiggler with an exponential gain section followed by a tapered section for high extraction efficiency.   Proper estimation of the FEL performance should really be done on the specific 6-D electron phase space as produced by the injector and beam transport.  For the purposes of this note, we use a popular scaling developed by MIng Xie [\cite{xie1995design}].   From the FEL resonance condition at 300 MeV we get a wiggler wavelength of 72 mm at a $K_{rms}$ of 1.95.  This yields an easily achievable peak field of 4.1 kG in the permanent magnet wiggler.  The wiggler gap can be 35 mm; it is good to provide a generous aperture with such a high average current though it should be noted that the wiggler beam pipe will still need active cooling to control resistive wall heating.  The wiggler is built in roughly 2 m sections with beam focusing and diagnostics in between.  The last few wiggler sections need to be capable of tapering by adjusting the wiggler gap.  

We assume a slice energy spread of 0.1\% or 300 keV. At the assumed bunch charge of 440 pC a normalized emittance as preserved from the injector is projected to be 1.6 mm-mrad.   Both the emittance and energy spread are sufficiently low as to only minimally affect the gain and saturation power; the FEL would be tolerant of degradation in either.  The gain length from the [\cite{xie1995design}] scaling is 0.64 m and the power grows from 1 W of noise to saturation in 15.3m of wiggler.  The untapered FEL saturates at an extraction of 0.65\%, a peak power of 6.5 GW in the micropulse.  We begin the wiggler taper slightly before saturation, perhaps with 1 GW peak power in an optical mode waist of 240 microns (Rayleigh range 0.36m).  The KMR scaling [\cite{kroll1980free}] for the tapered section predicts an extraction of .081/m (8.1\%/m) of the trapped electrons.  If we assume 50\% of the electrons participate in the trapping then we can achieve 5\% extraction (5 MW) in 1.2 m of tapered wiggler, 10\% (10 MW) in 2.4 m.   As mentioned above, these numbers are only rough estimates.  A study utilizing a modern FEL simulation would be required to get more exacting answers, as well as give guidance on the induced energy spread.

\subsubsection{Beam dump}

After the electrons are successfully decelerated down to below 10 MeV while compressing their energy spread, the beam is sent into a beam dump consisting of a water-cooled plate.  In order to prevent burn-through the beam is actively rastered on the dump face. The 3.3 MW+ of beam power needs to be distributed over an area exceeding 0.5 \si{m^2} to safely dispose of the power.  A significant level of X-rays are produced which are managed by local shielding with minimal activation.  Figure \ref{fig:JLab_beam_dump} shows the much smaller beam dump of the Jefferson Laboratory FEL.

\begin{figure}
\begin{centering}
\includegraphics[width=\textwidth]{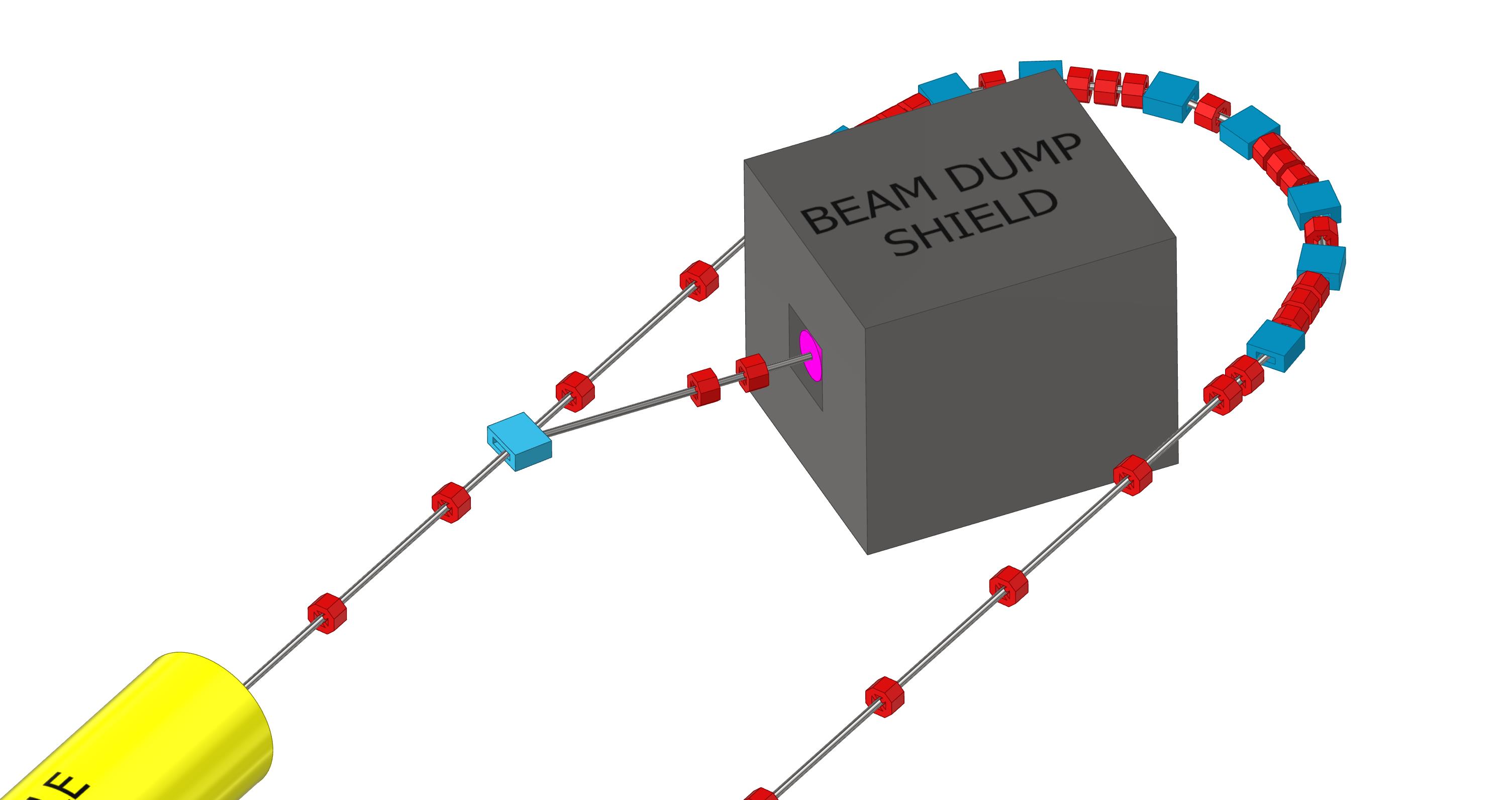}    
\caption{\label{fig:FEL-03} Closeup of the beam dump surrounded by 200 tons of steel shielding.  The dump is surrounded by 1 meter thick steel.}
\end{centering}
\end{figure}

\begin{figure}
\begin{centering}
\includegraphics[width=.8\textwidth]{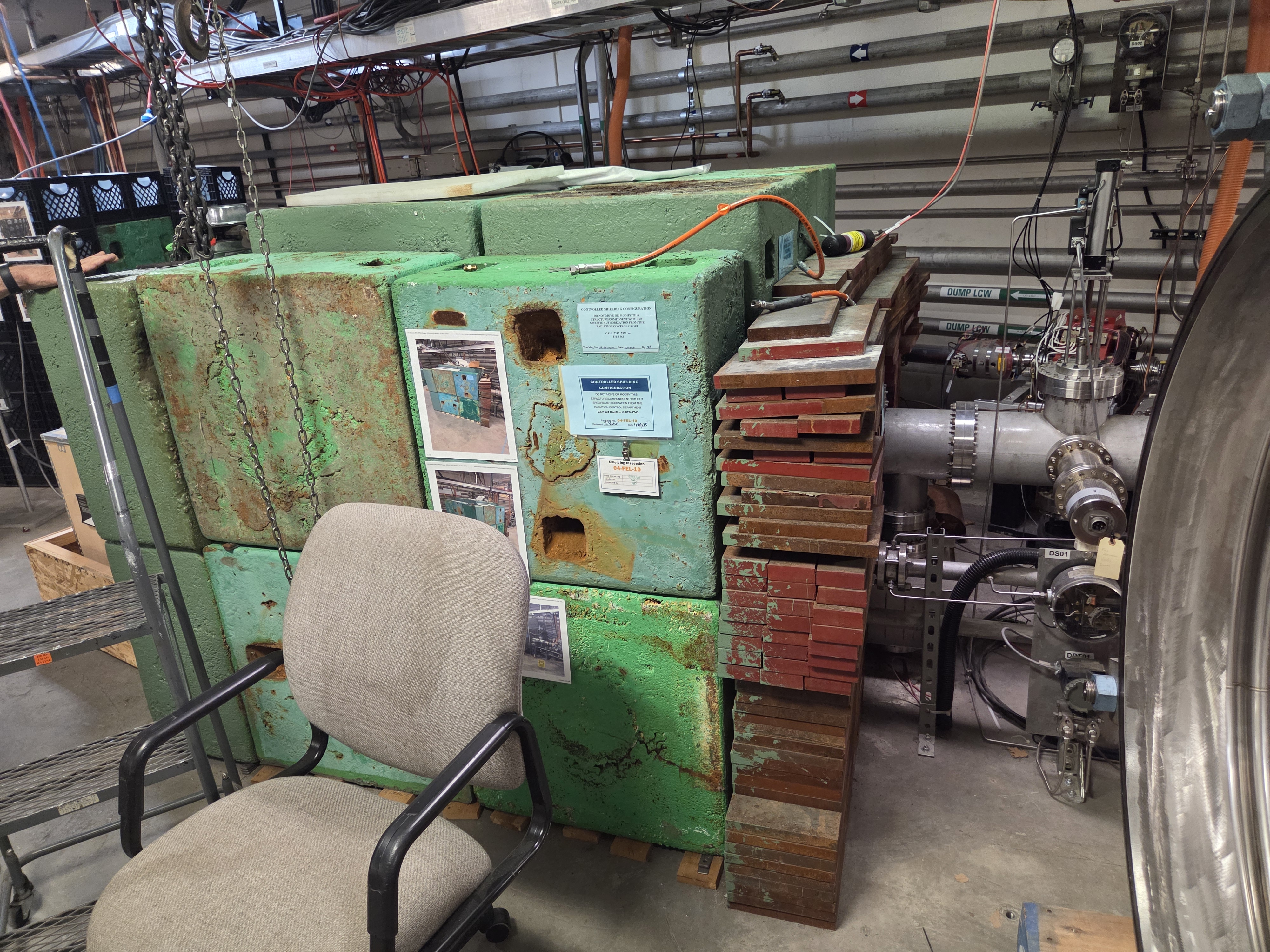}    
\caption{\label{fig:JLab_beam_dump} Photo of the beam dump in the Jefferson Laboratory FEL.  This FEL was much lower in energy than the conceptual design presented here.}
\end{centering}
\end{figure}
\subsubsection{Optical transport}

The optical mode of the FEL will begin at 0.24 mm and grow as z/0.36m.   Losses on high-power coated optics at 500 nm might be \num{5E-5} and absorbed fluxes on cooled mirrors need to be less than 100 \si{W/cm^2} so 2 \si{MW/cm^2} is the maximum incident average normal fluence.  This suggests on the order of 37 Rayleigh ranges or 14 m will be needed before the first normal optic.  Perhaps grazing mirrors could shorten this distance, though absorption for metal coatings is higher. 

\begin{figure}
\begin{centering}
\includegraphics[width=\textwidth]{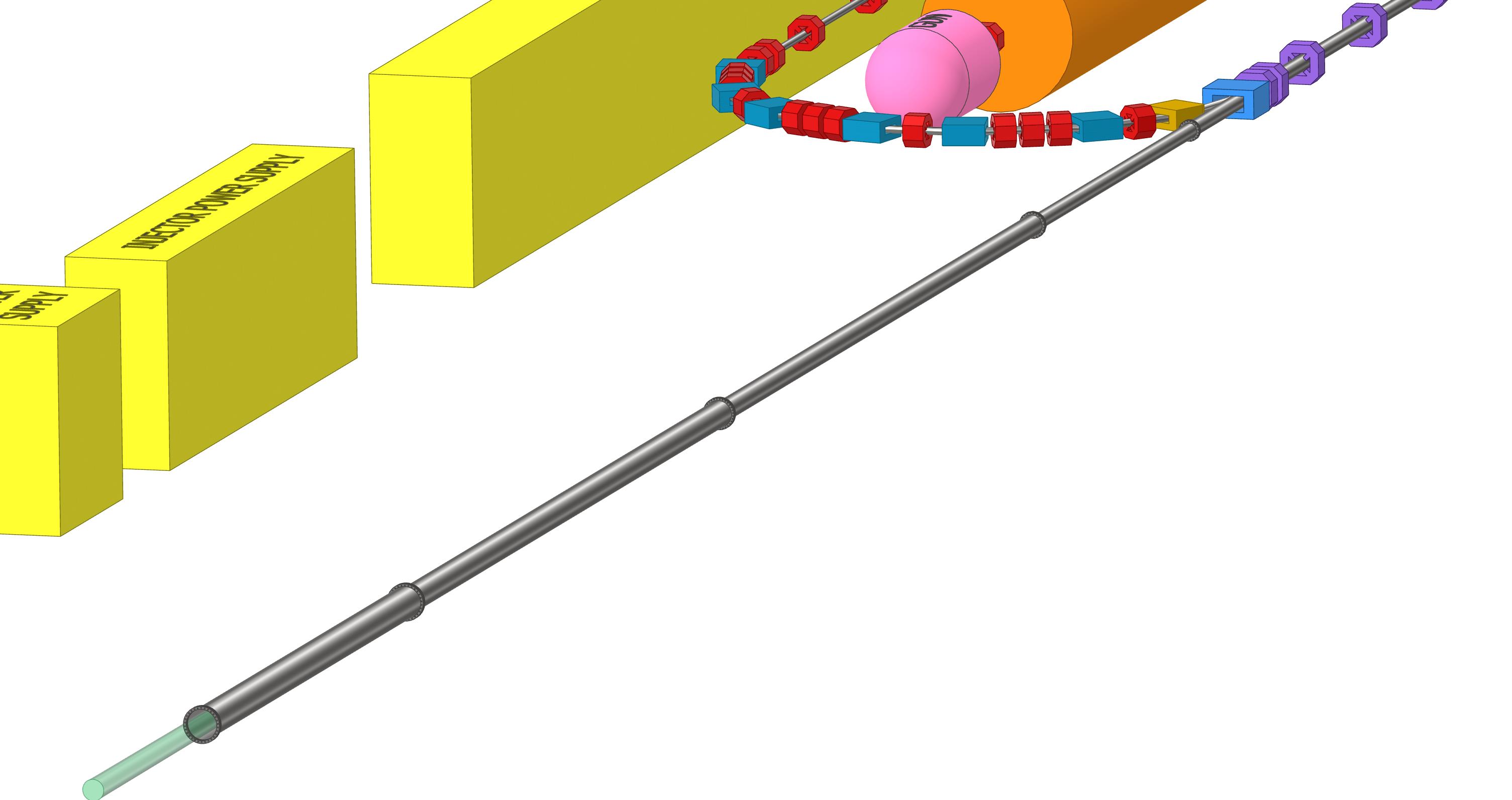}    
\caption{\label{fig:FEL-04} View of the vacuum pipe that is the first stage of laser beam expansion.  The beam must be at least 19.6 cm in diameter to have an energy density low enough to pass through the sapphire window without damaging it.  The sapphire window needs cryogenic cooling to dissipate the heat absorbed from the beam.  (Cooling not shown.)}
\end{centering}
\end{figure}

\begin{figure}
\begin{centering}
\includegraphics[width=\textwidth]{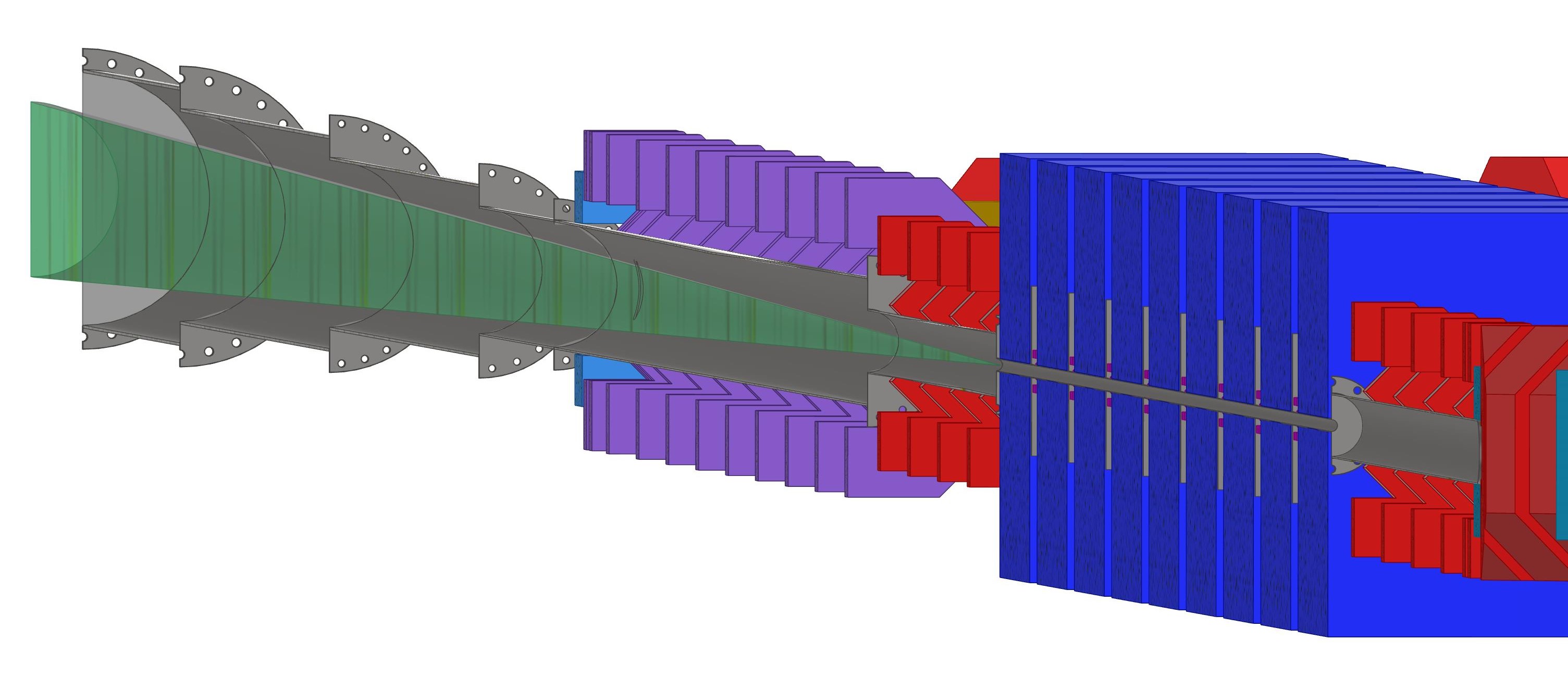}    
\caption{\label{fig:FEL-05} Oblique section view showing the beam diverging toward the sapphire window.  Note the need for increased aperture quadrupoles as the laser diverges and the beam pipe is made larger to accommodate the laser beam.}
\end{centering}
\end{figure}
\subsubsection{Diagnostics and controls}

The primary beam diagnostics and controls are generally available commercially.  One exception is the beam position monitors (BPMs) used in the accelerator section of the machine.  Since these see both accelerating and decelerating beams, a means of phase/timing detection will be necessary to simultaneously monitor the accelerating and decelerating beams.   One other item of note regarding BPMs: the alignment tolerances in SASE wigglers are generally quite tight and standard practice in X-ray SASE FELs is to use cavity BPMs to achieve the desired resolution.  This is not recommended in our high average current CW system as the resonant cavity power would lead to catastrophic heating of the diagnostic.  Button BPMs should provide sufficient resolution for this application given the longer wavelength operation and high bunch charge in CW operation. 

Beam loss monitors will provide for rapid shutoff of the beam should any accelerator component fail and direct the electron beam into the vacuum pipe wall.   At this power level the beam would burn through any beampipe in much less than a microsecond so the beam loss system must be designed to provide very rapid and redundant shutdown of the electron beam.

The development of virtual machines and AI guided tuning of the accelerator means that little operator intervention should be required once commissioning is complete.  One operator should be sufficient to bring the system into lasing and provide the necessary safety and machine health oversight of standard operations.  Modern RF systems and many DC power units are “fail soft”, that is, they gradually degrade in performance usually providing early warning of imminent failure allowing repairs to be performed either online or during scheduled maintenance periods.

\subsubsection{Facility}

The facility could be placed below ground to take advantage of earth radiation shielding, but this depends on the location of the laser power transmitter site.  Under normal operation, the radiation generated should be at a comparatively low level but personnel must be protected with four or more meters of dense material to handle the possibility of accidental beam loss. Packed earth is an inexpensive way to accomplish this.  Personnel access during operation will be excluded through use of a Personnel Safety System locking out passages into the vault.  Radiation levels at site boundaries will be actively monitored.

\begin{figure}
\begin{centering}
\includegraphics[width=\textwidth]{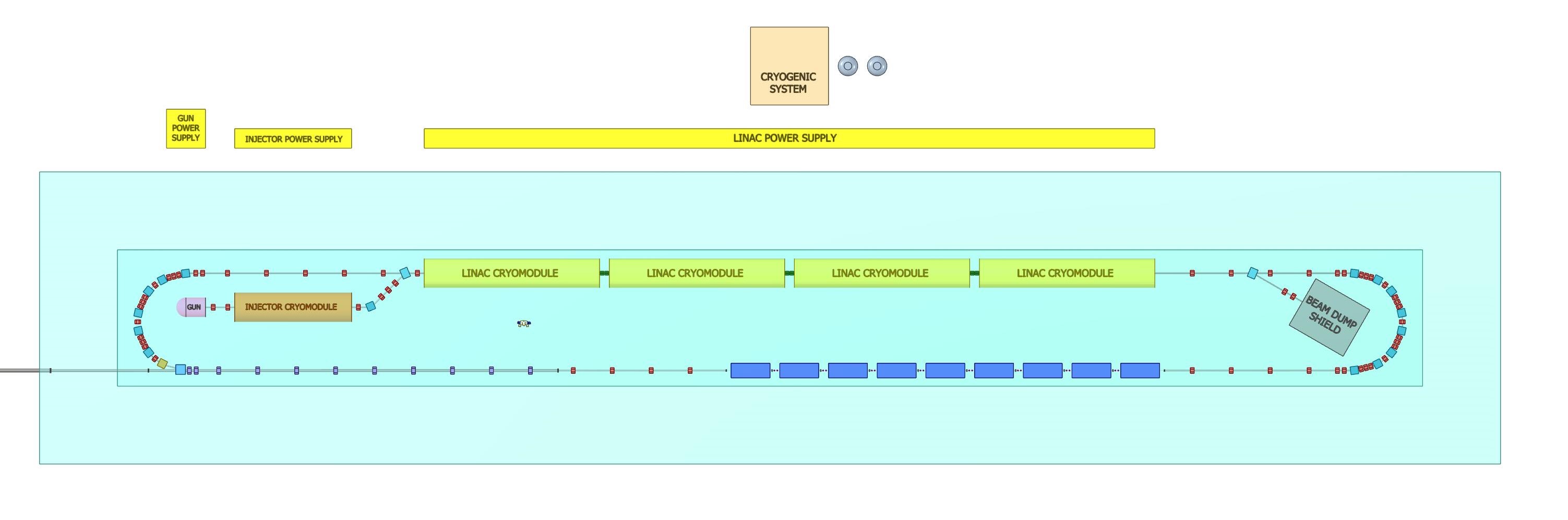}    
\caption{\label{fig:FEL-mm001} Top view of the FEL with 4 m of shielding all around (shown transparent blue.)  This thickness of shielding is required for personnel in close proximity to the accelerator.}
\end{centering}
\end{figure}

\begin{figure}
\begin{centering}
\includegraphics[width=\textwidth]{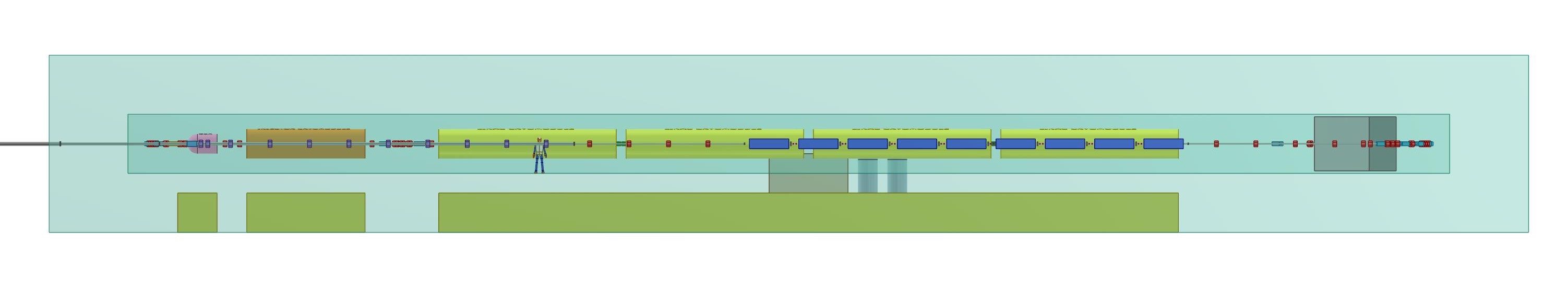}    
\caption{\label{fig:FEL-mm002} Elevation view of the electron accelerator showing 3 m thick shielding top and bottom.  The accelerator needs more shielding in the plane of the beam than it does above and below it.}
\end{centering}
\end{figure}

\begin{figure}
\begin{centering}
\includegraphics[width=\textwidth]{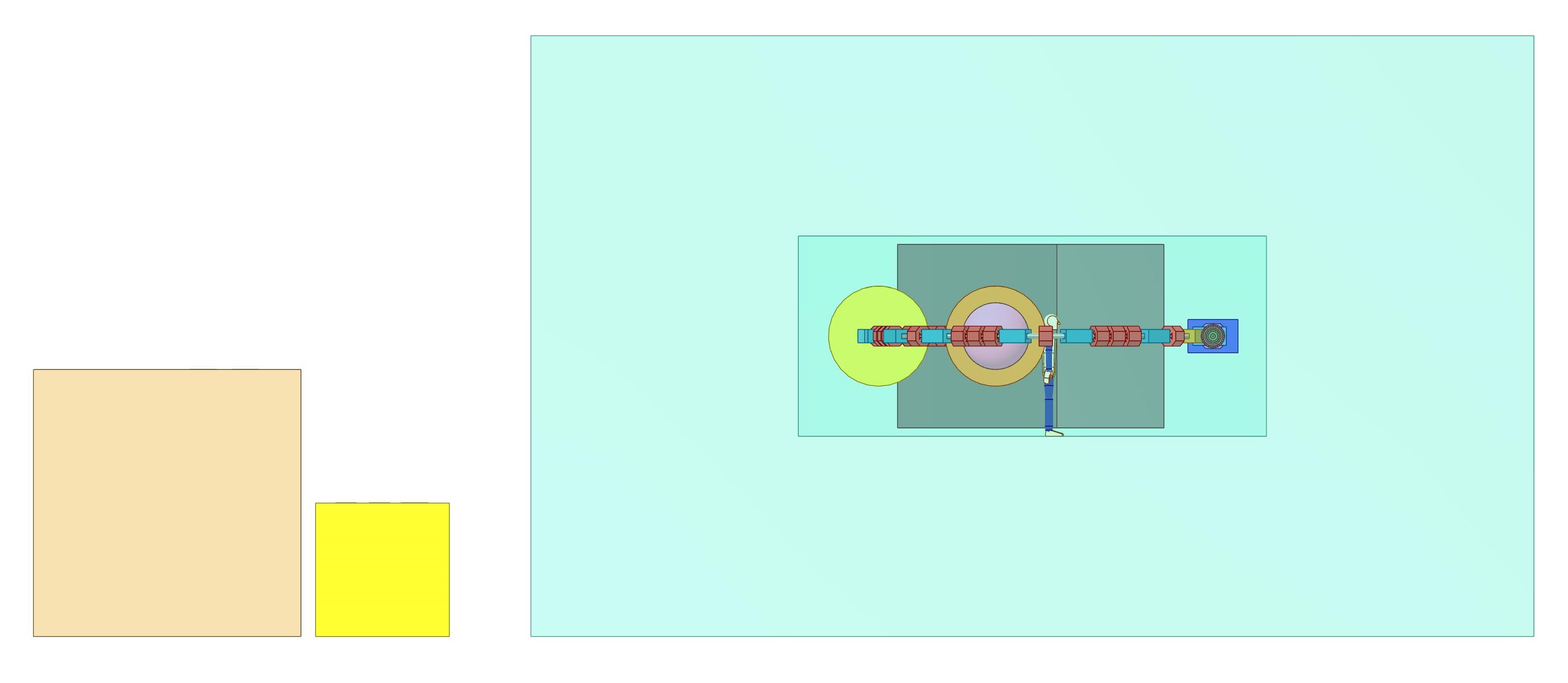}    
\caption{\label{fig:FEL-mm003} End view of the accelerator with shielding looking upstream (in the laser beam sense,) toward the wigglers.}
\end{centering}
\end{figure}

It is estimated that a below-ground accelerator vault of approximately 15 m by 75 m by 9 m high would hold the machine with a 200 mm optical beam pipe extending from the vault to the separately located optical beam director for expansion of the light beam and projection to the target.  Figure~\ref{fig:FEL-mm001} is a plan view of the accelerator showing the 4 m thick walls rendered transparent. Figure~\ref{fig:FEL-mm002} shows an elevation view with thinner (3 m) shielding above and below the accelerator. Figure~\ref{fig:FEL-mm003} shows an end elevation view of the accelerator.  Thick walls will provide not only shielding but also stability of the structure to meet the stringent alignment requirements.  
If the shielding were made from water tanks, the volume of water shown in the conceptual design would weigh 9,594 tons.  If the shielding is concrete, the volume shown would weigh 23,134 tons.  These numbers are relevant if the FEL is mounted on a floating platform that must carry that mass.  For example, a modern large cargo ship can float a payload of at least 220,000 tons, so the shielding is well within the capability of large ships.

It will be desirable to place power supplies etc above the accelerator vault for easy access to repair and feed the power and signals down to the vault.  It may also be desirable to have a pair of injectors for redundancy to improve reliability.  They could be located in separate vaults to allow access to one for photocathode regeneration while the other gun provides a beam. 

The cryoplant will be in a modest separate building to help isolate ground vibration transmission to the sensitive SRF cavities of the accelerator.  Cryogenic helium is fed closed-cycle to the linac in vacuum-insulated piping. 
\newpage
\subsubsection{System cost}

Table \ref{tab:FELcost} provides a ROM cost in 2025 US dollars for the major required components.  These estimates do not include any necessary R\&D for systems such as the high average current gun and high power RF couplers for the SRF cavities.   Many of the other components are available either off the shelf or with modest NRE costs: cavities and cryomodules, cryogenic system, focusing, bending, and trim magnets, wiggler.

\begin{table}
\begin{tabular}{@{}llll@{}}
\toprule
\textbf{Subsystem} & \textbf{Component}   & \textbf{Cost(\$M)} & \textbf{Comment} \\ 
\midrule
Injector  &             &           &                     \\
          & Drive Laser & \multicolumn{1}{r}{2.00}     &                     \\
          & DC Power    & \multicolumn{1}{r}{1.45}     &                     \\
          & RF          & \multicolumn{1}{r}{43.56}    & at \$12/watt        \\
          & Cryomodule  & \multicolumn{1}{r}{22.00}    & at \$22M/cryomodule \\
Linac     &             &           &                     \\
          & Linac RF    & \multicolumn{1}{r}{72.00}     & at \$12/watt        \\
          & Cryomodule  & \multicolumn{1}{r}{88.00}     & at \$22M/cryomodule  \\
Transport & Magnets     & \multicolumn{1}{r}{4.00}     &                     \\
Dump      &             & \multicolumn{1}{r}{2.00}     &                     \\
Wiggler   &             & \multicolumn{1}{r}{7.20}     & at\$400k/m          \\
Cryoplant &             & \multicolumn{1}{r}{25.00}     &                     \\
Controls and Diagnostics &     & \multicolumn{1}{r}{10.00}  &                 \\
Building  &             & \multicolumn{1}{r}{25.00}     &                     \\
Cooling   &             & \multicolumn{1}{r}{2.50}     & LCW                 \\
          &             &           &                     \\
\textbf{Total}     &             & \multicolumn{1}{r}{\textbf{304.71}}    &   \\ 
\bottomrule
\end{tabular}
\caption{Rough estimate of the cost of a 5 MW FEL.}
\label{tab:FELcost}
\end{table}

\subsubsection{System performance}

Table \ref{tab:FELeff} provides an estimate of the overall wall-plug efficiency.   While the FEL is designed to convert 5\% to 10\% of the beam power to light the wall-plug efficiency is enhanced through the reuse of spent beam.  Overall system efficiency is calculated by dividing the output power by the sum of all systems drawing power.  It is evident that the main hit to the wall-plug efficiency is in the conversion of AC to RF power presently assumed to be around 55\%.  There is progress in this area so potential improvements in performance may be on the horizon.  Other possibilities for improvement included the above-mentioned use of 4K operation with coated Nb cavities (note that high-temperature superconductors at nitrogen temperatures used for DC transmission are not appropriate for RF cavity use).  This would cut the cryogenic plant draw by a factor of five or so.   High-temperature superconductors might be used for the bending magnets, though the gain in overall system performance is not large.

\begin{table}
\begin{tabular}{@{}llccl@{}}
\toprule
\textbf{Subsystem} & \textbf{Component}   & \textbf{Power (kW)} & \textbf{AC Power (kW)} & \textbf{Comment} \\ \midrule
Injector  & Gun         & \multicolumn{1}{r}{145}    & \multicolumn{1}{r}{150}  &   \\
          & RF          & \multicolumn{1}{r}{3,630}  & \multicolumn{1}{r}{6,600} & at 55\% efficiency \\
Linac     & RF          & \multicolumn{1}{r}{6,000}  & \multicolumn{1}{r}{10,909}  & at 55\% efficiency \\
          & Cryo        & \multicolumn{1}{r}{600 W @2K}  & \multicolumn{1}{r}{600} & at 1000W/W @2K \\
Magnets   & DC          & \multicolumn{1}{r}{500}      & \multicolumn{1}{r}{550}  & WAG       \\
Hotel Power & AC        &                              & \multicolumn{1}{r}{400}  & WAG        \\
LCW       & Pumps       &                              & \multicolumn{1}{r}{200}  & WAG  \\
          &             &                              &                          &         \\
\textbf{Total AC}       &   &  & \multicolumn{1}{r}{\textbf{19,409}} & \multicolumn{1}{l}{\textbf{kW}}  \\
          &             &                              &                          &         \\
\textbf{FEL Power Out}  &  &   & \multicolumn{1}{r}{\textbf{5,000}} & \multicolumn{1}{l}{\textbf{kW}}  \\
\textbf{FEL Efficiency}  &  &   & \multicolumn{1}{r}{\textbf{5\%}}  \\
\textbf{System Efficiency}  &  &   & \multicolumn{1}{r}{\textbf{26\%}}    \\ \bottomrule
\end{tabular}
\caption{Power performance of a 5 MW FEL.}
\label{tab:FELeff}
\end{table}

\subsubsection{Single FEL power limits}
\label{sec:FELplimit}

Given the significant power that a climber will require, a question arises as to the ultimate power limit achievable from an FEL.   The answer is necessarily speculative since the highest power demonstration so far has been only 14 kW.  To be sure, there are no hard limits, but rather as the power increases, some parts of the system become increasingly difficult or unattractive.   

Recall that the power out of an FEL is equal to the electron beam power times the FEL conversion efficiency.  The electron beam power in turn is simply the average current times the voltage.  Adding voltage (beam energy in MeV) to an electron beam is straightforward and the limit is only how much accelerator one is willing to buy.  Unfortunately, the FEL gain and conversion efficiency is output wavelength dependent and falls as (1/beam energy). The result is a broad performance optimum in operating beam energy for a given wavelength desired; in our case of 0.5 micron output this optimum occurs at several hundred MeV. We chose 300 MeV for our operating point.   Above this range, any increase in power from the higher electron beam power is offset by a reduction in the conversion efficiency.   The conversion efficiency projected is 5\%, near the maximum demonstrated of any FEL operating in the near IR or visible.  The induced energy spread on the beam is roughly six times this conversion efficiency so even at this level electron transport for energy recovery will be difficult.   

That leaves only the average beam current to work with.  The electron (and thus optical) beam is provided in a continuous stream of micropulses at a rate up to the accelerating RF cavity frequency.  Higher beam currents demand lower RF cavity frequencies because larger volumes of RF field are required to accelerate in a stable fashion.   We chose 750 MHz as the highest RF frequency capable of accelerating the high bunch charges required.  Thus, for the average current of 0.33 Amp we require a bunch charge of 0.33/7.5e8 or 440 pCoul/bunch.   This is quite a high bunch charge to produce at high quality from the injector gun due to the high self space charge fields felt at low beam energy.  These self fields cause the beam to blow up before it can be accelerated to higher energy where space charge forces become less significant.  There are also other charge-dependent instabilities, such as coherent synchrotron emission, which work to ruin beam quality during the acceleration process.  As a result, most present FELs utilize bunch charges in the 0.1 nCoul range.   Because we are operating in the visible region where tolerances are more forgiving, we have allowed the bunch charge to increase by 4X beyond the present common value of around 100 pCoul and paid the price of 20\% increased gain length due to the expected increase in emittance to 1.6 mm mrad.   Further increases in bunch charge would result in lower electron beam quality due to increased emittance, leading to increasingly significant reductions in FEL performance.

The bottom line of this discussion is that the performance values we have chosen are already aggressive.  While arguments could be made for modest increases in some performance parameters, the projected output is near the limits of what can be achieved in a single FEL.   For larger output powers, multiple FELs could be utilized with the additional benefit of increased reliability (at reduced but still usable powers) due to system redundancy.

\subsubsection{Turning down the power for higher altitude climbers}

Table \ref{tab:sevenclimb} shows that climbers at the lowest altitudes need the highest power to climb.  The higher climbers have much lower power requirements, but they still need the large waist beam and high beam quality to preserve the power delivered.  Turning the power down on a 5 MW FEL is easy to do.  It is accomplished at the injector gun by controlling the pulses of the drive laser that strike the gun cathode.  Instead of providing pulses at 750 MHz, 1 MHz can be left out which gives pulses at 749 MHz, leading to 749/750 times the laser power. This can be regulated down to 1 pulse per second as desired.  The typical process is to set the controls to drop/increase the beam current by 1\% each time.

The AC power draw decreases along with the current reduction except for the cryogenic refrigerator, which is a constant draw.

\subsubsection{Summary}

We have examined technologies for a high-power FEL system in the visible for use in beaming power.  Except for the substantial average beam current required, most aspects of the technology are in hand today and provide an attractive system for high-power operation.   The estimated 26\% conversion of AC power to light offers the possibility of improvement through foreseen near-term technology advances.  By pushing each of the component developments and adding energy recycling components it might be possible to push the overall efficiency to 33\%. Key areas for focus would be the operation of the RF cavities at 4K or above, and improvements in the RF generation efficiency.

\newpage
\subsection{Conceptual design of a beam expander/transmitter on the ground--the 7.5 m diameter telescope for launching the 5 m diameter beam}
\label{sec:beamexpand}
\begin{figure}[hb]
\begin{centering}
\includegraphics[width=.8\textwidth]{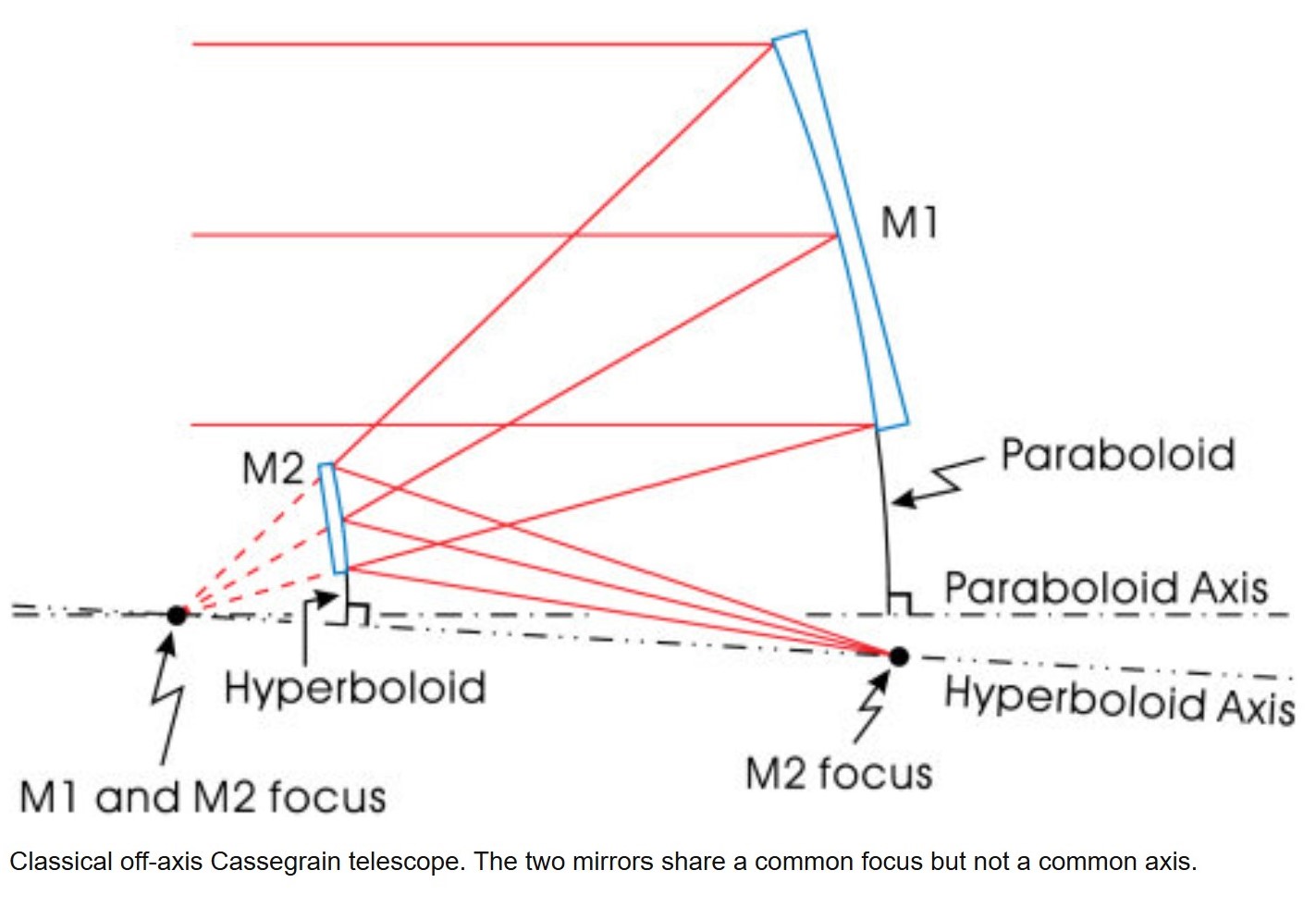}    
\caption{\label{fig:Cassegrain} Schematic diagram of offset Cassegrain optics from [\cite{Chang}].}
\end{centering}
\end{figure}

Once the 5 MW laser beam has passed through the cryogenically cooled sapphire window at the end of the FEL beam pipe shown in Figure \ref{fig:FEL-04}, it must be expanded so that the beam waist at the transmitter is approximately 2.5 m, or 5 m in diameter.  The conceptual design for the beam expander is an offset Cassegrain telescope.  (See Figure \ref{fig:Cassegrain}.) This design was chosen to avoid mirror and mirror support material in the outgoing beam, as would be the case in a typical Cassegrain telescope. The dominant aberration of offset Cassegrain telescopes is linear astigmatism, but reference [\cite{Chang}] shows an equation to eliminate linear astigmatism.  The M1 or primary mirror has a concave paraboloid surface.  The M2 or secondary mirror has a convex hyperboloid surface.

\begin{figure}
    \begin{minipage}[t]{.48\textwidth}
        \centering
        \includegraphics[width=\linewidth]{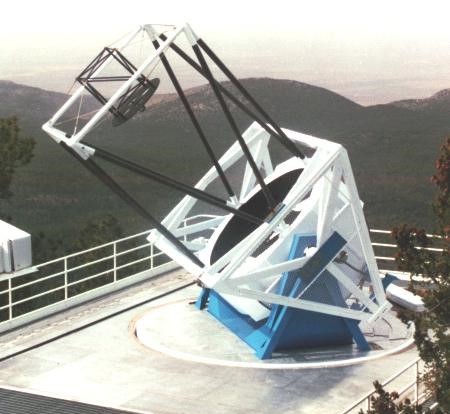}
        \captionsetup{justification=centering}
        \caption{The Sloan Telescope without the wind baffle on showing the structure that supports the optics. From \url{https://classic.sdss.org/tour/main_telescope.php}}
        \label{fig:Sloannaked}
    \end{minipage} \hfill
    \begin{minipage}[t]{.48\textwidth}
        \centering
        \includegraphics[width=\linewidth]{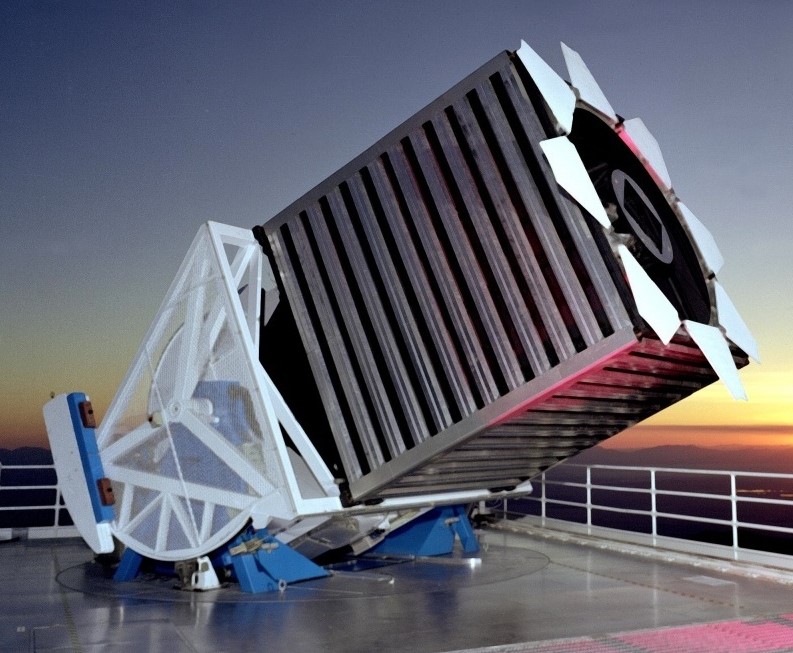}
        \captionsetup{justification=centering}
        \caption{View of the Sloan telescope with the wind baffle on.  From \url{https://www.planetary.org/space-images/sloan-foundation-telescope}}
        \label{fig:Sloanfull}
    \end{minipage}
\end{figure}

The design of the beam expander mounting was inspired by the 2.5 m Sloan Digital Sky Survey telescope at Apache Point Observatory in NM.  (See figure \ref{fig:Sloannaked} and \ref{fig:Sloanfull}.)

\begin{figure}
\begin{centering}
\includegraphics[width=\textwidth]{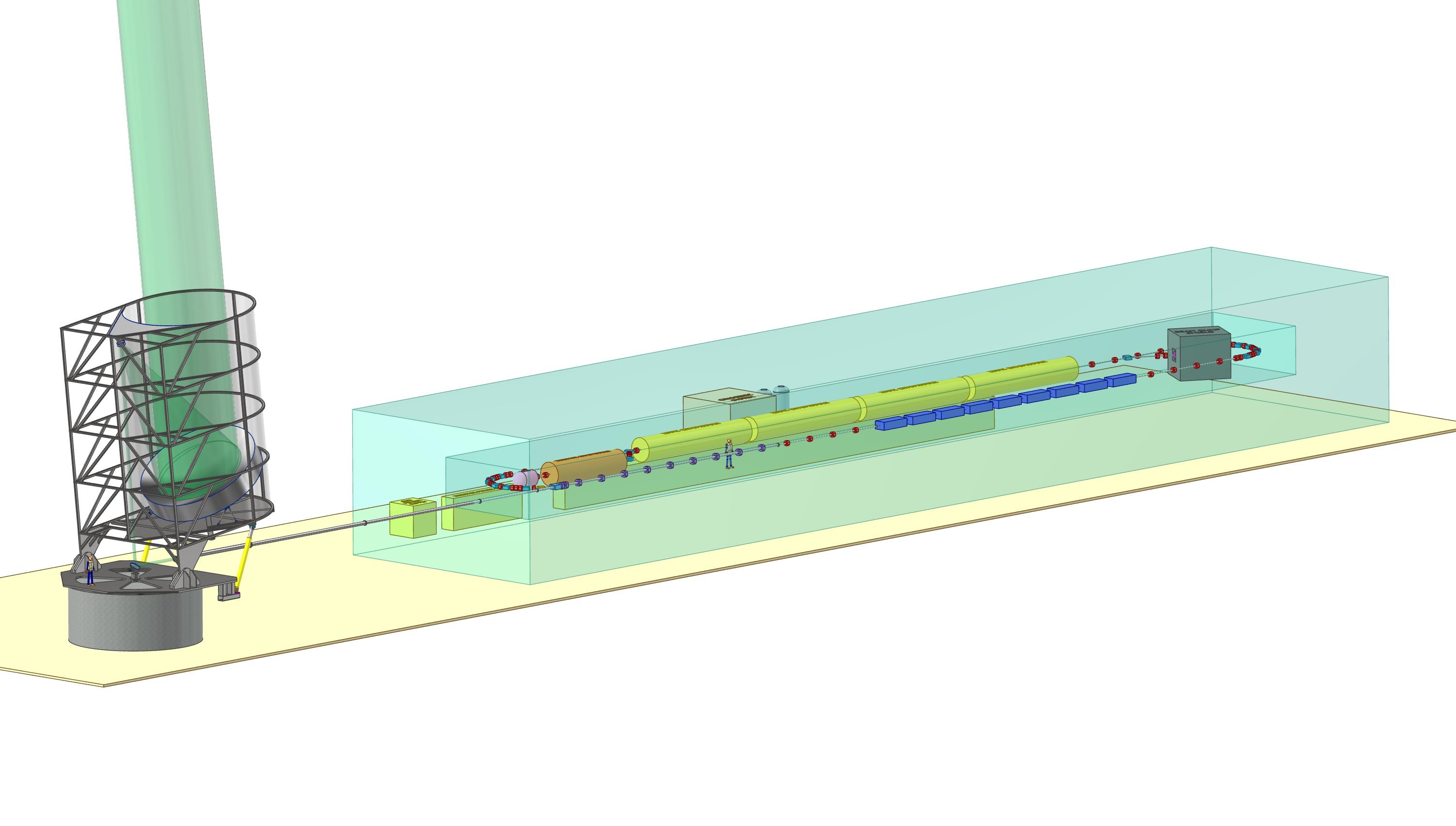} 
\caption{\label{fig:FEL-BE-01} An oblique view of the entire FEL connected to the Beam Expander.  The outer shell of the Beam Expander is rendered transparent to see the mirrors in the optical path.}
\end{centering}
\end{figure}

\begin{figure}
\begin{centering}
\includegraphics[width=\textwidth]{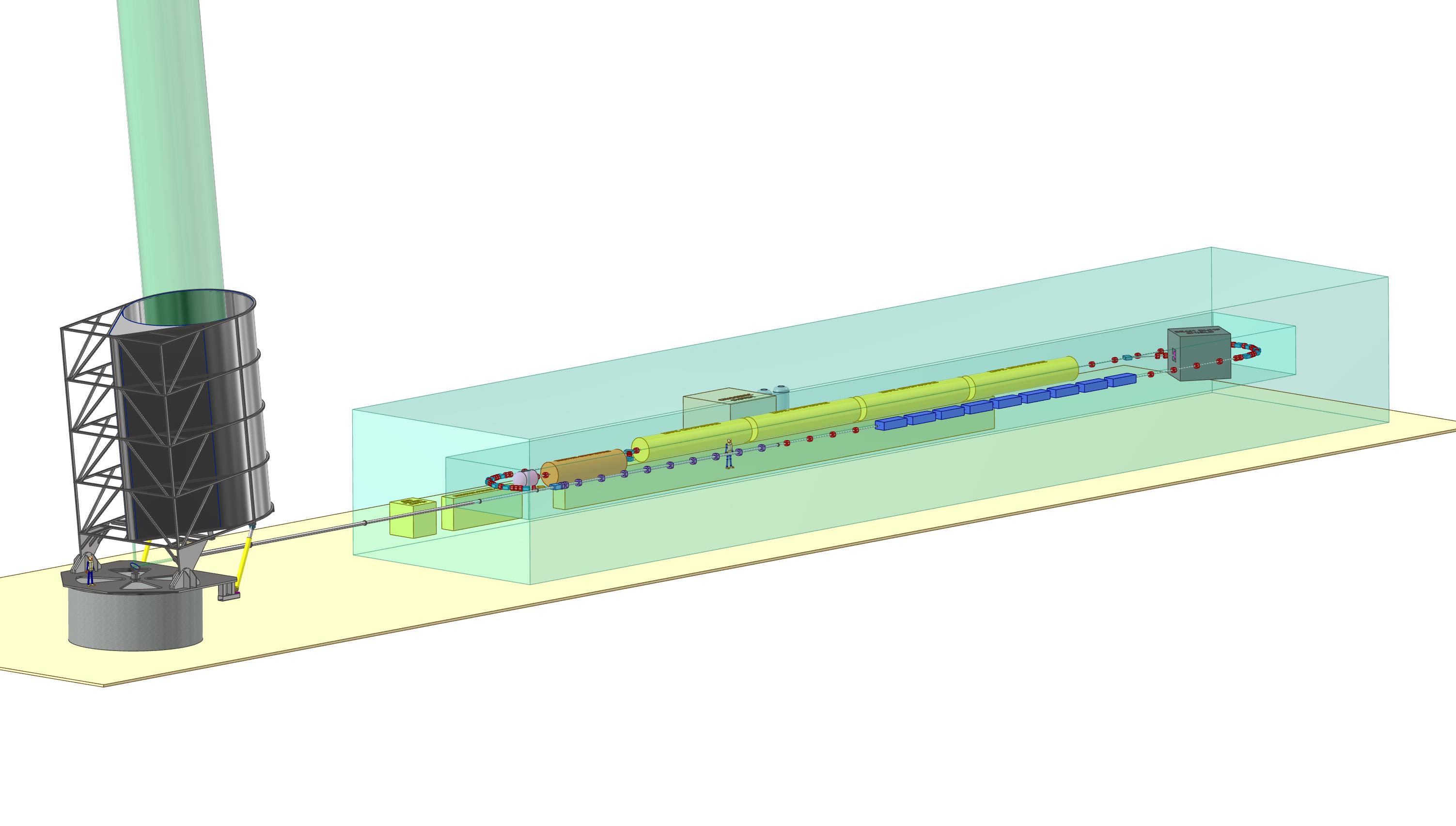} 
\caption{\label{fig:FEL-BE-02} Another oblique view of the FEL and Beam Expander.  The shell around the optical path is rendered opaque as the aluminum would be in reality.  Additional optical shielding will be needed around the base of the Beam Expander for human eye protection.}
\end{centering}
\end{figure}

Figure~\ref{fig:FEL-BE-01} shows an oblique view of the FEL and the Beam Expander downstream of it.  The shell of the beam expander has been rendered transparent to see the primary and secondary mirrors in the optical path. Figure~\ref{fig:FEL-BE-02} shows the shell rendered opaque, as it would be in reality.

\begin{figure}
\begin{centering}
\includegraphics[width=\textwidth]{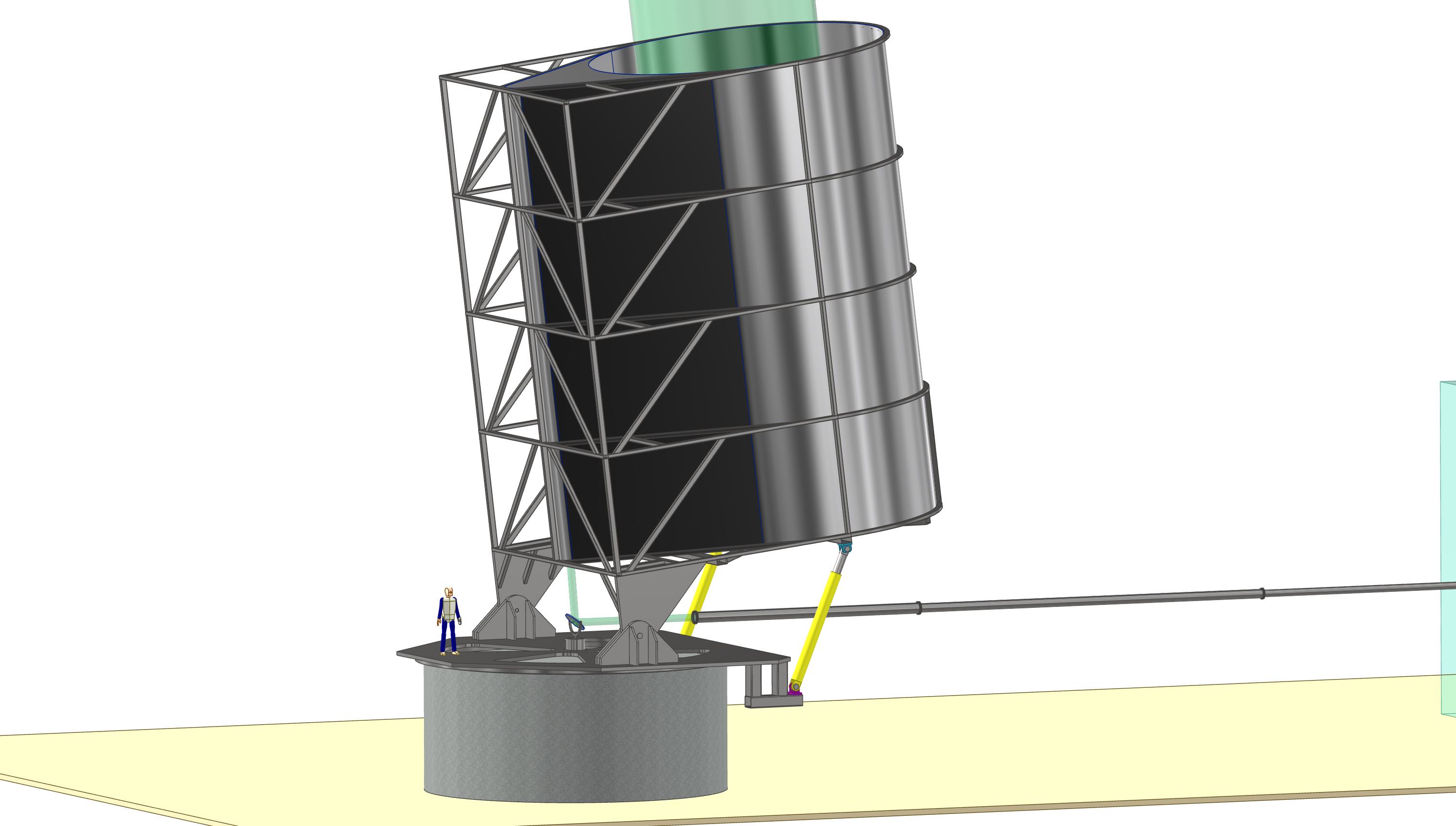} 
\caption{\label{fig:FEL-BE-03} Close-up of the offset Cassegrain Beam Expander showing the alt-az mount.  The outer shell is rendered opaque.  A six foot man is shown for scale.}
\end{centering}
\end{figure}

\begin{figure}
\begin{centering}
\includegraphics[width=\textwidth]{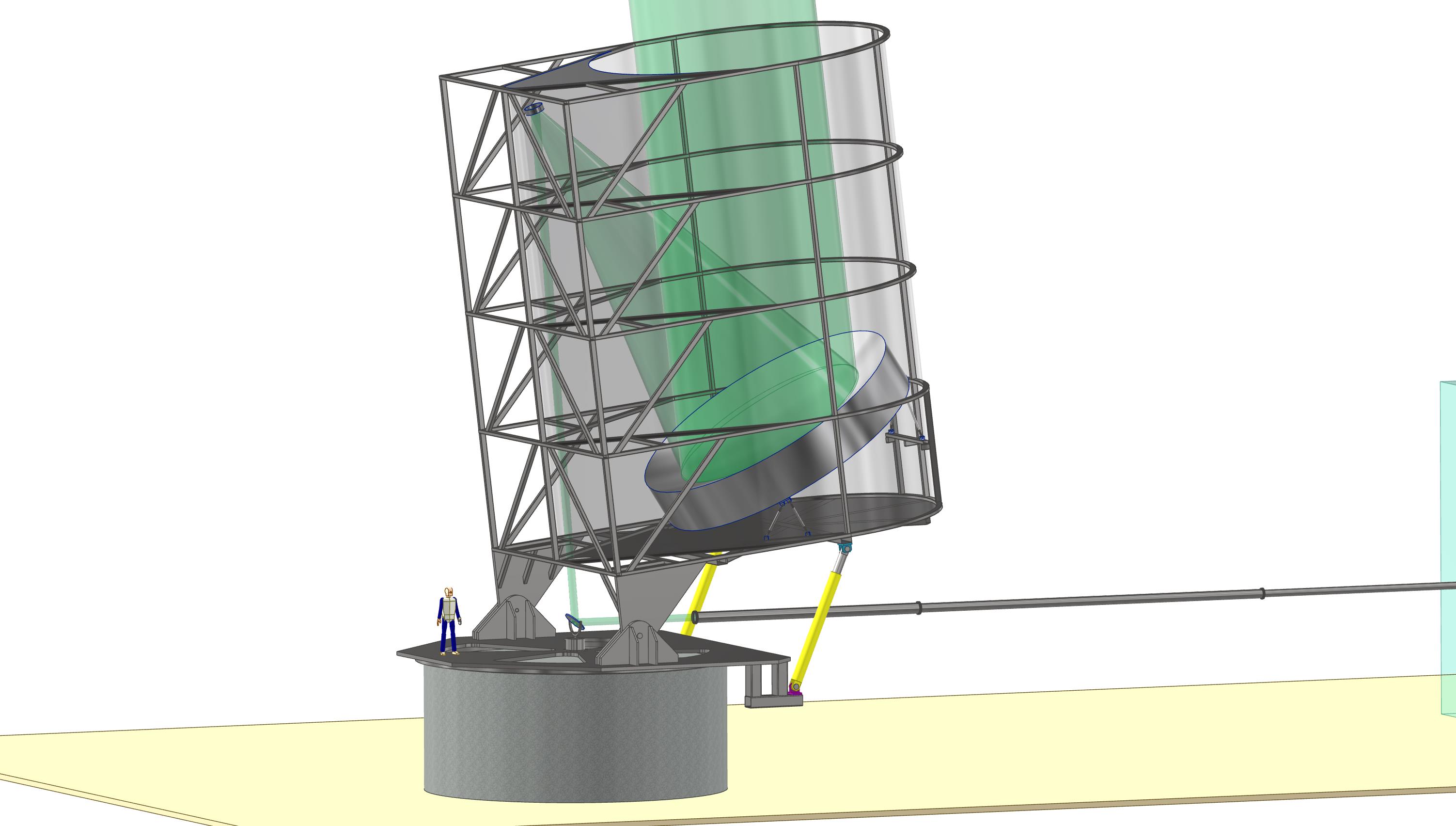} 
\caption{\label{fig:FEL-BE-04} Close-up of the Beam Expander showing the outer shell rendered transparent to see the primary and secondary mirrors.}
\end{centering}
\end{figure}

Figures \ref{fig:FEL-BE-03} and \ref{fig:FEL-BE-04} show close-up views of the beam expander and its alt-az telescope mounting.  The design concept is a pitch (alt, short for altitude,) degree of freedom actuated by hydraulic cylinders shown in yellow.  The entire structure rotates around a vertical axis (the az, or azimuth axis).   

One significant difference between this beam expander and the Sloan telescope is that the Sloan telescope has a wind baffle that completely surrounds the optical telescope to protect it from winds up on the mountain.  The wind baffle is mounted on a separate structure that is vibrationally isolated from the foundation of the optical telescope.  The wind baffle has its own alt-az mount that follows the optical mount.  It remains for final design and site selection to determine whether the FEL beam expanders need a wind baffle.

\begin{figure}
    \begin{minipage}[t]{.42\textwidth}
        \centering
        \includegraphics[width=\textwidth]{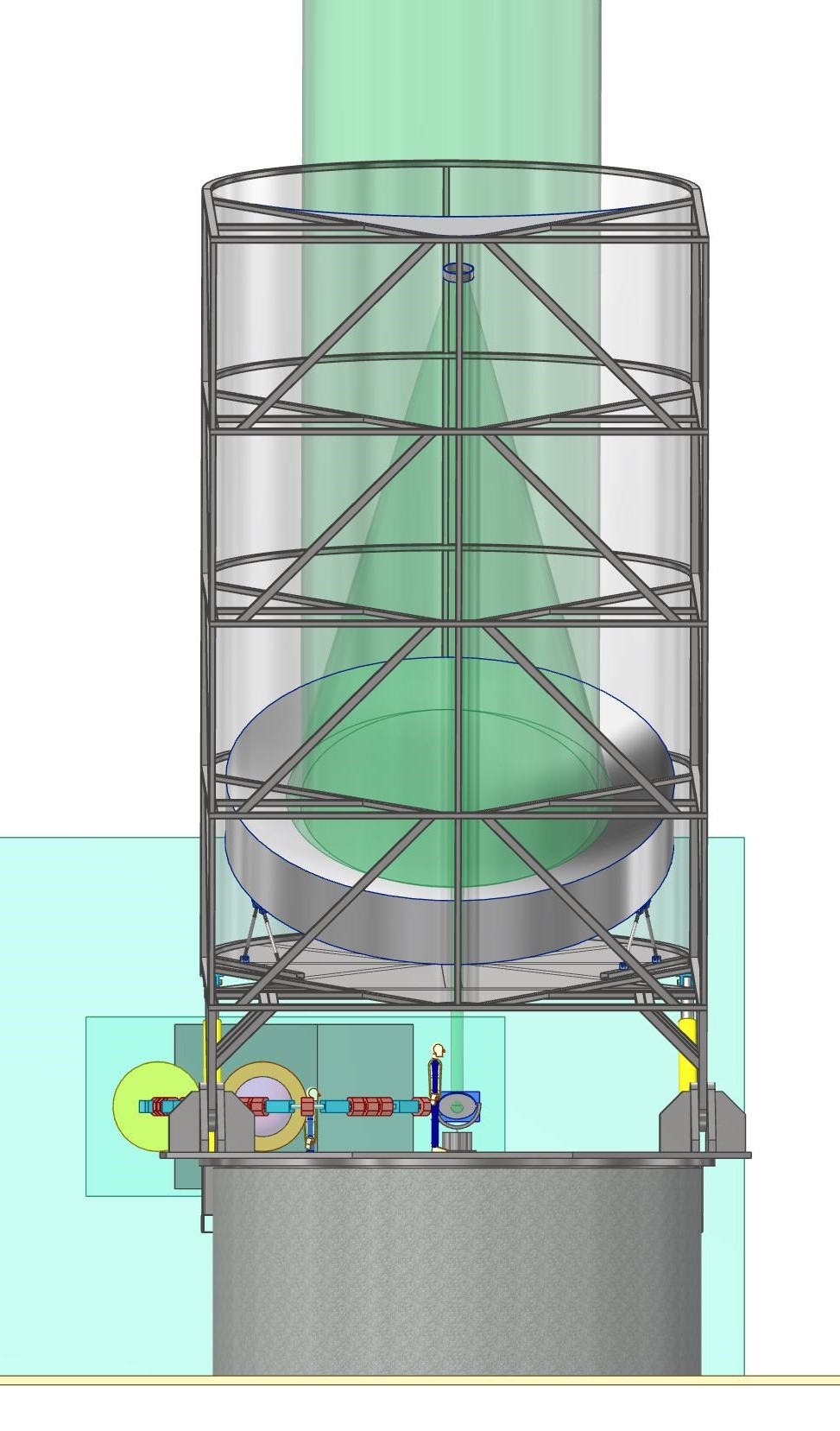} 
        \caption{\label{fig:FEL-BE-06} A view of the Beam Expander looking upstream, shell transparent.}
   \end{minipage} \hfill
   \begin{minipage}[t]{.50\textwidth}
        \centering
        \includegraphics[width=\textwidth]{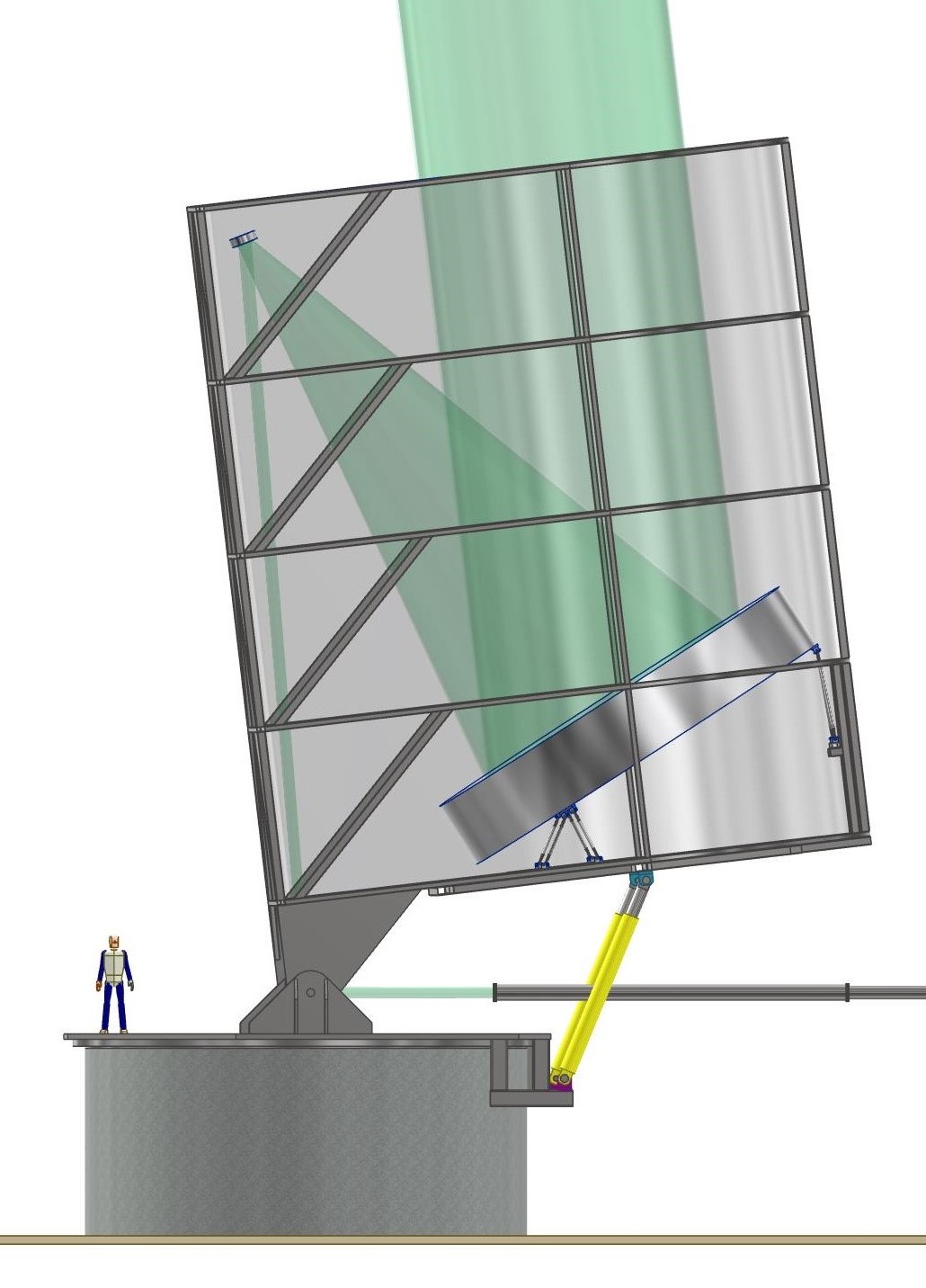} 
        \caption{\label{fig:FEL-BE-05} The primary mirror is shown on a conceptual hexapod mounting.  The mounting is comprised of turnbuckles to create the initial alignment.  Access hatches have not been designed to reach these adjusters.  The primary mirror may be segmented to facilitate the adaptive optics needed to keep the beam pointing at the climbers.  Further development may show the need for motorized actuation of the hexapod.}
    \end{minipage}
\end{figure}

Figures \ref{fig:FEL-BE-05} and \ref{fig:FEL-BE-06} show orthogonal elevations of the beam expander looking transverse to the FEL beam pipe and along it.  The aluminum shell of the expander is rendered transparent to show the turnbuckle hexapod support concept for the primary mirror. No supports are shown for the FEL beam pipe in Figure \ref{fig:FEL-BE-05}, but they are certainly required.  The cryogenic system required to cool the sapphire window at the end of the beam pipe is not shown in any figure.

Figure \ref{fig:FEL-BE-08} shows a section view through the turret bearing that allows the azimuth rotation.  Since this bearing is larger than 8 m in diameter, it would be made from segments.  No azimuthal rotation mechanism is shown, but a common way to actuate this type of axis is to put an external gear on the outer race of the bearing and use a motor and gear drive to rotate the outer race and structure.  The inner race would be bolted solidly to a flat base structure.  

\begin{figure}
\begin{centering}
\includegraphics[width=\textwidth]{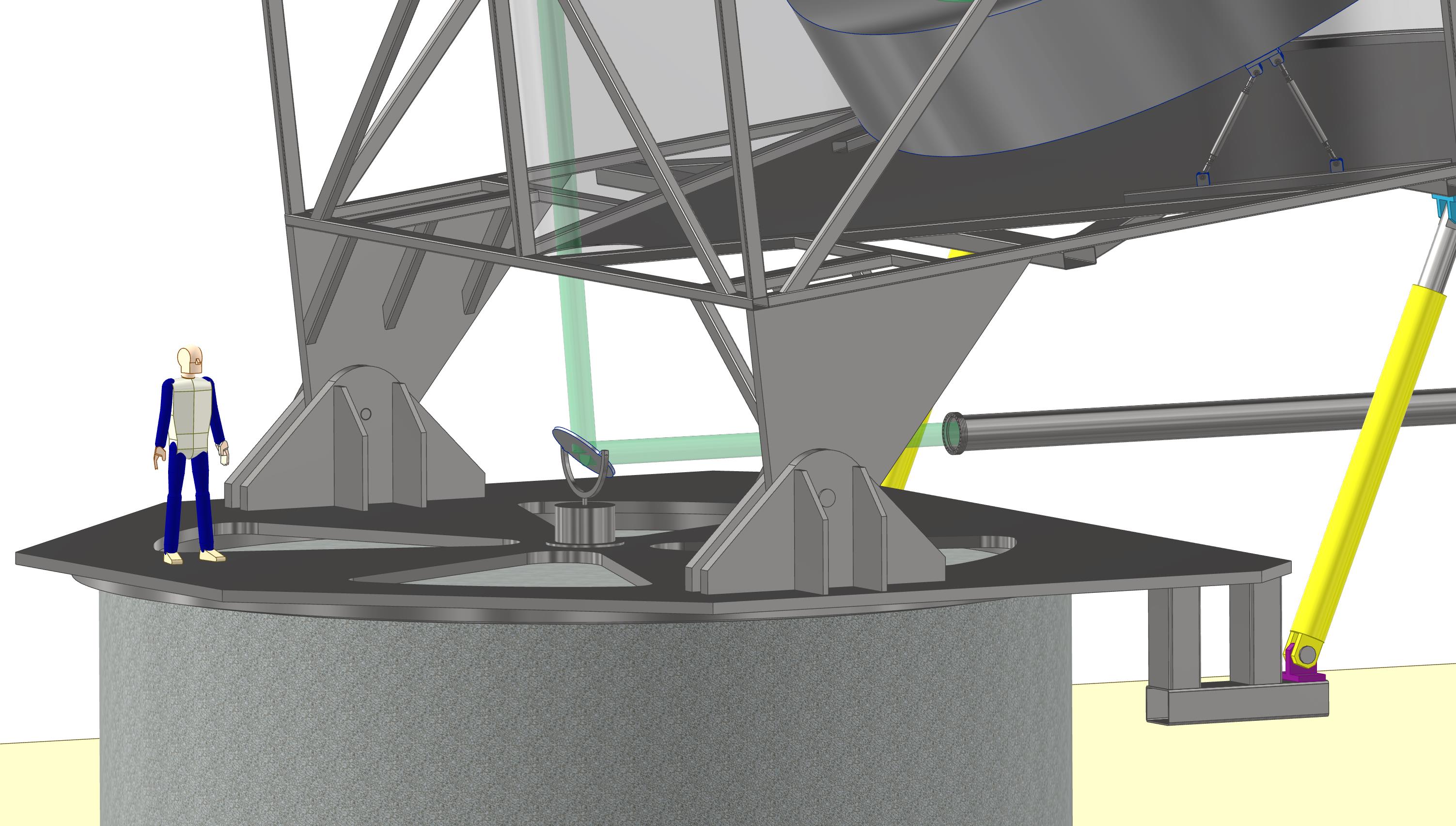} 
\caption{\label{fig:FEL-BE-07} This view shows the articulated mirror needed to direct the FEL beam into the Beam Expander.  No actuators on this mirror are shown.}
\end{centering}
\end{figure}

\begin{figure}
\begin{centering}
\includegraphics[width=\textwidth]{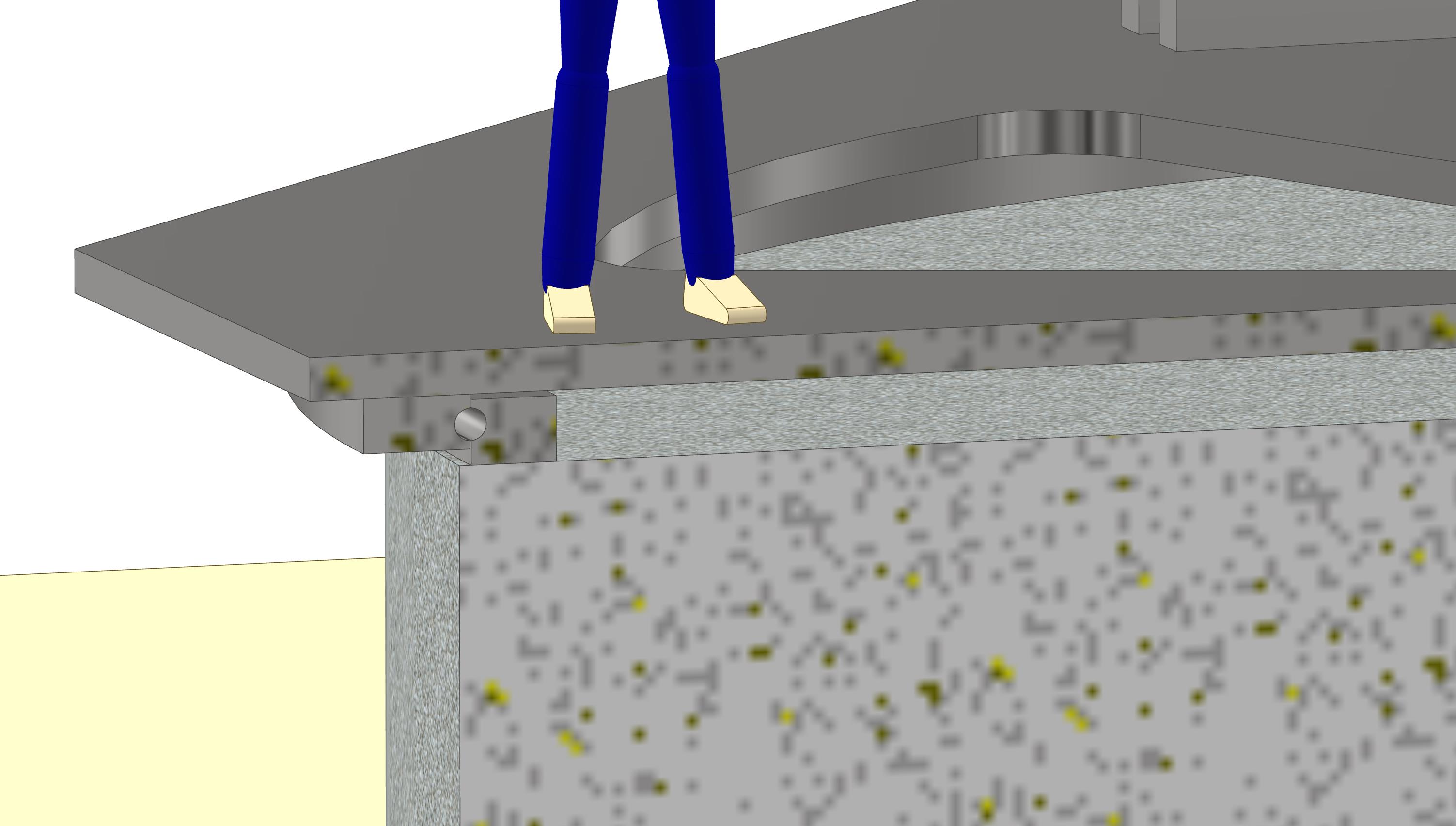} 
\caption{\label{fig:FEL-BE-08} Section view of the azimuth mount showing the turret bearing required to support the Beam Expander and change its yaw position.  Ascending climbers will deflect the tether West because of Coriolis forces and the beam expander has to follow the climber no matter where it is in the sky.}
\end{centering}
\end{figure}

Several styles of turret bearings can absorb moment and radial forces. The rolling elements may be balls with the races having a small ellipticity such that the balls only contact the races in an X configuration.  A crossed roller bearing is another possibility.  The alt axis rotates on a high-load tapered roller bearing pair on each side.
\newpage
The small flat mirror that bounces the laser beam from the FEL into the beam expander, shown in close-up in Figure \ref{fig:FEL-BE-07}, is articulated to keep the beam entering the expander at every altitude and azimuth angle of the expander.  No articulation mechanisms are shown in the figures.  This mirror is also mounted in an alt-az configuration with its az axis matching the az axis of the whole expander.  Its altitude axis is at the same elevation as the altitude axis of the whole expander, but it only becomes colinear with the larger axis when the azimuth rotation is set to zero degrees off the FEL beam axis.  

\begin{figure}
\begin{centering}
\includegraphics[width=\textwidth]{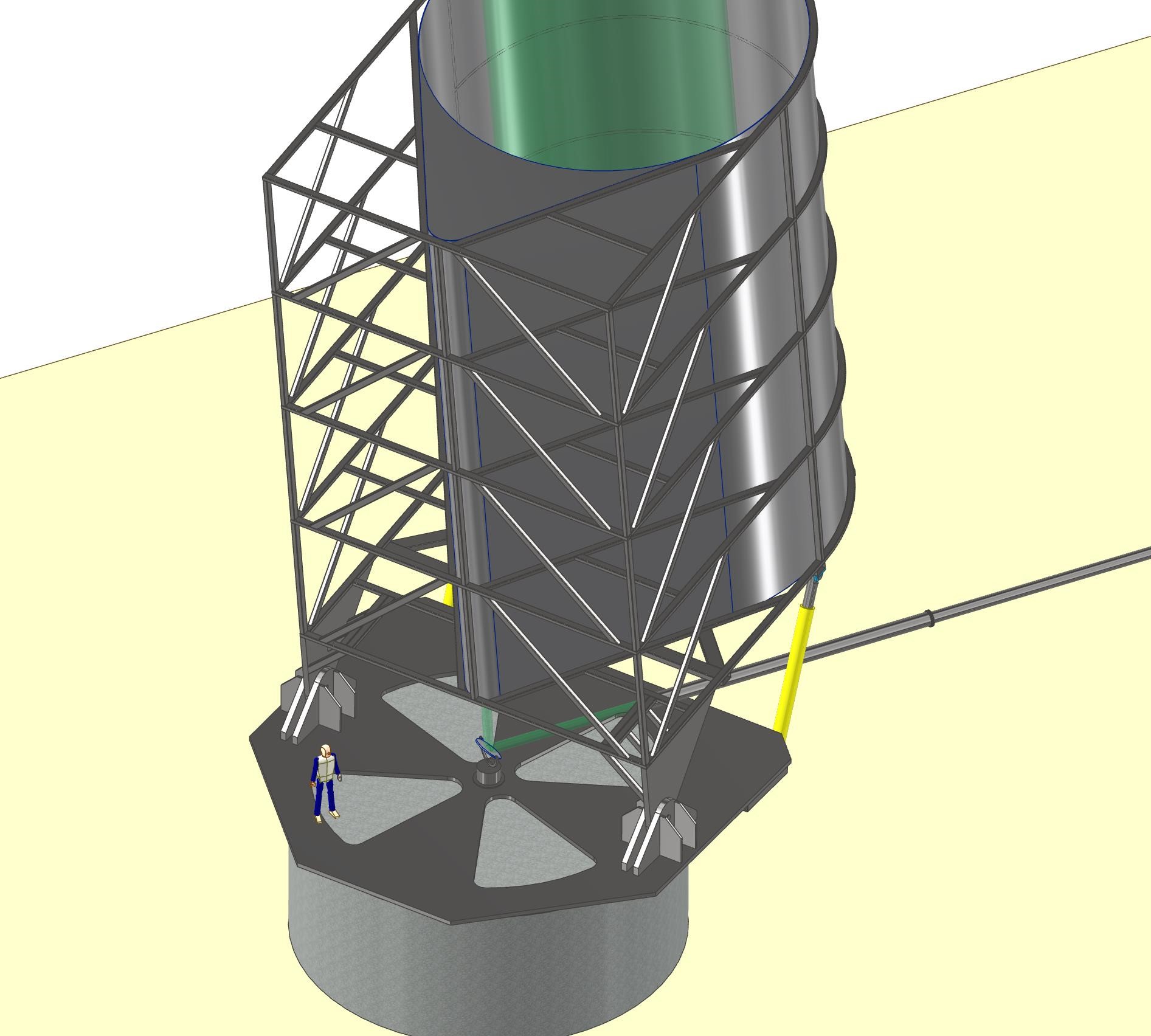} 
\caption{\label{fig:FEL-BE-09} View showing the Beam Expander rotated approximately 36 degrees in azimuth, point the beam to FEL beam left.  The azimuth range is +/- 36 degrees with the pitch hydraulic cylinders as shown.  The beam is pointing the 5 m diameter output beam to beam-left with respect to the FEL beam.}
\end{centering}
\end{figure}

\begin{figure}
\begin{centering}
\includegraphics[width=\textwidth]{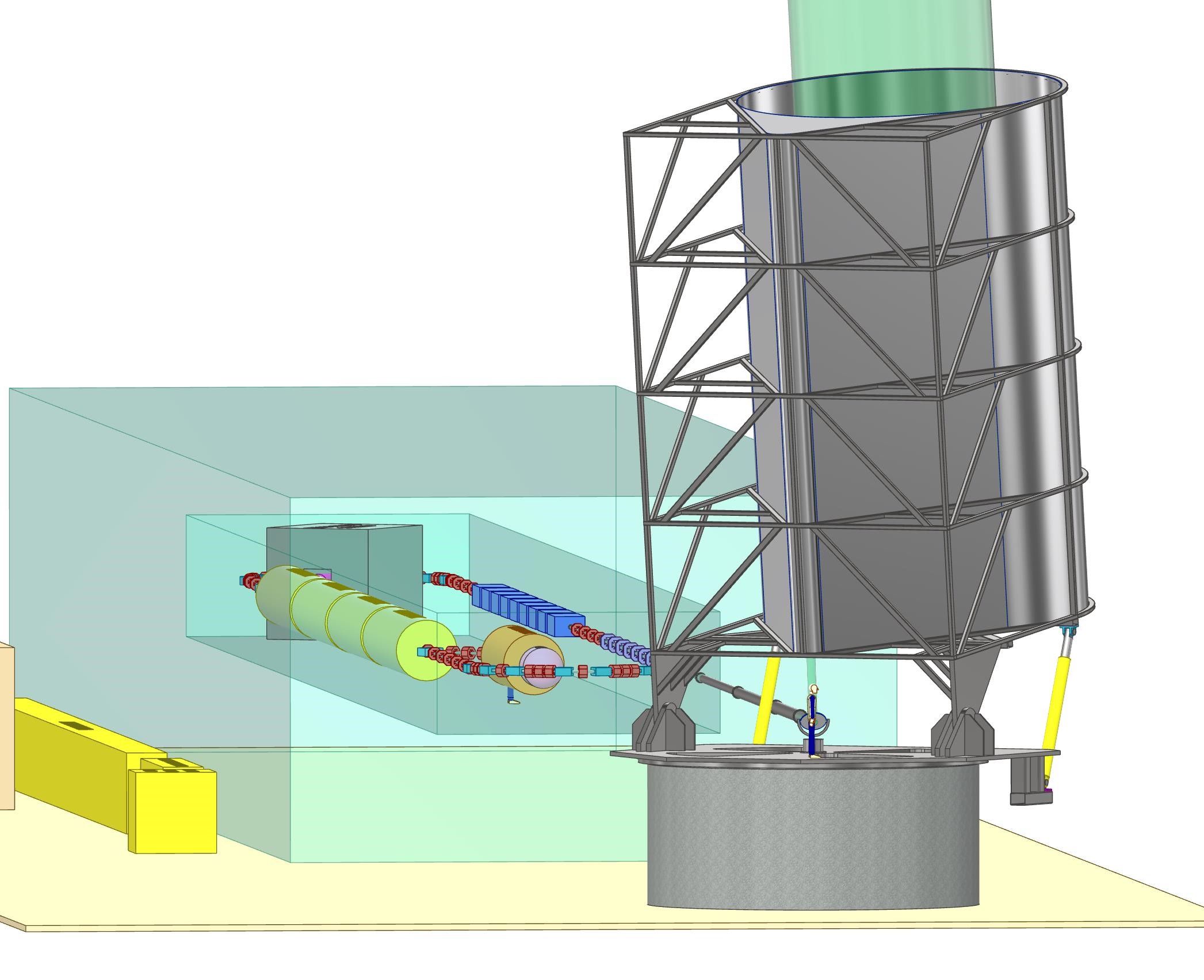} 
\caption{\label{fig:FEL-BE-10} View showing the Beam Expander rotated all the way around to its other azimuth stop near the FEL beam line at beam-right.}
\end{centering}
\end{figure}

\begin{figure}
\begin{centering}
\includegraphics[width=\textwidth]{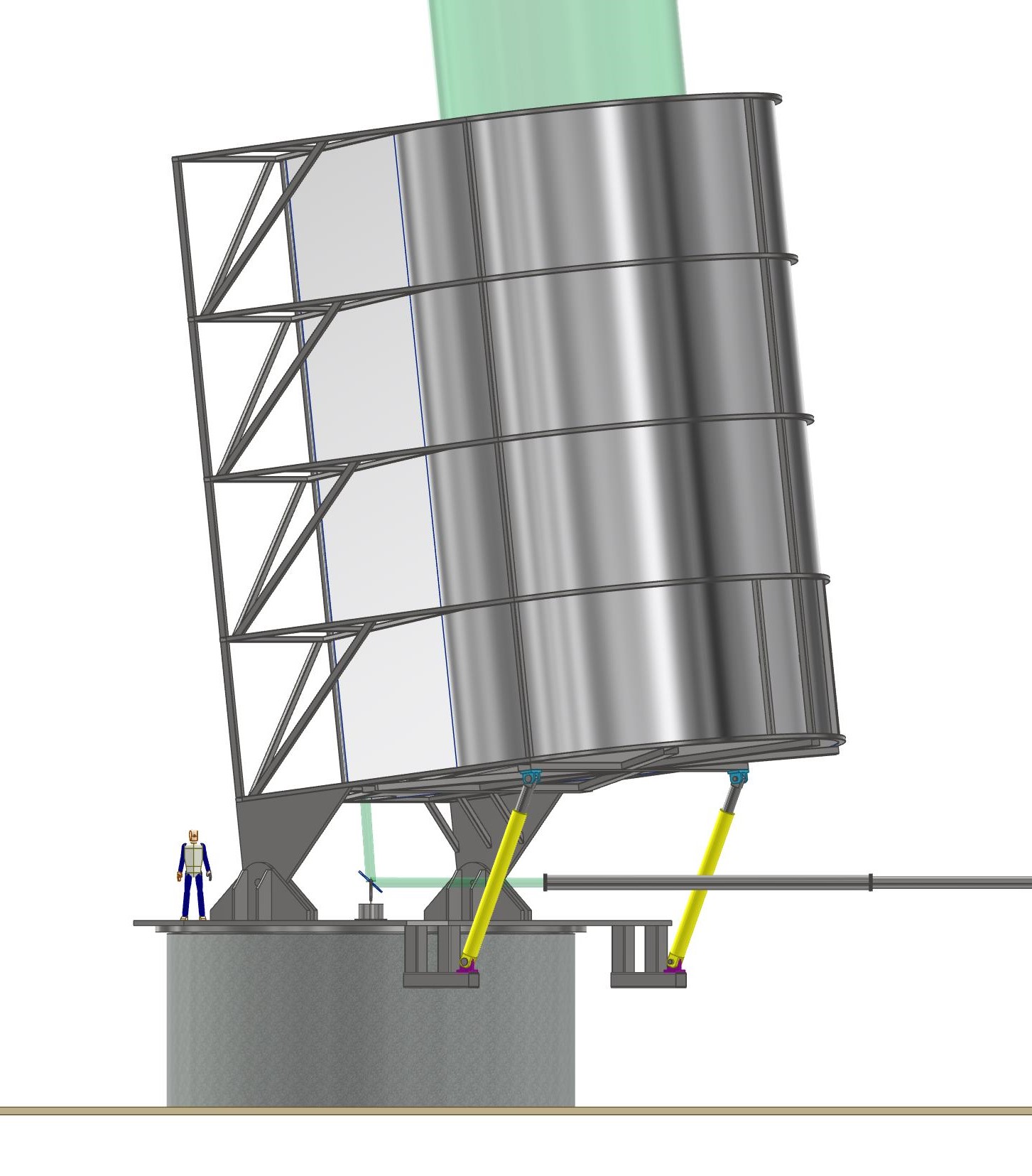} 
\caption{\label{fig:FEL-BE-11} Another view of the Beam Expander rotated to point the 5 m beam 36 degrees to beam-right.}
\end{centering}
\end{figure}

Figures \ref{fig:FEL-BE-09}, \ref{fig:FEL-BE-10}, and \ref{fig:FEL-BE-11} show the mechanical limits of the azimuth axis at +/- 36 degrees off the FEL beam axis.  This is when the pitch hydraulic cylinders come close to the FEL beam.  A different configuration of cylinder actuators, or a totally different actuation mechanism, could allow for a larger azimuthal range.  The range required will also be determined by what the beam expander is mounted on.  If it is mounted on an aircraft carrier, as shown in the cover art image, then the carrier could provide additional motion.

\begin{figure}
    \begin{minipage}[t]{.49\textwidth}
        \centering
        \includegraphics[width=\textwidth]{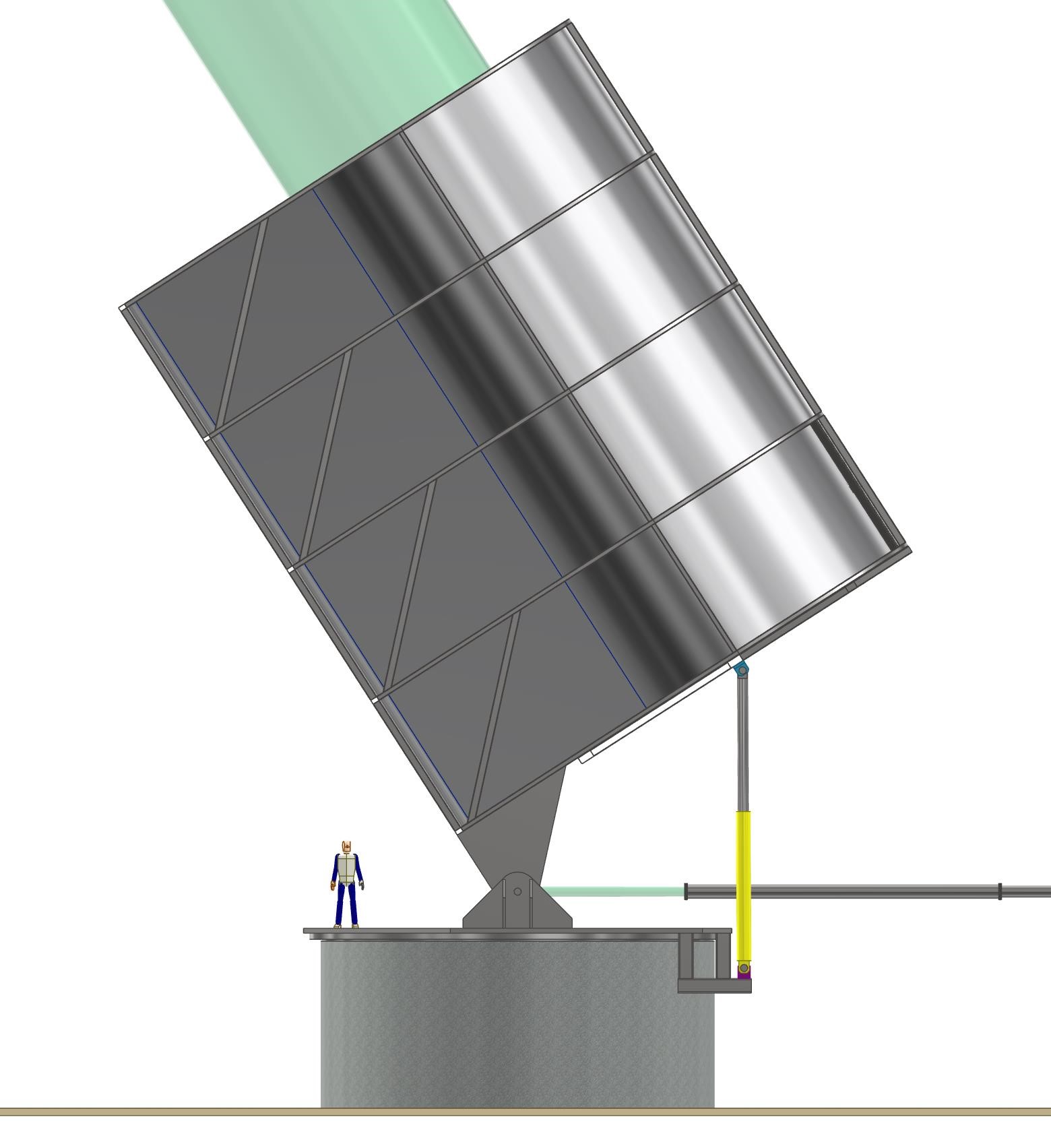} 
        \caption{\label{fig:FEL-BE-12} This view shows the maximum pitch angle from vertical, 30 degrees, created by the length of hydraulic cylinders shown.  An increased pitch angle requires longer hydraulics and lower attachment points to the azimuth assembly; however, larger angles become problematic for the mirror that directs the light into the expander.  The hydraulic cylinders are double-acting.  No hydraulic plumbing is shown on the alt/pitch cylinders.}
    \end{minipage} \hfill
    \begin{minipage}[t]{.49\textwidth}
        \centering
        \includegraphics[width=\textwidth]{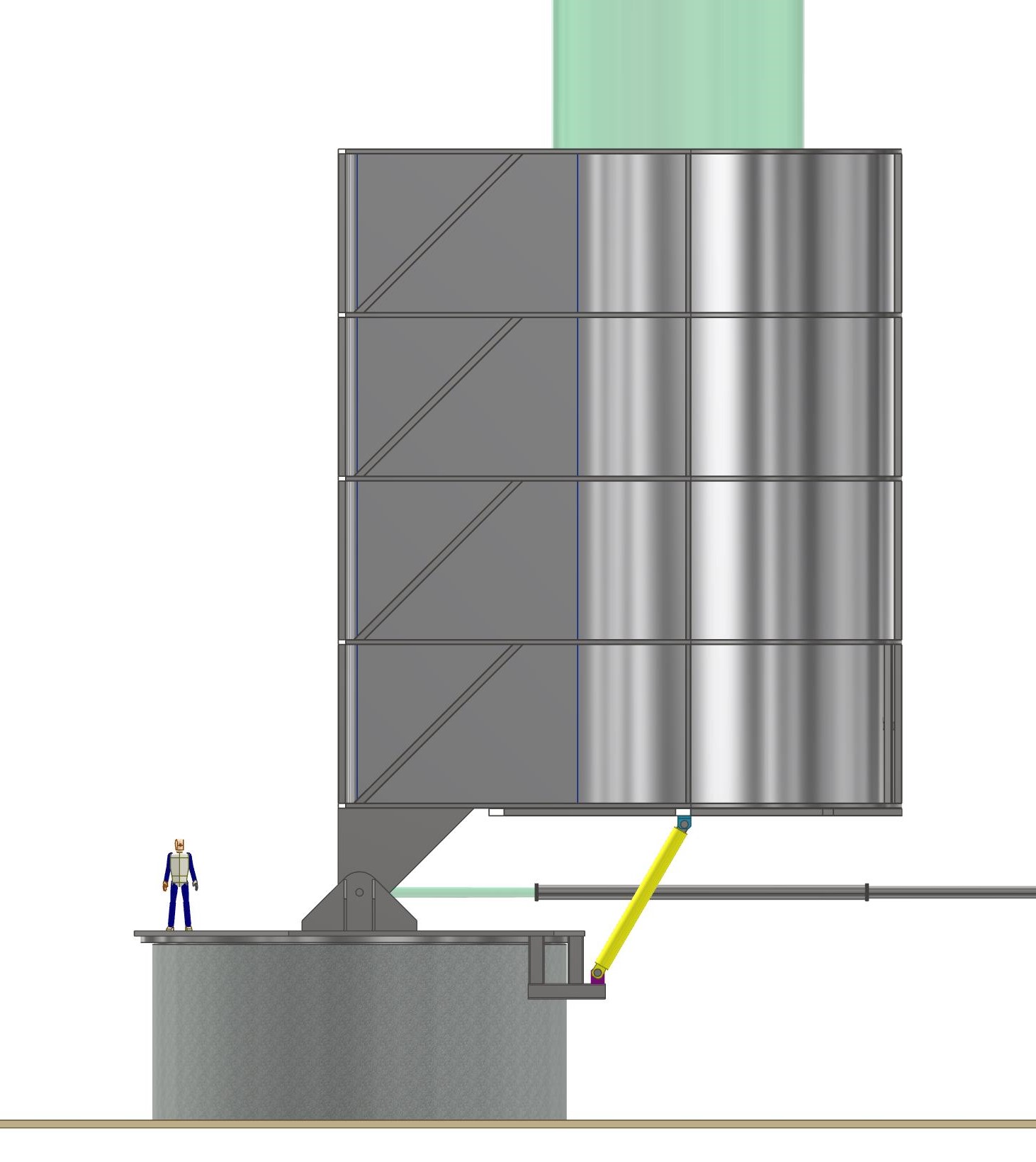} 
        \caption{\label{fig:FEL-BE-13} The Beam Expander is capable of pointing straight upwards if needed.  A larger pitch angle range would require a reconfiguration of the hydraulic cylinders.}
    \end{minipage}
\end{figure}

Figures \ref{fig:FEL-BE-12} and \ref{fig:FEL-BE-13} show the altitude motion through its range from 30 degrees off vertical to straight up.  Again, a different configuration of hydraulic cylinders could extend the range of motion to beyond vertical in the direction of the FEL, but this is probably not necessary given that the beam expanders will probably be set at a significant distance away from the base of the tether.

One aspect of the beam expander that has not been developed yet is where in the optical chain the adaptive optics would occur.  Adaptive optics (AO) is a feedback mechanism where information coming back from a reference beam in the atmosphere allows actuated mirror surfaces to adjust to turbulence in the atmosphere that would deflect the beam from illuminating the PV arrays on the climber.

\begin{figure}
\begin{centering}
\includegraphics[width=\textwidth]{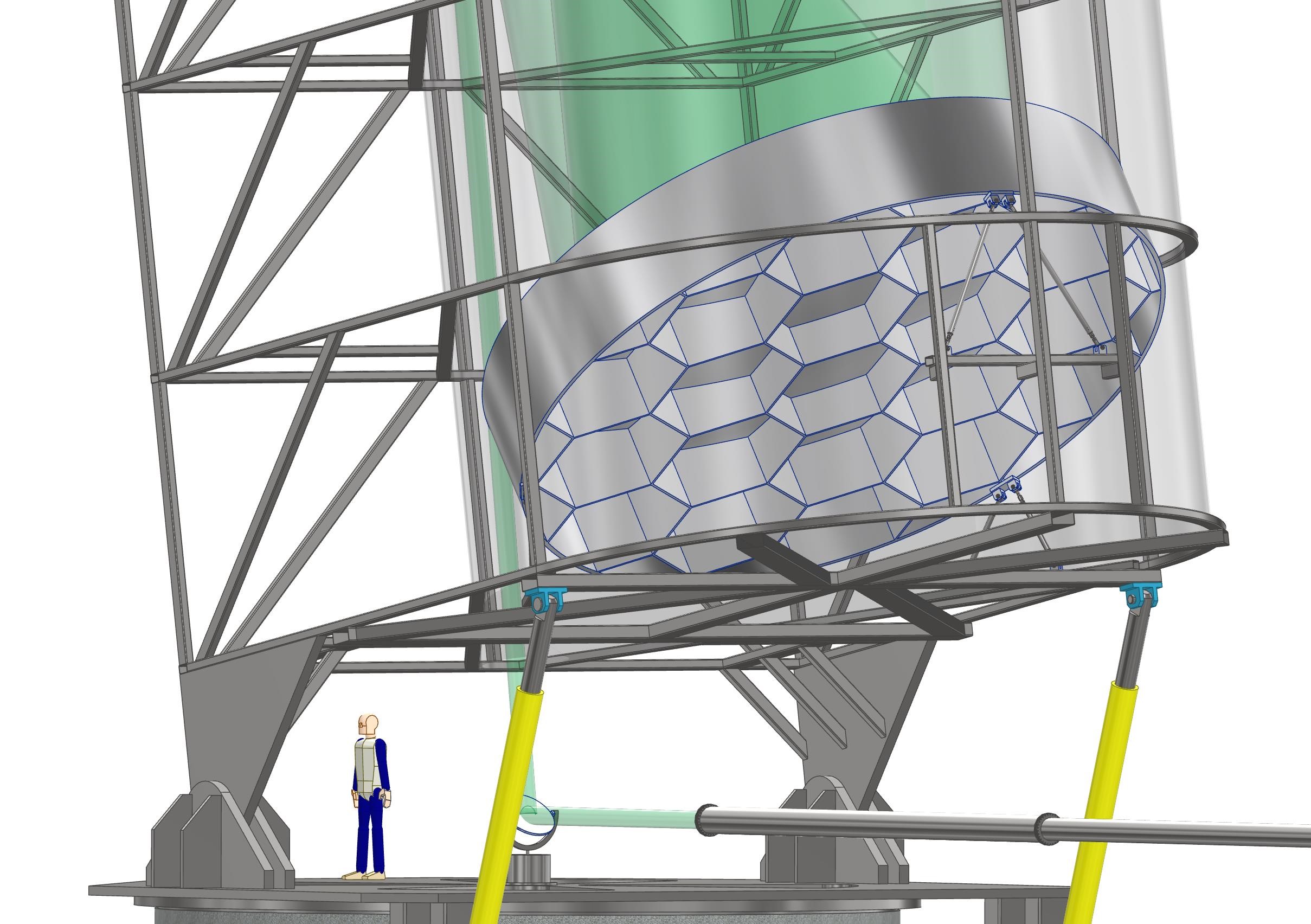} 
\caption{\label{fig:FEL-BE-14} A view of the Beam Expander showing the hexagonal hollows modeled in the primary mirror.  These hollows could be where actuators are located that position hexagonal mirrors, turning the primary mirror into an array of smaller hexagonal mirrors for adaptive optics.}
\end{centering}
\end{figure}

Figure \ref{fig:FEL-BE-14} shows a view of the beam expander where the shell is rendered transparent, exposing the bottom side of the primary mirror.  This mirror was modeled with hexagonal pockets in the backside that could represent multiple things.  First, any mirror like this will be hollowed out as much as possible to reduce weight.  Second, the primary mirror could be made of many individual hexagonal segments.  The pockets provide space behind the mirrored surface where actuators could be mounted to provide for the adaptive optics.

Any of the figures with a transparent shell around the beam expander, such as Figure \ref{fig:FEL-BE-14}, also show the conceptual design of the primary mirror mounting as a hexapod made with turnbuckles.  Final design will determine if these turnbuckles also need to be actuated remotely. 

\newpage
\subsubsection{Alternative design for the beam expander}

\begin{figure}
\begin{centering}
\includegraphics[width=\textwidth]{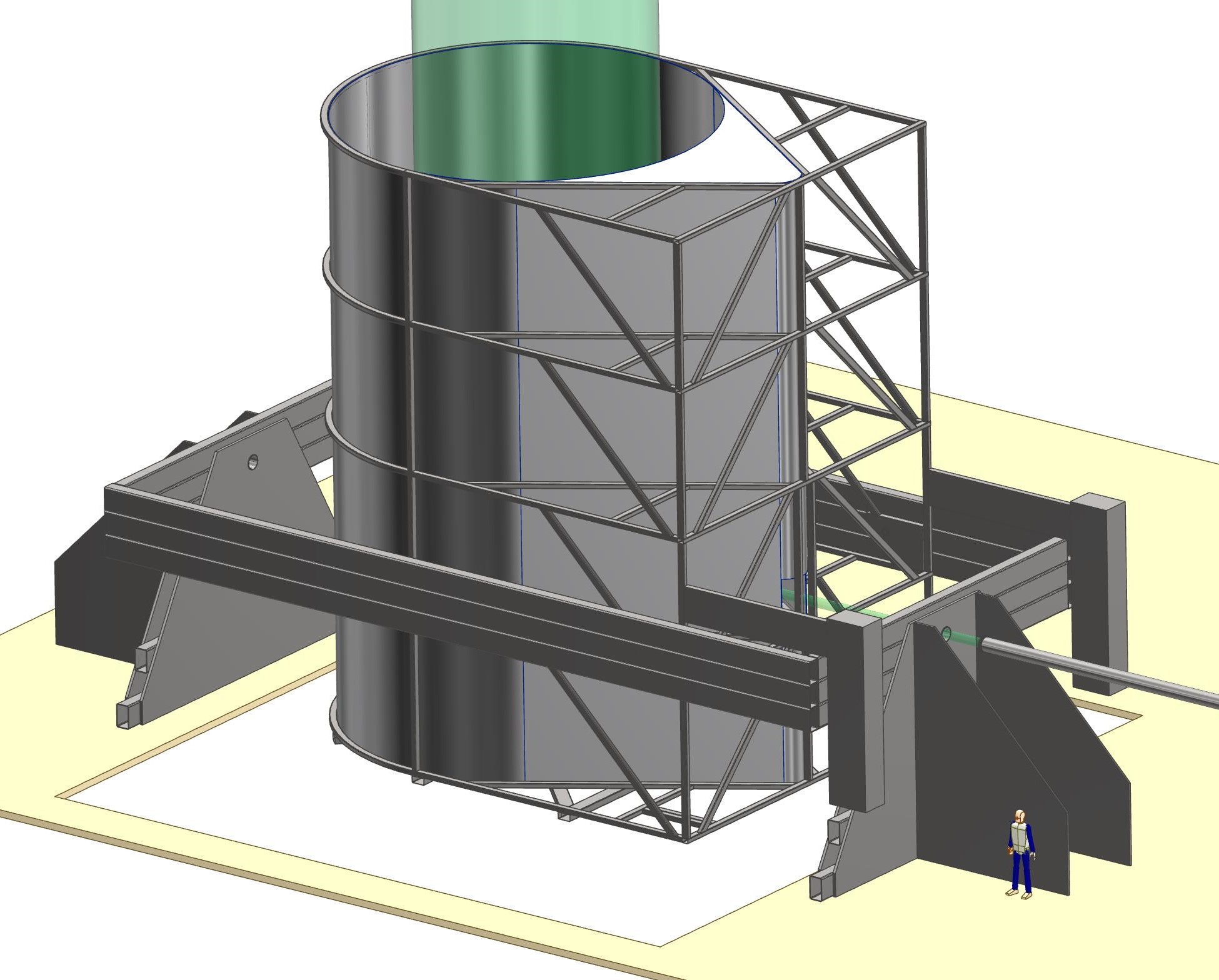} 
\caption{\label{fig:FEL-BG-04} A view of an alternative Beam Expander suggested by Blaise Gassend.}
\end{centering}
\end{figure}

\begin{figure}
\begin{centering}
\includegraphics[width=\textwidth]{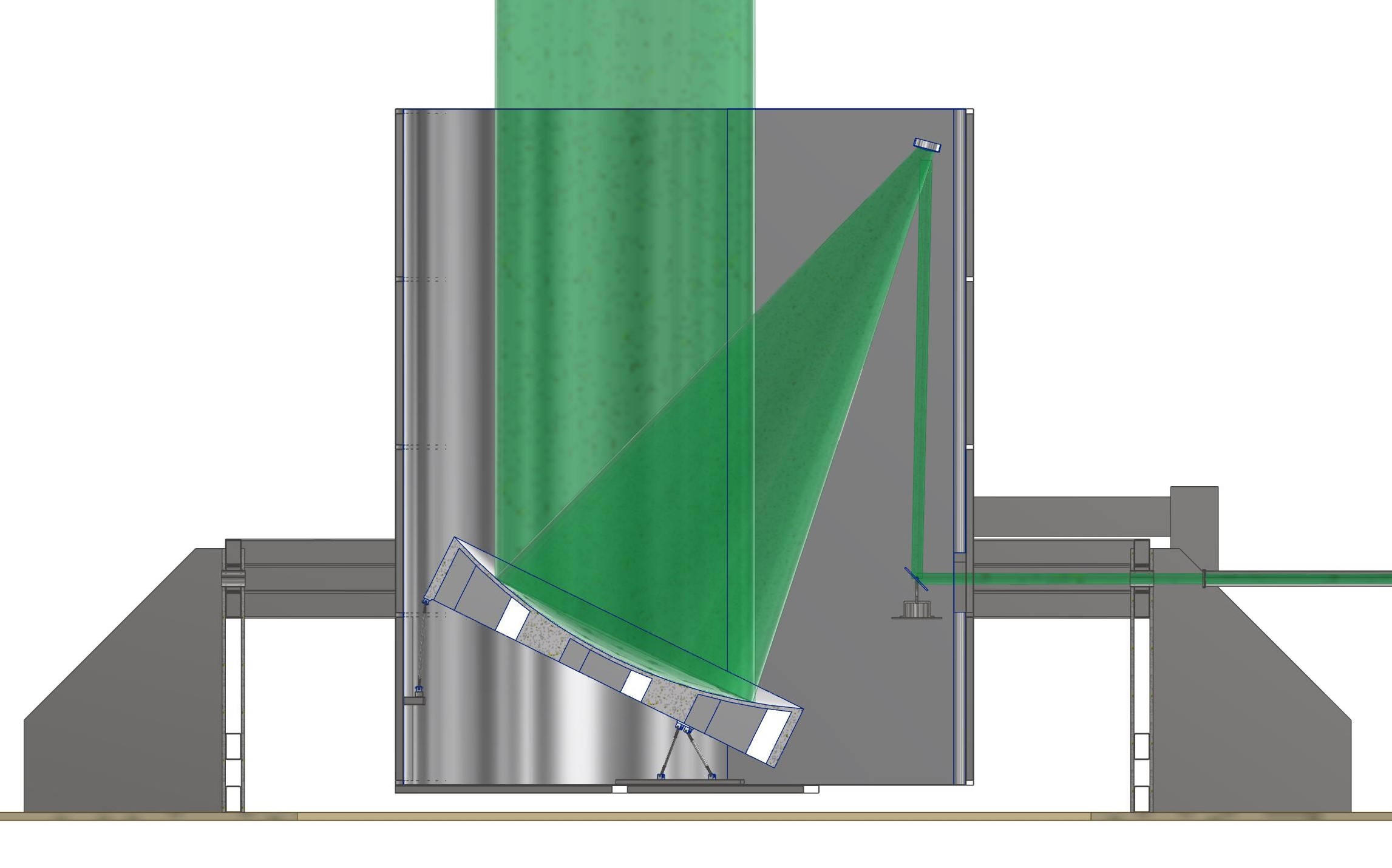} 
\caption{\label{fig:FEL-BG-03} A section view of an alternative Beam Expander showing the light path.}
\end{centering}
\end{figure}

\begin{figure}[t]
\begin{centering}
\includegraphics[width=\textwidth]{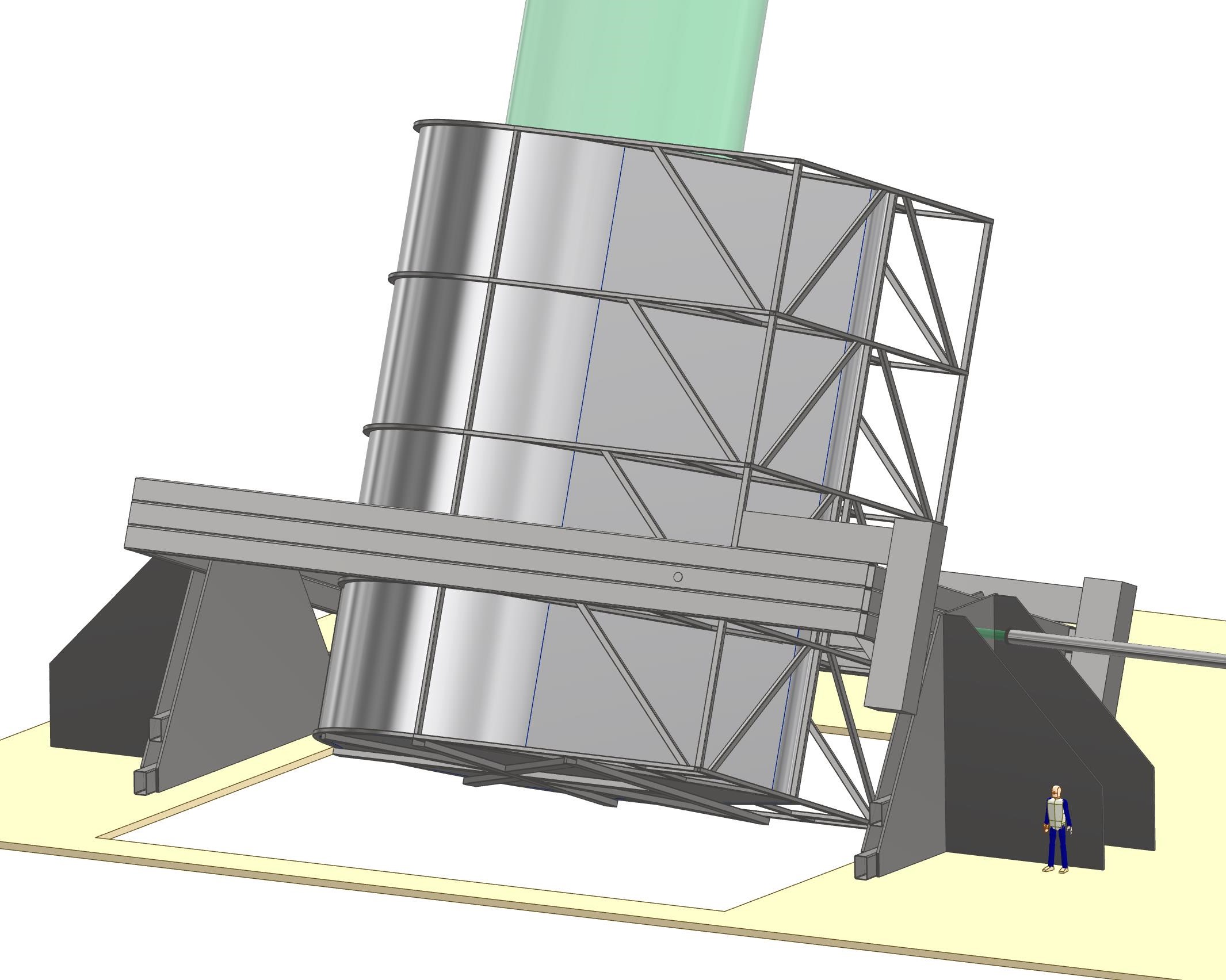} 
\caption{\label{fig:FEL-BG-02} A view of an alternative Beam Expander with the roll axis rotated around the FEL beam.  The ranges of roll and pitch rotation are very similar to those of the alt-az design.}
\end{centering}
\end{figure}

\begin{figure}[b]
\begin{centering}
\includegraphics[width=\textwidth]{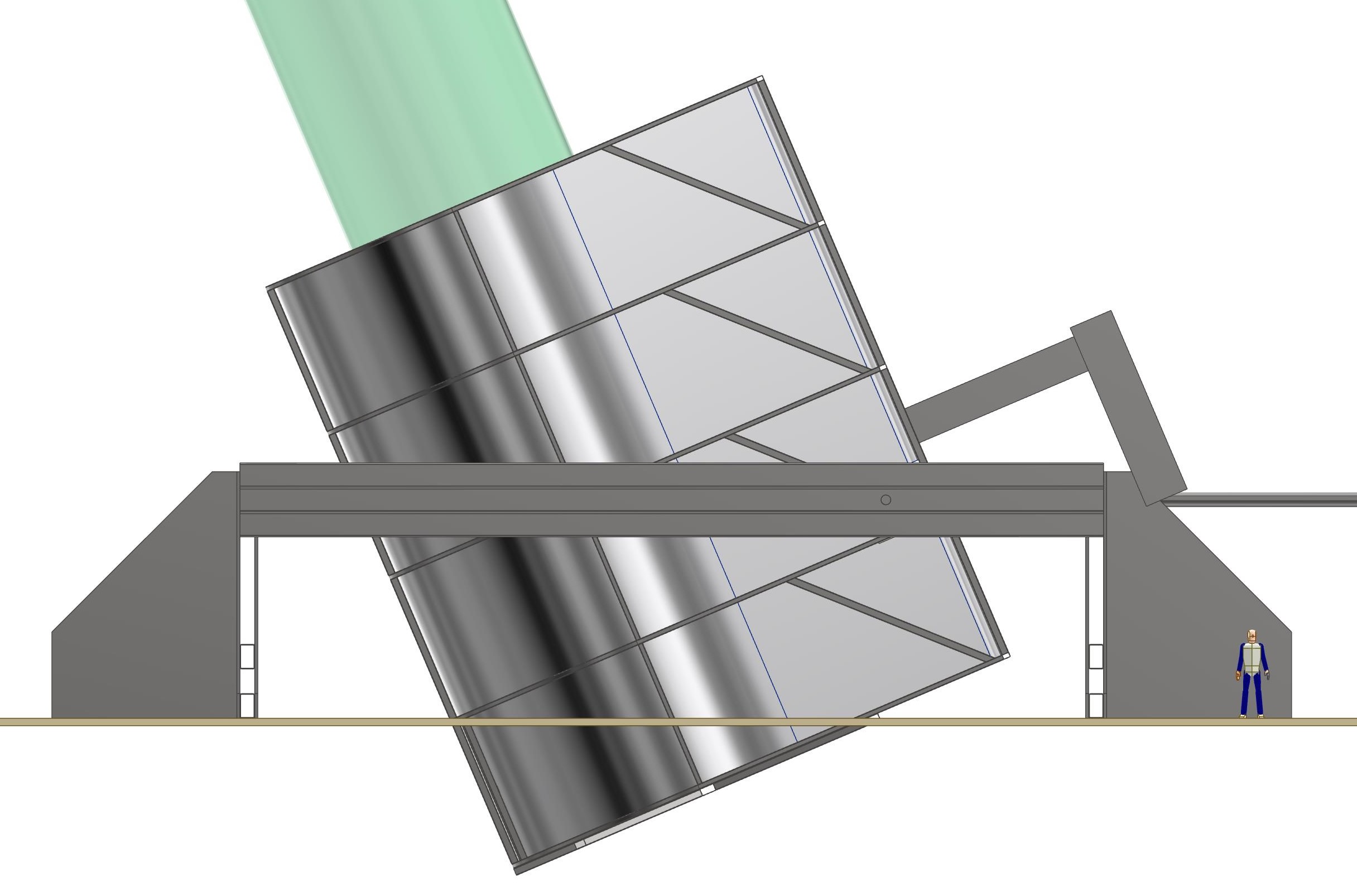} 
\caption{\label{fig:FEL-BG-05} A view of an alternative Beam Expander with the pitch axis rotated. There is a hole cut in the deck below the beam expander to allow for the pitch rotation, given the constraint of the height of the FEL beam from the floor.}
\end{centering}
\end{figure}

A criticism of the alt-az beam expander shown is that the pitch rotation axis is not balanced around its rotation point, requiring large forces from the hydraulic cylinders to rotate the expander.  Blaise Gassend proposed the alternative geometry shown in Figures \ref{fig:FEL-BG-04}, \ref{fig:FEL-BG-03}, \ref{fig:FEL-BG-02}, and \ref{fig:FEL-BG-05}.  No actuating mechanisms are shown in these figures.  The structure has not been fully analyzed for stress and deflection, but the stacked rectangular box beam weldment was given a preliminary analysis to size the cross-section of the weldment for deflection.

This beam expander requires much more structural steel than the alt-az design.  The housing of the offset Cassegrain expander is very similar in design to the alt-az model, except that making it counterweighted to be balanced around the pitch axis causes it to weigh 102 tons, instead of half that for the alt-az design.

Another difference between this design and the alt-az design is the additional end mounting structures that allow the roll rotation around the FEL beam axis.  These structures have not been analyzed for any of the lateral forces they may be required to absorb and could become larger.

Figure \ref{fig:FEL-BG-04} shows the overall concept in which one axis rolls around the axis of the FEL beam.  The other axis is orthogonal and pitches around pins in the rectangular weldment of tubular beams.  

This design is shown as one of several possible alternatives to the alt-az design presented.  Final design will determine the structure based on the final requirements and cost comparisons between different designs.

\FloatBarrier

\newpage
\subsubsection{Beam expander implications for the Earth anchor design}

The cover art for this report shows two FELs and two beam expanders mounted on each of two aircraft carriers.  This image was meant to convey the idea that the platforms for the lasers may not be land-based, but could be sea-based.  The distance between the beam expanders and the base of the tether is determined by two things: the altitude of the climber on the tether at which the lasers are turned on, and the altitude angle the beam expander is capable of rotating through.  A smaller altitude angle range, making the beam closer to vertical, forces the beam expander to be closer to the base of the tether than if it can point further away from vertical.  The higher the altitude at which the lasers are turned on, the further away the beam expander can be from the base of the tether.  Climbers closer to GEO can be powered from lasers that are kilometers away from the base of the tether.

\begin{figure}
\begin{centering}
\includegraphics[width=\textwidth]{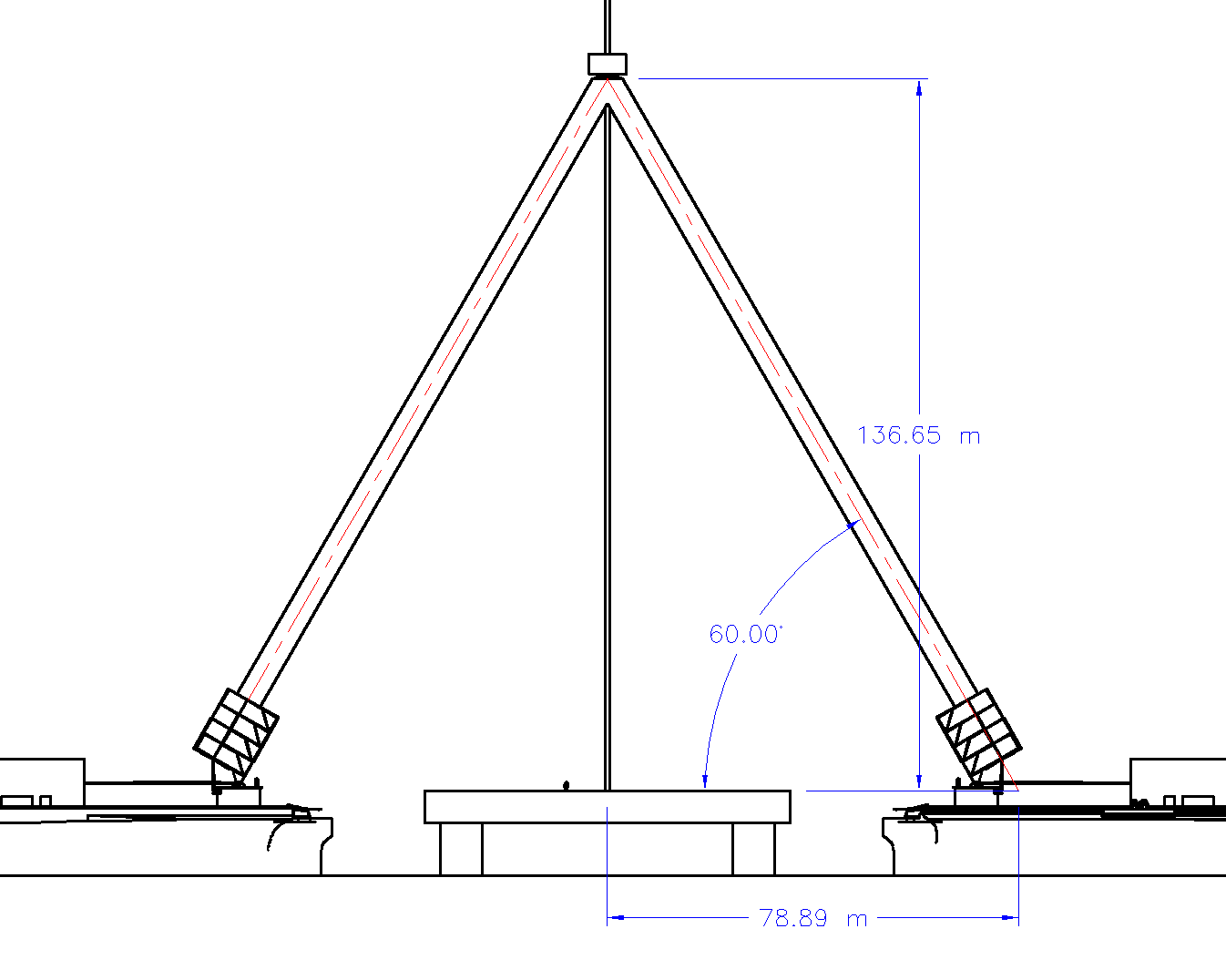} 
\caption{\label{fig:FEL-02-06} Simplified elevation view drawing of the Earth Anchor platform with aircraft carriers to either side of it beaming power to a climber.}
\end{centering}
\end{figure}

Figure \ref{fig:FEL-02-06} shows a line drawing taken from the cover art model giving the dimensions between the beam expanders and the anchor platform for beam expanders that can point $30 \degree$ away from vertical.  This configuration is not realistic as the aircraft carriers are too close to the platform for rough sea conditions.  (The only way this configuration could work is if the tether anchor, FELs, and beam expanders were all on the same platform.) But it can be used to scale the numbers for more realistic configurations.  For example, if the climber starts receiving beam at 1 km up the tether, the horizontal distance completing the triangle would be:

$$x = \frac{y}{\mathrm{tan}(60\degree)} = \frac{1 \ \si{km}}{1.7321} = 0.577 \ \si{km}$$

The beam expanders would be based almost .6 km away from the tether base for a climber starting to receive power 1 km up and with beam expanders that can rotate $30 \degree$ away from vertical.

Setting power beaming platforms such that their angular ranges of motion cover single climbers at the altitudes shown in Table \ref{tab:8climbFEL} gives the horizontal distances between the base of the tether and the power platforms.

One aspect of a sea-based platform that is very problematic is the ubiquitous presence of salt spray in the atmosphere getting on all of the optical mirror surfaces.  The beam expander mirrors must be kept clean to preserve the beam quality.  Smaller mirrors such as the flat mirror at the base of the expander and the M2 mirror also see a much higher power density and dust and other imperfections on their surfaces will lead to burning and mirror damage.

Perhaps the biggest driver of the distance between the beam expanders and the base of the tether is the range of motion the base of the tether needs to move through to avoid collisions with satellites in LEO and MEO.  The strategy for avoiding these collisions between the tether and satellites has been to induce transverse motion of the tether by moving the base. Base motion will translate the tether in LEO away from the approaching satellite.  ISEC has not been able to determine whether the base of the tether needs to move through +/- 100 m, or +/- 1 kilometer or more to avoid orbital collisions.  A larger movement of the base would either require close coordination of the laser platforms with the tether base platform, or enough distance between them that the tether base motion can occur without affecting the laser platforms.

\newpage
\subsection{Conceptual design of the PV arrays on the climber to receive the laser light}
\subsubsection{Matching the photovoltaic (PV) technology of the receivers on the climber to the laser frequency}
\label{sec:PVarray}

The problem of collecting laser light and converting it into electricity is very different from collecting sunlight and converting that into electricity.  Sunlight contains a large range of wavelengths and provides a uniform energy distribution on PV surfaces.  Lasers are monochromatic, emitting a single or very narrow band of wavelengths.  The goal of solar cell manufacturers is to design panels that can absorb and convert to electricity as many different wavelengths of light as possible.  The goal of the designers of the PV arrays on the climber is to maximize the efficiency of converting the monochromatic light of the laser to electricity and creating the lightest, smallest receiver size possible.

There are three main categories of photovoltaic cell in use today: conventional mono- and poly-crystalline silicon (c-Si) cells, thin film solar cells (a-Si, CIGS and CdTe), and multi-junction (MJ) solar cells.  The oldest and cheapest technology in common use is conventional mono- and poly-crystalline silicon cells.  This material has a peak light conversion efficiency in the near infrared and so is unsuitable for 500 nm laser light.

\begin{table}[hb]
    \centering
    \begin{tabular}{|l|l|c|c|c|c|c|}
        \hline
        Categories & Technology & $\eta$ (\%) & $V_{\mathrm{OC}}$ (V) & $I_{\mathrm{SC}}$ (A) & W/m$^{2}$ & $t$ ($\mu$m) \\
        \hline
        \multirow{2}{*}{Crystalline silicon cells} 
            & Monocrystalline & 24.7 & 0.5   & 0.8  & 63  & 100 \\ \cline{2-7}
            & Polysilicon     & 20.3 & 0.615 & 8.35 & 211 & 200 \\ 
        \hline
        \multirow{3}{*}{Thin film solar cells}
            & Amorphous silicon & 11.1 & 0.63 & 0.089 & 33 & 1 \\ \cline{2-7}
            & CdTe              & 16.5 & 0.86 & 0.029 & -- & 5 \\ \cline{2-7}
            & CIGS              & 19.5 & --   & --    & -- & 1 \\
        \hline
        Multi-junction cells & MJ & 40.7 & 2.6 & 1.81 & 476 & 140 \\
        \hline
    \end{tabular}
    \caption{Categories of photovoltaic technologies in use now with typical properties.  $\eta$ is the efficiency of the technology.  $V_{OC}$ (Open-Circuit Voltage) is the maximum voltage a solar cell or panel can produce when no current is flowing.  $I_{SC}$ (Short-Circuit Current) is the maximum electrical current a solar cell or panel can produce when its positive and negative terminals are directly connected (shorted), resulting in zero voltage across them, and it indicates the cell's maximum current-producing capability under specific light conditions.  $\si{W/m^2}$ is the nominal power of PV devices measured under standard test conditions.  The power is measured while varying the resistive load on the module between an open and closed circuit (between maximum and minimum resistance). The highest power thus measured is the 'nominal' power of the module in watts.  t is the thickness of the photocell.  From [\cite{multijunc}].}
    \label{tab:PVtech}
\end{table}

Table \ref{tab:PVtech} shows that the efficiency of thin film solar cells is far less than either the silicon cells or the multi-junction cells.  They are primarily used in small products where flexibility of the array is important.  The efficiency of multi-junction cells is better than that of silicon.  This is because multi-junction photocells are made of multiple layers of semiconductor materials sensitive to different wavelength bands.

Multi-junction solar cells are more expensive and harder to make than the other types of cells, but their high efficiency and high power-to-weight ratio makes them ideal for aerospace applications.

\begin{figure}
\begin{centering}
\includegraphics[width=\textwidth]{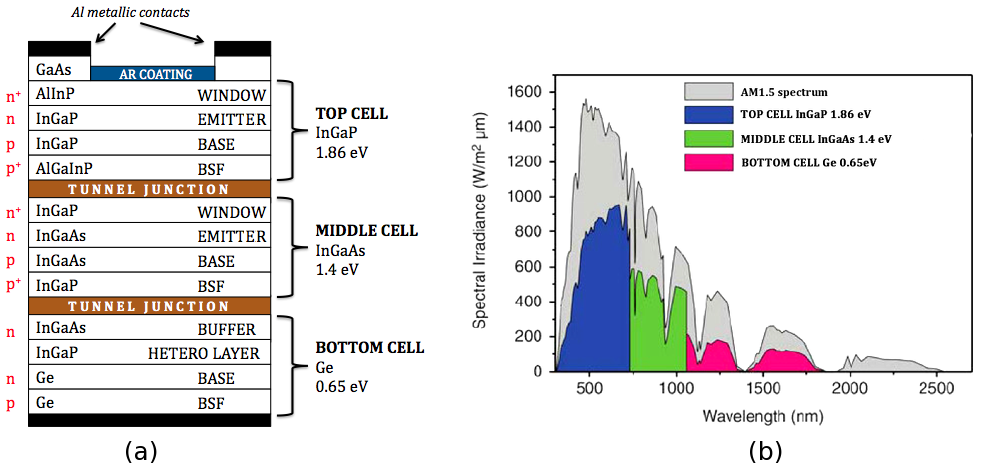} 
\caption{\label{fig:multij} The structure of a multi-junction photocell showing the sensitivities of the different layers to different wavelengths of light.  AR stands for antireflective coating.  From [\cite{multijunc}].}
\end{centering}
\end{figure}

Figure \ref{fig:multij} shows a multi-junction solar cell with three layers sensitive to three different bands of light.  The band of interest is the top layer, the Indium Gallium Phosphide (InGaP) layer.  This semiconductor has its peak collection efficiency near 500 nm.  In designing custom PV arrays for the climbers, the other layers of different materials can be eliminated and replaced with multiple layers of the same optimum material.

\begin{table}
    \centering
    \begin{tabular}{lccc}
        \toprule
        Material & band gap type & band gap energy & bandgap wavelength \\
        \midrule
        lead selenide (PbSe)         & direct  & 0.27 eV  & 4.57 $\mu$m \\
        lead telluride (PbTe)        & direct  & 0.32 eV  & 3.86 $\mu$m \\
        indium arsenide (InAs)       & direct  & 0.36 eV  & 3.43 $\mu$m \\
        lead sulfide (PbS)           & direct  & 0.37 eV  & 3.34 $\mu$m \\
        germanium (Ge)               & indirect& 0.67 eV  & 1.84 $\mu$m \\
        gallium antimonide (GaSb)    & direct  & 0.726 eV & 1.70 $\mu$m \\
        silicon (Si)                 & indirect& 1.12 eV  & 1.1 $\mu$m  \\
        indium phosphide             & direct  & 1.35 eV  & 915 nm      \\
        gallium arsenide (GaAs)      & direct  & 1.43 eV  & 873 nm      \\
        cadmium telluride (CdTe)     & direct  & 1.5 eV   & 828 nm      \\
        cadmium selenide (CdSe)      & direct  & 1.74 eV  & 712 nm      \\
        aluminum arsenide (AlAs)     & indirect& 2.12 eV  & 585 nm      \\
        gallium phosphide (GaP)      & indirect& 2.24 eV  & 551 nm      \\
        cadmium sulfide (CdS)        & direct  & 2.42 eV  & 513 nm      \\
        gallium nitride (GaN)        & direct  & 3.2 eV   & 389 nm      \\
        cubic zinc sulfide (ZnS)     & direct  & 3.54 eV  & 349 nm      \\
        hexagonal zinc sulfide (ZnS) & direct  & 3.91 eV  & 318 nm      \\
        aluminum nitride (AlN)       & direct  & 6.015 eV & 205 nm      \\
        \bottomrule
    \end{tabular}
    \caption{Type and energy for some common semiconductor materials ordered by increasing band gap energy. From [\cite{bandgap}]}
    \label{tab:bandgaps}
\end{table}

Table \ref{tab:bandgaps} shows the band gap energy and associated bandgap wavelength for some common semiconductors.  To see why the InGaP photocell converts efficiently at the 500 nm wavelength incident light, we calculate the energy of a 500 nm photon.  Given the relation $E=hc/\lambda$, where h = Planck's constant, c is the speed of light, and $\lambda$ is the wavelength, we find that the energy of a 500 nm photon is 2.48 eV.  The GaP semiconductor's band gap energy is 2.24 eV.  (Indium is added to tune the band gap energy.)  If a 500 nm photon hits a semiconducting material whose band gap is 2.48 eV or less, electrons and other charge carriers are excited in the material.  This is the basis of the photovoltaic effect.  This excitation causes the electric field that leads to the conversion of the incoming light to electricity.  When photons interacting with the semiconductor material have higher or lower energies than the bandgap energy, they contribute to waste heat in the semiconductor.  Temperature has a large effect on the efficiency of the photocell.  Higher temperatures lead to lower efficiencies.

Reference [\cite{york2017high}] describes the modeling and actual performance of Laser Power Converters (LPC) made from a novel vertical epitaxial heterostructure architecture (VEHSA).  Unfortunately for this application, all of the work done in this paper involved multijunctions of GaAs semiconductor whose peak conversion efficiency is from 810 nm 850 nm, in the near infrared spectrum.  That peak efficiency is much higher than for multijunction photocells converting the solar spectrum.  Quoting [\cite{york2017high}], Recent progress with VEHSA devices has demonstrated the following performance attributes:

\begin{itemize}
    \item The highest optical to electrical efficiency ever achieved
    \item The highest output powers ever reported for a high-efficiency monolithic PV cell with 5.87 W of converted output.
    \item The highest efficiencies ever reported for any types of optical to electrical power conversion devices simultaneously combining high photovoltage and output powers (\textgreater 5 W at \textgreater 7 V with \textgreater 60\% efficiency and \textgreater 3 W at \textgreater 14 V with \textgreater 60\% efficiency).
    \item Unique experimental evidence of the significant impact of photon recycling in these photovoltaic devices.
    \item The highest efficiency ever reported of 61.8\% with a significantly detuned optical input.
\end{itemize}

\begin{figure}
\begin{centering}
\includegraphics[width=.7\textwidth]{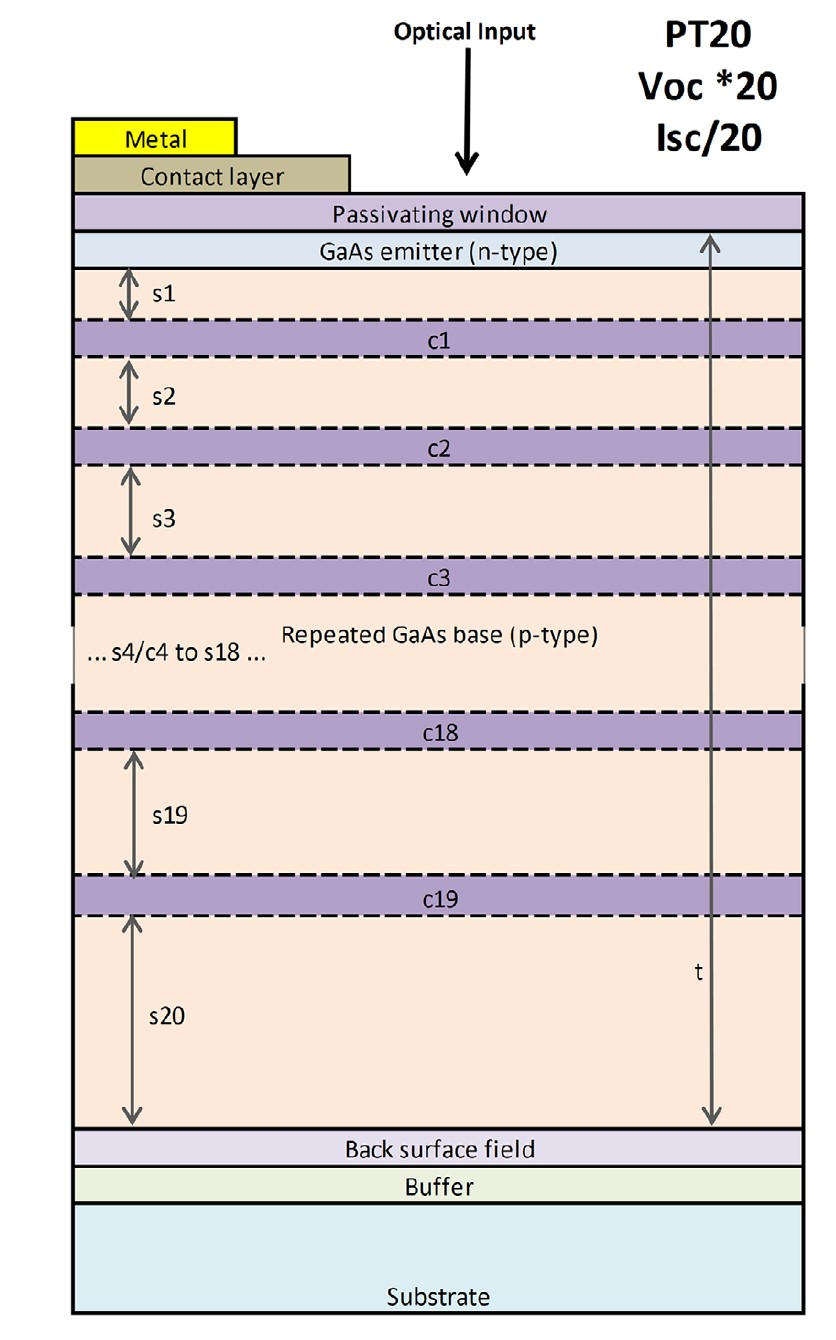} 
\caption{\label{fig:VEHSA} The structure of a VEHSA multi-junction photocell made from 20 layers of GaAs cells, as distinct from the multi-junction cells for solar spectrum conversion shown in Fig \ref{fig:multij}.  From [\cite{york2017high}].}
\end{centering}
\end{figure}

\begin{figure}
\begin{centering}
\includegraphics[width=\textwidth]{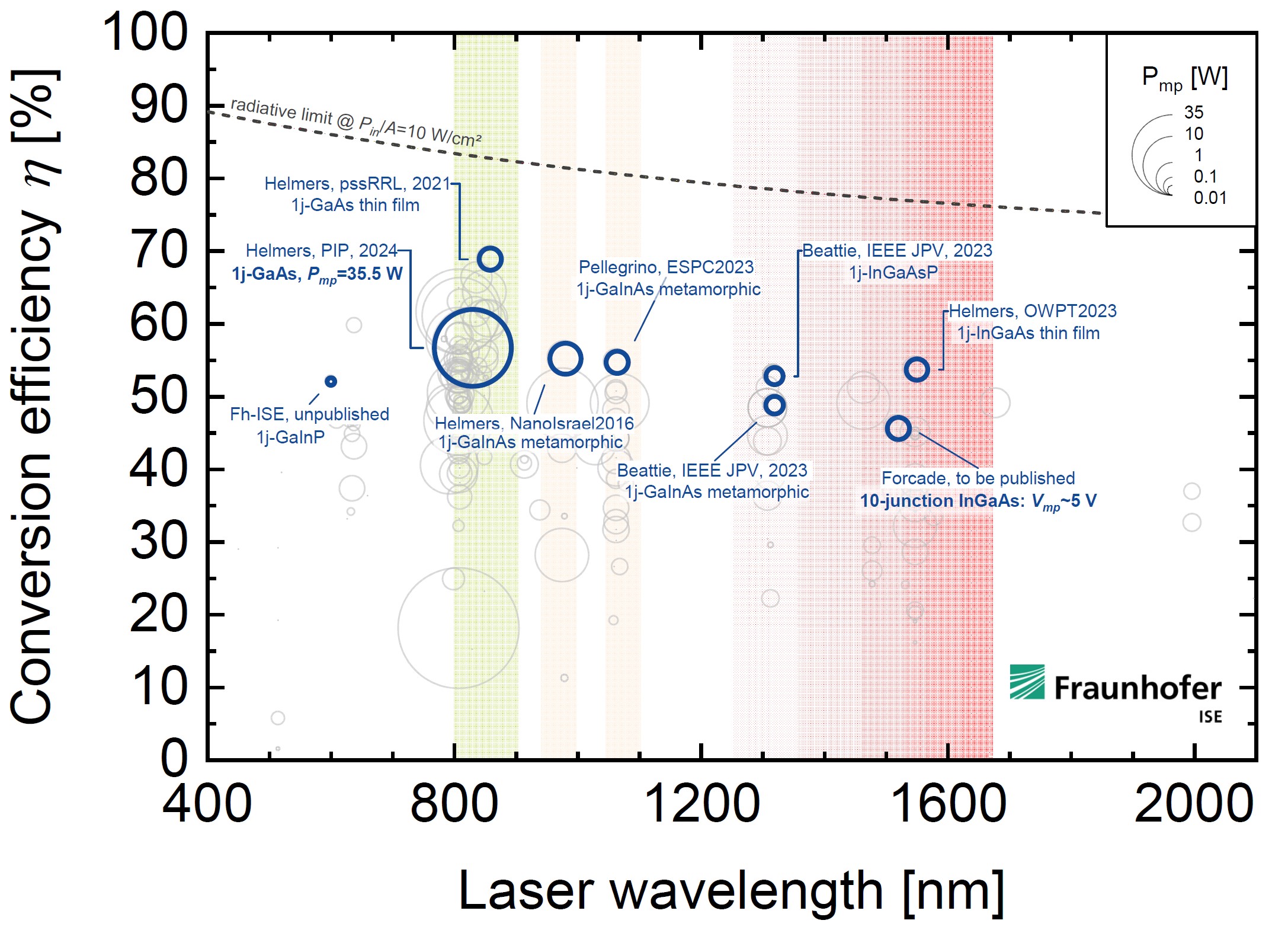} 
\caption{\label{fig:laserconv} Measured conversion efficiencies of photonic power converters for various wavelengths. The size of the circle symbols scales with electrical output power $P_mp$. Blue symbols represent selected PPC devices fabricated at Fraunhofer ISE; the captions state absorber materials and device-specific remarks. Gray symbols represent literature data of various research groups.  From [\cite{helmers2024advancing}].}
\end{centering}
\end{figure}

One of the challenges of this study has been to project the efficiency of the LPC PV arrays on the climber. The highest conversion efficiency of solar spectrum conversion at 47.6\% (shown in [\cite{helmers2024advancing}]), seems inappropriate given the development of the VEHSA architecture.  As [\cite{helmers2024advancing}] shows though, most of the work in VEHSA architecture has been done in GaAs for the near infrared region.  Figure \ref{fig:laserconv} shows the conversion efficiencies as a function of wavelength for various devices made at the Fraunhofer Institute and elsewhere.  There is a paucity of data for devices that convert in the 500 nm wavelength.  The two gray circles above the 500 nm tick mark in the graph show very low efficiencies, while the devices in the 600-700 nm range go all the way up to 60\% efficiency.  It is not known at this point whether concentrated research effort in VEHSA architecture (or others) using InGaP will yield conversion efficiencies above 60\%.  Reference [\cite{klitzke2025rear}] is encouraging because it reports 59\% efficiency at 590 nm. This remains an area of research that will need to be done in the development of climber PV arrays.  The final efficiency of these arrays has two effects:

\begin{enumerate}
    \item Lower efficiency PV arrays will be hotter from waste heat which makes the temperature management problem of the PV arrays more difficult.
    \item Lower efficiency PV arrays require larger ground wall-plug powers, which makes the ground infrastructure more expensive.
\end{enumerate}

As noted in [\cite{cichelli2025photovoltaic}], laser light is monochromatic and illuminates a PV surface with a cylindrical beam whose power density varies across the array as a Gaussian function.

\begin{equation}
    I(r,z) = I_0 \left( \frac{w_0}{w(z)} \right)^2 \exp\left( -\frac{2r^2}{w(z)^2} \right)
    \label{eq:intensity}
\end{equation}
\\
where the peak intensity in the middle of the circular PV array is $I_0 = 2P/(\pi w_o^2)$.  $w_0$ is the beam radius at the waist, as defined in Section \ref{sec:diff}.  P is the optical power of the beam, and w(z) is defined in \ref{eq:beamradius}.  Considering the four 5 MW laser configuration, near the surface of the Earth, P = 10 MW on a single array and $w_o$ = 2.5 m.  This gives a maximum beam intensity in the center of the PV array as 1.02 \si{MW/m^2}. The intensity of sunlight at noon at ground level is about 1 \si{kW/m^2}, so the laser is illuminating the center of the PV array near the ground with more than 1,000X the intensity of sunlight.  Concentrated photovoltaics (CPV) have sustained an intensity of 1,000 suns, but the heat management problem on the PV cells is an engineering challenge.

At GEO, the waist of the beam is 3.38 m, and with a single 5 MW laser beaming from the ground at full power (as shown in Figure \ref{fig:1FEL25m}), the power falling on a single climber PV array would be about 1 MW.  This gives the intensity in the center of an array as 56 \si{kW/m^2}.  Since at GEO the insolation is 1,368 \si{W/m^2}, the intensity of the lasers would be 41X that of sunlight.  

To quote [\cite{cichelli2025photovoltaic}], "Space-dependent intensity makes the design much more complicated compared to the case of uniform solar distribution, because the single cells over the surface of the PV panel are operated at different incident intensities, hence different operating points along the I–V curve. This requires optimization of the panel geometry, specifically the arrangement of series and parallel connections between cells, to avoid bottle-necking effects on the photocurrents. In addition, temperature variations between different parts of the panel must be taken into proper account for an accurate modeling of system performance."

For a given distance between source and PV array, we can simplify the equation for the variable intensity across the PV array as:

\begin{equation}
    I(r) = \frac{2P}{\pi w^{2}} \exp\left(-\frac{2r^{2}}{w^{2}}\right)
    \label{eq:arrayintensity}
\end{equation}

Reference [\cite{cichelli2025photovoltaic}] proposes and analyzes two different layouts of photocells into arrays, a ring structure and a tree structure.  The paper assumed vertically-stacked InGaAsP triple-junctions illuminated by 1064 nm laser light.  The goal was to optimize energy conversion efficiency while complying with the mechanical and electrical constraints required by space applications.  The authors were primarily modeling power beaming into permanently shadowed craters on the moon over much shorter distances than from Earth to GEO, but many of their conclusions about panel structuring apply to the climber PV arrays.

\begin{figure}
\begin{centering}
\includegraphics[width=.88\textwidth]{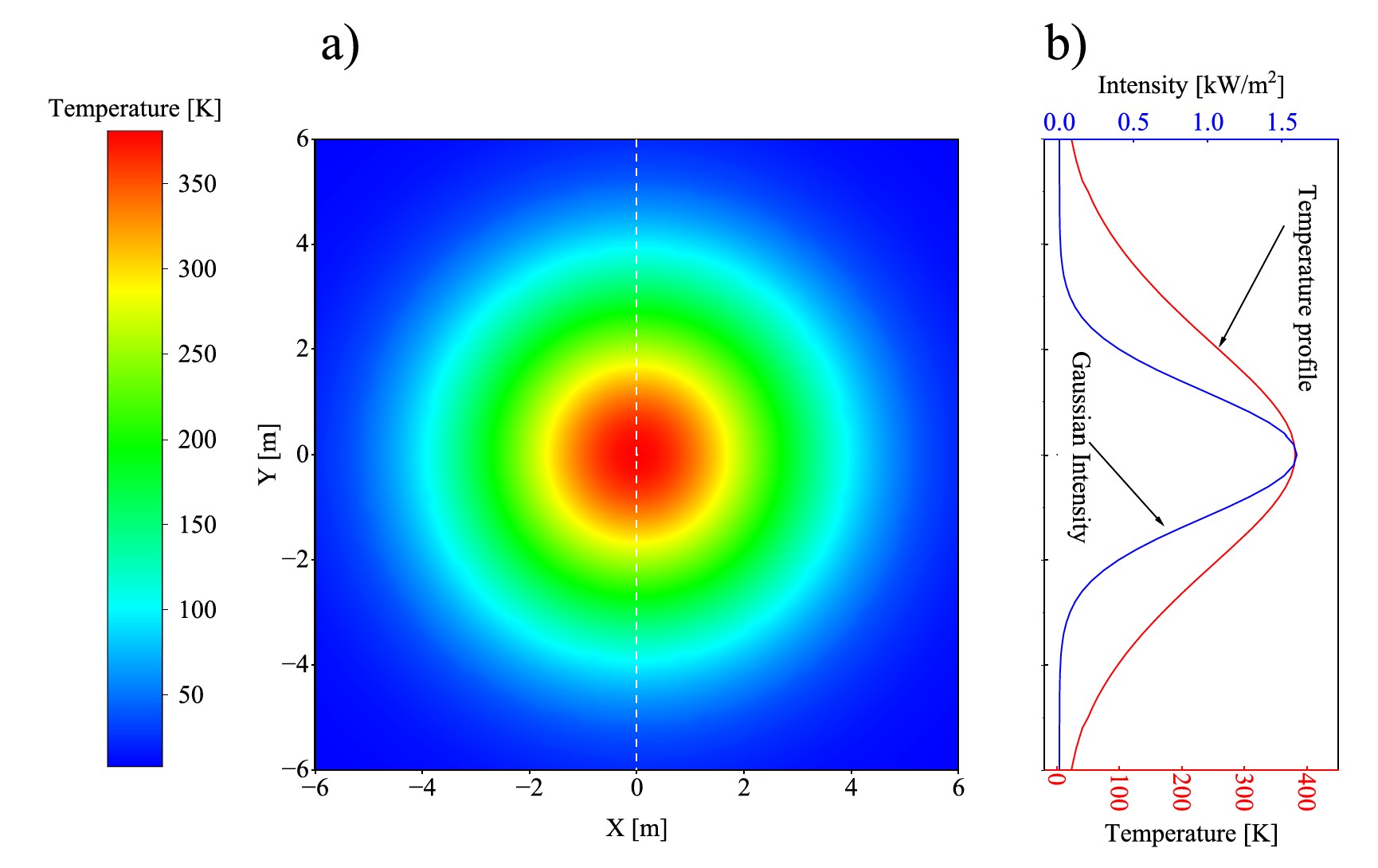} 
\caption{\label{fig:gausstemp} (a)Two-dimensional thermal map of the panel exposed to a Gaussian beam with 1600 \si{W/m^2} peak intensity. (b) Comparison between the temperature profile and the Gaussian intensity distribution across a vertical section of the panel. The thermal spatial profile of the two-dimensional InGaAsP PV panel is modeled by finite element analysis.  From [\cite{cichelli2025photovoltaic}].}
\end{centering}
\end{figure}

Figure \ref{fig:gausstemp} motivates the reasoning behind laying out the PV cells in a ring array.  Because of the non-uniform distribution of energy across the array, the temperature of the cells will be higher in the center where the intensity is the maximum.  Temperature is another factor in the performance of the photocells.  Higher temperatures degrade the efficiency of the photocells.

\begin{figure}
\begin{centering}
\includegraphics[width=.9\textwidth]{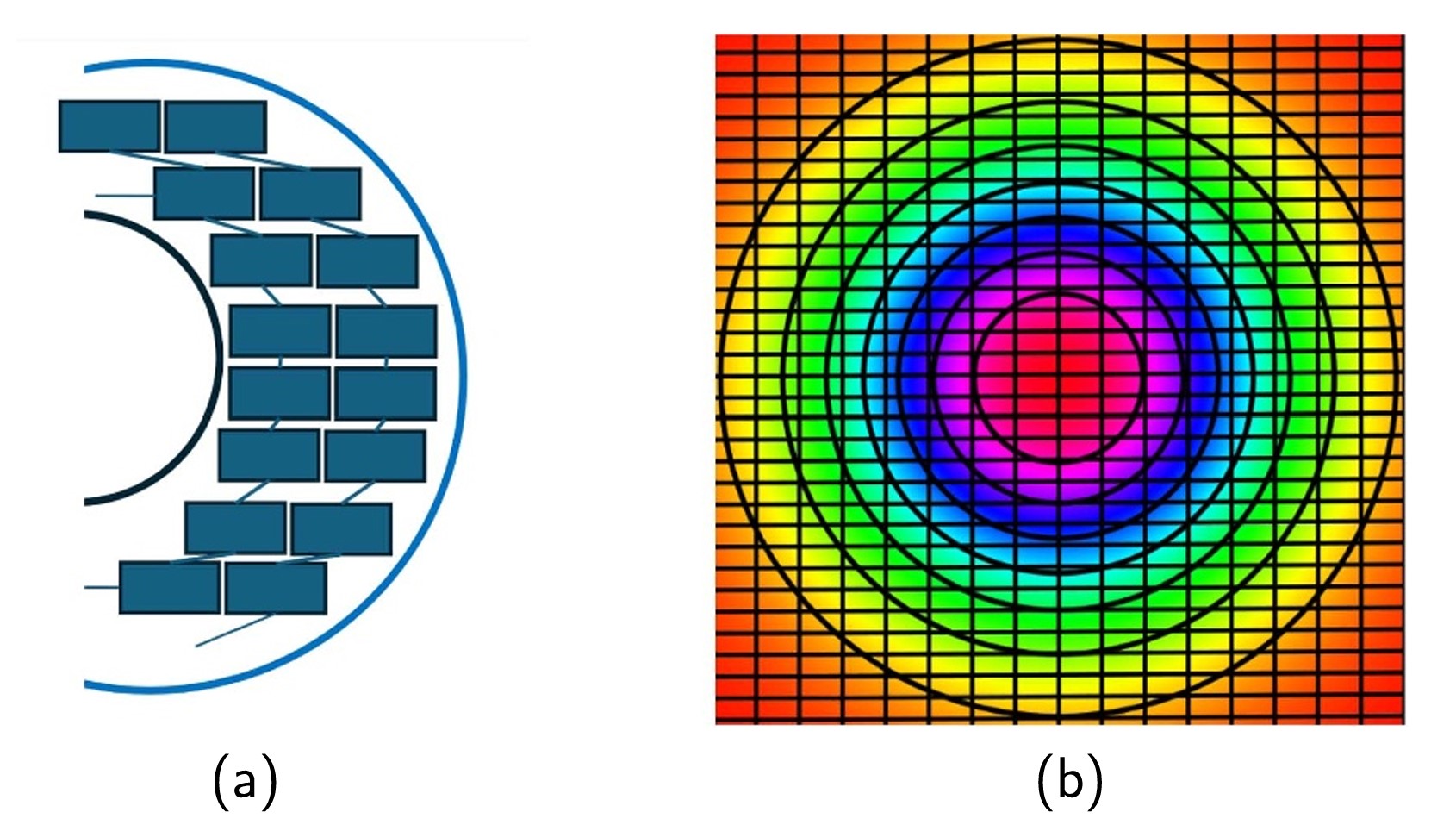} 
\caption{\label{fig:ringarray} Ring-Shaped design. (a) 8-cell strings in Ring-Shaped design interconnection scheme. (b) Optical field intensity across the cells.  From [\cite{cichelli2025photovoltaic}].}
\end{centering}
\end{figure}

Figure \ref{fig:ringarray} shows the connection pattern between photocells that occupy the same annular area, so that they would be at similar temperatures and electrical output performance.

\begin{figure}
\begin{centering}
\includegraphics[width=.9\textwidth]{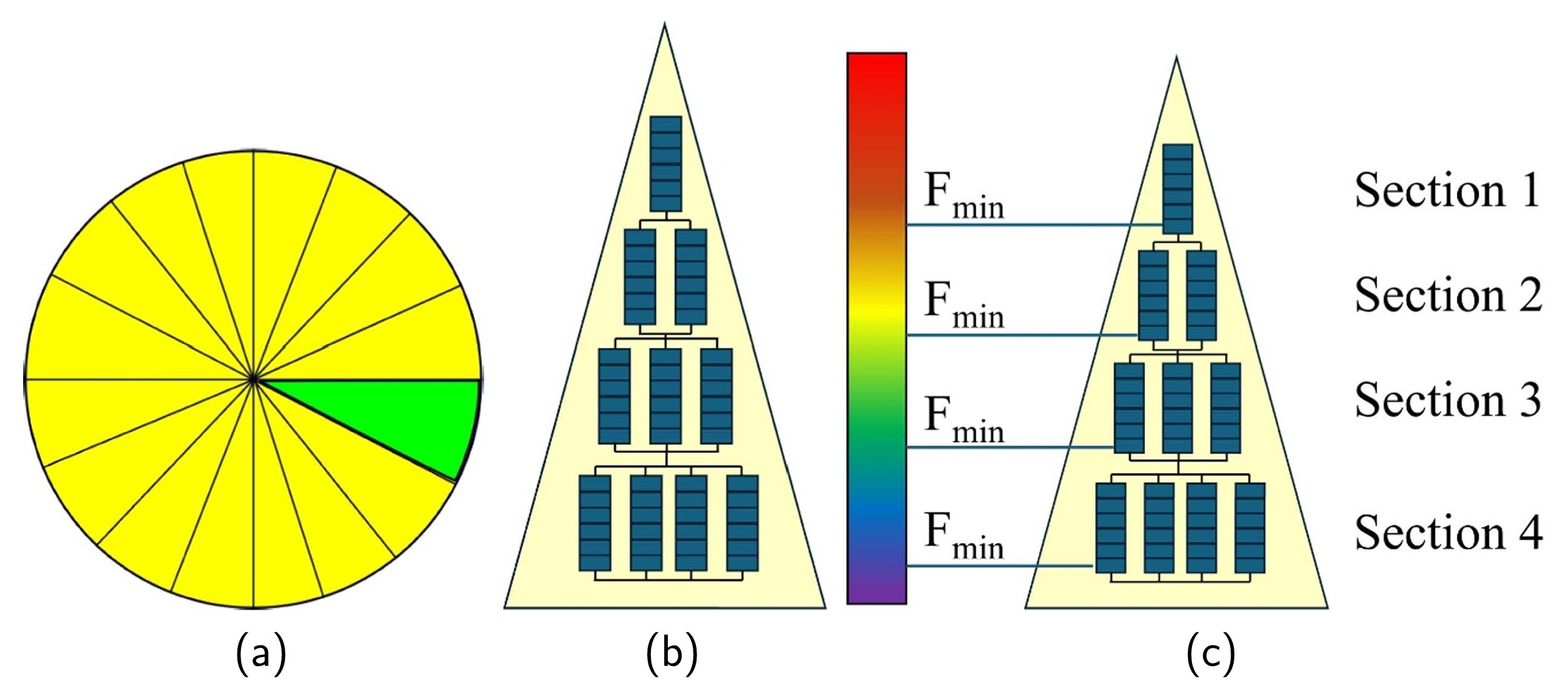} 
\caption{\label{fig:treearray} Tree-Shaped design. (a) Circular panel. (b) Each circular sector is composed of a certain number of sections connected in series. (c) Circular sector with sections of varying lengths. The gradient bar indicates the intensity at the cell locations.  From [\cite{cichelli2025photovoltaic}].}
\end{centering}
\end{figure}

Figure \ref{fig:treearray} shows the other connection pattern between photocells analyzed in the paper, referred to as the tree-shaped design. [\cite{cichelli2025photovoltaic}] concluded that the ring-shaped array was more efficient than the tree-shaped array by a few percent.  However, the tree-shaped design exhibited a more compact surface area.  The paper also concluded that a future iteration of the research could be based on a hybrid configuration, combining the Tree-Shaped design for $r<0.9w$ and the Ring-Shaped design for $0.9w<r<1.5w$. This would effectively combine the scalability of the second design with the efficiency of the first.  They quote an efficiency for the ring-shaped triple junction photocells as 37\% at 500 \si{W/m^2}, and 35.6\% for the tree-shaped configuration.  

Another point brought out in the paper is that the bus for these photocells operated at 100 V.  The primary motors of the climber operate at a voltage of 800 V.  This would require a step-up DC/DC converter but it also means that the current in the 100 V bus would be eight times the current in the 800 V bus.  Since the 800 V bus is designed to operate at 7.5 kA at full power, the current in the PV bus would be 60 kA.  The $I^2 R$ losses in the PV bus would become much larger.  It remains for final design by electrical engineers familiar with space-based photovoltaics to determine whether the voltage of the PV array can be increased, and exactly which semiconductor junction to use.

\newpage
\subsubsection{Designing the arms that support the PV arrays from the climber for 1 G load}
\label{sec:PVin1G}

The 2023 study report, [\cite{wright2023climber}], shows a mass budget for the 20 tonne climber that included a placeholder mass for the PV arrays of 484.6 kg.  At the time of writing, very little was known about the size or structure of the PV arrays and their supports.  The arrays were far too small, given what has been presented in this report.

\begin{figure}
\begin{centering}
\includegraphics[width=\textwidth]{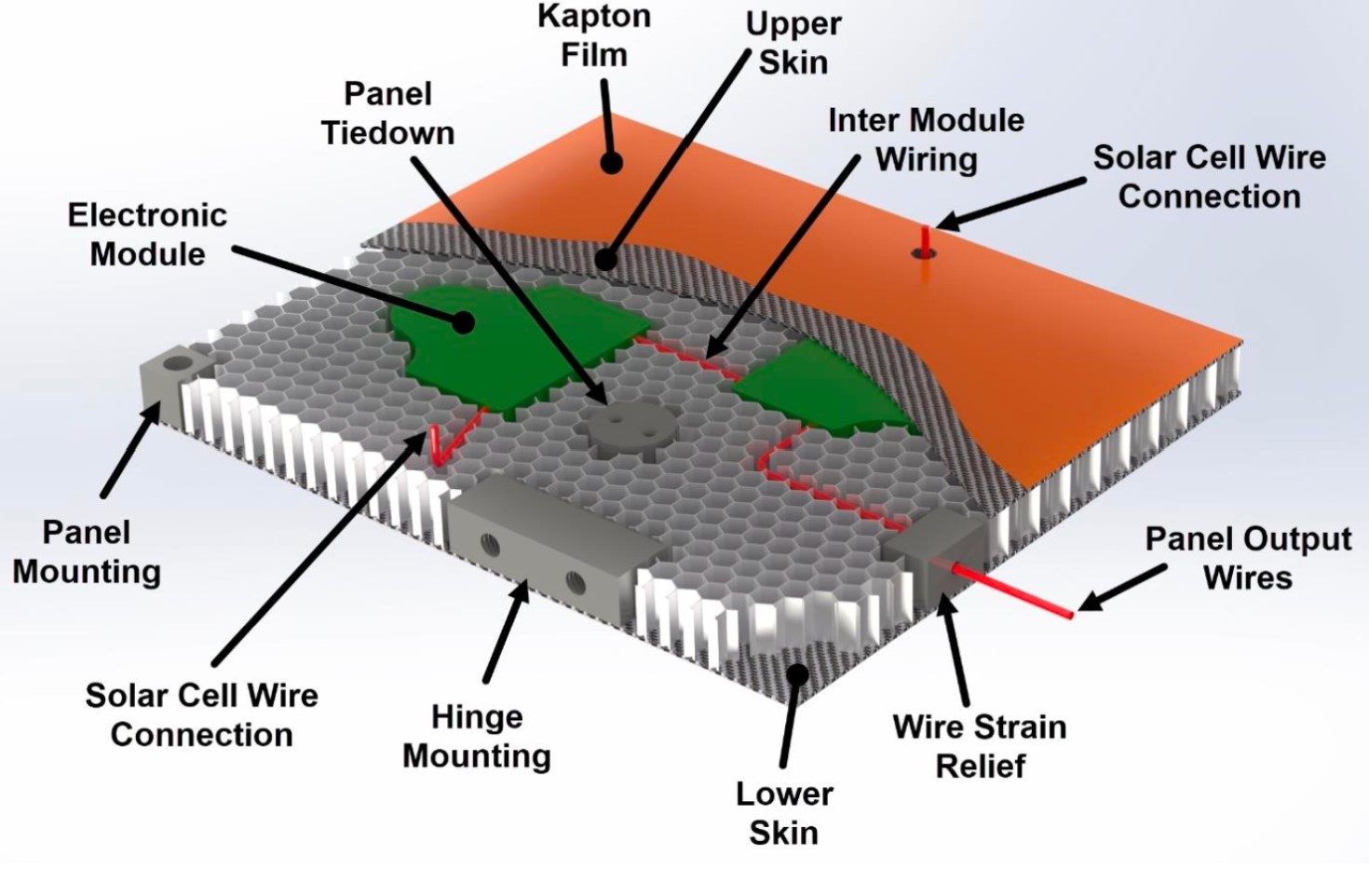} 
\caption{\label{fig:CSA_PV} Concept rendering of a PV array backing for space solar power that has electronics embedded in an aluminum honeycomb. [\cite{CSAPV}]}
\end{centering}
\end{figure}

The Canadian Space Agency posted a new space technology development program in 2025 at [\cite{CSAPV}] to develop PV arrays with electronics embedded in aluminum honeycomb panels.  The concept for this type of panel is shown in Figure \ref{fig:CSA_PV} and was used as the basis for the finite element analyses of the PV arms attached to the climber.  The construction of the arms was assumed to be Carbon Fiber Reinforced Polymer (CFRP) to maximize strength and minimize mass.

\begin{figure}
\begin{centering}
\includegraphics[width=\textwidth]{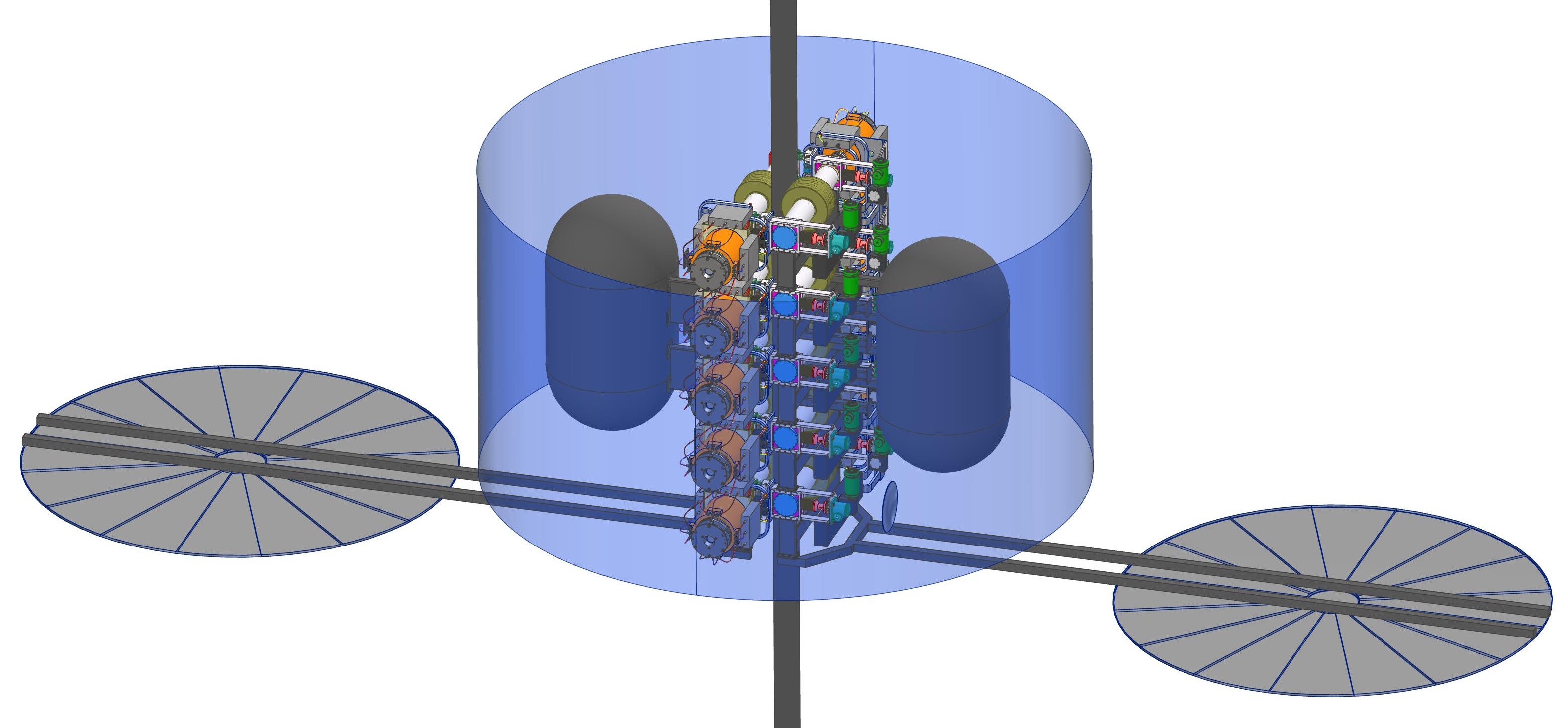} 
\caption{\label{fig:BE_climb1} Model of PV supports made from CFRP and bolted to the bottom of the climber interface structure.}
\end{centering}
\end{figure}

\begin{figure}
\begin{centering}
\includegraphics[width=\textwidth]{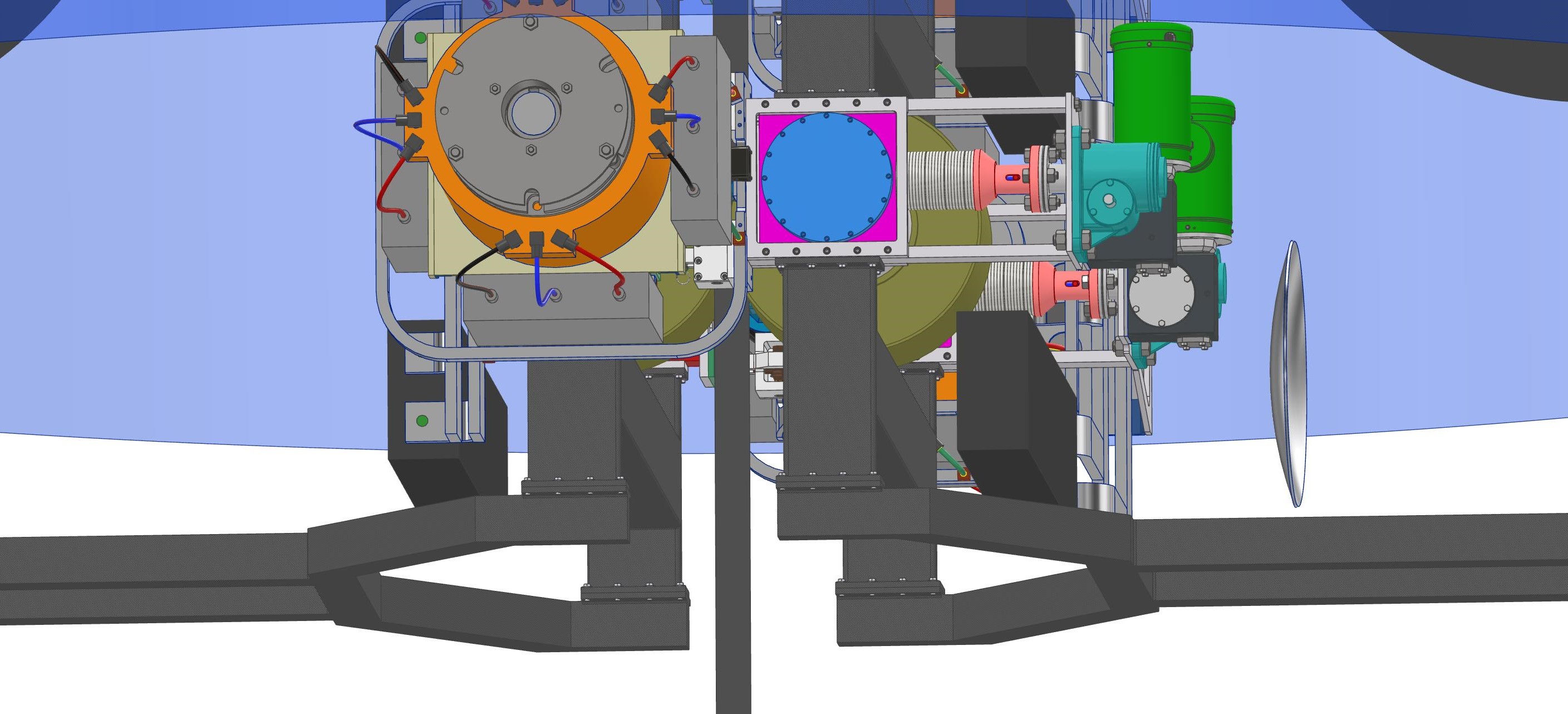} 
\caption{\label{fig:BE_climb2} Close-up of the connection between the PV support arms and the Climber structure.  Not shown are bus bar connections to the climber or conditioning electronics.}
\end{centering}
\end{figure}

The conceptual design of the arms supporting the PV is shown in Figures \ref{fig:BE_climb1} and \ref{fig:BE_climb2}.  The PV cells are supported by a .25 in (6.35 mm) thick aluminum honeycomb panel.  The PV cells themselves are only a few hundred microns thick, and add almost nothing to the overall mass or stiffness of the array structure.  They are neglected in the FEA of the array structure.  The arms are 4 in x 4 in x .125 in (101.6 mm x 101.6 mm x 3.18 mm) square tube of CFRP with flanges that allow them to be bolted to the climber half.  No clips or connections between the CFRP tubes and the aluminum ribs of the array are shown.

\begin{figure}
\begin{centering}
\includegraphics[width=\textwidth]{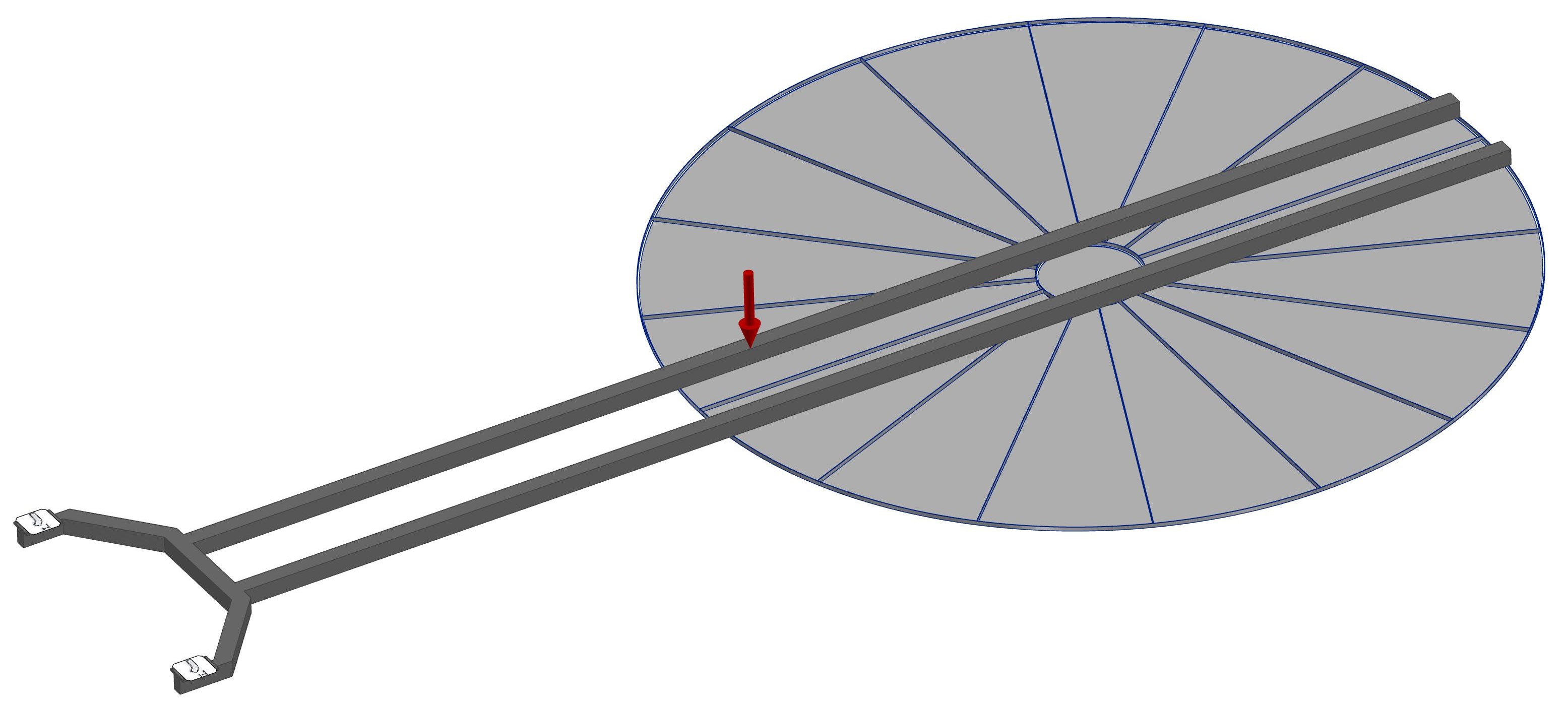} 
\caption{\label{fig:FEAPV-01} The CAD model used for the FEA of the PV array support arms. The white symbols on the left side of the arm structure are fixed boundary conditions where the arm bolts to the climber.  The red arrow represents gravity.  The first load step was just the self weight of the array and arms.}
\end{centering}
\end{figure}

The FEA model is shown in Figure \ref{fig:FEAPV-01}.  The arms were given fixed boundary conditions at the structural connection to the climber.  The red arrow in the figure represents the load of one G of gravity on the structure.

\begin{figure}
\begin{centering}
\includegraphics[width=\textwidth]{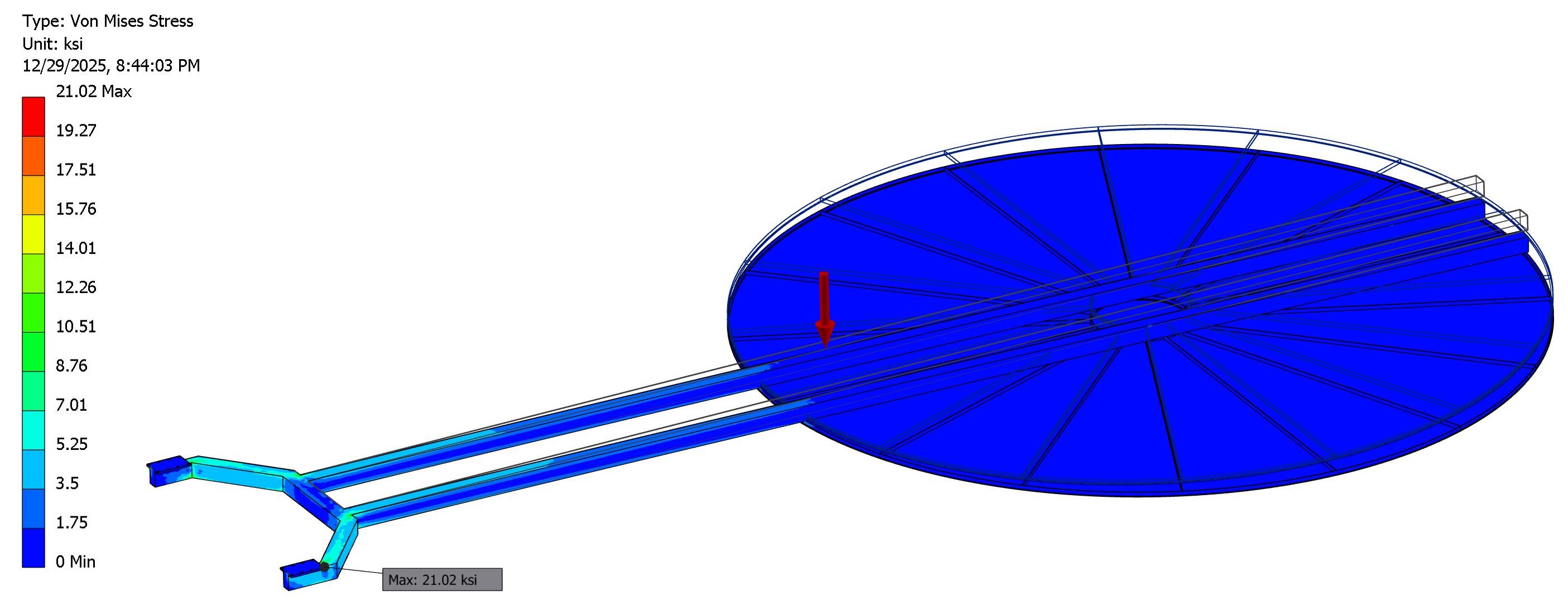} 
\caption{\label{fig:FEAPV-VM-01} A plot of Von Mises stress in the structure showing the actual deflection under load.  Maximum Von Mises stress is 21.02 ksi (144.9 MPa).}
\end{centering}
\end{figure}

\begin{figure}
\begin{centering}
\includegraphics[width=\textwidth]{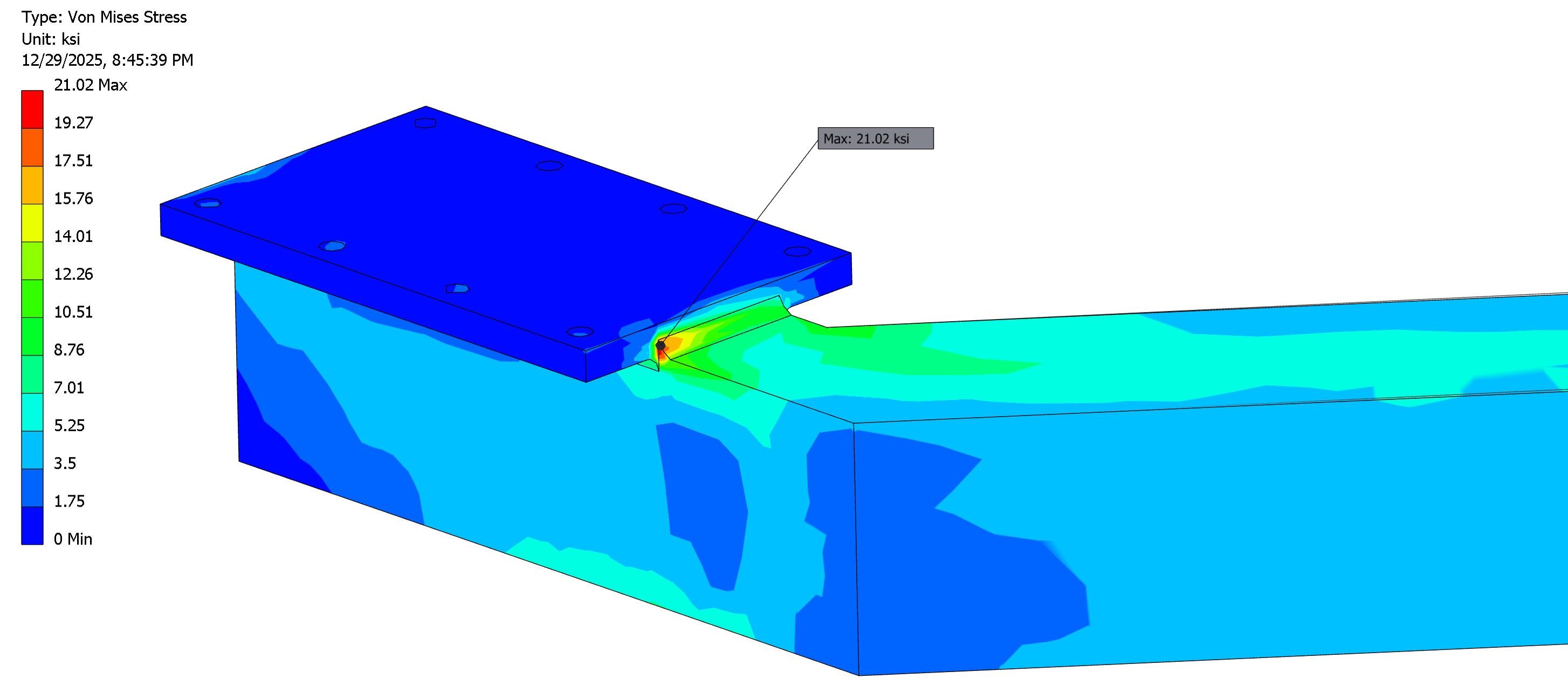} 
\caption{\label{fig:FEAPV-VM-02} Close-up plot of Von Mises stress in the structure showing the stress concentration that causes the maximum stress.  This stress occupies a very small volume and can be either a meshing artifact or an actual stress riser.  More detailed design work is necessary, but there are many ways to reduce this stress.  The important thing is to notice that the average stress in the material at the base of the cantilever is much less.}
\end{centering}
\end{figure}

Figure \ref{fig:FEAPV-VM-01} shows a Von Mises stress plot of the cantilevered arms.  Maximum stress is 21.02 ksi (144.9 MPa), but Figure \ref{fig:FEAPV-VM-02} shows that this stress occurs in a small volume.  This stress can be either a mesh artifact, or a real stress concentration.  More design and analysis work is needed.  If the arms were made of a metal like aluminum or steel, stress concentrations like this are ignored in static load situations.  Metals can locally deform to relieve stresses like this.  The main concern is in fatigue because the arms will oscillate as cantilevers and be buffeted by wind.  This cyclic stress may be of greater concern in CFRP construction as composites may delaminate and crack with cyclic loads.

\begin{figure}
\begin{centering}
\includegraphics[width=\textwidth]{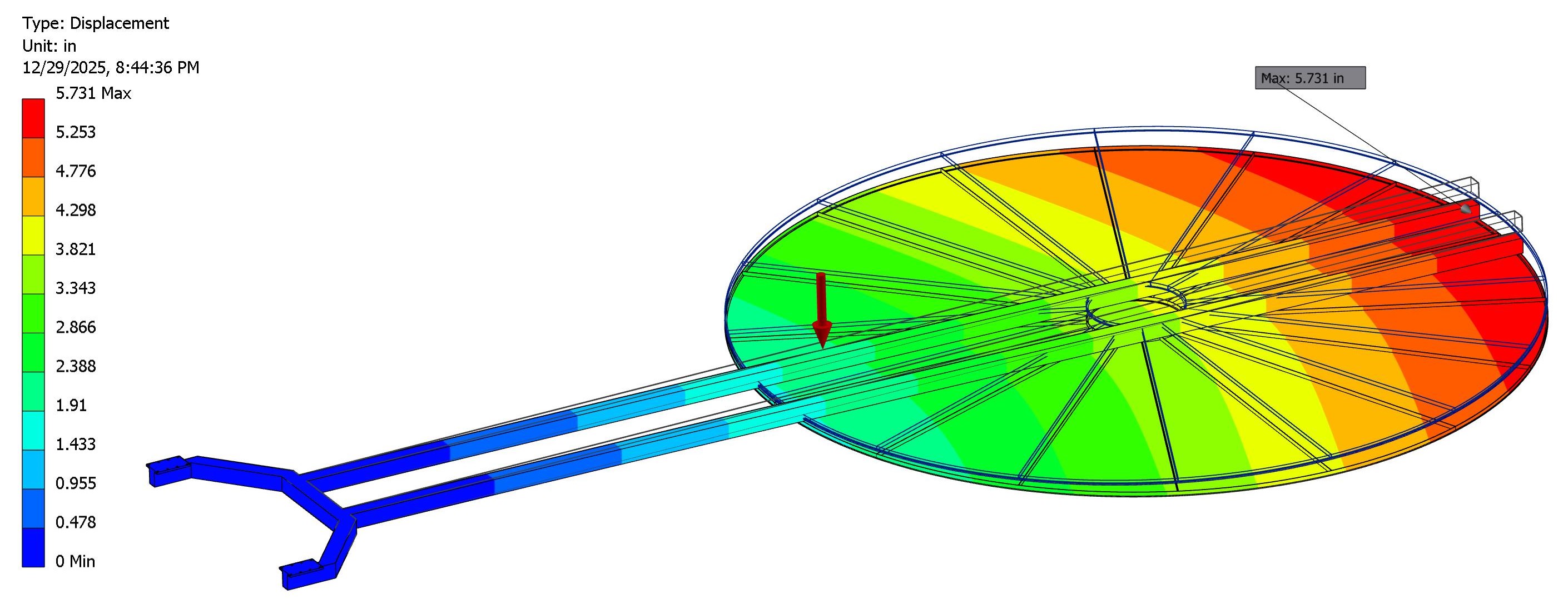} 
\caption{\label{fig:FEAPV-defl-01} A plot of deflection in the structure showing the actual deflection under load.  Maximum deflection at the end of the array is 5.7 inches (144.8 mm).}
\end{centering}
\end{figure}

Figure \ref{fig:FEAPV-defl-01} shows the deflection of the structure.  Maximum deflection is at the end of the cantilevered array and is 5.7 inches (144.8 mm).  This deflection on a structure of this size is acceptable, but is also easy to reduce with more design work.

\begin{figure}
    \begin{minipage}[t]{.48\textwidth}
        \centering
        \includegraphics[width=\linewidth]{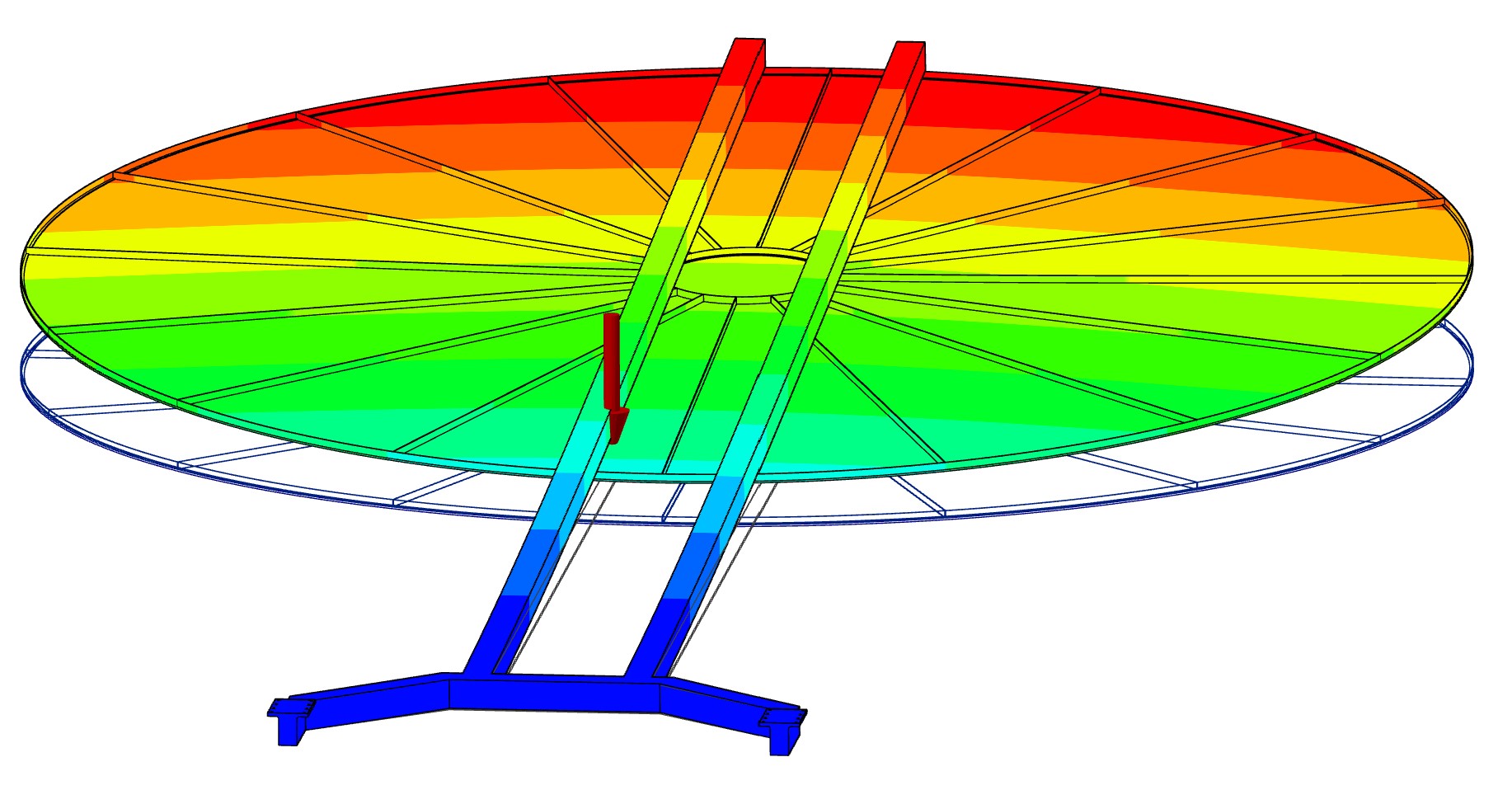}
        \captionsetup{justification=centering}
        \caption{An exaggerated deflection plot of the lowest natural frequency of the PV array support, 1.67 Hz cantilever oscillation.}
        \label{fig:FEAPV-1.67Hz}
    \end{minipage} \hfill
    \begin{minipage}[t]{.48\textwidth}
        \centering
        \includegraphics[width=\linewidth]{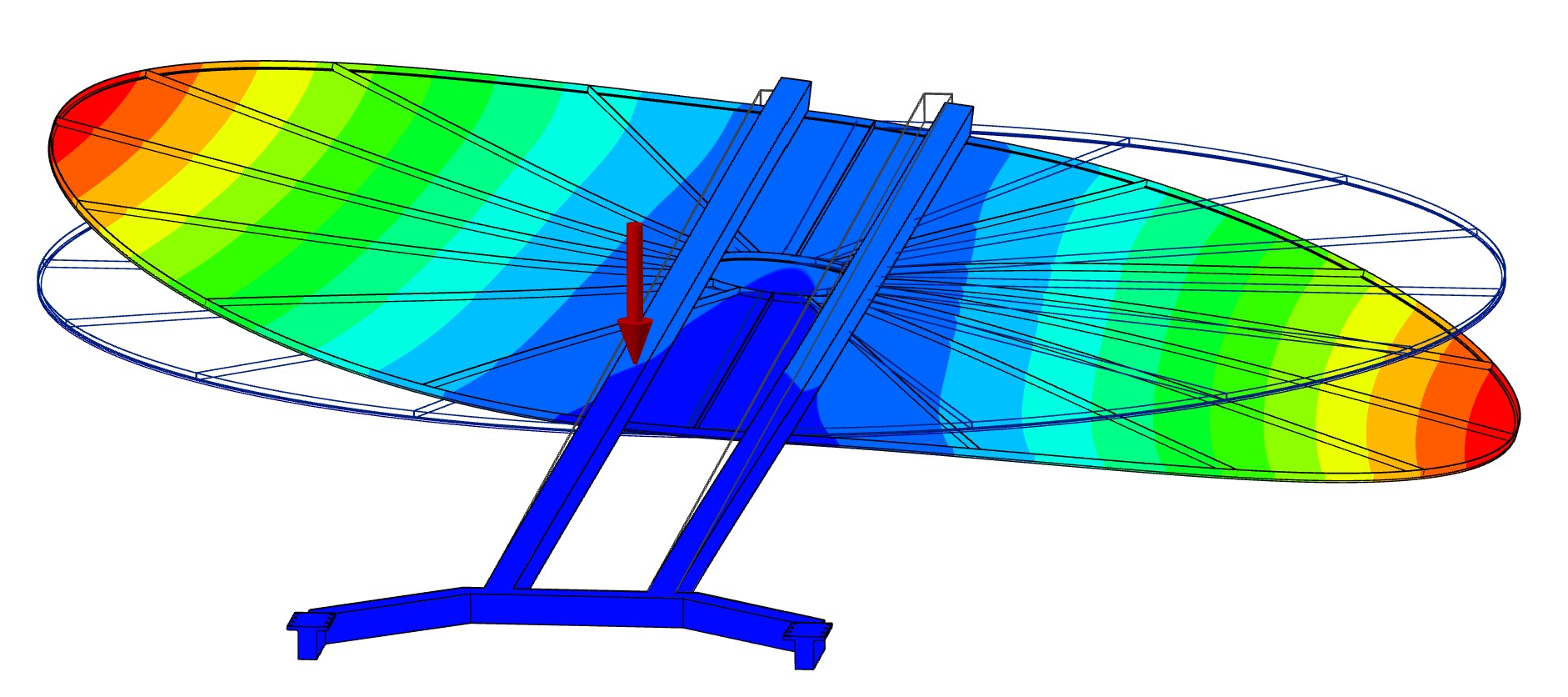}
        \captionsetup{justification=centering}
        \caption{Exaggerated deflection plot of the second natural frequency of the PV array support, 4.69 Hz torsional oscillation.}
        \label{fig:FEAPV-4.69Hz}
    \end{minipage}
    \begin{minipage}[t]{.48\textwidth}
        \centering
        \includegraphics[width=\linewidth]{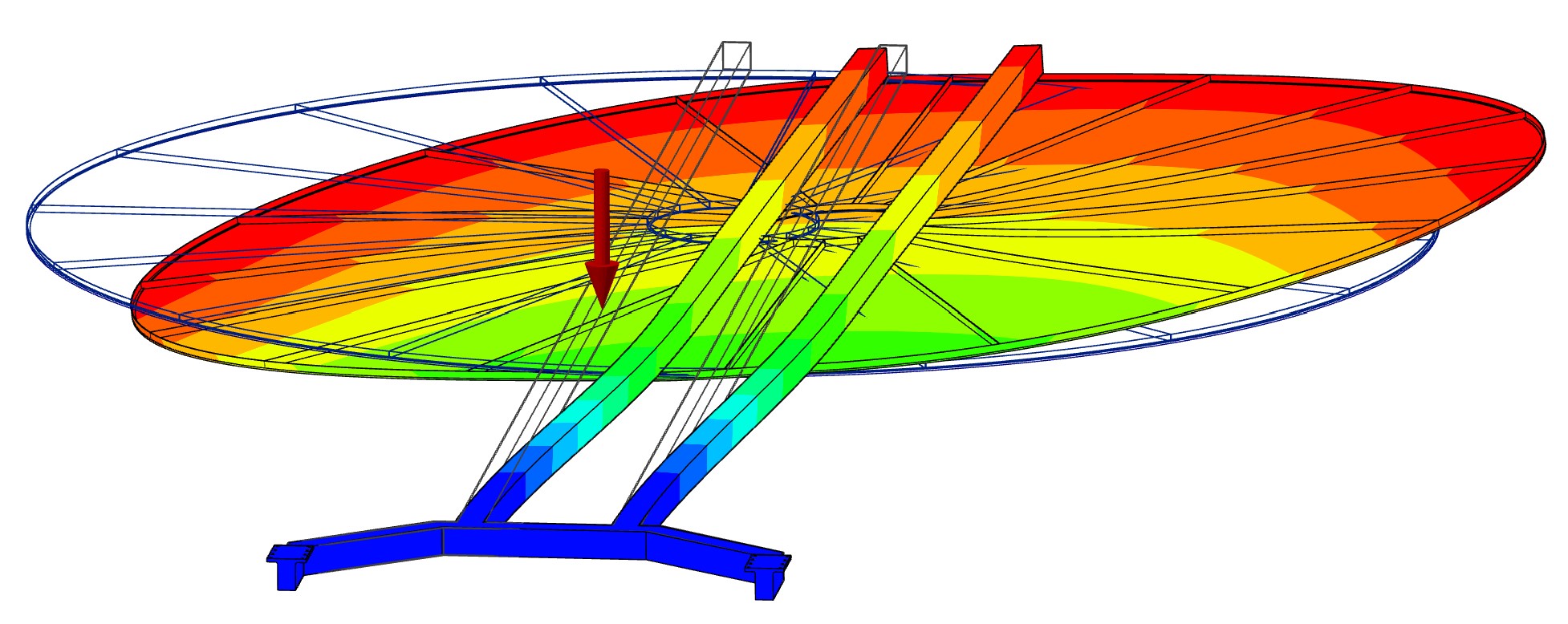}
        \captionsetup{justification=centering}
        \caption{Exaggerated deflection plot of the third natural frequency of the PV array support, 5.92 Hz, a mixed mode of yaw of the arms and torsion of the array.}
        \label{fig:FEAPV-5.92Hz}
    \end{minipage}
    \begin{minipage}[t]{.48\textwidth}
        \centering
        \includegraphics[width=\linewidth]{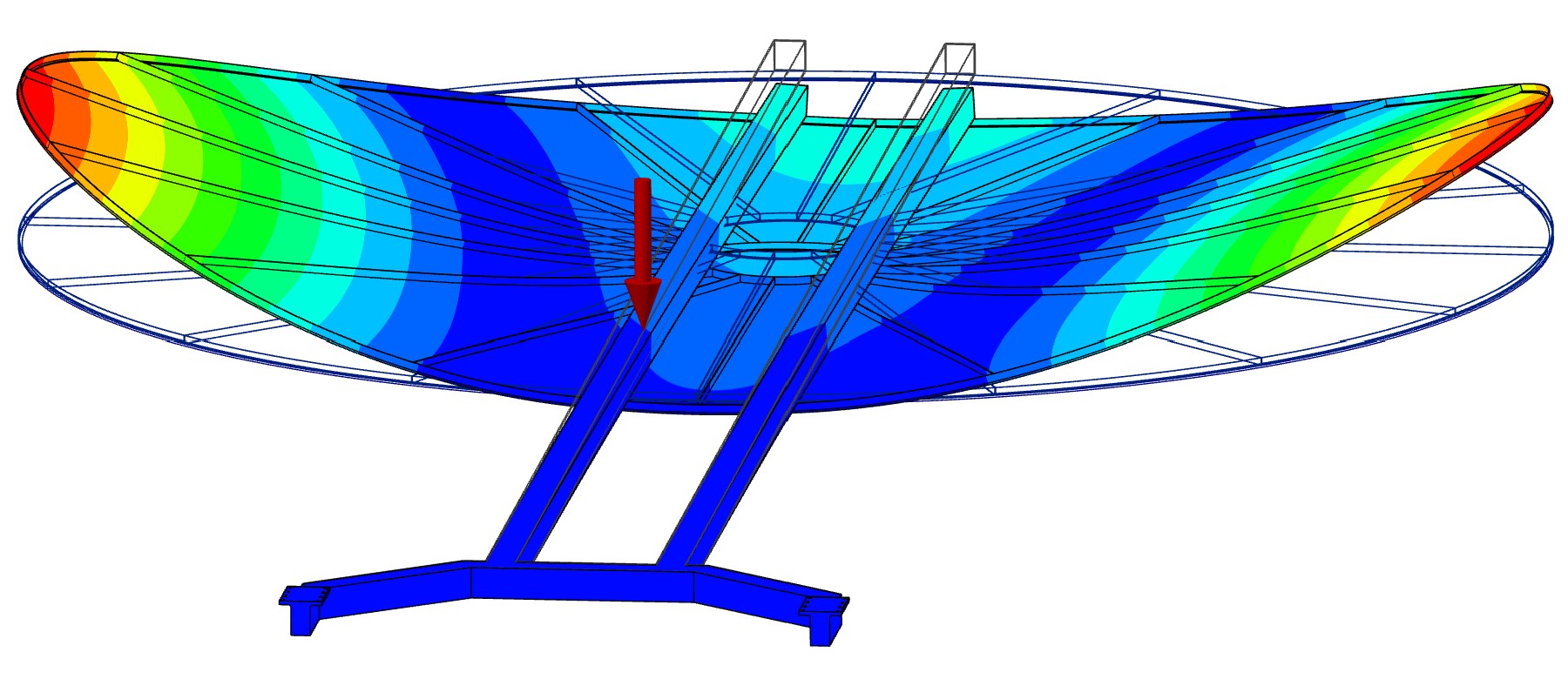}
        \captionsetup{justification=centering}
        \caption{Exaggerated deflection plot of the fourth natural frequency of the PV array support, 6.72 Hz, the PV array is flapping like a bird's wings.}
        \label{fig:FEAPV-6.72Hz}
    \end{minipage}
    \begin{minipage}[t]{.48\textwidth}
        \centering
        \includegraphics[width=\linewidth]{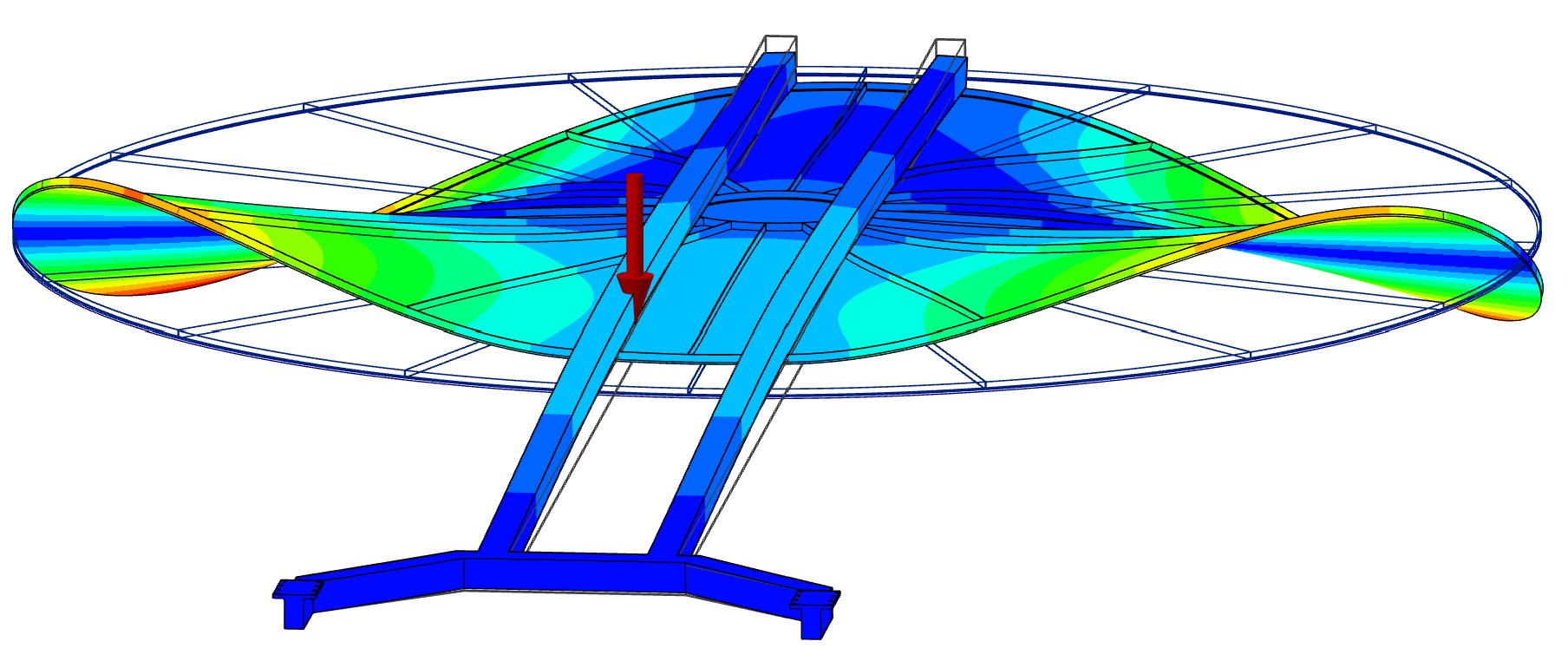}
        \captionsetup{justification=centering}
        \caption{Exaggerated deflection plot of the fifth natural frequency of the PV array support, 10.85 Hz, a diaphragm mode of the array mixed with flexing of the arms.}
        \label{fig:FEAPV-10.85Hz}
    \end{minipage}
    \begin{minipage}[t]{.48\textwidth}
        \centering
        \includegraphics[width=\linewidth]{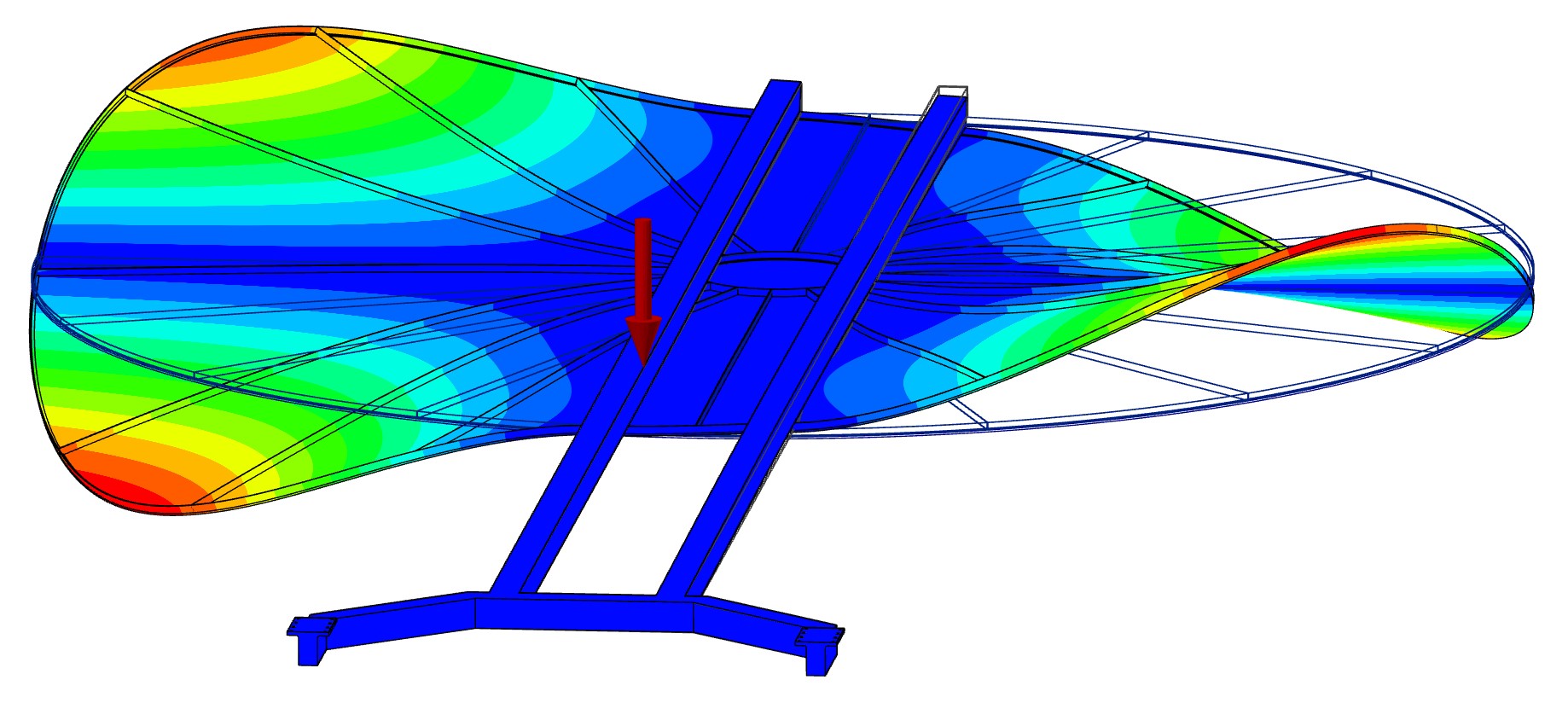}
        \captionsetup{justification=centering}
        \caption{Exaggerated deflection plot of the sixth natural frequency of the PV array support, 11.06 Hz, a diaphragm mode of the array without contribution from the arms.}
        \label{fig:FEAPV-11.06Hz}
    \end{minipage}
\end{figure}

A modal analysis was also performed on this model to determine the natural frequencies and normal modes.  Figures \ref{fig:FEAPV-1.67Hz} to \ref{fig:FEAPV-11.06Hz} show the lowest six natural frequencies and normal modes of the PV array support structure.  As expected, the fundamental mode of the structure is a cantilever oscillation of the arms at 1.67 Hz.  A natural frequency this low is uncomfortable for people to walk on (like a plank between two ladders) as it induces bouncing from walking.  Solar arrays in space are acceptable if their lowest natural frequency is between 0.04 to 0.5 HZ, but this is for objects in orbit to avoid exciting structural vibrations from spacecraft maneuvers.  Higher stiffness typically means higher mass, always a difficult trade-off for spacecraft.  It remains for further engineering analyis to determine if this structure is acceptably stiff without being too heavy. 

The climber requires two PV arrays and support structures and each of the arms and arrays shown in the FEA model weighs 127.5 lbs (57.8 kg).  This gives a total mass for two arrays as 255 lbs (115.6 kg).  Even if they were to get heavier by a factor of 7, they would still be lighter than any other power receiver shown in this report.

\subsubsection{Calculating the effect of wind drag on the PV arrays}

The calculation that follows is a worst-case condition because it is predicated on the climber moving at 200 km/h very close to sea level.  We also assume that the wind acting on the horizontal PV arrays comes from the motion of the climber.  To determine the drag coefficient on the PV arrays, we first calculate the Reynolds number, $Re$ for the moving air over the array.

$$Re = \frac{VL}{\nu}$$ \par
\bigskip
\noindent where: \\
V is the speed of the climber, 200 km/h = 55.6 m/s \\
L is a characteristic length which is 5 m for the diameter of the array \\
$\nu$ is the kinematic viscosity of air, 1.51 $\times 10^-5$ \si{m^2/s} (air at 1 atm and $20 \degree$C)
\bigskip

\begin{figure}
\begin{centering}
\includegraphics[width=\textwidth]{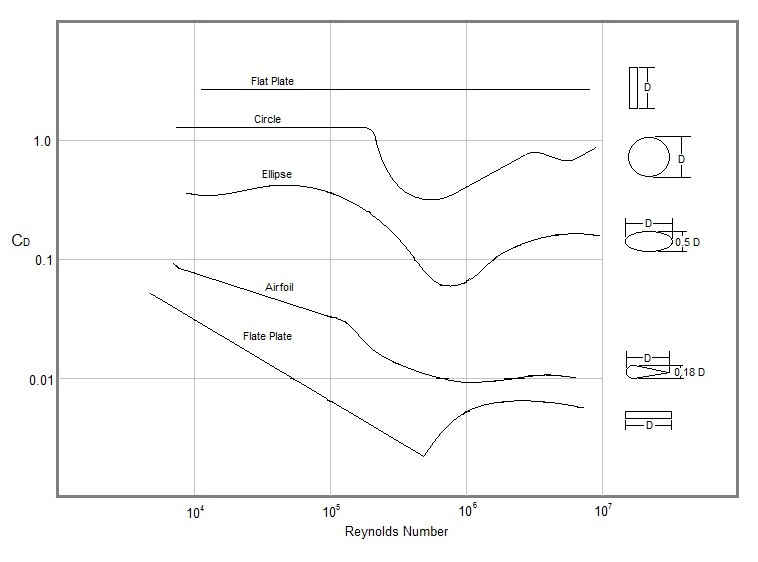} 
\caption{\label{fig:dragvsre} Graph of the drag coefficient for different shapes versus Reynolds number.  The drag coefficient for a flat plate does not vary with Reynolds number. [\cite{munson1995fundamentals}]}
\end{centering}
\end{figure}

\begin{figure}
\begin{centering}
\includegraphics[width=\textwidth]{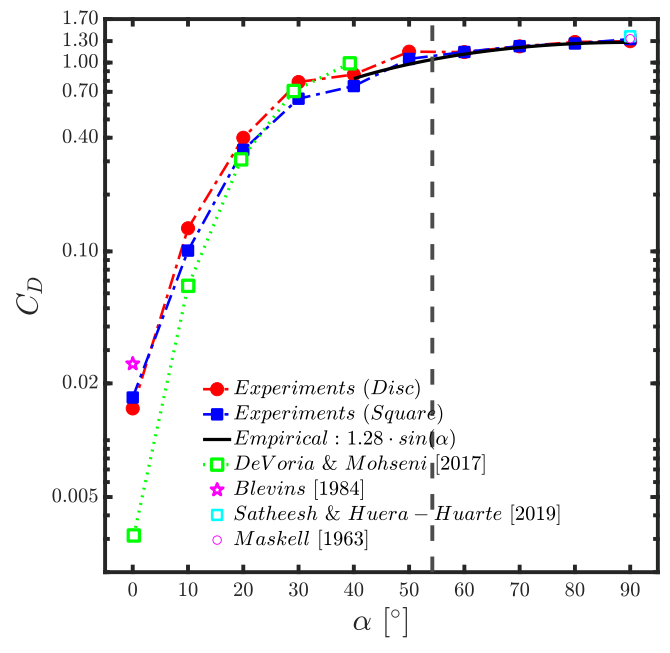} 
\caption{\label{fig:draggraph} Graph of the drag coefficient on a disk at an angle $\alpha$ to the flow.  $90\degree$ is the angle of the PV array to the direction of air flow when the array is horizontal and the climber is moving upward. [\cite{satheesh2022drag}]}
\end{centering}
\end{figure}

The Reynolds number calculates to 1.84 $\times 10^7$.  Figure \ref{fig:dragvsre} shows that the drag coefficient for a disk perpendicular to the fluid flow doesn't vary with Reynolds number.  Figure \ref{fig:draggraph} shows how the drag coefficient varies when the angle of the disk is varied with respect to the flow direction.  The PV arrays are at an angle of $90\degree$ to the air flow when they are horizontal and the climber is moving upwards. The drag coefficient $C_D$ used in the FEA was 1.3.  The drag coefficient for a flat disk perpendicular to the flow does not vary significantly with Reynolds number because it is a 2D shape with respect to the flow.  All of the drag comes from the frontal resistance.

The force F on the PV array from the wind speed caused by the relative motion of the climber is given as:

$$F = C_D\left(\frac{1}{2}\rho V^2A\right)$$ \par
\bigskip
\noindent where: \\
$C_D$ is the drag coefficient, in this case 1.3 \\
$\rho$ is the density of air, 1.2 \si{kg/m^3} \\
V is the speed of the climber, 200 km/h = 55.6 m/s \\
A is the area of the array disk, $\pi(2.5 \,\mathrm{m})^2 = 19.635$ \si{m^2} \\
\bigskip

An equivalent pressure on the array disk can be calculated by moving the area of the disk over to the other side of the equation:

$$P_{eq} = \frac{F}{A} = C_D\left(\frac{1}{2}\rho V^2\right)$$

The force is calculated as 47,345.1 N (10,643.6 lbs), and the pressure is found to be 2,411.2 Pa (0.35 psi). This force is far too large to be supported by the array, and this was confirmed by the FEA.

The conclusion from this calculation is that even if we could push the amount of power necessary to climb at 200 km/h near sea level, this design of PV array and support cannot support the drag force.

The 2023 study [\cite{wright2023climber}] gave an acceleration profile for the climber starting at sea level such that the climber was traveling at 73.5 km/h at an altitude of 970 m.  The density of dry air at 970 m is approximately 1.12 \si{kg/m^3}, a 6.7\% reduction from the density at sea level.  Plugging in a speed of 73.5 km/h (20.42 m/s) and a density of 1.12 \si{kg/m^3}, we get a drag force of 5,960.4 N (1,340 lbs) or a pressure of 303.6 Pa (.044 psi).

\begin{figure}
\begin{centering}
\includegraphics[width=\textwidth]{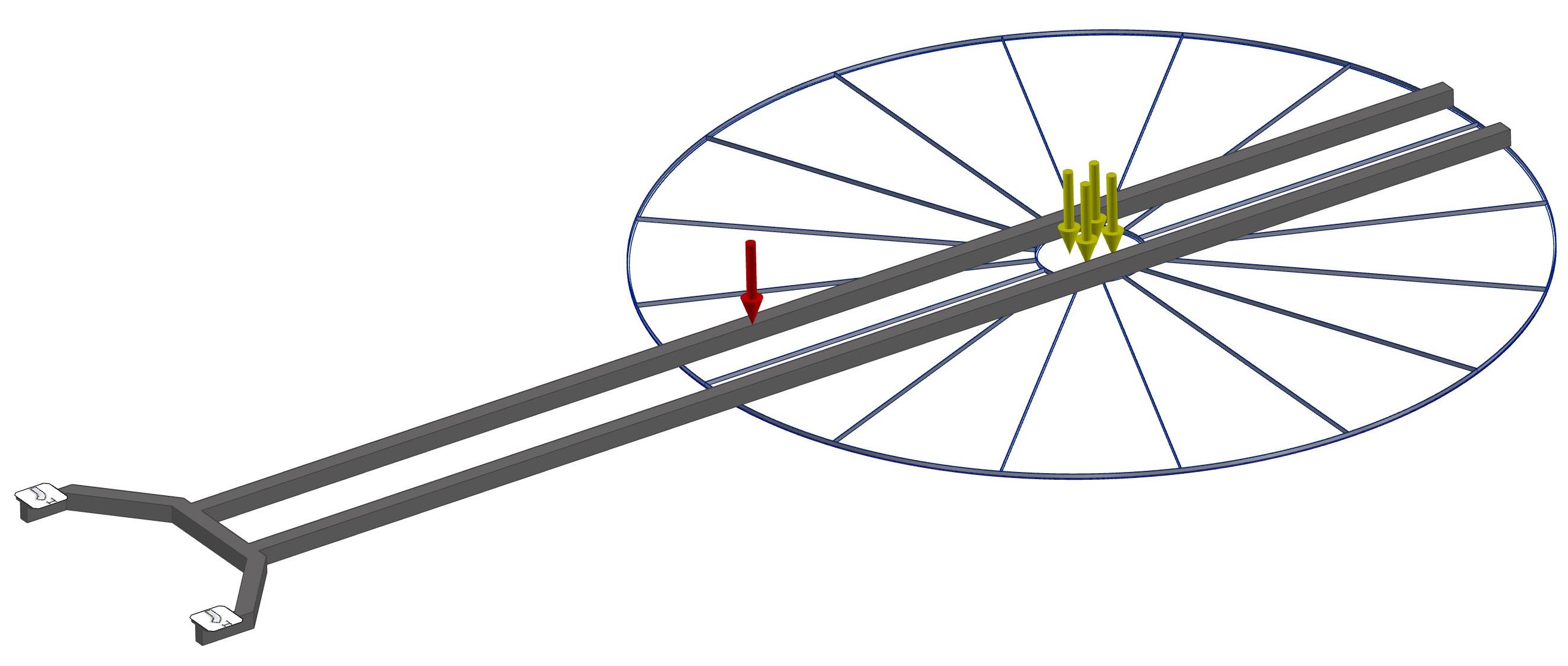} 
\caption{\label{fig:FEAPV-drag} The CAD model with an additional symbol on the PV array indicating the pressure applied from drag. The white symbols on the left side of the arm structure are fixed boundary conditions where the arm bolts to the climber.  The red arrow represents gravity.}
\end{centering}
\end{figure}

\begin{figure}
\begin{centering}
\includegraphics[width=\textwidth]{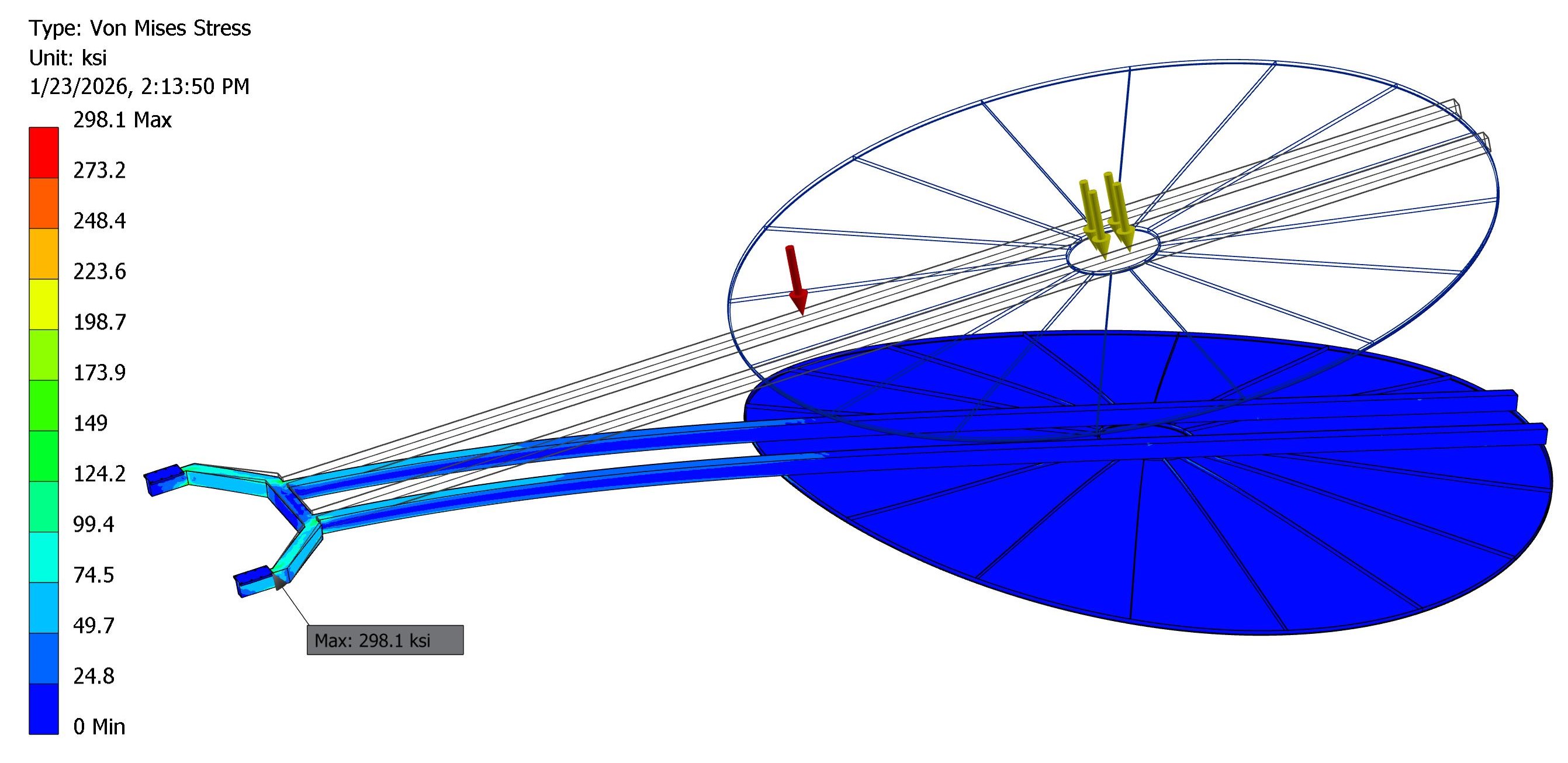} 
\caption{\label{fig:FEAPV-dragVM} A Von Mises stress plot of the array with the drag pressure applied. Maximum stress is 298.1 ksi (2.1 GPa). This is beyond the strength of CFRP.}
\end{centering}
\end{figure}

\begin{figure}
\begin{centering}
\includegraphics[width=\textwidth]{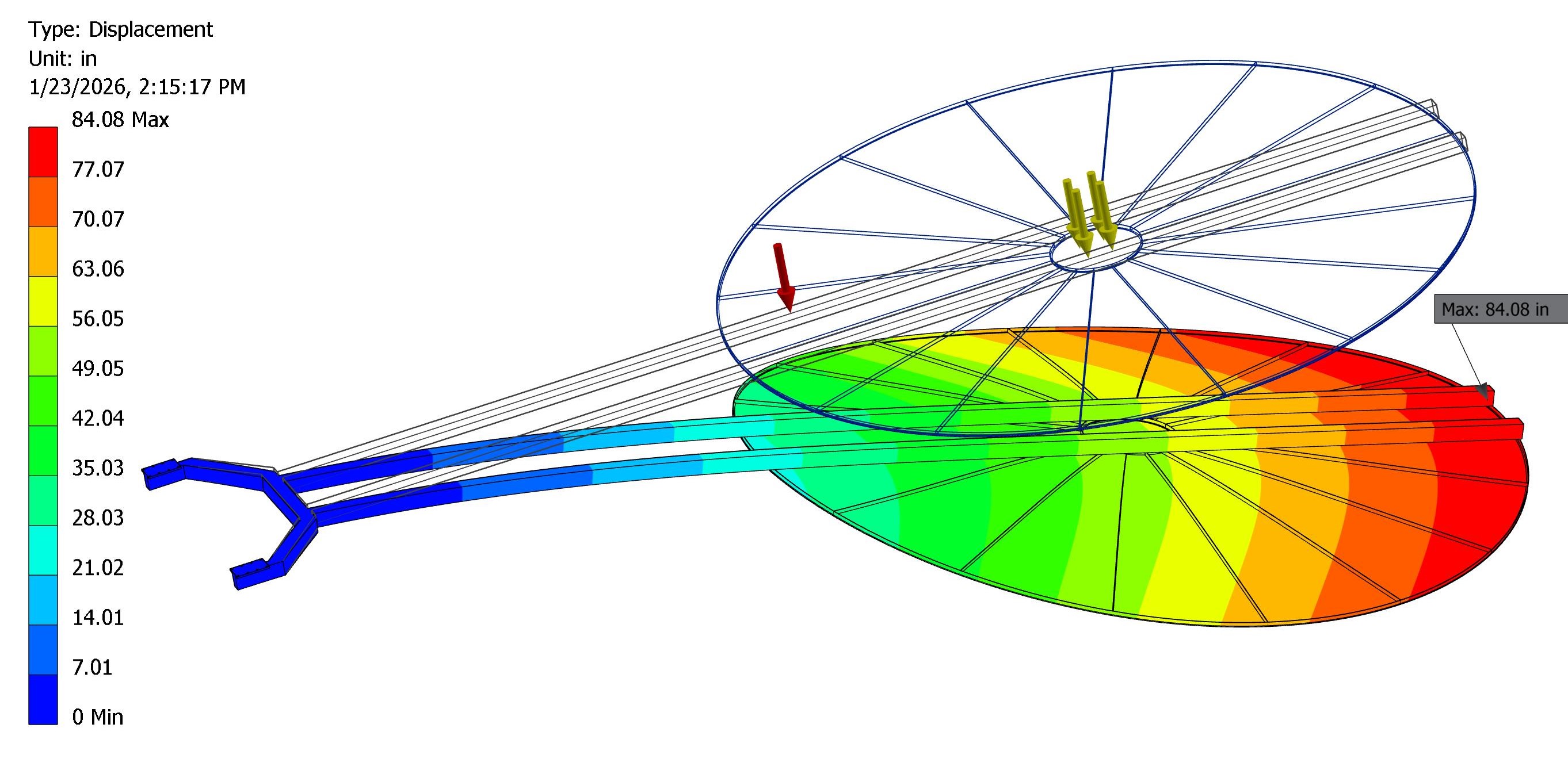} 
\caption{\label{fig:FEAPV-dragdefl} A deflection plot of the array with the drag pressure applied. Maximum deflection is 84.1 in (2.14 m). This deflection is too large for this analysis to be correct. The deflection violates the assumption of small deflection with respect to the size of the model.  This would have to be analyzed with non-linear FEA.}
\end{centering}
\end{figure}

Figure \ref{fig:FEAPV-drag} shows the FEA CAD model with the additional yellow arrows that indicate the pressure of .044 psi on the array.  Figure \ref{fig:FEAPV-dragVM} shows a contour plot of Von Mises stress on the actual deflection scaling.  The deflection in this model is too large and violates the assumptions of static, linear, small displacement FEA.  The stress is significantly over the allowable stress for CFRP.  Figure \ref{fig:FEAPV-dragdefl} shows the calculated deflection.  The stress and deflection in a static, linear FEA cannot be trusted when they violate the small displacement assumption.  The proper way to analyze the structure is to use non-linear FEA.

The conclusion is that the PV arrays as designed cannot withstand the drag forces in the atmosphere even following an acceleration profile as outlined in [\cite{wright2023climber}].

This is a very strong motivation to use a launch technique such as the reel-out method described in Section \ref{sec:reelout}.  That method can raise climbers above 20 km where the air is too thin to offer significant drag.  The PV arrays can be stowed in the vertical position where they would offer far less drag and be raised to the horizontal position when the air is thin enough to allow it.  This also motivates the next section about how to design a flexible coupling that can carry kiloamps of current.  Such a coupling is needed if the PV arrays are to be rotated into position.  Additional hardware and design effort is needed to modify the fixed support structure shown to accommodate this array deployment.

\FloatBarrier

\subsubsection{Designing kiloamp power couplings that can cross rotational linkages to allow for PV array articulation}

There are two basic ways to make a flexible joint that can carry kiloamps of current.  One method involves oxygen-free high-conductivity (OFHC) copper wire braids that are compressed into terminal blocks at the ends.  Two companies that make custom versions of these are Thermal Space, \url{https://thermal-space.com/copper-thermal-straps/} and Technology Applications Incorporated, 
\\ 
\url{https://www.techapps.com/copper-thermal-strap-assemblies}.  These straps are usually made as thermal connections in dilution refrigerators and cryo-coolers because of their extremely low thermal resistivity, but they also have the characteristic of very high electrical conductivity as well.

\begin{figure}[hb]
\begin{centering}
\includegraphics[width=.8\textwidth]{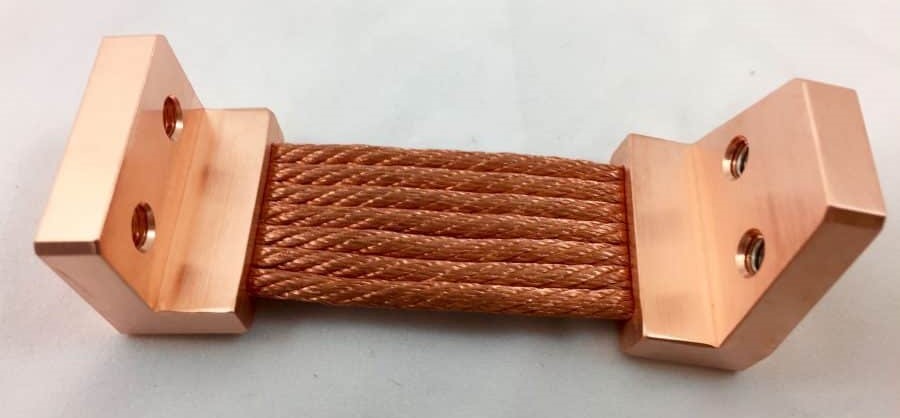} 
\caption{\label{fig:thermalspace} Photo of a custom strap made from copper braids that can be used to carry high currents and is flexible. This one is available from Thermal Space, Ltd.}
\end{centering}
\end{figure}

\begin{figure}[t]
\begin{centering}
\includegraphics[width=.7\textwidth]{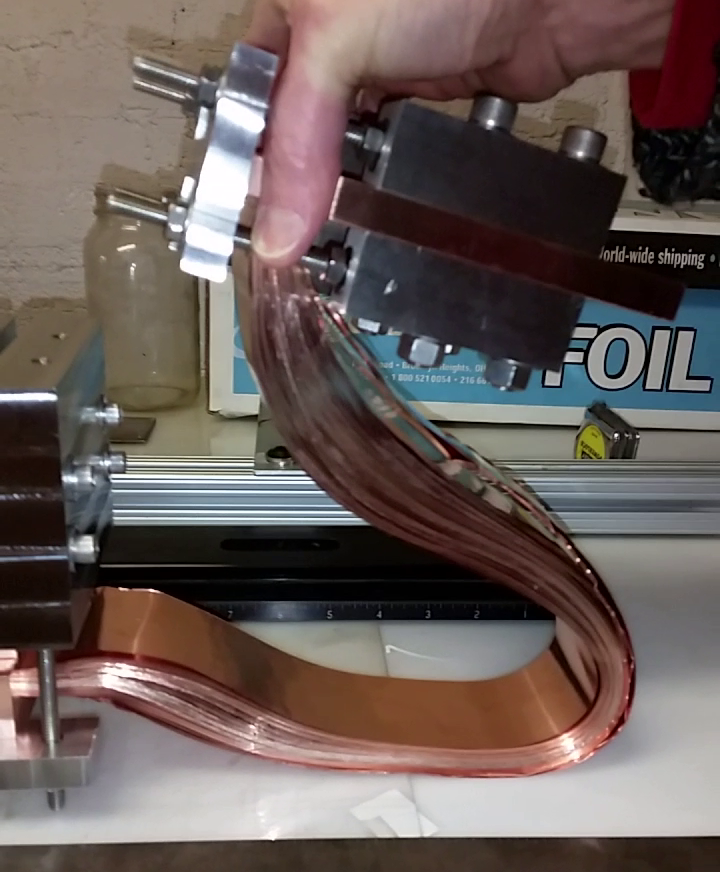} 
\caption{\label{fig:UIUCstrap} Photo of a custom strap made from copper foils that can be used to carry high currents and is flexible. This one was made at the University of Illinois, Urbana-Champaign, for the nEDM@SNS experiment.}
\end{centering}
\end{figure}

The other common method involves thin copper strips that are extremely electrically conductive and very flexible. References [\cite{steyert1978high}] and [\cite{rosenblum1977simple}] describe the technique of heat treating .005 inch thick foils in a partial vacuum with a dry air leak to drive impurities to the surface of the foil.  Figure \ref{fig:UIUCstrap} shows a strap that was made at the University of Illinois, Urbana-Champaign (UIUC), for the Neutron Electric Dipole Moment experiment planned for the Spallation Neutron Source in Oak Ridge, TN, (nEDM@SNS experiment). By stacking strips and electron beam welding the ends to connecting blocks, flexible kiloamp-carrying connections can be made between two parts that move with respect to each other.

Copper can carry 2 $\si{amp/mm^2}$ of cross-sectional area in air.  Careful design will have to be done to size the copper braids or foils for high currents in vacuum.  A braided cable able to carry 7.5 kA in air would require a minimum copper cross-sectional area of 3,750 $\si{mm^2}$.  This area of solid copper would fit in a square 61 mm on a side, larger for braid given the packing factor of wires.  To be flexible and allow the braids to radiate heat into space, a cross-section shape of a thin, wide rectangle would be preferred over a square cross-section.  This much copper is a significant mass over several meter lengths so careful design is required to achieve the necessary flexibility at the minimum length.  The PV arrays on the climber shown in this report do not currently have a rotary joint that allows them to be stowed in a vertical position during lift-off. This will need to be designed in future. They are also not currently planned to have active control in yaw and pitch rotation.
\FloatBarrier
\newpage
\subsection{The total number of FELs required to power eight climbers}

Multiple FELs are needed to power eight climbers on the tether simultaneously.  The plan is assumed to be to switch between lasers from day to day as climbers achieve given altitudes.  A single 5 MW FEL is degraded to 68.4\% of its nominal power by the atmospheric absorption, then another 40\% for the efficiency of the PV arrays.  This gives the power available from a single FEL as 1.37 MW before derating it for divergence loss.  Figure \ref{fig:1FEL25m} shows that a single 5 MW FEL with a 2.5 m beam waist can power climbers above 14,400 km up.  An option for the higher climbers is to have dedicated FELs with different configurations, optimized for lower power.  

The climber is designed to split into two separate machines that assemble to each other and surround the tether.  The two halves are as independent as possible for redundancy and so that a minimum number of connections need to be made when the two halves are mated around the tether.  This style of climber has two PV arrays to supply each half of the climber independently,  so the beam from a single FEL must be passed through a beam splitter, then sent to the beam expanders. Restating Table \ref{tab:sevenclimb} but adding a column for the number of lasers required gives Table \ref{tab:8climbFEL}.

We see from the table that a total of 14 Free Electron Lasers are needed to power all the climbers on the tether at one time.  Redundancy is not factored into this number.

\begin{table}[hb]
    \begin{centering}
    \begin{tabular}{cccc}
    \toprule
        Climber number & Altitude, km & Power Required & No. of FELs\\
        \midrule
        1 & 100 & 10.9 MW & 4 \\
        2 & 4,800 & 3.5 MW & 3 \\
        3 & 9,600 & 1.6 MW & 2 \\
        4 & 14,400 & 904 kW & 1 \\
        5 & 19,200 & 526 kW & 1 \\
        6 & 24,000 & 301 kW & 1 \\
        7 & 28,800 & 150 kW & 1 \\
        8 & 33,600 & 41 kW & 1 \\
        \midrule
         & & Total no. of FELs: & 14 \\
        \bottomrule
    \end{tabular}
    \caption{Table of the power required at the wheels of eight 20 tonne climbers on the tether as shown in Figure \ref{fig:multiclimber}, along with the number of FELs needed to provide the power required at each altitude. Climber \#1 gets laser power at 100 km up after reel-out launch as described in Section \ref{sec:reelout}.}
    \label{tab:8climbFEL}
    \end{centering}
\end{table}
\FloatBarrier

\subsection{The transfer function and overall efficiency of the laser power beaming system}
\label{sec:transfer}
To calculate the amount of wall-plug power needed on the ground to drive eight climbers on the tether simultaneously, we need a transfer function that includes all of the inefficiencies that happen between the transmitter and the receiver.  Equation \ref{eq:syseffic} shows the relationship between all of those inefficiencies.

\begin{equation}
    P_{\text{out}} = \varepsilon_1 \cdot \varepsilon_2(z) \cdot \varepsilon_3(z) \cdot \varepsilon_4 \cdot \varepsilon_5(z) \cdot \varepsilon_6(z) \cdot P_{\text{in}}
    \label{eq:syseffic}
\end{equation}

\noindent where $P_{out}$ is the shaft power at the wheels of the climber\\
$\varepsilon_1$ is the laser efficiency \\
$\varepsilon_2(z)$ is the atmospheric transmittance as a function of altitude\\
$\varepsilon_3(z)$ is the laser power efficiency after beam divergence as a function of altitude\\
$\varepsilon_4$ is the PV efficiency \\
$\varepsilon_5(z)$ is the drive electronics efficiency as a function of altitude\\
$\varepsilon_6(z)$ is the motor efficiency as a function of altitude\\
$P_{in}$ is the wall power input to the system\\

Figure \ref{fig:transfer} is a graphical representation of the transfer function as a block diagram.  Each block is a step in the transmission process that degrades the wall-plug power until the shaft power at the wheels remains.  Some of the elements of the transfer function are functions of the altitude of the climber.  For example, at sea level, the atmospheric transmittance would be close to 1.0 because the length of beam passage in the atmosphere is very short.  At the top of the atmosphere, roughly 100 km up, the full effect of atmospheric absorption would have happened to the laser light, leaving 68.4\% of the original 5 MW output. 

The drive electronics efficiency and motor efficiency are also shown as functions of altitude because the efficiency is low at startup (low altitude) and maximizes as the motor delivers its rated torque and power.  These efficiencies may decline as rated torque and power are not required at higher altitudes.  These functions are not currently known.  Numbers shown in Table \ref{tab:altsyseffic} are estimates.

\begin{figure}
\begin{centering}
\includegraphics[width=.9\textwidth]{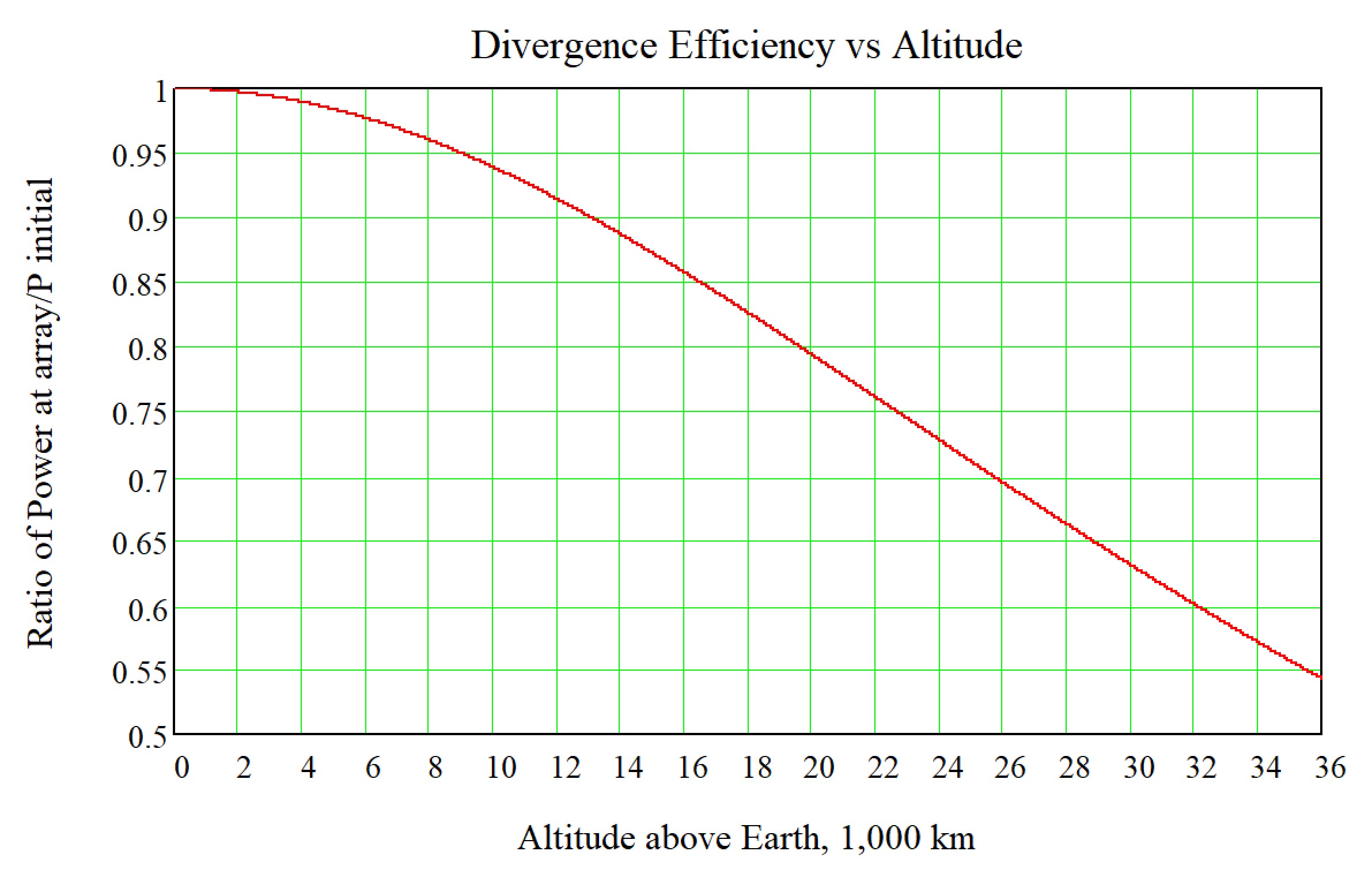} 
\caption{\label{fig:diveffic} Graph of the efficiency associated with the divergence of the 0.5 micron laser beam with a beam waist of 2.5 m as a function of altitude.}
\end{centering}
\end{figure}

\begin{figure}
\begin{centering}
\includegraphics[width=.6\textwidth]{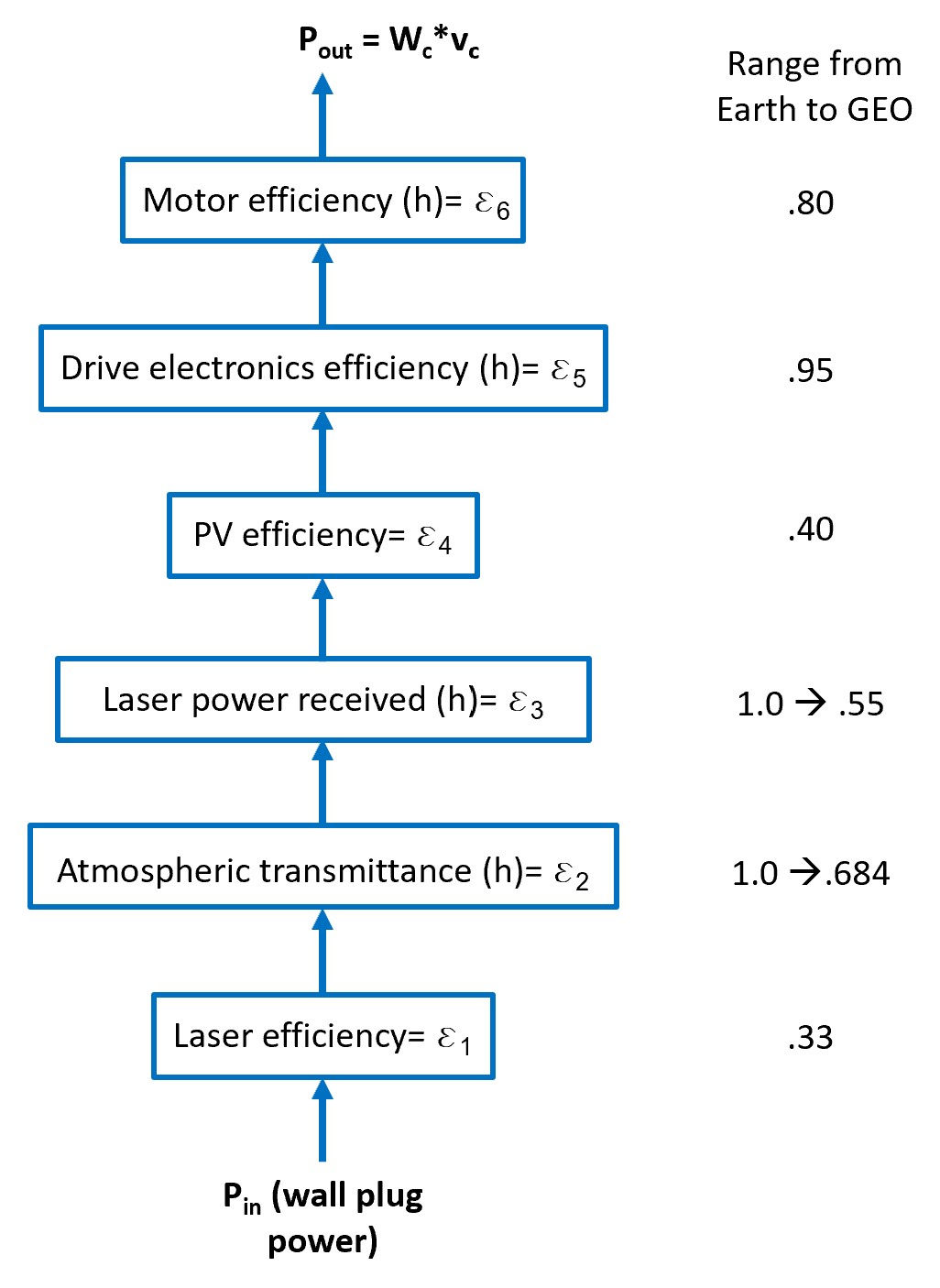} 
\caption{\label{fig:transfer} Block diagram of the transfer function that describes the stacked inefficiencies of the laser power beaming system.}
\end{centering}
\end{figure}

Some of the efficiency functions with altitude are complex processes that are beyond the scope of this report.  The function to calculate $\varepsilon_3(z)$ was described in section \ref{sec:diff} and is shown in Figure \ref{fig:diveffic}. The main focus is to understand how much power is needed at the ground to power the 14 lasers needed for all of the climbers on the tether at one time.

Each climber moves 4,800 km in 24 hours traveling at 200 km/h.  The amount of power required varies the most in the early part of the trip to GEO so we can create a table that shows the power required at the beginning and end of each 24 hour period for each climber moving at 200 km/h.  (The exception is always the first climber as it doesn't get enough power to move at 200 km/h until 3,000 km up.) At the end of a 24 hour period, climber 1 becomes climber 2, and so on.  The lasers powering the climber at the beginning of the 24 hour period have to turn off and rotate downward to acquire the next climber down, then repeat the power transmission they performed the day before.  Table \ref{tab:powerbyday} shows the shaft power required at the climber at the beginning and ending of each 24-hour period of climbing.

\newcolumntype{C}[1]{>{\centering\arraybackslash}p{#1}}
\begin{table}[h]
    \centering
    \begin{tabular}{C{2cm} C{2cm} C{3.5cm} C{3.5cm}}
        \toprule
        Altitude start, km & Altitude end, km & Start Shaft Power Required, MW & End Shaft Power Required, MW \\
        \midrule
        100   & 4800  & 10.900 & 3.500 \\
        4800  & 9600  & 3.500  & 1.600 \\
        9600  & 14400 & 1.600  & 0.904 \\
        14400 & 19200 & 0.904  & 0.526 \\
        19200 & 24000 & 0.526  & 0.301 \\
        24000 & 28800 & 0.301  & 0.150 \\
        28800 & 33600 & 0.150  & 0.041 \\
        33600 & 36000 & 0.041  & 0.000 \\
        \bottomrule
    \end{tabular}
        \caption{The shaft power required at the beginning and end of each 24 hr period of climbing for each climber.}
    \label{tab:powerbyday}
\end{table}

Calculating the system efficiency at each altitude is a matter of multiplying all of the efficiencies shown in Equation \ref{eq:syseffic}.  Table \ref{tab:altsyseffic} shows the efficiencies at different altitudes and how they combine to give the system efficiency.  This is in line with statements made at the beginning of Section \ref{sec:microwave} on microwave power beaming that optical efficiencies will be lower than microwave efficiencies.  Efficiency is not the most important parameter here as it is clear that optical receiver arrays are much smaller than microwave receivers.  Mass of the receiver determines its viability for the climber.

\newcolumntype{D}[1]{>{\centering\arraybackslash}p{#1}}
\begin{table}[h]
    \centering
    \begin{tabular}{D{1.5cm} D{1.5cm} D{1.7cm} D{1.7cm} D{1.7cm} D{1.8cm} D{1.5cm} D{1.5cm}}
        \hline
            Altitude start, km
            & $\varepsilon_1$, Laser efficiency 
            & $\varepsilon_2$, Atmospheric transmittance 
            & $\varepsilon_3$, Divergence Efficiency 
            & $\varepsilon_4$, PV efficiency 
            & $\varepsilon_5$, Drive electronics efficiency 
            & $\varepsilon_6$, Motor Efficiency 
            & $\varepsilon_S$, System Efficiency \\
        \hline
        100   & 0.33 & 0.684 & 1.000 & 0.40 & 0.95 & 0.80 & 6.86\% \\
        4800  & 0.33 & 0.684 & 0.985 & 0.40 & 0.95 & 0.96 & 8.11\% \\
        9600  & 0.33 & 0.684 & 0.944 & 0.40 & 0.95 & 0.96 & 7.77\% \\
        14400 & 0.33 & 0.684 & 0.881 & 0.40 & 0.95 & 0.96 & 7.25\% \\
        19200 & 0.33 & 0.684 & 0.807 & 0.40 & 0.95 & 0.96 & 6.65\% \\
        24000 & 0.33 & 0.684 & 0.728 & 0.40 & 0.95 & 0.96 & 5.99\% \\
        28800 & 0.33 & 0.684 & 0.650 & 0.40 & 0.95 & 0.96 & 5.35\% \\
        33600 & 0.33 & 0.684 & 0.577 & 0.40 & 0.95 & 0.96 & 4.75\% \\
        \hline
    \end{tabular}
    \caption{Combining all the inefficiencies to calculate the system efficiency at each altitude. Motor efficiency as a function of altitude requires more realistic analysis, as they do not operate at full rated power or torque at higher altitudes.}
    \label{tab:altsyseffic}
\end{table}

\newcolumntype{L}[1]{>{\centering\arraybackslash}p{#1}}
\begin{table}[h]
    \centering
    \begin{tabular}{L{1cm} L{1.2cm} L{1.7cm} L{1.7cm} L{1.7cm} L{1.5cm} L{1.5cm}}
        \toprule
        Day & Number of FELs & Max FEL power, MW & Start Max Shaft Power, MW & End Max Shaft Power, MW & Start Climber speed, km/h & End Climber speed, km/h \\
        \midrule
        1 & 4 & 20 & 4.159 & 4.096 & 76.3  & 234.1 \\
        2 & 3 & 15 & 3.687 & 3.533 & 210.7 & 441.7 \\
        3 & 2 & 10 & 2.355 & 2.198 & 294.4 & 486.4 \\
        4 & 1 & 5  & 1.099 & 1.007 & 243.2 & 382.8 \\
        5 & 1 & 5  & 1.007 & 0.908 & 382.8 & 603.5 \\
        6 & 1 & 5  & 0.908 & 0.811 & 603.5 & 1,081.3 \\
        7 & 1 & 5  & 0.811 & 0.720 & 1,081.3 & 3,511.6 \\
        8 & 1 & 5  & 0.720 & 0.000 & 3,511.6 & 0.0 \\
        \bottomrule
    \end{tabular}
    \caption{Calculating the speed of the climbers at each altitude if the FELs were all run at their full power of 5 MW. It is clear that the FELs do not need to run at full 5 MW for the higher-altitude climbers, as the climbers cannot climb at those speeds.}
    \label{tab:maxspeed}
\end{table}

Almost all of the graphs shown in Section \ref{sec:powerlossreq} show that once the climber gets enough power to climb at 200 km/h at the lowest altitude possible, there is substantial power available at higher altitudes if the 5 MW FELs are run at full power.  Table \ref{tab:maxspeed} shows that this extra power translates to possible climber speeds that are way beyond the mechanical limits of bearings and other parts of the drive train.  More detailed engineering of the climber drive train in future may allow speeds of 300 or 400 km/h eventually, but for the purposes of this study the maximum speed of a climber is limited to 200 km/h.  Given this speed limit, it then makes sense to lower the power in the higher altitude FELs to preserve the speed limit and run the elevator in the most efficient way by not beaming power up where it is not needed.

\newcolumntype{E}[1]{>{\centering\arraybackslash}p{#1}}
\begin{table}[h]
    \centering
    \begin{tabular}{E{1cm} E{2cm} E{2cm} E{2cm} E{2cm} E{2cm} E{2.2cm}}
        \hline
        Day & Number of FELs & Start FEL Power derating & End FEL power derating & Max Shaft Power, MW & Total FEL Power, MW & Max Wall-plug power, MW \\
        \hline
        1 & 4 & 1.00 & 0.85 & 4.16 & 20.00 & 60.61 \\
        2 & 3 & 0.95 & 0.45 & 3.50 & 14.24 & 43.15 \\
        3 & 2 & 0.68 & 0.41 & 1.60 & 6.79 & 20.58 \\
        4 & 1 & 0.82 & 0.52 & 0.90 & 4.11 & 12.46 \\
        5 & 1 & 0.52 & 0.33 & 0.53 & 2.61 & 7.92 \\
        6 & 1 & 0.33 & 0.18 & 0.30 & 1.66 & 5.02 \\
        7 & 1 & 0.18 & 0.06 & 0.15 & 0.92 & 2.80 \\
        8 & 1 & 0.06 & 0.00 & 0.04 & 0.28 & 0.86 \\
        \hline
        \multicolumn{6}{r}{Total Wall-Plug Power:} & 153.41 \\
        \hline
    \end{tabular}
        \caption{Table of the wall-plug power required to power eight 20 tonne climbers moving at 200 km/h upwards on the tether simultaneously as shown in Figure \ref{fig:multiclimber}. Columns are shown for the number of FELs and the amount of derating each FEL platform needs at the beginning and end of each day to keep climbers moving at only 200 km/h.  Climber 1 is getting the full output of four 5 MW lasers at the beginning of its climb, but cannot move at 200 km/h because the lasers are reduced to 13.68 MW just above the atmosphere and that is reduced to 4.2 MW at the wheels by the remaining inefficiencies, not the 10.9 MW required for 200 km/h travel.}
        \label{tab:wallplug}
\end{table}

Table \ref{tab:wallplug} compiles the shaft power delivered at each of the climbers rolling upwards at 200 km/h at each of their altitudes, and factors in all of the inefficiencies to give a total wall-plug power for the system of 153.4 MW.  The derating factors at the higher altitudes show that climbers 4 through 8 don't need a 5 MW FEL, but can use a lower-power laser system.  Even if the laser source is not an FEL, the output beam still needs to be of high quality and passed through the beam expander to keep its divergence as required.

The ground power system has to be sized for the maximum load, so the FEL power at the beginning of each day is used to compute the wall-plug power required.  How this power is transmitted to each platform will be a function of the location of each FEL platform and the power it requires for the altitudes it serves.

\FloatBarrier

\clearpage
\section{Microwave Power Beaming to the Climbers}
\label{sec:microwave}
\subsection{Introduction}

Powering the space elevator climber has many challenges.  The power required is immense and the mechanism must be kept lightweight.  If not done right, the climber propulsion would be almost as inefficient as the rocket equation.  Microwave power beaming offers a possible means of wirelessly transferring electrical power to the climber for propulsion.  
The basic physics of microwave power beaming will be presented.  The general challenges of applying the physics to the SE climber will be analyzed and possible approaches will be proposed.  Further research needed will be described.

This section summarizes analyses and concepts for using microwave power beaming to power the space elevator.  It concentrates on physics, which impacts scientific readiness level, and does not dwell on technology readiness.  Many of the references do present technological details.

\subsection{Assumptions and inputs}
\label{sec:microwave-assumptions}

Assumptions and inputs are based on [\cite{Robinson2022}]:
\begin{enumerate}
    \item Gross vehicle weight of a loaded climber = 20 tonnes
    \item Climber shaft power is limited to 4 MW. To compensate for inefficiencies, the power received at the climber must be more, and the source power delivered to the transmitter may be much more.
    \item Climber speed is limited to 200 km/h
    \item Climber trip on a tether from ground to geosynchronous orbit is a distance of about 36,000 km
    \item The climber will be equipped with microwave power receiver system which converts microwave energy into direct current [\cite{rectenna}]
        \begin{enumerate}
            \item Commonly described as a rectenna for rectifying antenna.
            \item Rectenna power conversion efficiency is 85\%.
            \item Climber motor drive efficiency is 95\%, for combined climber efficiency of 81.6\%
            \item Rectenna mass is 1 g/W [\cite{ozawa2015study}] or 1 kg / kW, or 1 tonne/MW, thus a rectenna capable of receiving 4~MW would weigh 4 tonnes or 4,000 kg.
        \end{enumerate}
    \item A microwave transmitter will be focused at the rectenna.
        \begin{enumerate}
            \item Different transmitter locations will be considered
            \item Transmitter power will be controlled based on telemetry from the climber to keep shaft power below 4~MW, or just what is needed for 200 km/h, whichever is less.
        \end{enumerate}
    \item A 20 tonne climber module uses pairs of rollers that pinch the tether, each wheel powered by an electric motor. A microwave rectenna provides electrical power for the motors.
\end{enumerate}

It is desirable to limit beam power density to 230 \si{W/m^2} in the ionosphere, 100 km to 400 km altitude [\cite{jaffe2024power}, pp. 285], but this figure assumes a 10 km wide beam from a Solar Power Satellite beaming power to a 10 km wide rectenna on the ground delivering up to 5 GW.  A space elevator climber will have a much smaller rectenna with a much smaller beam width and much lower total power in the beam, so it will be unlikely to excite undesirable effects in the ionosphere based on this excerpt: 

"Finally, it is also essential to carry out further experimental studies of microwave-plasma interactions. The single small-scale MINIX (Microwave Ionosphere Nonlinear Interaction Experiment) study reported by Kaya et al., 1986 is a rather inadequate basis for inferring the behavior of a full-scale SPS system for various reasons indicated in section 3 above. The MINIX results did indicate that plasma waves were excited by the stimulated Raman scattering mechanism, but other important effects such as stimulated Brillouin scattering, plasma heating and nonlinear self-focusing, which are expected from theory and which could have important consequences for both the operating efficiency of the SPS system and the impact on the environment, were not detected. The reasons for this have been discussed above. In particular, the narrow microwave beams used on the MINIX would also be unlikely to excite self-focusing effects. Much broader beams in real SPS systems would not necessarily behave in the same manner."  [\cite{robinson2004environmental}, pp.20]

The following analysis will show how to comply with the SPS PD limit, but will also consider higher power density approaches.

\subsection{NASA comparison of microwave and optical wireless power transmission}

One motivation for considering microwave power beaming for the space elevator is an old NASA comparison with optical power beaming [\cite{laserMWcomp}]. From the NASA comparison: 

Microwave efficiency: Rectenna RF-DC conversion 85\%, Overall ~50\%

Optical efficiency: DC conversion 70\%, Overall ~8\%

Microwave power beaming will likely be much more efficient than optical power beaming.

Microwave power beaming will likely be much more efficient than wireless communications transmission because the power beaming is beamcast, not broadcast.  The power beaming transmitter is focused on the receiver so it is limited by diffraction, not the inverse square law, and very little power is lost in transmission.

\begin{figure}
\begin{centering}
\includegraphics[width=\textwidth]{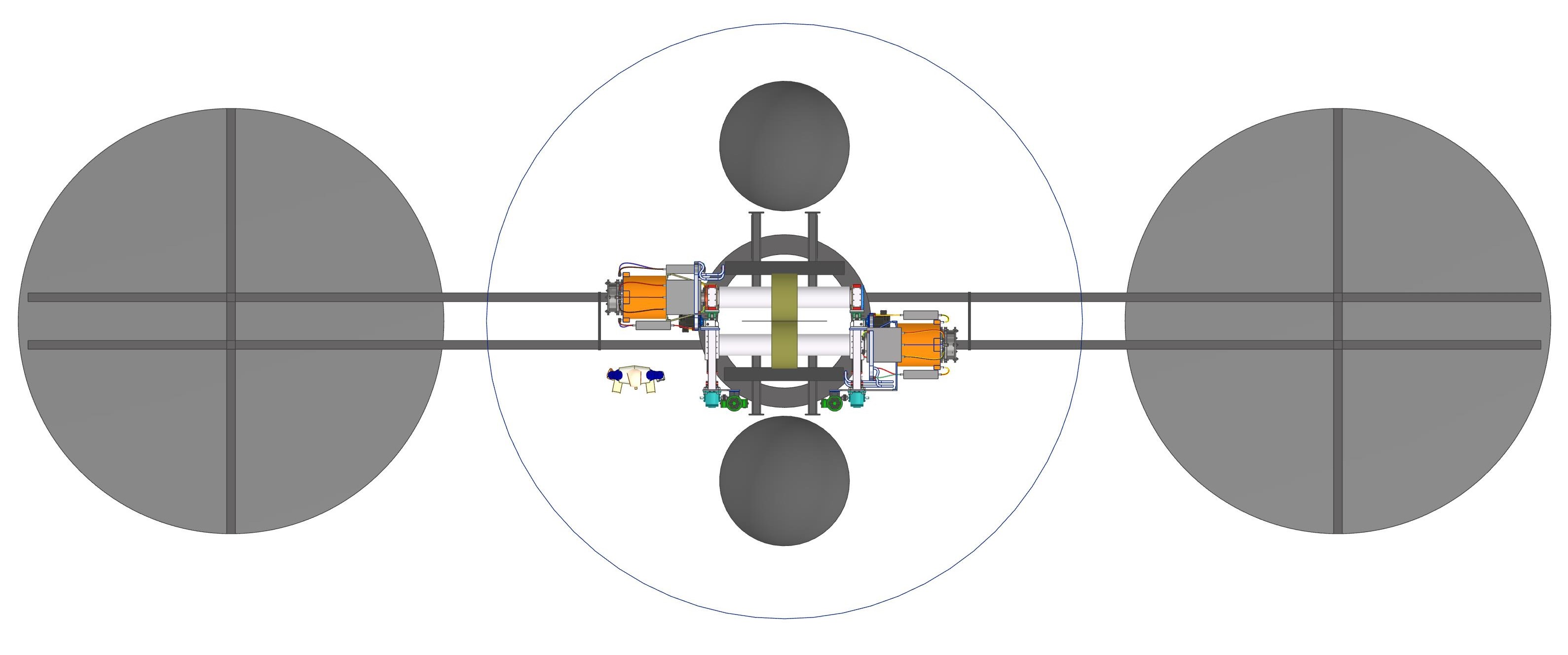}    
\caption{\label{fig:Fig1climber} Climber top view rendering with two LOX tanks as payload (small dark gray circles above and below the wheels) and two notional 5 m diameter rectennas on the left and right sides to receive power. This was a conceptual sketch comparing the optical PV arrays and a possible rectenna configuration, but it may not represent a workable rectenna design.}
\end{centering}
\end{figure}

\subsection{SE power and speed versus altitude}

\begin{figure}
\begin{centering}
\includegraphics[width=.9\textwidth]{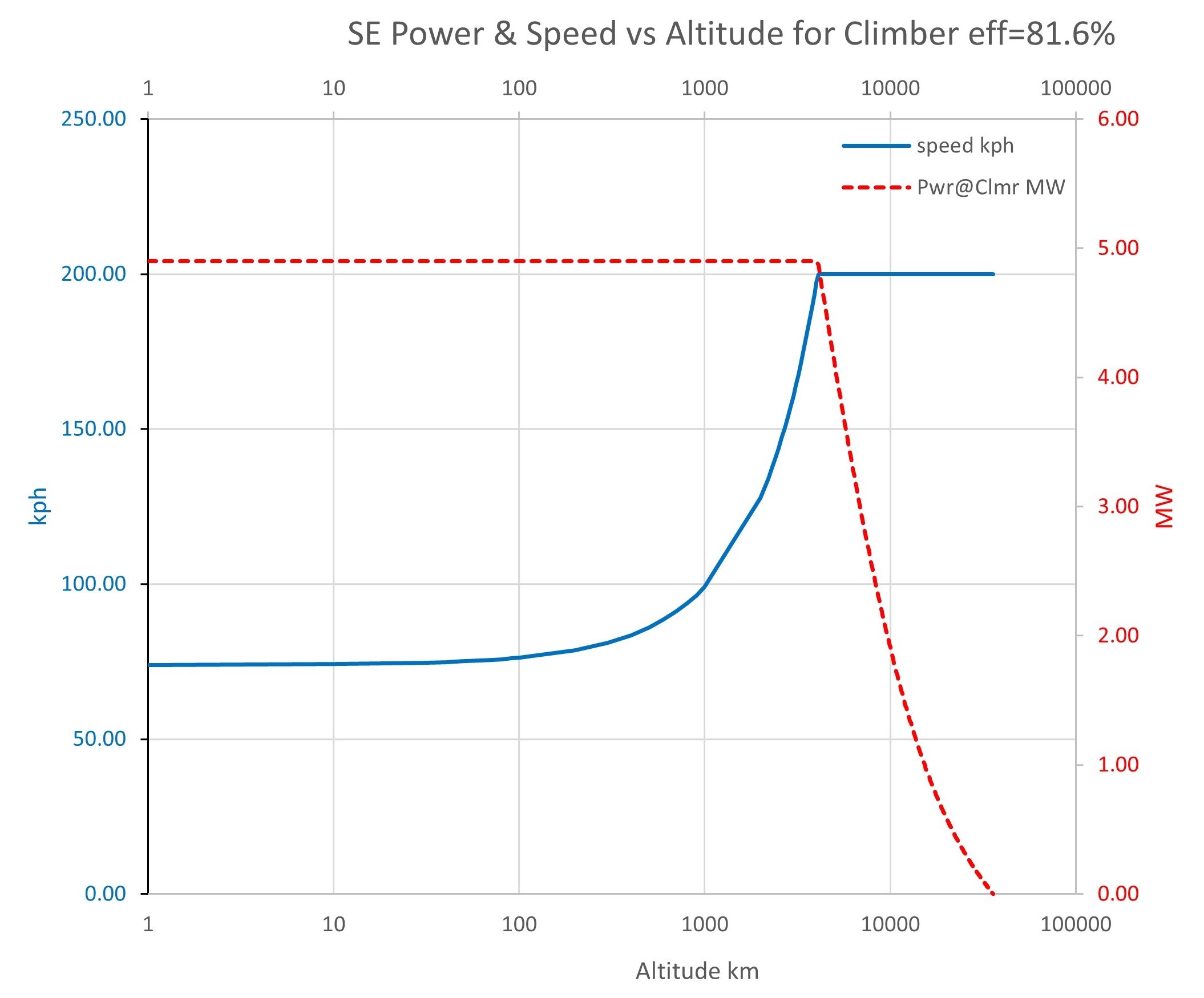}
\caption{\label{fig:Fig2power} Climber power and speed versus altitude with limits and some inefficiencies. Pwr@Clmr is the power incident at the rectenna before inefficiencies.}
\end{centering}
\end{figure}

The power required to operate a space elevator increases with speed, like any terrestrial elevator. Increasing altitude diminishes the force of gravity by the inverse square law, and, since the elevator rotates with the Earth, increasing altitude increases centrifugal force, which counters gravity.  Ultimately, at the altitude designated the geosynchronous orbit, these forces balance out and produce a microgravity environment. [\cite{Robinson2022}] wrote a nice analysis of this, but others may be helpful [\cite{earthgrav}], [\cite{elforce}], [\cite{orbits}], [\cite{tangentsp}].  Figure \ref{fig:Fig2power} shows how the incident rectenna power needed to deliver 4 MW shaft power limit, with inefficiencies, and the 200 km/h speed limit interact for a 20 tonne climber as it ascends in altitude. PWR@Clmr is the power incident at the rectenna before inefficiencies.

Microwave exposure power density examples: 

10 \si{mW/cm^2} = 100 \si{W/m^2} generally considered maximum safe for personnel exposure

230 \si{W/m^2} = maximum to prevent adverse ionospheric effects

1000 \si{W/m^2} = sunlight

10,000 \si{W/m^2} = microwave oven = 10 times sunlight

\subsection{Approach to microwave power beaming analysis}

Efficiency calculations are based on the Goubau relationship [\cite{jaffe2024power} pp. 167].  This is a focused system, deliberately violating far-field criteria so the inverse square law does not apply.  The transmitter utilizes technology from Active Electronically Scanned Arrays (AESA) for the microwave phased array antenna, which is focused like a virtual hologram of a Fresnel lens forming a retrodirective array using phase differences from a pilot signal sent from the receiver [\cite{jaffe2024power}, pp. 266] to keep the power beam directed at and focused on the rectenna.

“In the AESA, each antenna element is connected to a small solid-state transmit/receive module (TRM) under the control of a computer, which performs the functions of a transmitter and/or receiver for the antenna. \dots AESAs add many capabilities of their own to those of the PESAs. Among these are: the ability to form multiple beams simultaneously \dots” [\cite{AESA}]
\newpage

This capability to form multiple beams simultaneously will be employed to form power beams for multiple SE climbers.  Multiple beams may be produced simultaneously if the power amplifiers at each element are linear so that superposition holds.  Thus, one transmitter array may produce multiple focused beams simultaneously to power multiple climbers.  It will just need to be designed to handle the worst-case sum of power for multiple beams.

\begin{figure}
\begin{centering}
\includegraphics[width=.7\textwidth]{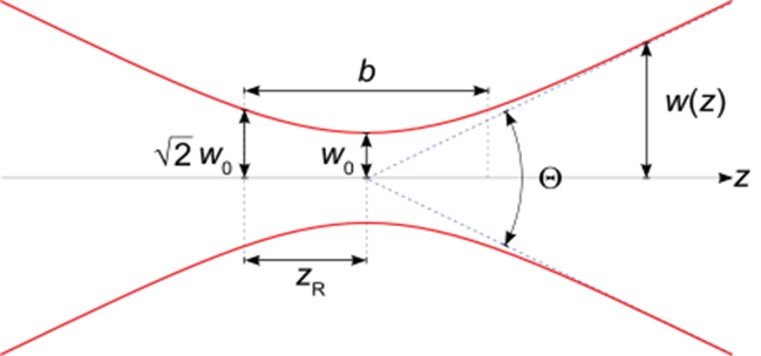}    
\caption{\label{fig:Fig4gaussian} Gaussian beam width w(z) as a function of the axial distance z}
\end{centering}
\end{figure}

Like all focused systems, the smallest spot size is limited by diffraction, so that the inverse square law for far field antenna systems does not apply (see Figure \ref{fig:Fig4gaussian}). Microwave power beaming is based on this for the best efficiency.

A focused system is diffraction-limited to a Gaussian beam width as depicted in Figure \ref{fig:Fig4gaussian} [\cite{beamdiv}], with these parameters: 
\bigskip

$w_0$: beam waist; 

b: depth of focus; 

$z_R$: Rayleigh range; 

$\Theta$: total angular spread
\bigskip

Derivation of the Goubau relationship for power-link efficiency; subscripts r for receiver, t for transmitter:
\begin{equation}  
\mathrm{\Theta}_r=2\theta\simeq\frac{4\lambda}{\pi D_r}
\label{eq:Thetatheta}
\end{equation}

\begin{equation} 
\mathrm{\Theta}_t=\frac{D_t}{f}=\frac{D_t}{R}
\end{equation}

$\Theta$ = total angular spread 

D = Diameter of aperture; D= 2r (r is radius)

A = area of aperture; $A = \pi r^2 = (\pi D^2)/4$

$\lambda$  = wavelength

f = focal length of $D_t$; set equal to R

R = distance between source aperture and receive aperture

$\tau$ = power-link parameter

P = power

$\varepsilon$ = power efficiency, proportion of transmitted beam within receiver aperture~\label{eq:GEpsilon}

\begin{equation}
\tau=\frac{\mathrm{\Theta}_t}{\mathrm{\Theta}_r} =\frac{{\pi D_rD}_t}{4\lambda R} =\frac{\sqrt{{A_rA}_t}}{\lambda R}
\end{equation}

\begin{equation}
\varepsilon=\frac{P_r}{P_t}\simeq1-e^{-\tau^2}
\end{equation}

This is the Goubau relationship for power efficiency, which “can provide a quick, straightforward approximation of beam collection efficiency for power beaming links. It addresses geometric factors only and thus neglects inefficiencies of conversion or sources of loss in the transmission medium” [\cite{jaffe2024power} pp.167].  These formulas imply a Gaussian taper for the beam spot at the receiver.  Beam shaping is commonplace in the optics and microwave industries, and these techniques are applicable to WPT.  A custom illumination function at the transmitter may be used to produce a flat, “top-hat”, profile at the receiver at the expense of a larger and more complex transmitter aperture.  “The use of diverse forms of phased arrays makes both the range of possible apertures and possible beam profiles effectively limitless.” [\cite{jaffe2024power} pp.152-154]

The essence of the Goubau relationship is that a smaller aperture on one side requires a larger aperture on the other side for good efficiency. Beam efficiency is a measure of how much of the transmitted beam is collected by the receiver aperture; the rest misses it and disperses into space past the receiver.  

\subsection{Proposals for microwave power beaming to the SE climber}

\subsubsection{Ground power only for microwave power transmission to the SE climber}

The anticipated initial operating capability for the microwave-powered space elevator will utilize power from a ground-mounted transmitter only. A spreadsheet analysis has been done which combines the above equations with the power and speed versus altitude profile in Figure \ref{fig:Fig2power}, for different parameters.  See Table \ref{tab:groundpowersummary} for the cases analyzed for ground power only to the climber. Only cases deemed optimal or illustrative will have full graphs presented.

The 10 m receive aperture on the climber cases (rows 1 and 2 in Table \ref{tab:groundpowersummary}) are optimal from a climber mechanical perspective by having a smaller receive array, but require a very large transmitter diameter (Dt) for high beam efficiency at high altitudes.  Even with Dt = 10 km, the transmitter is required to greatly increase power to follow the power requirements of the SE climber, indicating a poor WPT link efficiency at higher altitudes.  Additionally, the power density greatly exceeds the desired power density (PD) limit for the ionosphere, but the beam width is very small, which should mitigate undesirable effects in the ionosphere.

\newcolumntype{M}[1]{>{\centering\arraybackslash}p{#1}}
\begin{table}[h]
    \centering
    \begin{tabular}{ccc M{2.5cm} M{3cm} cc}
        \toprule
        Dr & Dt & ETA & Pk PD@RX & PWR 8 CLMRS & Energy/CLMR & Cost/CLMR \\
        (m) & (m) & (days) & (\si{W/m^2} at 100 km) & (MW) & (kWh) & (\$/kg) \\
        \midrule
        10  & 10,000 & 8.09 & 62,413.70 & 226.79 & 5.43E+06 & \$271.39  \\
        10  & 30,000 & 8.09 & 62,413.70 & 36.69  & 8.93E+05 & \$44.63   \\
        30  & 3,000  & 8.09 & 6,934.86  & 277.21 & 6.64E+06 & \$331.83  \\
        30  & 10,000 & 8.09 & 6,934.86  & 36.69  & 8.93E+05 & \$44.63   \\
        50  & 1,000  & 8.09 & 2,496.55  & 871.98 & 2.09E+07 & \$1{,}045.75 \\
        50  & 2,000  & 8.09 & 2,496.55  & 226.79 & 5.43E+06 & \$271.39  \\
        165 & 1,000  & 8.09 & 229.25   & 90.96  & 2.18E+06 & \$109.06  \\
        165 & 2,000  & 8.09 & 229.25   & 32.76  & 7.99E+05 & \$39.93   \\
        165 & 10,000 & 8.09 & 229.25   & 17.01  & 4.22E+05 & \$21.09   \\
        \bottomrule
    \end{tabular}
    \caption{Summary of ground power only microwave beaming energy and cost for 20 tonne climber with 50\% payload (10 tonnes), for various transmitter and receiver diameters. Total transmitter wall power is summarized for eight simultaneous climbers on the tether at the altitudes listed in Table \ref{tab:sevenclimb} and depicted in Figure \ref{fig:multiclimber}, designated as “PWR 8 CLMRS”.  “ETA" is estimated time of arrival, or trip time.  “Pk PD@RX” is the peak power density at the receiver (climber rectenna).  “Energy/CLMR” is the energy consumption for the ascent by a single climber.  Cost/CLMR” is the estimated cost of the energy consumed for the ascent by a single climber using a cost of \$0.50/kWh.}
    \label{tab:groundpowersummary}
\end{table}
\FloatBarrier

\begin{figure}
\begin{centering}
\includegraphics[width=\textwidth]{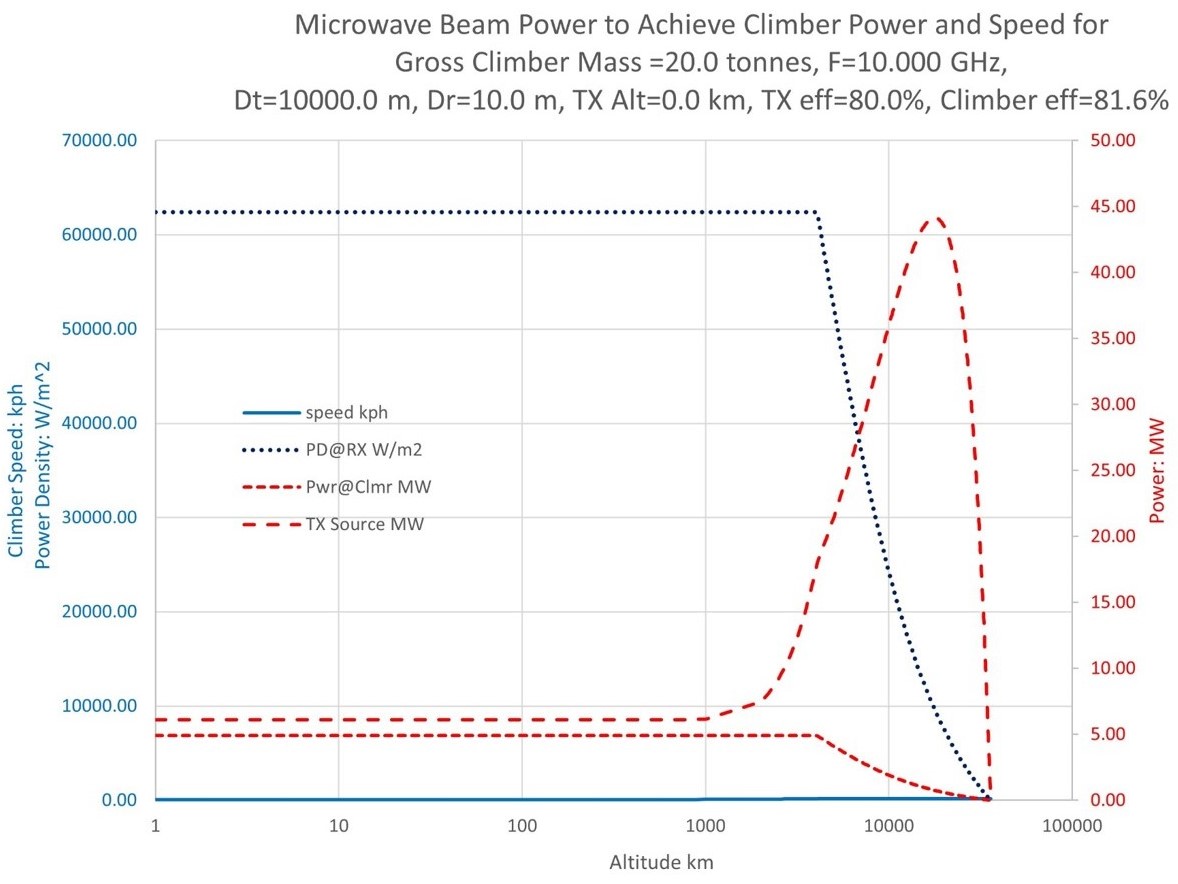}    
\caption{\label{fig:Dr10mDT10km} Climber Power and Speed for ground power only at Dr=10 m, Dt=10 km (Row 1 of Table \ref{tab:groundpowersummary}.)}
\end{centering}
\end{figure}

The Dr=30 m cases (rows 3 and 4 of Table \ref{tab:groundpowersummary}) require the transmitter to have much greater power to follow the propulsion power requirements of the SE climber.  A very large Dt=10 km is required for good beam efficiency.  With Dt=3 km, the high cost indicates poor WPT link efficiency.  Additionally, the power density exceeds the desired PD limit for the ionosphere, but the beam width is pretty small, which should mitigate undesirable effects in the ionosphere.

The Dr=50 m cases (rows 5 and 6 of Table \ref{tab:groundpowersummary}) require the transmitter to also have much greater power to follow the propulsion power requirements of the SE climber, indicating poor WPT link efficiency.  Additionally, the power density exceeds the desired PD limit for the ionosphere, but the beam width is small which should mitigate undesirable effects in the ionosphere.  

The Dr=165 m cases (rows 7 to 9 of Table \ref{tab:groundpowersummary}) attempt to limit power density to 230 \si{W/m^2} in the ionosphere and still provide 4 MW of power after inefficiencies by increasing the size of the rectenna to 165 m diameter.  The rectenna area is estimated by solving:  4 MW/(inefficiencies*230 \si{W/m^2}) for area, which gives a disc diameter of about 165 m for the rectenna.  The actual power transmitted will need to change depending on distance and transmitter diameter in order to maintain this power density.  The WPT efficiency is much greater than the first two cases as indicated by a reduced launch cost.  The full 4 MW shaft power is available to the climber at low altitudes.  Improving transmission power link efficiency will require a larger transmitter, such as 2 km in diameter, which provides a mostly level power output from the transmitter.  While poor efficiency is estimated for higher altitudes, the dropping power requirement for climber shaft power keeps the transmitter power reasonable.  The Dr=165 m and Dt=2 km case shows better efficiency than the previous cases.  The full 4 MW power limit is available to the climber at low altitudes.

\begin{figure}
\begin{centering}
\includegraphics[width=\textwidth]{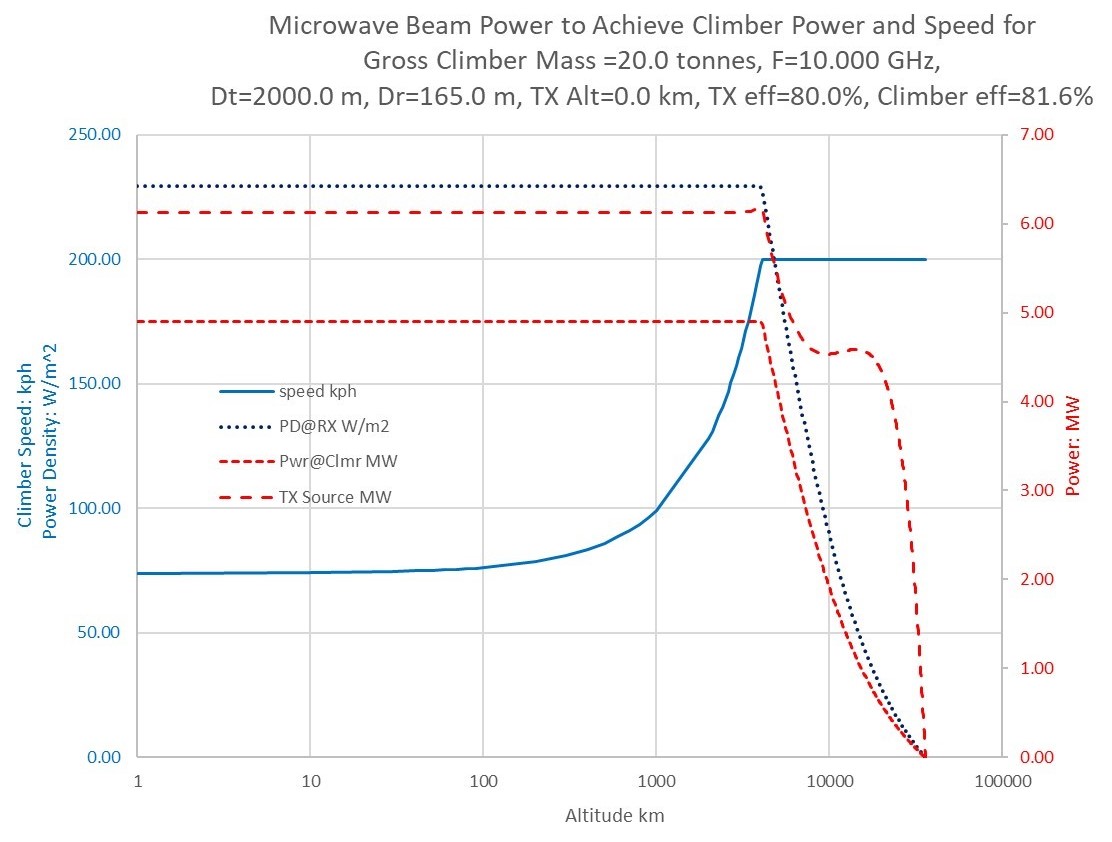}    
\caption{\label{fig:4p77} Climber Power and Speed for ground power only at Dr=165 m Dt=2 km (Row 8 of Table \ref{tab:groundpowersummary}.)}
\end{centering}
\end{figure}

\FloatBarrier

\subsubsection{Geosynchronous solar power satellite only powering a climber}

An early use of a space elevator will be to carry components to construct Solar Power Satellites (SPS) in Geosynchronous Orbit.  These SPS may then be used to power space elevator climbers.  In the SPS scenario with the transmitter at 36,000 km, several cases were analyzed (see Table \ref{tab:SPSpowersummary}).

The TX diameter was changed from 1,000 m to 30 km, and the RX diameter from 10 m to 165 m.  Only the 10,000 m TX and 165 m RX diameters provided good efficiency and kept the power density in the ionosphere constrained to 230 \si{W/m^2}.  However, these large aperture diameters may be impractical. 

\begin{table}[h]
    \centering
    \begin{tabular}{ccccccc}
        \toprule
        Dr & Dt & ETA & Pk PD@RX & PWR 8 CLMRS & Energy/CLMR & Cost/CLMR \\
        m & m & days & W/m$^2$ at 100km & MW & kWh & \$/kg \\
        \midrule
        10  & 10,000 & 8.09 & 62,413.70 & 2,310.39 & 5.63E+07 & \$2{,}814.33 \\
        10  & 30,000 & 8.09 & 62,413.70 & 264.40  & 6.44E+06 & \$322.24   \\
        30  & 3,000  & 8.09 & 6,934.86  & 2,850.33 & 6.94E+07 & \$3{,}472.00 \\
        30  & 10,000 & 8.09 & 6,934.86  & 264.40  & 6.44E+06 & \$322.24   \\
        50  & 1,000  & 8.09 & 2,496.55  & 9,216.06 & 2.25E+08 & \$11{,}225.70 \\
        50  & 2,000  & 8.09 & 2,496.55  & 2,310.39 & 5.63E+07 & \$2{,}814.33 \\
        165 & 1,000  & 8.09 & 229.25   & 854.04  & 2.08E+07 & \$1{,}040.43 \\
        165 & 2,000  & 8.09 & 229.25   & 220.04  & 5.36E+06 & \$268.21   \\
        165 & 10,000 & 8.09 & 229.25   & 19.95   & 4.91E+05 & \$24.56    \\
        \bottomrule
    \end{tabular}
    \caption{Summary of GEO SPS power only microwave beaming energy and cost for 50\% payload fraction (10 tonnes) and \$0.50/kWh}
    \label{tab:SPSpowersummary}
\end{table}

\begin{figure}
\begin{centering}
\includegraphics[width=\textwidth]{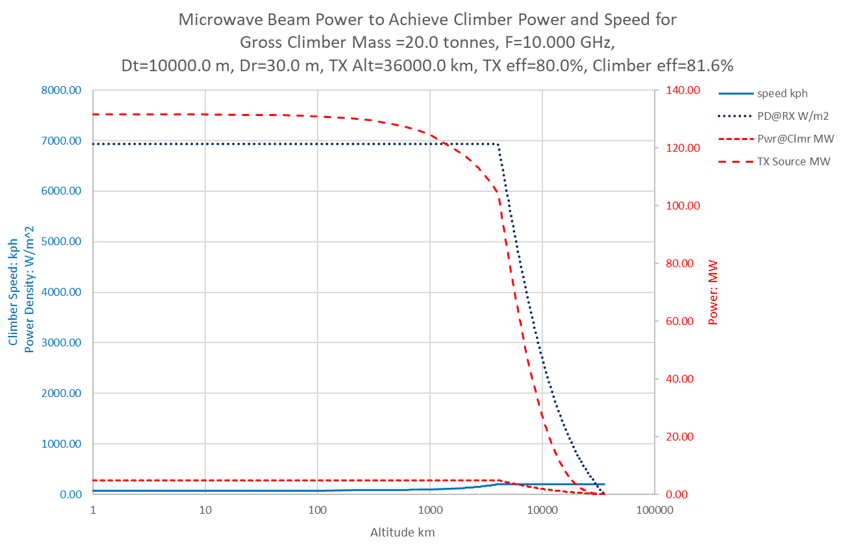}    
\caption{\label{fig:SPSDr30Dt10km} Climber power and speed for GEO SPS power only at Dr=30 m Dt=10 km (Row 4 of Table \ref{tab:SPSpowersummary}.) The power at the climber is a small fraction of the power at the transmitter.}
\end{centering}
\end{figure}

\FloatBarrier

\subsubsection{Solar power satellite suspended from tether at arbitrary altitude powering a climber}

\begin{figure}[h]
\begin{centering}
\includegraphics[width=\textwidth]{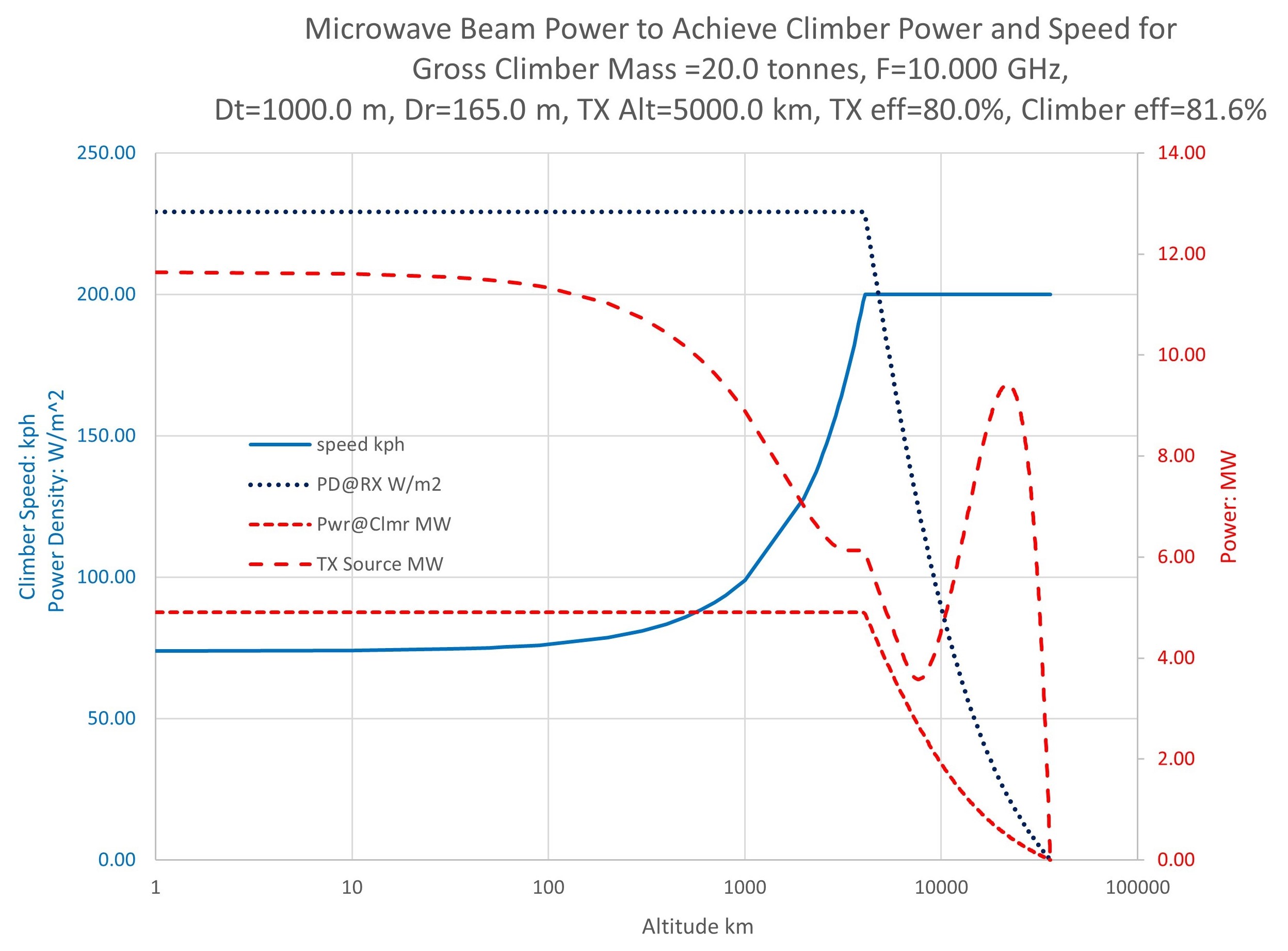}    
\caption{\label{fig:Fig8Tx5000km} Climber Power and speed for a Solar Power Station on a secondary tether at an altitude of 5,000 km}
\end{centering}
\end{figure}

Another alternative considered is a Solar Power Satellite suspended from a different tether at 5,000 km altitude, providing wireless power transfer for a climber with a 165 m diameter rectenna to keep power density within limits for the ionosphere.  This configuration seems to provide poor efficiency with a transmitter aperture of 1,000 m diameter, but at 2,000 m diameter, the efficiency is pretty good.  Plus, the full 4 MW power limit is available to the climber at low altitudes. Figure \ref{fig:Fig8Tx5000km} shows the curves for this configuration.  (Suspending an SPS from the tether that the climber climbs on presents formidable engineering challenges.)

\FloatBarrier

\subsubsection{Geosynchronous solar power satellite beaming power to a ground receiver array with an SE climber parasitically tapping into the beam}

Many designs for SPS providing power to a ground station have a nominally 1 km diameter transmitter feeding a 10 km diameter rectenna on the ground, with the power adjusted to provide a receiver power density of 230 \si{W/m^2} at the ground.  In this alternative, the diameters of the transmitter and receiver are made the same so the power density is constant with altitude, and the power is adjusted for 230 \si{W/m^2}. Reversing the equations for beam efficiency and tau, at 10.0 GHz and 36,000 km distance, this will require SPS and ground diameters of about 2,000 m for 99.9\% efficiency.  The SPS and ground receiver arrays would be positioned so that the tether goes through their centers, and the climber is in the power beam.  This will provide about 650 MW on the ground receiver, while a 165 m diameter rectenna on the climber will provide about 4 MW shaft power after inefficiencies from that same beam.  This should allow multiple simultaneous climbers, and for higher speeds at higher altitudes if the mechanical issues could be resolved.

\subsubsection{Energy harvesting from a descending climber}

The concept in the previous section of a climber parasitically tapping into a beam from an SPS also enables the exporting of regenerative braking energy in a descending climber by radiating it as microwaves to a big rectenna on the ground.  At lower altitudes, this should be fairly efficient in recovering the gravitational energy generated by a descending climber.  See Section \ref{sec:mwavebeamoff} for more details.

\subsubsection{Simultaneous ground and GEO SPS power for the climber with power density limited just in the ionosphere}

This section will make an assumption that a bidirectional rectenna will be developed so that beamed power may be received simultaneously from both sides, that is, from the ground and from an SPS in GEO.  Analysis will include limiting the power density while in the ionosphere at 230 \si{W/m^2} from above and below, assuming the rectenna array will shadow the power density from one side to the other.  This will keep the total power density while traversing the ionosphere controlled to about 460 \si{W/m^2} total power density.  Analysis will also include unlimited power density at the climber.  Several different rectenna and transmitter aperture diameters are analyzed to produce 4 MW maximum shaft power for climber propulsion with inefficiencies.  

For a limited power density, the speed of the SE climber may be limited by the power received depending on the receiver aperture diameter.  Both the ground-based and GEO SPS power transmitters will be controlled to produce either power required for the climber, or the specified limited power density at the climber.

A spreadsheet analysis was used to produce plots similar to the previous section but for dual transmitters at two different altitudes that simultaneously provide power to the climber.  For some cases, the power density is limited at the altitudes of the ionosphere to 230 \si{W/m^2} for each transmitter.  Table \ref{tab:dualrectenna2} has a summary of these cases with the ground transmitter, Dt1, only active below 400 km altitude, and the GEO SPS transmitter, Dt2, only active above 100 km.  Two columns are presented for the climber receive aperture because the spreadsheet allows making the aperture to receive ground power different from the aperture to receive SPS power, which is explored in the next section. 

\begin{table}[h]
    \centering
    \begin{tabular}{ccccc M{2cm} M{2cm} M{2cm} M{2cm}}
        \hline
        Dr1 & Dr2 & Dt1 & Dt2 & ETA & Pk PD@RX & PWR 8 CLMRS & Energy /CLMR & Cost /CLMR \\
        (m) & (m) & (m) & (m) & (days) & (W/m\textsuperscript{2} at 100\,km) & (MW) & (kWh) & (\$/kg) \\
        \hline
        10  & 10  & 300 & 10,000 & 15.15 & 230.00 & 1,161.01 & 2.40E+08 & \$11{,}976.43 \\
        10  & 10  & 300 & 30,000 & 15.15 & 230.00 & 134.00  & 2.73E+07 & \$1{,}362.85 \\
        30  & 30  & 300 & 10,000 & 8.95  & 230.00 & 138.02  & 8.11E+06 & \$405.66   \\
        30  & 30  & 300 & 30,000 & 8.95  & 230.00 & 21.33   & 1.16E+06 & \$58.06    \\
        50  & 50  & 300 & 1,000  & 8.46  & 230.00 & 5,035.39 & 2.31E+08 & \$11{,}546.54 \\
        50  & 50  & 300 & 2,000  & 8.46  & 230.00 & 1,263.55 & 5.79E+07 & \$2{,}895.09 \\
        50  & 50  & 300 & 10,000 & 8.46  & 230.00 & 56.89   & 2.54E+06 & \$127.10   \\
        165 & 165 & 300 & 1,000  & 8.23  & 114.63 & 644.25  & 2.04E+07 & \$1{,}019.64 \\
        165 & 165 & 300 & 2,000  & 8.23  & 114.63 & 168.73  & 5.28E+06 & \$263.93   \\
        165 & 165 & 300 & 10,000 & 8.23  & 114.63 & 19.01   & 5.10E+05 & \$25.51    \\
        \hline
    \end{tabular}
    \caption{Climber power and speed for dual ground and GEO sources with limited PD in the ionosphere for 50\% payload fraction (10 tonnes) and \$0.50/kWh}
    \label{tab:dualrectenna2}
\end{table}

The 10 m receive aperture cases have poor efficiency and long trip time, even with an extremely large transmit aperture at the GEO SPS.  The ground transmitter aperture diameter of 300 m is adequate up to 400 km altitude when it will be shut off. 

\begin{figure}
\begin{centering}
\includegraphics[width=\textwidth]{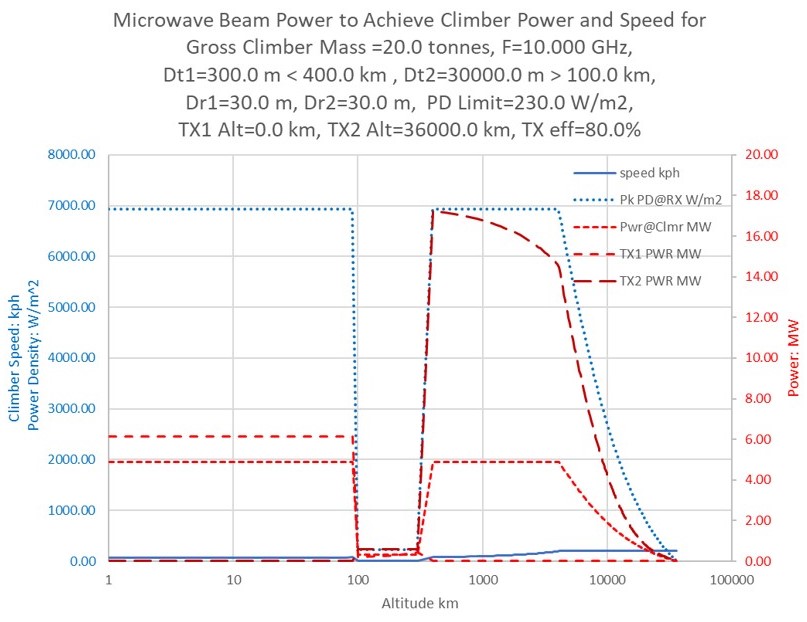}    
\caption{\label{fig:GNDGEO-2} Climber power and speed for Dr=30 m diameter receiver on the climber with a ground transmitter, Dt1=300 m, and a GEO transmitter, Dt2=30 km (Row 4 of Table \ref{tab:dualrectenna2}). Power density in ionosphere limited to 230 \si{W/m^2}.}
\end{centering}
\end{figure}

Row 4 of Table \ref{tab:dualrectenna2} and Figure \ref{fig:GNDGEO-2} shows that having a 30 m diameter rectenna on the climber with the transmitter at GEO being 30 km diameter provides good efficiency.  The trip time becomes nearly 9 days because of the slow speed in the ionosphere due to power density limits.

\begin{figure}
\begin{centering}
\includegraphics[width=\textwidth]{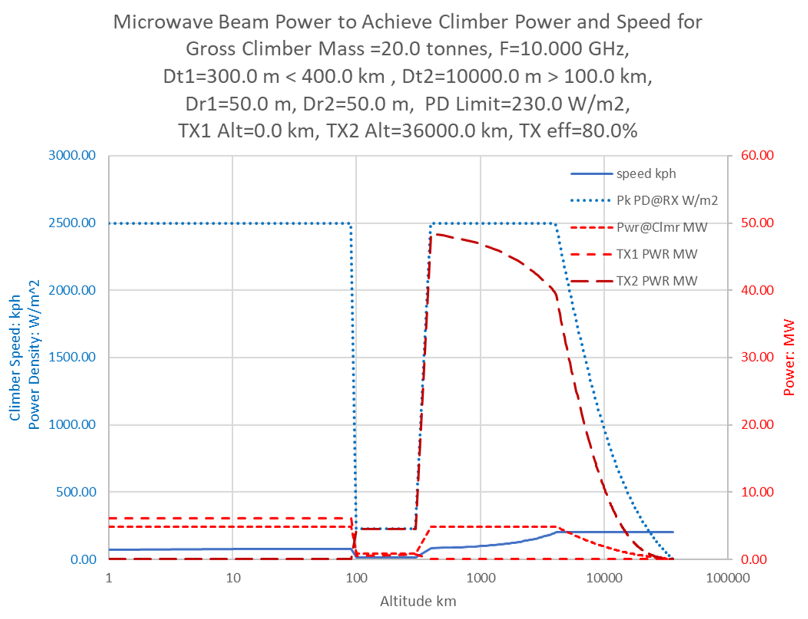}    
\caption{\label{fig:Dr50Dt1-300Dt2-10km} Climber power and speed for Dr=50 m diameter receiver on the climber, with a ground transmitter, Dt1=300 m diameter, and Dt2= 10 km diameter GEO power. (Row 7 of Table \ref{tab:dualrectenna2})}
\end{centering}
\end{figure}

Row 7 of Table \ref{tab:dualrectenna2} and Figure \ref{fig:Dr50Dt1-300Dt2-10km} shows that increasing the rectenna on the climber to 50 m diameter, the GEO transmitter may be reduced to 10 km diameter, and efficiency is still fair.  The trip time is about 8.5 days.

The highest efficiency was obtained with the climber receive aperture enlarged to 165 m diameter, which is the diameter that supplies 4 MW at a power density of 230 W/m2.  With the GEO transmitter at 10 km diameter, the trip time drops to about 8.2 days and the energy efficiency is very good.

\begin{figure}
\begin{centering}
\includegraphics[width=\textwidth]{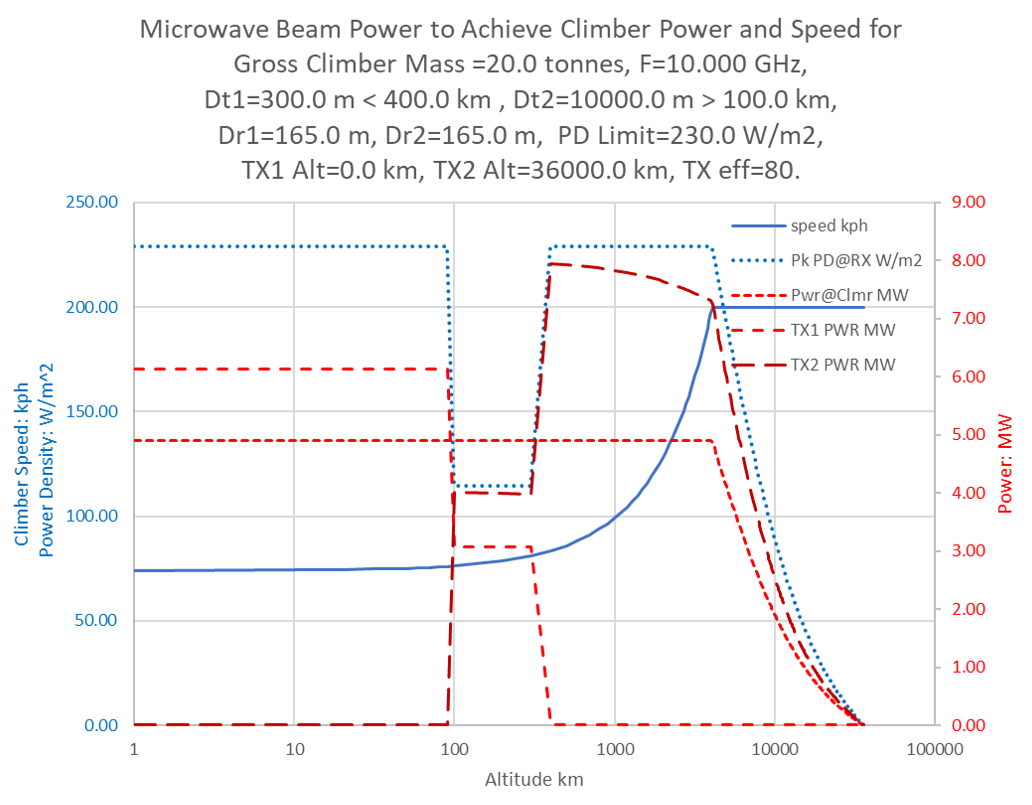}    
\caption{\label{fig:Dr165Dt1-300Dt2-10km} Climber power and speed for Dr=165 m and PD limit = 230 \si{W/m^2} with a ground transmitter, Dt1=300 m diameter, and Dt2=10 km diameter GEO power (Row 10 of Table \ref{tab:dualrectenna2})}
\end{centering}
\end{figure}

\bigskip

Next, the power density limit is removed.  Table \ref{tab:dualrectenna-noPDlimit} has a summary of these cases with the ground transmitter, Dt1, only active below 400 km altitude, and the GEO SPS transmitter, Dt2, only active above 100 km.

\begin{table}[h]
    \centering
    \begin{tabular}{ccccc M{2cm} M{2cm} M{2cm} M{2cm}}
        \hline
        Dr1 & Dr2 & Dt1 & Dt2 & ETA & Pk PD@RX & PWR 8 CLMRS & Energy /CLMR & Cost /CLMR \\
        (m) & (m) & (m) & (m) & (days) & (W/m\textsuperscript{2} at 100 km) & (MW) & (kWh) & (\$/kg) \\
        \hline
        10  & 10  & 300 & 10,000 & 8.23 & 31,206.85 & 1,740.19 & 5.53E+07 & \$2{,}766.21 \\
        10  & 10  & 300 & 30,000 & 8.23 & 31,206.85 & 205.58 & 6.55E+06 & \$327.42 \\
        30  & 30  & 300 & 10,000 & 8.23 & 3,467.43  & 202.01 & 6.35E+06 & \$317.60 \\
        30  & 30  & 300 & 30,000 & 8.23 & 3,467.43  & 32.37  & 9.56E+05 & \$47.80  \\
        50  & 50  & 300 & 1,000  & 8.23 & 1,248.27  & 6,916.36 & 2.20E+08 & \$10{,}987.20 \\
        50  & 50  & 300 & 2,000  & 8.23 & 1,248.27  & 1,736.61 & 5.51E+07 & \$2{,}755.72 \\
        50  & 50  & 300 & 10,000 & 8.23 & 1,248.27  & 79.44  & 2.44E+06 & \$122.12 \\
        165 & 165 & 300 & 1,000  & 8.23 & 114.63   & 644.25 & 2.04E+07 & \$1{,}019.64 \\
        165 & 165 & 300 & 2,000  & 8.23 & 114.63   & 168.73 & 5.28E+06 & \$263.93 \\
        165 & 165 & 300 & 10,000 & 8.23 & 114.63   & 19.01  & 5.10E+05 & \$25.51  \\
        \hline
    \end{tabular}
    \caption{Climber power and speed for dual ground and GEO sources with unlimited PD in the ionosphere for 50\% payload fraction (10 tonnes) and \$0.50/kWh}
    \label{tab:dualrectenna-noPDlimit}
\end{table}

\begin{figure}
\begin{centering}
\includegraphics[width=\textwidth]{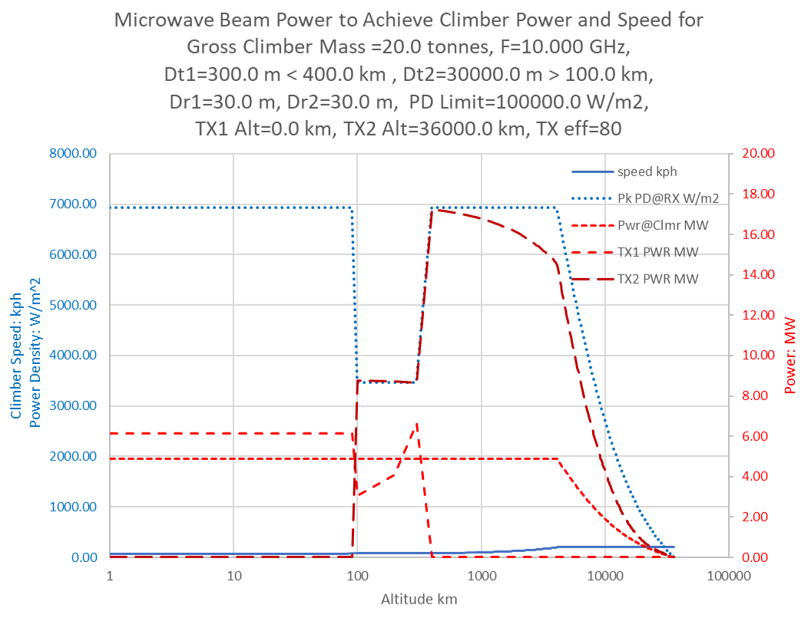}    
\caption{\label{fig:Dr30Dt1-300Dt2-30kmnoPDlimit} Climber power and speed for Dr=30 m diameter receiver on the climber with a ground transmitter, Dt1=300 m, and a GEO transmitter, Dt2=30 km (Row 4 of Table \ref{tab:dualrectenna-noPDlimit}). Power density in the ionosphere is not limited.}
\end{centering}
\end{figure}

Removing the power density limit for a climber with a constant diameter receiver aperture does not greatly improve efficiency, but it does improve climber travel time.  Likely the best compromise is with the 30 m climber receive aperture, see Figure \ref{fig:Dr30Dt1-300Dt2-30kmnoPDlimit}.

\FloatBarrier

\subsubsection{Dual stage rectenna on climber}

In this analysis, a smaller rectenna is attached below the climber and used for launch powered by a ground transmitter.  Later, when entering the ionosphere and above, where wind loading would be insignificant, a larger flexible rectenna is unfurled like a sail or umbrella and used for the rest of the climb.  Power is supplied by both the ground and GEO transmitters while traversing the ionosphere, and by the GEO transmitter while above the ionosphere.  Table 4-12 shows several examples of the dual-stage rectenna on the climber

\begin{table}[h]
    \centering
    \begin{tabular}{ccccc M{2cm} M{2cm} M{2cm} M{2cm}}
        \toprule
        Dr1 & Dr2 & Dt1 & Dt2 & ETA & Pk PD@RX & PWR 8 CLMRS & Energy /CLMR & Cost /CLMR \\
        m & m & m & m & days & W/m\textsuperscript{2} at 100km & MW & kWh & \$/kg \\
        \midrule
        10 & 50  & 300 & 10,000 & 8.23 & 31,206.85 & 79.44  & 2.44E+06 & \$122.21 \\
        10 & 50  & 300 & 30,000 & 8.23 & 31,206.85 & 20.00  & 5.47E+05 & \$27.36  \\
        10 & 165 & 300 & 10,000 & 8.23 & 31,206.85 & 19.01  & 5.12E+05 & \$25.60  \\
        30 & 50  & 300 & 2000  & 8.23 & 3,467.43  & 1736.61 & 5.51E+07 & \$2,755.72 \\
        30 & 50  & 300 & 10,000 & 8.23 & 3,467.43  & 79.44  & 2.44E+06 & \$122.12 \\
        30 & 50  & 300 & 30,000 & 8.23 & 3,467.43  & 20.00  & 5.45E+05 & \$27.27  \\
        30 & 165 & 300 & 1,000  & 8.23 & 3,467.43  & 644.25 & 2.04E+07 & \$1,019.64 \\
        30 & 165 & 300 & 2,000  & 8.23 & 3,467.43  & 168.73 & 5.28E+06 & \$263.93 \\
        30 & 165 & 300 & 10,000 & 8.23 & 3,467.43  & 19.01  & 5.10E+05 & \$25.51  \\
        50 & 165 & 300 & 10,000 & 8.23 & 1,248.27  & 19.01  & 5.10E+05 & \$25.51  \\
        \bottomrule
    \end{tabular}
    \caption{Climber power and speed for dual stage rectenna for ground and GEO sources with unlimited PD in the ionosphere for 50\% payload fraction (10 tonnes) and \$0.50/kWh}
    \label{tab:2diffrectennas-noPDlimit}
\end{table}

\begin{figure}
\begin{centering}
\includegraphics[width=\textwidth]{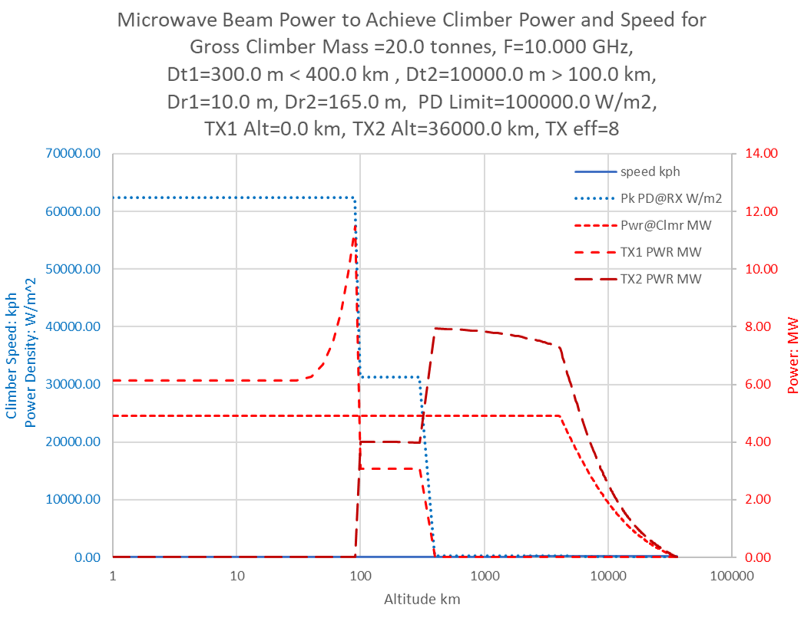}    
\caption{\label{fig:Dr1=10Dr2=165noPDlimit} Climber power and speed for Dr1=10 m, Dr2 = 165 m and unlimited PD with a ground transmitter, Dt1=300 m, and a GEO transmitter, Dt2=10 km.   (Row 3 of Table \ref{tab:2diffrectennas-noPDlimit}).}
\end{centering}
\end{figure}

Likely the best compromise starts with a 10 m climber aperture illuminated by a 300 m ground transmitter aperture, then at 100 km altitude, a 165 m aperture is deployed which is illuminated by a 10 km transmitter aperture SPS at GEO, see Figure \ref{fig:Dr1=10Dr2=165noPDlimit}.

\FloatBarrier

\subsection{Space power grid}

[\cite{jaffe2024power} pp.348] describes an example of a Solar Power Satellite (SPS) operating in a Medium Earth Orbit (MEO) “between 2,000 and 35,786 km (1,243 and 22,236 mi) above sea level” [\cite{MEO}]. While this was intended to supply power to the ground, conceptually, this could also be useful for powering the space elevator.  While Jaffe’s example is for an optical beaming system, conceptually, this approach should work for microwave power beaming as well.

“For a medium Earth orbit (MEO) … [a] constellation of satellites would be needed for continuous receiver coverage for the ground receiver…” [\cite{jaffe2024power} pp.348] or for other spacecraft.  Establishing this constellation would be time-consuming and expensive.  However, a large constellation of SPS at MEO would provide a space power grid (SPG) [\cite{komerath2006space}], or space-based energy grid, for many customers and could see break-even on investment in roughly 10 to 15 years.  The SPG would drive energy prices down to commodity levels, similar to what large constellations of low earth orbit (LEO) satellites have done for space-based communications.  A large constellation of MEO SPS in a space power grid would allow for operational flexibility in that multiple SPS could illuminate a single receiver for greater power received, although there may still be a power density limit for a spacecraft receiver within the ionosphere.  A large constellation would also offer redundancy for operational reliability.

Shorter transmission distances from SPG SSP to customer power collectors versus GEO SPS may improve efficiencies for smaller aperture diameters for the receiver and/or the transmitter.
Tables \ref{tab:SPGmicrowave} and \ref{tab:SPGinfrared} gives some examples of $D_t$ and $D_r$ for 90\% beam efficiency at 2000 km distance using the Goubau relationship.  An optical wavelength allows smaller $D_t$ and $D_r$ versus microwave beaming.

\begin{table}[h]
    \begin{minipage}[t]{.48\textwidth}
    \centering
    \begin{tabular}{ccc}
        \hline
        Dt (m) & Dr (m) & TX Eff (\%) \\
        \hline
        100  & 1158.43  & 90.0\% \\
        200  & 579.213  & 90.0\% \\
        400  & 289.607  & 90.0\% \\
        800  & 144.803  & 90.0\% \\
        1000 & 115.843  & 90.0\% \\
        2000 & 57.9213  & 90.0\% \\
        4000 & 28.9607  & 90.0\% \\
        \hline
    \end{tabular}
    \caption{Space power grid examples for 10 GHz microwave beaming, 2000 km distance}
    \label{tab:SPGmicrowave}
    \end{minipage} \hfill
    \begin{minipage}[t]{.48\textwidth}
    \centering
    \begin{tabular}{ccc}
        \hline
        Dt (m) & Dr (m) & TX Eff (\%) \\
        \hline
        0.05 & 77.2819 & 90.0\% \\
        0.1 & 38.641 & 90.0\% \\
        0.2 & 19.3205 & 90.0\% \\
        0.4 & 9.66024 & 90.0\% \\
        0.8 & 4.83012 & 90.0\% \\
        1 & 3.8641 & 90.0\% \\
        2 & 1.93205 & 90.0\% \\
        4 & 0.96602 & 90.0\% \\
        8 & 0.48301 & 90.0\% \\
        10 & 0.38641 & 90.0\% \\
        \hline
    \end{tabular}
    \caption{Space power grid examples for 1 micron wavelength infrared power beaming, 2000 km distance}
    \label{tab:SPGinfrared}
    \end{minipage}
\end{table}

An SSP in LEO or MEO based on optical power beaming using a high-energy laser could also be employed for orbital debris remediation [\cite{fox2025space}], which could be another customer base for the SPG.

Operating in the magnetosphere enables electrodynamic tether (EDT) propulsion from about 640 km to about 36,000 km altitude.  An SPS would use the electrical power it produces to energize an EDT [\cite{EDT}] with propellent-less propulsion for indefinite mission duration, and inexpensively self-deploy from LEO to MEO or GEO.  This would also be used for both orbital modification and attitude control which would use no fuel, providing an indefinite operational lifetime.  Additionally, an SPS delivered into LEO could self-deploy to MEO or even GEO with no propellent, greatly reducing the cost of launch to MEO or GEO.

Combining an EDT with an SPS suggests another use: that of space tug, moving other satellites to new orbits.  This could be an additional market for SSP owners and operators.

Operating in MEO has the disadvantage of operating within the Van Allen radiation belts [\cite{vanallen}], requiring hardening which may increase weight and cost, and may shorten spacecraft lifetime.  However, an SPS operating within the Van Allen radiation belts could use some of its energy to remediate the radiation.  High Voltage Orbiting Long Tether, or HiVOLT, is a concept proposed by Russian physicist V. V. Danilov and further refined by Robert P. Hoyt and Robert L. Forward for draining and removing the radiation fields of the Van Allen radiation belts that surround the Earth [\cite{hudoba2016using}].  The trick here is developing a market so that this activity could turn a profit for SSP owners and operators.  Governments could fund this as a way to mitigate the hazard to navigation posed by these radiation fields.  Operators of other spacecraft could also pay for this service.  

\subsubsection{Space power grid companies}

Aetherflux

\url{https://www.aetherflux.com/}

\url{https://dig.watch/updates/aetherflux-raises-50-million-for-space-solar-power-project}
\\
\\
Space Solar

\url{https://www.spacesolar.co.uk/}
\\
\\
Star Catcher orbital energy grid

\url{https://spacenews.com/star-catcher-banks-12-25-million-for-orbital-energy-grid/}

\url{https://star-catcher.com/}
\\
\\
Virtus Solis

\url{https://virtussolis.space/}

\url{https://spacenews.com/orbital-composites-and-virtus-solis-announce-space-based-solar-power-demonstration/}

\subsection{Conclusions on microwave power beaming}

\begin{figure}
\begin{centering}
\includegraphics[width=\textwidth]{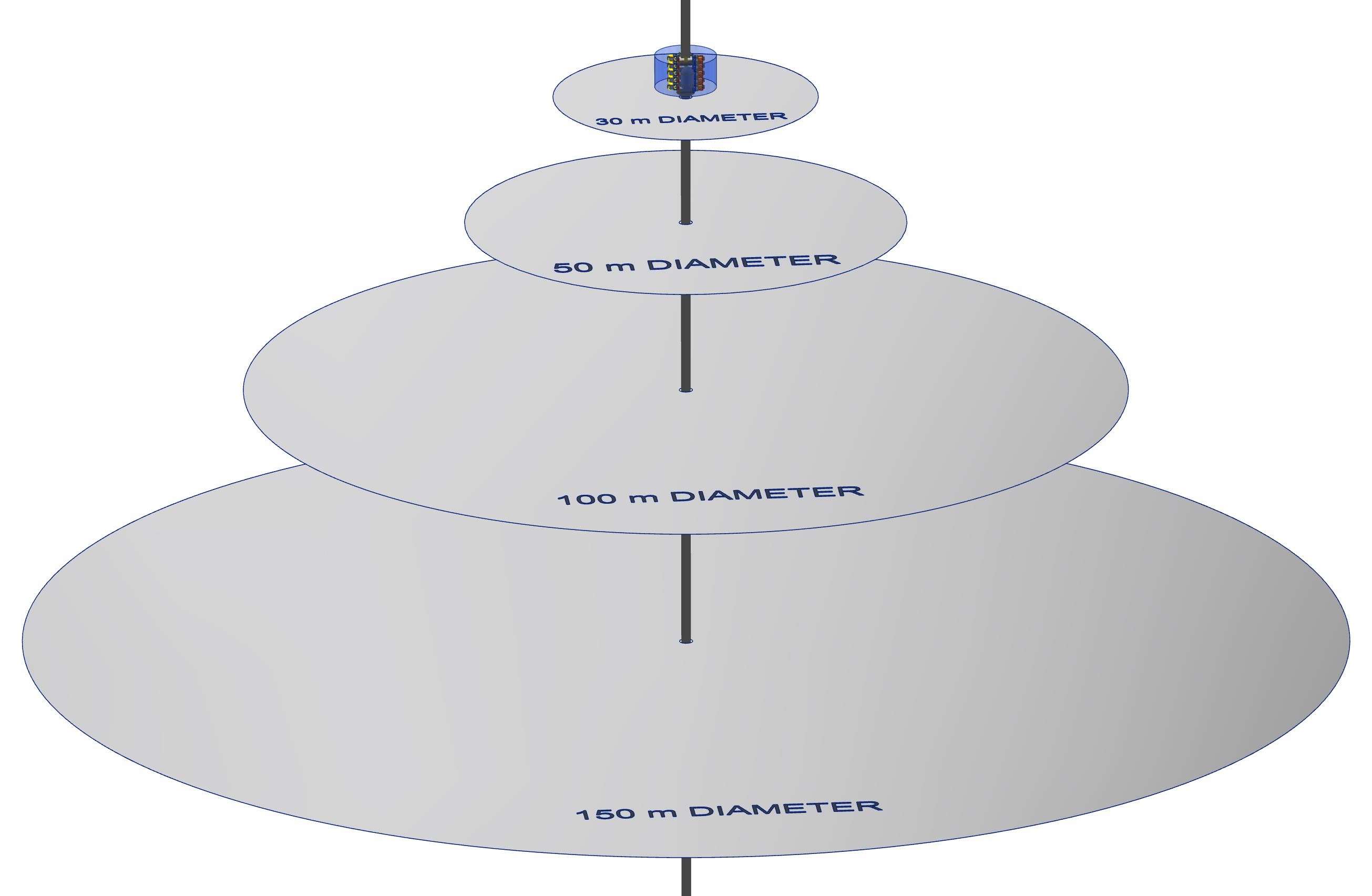}
\caption{\label{fig:Fig14rectennasizes} The various rectenna sizes discussed relative to the baseline 20 tonne climber}
\end{centering}
\end{figure}

\begin{enumerate}
    \item Larger transmitter diameters on the ground and in space along with higher frequencies will support smaller rectennas on the climber, but there will be technology challenges with higher frequencies.
    \item Keeping the rectenna size at the climber reasonable requires more than one transmitter in order to achieve the best efficiency, but the larger the better, such as 165 m diameter. 50 m diameter is not too terrible, and even 30 m could be made to work.
    \item For a space elevator, a ground-based power beaming system would be a useful aid for launching a climber.  Unless the receiver diameter is very large for microwave power beaming, on the order of 165 m diameter, the SE climber would have to move slowly through the ionosphere if it is decided that the microwave power density must be limited.  Once above the ionosphere at 400 km, the received power density from a space power satellite would be increased to support higher speeds.
    \item A ground transmitter likely could have a reasonable size (100 m diameter) when used in conjunction with a GEO transmitter, but then the GEO transmitter will be very large (up to 10 km diameter) to obtain good efficiency.  The energy cost for the two-transmitter configuration is lower than the single-transmitter configurations analyzed.
    \item Allowing the climber speed to drop while traversing the ionosphere to limit power density therein increases the time to GEO by only a day or two.
    \item This analysis is based on the Goubau relationship for the physics of a power beaming link, and while this section has glossed over technology challenges, they are addressed in some of the references.  Addressing the technology specifics will require considerable additional analysis.
    \item A large constellation of SPS in MEO forming a Space Power Grid could have great advantages for powering the space elevator, while potentially providing other benefits such as mitigating space debris and remediating the Van Allen radiation belts.
    \item Construction challenges:  The antenna and rectenna sizes discussed are very large.  However, the solar power satellite community and other organizations have been working on the technology of building large structures in space since the 1960s. They still have much work to do.  The largest single such structure this author has discovered was the Sunjammer solar sail demonstrator, which was 38 m on a side.  However, it was canceled before it actually flew in space. [\cite{sunjam}] Figure \ref{fig:Fig14rectennasizes}  shows the various rectenna sizes discussed relative to the baseline 20 tonne climber.
    \item Further research and experimental studies are needed of microwave-plasma interactions in the ionosphere.  
“It would seem essential therefore, that experiments should be conducted on spatial scales which better match those that will be encountered in any future SPS system. It may be possible to carry out such experiments with suitable ground-based high-power radio transmitters in the first instance. Otherwise some large space platform would have to be utilised for large scale experimental studies in space.”  [\cite{robinson2004environmental}, pp.20]

\end{enumerate}

\clearpage

\chapter{Power Sources on the Ground for Power Transmission}

The possible power sources that could power the laser, microwave, or the conductive tether power transmission systems on the ground are as follows:

\begin{itemize}
    \item Coal or gas-fired power plants
    \item Nuclear fission reactors
    \item Solar farms
    \item Wind farms
    \item Wave farms
    \item Fusion reactors
\end{itemize}

The maximum power required for a single platform would be to power two FELs.  From Table \ref{tab:wallplug}, the wall-plug power required for four FELs is 60.61 MW, so the power required for two FELs on the same platform would be 30.3 MW.

This section is intended to be a comprehensive survey, but a key goal is to power the space elevator in an environmentally friendly manner. An example of a fossil fuel power plant in the right power range is shown, but fossil fuel plants would not be the first choice on environmental grounds.

\section{Coal or gas-fired power plants}

\begin{figure}[ht]
\begin{centering}
\includegraphics[width=.8\textwidth]{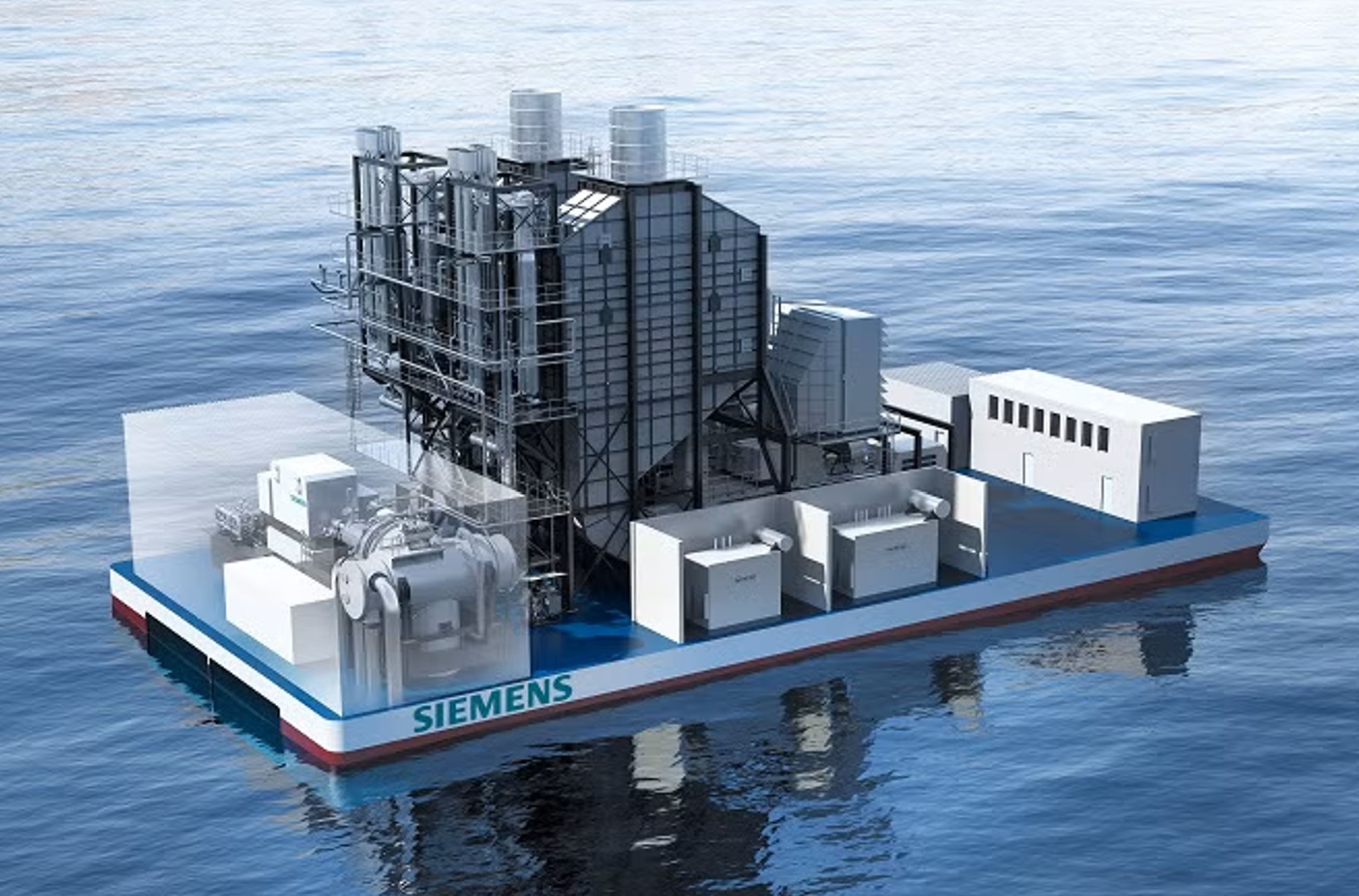} 
\caption{\label{fig:Siemenspbarge} An example of a floating LNG power barge capable of outputting 145 MW. [\cite{Siemensbarge}]}
\end{centering}
\end{figure}

Figure \ref{fig:Siemenspbarge} shows an example of an LNG-fueled marine power plant intended to bring power to remote locations in the Dominican Republic. It can deliver 145 MW of power, making it large enough for the dual FEL platform. The technology is mature, and fossil fuel power plants are available in a variety of capacities.

One drawback to fossil-fueled platforms is that they require regular refueling.  The power-beaming platform may be in a remote location, so ensuring the supply line poses a logistics challenge for the SE company.

\FloatBarrier
\section{Nuclear fission reactors}

Quoting reference [\cite{nuclearland}], "The R. E. Ginna Nuclear Power Plant in New York is the smallest nuclear power plant in the United States, and it has one reactor with an electricity generating capacity of 582 megawatts (MW)."  A small land-based reactor could be used to power the FELs and an entire associated Earth Port facility.

GE Vernova Hitachi is building small modular boiling water reactors that can deliver 300 MW(e) power [\cite{smallnukes}]. These can be delivered to wherever needed and are small enough to be mounted on movable platforms.

Rolls-Royce in the UK is also planning small modular reactors (SMR) that use low-enriched uranium (LEU).  Their SMRs will produce 470 MW(e) of power.  They are not mobile, but occupy one-tenth the area of traditional nuclear reactors [\cite{rollsroyce}].

Naval nuclear reactors are in the range needed for power beaming.  Reactor sizes in US naval vessels range up to ~500 MWt (about 165 MWe) in the larger submarines and surface ships. [\cite{nukeships}] US naval nuclear vessels all use pressurized water reactors (PWR) and burn highly enriched uranium (HEU). HEU allows for compact, long-lasting reactors without refueling for decades.  Submarine nuclear reactors only need refueling every twenty years, once in the life of the vessel.  Either a submarine or an aircraft carrier could supply the power needed to run two FELs or more.

The French have switched their naval reactors from highly enriched uranium to low-enriched uranium because LEU is more available and less of a threat to nuclear proliferation.

An arrangement with a government that has nuclear-powered naval vessels would have to be worked out to allow the space elevator to be powered by one.

\begin{figure}
\begin{centering}
\includegraphics[width=\textwidth]{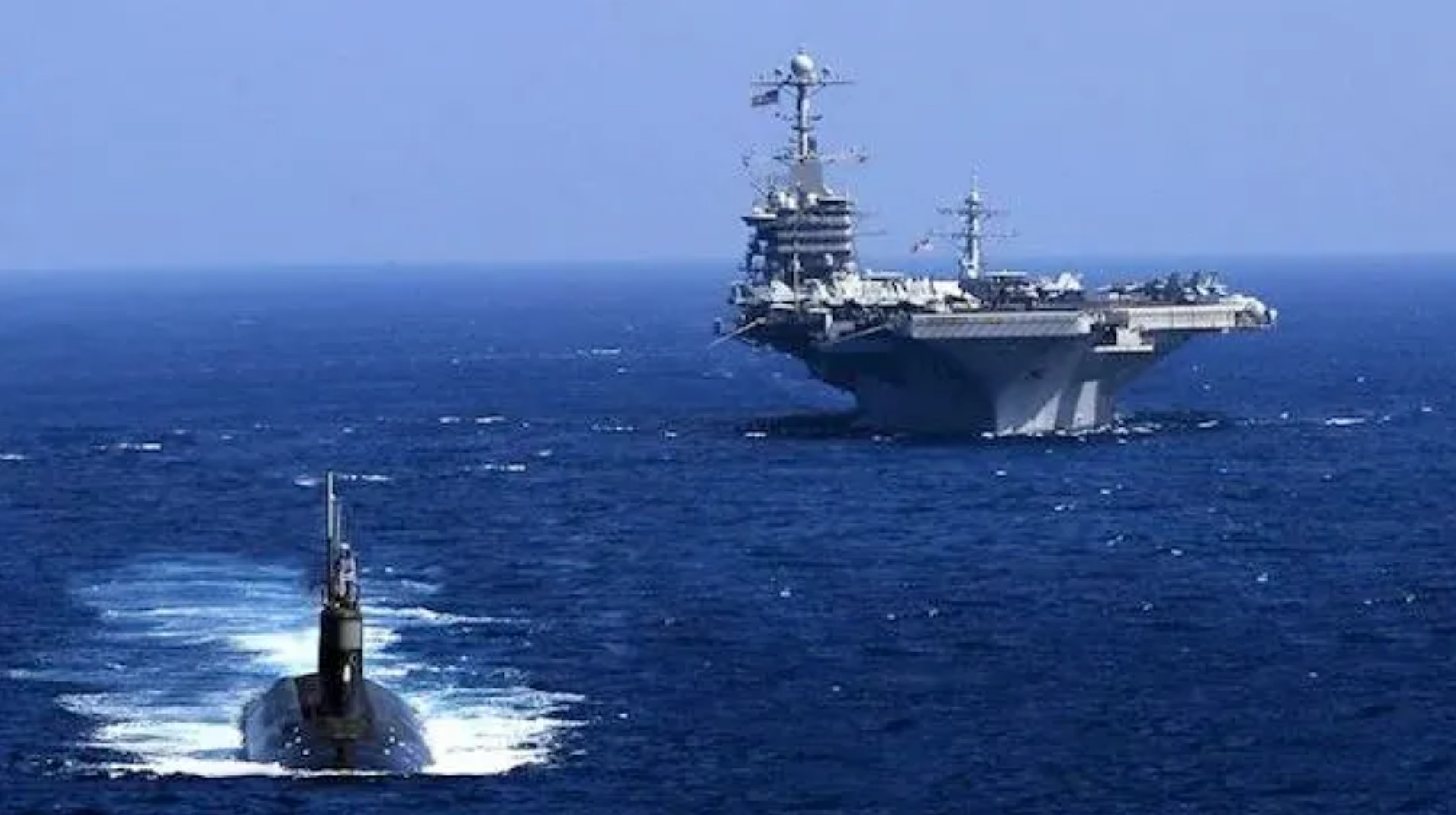} 
\caption{\label{fig:subcarrier} A US naval nuclear submarine and aircraft carrier. [\cite{navynukes}]}
\end{centering}
\end{figure}

There are six fourth-generation nuclear reactor types recognized by the Generation IV International Forum (GIF) [\cite{GIF}].  These are intended to be the successors of today's fission technology.  The goal is to bring online some of these alternative fission technologies by 2030.

\begin{figure}
\begin{centering}
\includegraphics[width=.8\textwidth]{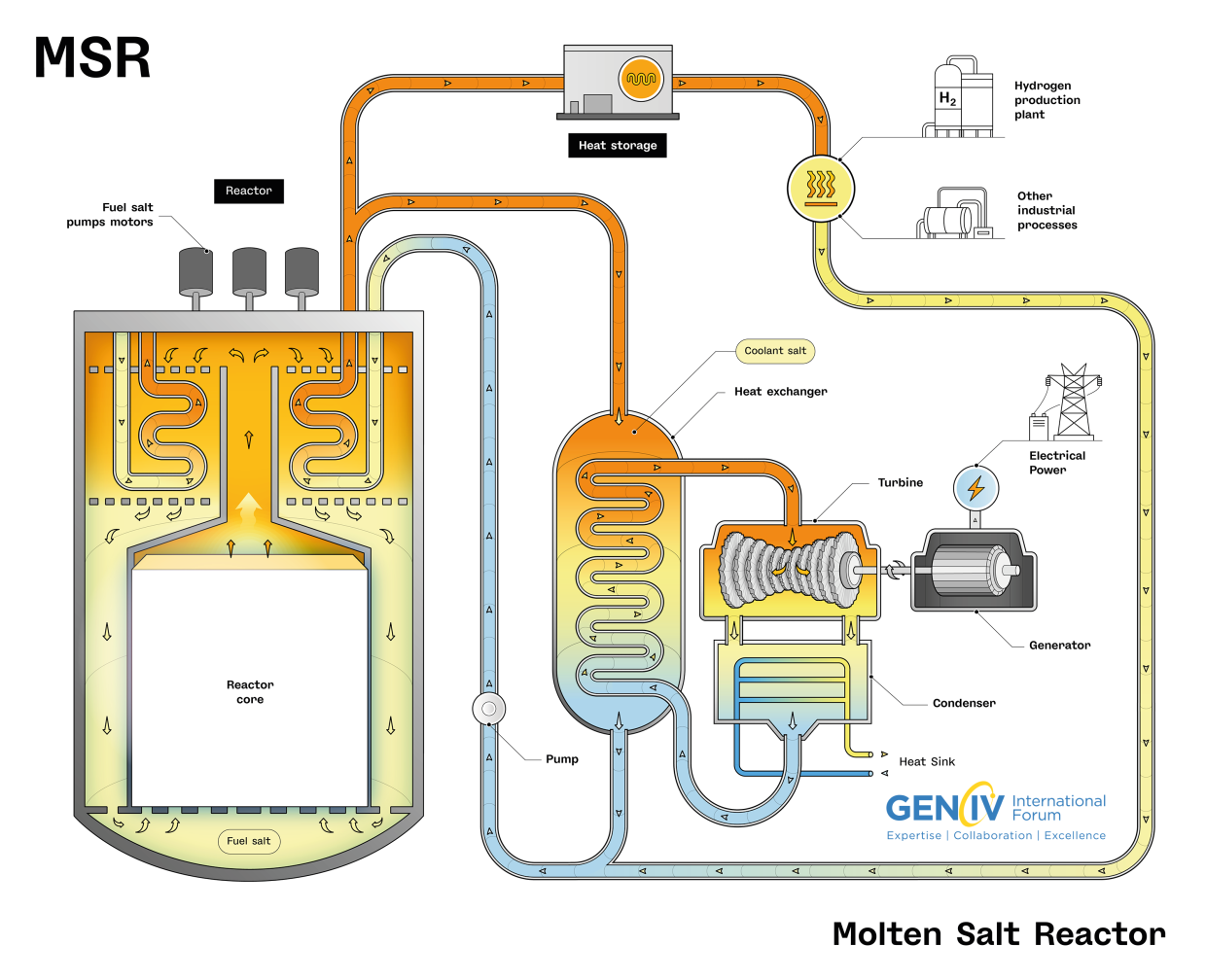} 
\caption{\label{fig:MSR} Schematic diagram of a Molten Salt Reactor (MSR). [\cite{MSR}]}
\end{centering}
\end{figure}

One of the six technologies is Molten Salt Reactors (MSRs), which include thorium reactors.  Thorium reactors have all of the fissile material dissolved in molten salt such as Flibe.  These reactors can't melt down because they're already in the molten state and fissile material has to be added to keep them hot.  MSRs can burn a variety of fissile materials.  Thorium is more readily available in many parts of the world and can be used to breed more fissile material for other reactors.  They were initially developed in the 1960s, but were considered too complicated and there were few incentives for investors to develop them.  There was already significant investment in light-water reactor (LWR) technologies.  Thorium reactors may still come into their own in the near future. Figure \ref{fig:MSR} shows a schematic diagram of a molten salt reactor.

\begin{figure}
\begin{centering}
\includegraphics[width=.8\textwidth]{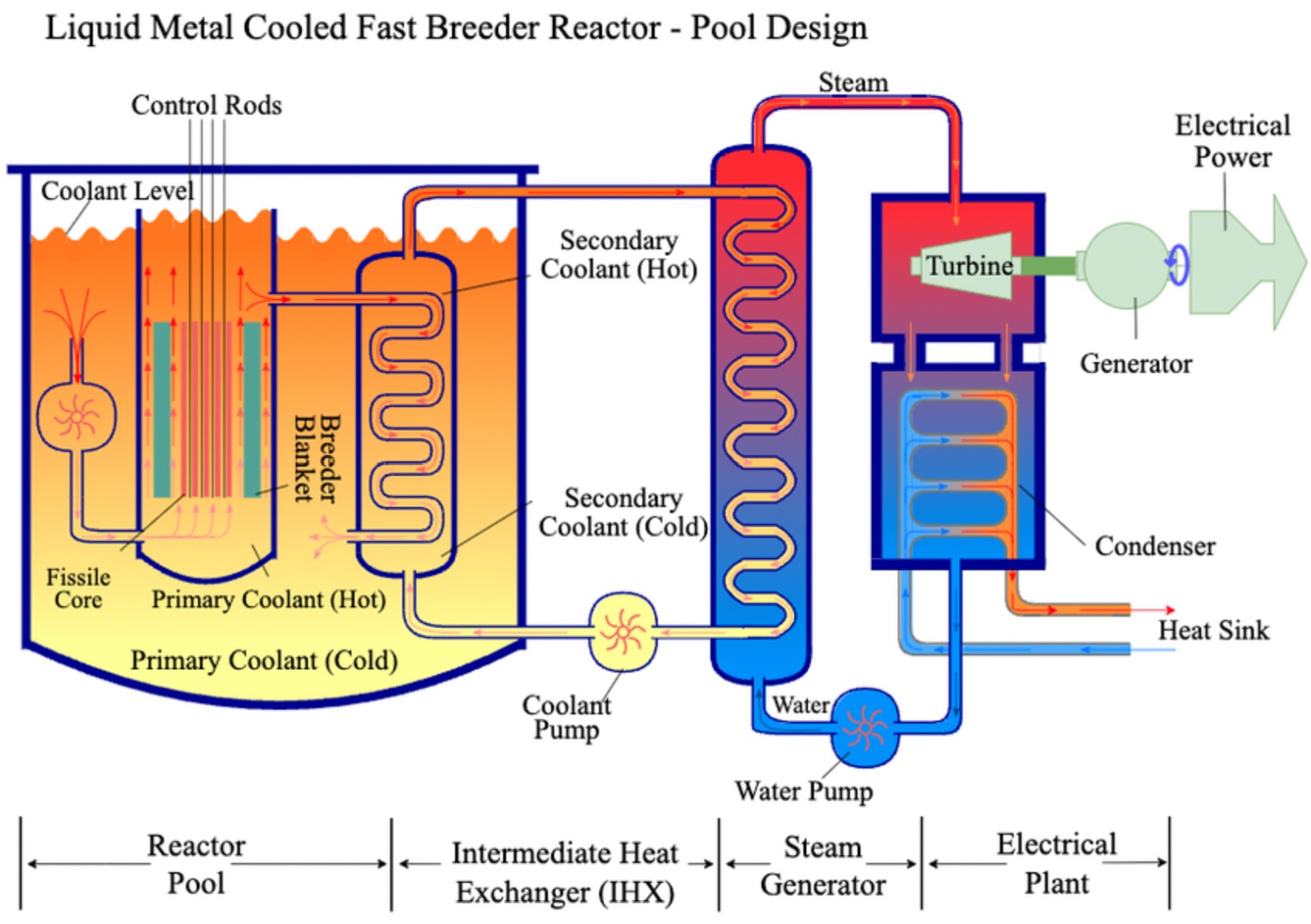} 
\caption{\label{fig:LMRbreeder} Schematic diagram of a Liquid Metal breeder reactor, pool design. [\cite{LMR}]}
\end{centering}
\end{figure}

A reactor design similar to the MSR is the Liquid Metal-cooled Reactor (LMR).  GIF recognizes two different LMR reactor designs among the six fourth-generation technologies.  LMRs differ from MSRs in the coolant circulated around the fissile material.  LMRs use sodium (SFR) or lead (LFR) to cool solid fuel, while MSRs have the fissile material dissolved in molten salt.  Figure \ref{fig:LMRbreeder} shows a schematic for one design of a liquid metal cooled fission reactor (LMR).  Both MSRs and LMRs are SMRs.  

\FloatBarrier

\section{Solar farms}

\begin{figure}[hb]
\begin{centering}
\includegraphics[width=\textwidth]{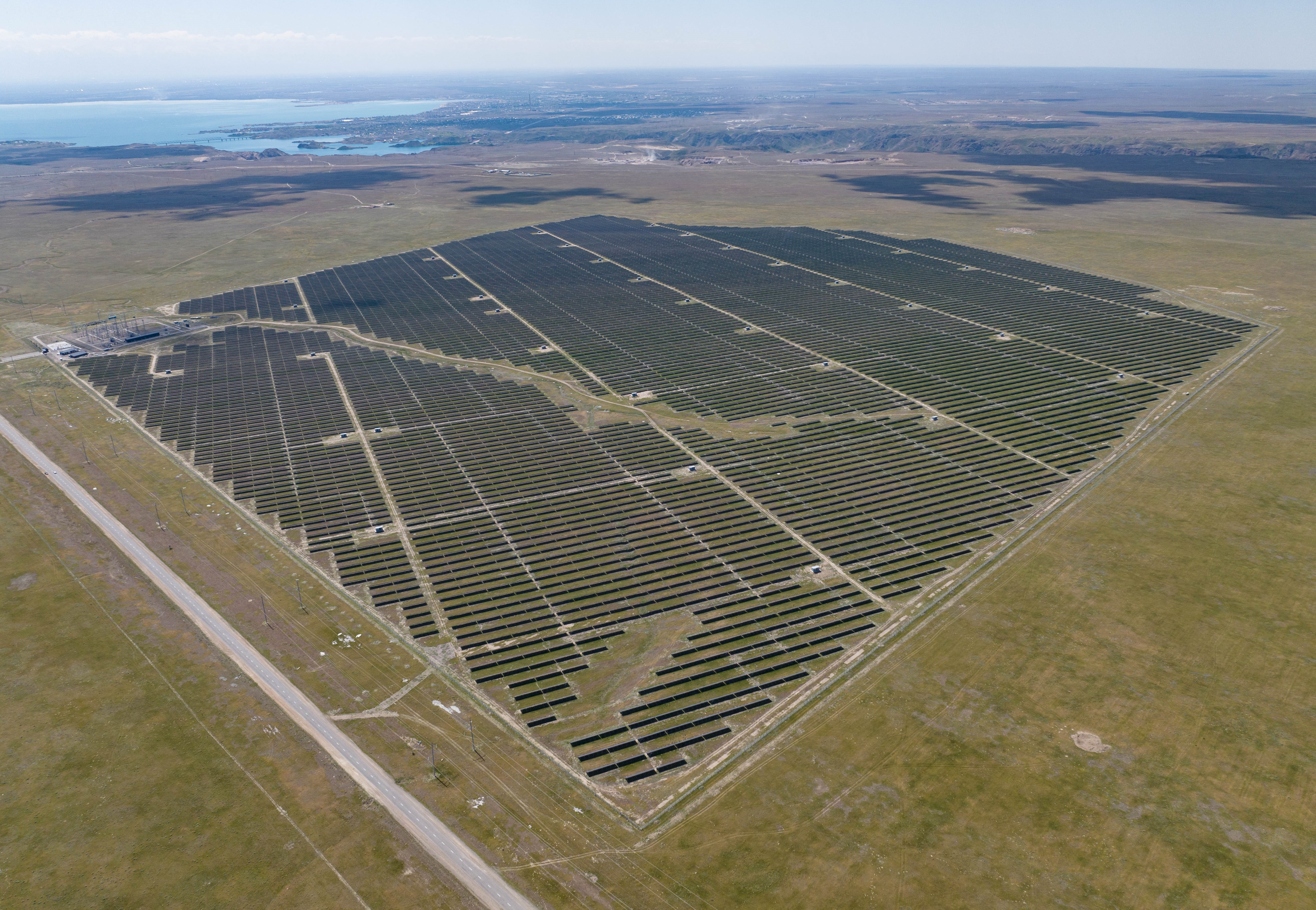} 
\caption{\label{fig:solarfarm} The Nurgisa Solar Farm, rated for 100 MW capacity.  It is situated on 270 hectares of land in Kazakhstan. [\cite{solarfarm}]}
\end{centering}
\end{figure}

Solar farms have been built in a range of capacities.  Figure \ref{fig:solarfarm} shows one rated for 100 MW in Kazakhstan.  It occupies 270 hectares of land (a square 1.643 km on each side).  Solar farms are primarily useful at middle latitudes that have sufficient average sunlight.  This is consistent with the space elevator being on the equator.  Solar farms of sufficient capacity are not mobile power sources.  They also receive no power at night, so a battery farm or alternative power method is also necessary.

\begin{figure}
\begin{centering}
\includegraphics[width=\textwidth]{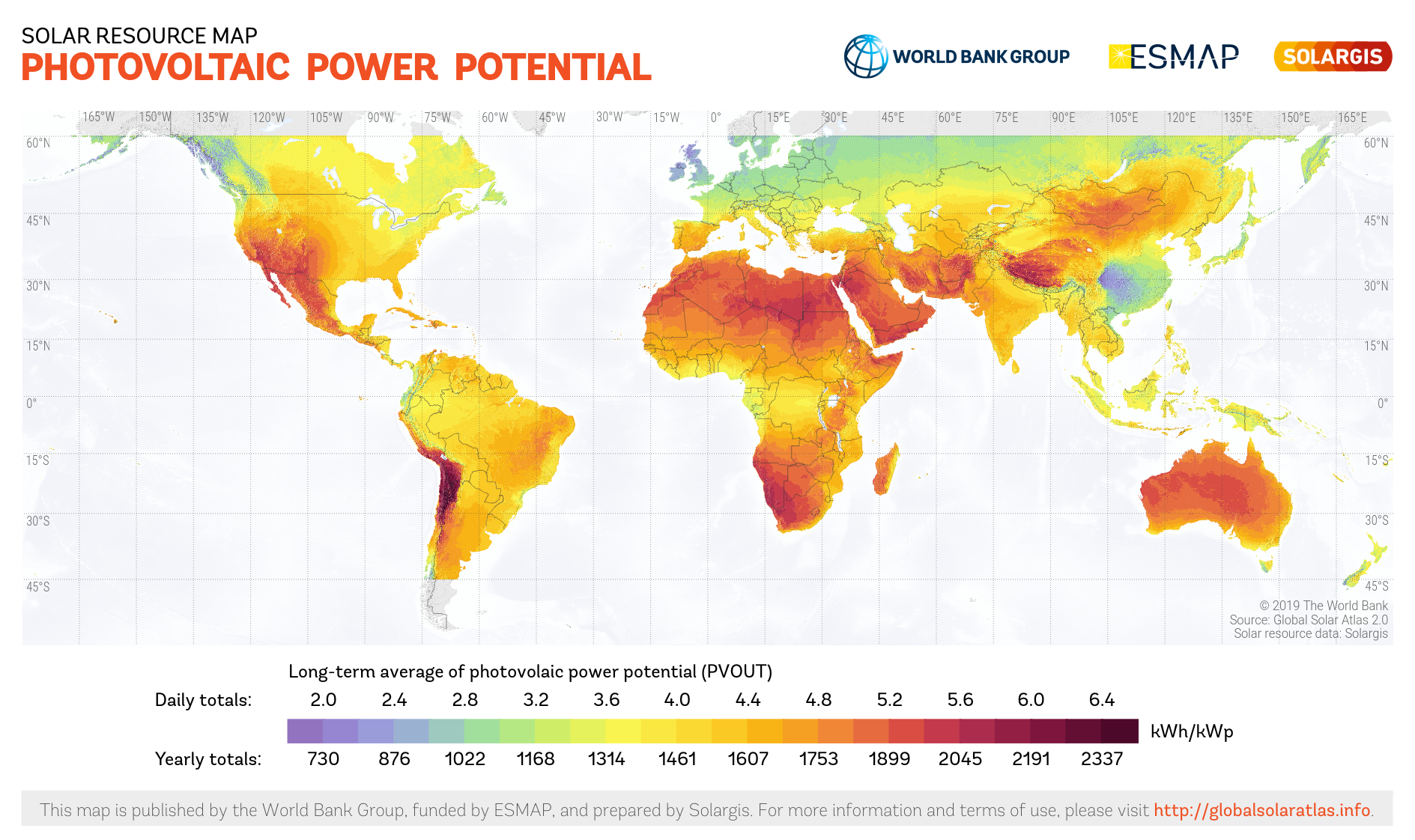} 
\caption{\label{fig:worldsolar} Map of solar installations around the world.  [\cite{worldsolar}]}
\end{centering}
\end{figure}

Figure \ref{fig:worldsolar} shows the distribution of solar power in the world.  It is interesting to note that the highest concentrations of solar power installations are not on the equator, but above and below it.  To use solar farms to power the space elevator, the space elevator company will either have to build its own or run long HV transmission lines from existing facilities.  It may also be possible to have power beaming stations at higher latitudes for climbers at higher altitudes to take advantage of existing solar farms.

\FloatBarrier

\section{Wind farms}

\begin{figure}[hb]
\begin{centering}
\includegraphics[width=.9\textwidth]{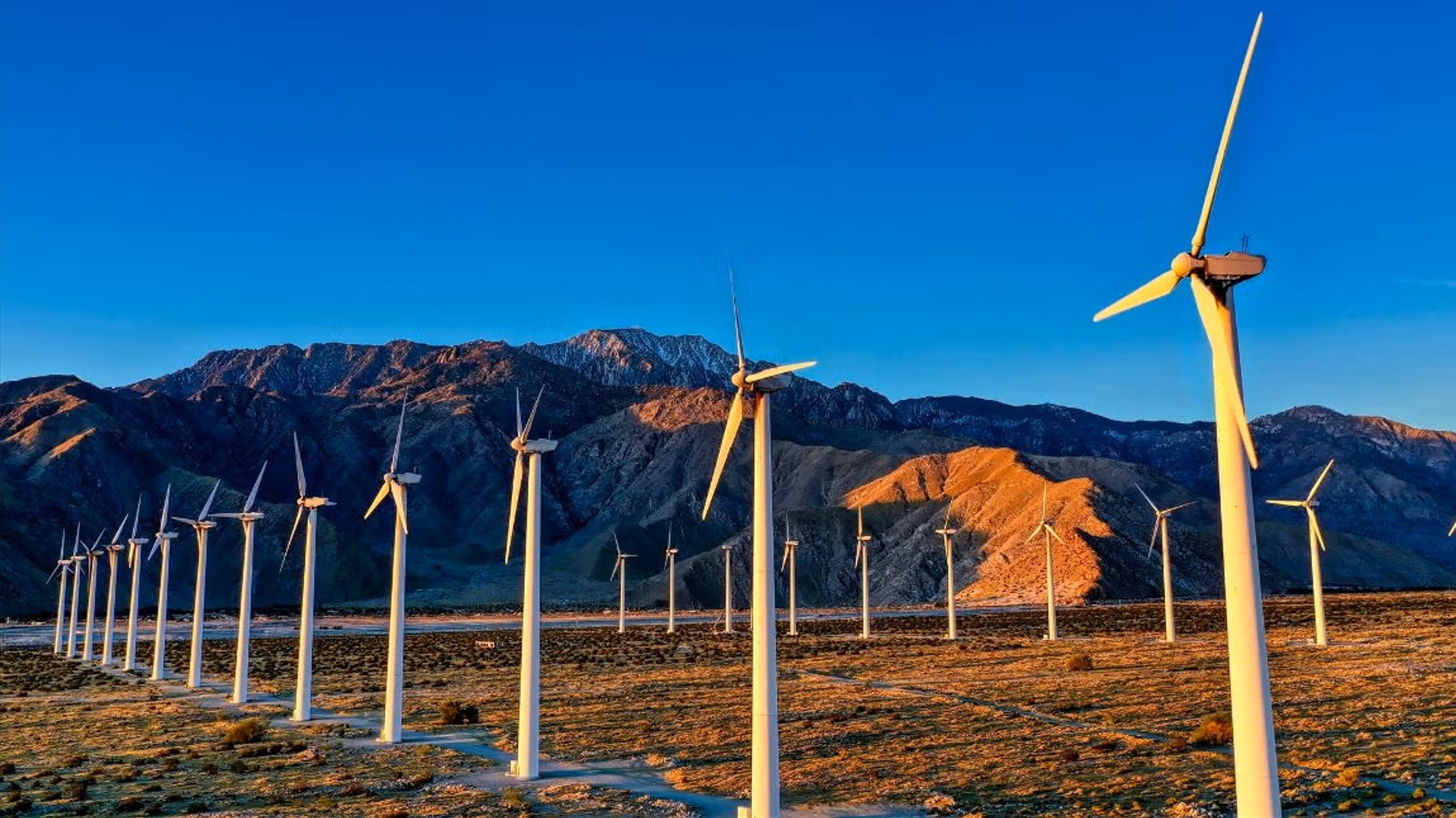} 
\caption{\label{fig:100MWwind}Buckthorn Wind, a 100.5 MW wind farm located near Dallas, Texas. [\cite{windfarm}]}
\end{centering}
\end{figure}

Another renewable source of electrical energy is wind farms.  These can be located on land or offshore.  Figure \ref{fig:100MWwind} shows a land installation near Dallas, Texas. Buckthorn Wind Farm is equipped with 29 Vestas Wind Systems turbines, each with a 3.45 MW nameplate capacity.

A wind farm needs roughly 30-60 acres (12.14-24.28 hectares) per megawatt (MW) for spacing, so a 100 MW farm needs 3,000 to 6,000 acres (1,214-2,428 hectares) total, though some designs need much more for optimal wind flow.  This is another example of a non-mobile power source.

\begin{figure}
\begin{centering}
\includegraphics[width=\textwidth]{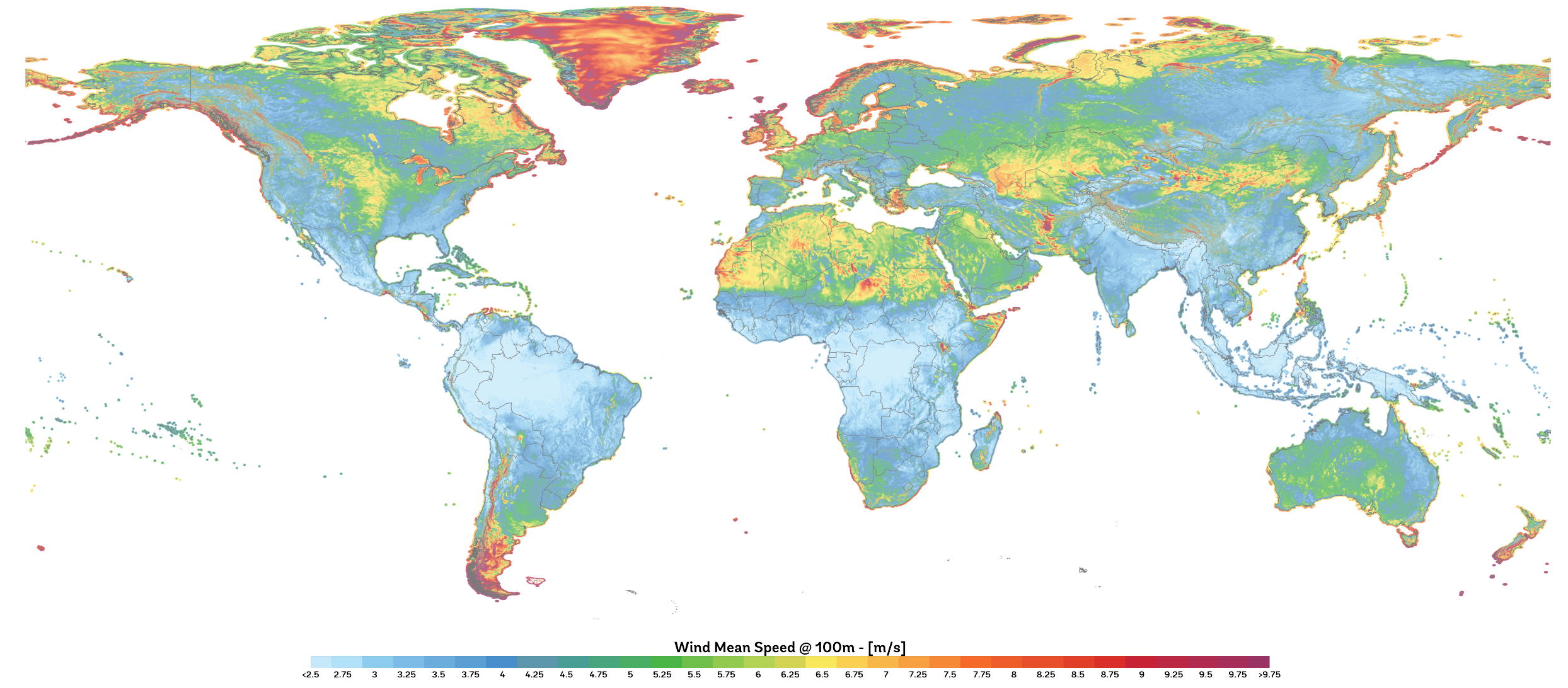} 
\caption{\label{fig:globalwinds}Map of wind speed around the world. [\cite{globalwind}]}
\end{centering}
\end{figure}

\begin{figure}
\begin{centering}
\includegraphics[width=\textwidth]{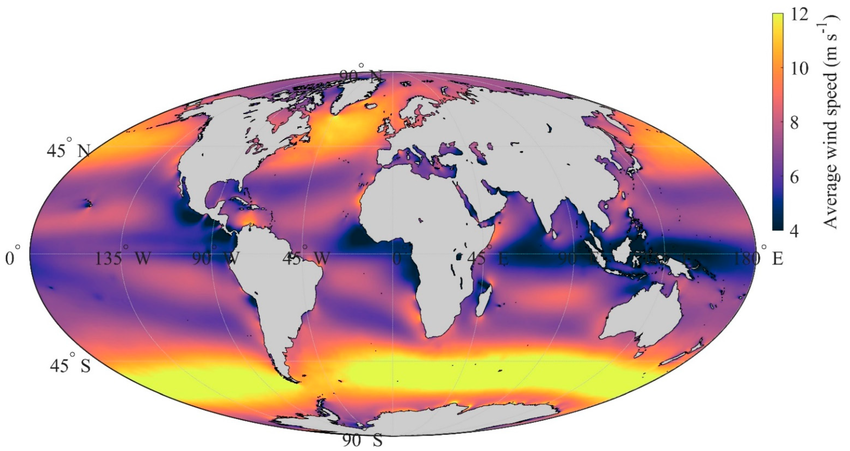} 
\caption{\label{fig:offshorewinds}Global offshore hourly mean wind speed at 100 m above sea level computed using the ERA5 reanalysis database from 2010 to 2020 [\cite{offshorewind}].}
\end{centering}
\end{figure}

Figure \ref{fig:globalwinds} shows the distribution of wind speed at 100 m above the ground over the continents.  Figure \ref{fig:offshorewinds} shows wind speeds at 100 m above sea level over the oceans.  The space elevator needs to be located on the equator (or very close), but as the maps show, winds are at their global minima all around the equator.  Wind farms are not a likely source of power for the space elevator.

\FloatBarrier

\section{Wave farms}

Reference [\cite{waveenergy1}] and [\cite{waveenergy2}] give the history and current survey of technologies in harvesting energy from the wave motion of the oceans.  The references list at least 25 companies actively pursuing electrical generation from water wave action.  Wave Energy Converters (WEC) generally refer to devices that harvest wave energy anywhere, but there are specific devices to harvest current and tidal energy.  These are a different category of energy conversion devices and are specific to locations with tidal activity.  There is extreme differentiation in the WEC marketplace.  The second reference shown lists six different types of WECs:

\begin{itemize}
    \item Point Absorbers
    \item Rotating Mass
    \item Submerged Pressure Differential Devices
    \item Terminators
    \item Oscillating Water Column Devices
    \item Attenuators
\end{itemize}

The technology of these devices varies widely with some working on the surface of the water and some submerged.  Some WECs can absorb wave energy from any direction, while others are oriented to the waves and intended to capture energy from wave pitch, roll, or yaw.  Some are anchored to the seabed, others not.

\begin{figure}
\begin{centering}
\includegraphics[width=\textwidth]{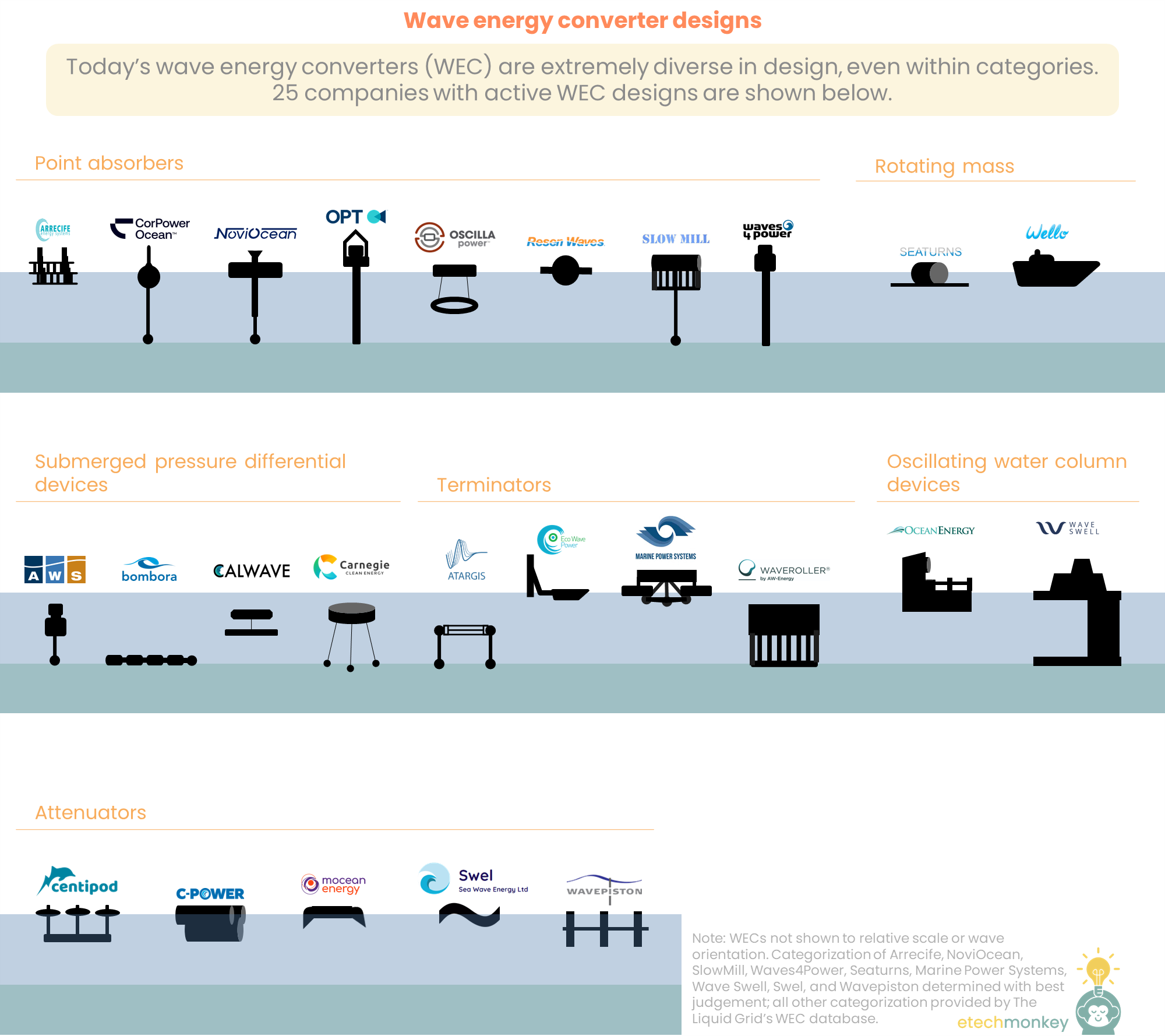} 
\caption{\label{fig:wavedesign} Graphic of the 25 companies currently pursuing wave energy converter development and marketing. [\cite{waveenergy2}]}
\end{centering}
\end{figure}

CalWave, \url{https://calwave.energy/}, plans scalable installations that can provide megawatts of power, but because of the extreme variability in location environments, it was not possible to get solid numbers on the size of wave farm needed to provide 100 MW of power.  Calwave claims that wave power is 20 to 60 times denser than other renewable energy sources.  A 100 MW wave farm would likely occupy several square kilometers of ocean.

Wave energy conversion may be one of the sources of power for future space elevator power beaming platforms.
\FloatBarrier
\section{Fusion reactors}

While it is not currently available to power the space elevator, nuclear fusion is a clear possibility for the future.  A non-exhaustive list of companies trying to develop fusion power is:

\begin{itemize}
    \item Avalanche Energy, \url{https://www.avalanchefusion.com/}
    \item Commonwealth Fusion Systems, \url{https://cfs.energy/}
    \item First Light Fusion, \url{https://firstlightfusion.com/}
    \item Focused Energy, \url{https://www.focused-energy.co/}
    \item General Fusion, \url{https://generalfusion.com/}
    \item Helion Energy, \url{https://www.helionenergy.com/}
    \item Marvel Fusion, \url{https://marvelfusion.com/}
    \item Realta Fusion, \url{https://realtafusion.com/}
    \item TAE Technologies, \url{https://tae.com/}
    \item Tokamak Energy, \url{https://tokamakenergy.com/}
    \item Type One Energy, \url{https://typeoneenergy.com/}
    \item Zap Energy, \url{https://www.zapenergy.com/}
\end{itemize}

\begin{figure}
\begin{centering}
\includegraphics[width=\textwidth]{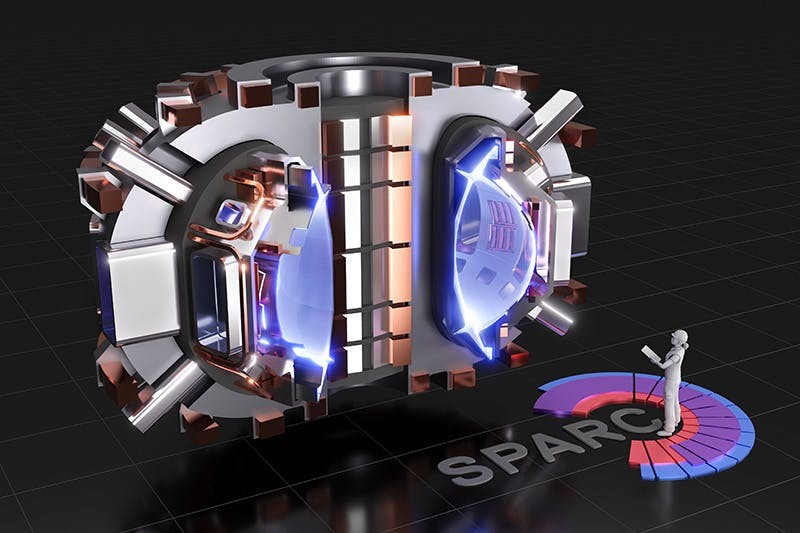} 
\caption{\label{fig:SPARC}Rendering of the SPARC tokamak fusion reactor under construction by Commonwealth Fusion Systems. High-temperature superconducting (HTS) tape enables smaller magnets to operate at higher temperatures and magnetic fields than in previous tokamak cryogenic systems. SPARC is a technology demonstrator and will not generate electricity, but it is the prototype for ARC, a utility-scale fusion reactor. [\cite{SPARCCFS}]}
\end{centering}
\end{figure}

Once developed and commercialized, fusion reactors will come in a range of sizes from microreactors to large-scale utilities.  They may be as portable as fission reactors, but that remains to be seen.

\chapter{Solar Powered Climbers}\label{chapt:Solar}

One method that has been proposed several times is to use solar power for the motors on a climber, [\cite{Assess}, especially Figure 4-3]. This method avoids the difficulties of aiming laser beams over long distances, and it eliminates the need for large and expensive power infrastructure needed on Earth, such as that of the power beaming designs. It has the drawbacks that no solar power is available at night, and the receiver needs to be very large.

The most promising material for the solar arrays appears to be perovskite. Although it is not in general use yet, there is much activity developing it for use terrestrially. It is under consideration for use on the Moon because its optimum thickness is only about 400 nanometers (nm).[\cite{tu2021perovskite}, \cite{click}, \cite{NASAunusual}] This means that large solar arrays can be constructed with only small quantities of perovskite. On Earth, it has to be protected from moisture, but in space and on the moon that is not an issue. In the vicinity of Earth, the Sun's light power (including visible light, infra-red and ultra-violet) is typically in the range 1,361 \si{W/m^2} to 1,366 \si{W/m^2}, and perovskite is able to produce about 400 \si{W/m^2} from this insolation. For a 4 MW climber an area of 12,500 square meters is needed to generate 5 MW, which allows for some losses.

A mechanical design could be for a flat plate that has to be steered to face the sun, but this presents difficulties because the solar arrays must withstand Earth’s gravity near Earth, which is quite different from satellites that are in orbit. Hence, they will need structural support such as reinforcing struts, which will add to their weight. A more promising design is to use a sphere that is folded at first and then opens rather like an umbrella or pair of umbrellas when it is above the atmosphere.

A circular plate of 12,500 \si{m^2} area will have a radius of 63 meters (a diameter of about 400 feet). If we use a sphere of that radius, it will present to the sun an effective surface of 12,500 \si{m^2}, but its surface area will be 50,000 \si{m^2} $(4\pi R^2)$. Of course, if more power is needed for a faster climb or a bigger climber, a greater radius will be required. Doubling the radius will give four times the power.

\section{Designs of perovskite solar cells}

\begin{figure}
\begin{centering}
\includegraphics[width=.5\textwidth]{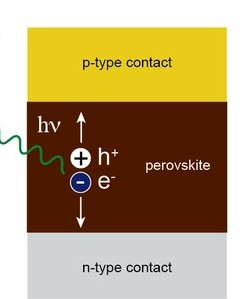}    
\caption{\label{fig:fig1} Structure of a perovskite cell (from \cite{tu2021perovskite}).}
\end{centering}
\end{figure}

The basic structure of one type of perovskite cell is shown in Figure \ref{fig:fig1}. The p-type contact is also known as the hole-transporting layer (HTL), and the n-type contact is known as the electron-transporting layer (ETL).
In addition to these three layers, two more layers are needed to conduct the current away and to form a substrate. A thin layer (100nm) of silver or gold may be used above the HTL to carry the positive charges away, and a somewhat thicker (150nm) layer below the ETL may use indium tin oxide, which is a common electronic substrate, to conduct the negative charges away.

Perovskite was discovered in the Ural mountains and is named after a mineralogist called Lev Perovsky. The term refers to calcium titanium oxide $(CaTiO_3)$ and also more generally to a class of materials that have a similar crystalline structure. One form of perovskite used in solar cells has the formula $CH_3NH_3PbI_3$.The elements here are carbon, hydrogen, nitrogen, lead and iodine.

\begin{table}[h]
    \centering
    \begin{tabular}{l l l}
        \toprule
        Name & Composition & Thickness \\
        \midrule
        Conductive layer & Ag or Au (silver or gold) & 100 nm \\
        p-type contact (HTL) & PCBM=Phenyl-C61-butyric acid methyl ester & 100 nm \\
        Perovskite & CH$_3$NH$_3$PbI$_3$ & 300 nm \\
        n-type contact (ETL) & PEDOT=Poly(3,4-ethylenedioxythiophene) & 50 nm \\
        Conducting substrate & ITO=Indium tin oxide & 150 nm \\
        \bottomrule
    \end{tabular}
    \caption{Layers of a typical perovskite photo-voltaic cell}
    \label{tab:perovskite}
\end{table}

The typical layers of a perovskite photo-voltaic cell are listed in Table \ref{tab:perovskite}. The total thickness including the substrate is 700 nm (0.7 micron). The density of perovskite is approximately 4000 \si{kg/m^3}.

Indium tin oxide is a widely used substrate for electronic applications. Adapting this layout to space elevators, aluminium is the preferred substrate because it is much lighter (2700 vs 7120 \si{kg/m^3}). Silver is lighter than gold, so it is preferred as the top layer for this application. 

\begin{table}
    \centering
    \begin{tabular}{l l p{1.8cm} p{1.8cm} p{2cm}}
        \toprule
        Layer & Material & Density (kg/m\textsuperscript{3}) & Thickness (nm) & Mass per unit area (kg/m\textsuperscript{2}) \\
        \midrule
        Conductive layer & Silver & 10490 & 100 & $10.5 \times 10^{-4}$ \\
        p-type contact (HTL) & PCBM & 1630 & 100 & $1.63 \times 10^{-4}$ \\
        Perovskite & CH\textsubscript{3}NH\textsubscript{3}PbI\textsubscript{3} & 4000 & 300 & $12 \times 10^{-4}$ \\
        n-type contact (ETL) & PEDOT & 1331 & 50 & $0.67 \times 10^{-4}$ \\
        Conducting substrate & Aluminium & 2700 & 150 & $4.1 \times 10^{-4}$ \\
        \midrule
        TOTAL & & & 700 & $2.90 \times 10^{-3}$ \\
        \bottomrule
    \end{tabular}
    \caption{Perovskite cell layers adapted for space elevators}
    \label{tab:pskiteSE}
\end{table}

From Table \ref{tab:pskiteSE} we take an approximate mass density of \unit{\num{3E-3}} \si{kg/m^2}, leading to a total mass over 50,000 \si{m^2} of 147 kg. This mass is included in Section \ref{sec:provmass}, Provisional mass budget.

\section{Structure of the sphere}

\begin{figure}
\begin{centering}
\includegraphics[width=\textwidth]{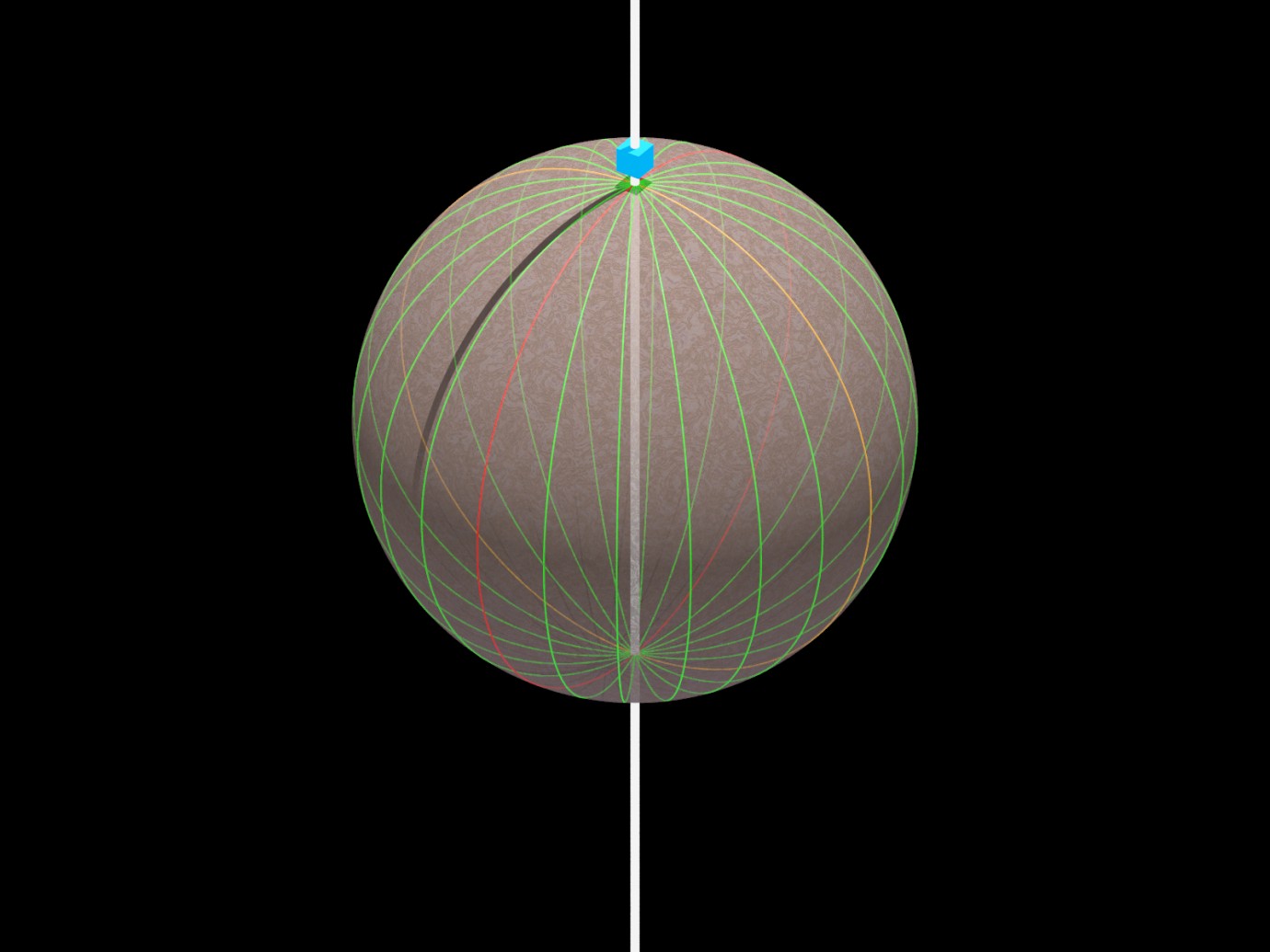}    
\caption{\label{fig:fig2} The solar array on a wireframe sphere below the climber; the plane of the orange circle is parallel to the plane of the Earth's equator; the climber is at the top in light blue; the red rods define a circle whose plane is orthogonal to the orange circle/equator.}
\end{centering}
\end{figure}

It is assumed that we can maintain a stable orientation of the sphere using reaction wheels or a similar mechanism. The sphere will consist of a graphene membrane stretched over a frame of 24 flexible rods made of carbon fiber reinforced polymer (CFRP) as in Figure \ref{fig:fig2}. The membrane will support the solar cells and will contribute to the stability of the sphere. Then the sun follows a path within ±23.5° of a circle parallel to the Earth’s equator. This circle is marked in orange in Figure \ref{fig:fig2}, not to be confused with the red circle that is orthogonal to it.

Each perovskite cell delivers a voltage of 1.25 V, so a string of 80 cells connected in series will produce 100 V. It will occupy 0.76 \si{m^2} and will deliver 3 Amps and thus 300 Watts. All the cells in one string will receive sunlight at the same time of day. There will be little or no sunlight near the north and south poles of the sphere, and so the solar strings there will produce less current, depending on the season; at some point near the poles it may not be worth placing solar cells at all.

\section{Power conversion}

\begin{figure}
\begin{centering}
\includegraphics[width=0.8\linewidth]{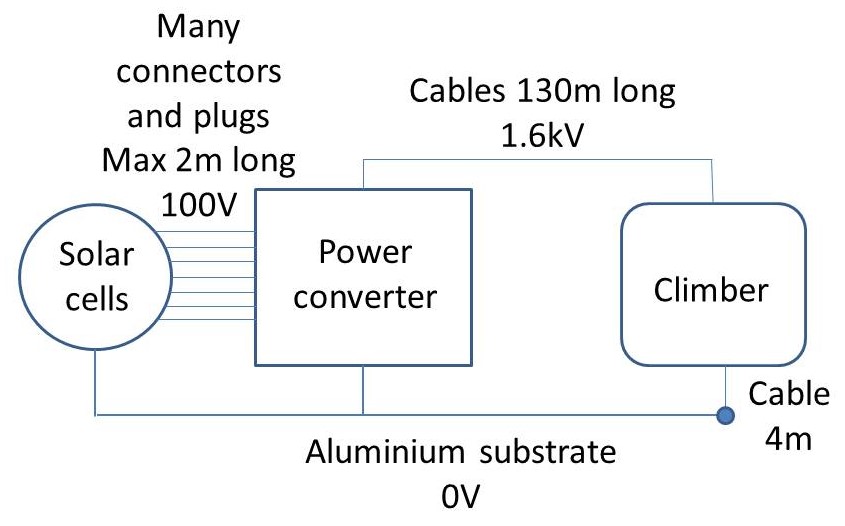}
\caption{Conceptual diagram of the main power components}
\label{fig:ElecDiagram}
\end{centering}
\end{figure}

To get power from the solar cells to the climber, a power converter is needed. Figure \ref{fig:ElecDiagram} shows the main elements. Strings of solar cells are connected in series to produce output at 100 Volts. Many short connectors and plugs carry power from the solar cells to a power converter that changes the voltage to 1.6 kV in order to minimize transmission losses in the long cables that carry power to the climber. The short return cables connect to the aluminium substrate to provide the neutral return connection. This and the following two sections go into detail on these components.

Current from many of the solar cells will have to travel to the climber. In early morning and late evening, the sun will be shining on the lower parts of sphere. We may want to allow for laser light to shine onto the bottom at night, providing a hybrid solar and laser power system. Hence current may need to flow as far as 2R plus the distance to the climber, where R is the radius of the sphere. The total distance could be 130 meters, and so to avoid heavy losses and excessive heating we will need power conversion from 100 V to at least 800 V, and preferably 1.6 kV to drive a pair of 800 V motors in series.

It is tempting to think about connecting solar cells in series above 100 V, but experience has shown that this causes problems with flashover and corona discharges. [\cite{OsborneEE}] There are published standards showing how to protect high-voltage circuits in space. [\cite{NASA-HDBK-4007}]

\begin{figure}
\begin{centering}
\includegraphics[width=.8\textwidth]{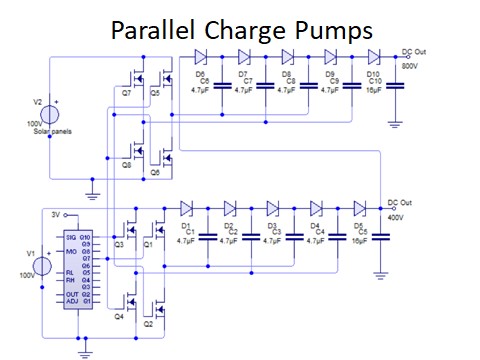}    
\caption{\label{fig:fig3} Circuit diagram of a charge pump}
\end{centering}
\end{figure}

Air or vacuum-core transformers are a possible technology. An alternative method is to use charge pumps such as the one illustrated in the circuit diagram in Figure \ref{fig:fig3}. The distribution of mass across the sphere would be similar.

One manufacturer claims to achieve 35 kW/kg, [\cite{HV}], but this applies to voltage doubling, whereas we need multiplication by a factor of at least eight. One study has claimed a power-to-weight ratio for a similar method of 27 kW/kg, [\cite{gemin2015arch}, especially Figure 35], delivering a voltage of several kilovolts for use in turbo-electric aircraft; this would imply a total mass for 5 MW of 185 kg. This mass needs to move inside the sphere, following the sun. Otherwise, the power converters would need to be on all parts of the sphere to cope with the diurnal variations, including the possibility of using laser power from Earth at night, and so the total mass would be 740 kg. To reduce this mass, a mechanism is needed to move the power converter.

\begin{figure} [!h]
    \begin{minipage}[t]{.5\textwidth}
        \centering
        \includegraphics[width=.95\linewidth]{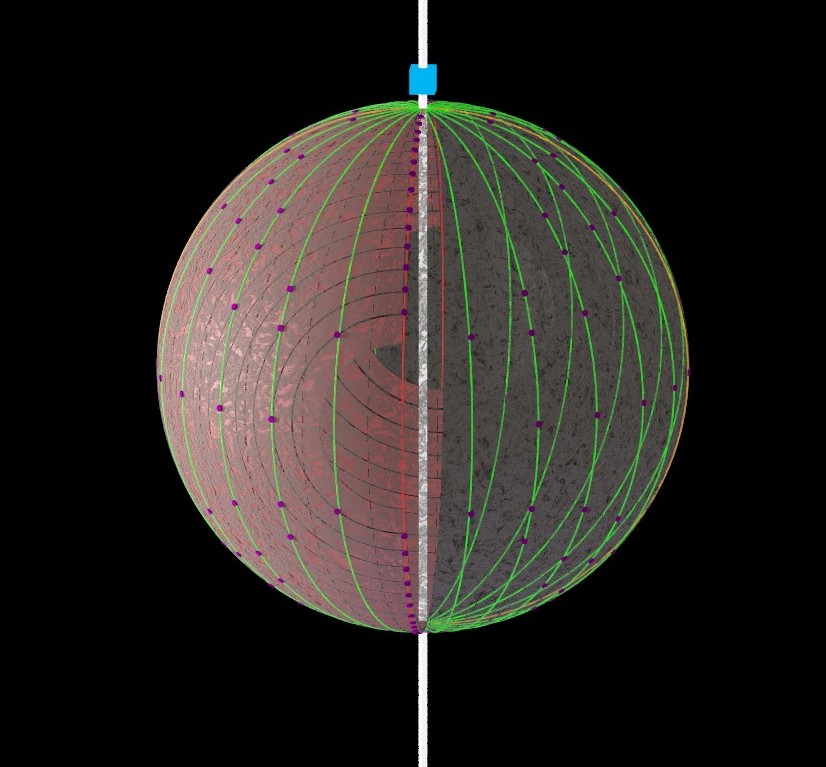}
        \captionsetup{justification=centering}
        \caption{The power converter is in reddish brown; it travels inside the sphere and follows the sun; this is the position at 6am.}
        \label{fig:fig4}
    \end{minipage}
    \begin{minipage}[t]{.5\textwidth}
        \centering
        \includegraphics[width=.95\linewidth]{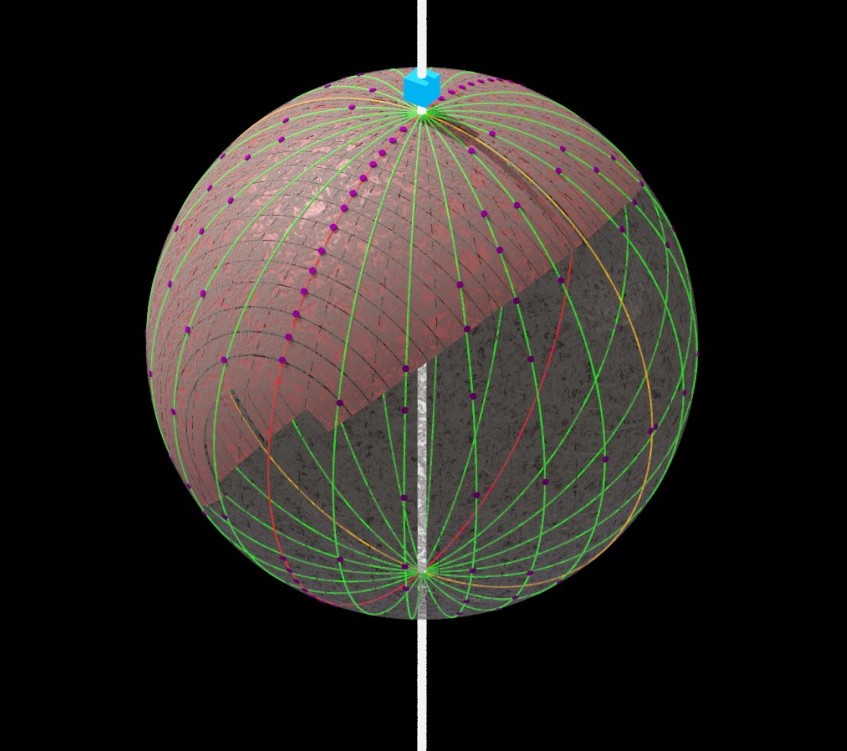}
        \captionsetup{justification=centering}
        \caption{Position at 9am}
        \label{fig:fig5}
    \end{minipage}
    \begin{minipage}[t]{.5\textwidth}
        \centering
        \includegraphics[width=.95\linewidth]{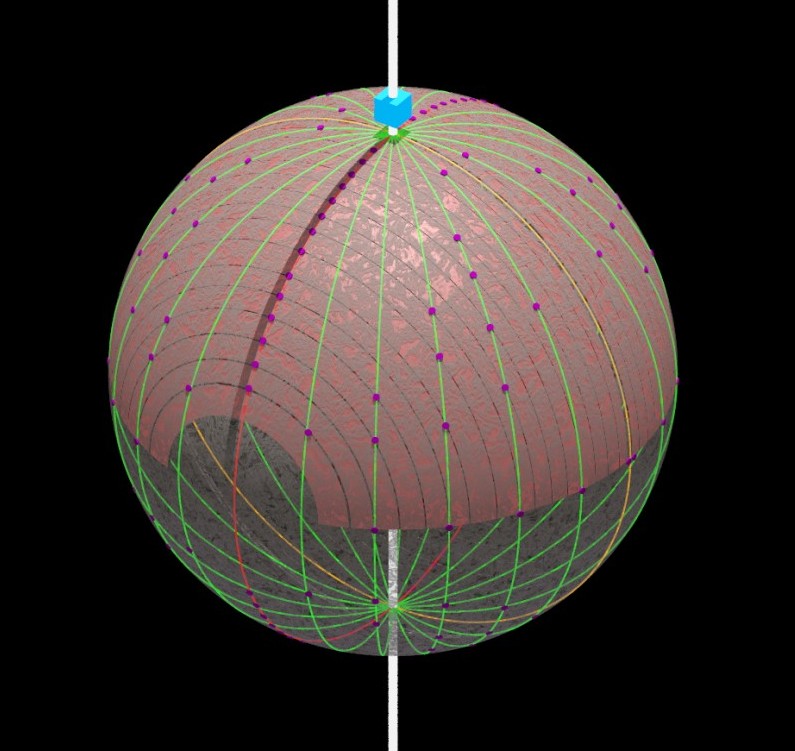}
        \captionsetup{justification=centering}
        \caption{At noon}
        \label{fig:fig6}
    \end{minipage}
    \hfill
    \begin{minipage}[t]{.5\textwidth}
        \centering
        \includegraphics[width=.95\linewidth]{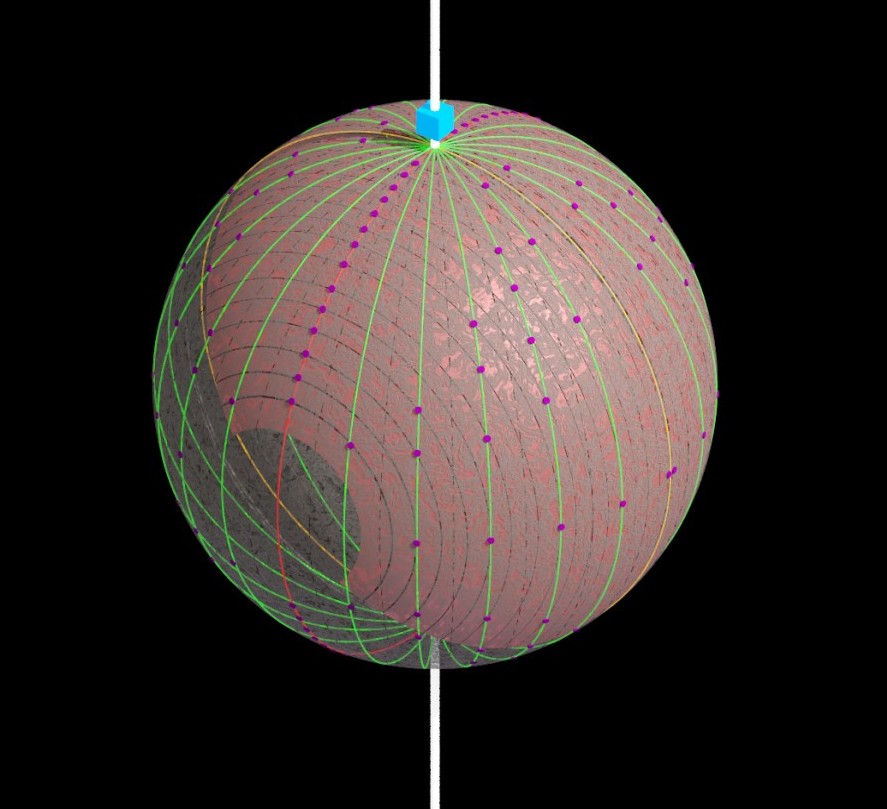}
        \captionsetup{justification=centering}
        \caption{At 3pm; the magenta-colored motors are most clearly visible here.}
        \label{fig:fig7}
    \end{minipage}
\end{figure}

The movement required is illustrated in Figures \ref{fig:fig4} to \ref{fig:fig7}. Motors on the rods are shown in magenta. There are 256 of them, and they drive the power converter around the inside of the sphere. In these figures they appear as small magenta objects on the rods. They are inside the membrane, and they are placed so that each of the toothed tracks marked as gray lines on the power converter engage with three motors at any moment. The motors will need to drive the power converter at the speed of a 24-hour clock with a circumference of $2\pi R$, which amounts to 4.6 mm/s.

The power converter is split into two sides, north and south, so that each side can move past the meter-wide tether, leaving a gap between them. The solar cells are still active in this gap; their output is collected by the nearest part of the power converter. It is easiest to see the gap in Figure \ref{fig:fig7}; it coincides with the orange rod that is in a plane parallel to Earth’s equator.

Moving the power converter around the sphere each day will cause an imbalance in weight. As a result, the sphere will tilt, an effect partly balanced by tension in the tether. 

One or more cables are needed to carry the power from the converter to the climber, and they will need to be long enough to reach from the lowest point of the sphere. 5 MW of power at 800 V requires a total current of I=6.25 kA, and this will be carried through several aluminium conductors of length 2R plus 4 meters to reach the climber $(l=2R+4\approx130 \, \si{m})$. If the conductors’ total area of cross section is A and their resistivity is $\rho_c$, the resistance $R_c$ is given by $R_c=(2R+4)\rho_c/A$. The loss will be $L_c=I^2R_c$. Therefore 
$$A=(2R+4)\rho_cI^2/L_c$$

Now $\rho_c=2.65\times{10}^{-8}$ $\si{\Omega\cdot\m}$ for aluminium. If we want to keep the loss to $L_c=200$ $\si{kW}$ then
$$A=130\times2.65\times{10}^{-8}\times39\times{10}^6/2\times{10}^5\approx6.72\times{10}^{-4} \, \si{m^2}$$

This is 6.72 \si{cm^2}. The mass of the conductor is $2RA\rho_a$, where $\rho_a=2.70\times{10}^3$ $\si{kg/m^3}$, the mass density of aluminium. This comes to $130\times6.72\times{10}^{-4}\times2.70\times{10}^3\approx236$ \si{kg}. We will also need insulation probably amounting to 20 kg, giving a total of 256 kg. There will also need to be a return conductor, which may take the form of a  conductive layer on the membrane or another cable or set of cables. In practice there will be several cables coming from different parts of the power converter, but the total mass will be similar.

If we had no power conversion and kept the voltage at 100 V, the cable mass would be 15 tonnes. We would also need return cables to deal with the large current, totaling 30 tonnes. On the other hand, if we raise the voltage to 1.6 kV, the total conductor mass will only need to be 59 kg each way plus 20 kg insulation, totaling 138 kg. This is the recommended configuration. A cited study describes a power converter that will take several kV so it is reasonable to assume 1.6 kV. [\cite{gemin2015arch}]

The cable carrying the power to the climber will pass through the hole at the top of the sphere which also accommodates the one-meter-wide tether.

\section{Power to the converter}

Power from the solar cells must be carried to the nearest part of the power converter. The connections must be kept short to minimize the current and hence minimize the transmission losses. Since the power converter is inside the membrane, a large number of holes in the membrane will be needed, and these will be filled with so-called plugs. They are spaced one meter apart, so that there will be about 50,000 of them over the whole membrane. They must be positioned so as not to compromise the strength of the membrane significantly as it is stretched over the rods, but they must also allow for the variation in angle of movement of the power converter at different parts of the sphere.

\begin{figure}
\begin{centering}
\includegraphics[width=\textwidth]{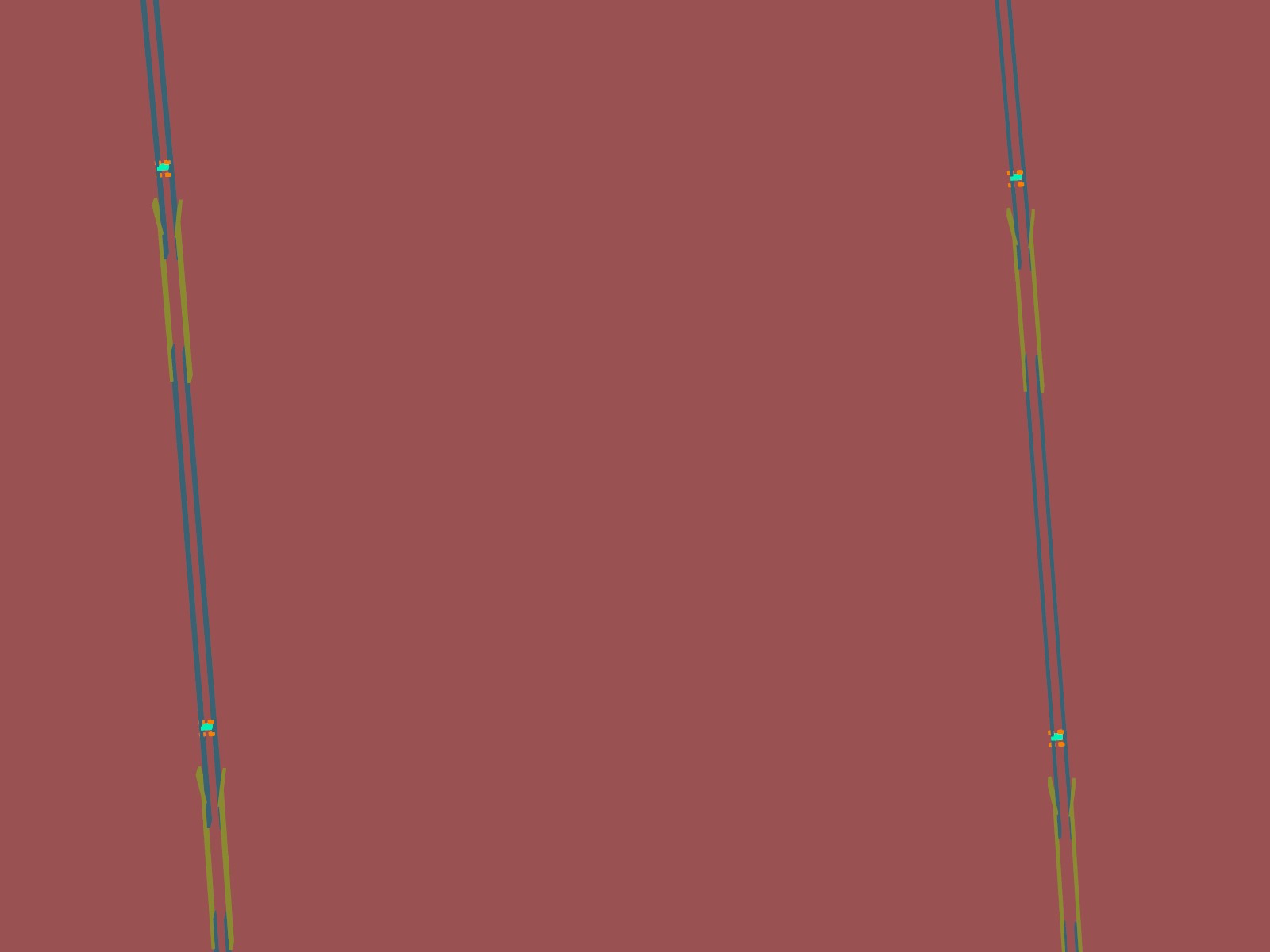}    
\caption{\label{fig:fig8} Tracks on the power converter showing connecting rails in yellow and plugs in green}
\end{centering}
\end{figure}

The plugs engage with tracks on the power converter, as in Figure \ref{fig:fig8}. They provide both mechanical and electrical connections. The circuitry is on the inside and hence not visible in the figure. The power converter is based on a flexible printed circuit to enable it to be folded while in the atmosphere, along with the rest of the sphere. To permit folding, most of the tracks are in short segments of 78 cm with connecting rails 30 cm long between them, overlapping by 4 cm at each end. However, near to rods and folds, the geometry is sometimes different, depending on the relative angles. The gauge of the track is 2.5 cm, a little wider than a typical model train set.

\begin{figure}
\begin{centering}
\includegraphics[width=\textwidth]{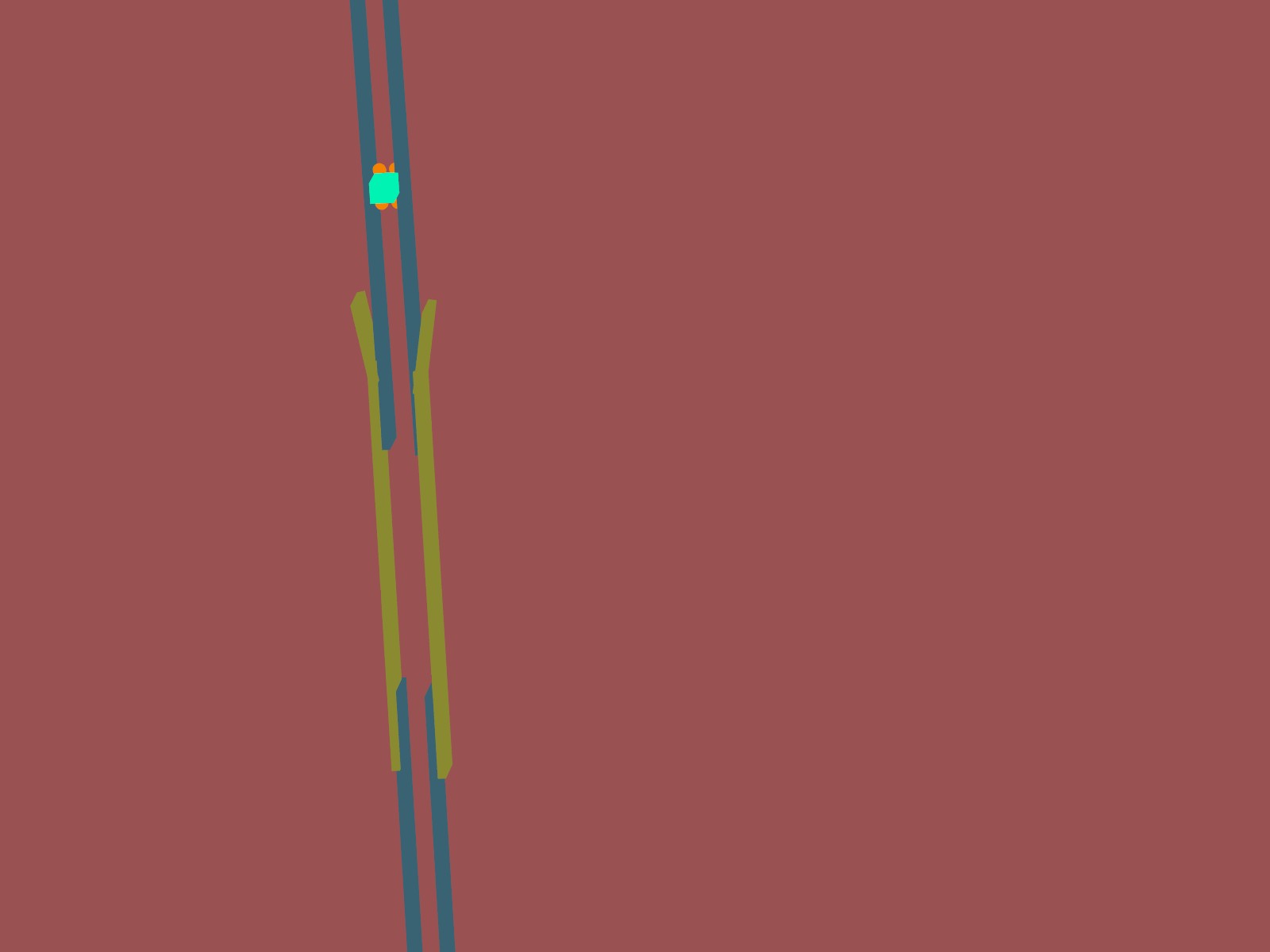}    
\caption{\label{fig:fig9} Plug detail (green) with orange wheels}
\end{centering}
\end{figure}

The plugs have wheels, shown in orange in Figure \ref{fig:fig9}, which fit into grooves in the rails, thus providing support to the rails and hence to the power converter. The wheels are spring loaded to allow for the wider sections of track at the connecting rails.

As the plugs are typically a meter apart, they will receive power from a square meter of solar cells yielding a maximum of 400 Watts depending on the angle of the sun, which is known by design. At 100 V, the current in a plug will be up to 4 Amps. The circuitry on the power converter must be distributed to minimize losses, keeping the path length of current flow to a maximum of 2 meters, including the rails and distance from the solar cells to the plug. 

\begin{figure}
\begin{centering}
\includegraphics[width=\textwidth]{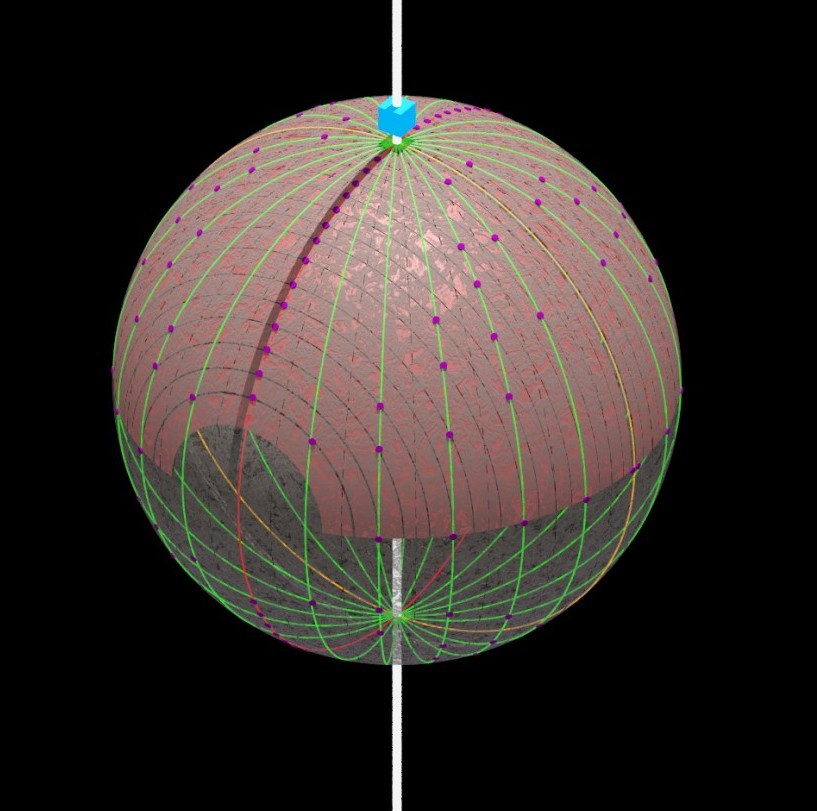}    
\caption{\label{fig:fig10} Drive motors (in magenta) spread strategically over the rods}
\end{centering}
\end{figure}

\begin{figure}
\begin{centering}
\includegraphics[width=.7\textwidth]{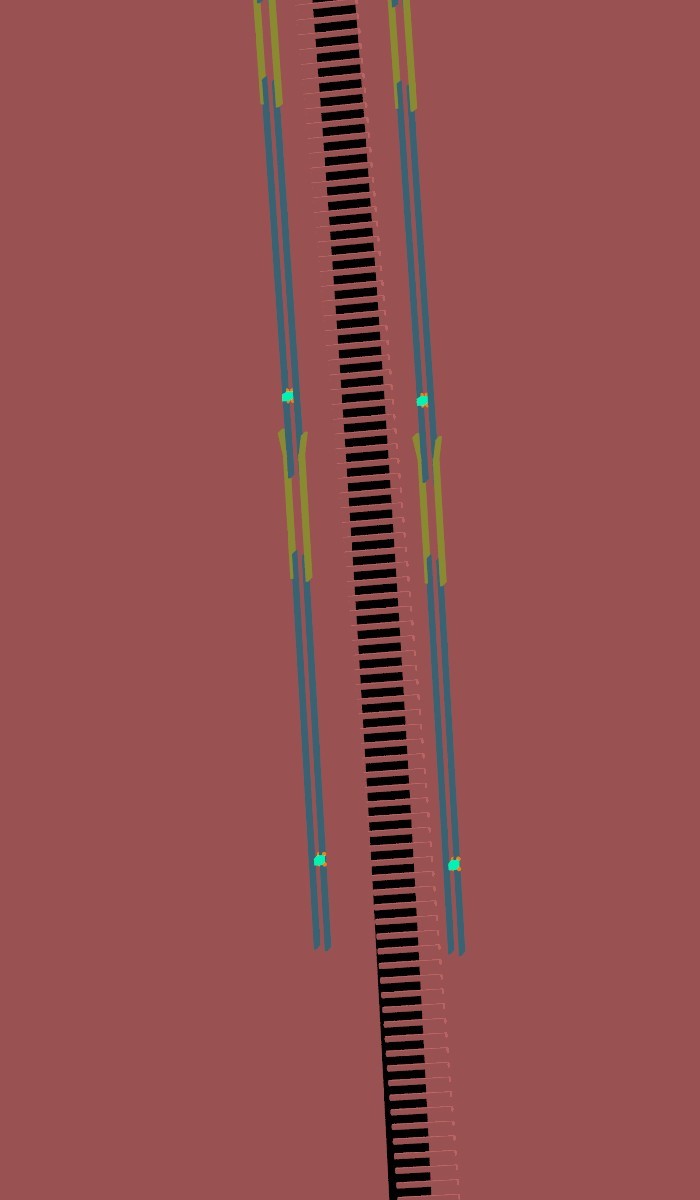}    
\caption{\label{fig:fig11} Toothed track for a motor to engage with; plugs and tracks on each side maintain the connection between the motor and the power converter}
\end{centering}
\end{figure}

\begin{figure}
\begin{centering}
\includegraphics[width=.75\textwidth]{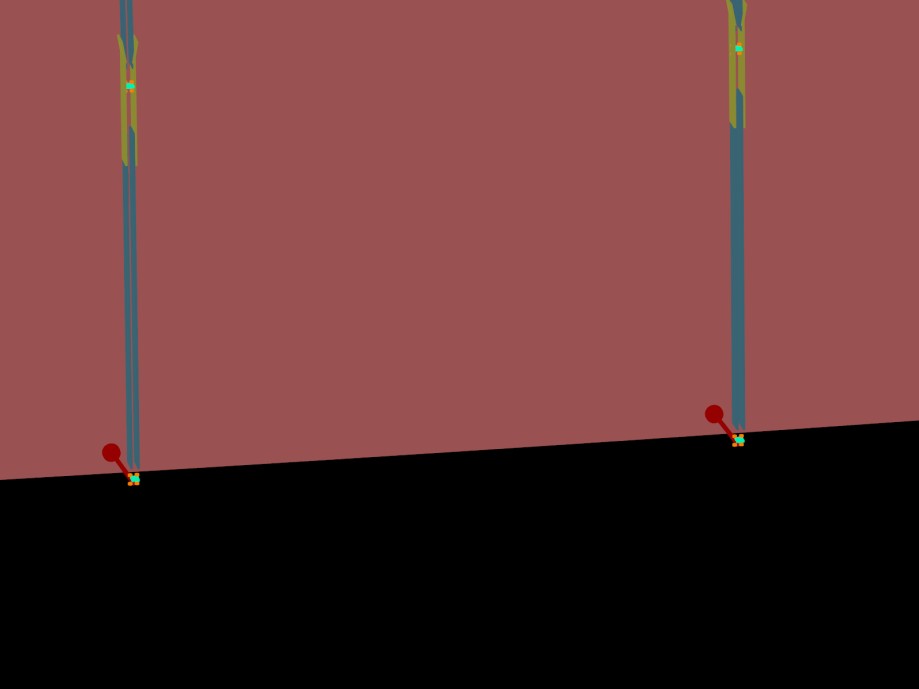}    
\caption{\label{fig:fig12} Grabbers (shown in red) at the leading edge catching plugs (green)}
\end{centering}
\end{figure}

\begin{figure}
\begin{centering}
\includegraphics[width=.75\textwidth]{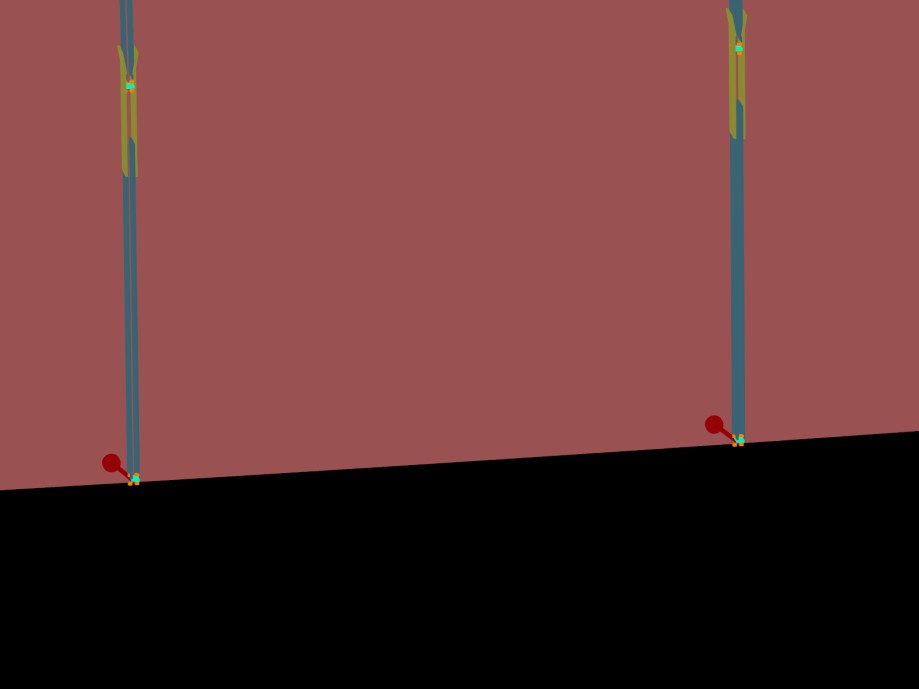}    
\caption{\label{fig:fig13} Grabbers pull the plugs into the track as the power converter moves}
\end{centering}
\end{figure}
\FloatBarrier

Referring back to the diagram in Figure \ref{fig:ElecDiagram}, we need to calculate the combined mass of all the connectors and plugs between the solar cells and the power converter. First look at the power to be carried, which is $p=VI$ for current $I$ and voltage $V$. Limit the loss to a fraction $f_p$, so that the loss is $pf_p$. Then $pf_p=I^2R_c$, where $R_c$ is the combined resistance of all these conductors. Now $R_c=l\rho_c/A$ for conductor length $l$, resistivity $\rho_c$ and combined cross-sectional area $A$. Hence
$$A=\frac{l\rho_c}{R_c}=\frac{l\rho_cI^2}{pf_p}=\frac{l\rho_cp^2}{pf_pV^2}=\frac{l\rho_cp}{f_pV^2}$$

The mass $m=lA\rho$, where $\rho$ is the mass density. Therefore
$$m=\frac{l^2\rho_c\rho p}{f_pV^2}$$
Take l=2 meters, $\rho_c=2.65\times{10}^{-8}$ \si{\Omega \cdot m} and $\rho=2700$ \si{kg/m^3} for aluminium, $p=5\times{10}^6$ (5 MW), $f_p=0.02$ (2\% power loss equivalent to 100 kW), and V=100 Volts. Then

$$m=\frac{2^2\times2.65\times{10}^{-8}\times2700\times5\times{10}^6}{0.02\times{10}^4}=22980\times{10}^{-4} \approx2.3 \, \si{kg}$$

The combined mass of all these conductors is 7.2 kg and the same again for the return current, making 14.4 kg. It probably does not need insulation. If the path length increases to 3 meters, the mass will be 32 kg. 

In the power converter, we can take advantage of the lower current being delivered near the leading and trailing edges to relax the path lengths, even though the solar cells will have the same packing density. Fewer conversion circuits or transformers will be needed there.

\section{Power converter drive motors}

The drive motors are responsible for moving the power converter to follow the sun, as shown in Figure \ref{fig:fig4} to Figure \ref{fig:fig7}. The dark lines indicate the tracks whereby the drive motors engage with the power converter to provide traction. Toothed tracks embedded in the substrate of the power converter are proposed; these tracks are shown as dark lines in Figure \ref{fig:fig10}. They run in the direction in which the power converter moves on its daily path, and so they remain in line with the motors. The advantage of toothed tracks is that they can be embedded in the power converter’s substrate and can bend with it. Figure \ref{fig:fig11} shows a toothed track in more detail with the plugs and their tracks beside them, which are placed there to maintain a connection between each motor and its associated toothed track at all times.

The disposition of the drive motors is designed to provide at least three motors in contact with each toothed track at all times. They are positioned next to the rods.

\section{Leading edge}

At any time of day, only half the plugs on the membrane will be connected to the power converter. As it moves to follow the sun, some plugs will disengage at the trailing edge and others will need to be connected at the leading edge. To make sure that every plug engages successfully, we need a grabbing mechanism. Otherwise, there is a danger that some plugs would be missed due to the possibility that the power converter or the membrane may sag, depending on the alignment with respect to gravity. In Figure \ref{fig:fig12} the power converter has moved near the next set of plugs, at which point the grabbers must find and grab them. The grabbers will need sensors and microprocessors to detect the position of the plugs and to control the mechanism. Figure \ref{fig:fig12} shows the position as the grabbing takes place. In Figure \ref{fig:fig13} the power converter has moved enough for the plugs to be in contact with the rails, and the grabbers can release them and wait for the next plugs, which come every 108 seconds.

Note that the grabber may be lifting the edge of the power converter or pulling the plug towards it or both, depending on their positions with respect to gravity. 

\section{Deployment}

\begin{figure} [ht]
    \begin{minipage}[t]{.5\textwidth}
        \centering
        \includegraphics[width=.9\linewidth]{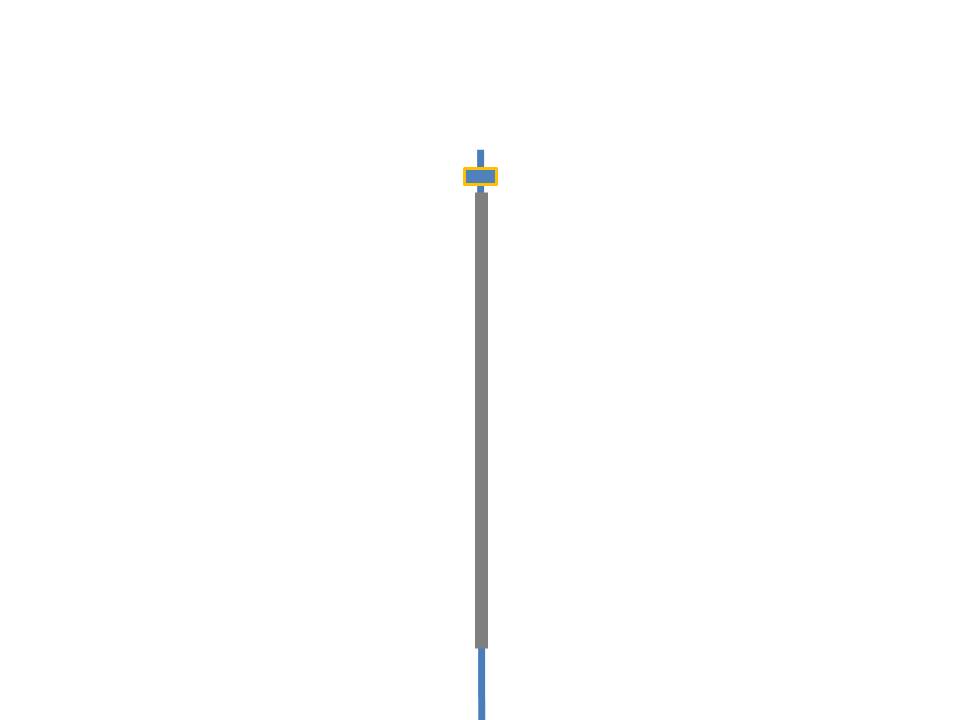}
        \captionsetup{justification=centering}
        \caption{Shape of solar power system in the atmosphere}
        \label{fig:fig14}
    \end{minipage}
    \begin{minipage}[t]{.5\textwidth}
        \centering
        \includegraphics[width=.9\linewidth]{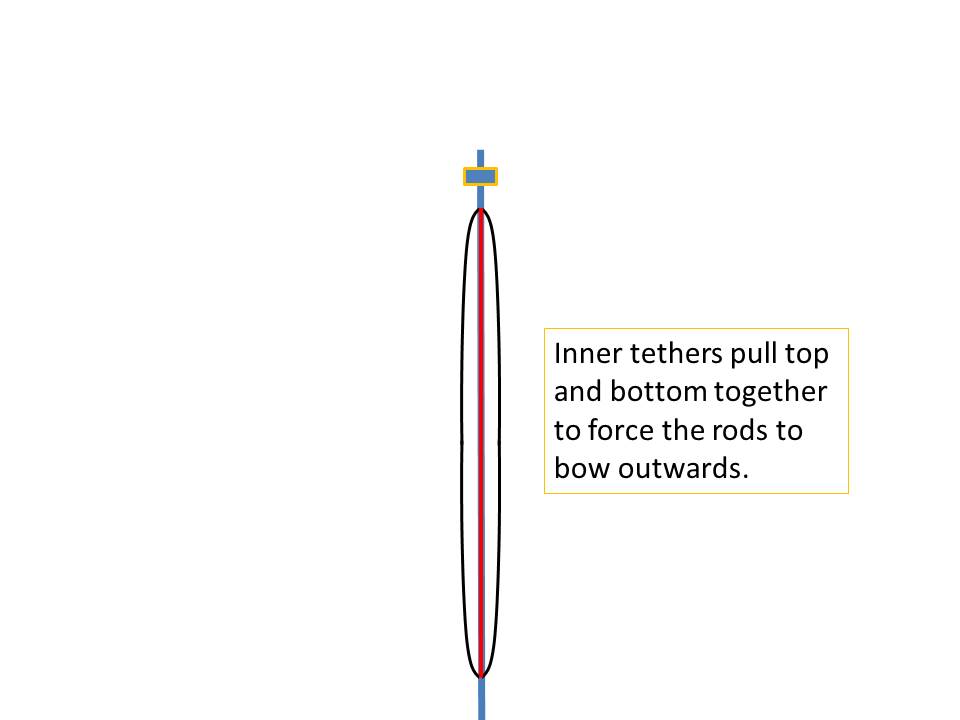}
        \captionsetup{justification=centering}
        \caption{Tethers are required to pull the top and bottom together and hold them in place.}
        \label{fig:fig15}
    \end{minipage}
    \begin{minipage}[t]{.5\textwidth}
        \centering
        \includegraphics[width=.9\linewidth]{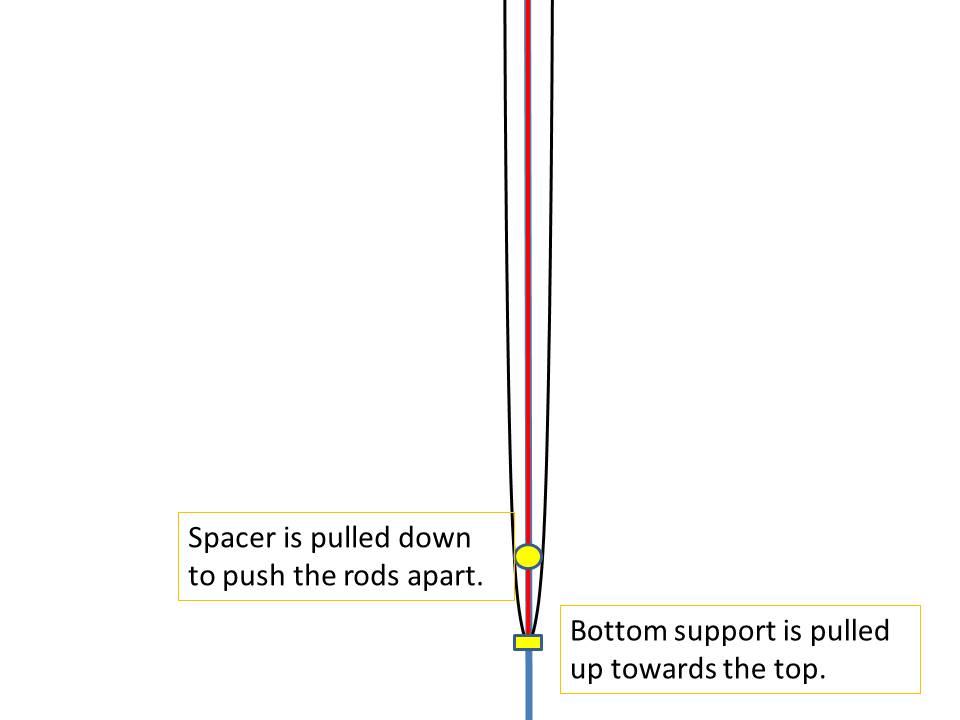}
        \captionsetup{justification=centering}
        \caption{A spacer pushes the rods apart as the bottom support is pulled up.}
        \label{fig:fig16}
    \end{minipage}
    \hfill
    \begin{minipage}[t]{.5\textwidth}
        \centering
        \includegraphics[width=.9\linewidth]{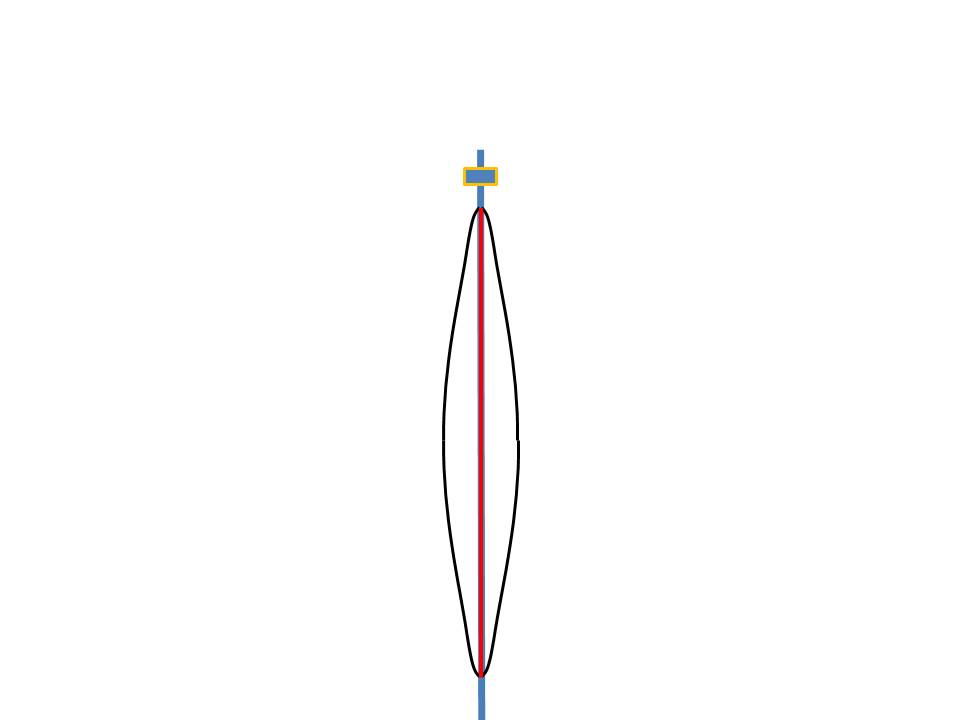}
        \captionsetup{justification=centering}
        \caption{The rods form a bow shape.}
        \label{fig:fig17}
    \end{minipage}
    \begin{minipage}[t]{.5\textwidth}
        \centering
        \includegraphics[width=.9\linewidth]{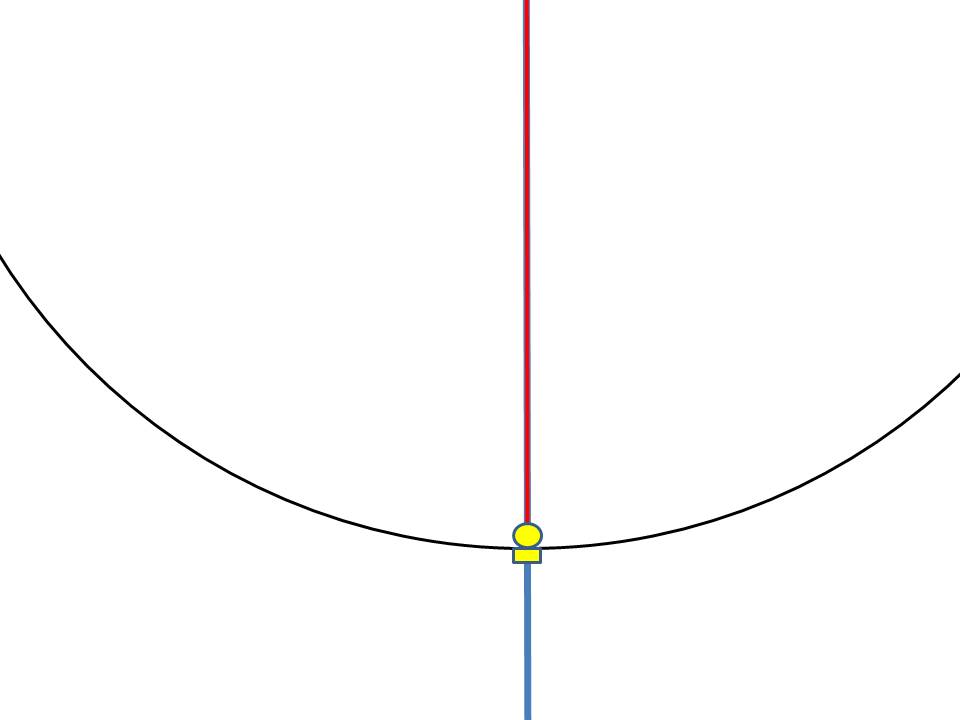}
        \captionsetup{justification=centering}
        \caption{Detail of bottom deployed}
        \label{fig:fig18}
    \end{minipage}
    \begin{minipage}[t]{.5\textwidth}
        \centering
        \includegraphics[width=.9\linewidth]{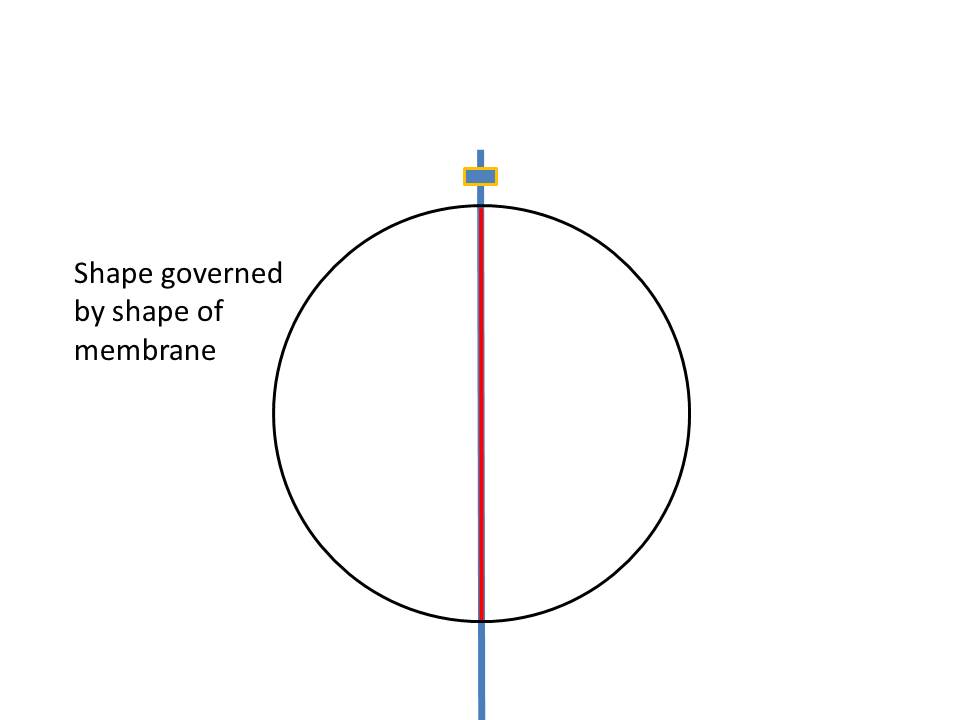}
        \captionsetup{justification=centering}
        \caption{2D diagram of deployment of sphere}
        \label{fig:fig19}
    \end{minipage}
\end{figure}

To initiate deployment, the rods must first be bent into a spherical shape. When the climber leaves the atmosphere the solar power system is narrow (Figure \ref{fig:fig14}). The first step in deployment is to pull the bottom towards the top (Figure \ref{fig:fig15}) using inner tethers. At the same time, a spacer is pulled down to start the rods’ bending (Figure \ref{fig:fig16}). A similar spacer is needed at the top. The actions cause the rods to bend (Figure \ref{fig:fig17}). The spacers end up next to the bottom (Figure \ref{fig:fig18}) or top. The membrane causes the rods to line up evenly spaced and the inner tethers maintain their tension to keep the rods in their desired shape of approximate semicircles, thus resulting in an overall shape close to a sphere (Figure \ref{fig:fig19}).

It should be possible to perform these movements by clamping the relevant parts to the main tether and inner tethers and moving the whole climber up or down; this avoids having to incorporate extra motors for deployment. A spool with a small electric motor will be needed to tidy away excess lengths of the inner tethers to avoid them entangling with the main tether.

\begin{figure}
\begin{centering}
\includegraphics[width=.75\textwidth]{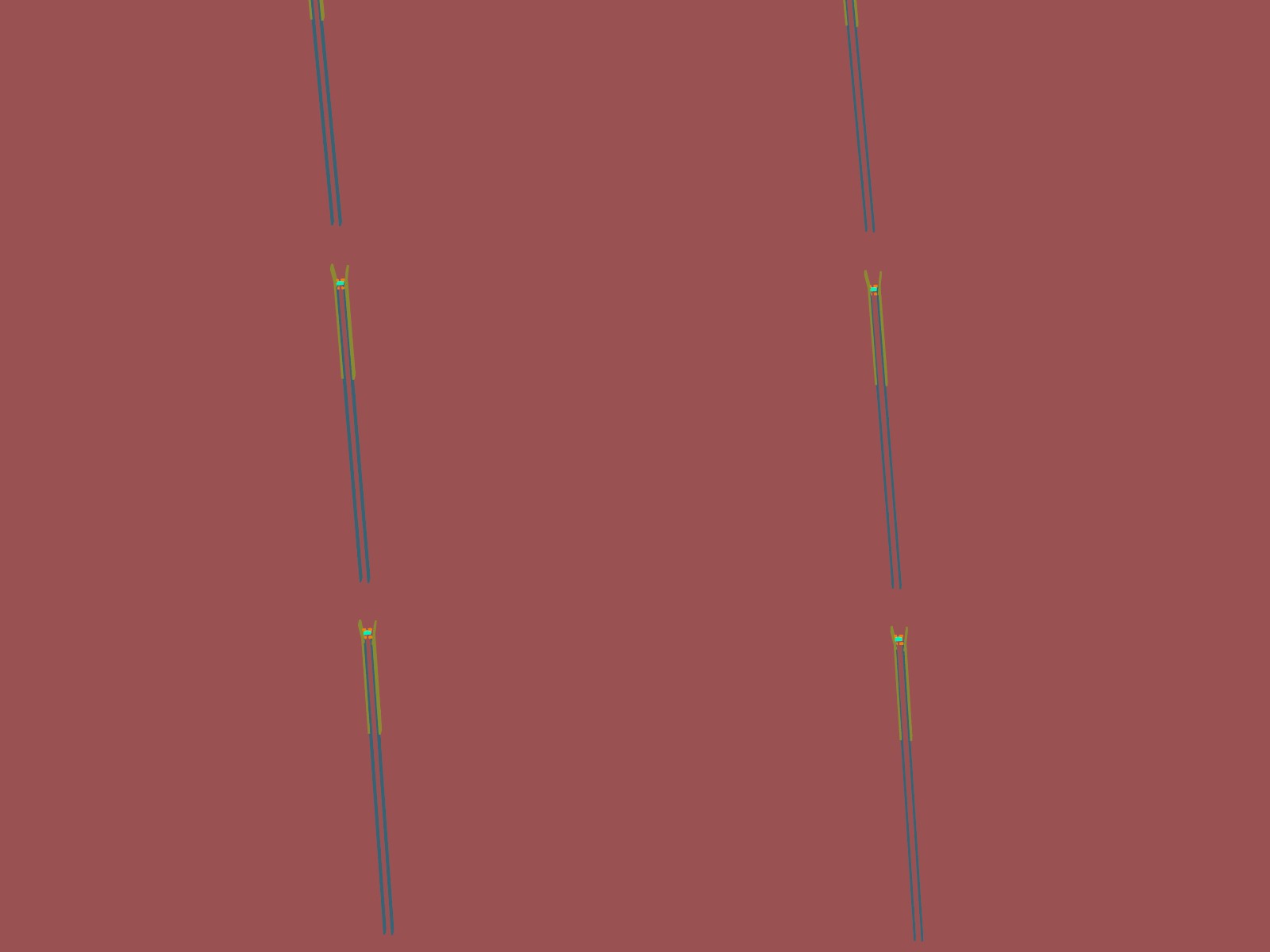}    
\caption{\label{fig:fig20} Connecting rails when the power converter has been unfolded}
\end{centering}
\end{figure}

In the atmosphere, the solar power system, consisting of solar cells, membrane, rods, power converter and other equipment, will be folded below the climber when it starts its ascent. The power converter will use flexible printed circuits rather than boards so that it can be folded. This means that the tracks that are engaged with the plugs must be segmented as in Figure \ref{fig:fig8} and Figure \ref{fig:fig9}. Initially the connecting rails are placed over the track segments as shown in Figure \ref{fig:fig20}. This is part of the setup that has to take place on Earth before the ascent commences. 

\begin{figure}
\begin{centering}
\includegraphics[width=.75\textwidth]{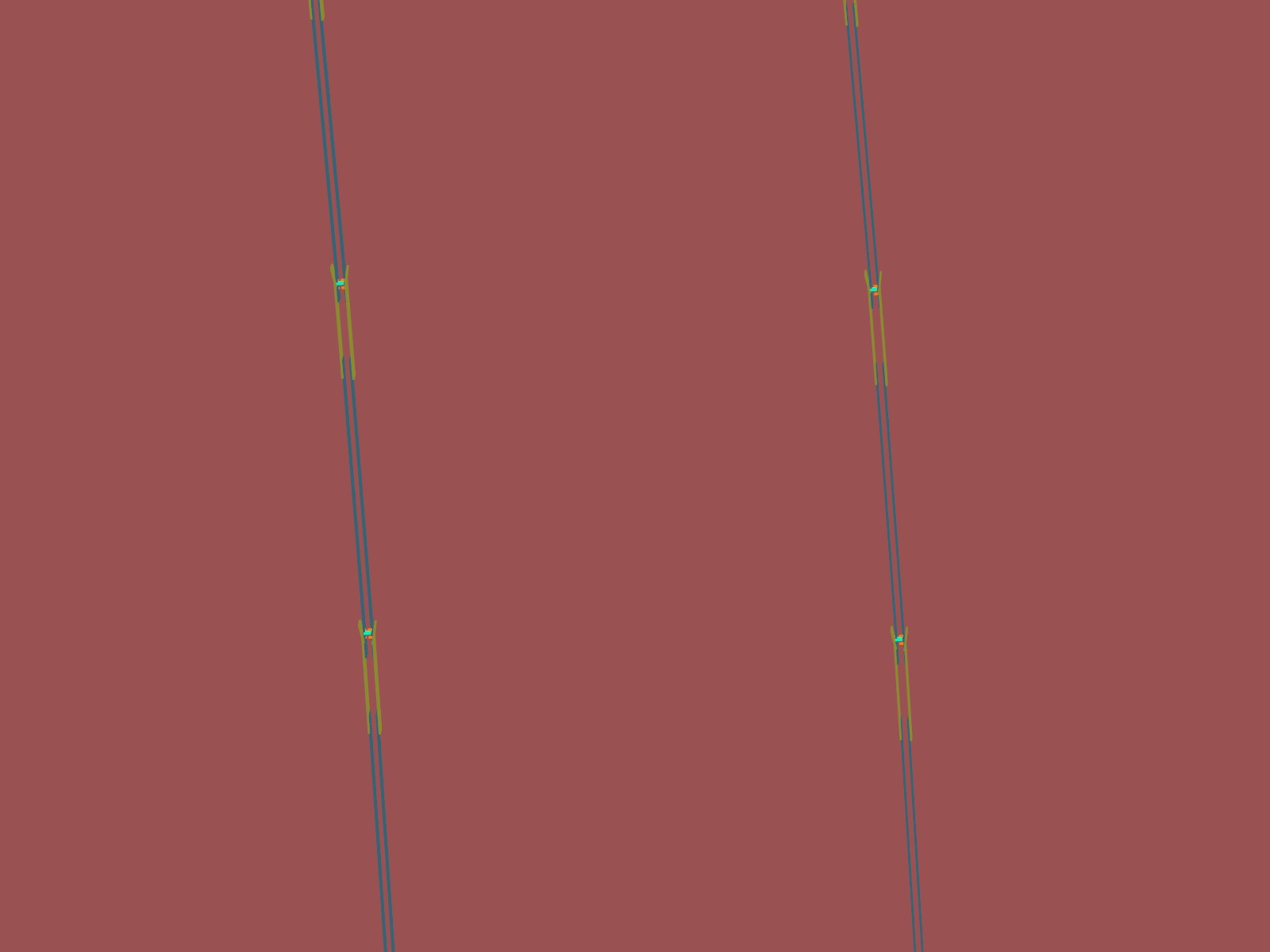}    
\caption{\label{fig:fig21} Connecting rails joined to the adjacent track}
\end{centering}
\end{figure}

\begin{figure}
\begin{centering}
\includegraphics[width=\textwidth]{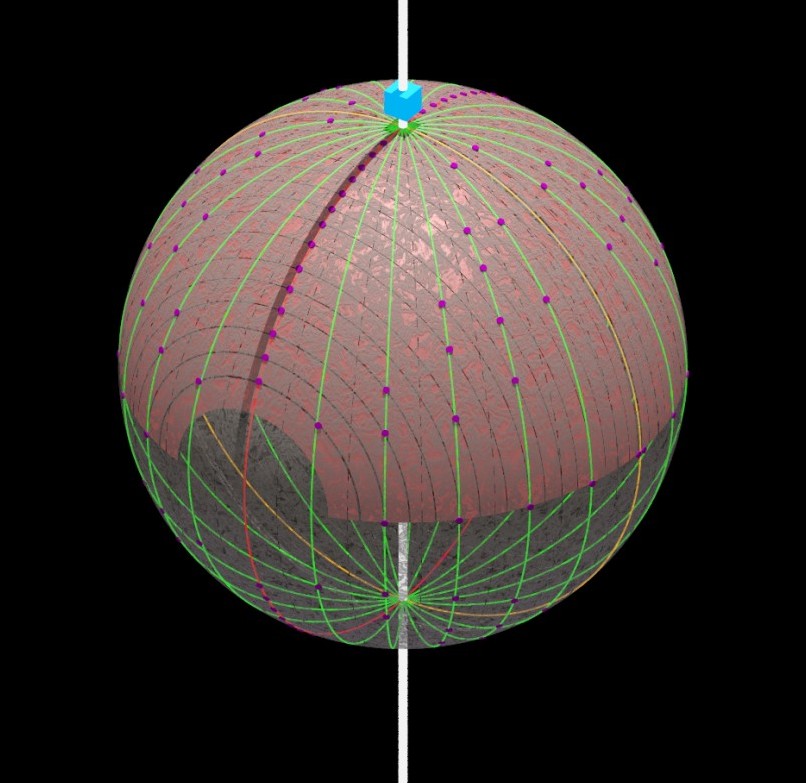}    
\caption{\label{fig:fig22} The dark lines running in the direction of motion are tracks to engage with the motors.}
\end{centering}
\end{figure}

About half the plugs will overlap with the power converter at any one time. On Earth they must be inserted into the tracks, and each plug should engage with one pair of the connecting rails. During deployment, each plug has the task of dragging its associated connecting rails until they join the adjacent segment of track as in Figure \ref{fig:fig21}. The plugs then release their hold on the connecting rails and proceed on as in Figure \ref{fig:fig8}. The movement is powered by the same motors that drive the power converter, shown in magenta in Figure \ref{fig:fig22}.

In Figure \ref{fig:fig22} (a copy of Figure \ref{fig:fig10}) there are dashed lines running midway between the rods. These are the fold lines. On Earth, each segment of the power converter, along with the membrane and solar cells, will be folded outward and to the right, thus making a concave fold at each rod and a convex fold along the fold lines. The motors (shown in magenta) are fixed to the rods. Plugs and their associated tracks will be needed adjacent to the toothed tracks and connecting rails to maintain the connection between the motors and the power converter, because otherwise the toothed wheels and tracks will tend to push each other apart, particularly when the power converter is hanging at the top of the sphere.

On each side of a convex fold midway between two rods, the tracks can be arranged so that, as the sphere opens out, the tracks come together and line up, closing the gap between them. On a concave fold at a rod, there will have to be space between the tracks, and the technique of dragging connecting rails to close the gap will be needed. On Earth, the solar power system (membrane, solar cells and power converter) will be wrapped around the rods, which will be straight and will surround the tether. Thus half of the solar power system’s area will be convex and the other half concave. 

\begin{figure}
\begin{centering}
\includegraphics[width=\textwidth]{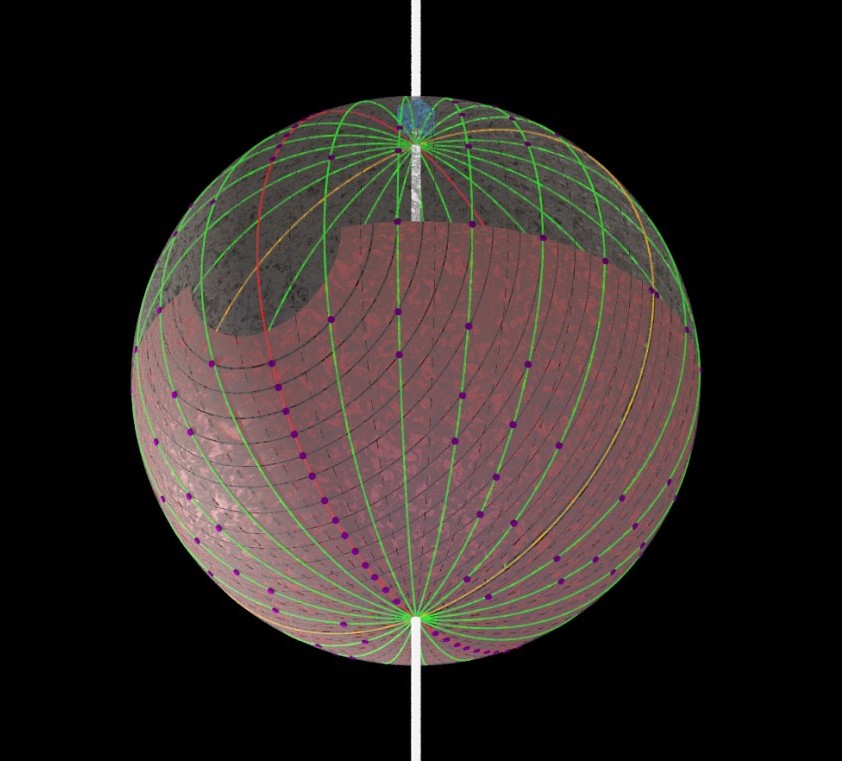}    
\caption{\label{fig:fig23} Position of power converter during deployment, viewed from below}
\end{centering}
\end{figure}

The power converter will start at the lowest position in the sphere – the midnight position – so that it is resting on the rods and membrane to ensure a good connection during the deployment process (Figure \ref{fig:fig23}). At dawn, half of it will receive sunlight, which will be more than enough to power its motors.

\FloatBarrier

\section{Initial preparation on Earth’s surface}

The solar power system will be folded below the climber when it starts its ascent. The folded, flattened sphere will be 198 meters high (about 650 feet), which presents logistical problems. A tower that high would be a substantial construction project in its own right, especially at sea. One idea to mitigate the problem is to pull down the tether by 200 meters or more and lay it out flat so that it can be threaded through the climber and the solar power system.

To do this, one method is to split the membrane that forms the sphere to allow it to be attached to the tether without having to thread the tether through it. This has already been proposed for the power converter. The split membrane will then need to be joined together again. One way of doing this is to use a Velcro fastener. Heavy duty brands on the market can hold as much as 15 lbs/sq in, which is over 100 \si{kN/m^2}. To hold a force of 10 kN/m, we need 10 cm of overlapping Velcro on each side of the join. At geostationary altitude, the fastening can be undone to remove the solar power system from the tether.

\begin{figure}[h]
    \begin{minipage}[t]{.5\textwidth}
        \centering
        \includegraphics[width=.85\linewidth]{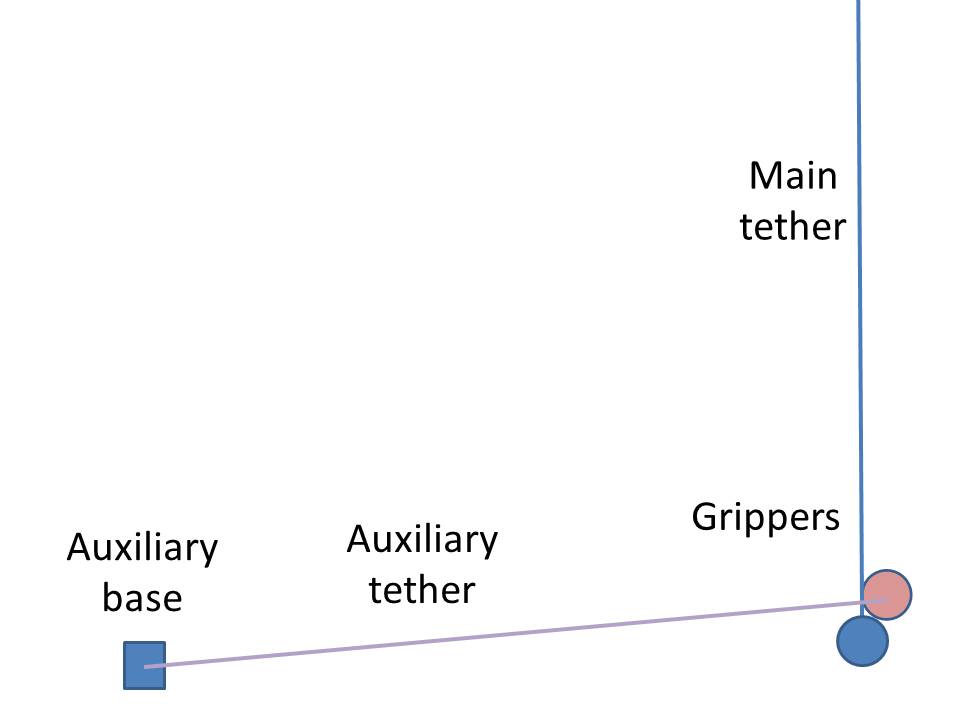}    
        \captionsetup{justification=centering}
        \caption{Auxiliary gripper climbs up the main tether}
        \label{fig:fig24}
    \end{minipage}
    \hfill
    \begin{minipage}[t]{.5\textwidth}
        \centering
        \includegraphics[width=.85\linewidth]{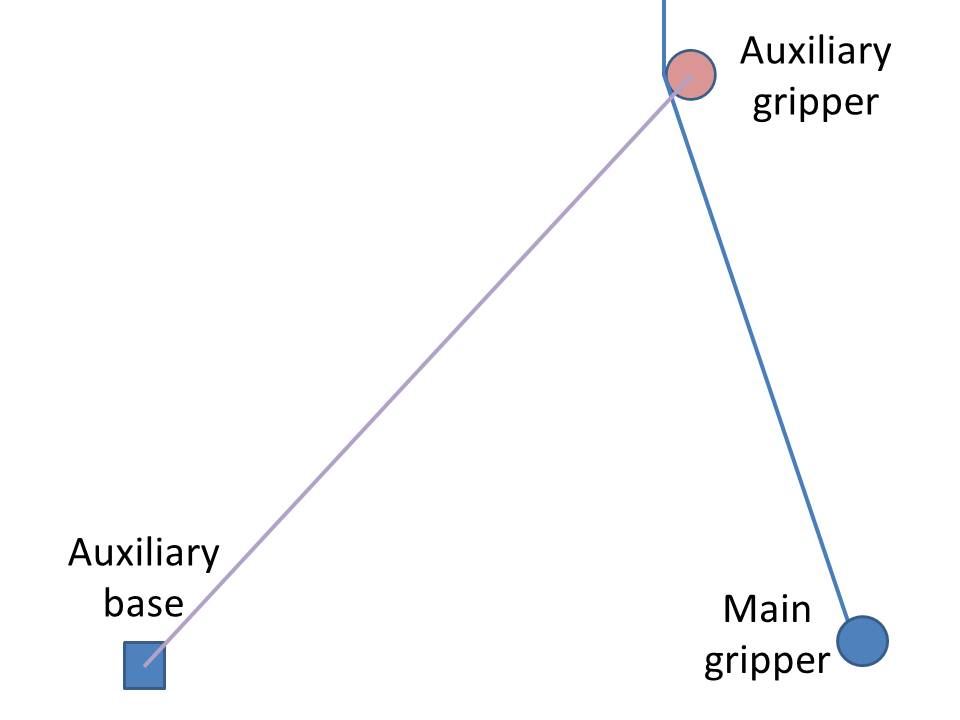}
        \captionsetup{justification=centering}
        \caption{The auxiliary base pulls the gripper and tethers}
        \label{fig:fig25}
    \end{minipage}
\end{figure}

\begin{figure}
    \begin{minipage}[t]{.5\textwidth}
        \centering
        \includegraphics[width=.85\linewidth]{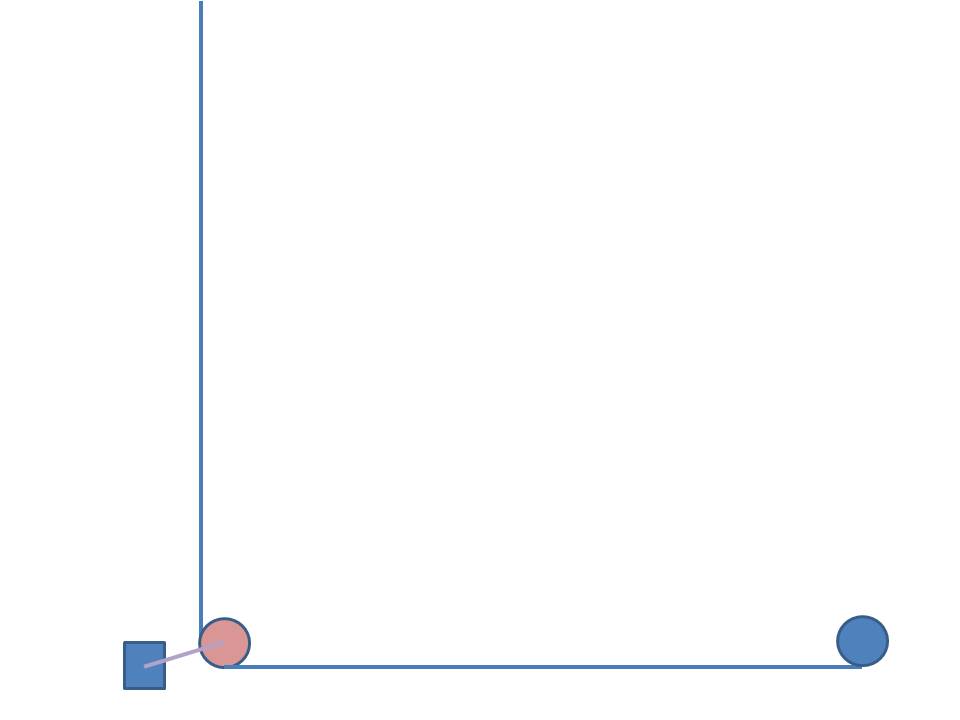}
        \captionsetup{justification=centering}
        \caption{More than 200 meters of main tether are horizontal}
        \label{fig:fig26}
    \end{minipage}
    \hfill
    \begin{minipage}[t]{.5\textwidth}
        \centering
        \includegraphics[width=.85\linewidth]{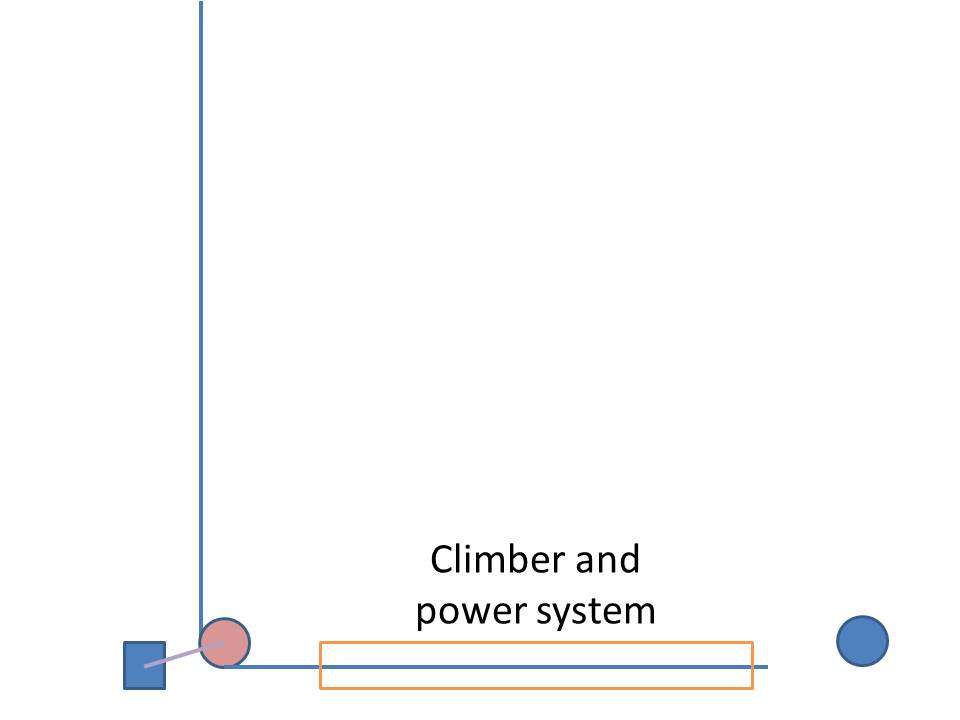}
        \captionsetup{justification=centering}
        \caption{Detach main tether from main gripper; auxiliary gripper holds it; thread climber and power system onto it}
        \label{fig:fig27}
    \end{minipage}
    \begin{minipage}[t]{.5\textwidth}
        \centering
        \includegraphics[width=.85\linewidth]{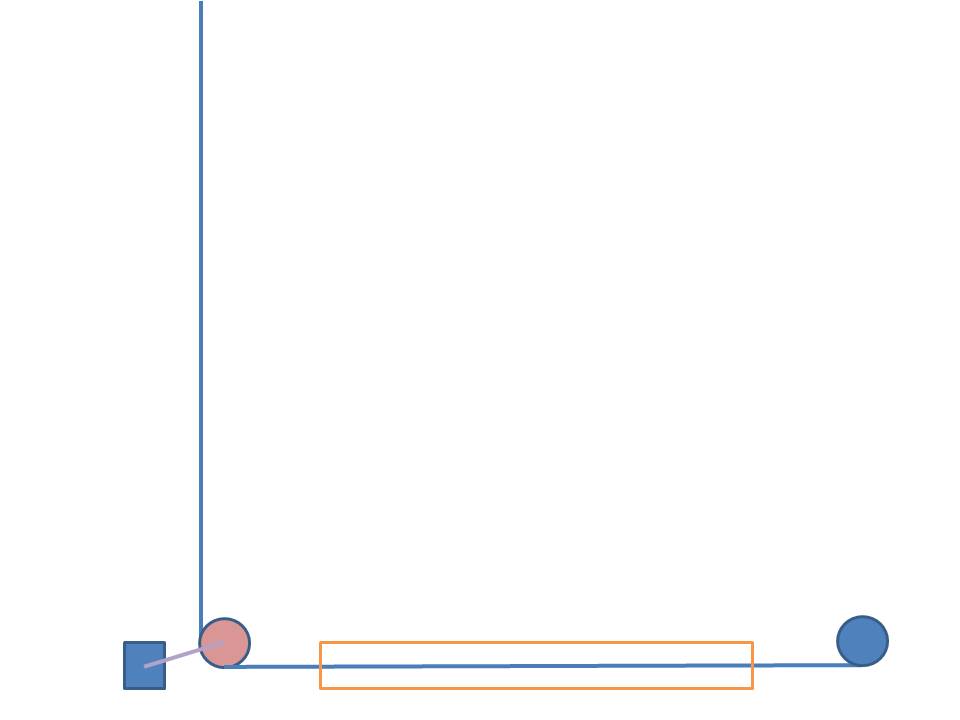}
        \captionsetup{justification=centering}
        \caption{Rejoin main tether to main gripper}
        \label{fig:fig28}
    \end{minipage}
     \hfill
    \begin{minipage}[t]{.5\textwidth}
        \centering
        \includegraphics[width=.85\linewidth]{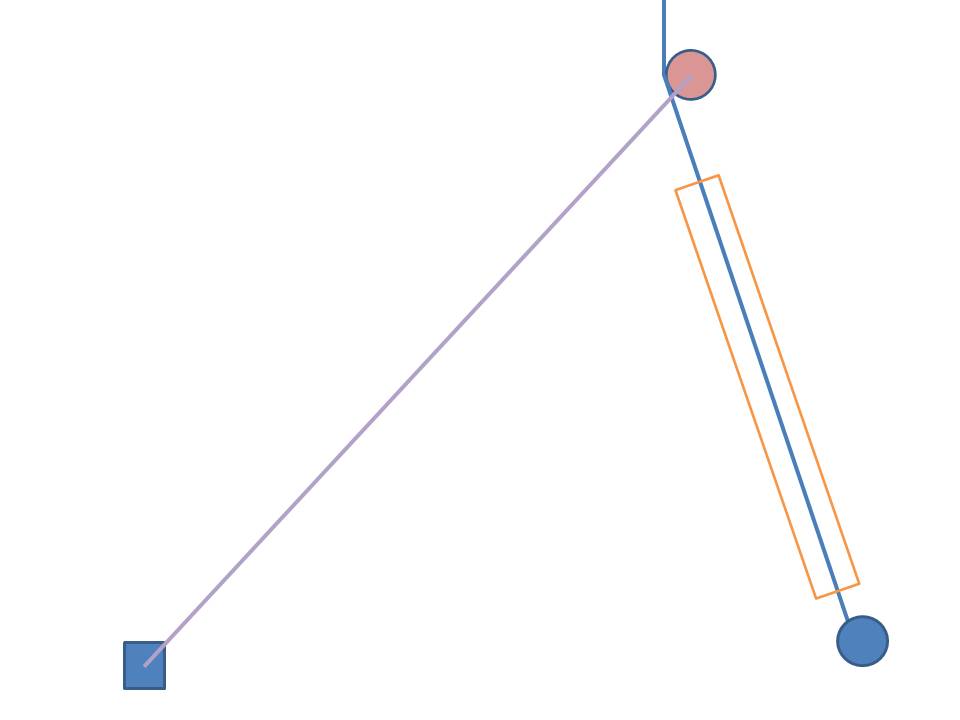}
        \captionsetup{justification=centering}
        \caption{Auxiliary base lets auxiliary gripper rise, which restores main tether near to vertical}
        \label{fig:fig29}
    \end{minipage}
    \begin{minipage}[t]{.5\textwidth}
        \centering
        \includegraphics[width=.85\linewidth]{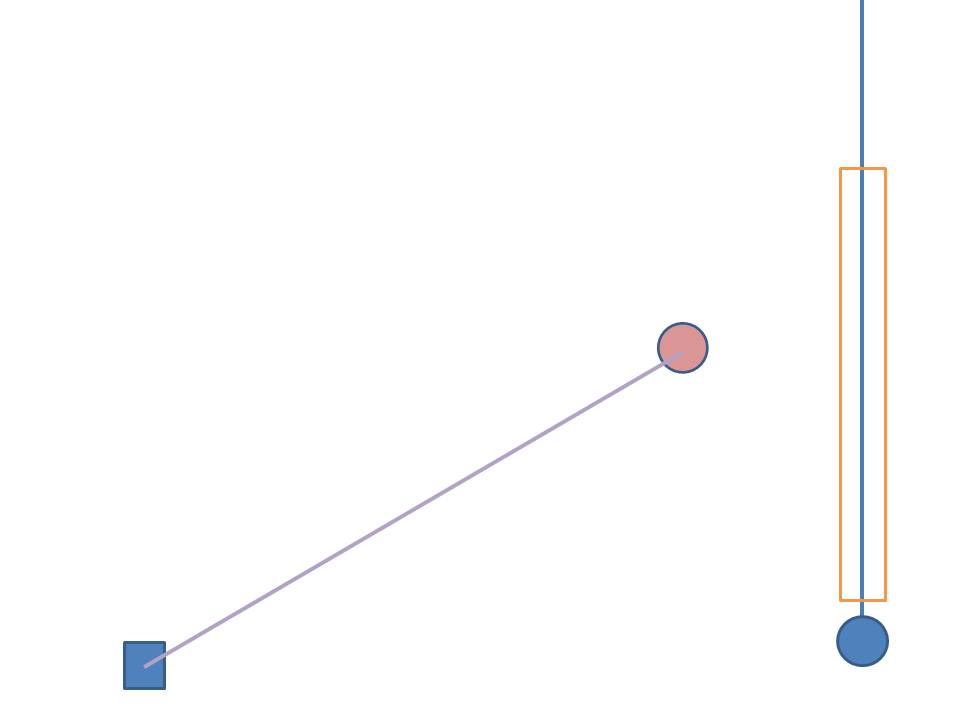}
        \captionsetup{justification=centering}
        \caption{Auxiliary gripper releases tether and falls; climber is free to climb}
        \label{fig:fig30}
    \end{minipage}
\end{figure}

An alternative method is to use an auxiliary gripper (see Figure \ref{fig:fig24}) that climbs at least 200 meters up the tether while attached to the auxiliary base. The auxiliary base (Figure \ref{fig:fig25}) pulls the auxiliary gripper until they are close together and a section of the main tether is horizontal (Figure \ref{fig:fig26}). The auxiliary gripper secures the tether so that the right-hand gripper can release it and allow the climber and solar power system to be threaded onto the tether (Figure \ref{fig:fig27}). Next the main tether is reconnected to the main gripper (Figure \ref{fig:fig28}). After that, the auxiliary base allows the auxiliary gripper to rise by slackening the auxiliary tether that connects them (Figure \ref{fig:fig29}). Finally the auxiliary gripper releases its grip on the main tether and falls back down (Figure \ref{fig:fig30}). The climber is now free to ascend with the folded power system.

\FloatBarrier

\section{Other shapes}

The obvious shape for solar panels is a flat plate, which is normal for satellites in microgravity. However, we have to deal with almost full Earth gravity, and so strong struts would be needed to support them, and they have to tilt to follow the sun. Consider a membrane stretched over a ring. It would need a diameter of 126 meters to garner 5 MW of power. A calculation of Euler’s buckling law shows that the ring would need a mass of more than 100 tonnes. [\cite{mascolo2021buckling}]

A cylinder is a somewhat more efficient shape for following the sun than a sphere (by a factor $4/\pi$), but it has the drawback that a heavy beam would be needed to maintain the straight horizontal axis.

An umbrella has a set of rods supporting a membrane under tension, but they typically have struts under compression that push the rods out. A calculation shows that each such strut would need a mass of 1.3 tonnes, whereas the rods rely on the main tether and inner tethers under tension to achieve and maintain their shape. The tension is less than one tonne; tethers with a mass less than 1 kg can sustain that tension.

\section{Calculations for the membrane and rods}

The key idea is to bend the rods while constraining them to conform to the shape of a sphere by appropriate design of the membrane. First we calculate the maximum compression the rods can take without failing and the maximum tension the membrane can take. Then we calculate the minimum tension in the membrane needed to achieve the spherical shape. However, this varies with the distribution of weight, especially the weight of the moving power converter. The next step is to apply very much more tension in the membrane and compression in the rods (within the maxima calculated) so that the diurnal and other variations are very small by comparison, which means that they will distort the sperical shape by only a small amount.

The following calculations are probably not optimal. Although the Euler-Bernoulli theory works well for bent beams, it assumes the bending is slight compared to the length of the beam. In our case, the rods are bent into a semicircle. Kirchhoff’s rod theory dates from the 19th century, but it appears that its application to all but the simplest examples is still the subject of active research. [\cite{batista2017large}] It would be useful to build a simulation of the forces on a rod.

When a rod with radius of cross-section $r$ is bent into a semicircle of radius $R$, the inner half of it (that half nearer to the center of the sphere) is compressed while the outer half is under tension. We can imagine these two halves separated by a semi-cylinder that is neither compressed nor stretched. The innermost part undergoes a compressive strain of $r/R$. The corresponding stress is $Er/R$, where $E$ is Young's modulus of the rod's material. If we add a longitudinal compressive force $C_l$ this is added to the other forces, and so the stress in the innermost part is

$$\sigma=\frac{Er}{R}+\frac{C_l}{\pi r^2}$$

The maximum load $C_l=C_{max}$ that a rod can bear is

$$C_{max}=\pi\ r^2\left(\sigma_{max}-\frac{Er}{R}\right)$$

For CFRP, $\sigma_{max}=570$ \si{MPa} and E=70 GPa. Take $r=3$ mm and $R=63$ meters. Then

$$C_{max}=28.27\times{10}^{-6}\left(5.7\times{10}^8-3.33\times{10}^6\right)\approx161\times{10}^2 \ \si{N}$$
This is 16.1 kN, so each rod can bear a load of 1.6 tonne force, much more than is needed for this application. Each rod is $\pi R$ long, so its mass is $\rho_r\pi r^2R$, where the density $\rho_r=1.6\times{10}^3$ $\si{kg/m^3}$. This comes to $8954\times{10}^{-3} \ \si{kg}$, which is approximately 9 kg. 24 rods will have a mass of 215 kg.

 The maximum tension in the membrane per unit length along a rod must satisfy the constraint $T_{max}\le\sigma_m\tau$, where $\sigma_m$ is the tensile strength and $\tau$ is the thickness. Taking $\sigma_g=8\times{10}^{10}$ \si{Pa} and $\tau={10}^{-6}$ \si{m} for the graphene layer, we require $T_{max}\le8\times{10}^4$, about 80 kN per meter. This is far more than the weight per meter on a rod, and so it is reasonable to conclude that the membrane’s strength is more than adequate to maintain the shape very close to spherical.

 Having established the maximum forces, we examine what is needed to support the weights of the rods, membrane, photo cells and power converter.

A rod will be subject to forces at each end, $F_t$ pulling up (positive sign) or pulling down (negative sign) from the top and $F_b$ pushing up from below. As well as causing bending, these forces must support the weight of the rod and of the adjacent part of the membrane, including the perovskite cells. The power converter will also need to be supported. Finally, the tension in the membrane will affect the bending of the rod. We will be able to control the tension in the membrane by designing it to fit more loosely or tightly in different places, and this will give some control over the ultimate shape of the rod, which needs to be close to a sphere.

We can calculate the weight of a rod per unit length as $\rho_rg\pi r^2$, where $g$ is the acceleration due to gravity. We find $\rho_rg\pi r^2=1.6\times{10}^3\times9.81\pi9\times{10}^{-6}\approx444$ \si{mN} near Earth where $g\approx9.81$ $\si{m/s^2}$.

From Table \ref{tab:pskiteSE} we see that the solar cells have a mass per unit area of approximately $3\times{10}^{-3}$ \si{kg/m^2}, and the graphene membrane is similar, making a total density of $\rho_M=6\times{10}^{-3}$ $\si{kg/m^2}$. The width of membrane supported by a rod is $2\pi R/n=\sfrac{396}{24}=16.5$ m at the widest point, where n=24 is the number of rods. Elsewhere along the rod it follows a sine rule, and so it is $(2\pi R/n) \sin{(s/R)}$, where $s$ is the displacement along the rod, $0\le s\le\pi R$. The weight due to the membrane and solar panels on a rod per unit distance along the rod is  $\rho_Mg(2\pi R/n)\sin{(s/R)}\approx(0.97)\sin{(s/R)}$.

Thirdly we have the moving weight of the power converter. The function $V(s,t,\varphi)$ represents the part of its weight that one rod supports; the weight moves inside the sphere during the day. It is a function of position $s$ on the rod, time of day $t$ in hours and the angle $\varphi$ of the rod with respect to the afternoon side of the plane in the sphere parallel to Earth's equator (shown in orange in Figure 10). On the afternoon side of the sphere we find

$$V_a\ (s,t,\varphi)=\Pi\left(\frac{s}{\pi R}+\frac{(12-t)}{12}\right)\Pi\left(\frac{\varphi}{\pi}\right)W_C\sin{\frac{s}{R}}\ \ \ \ \ \ \ \ \ \ \ \ \ \ \ \ for\ -\frac{\pi}{2} \leq \varphi \leq \frac{\pi}{2} $$

Here $\Pi\left(x\right)$ is the rectangle function, also known as Dirichlet’s discontinuous factor, such that $\Pi\left(x\right)=1$ when $-{\frac{1}{2}}<x<{\frac{1}{2}}$ and $\Pi\left(x\right)=0$ otherwise. $W_C$ is the weight per rod per meter of the power converter at the middle circle, i.e., the circle midway between the top and bottom where $s=\pi R/2$. $W_C$ will be maximal near the rods that align with Earth's equator (orange in the figures) and will diminish nearer the line of the north-south poles. For simplicity, it is taken here as a constant such that it sums to the overall weight of the power converter, which is 240 kg force, about $w_C=2.35$ kN. The weight per rod per meter for any value of $s$ is $W_C\sin{\frac{s}{R}}$, with $W_C=2w_C/n\pi R$. The factor of 2 arises because only half the sphere has this weight at any time. Thus $W_C\approx1.0$.

The expression for the morning side is

$$V_m\left(s,t,\varphi\right)=\Pi\left(\frac{s}{\pi R}+\frac{t-12}{12}\right)\Pi\left(\frac{\varphi}{\pi}-1\right)W_C\sin{\frac{s}{R}}\ \ \ \ \ \ \ \ \ \ \ \ \ \ \ \ for\ \frac{\pi}{2}\le\varphi\le\frac{3\pi}{2}$$

We can consolidate these into a single expression for the weight distribution of the power converter
\begin{equation}
V\left(s,t,\varphi\right)=\left[\Pi\left(\frac{s}{\pi R}+\frac{12-t}{12}\right)\Pi\left(\frac{\varphi}{\pi}\right)+\Pi\left(\frac{s}{\pi R}+\frac{t-12}{12}\right)\Pi\left(\frac{\varphi}{\pi}-1\right)\right]W_C\sin{\frac{s}{R}}
\label{eq:vstphi}
\end{equation}
The sum of the rectangle functions in square brackets is either zero or one depending on position and time of day.

In summary, the distributed forces on each rod due to weight are

\begin{equation}
w\left(s,t,\varphi\right)=\rho_rg\pi\ r^2+\ \rho_Mg\frac{2\pi R}{n}\sin{\frac{s}{R}}+V(s,t,\varphi)
\label{eq:wstphi}
\end{equation}
The sum of these weights per unit length of a rod is greatest when $s=\pi R/2$, i.e., at the middle circle. It varies between $0.444+0.971\approx1.5$ and $0.444+0.971+1.0\approx2.5$ \si{N/m}.
\\

The tension in the membrane also acts as a distributed force on each rod, but it acts in the direction normal to the rod, whereas the weights act vertically. The membrane will be designed to keep the curvature of the rod close to a constant $1/R$. We can control the tension by arranging for different parts of the membrane to be slightly slacker or tighter.

The equations for a slender rod with a large bend are below. [\cite{lautrup2011physics}]

$$EI\frac{d\theta}{ds}=-M$$

Here $E$ is Young's modulus of the rod material, $I=\pi r^4/4$ is the second moment of area, $r$ is the rod's radius, $\theta$ is the slope of the rod with respect to the z axis (the vertical) as a function of the displacement $s$ along the rod, $M$ is the bending moment, and $\frac{d\theta}{ds}$ is the curvature. This equation is the equivalent of the conventional equation for bending a beam $EI\frac{d^2x}{dy^2}=M$
However, that equation assumes that the gradient $\frac{dy}{dx}$ is small, which is not true in the case of a rod bent into a semicircle. Differentiating gives the following.

\begin{equation}
EI\frac{d^2\theta}{ds^2}=-F_y\cos{\theta}+F_z\sin{\theta}
\label{eq:EI}
\end{equation}

\begin{figure}[h]
    \begin{minipage}[t]{.35\textwidth}
        \centering
        \includegraphics[width=\textwidth]{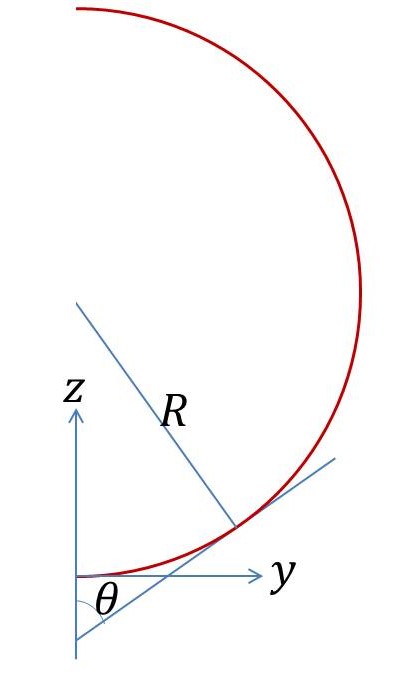}   
        \captionsetup{justification=centering}
        \caption{Angle $\theta$ with respect to the z coordinate}
        \label{fig:thetaz}
    \end{minipage}
    \hfill
    \begin{minipage}[t]{.5\textwidth}
        \centering
        \includegraphics[width=\textwidth]{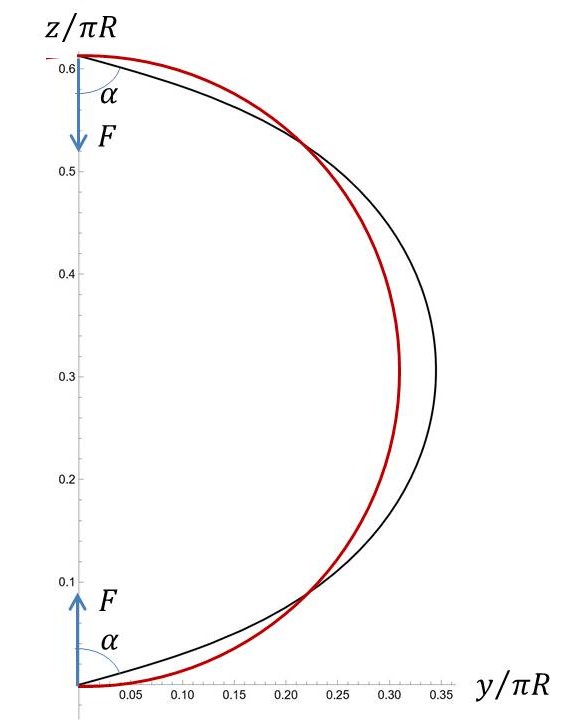}   
        \captionsetup{justification=centering}
        \caption{Plot of the shape of the rod under bending due to forces top and bottom}
        \label{fig:rodshape}
    \end{minipage}
\end{figure}

$F_y$ and $F_z$ are forces applied in the y direction and z direction, respectively.  See Figure \ref{fig:thetaz}. In the case of symmetrical forces $F$ at the top and bottom with $F_z=-F$ and $F_y=0$, this equation has the solution

\begin{equation}
y=\sqrt{\frac{EI}{F}2\left(\cos{\theta}-\cos{\alpha}\right)}\ \ \ \ \ \ \ \ \ \ \ \ \ \ \ \ \ \ \ z=\sqrt{\frac{EI}{F}}\int_{\theta}^{\alpha}{\frac{\cos{\theta\prime}}{\sqrt{2\left(\cos{\theta\prime}-\cos{\alpha}\right)}}d\theta\prime}
\label{eq:yz}
\end{equation}

It is shown in Figure \ref{fig:rodshape} with $\alpha=74.5 \degree$. The desired semi-circular shape is in red. We rely on the tension in the membrane to achieve a close approximation to it.

The force $F$ is provided by a tether connecting the top of each rod to the bottom. From equation \ref{eq:yz} we can see that it is approximately given by $F=EI/z^2$. If z=63 meters, $E={10}^{10} \, \si{Pa}$, $I=\pi r^4/4$, and r=3 mm then

$$F=\frac{{10}^{10}\pi81\times{10}^{-12}}{4\times{63}^2}\approx0.016 \, \si{N}$$

This is a tiny force, dwarfed by the weight of the rod, membrane, solar cells and power converter, represented in the function $w\left(s,t,\varphi\right)$ given in equations \ref{eq:vstphi} and \ref{eq:wstphi}. These weights will tend to squash and buckle the rod, an effect which is opposed by the tension in the membrane. The idea of using membranes this way is not new [\cite{alpermann2013restraining}], but the bending here is more extreme. For equilibrium, the sum of the forces at the top and bottom must be equal to the sum of the weights, and their difference will be similar to this calculated value of $F$. In the desired case of a semicircle, we have $\theta=s/R-\frac{\pi}{2}$. In addition, the tension in the membrane applies orthogonally to the rod. It should balance the tendency of the weights to cause too much bending. For this reason, the force $T\left(s,t,\varphi\right)$ on the rod due to tension should also be a function of the same variables as $w\left(s,t,\varphi\right)$.

Both $T\left(s,t,\varphi\right)$ and $w\left(s,t,\varphi\right)$ are distributed forces, whereas the forces in equation \ref{eq:EI} are point forces. Rather than differentiate again we introduce $T^\prime$ such that
\begin{equation}
\frac{dT^\prime}{ds}=T\left(s,t,\varphi\right)
\label{eq:eqn5}
\end{equation}

Later we will do something similar for $w\left(s,t,\varphi\right)$. $T^\prime$ acts towards the center of curvature. From equation \ref{eq:EI} we have

\begin{equation}
EI\frac{d^2\theta}{ds^2}=-F_y\cos{\theta}+F_z\sin{\theta}-T^\prime
\label{eq:EIdtheta}
\end{equation}

$F_y$ represents a lateral force exerted on the rod at the top and bottom. It balances the lateral component of the tension. The vertical components of the tension are nearly equal and opposite. They are much greater than the weight function $-w\left(s,t,\varphi\right)$ which is the weight per unit distance along the rod, defined in equations \ref{eq:vstphi} and \ref{eq:wstphi}. Similarly to $T^\prime$ it can be written

\begin{equation}
\frac{dF_z}{ds}=-w\left(s,t,\varphi\right)=A_w+B_w\sin{\frac{s}{R}}
\label{eq:dFds}
\end{equation}

$A_w=-\rho_rg\pi r^2\approx-0.45$, and $-1.0\gtrsim B_w\gtrsim-2.0$ incorporates the weight of the power converter.

\begin{equation}
B_w=-\rho_Mg\frac{2\pi R}{n}-W_C\left[\Pi\left(\frac{s}{\pi R}+\frac{12-t}{12}\right)\Pi\left(\frac{\varphi}{\pi}\right)+\Pi\left(\frac{s}{\pi R}+\frac{t-12}{12}\right)\Pi\left(\frac{\varphi}{\pi}-1\right)\right]
\label{eq:B}
\end{equation}

The sine in equation \ref{eq:dFds} reflects the varying width of the membrane along the rod’s length. Integrate equation \ref{eq:dFds}.

\begin{equation}
F_z=\int\left(A_w+B_w\sin{\frac{s}{R}}\right)ds=A_ws-RB_w\cos{\frac{s}{R}}+C
\label{eq:Fzint}
\end{equation}

$C$ is a constant of integration. See “Integration by Parts” below for the justification for treating $B_w$ as constant in the case of a movable power converter. It is obviously constant for a fixed power converter. When $s=0$, $F_z=F_t$, which is the force at the top, and $F_t=-RB_w+C$. Thus $C=F_t+RB_w$. Hence 

\begin{equation}
F_z=A_ws+RB_w\left(1-\cos{\frac{s}{R}}\right)+F_t
\label{eq:FzAs}
\end{equation}

The differential equation \ref{eq:EIdtheta} can be solved numerically if we simplify the tension function so that

$$\frac{dT^\prime}{ds}=T_c\frac{d\theta}{ds}-F_c$$

Here $T_c$ and $F_c$ are constant. We can integrate to obtain

$$T^\prime=T_c\theta-F_cs+D$$

$D$ is a constant of integration. Hence, from equations \ref{eq:EIdtheta} and \ref{eq:FzAs} the differential equation is

$$EI\frac{d^2\theta}{ds^2}=-F_y\cos{\theta}+\left[\left(A_ws+RB_w\left(1-\cos{\frac{s}{R}}\right)+F_t\right)\right]\sin{\theta}+F_cs-T_c\theta-D$$

To determine the constant of integration $D$, note that, when $s=0$, $\theta=\frac{\pi}{2}$ and $\frac{d^2\theta}{ds^2}=0$. Thus $D=F_t-\frac{\pi}{2}T_c$, and the equation becomes

\begin{equation}
EI\frac{d^2\theta}{ds^2}=-F_y\cos{\theta}+\left[\left(A_ws+RB_w\left(1-\cos{\frac{s}{R}}\right)+F_t\right)\right]\sin{\theta}+F_cs-T_c\left(\theta-{\frac{\pi}{2}}\right)
\label{eq:eqn12}
\end{equation}

In equation \ref{eq:eqn12}, the force $F_y$ comes from the top of the rod where it meets the ring surrounding the tether. This ring will be subject to compressive forces due to all of the rods but will also have compensating tension from the membrane. The force $F_c$ represents compressive forces in the rod due to tension in the membrane. 

\begin{figure}[h]
    \begin{minipage}[t]{.5\textwidth}
        \centering
        \includegraphics[width=\textwidth]{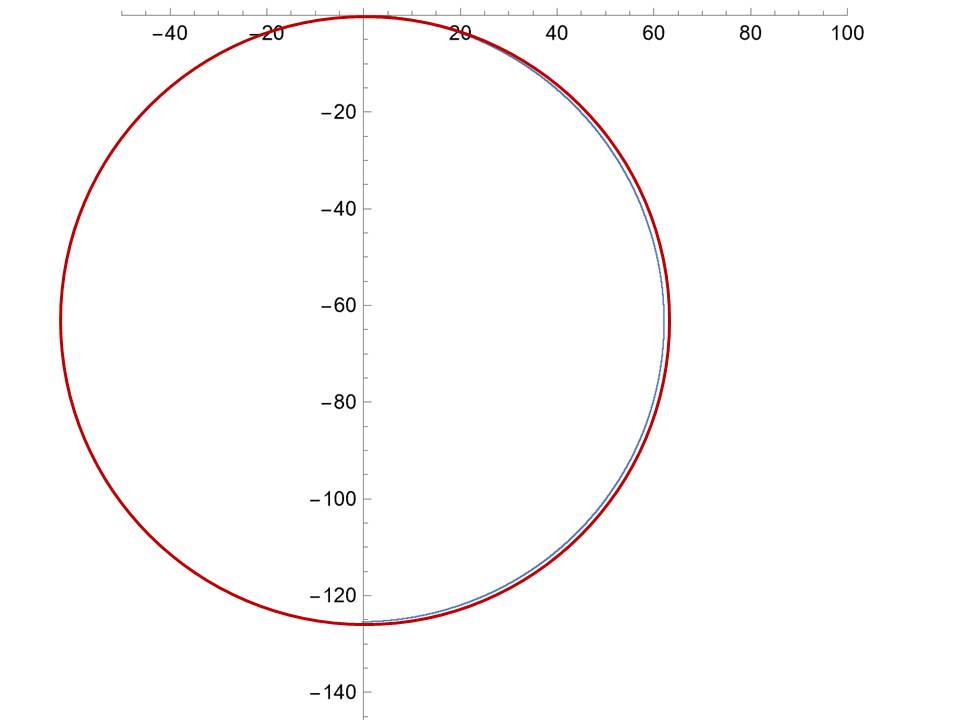}   
        \captionsetup{justification=centering}
        \caption{Shape of rod in blue with a reference circle in red}
        \label{fig:F1-new2}
    \end{minipage}
    \hfill
    \begin{minipage}[t]{.5\textwidth}
        \centering
        \includegraphics[width=\textwidth]{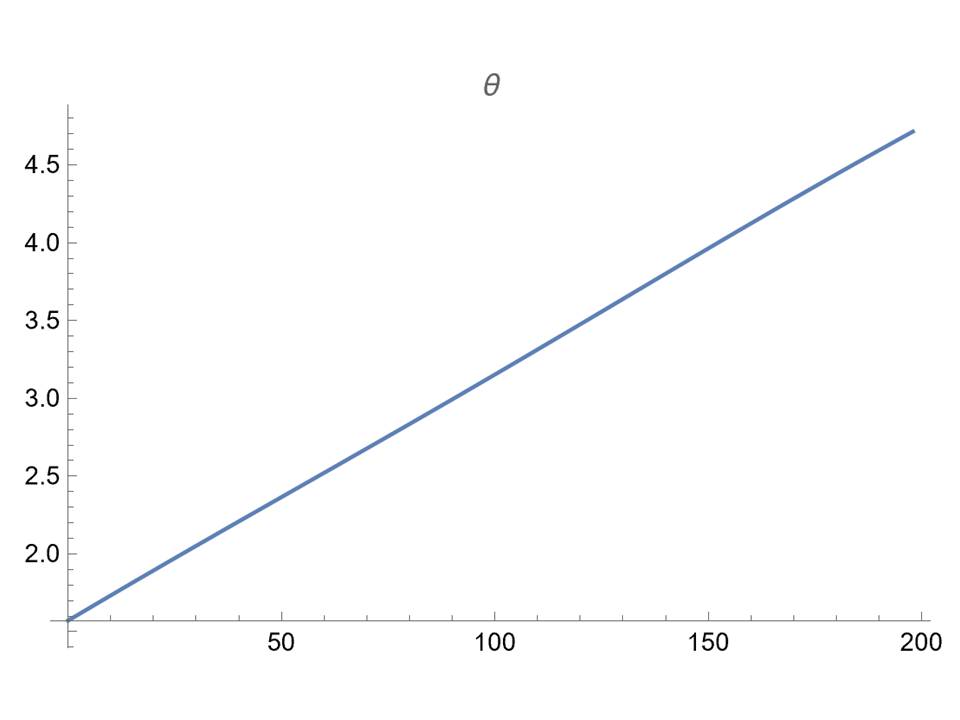}   
        \captionsetup{justification=centering}
        \caption{Slope of the rod (radians) against distance along it from zero to ${\pi R}\approx{198}$ meters}
        \label{fig:F2-new2}
    \end{minipage}
    \begin{minipage}[t]{.5\textwidth}
        \centering
        \includegraphics[width=\textwidth]{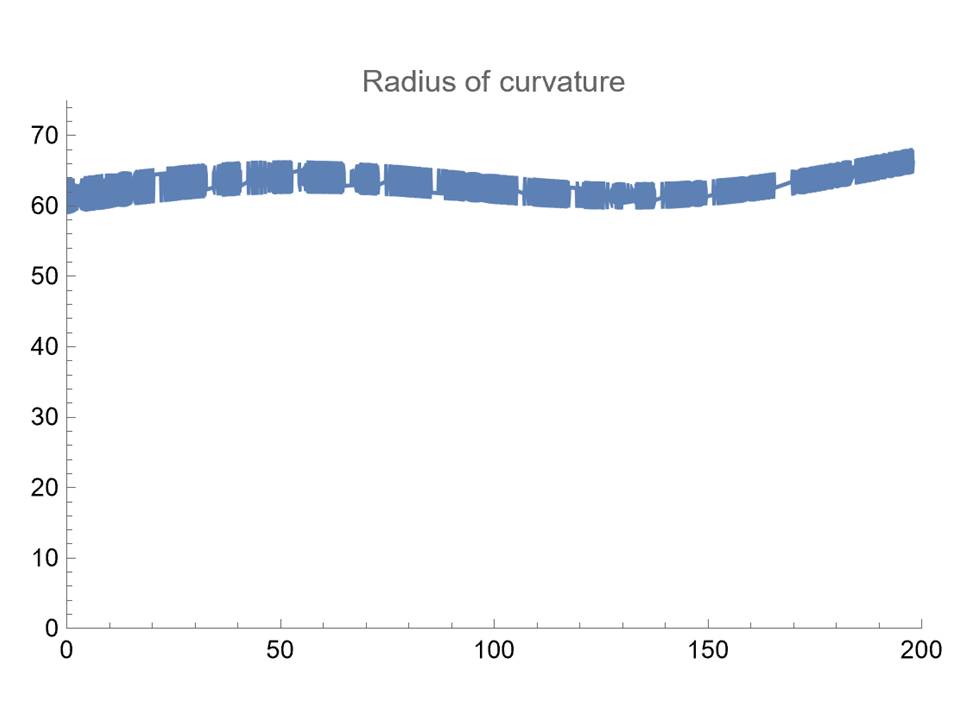}   
        \captionsetup{justification=centering}
        \caption{Radius of curvature $\frac{ds}{d\theta}$}
        \label{fig:F3-new2}
    \end{minipage}
    \hfill
    \begin{minipage}[t]{.5\textwidth}
        \centering
        \includegraphics[width=\textwidth]{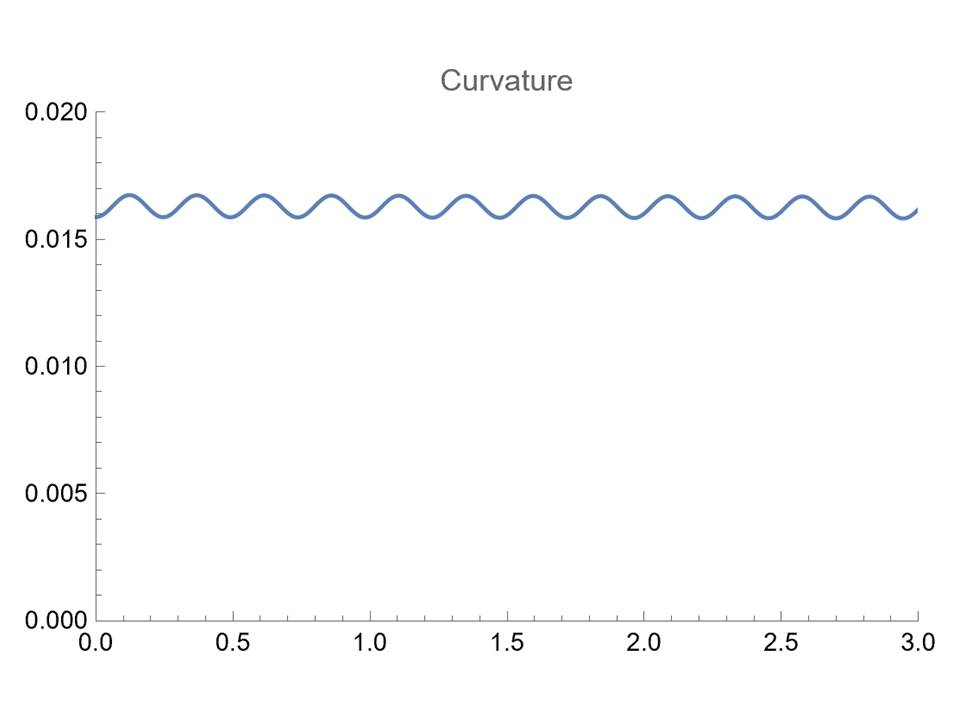}   
        \captionsetup{justification=centering}
        \caption{Detail of the curvature $\frac{d\theta}{ds}$}
        \label{fig:F4-new2}
    \end{minipage}
\end{figure}

This equation can be solved numerically with the result shown in Figure \ref{fig:F1-new2}. The ideal shape (in red) has a width of 63 meters, which is the target radius. The values used are $F_c=48$ N, $F_t=100$ N, $F_y=150$ N and T=3066.6 N. The shape plotted from equation \ref{eq:eqn12} has a slightly smaller width of 62.14 meters. The shape produced when the power converter is not present is indistinguishable at this scale. Its width is 61.92 meters, a difference of 22 cm. The reason is that the tension and other forces are considerably greater than the weight of the power converter.

Figure \ref{fig:F2-new2} shows the slope $\theta$ of the rod as a function of the distance $s$ along it. It goes through an angle of $\pi$ radians from $\frac{\pi}{2}$ to $\frac{3\pi}{2}$ (90° to 270°). This slope is used to calculate the plot in Figure \ref{fig:F1-new2}. Figure \ref{fig:F3-new2} shows the radius of curvature, which is the reciprocal of the curvature $\frac{d\theta}{ds}$. The average is 62.95 meters, very close to the desired radius of 63 meters. A small standing wave is present, and Figure \ref{fig:F4-new2} shows the detail of the curvature $\frac{d\theta}{ds}$ for the first three meters with the standing wave clearly visible. The wavelength is 25 cm.

\begin{figure}
\begin{centering}
\includegraphics[width=.7\textwidth]{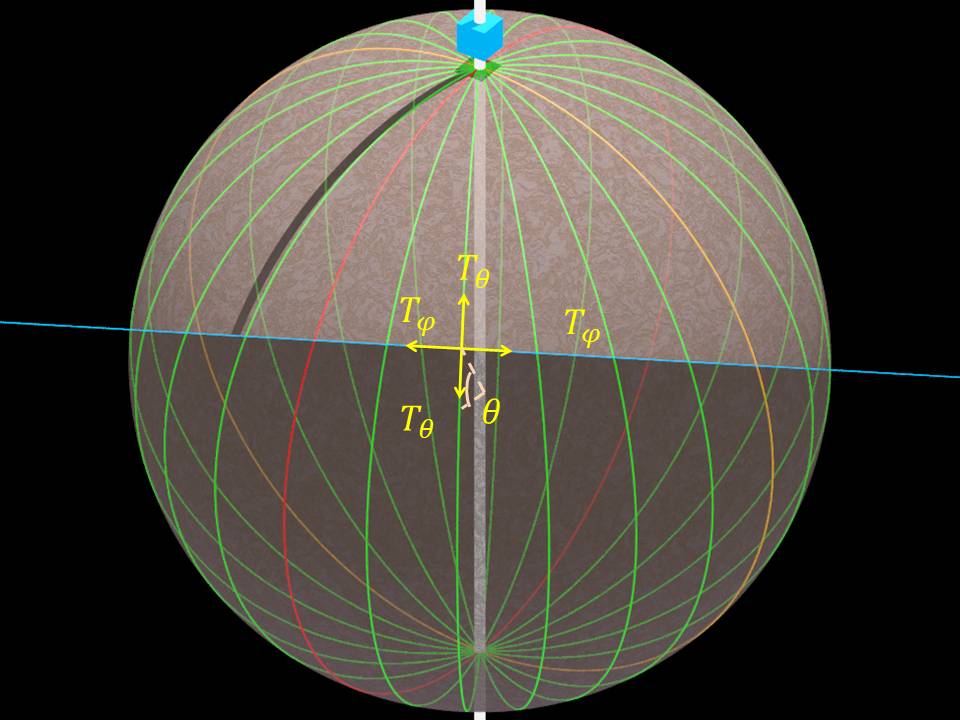}    
\caption{\label{fig:F5-new2} Sphere with a great circle}
\end{centering}
\end{figure}

Tension in a membrane is measured in Newtons per meter in two orthogonal directions. These components are $T_\theta$ along the length of a rod and $T_\varphi$ orthogonal to it. The result is an inward force proportional to the curvature at each point of a rod and directed towards the center of curvature. Provided we can maintain a shape close to a sphere, the center of curvature will be at the center of the sphere for both $T_\theta$ and $T_\varphi$. Thus the curvatures act towards the center of the great circles, one formed by the rod and its opposite rod and the other by the plane orthogonal to it as illustrated in Figure \ref{fig:F5-new2}.

The rods cause the membrane to bend at the points of contact rather than forming a continuous spherical surface, and so the action of $T_\varphi$ will be concentrated at each rod and will apply in proportion to the angle in the membrane there, which is $\frac{2\pi}{n}\sin{\frac{s}{R}}$ in radians, where $n=24$ is the number of rods. In degrees, this angle has a maximum value of 15°. Therefore the effective force of $T_\varphi$ on the rod will be $\frac{2\pi}{n}T_\varphi\sin{\frac{s}{R}}$. The action of $T_\theta$ applies only to the area of contact with the rod, which is $\frac{2\pi r}{n}\sin{\frac{s}{R}}$, $r$ being the radius of the rod. This causes an effective force of approximately $\frac{2\pi r}{nR}T_\theta\sin{\frac{s}{R}}$, which is proportional to the curvature $\sfrac{1}{R}$. The sum of these two quantities is $\frac{2\pi}{n}\left(\frac{r}{R}T_\theta+T_\varphi\right)\sin{\frac{s}{R}}$.

The term $T_c\left(\theta-{\frac{\pi}{2}}\right)$ in equation \ref{eq:eqn12} must be differentiated to give the distributed force per unit distance on a rod due to tension, as in the term $\frac{dT^\prime}{ds}$ which appears in equation \ref{eq:eqn5}. Therefore

\begin{equation}
\frac{dT^\prime}{ds}=T_c\frac{d\theta}{ds}=\frac{2\pi}{n}\left(\frac{r}{R}T_\theta+T_\varphi\right)\sin{\frac{s}{R}}
\label{eq:eqn13}
\end{equation}

Now $\frac{r}{R}\approx4.76\times{10}^{-5}$ is a very small quantity, so we depend on $T_\varphi$ to provide a useful force to counteract the forces in the rods. To a first order, ignoring Poisson’s ratio, if we want a compressive force of 3 kN, we need $\frac{2\pi}{n}T_\varphi=3\times{10}^3$. Therefore $T_\varphi=3\times{10}^3n/2\pi\approx11.46\times{10}^3$, which is just under 11.5 kN/m.

\begin{figure}[h!]
\begin{centering}
\includegraphics[width=.7\textwidth]{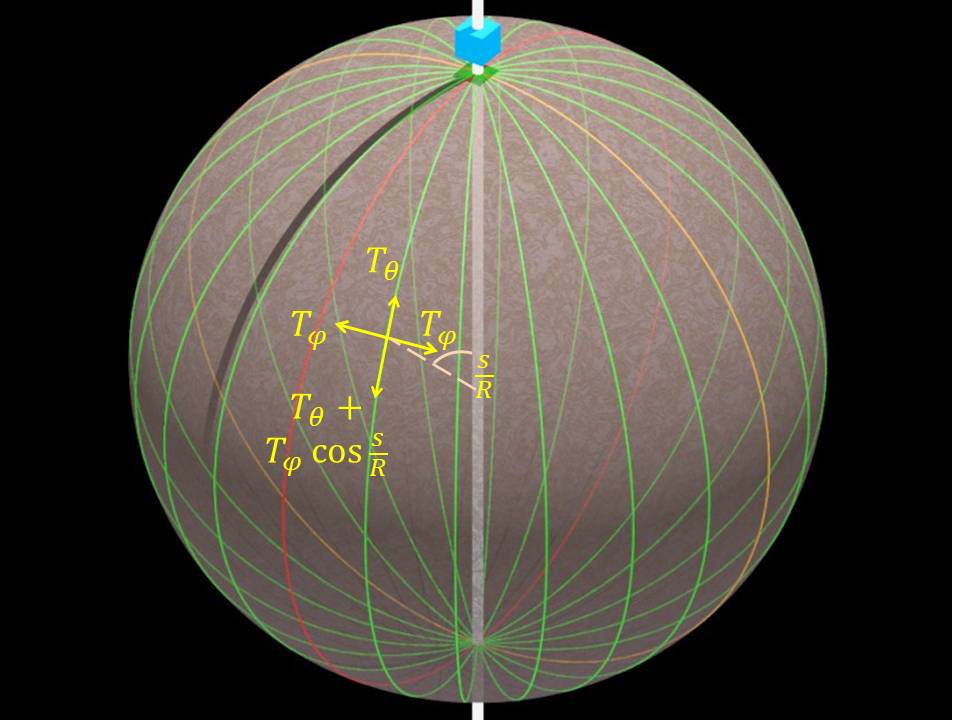}    
\caption{\label{fig:F6-new2} Geometrical relationship between $T_\varphi$ and $T_\theta$}
\end{centering}
\end{figure}

There is a geometric relationship between $T_\theta$ and $T_\varphi$ as illustrated in Figure \ref{fig:F6-new2}. 

The tension $T_\theta$ prevents the membrane from slipping towards the top and bottom. It acts across the circumference $c$ of a circle about the tether, where $c=2\pi R\sin{\frac{s}{R}}$. Thus $T_\varphi$ augments $T_\theta$ according to the following relationship: 

$$c\frac{dT_\theta}{ds}=T_\varphi\cos{\frac{s}{R}}$$

\begin{equation}
\therefore\ T_\varphi=2\pi R\frac{dT_\theta}{ds}\tan{\frac{s}{R}}
\label{eq:eqn14}
\end{equation}

Hence and from equation \ref{eq:eqn13}

$$T_c\frac{d\theta}{ds}=\frac{2\pi}{n}\left(\frac{r}{R}T_\theta+2\pi R\frac{dT_\theta}{ds}\tan{\frac{s}{R}}\right)\sin{\frac{s}{R}}$$

Approximate $\frac{d\theta}{ds}=\frac{1}{R}$.

$$T_c=\frac{2\pi}{n}\left(rT_\theta+2\pi R^2\frac{dT_\theta}{ds}\tan{\frac{s}{R}}\right)\sin{\frac{s}{R}}$$

\begin{figure}
    \begin{minipage}[t]{.5\textwidth}
        \centering
        \includegraphics[width=\textwidth]{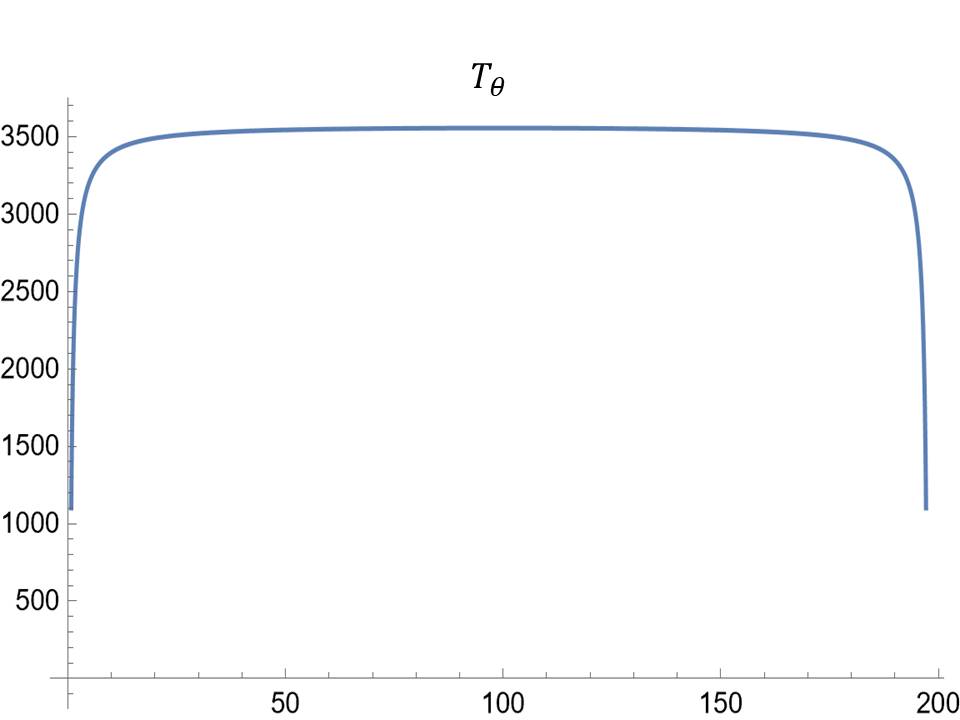}   
        \captionsetup{justification=centering}
        \caption{Plot of tension $T_\theta$, N, versus distance along a rod}
        \label{fig:F7-new2}
    \end{minipage}
    \hfill
    \begin{minipage}[t]{.5\textwidth}
        \centering
        \includegraphics[width=\textwidth]{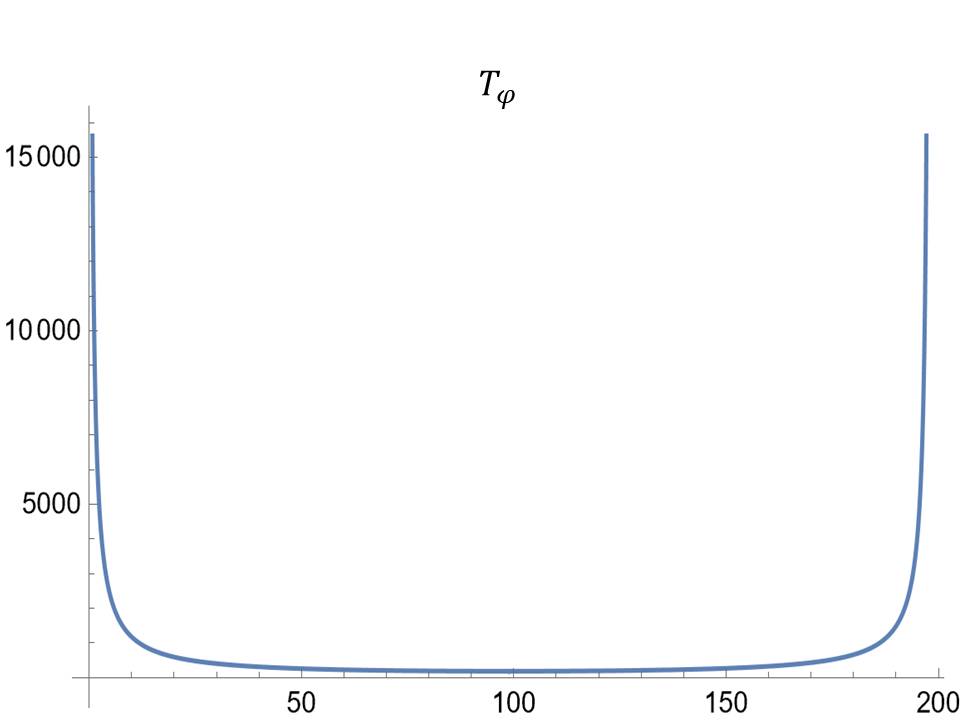}   
        \captionsetup{justification=centering}
        \caption{Plot of tension $T_\varphi$, N, versus distance along a rod}
        \label{fig:F8-new2}
    \end{minipage}
\end{figure}

We can plot $T_\theta$ from this equation and then use equation \ref{eq:eqn14} to obtain $T_\varphi$, with the results shown in Figure \ref{fig:F7-new2} and Figure \ref{fig:F8-new2}. Near the top and bottom, the sum of these orthogonal components of tension is nearly 17 kN. Near the middle the sum is 3.7 kN. These are well within the capability of the graphene membrane one micron thick, which can take 80 kN. This allows flexibility in the design of the sphere. More detailed design work may show that we can make parts or all of the graphene membrane even thinner or incorporate a conductive layer of aluminium as an alternative to cables for the return current plus an insulating layer between it and the perovskite cells. For example, we could reduce the graphene's thickness to 0.5 micron, which is still strong enough for 40 kN, more than double what is needed.

\FloatBarrier

\subsection{Integration by parts}

The justification for treating $B_w$ as constant comes from integration by parts. We have

$$B_w=-\rho_Mg\frac{2\pi R}{n}-W_C\left[\Pi\left(\frac{s}{\pi R}+\frac{12-t}{12}\right)\Pi\left(\frac{\varphi}{\pi}\right)+\Pi\left(\frac{s}{\pi R}+\frac{t-12}{12}\right)\Pi\left(\frac{\varphi}{\pi}-1\right)\right]$$

Clearly, the first term is constant. The derivative of $\Pi(s)$ is $\delta\left(s+{\frac{1}{2}}\right)-\delta\left(s-{\frac{1}{2}}\right)$. Generalize the critical terms by introducing $a=\pm1$ so that we need to integrate $\Pi\left(\frac{s}{\pi R}+a\frac{12-t}{12}\right)\sin{\frac{s}{R}}$. Then

\begin{equation*}
\begin{split}
\int{\Pi\left(\frac{s}{\pi R}+a\frac{12-t}{12}\right)\sin{\frac{s}{R}}ds=\left(\int{\sin{\frac{s}{R}}ds}\right)\Pi\left(\frac{s}{\pi R}+a\frac{12-t}{12}\right)} \\ -\int\left(\int{\sin{\frac{s}{R}}ds}\right)\frac{1}{\pi R}\left(\delta\left(\frac{s}{\pi R}+a\frac{12-t}{12}+\frac{1}{2}\right)-\delta\left(\frac{s}{\pi R}+a\frac{12-t}{12}-\frac{1}{2}\right)\right)ds
\end{split}
\end{equation*}

\begin{equation*}
\begin{split}
=R\cos{\frac{s}{R}}\Pi\left(\frac{s}{\pi R}+a\frac{12-t}{12}\right)- \\ \frac{1}{\pi}\int\cos{\frac{s}{R}}\left[\delta\left(\frac{s}{\pi R}+a+\frac{1}{2}-\frac{at}{12}\right)-\delta\left(\frac{s}{\pi R}+a-\frac{1}{2}-\frac{at}{12}\right)\right]ds*
\end{split}
\end{equation*}

$$=R\cos{\frac{s}{R}}\Pi\left(\frac{s}{\pi R}+a\frac{12-t}{12}\right)-\frac{1}{\pi}\int\left[\cos{\left(\pi\left(a+\frac{1}{2}-\frac{at}{12}\right)\right)}-\cos{\left(\pi\left(a-\frac{1}{2}-\frac{at}{12}\right)\right)}\right]ds$$

$$=R\cos{\frac{s}{R}}\Pi\left(\frac{s}{\pi R}+a\frac{12-t}{12}\right)+\frac{1}{\pi}\int{2\sin{\left(a\pi\left(1-\frac{t}{12}\right)\right)}\sin{\pi}}ds$$


But $\sin{\pi}=0$ and therefore this confirms that we can treat B as constant in the integration.

In the above equation, there is a step marked * in which $\int{\cos{(\pi x)}  \delta(x+b)ds}=\cos{\pi b}$ occurs, where $x=\frac{s}{\pi R}$ and $b=a\pm\frac{1}{2}-\frac{at}{12}$. It relies on the fact that $\int\delta\left(x+b\right)dx=H(x+b)$, where H(x+b) is the Heaviside function such that when $x\geq-b$,\ $H\left(x+b\right)=1$ and $H\left(x+b\right)=0$ otherwise. Now

\begin{equation*}
\begin{split}
\int{\cos{(\pi x)}\delta\left(x+b\right) ds}=\int{\cos{(\pi x)}\delta\left(x+b\right)\frac{ds}{dx}dx=\pi R\int{\cos{(\pi x)}\delta\left(x+b\right) dx}} \\ =\pi R(\int\delta\left(x+b\right)dx)\cos{\pi x}-\pi R\int{\left(\int\delta\left(x+b\right)ds\right)\pi\sin{\pi x}}dx \\ =\pi RH\left(x+b\right)\cos{\pi x}-\pi^2R\int{H\left(x+b\right)\sin{\pi x}dx} \\ =\pi RH\left(x+b\right)\cos{\pi x}-\pi^2R\int_{-b}^{x}{\sin{\pi x^\prime}dx^\prime} \\ =\pi RH\left(x+b\right)\cos{\pi x}-\frac{\pi^2R}{\pi}\left(\cos{\pi x}-\cos{\pi b}\right)=\pi R\cos{\pi b}, \ \text{since}\ x>-b
\end{split}
\end{equation*}

\subsection{Alternative solar power method}

A suggestion worth exploring in the future is to reflect and concentrate sunlight on the climbers from one or more satellites in geostationary orbit, or from a station connected to another space elevator. The main consideration to be studied is the design of the optics to deal with varying angles of the sun in the course of the day. The receiver at each climber would be similar to that needed for laser power. The maximum interruption to sunshine would occur during an equinox and would only last for 50 minutes per day; for much of the year there would be no interruption. Judicious choice of the satellite’s location should ensure that there is no danger to people on Earth.

\subsection{Resilience}

Impacts with space debris may occur on occasions, since the area presented by the sphere is 12,500 \si{m^2}. Objects larger than 10 cm are tracked. [\cite{spacedebris}] If an impact with a tracked object is predicted, the climber will need to stop short of the expected point of impact until the debris has passed.

Untracked objects are likely to make two holes in the membrane and also damage the power converter. This will weaken the membrane, but it is designed to sustain a tension as high as 80 kN/m, whereas the required tension is no greater than 355 N/m. This redundancy in strength could be exploited to reduce the overall mass of the membrane and still leave enough strength to cope with the holes, even in a bad case where the debris takes out a slice of membrane between two rods.

Damage to a rod is a potential hazard. In that case, we rely on the other rods to maintain the integrity of the sphere. Each rod can sustain a compressive force of 16.1 kN, whereas the actual weight placed on one is about 40 kgf or 390 N. This redundancy ensures that the neighboring rods will be able to share the load, albeit with some distortion to the spherical shape of the membrane.
The power converter’s electronics will need to be designed in a modular fashion to be able to cope with damage of this sort. If one of the drive motors is hit, there are 255 other motors to carry on. Damage to a track that engages with a plug or to a toothed track that engages with a drive motor will need some further study and careful design to make sure that it does not prevent the whole power converter from moving.
The sphere will heat up in sunlight just like any other spacecraft. 30\% of the incoming radiation will be converted to electrical power, most of which will be converted to mechanical power, and the rest will cause moderate heating. At the International Space Station, the temperature fluctuates between $\pm250 \degree$ F over the day-night cycle, which is $120 \degree$ C to $-155 \degree$ C. [\cite{spacecold}] We can expect something similar.

\FloatBarrier
\section{Provisional mass budget}
\label{sec:provmass}

Table \ref{tab:calcmassmovePC} is a summary of the various masses calculated for the different components of the solar power system, assuming a movable power converter. Table \ref{tab:calcmassfixPC} assumes a fixed power converter.

\begin{table}[h]
    \centering
    \begin{tabular}{l r}
        \toprule
        Component & Mass (kg) \\
        \midrule
        Perovskite cells & 147 \\
        Membrane under tension (50,000 m\textsuperscript{2}) & 150 \\
        Rods & 215 \\
        Power converter (5 MW) & 185 \\
        Tracks, connecting rails and plugs & 20 \\
        120 grabbers at leading edge @ 0.05kg & 6 \\
        256 motors to move power converter @ 0.028kg & 7 \\
        Conductors 1.6kV, 100 kW loss & 138 \\
        \midrule
        \textbf{TOTAL} & \textbf{868} \\
        \bottomrule
    \end{tabular}
    \caption{Summary of calculated masses for a movable power converter}
    \label{tab:calcmassmovePC}
\end{table}

\begin{table}[h]
    \centering
    \begin{tabular}{l r}
        \toprule
        Component & Mass (kg) \\
        \midrule
        Perovskite cells & 147 \\
        Membrane under tension (50,000 m\textsuperscript{2}) & 150 \\
        Rods & 215 \\
        Power converter (4 X 5 MW) & 740 \\
        Plugs & 10 \\
        Conductors 1.6kV, 100 kW loss & 235 \\
        \midrule
        \textbf{TOTAL} & \textbf{1,497} \\
        \bottomrule
    \end{tabular}
    \caption{Summary of calculated masses for a fixed power converter}
    \label{tab:calcmassfixPC}
\end{table}
\FloatBarrier

\newpage

\section{Journey times}

Assuming a maximum power of 4 MW and a maximum speed of 200 km/h, the climbing times during the hours of daylight are shown in Table \ref{tab:journeytime}. The worst case for solar power is at an equinox. Then the journey will take until 7:55 am on day 10. Nearer to the solstice the journey to reach GEO (35,786 km) will take nine days. It may be possible to attain a higher speed when the climber is beyond 13,000 km, because its weight will less than 6\% of its weight on Earth.

There is a benefit in projecting laser power to the lowest two climbers. Not only will it reduce the journey times, it will also reduce the overall weight of climbers on the tether.

\newcolumntype{P}[1]{>{\centering\arraybackslash}p{#1}}
\begin{table}[h]
    \centering
    \begin{tabular}{P{2.3cm} P{2cm} P{2.3cm} P{2cm} P{2.2cm}}
        \toprule
        \multicolumn{2}{c}{At equinox} & \multicolumn{2}{c}{At solstice} & \multirow{2}{*}{Day number} \\
        \cmidrule(r){1-2} \cmidrule(r){3-4}
        Altitude at midnight (km) & Time of arrival & Altitude at midnight (km) & Time of arrival & \\
        \midrule
        1375   & 20:20 & 1395   & 20:35 & 1  \\
        3960   & 21:28 & 4180   & 22:02 & 2  \\
        7900   & 22:15 & 8455   & 23:21 & 3  \\
        12,080 & 22:40 & 13,115 & 24:00 & 4  \\
        16,395 & 22:55 & 17,920 & 24:00 & 5  \\
        20,800 & 23:06 & 22,725 & 24:00 & 6  \\
        25,270 & 23:15 & 27,530 & 24:00 & 7  \\
        29,915 & 23:20 & 32,330 & 24:00 & 8  \\
        34,720 & 23:25 & 35,786 & 17:16 & 9  \\
        35,786 & 07:55 &        &       & 10 \\
        \bottomrule
    \end{tabular}
        \caption{Climbing times}
    \label{tab:journeytime}
\end{table}

There may be brief periods when one climber comes between the sun and another climber. The worst case is on day 1 when the higher climber may be as close as 1,365 km to the lower one. The effect is very small. The angle in radians subtended by the sun near Earth is the sun’s diameter (1.39 million km) divided by the distance (150 million km), which is $1.39/150\approx9.27\times{10}^{-3}$. The angle subtended by a climber with diameter 126 m at a distance 1,365 km is $9.23\times{10}^{-5}$. The proportion of sunlight blocked by occlusion is the ratio of the areas, which is $(9.27\times{10}^{-3}/9.23\times{10}^{-5})^2\approx{10}^{-4}$. This is 0.01\% of the incoming power, which is inconsequential.

Of course, each climber will cast a shadow over its own solar cells at some times of day. The effective area that the solar cells present to the sun is 12,500 \si{m^2}, and the greatest area of cross section of a climber is 30 \si{m^2}. Hence the 5 MW power will be diminished by at most 30/12,500, which is 0.24\%. Much as for solar cells on Earth, the power converter will have to deal with variable shadows in the region affected by the shadow; there are well-understood techniques for doing this. [\cite{seriesparPV}]

\chapter{Conducting Electrical Power Through the Tether}\label{chapt:ACDC}

The ISEC Power Study Group began a number of conceptual designs to deliver electrical power to the climbers by using the electrical conductivity of a graphene tether as a power conductor.  One of the biggest advantages of a tether-powered climber is that there would be no outrigger power receivers, such as the laser power converters or microwave rectennas.  This design could potentially lead to climbers with the smallest cross-sectional areas of any of the power-beaming or solar designs.  A smaller cross-section means a smaller target to orbital debris, or a higher payload ratio.

The concept focused on DC power only as AC power on the tether would be lost by RF radiation, as well as by joule heating.  The climber was conceived to collect electrical current off the tether by brushes, similar to the way electric buses and trains draw power from an overhead grid of wires in many cities and along high-speed rail lines.  Graphene seemed to be the ideal material as it was a good electrical conductor and one of only a few material candidates strong enough to support itself over the length of the tether.

\section{Graphene electrical conductivity}

The work was based on the fact that a single atomic layer of graphene is the world’s best conductor of electricity, with an in-plane electrical conductivity of $96 \times 10^6$ S/m, about 1.66 times higher than copper. [\cite{nobelgraphene}]  We had assumed that the electrical conductivity of tens of thousands of monolayers of graphene bonded into a van der Waals homostructure (graphene laminate) would be the same as that of the monolayer.

Samples of multilayered graphene have been made in the laboratory and their electrical conductivity has been measured.  The results clearly show that the electrical conductivity declines rapidly as sheets of graphene are stacked on top of one another in a van der Waals homostructure to form graphene laminate [\cite{wang2019ultrastiff}, \cite{amer2020graphene}].  Table \ref{tab:conductivity} shows the electrical conductivities of single and multilayer graphene in plane, and cross-plane, and some metals.

\begin{table}
    \centering
    \begingroup 
    \renewcommand{\arraystretch}{1.5} 
    \begin{tabular}{p{7cm} p{2.5cm} p{3.5cm}}
        \toprule
        Material & Electrical conductivity (S/m) & Reference \hfill\\
        \midrule
        Single crystal graphene (in-plane) & $96.0 \times 10^{6}$ & \cite{nobelgraphene} \\
        More than 20 layers of single crystal graphene (in-plane) & $2.2 \times 10^{5}$ &  \cite{wang2019ultrastiff}, \cite{amer2020graphene} \\
        More than 20 layers of single crystal graphene (cross-plane) & $1.4 \times 10^{4}$ & \cite{tsang1976c} \\
        The highest measured conductivity of single-walled carbon nanotube fibres (in-plane) & $10.0 \times 10^{6}$ & \cite{xiang2025electrical} \\
        Copper & $58.0 \times 10^{6}$ & \cite{matula1979electrical} \\
        Gold & $42.0 \times 10^{6}$ & \cite{matula1979electrical} \\
        \bottomrule
    \end{tabular}
    \endgroup 
    \caption{Electrical conductivities of graphene, CNTs and some metals}
    \label{tab:conductivity}
\end{table}

\begin{figure}
\begin{centering}
\includegraphics[width=\textwidth]{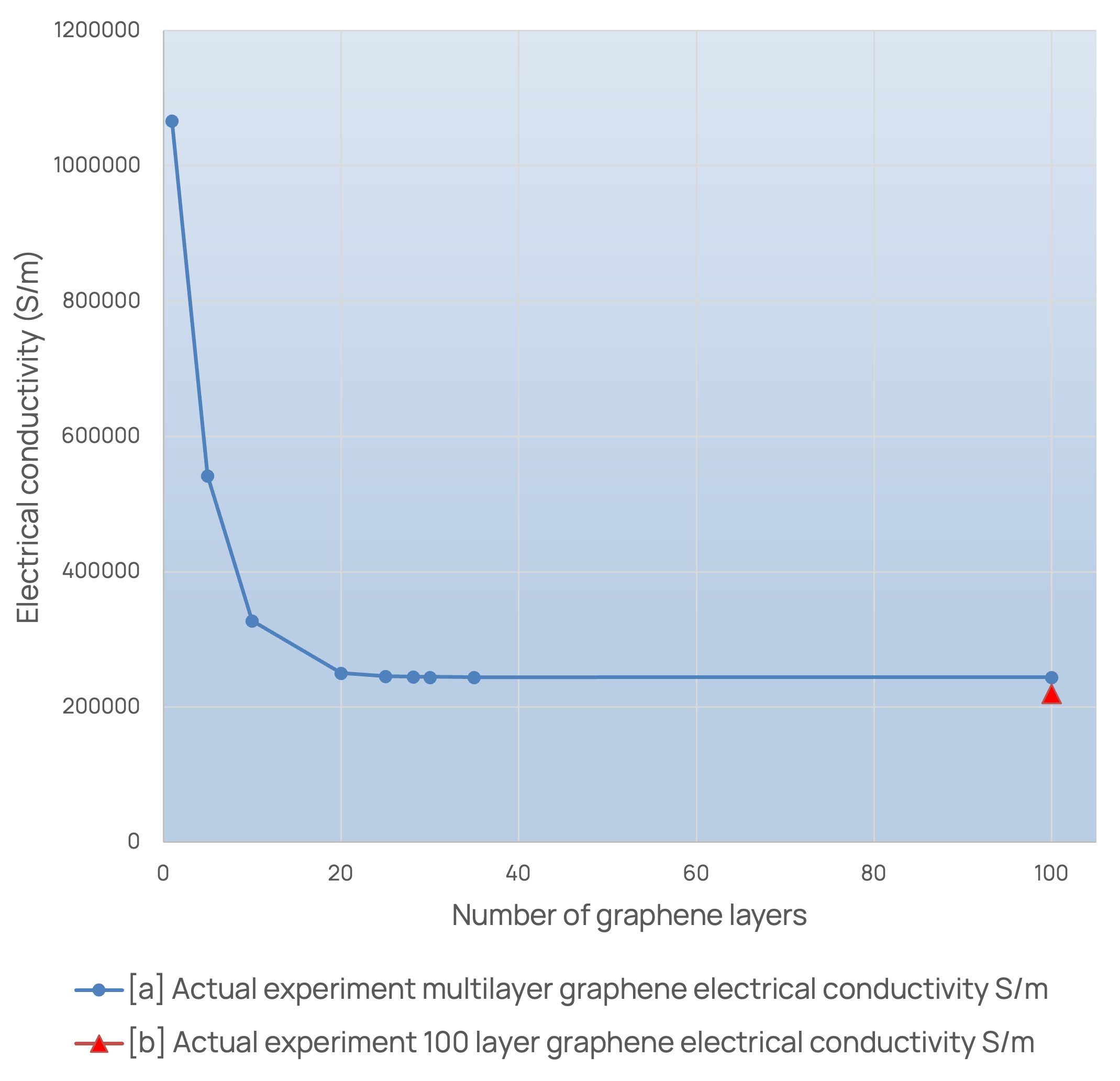}    
\caption{\label{fig:graphcond} How the in-plane electrical conductivity of graphene laminates changes with the number of layers}
\end{centering}
\end{figure}

This trend in Figure \ref{fig:graphcond} shows that the electrical conductivity of graphene declines rapidly as more layers are added.  After 20 layers, the electrical conductivity stabilizes at $2.2 \times 10^5$ S/m and no further decline has been reported.

The electrical conductivity of bundles of carbon nanotubes (CNTs) has also been measured, and the highest reported was found to be $10.0 \times 10^6$ S/m [\cite{xiang2025electrical}].  

The conductivity values for graphene and CNTs indicate that using these materials alone as electrical power conduits will create substantial resistive losses of power and, at present, cannot be recommended for power transmission for the space elevator.

\section{hBN interleaving}

A possible mitigation for the reduction in electrical conductivity is laminating monolayer graphene with an insulating material, such as hexagonal Boron Nitride (hBN). The idea is to separate individual graphene layers from each other so that their electrical conductivity is restored to the monolayer value.  hBN is a very high specific strength material and has been considered as a tether material.  Interleaving the two would not reduce the overall strength of a tether.  Two different tethers, one only graphene and one with mixed hBN and graphene, could have the same strength and be close to the same mass.

An experiment has already been done to examine the effect of interleaving graphene monolayers with hBN [\cite{overweg2020electronic}].  Two monolayers of graphene were separated by a 3.5 nm thick spacer of approximately 10 layers of hBN. The results showed that the spacing caused the decoupling of the graphene layers and the electrical conductivity of the graphene was restored to the monolayer value.  It is possible that fewer layers of hBN are needed to decouple the graphene electrical conductivity but this work has yet to be done.

An implication of interleaving hBN with graphene to recover the electrical conductivity of the graphene is that the total tether could become larger and heavier than a graphene-only tether, when adding both the graphene and hBN masses.  Detailed design studies and experiments would be needed to determine the optimal number of graphene layers to achieve the electrical conductivity required, and whether the laminate was strong enough to be a tether.

\begin{figure}
\begin{centering}
\includegraphics[width=\textwidth]{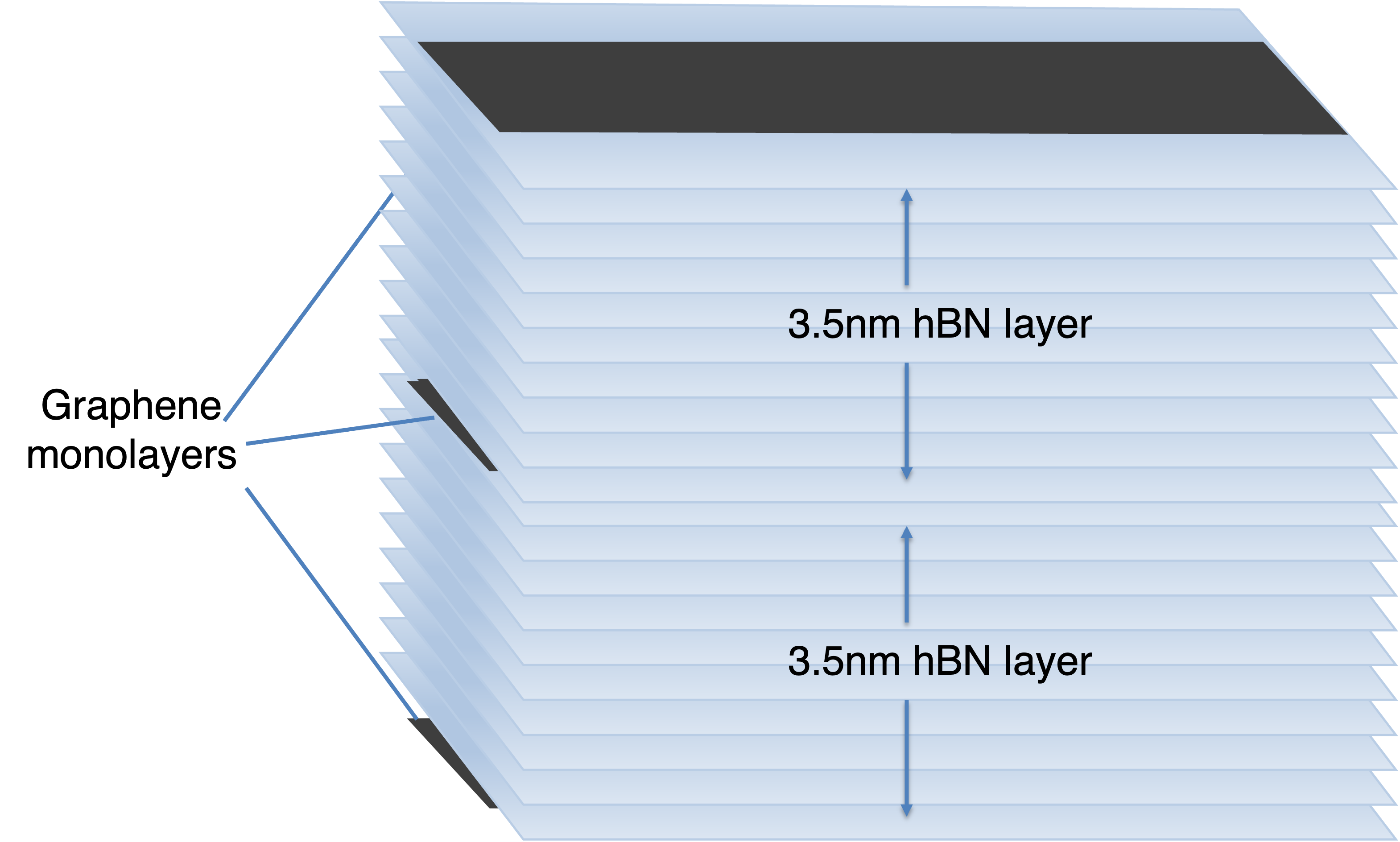}    
\caption{\label{fig:layerteth} Laminating alternating graphene monolayers with hBN multilayered insulating sheets restores the electrical conductivity of the graphene.}
\end{centering}
\end{figure}

\FloatBarrier

\section{Voltages and power losses}

Calculations showed that even with the electrical conductance of monolayer (single crystal) graphene, the tether would need to be energized with high voltage (hundreds of kV up to 1 MV) to reduce the resistive power losses of the current in the tether.  The worst case is powering the tether at the Earth anchor with the climber at about 4,000 km altitude, which is the altitude where the decreasing force of gravity starts reducing the power required to below 4 MW to maintain a constant speed of 200 km/h.  Assuming the tether contains two conductors with the conductivity of single-crystal graphene and dimensions of 1 m by 10 microns, the resistance of each is about 1 Ohm per km, so 2 Ohms per km for both conductors.  This calculates to a total of about 8,000 Ohms for 4,000 km of length along the tether.  Assuming that the power converter on the climber requires 10 kV, the load current for a 4 MW load will be about 400 A.  Using Ohm's law, the voltage drop on the tether would be 3,200,000 V (3.2 MV), and at 400 A, the Joule heating power loss ($I^2R$) is 1.28 GW, for a power efficiency of about 0.3\%.  Raising the efficiency would require changing any or all of several factors:  Reducing the tether resistance by making it larger (wider and/or thicker); increasing the tether conductivity; or reducing the length by having power sources closer to the climber. 

One method to reduce the voltage on the tether to 10 kV is to add power sources (such as solar panels) intermittently along the tether, spaced 100 km apart or more frequently.  Ideas have been put forth that do not yet reach the level of conceptual designs that incorporate solar panels suspended from the sides of the tether and are compatible with a climber that surrounds the tether.  What follows is a non-exhaustive list of the many engineering issues to be solved: 

\begin{itemize}
    \item Electrical connections between the panels and tether, given the interleaved layers of the graphene conductors.
    \item Design of the power converters that match the different voltages between the solar arrays and the tether.
    \item Compact packaging of the panels so that a surrounding climber does not need to be too much larger to clear them as it passes. See Section \ref{sec:passingproblem} for more discussion on what it means to be a surrounding climber.
    \item The additional weight of solar panels. They do not contribute to the tensile strength of the tether and so they subtract from its climber carrying capacity, or force the tether to be more massive.
    \item The power distribution from the solar panels.  The area of the panels varies as a function of altitude because climber power requirements decrease with altitude.
    \item The design of the panels to catch sunlight throughout the daily trajectory of the sun.
    \item The effect of the power variations caused by the moving position of the sun with respect to the tether panels and shadowing of the panels by the tether.
    \item Solving the deployment issues of a tether with solar panels attached to it.
\end{itemize}

\section{Ascenders, descenders, and the passing problem}
\label{sec:passingproblem}

Another idea proposed to add intermittent power sources to the tether is to interleave ascending climbers with descending climbers.  The descenders would dump power into the tether from their regenerative braking systems.  (See Chapter \ref{chapt:descclimb}).  Descenders would have to be within 100 km of ascenders to make the power transfer efficient enough.  The problems with this idea are several-fold.  The first is that a climber design would have to be found to allow descenders to pass ascenders on the same tether.  

The Reference Conceptual Design (RCD) climber surrounds the tether because this minimizes the mass of the structure that supports the wheels of the climber.  The coefficient of friction between the wheels and the tether is likely to be 0.1.  Such a low coefficient of friction means that the wheels of the climber need to be pressed together with 40 tons of compressive force to be able to get traction with the tether and allow the 20 tonne climber to roll upwards.  

This compressive force in the RCD climber is reacted in tension by the quick-disconnect mechanisms (described in Section \ref{sec:QDmech}) on both sides of the climber.  This is what is meant by the climber surrounding the tether.  This condition prevents two climbers that both surround the tether from passing each other, one going up and the other going down.  

Tension mechanisms are inherently lighter than mechanisms that absorb force in bending because materials are stronger in tension than in bending.  To make passing climbers, the climbers cannot completely surround the tether, but they must be open on one side to make room for a passing climber.  Removing the quick disconnect mechanism on one side of the climber forces all of the wheel compression force to be absorbed on one side of the climber, similar to the way the C in a C-clamp absorbs the compressive force of the screw by going into bending.  The RCD climber also has motors on both sides of the tether to power each wheel.  There are significant packaging problems to mount both motors on the same side of the climber, or to have one motor on the open side of the climber.  Passing climbers have to have their wheels in separate lanes on the tether to avoid collision, ascenders on one side, descenders on the other.  This means that the load of any climber will always be asymmetrically distributed on the tether.  While the climber may be counterbalanced to avoid torque on the tether from this asymmetry, it is not clear what the effect of the separate lanes would be on the stress in the tether.  The RCD climber was designed as it is to avoid all of these complications.  

Another problem is that a 100 km separation between climbers means that there would be 360 climbers on the tether at the same time between Earth and GEO.  The IOC for the space elevator allows eight climbers to be on the tether at the same time.  A tether to support 360 20 tonne climbers is 45 times heavier than a tether to support eight 20 tonne climbers.  A more massive tether means a more expensive tether and more difficult deployment of the tether.  20 tonne climbers would have to be mounted and launched every half hour to populate the tether with this many climbers.

As mentioned in Section \ref{sec:descentcon}, ascending climbers are forced West by Coriolis forces, and descenders are forced East.  If ascenders and descenders are present on the same tether, these forces would induce a torque on the tether and a tensile stress in the short direction of the tether.  This effect has not been simulated and is currently unknown.

\section{Other design challenges}

The ribbon shape of the tether causes the electric fields along the edges of the tether to far exceed those required to cause corona discharge in the atmosphere. This means that any design for an electrically conductive tether, made from multiple graphene monolayers interleaved with hBN, would need to contain the electrically conductive layers well within the electrically insulating hBN layers.  Electrical contacts at the ends of the tether would provide connection points for the power. Figure \ref{fig:layerteth} shows an example of this design.

Using the tether as the power conduit raises another design challenge: how to extract power from the tether for a moving climber?  If the edges of the ribbon are well insulated, then embedding electrical contacts in the tether is an alternative.  This would mean making holes in the insulating layer of the tether to anchor electrically conductive contacts.  Layered printed circuit board technology uses metal contacts termed vias [\cite{vias}].  It is possible that vias could be adapted to work with high voltages, although the additional mass of the metal connections would add weight to the tether that does not contribute to its tensile strength. This additional weight reduces the tether's capacity to carry climbers.  These holes would also present areas of stress concentration that could lead to potential points of mechanical failure, so they would need careful design.

One of the issues identified in the 2023 study report [\cite{wright2023climber}] is that multilayered graphene has an inherent weakness: very low shear strength in the interlayer bonds allowing the layers to slide over each other.  Climbing the tether with wheels pressed together from opposite sides applies tractive shear stresses to the surface layers.  If this stress is not distributed to the internal layers of the tether, the surface layers will fail and destruction of the tether would occur.  A method is needed to bond the layers of the tether together so that the tractive shear stress can be distributed to the internal layers and not just taken by the surface layers.  A recent hypothesis for cross-linking between the layers of graphene laminates, (spot welding) has shown that this is possible [\cite{spotwelding}].  Further work has demonstrated that bonds between graphene and hBN can be made forming a lateral heterostructure (butt joints between graphene and hBN) [\cite{HBNcrosslink}].  This means that spot welding between layers of hBN and graphene should also be possible, although this has yet to be demonstrated experimentally.

Studies showed that the tether would need to be energized at both the Earth anchor with base power and at GEO by a large solar PV array for best efficiency.  The high voltage at the source would be dropped by the tether resistance, and would be controlled by telemetry from the climber to keep the voltage at the climber at 10 kV.  The 10 kV voltage would have to be stepped down at the climber to operate the primary motors at 800 V.  Modern high-voltage solid-state DC-DC power converters have been demonstrated with mass power densities of 0.5 kW/kg to over 2 kW/kg, and high-performance/Aerospace/EV converters often achieve 10 to 25 kW/L volume density.  Thus, a 4 MW converter would likely weigh about 4 tonnes, and occupy 400 L of volume.  4 tonnes is 20\% of a 20 tonne climber.  Future development will be needed to reduce the mass of these converters.  This is also an issue for power beaming if the laser power converters (PV arrays) can only be operated at about 100 V and need to be stepped up for the 800 V motors.

High voltage applied to the tether becomes problematic as it would be prone to arc flash across the tether.  Mitigating this requires the positive and negative electrodes to be separated by distances wider than a practical tether.  Plasma conductivity in the ionosphere between 100 km and 400 km could conduct parasitic currents from exposed conductors, resulting in power loss, depending on the tether voltage, and the separation and width of the contacts.   

\section{Conclusion about the tether as a power conduit}

Given all of these factors, it is currently infeasible to power the climbers by using the tether as a power conduit.  Significant further study to mitigate these challenges is needed for this concept to be considered a technology ready for use with the space elevator.
\chapter{The Descending Climber Power Problem}\label{chapt:descclimb}
\section{Introduction to the difference between ascending and descending}

For twenty years, little attention has been paid to the problem of bringing payloads down to the surface of the Earth on climbers descending from GEO.  [\cite{Robinson2023}] discusses the problem of ascending from GEO to the Apex Anchor, which resembles descending to Earth from GEO in that the climber must dissipate power instead of receiving power to maintain speed.  However, the apparent weight of climbers above GEO is very small compared to their weight near Earth, so the power required to be dissipated is much smaller.

The assumption for decades has been that all the traffic on the elevator would be up from Earth to establish the cis-lunar infrastructure.  In 2025, Steven Griggs [\cite{Griggs}] gave a talk at the ISEC annual conference in which he presented that one of the few ways to pay back investors in space settlements or asteroid mining was to bring back to Earth precious metals and platinum group metals from asteroid mining.  He showed that aerobraking these payloads was impractical compared to lowering the payloads on the space elevator.  The time has come to consider the implications of descending on climber hardware and the power infrastructure of the elevator.  This report is only the beginning of the effort needed and is intended as an introduction to the technical problems of descent.

Ascending climbers consume large amounts of power to climb.  Looking at Figure \ref{fig:power}, we see that most of the power required is at or below a few thousand kilometers above the surface of the Earth.  Equation \ref{eq:energy} shows that a 20 tonne climber needs a total of almost a terajoule of energy to ascend to GEO from sea level.

\begin{figure}[hb]
\begin{centering}
\includegraphics[width=.8\textwidth]{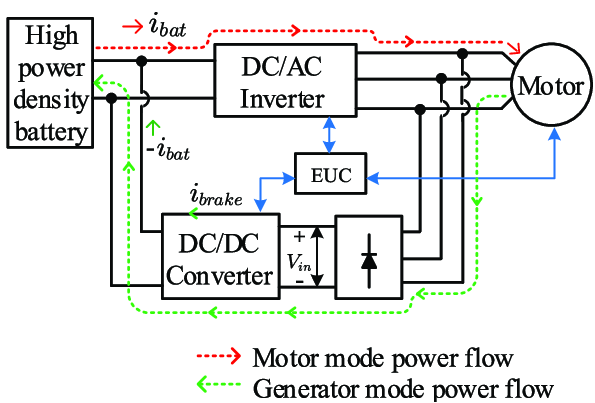}
\caption{\label{fig:block} Block diagram of a regenerative braking system. [\cite{nama2022design}]  In a typical EV, the battery that powers the motors also receives the regenerative braking power to recharge.  The descent from GEO produces about a terajoule of energy that the climber power system must manage.}
\end{centering}
\end{figure}

To keep a descending climber moving at only 200 km/h, (not freely falling), the climber must store or dissipate the same amount of power shown in the figure or more, depending on efficiencies.  With the wheels of the climber firmly clamped on the tether to get traction and the power supply to the motors removed, the wheels drive the motors of the climber.  As the climber rolls down the tether, the motors act as generators, producing electrical current that must be stored or dissipated.  This is a regenerative braking scheme to keep the climber moving at constant speed going down.  See Figure \ref{fig:block} for a block diagram of a regenerative braking system.  As the ascent requires almost a terajoule of energy, the descent will generate a terajoule of energy.

An ideal situation would be that there is an energy storage technology with sufficient capacity that a descender could charge this energy reservoir on the trip down, and then that climber could be "topped off" on the ground (to account for any losses or inefficiencies) and then use that energy for its next ascent.  Except for the first trip of a climber up, this would eliminate the need for any power beaming system.  Of course, that technology would have to be light enough to fit within the mass budget of the climber.

As shown in [\cite{wright2023climber}], the mass of the batteries currently assumed for the 5 wheel-pair climber is 827 kg, or about 4\% of the mass of the 20 tonne climber.  An energy storage technology that can store 1 terajoule of energy then needs to be able to store 1.21 GJ/kg (336.1 kWh/kg).

The next section is a survey of some possible storage technologies and their limitations.  As will be seen, current storage technology peaks at an energy density of less than 1 kWh/kg, making the state of the art today less than .3\% of what is needed. An important distinction to keep in mind is that all of the technologies shown can be characterized by both power density and energy density.  The power density is a measure of how quickly the storage can be charged or discharged.  Several of the technologies discussed below have sufficient power densities, but they all fall short in their energy density.  Without sufficient storage, the energy of descent must be dissipated off the climber.

\FloatBarrier

\section{Where to put the energy released by descent?}

\subsection{Survey of technologies for power storage on the climber}

This section introduces the principal storage options relevant to space elevator climbers:

\begin{itemize}
    \item Superconducting Magnetic Energy Storage (SMES): SMES systems store energy in the magnetic field generated by circulating superconducting currents. They offer very high power density and near-instantaneous charge and discharge capabilities with excellent efficiency. However, they require cryogenic cooling to maintain superconductivity, posing thermal management challenges. SMES systems can rapidly absorb regenerative braking energy with minimal losses, enhancing system efficiency.
    \item Supercapacitors: These devices provide very high power density and fast charge/discharge cycles, making them ideal for transient energy buffering and handling regenerative braking energy. While their energy density is lower than that of batteries, supercapacitors exhibit excellent thermal stability and longevity, supporting millions of cycles without significant degradation.
    \item Solid State Batteries: An emerging technology promising higher energy density and improved safety compared to traditional lithium-ion batteries. Solid state batteries feature superior thermal stability and potentially longer cycle life, making them attractive for space elevator applications where weight, volume, and operational safety are paramount.
    \item Fuel Cells: Standard fuel cells convert stored chemical energy into electrical energy with good energy density and scalability. They can complement batteries and supercapacitors, especially for long-duration energy supply. Standard fuel cells require fuel storage and regeneration strategies and are typically integrated with other storage systems to maintain performance during variable load profiles.  The climber use case is for reversible or regenerative fuel cells that use electricity to electrolyze water and produce hydrogen and oxygen, which are stored.
\end{itemize}

In the following sections, we will examine each of these energy storage technologies in detail, evaluating their strengths and limitations with respect to storage capacity, efficiency, thermal stability, scalability, and charge-discharge dynamics, including regenerative braking capabilities.  The total energy required to climb to GEO or stored by regenerative braking in a descent from GEO, is approximately 1 terajoule. The mass of each technology required to store this much energy is calculated to determine the adequacy of each technology.

\subsubsection{Superconducting Magnetic Energy Storage (SMES) System}

Superconducting Magnetic Energy Storage (SMES) systems store electrical energy in the magnetic field generated by a persistent current flowing through a superconducting coil. 

The stored magnetic energy E in an ideal superconducting coil is given by:

$$E = \frac{1}{2} L I^{2}$$

\noindent where:\\
L is the inductance of the coil (Henries)\\
I is the induced current in the coil (Amperes)\par
\bigskip
The inductance of a solenoid coil, possibly with an iron core of relative permeability $\mu_r$, is approximated as:

$$L = \mu_0 \mu_r \frac{N^2 A}{l}$$

\noindent where:\\
$\mu_0$ = 4$\pi \times 10^7$ H/m is the permeability of free space\\
$\mu_r$ is the relative permeability of a ferrous core\\
N is the number of turns\\
A is the cross-sectional area of the coil ($\si{m^2}$)\\
l is the length of the coil (m)\\

The magnetic field B inside the solenoid is:

$$B = \mu_0 \mu_r \frac{N}{l} I$$

The energy density u (energy per unit volume) stored is:

$$u = \frac{B^2}{2\mu_0}$$

\noindent Advantages of SMES:
\begin{itemize}
    \item Very high power density and fast response times, enabling rapid charging/discharging cycles suitable for regenerative braking scenarios.
    \item Exceptional efficiency with near-zero resistive losses due to superconducting coils.
    \item High power quality and durability with a virtually unlimited cycle life.
\end{itemize}

\noindent Limitations and Challenges:
\begin{itemize}
    \item Cryogenic Requirements: SMES coils must be maintained at very low temperatures (e.g., 20–90K for high-temperature superconductors) to sustain superconductivity, requiring complex and energy-intensive refrigeration systems.
    \item Energy Density and Scalability: Despite high power density, the \textit{energy density} (Wh/kg) of typical SMES systems is modest compared to chemical batteries and supercapacitors. Storing a high amount of energy demands bulky coils, greatly increasing weight.
    \item Quenching Risks: Exceeding critical current $I_c$, magnetic field $B_c$, or temperature $T_c$ causes quenching, loss of superconductivity, introducing resistance and thermal stress, potentially catastrophic for system stability.
    \item Mass Constraints: For space elevator climbers, the coil mass necessary to store energy comparable to the climber’s power demand exceeds the climber’s allowable mass budget.
\end{itemize}

Typically, superconducting magnetic energy storage (SMES) systems exhibit an energy density on the order of $1 - 10 \,\mathrm{Wh/kg}$, which is considerably lower than that of lithium-ion
batteries (generally around 100–250 Wh/kg) [\cite{burke2002high}], [\cite{Faradaybatt}].  Given an energy density of 10 Wh/kg, the mass required to store a terajoule of energy is 27.8 kilotonnes, well outside the mass budget of a 20 tonne climber.

In summary, while SMES offers exceptional power density and efficiency, the low energy density makes it nonviable as a storage system for space elevator climbers that need to store a terajoule of energy.

\subsubsection{Supercapacitors}

Supercapacitors, also known as ultracapacitors or electrochemical capacitors, offer a distinct form of energy storage characterized by a high power density and rapid charge-discharge capability. They operate by storing charge electrostatically at the electrode-electrolyte interface, unlike chemical energy storage in batteries.

The energy stored in a supercapacitor is governed by its capacitance and the applied voltage. The relationship is expressed as:

$$E = \frac{1}{2}CV^2$$

\noindent where:\\
C is the capacitance in farads (F)\\
V is the voltage across the capacitor (V)\par
\bigskip
The capacitance itself depends on the geometry and dielectric properties of the device. For a parallel-plate configuration, it is given by:

$$C = \varepsilon \frac{A}{d}$$

\noindent where:\\
$\varepsilon = \varepsilon_r \varepsilon_0$ is the permittivity of the dielectric material
between the plates\\
$\varepsilon_r$ is the relative permittivity or dielectric constant of the material\\
A is the effective surface area of the electrodes\\
d is the separation between electrodes\par
\bigskip
This relation highlights the strategies employed in supercapacitor design: maximizing electrode surface area (through porous nanostructures) and minimizing effective separation to achieve very high capacitance values. 

In practice, supercapacitors exhibit capacitances from a few millifarads (mF) up to tens of thousands of farads (F), while their maximum operating voltage is constrained by the electrolyte and material design.

\bigskip
\noindent Advantages of supercapacitors:
\begin{itemize}
    \item Very high power density (several kW/kg), enabling fast charge and discharge cycles.
    \item Long cycle life with minimal degradation over millions of cycles.
    \item Excellent thermal stability and operational reliability over a broad temperature range.
    \item Environmental friendliness and safer operation compared to many battery chemistries.
\end{itemize}

\noindent Limitations:
\begin{itemize}
    \item Lower energy density compared to electrochemical batteries, typically ranging from 3 to 10 Wh/kg [\cite{Faradaybatt}].
    \item Bulkier and heavier when scaled to meet the substantial energy storage needs of space elevator climbers.
    \item Higher self-discharge rates requiring frequent recharging intervals [\cite{Faradaybatt}].
    \item Limited voltage per cell necessitating complex series arrangements, affecting system design and reliability.
\end{itemize}

In the example of the 20 tonne climber, the voltage of the system is 800 V and the energy to be stored is 1 terajoule.  This gives a required capacitance of $3.13 \times 10^6$ farads, over 300 times larger than the largest currently in use.  800 V is also much higher than typical supercapacitor voltages, which are typically under 3 V.  

While supercapacitors provide excellent power delivery and regenerative braking capabilities, their low energy density leads to large size and weight for a given energy capacity. This fundamental scaling issue makes it impractical for primary energy storage in space elevator climbers, where minimizing mass and volume is critical.

\subsubsection{Solid state batteries}

Solid-state batteries (SSBs) replace the liquid electrolyte used in conventional lithium-ion batteries with a solid electrolyte, typically ceramic, glass, or sulfide-based materials. This substitution can offer significant improvements in thermal stability, safety, and potentially energy density [\cite{Faradaybatt}]. The fundamental energy storage in solid-state batteries is similar to lithium-ion cells:

$$E = Capacity \times Voltage$$

However, SSBs achieve higher volumetric and gravimetric energy densities by enabling:
– Thinner electrolytes due to higher ionic conductivity in solid materials.\\
– Compatibility with lithium metal anodes, which have greater theoretical specific capacity than graphite [\cite{Faradaybatt}].\par
\bigskip
\noindent Advantages of solid-state batteries include:

\begin{itemize}
    \item Enhanced safety by eliminating flammable liquid electrolytes.
    \item Potential for higher specific energy (up to 700 Wh/kg projected) and longer cycle life ($> 1,000$ cycles) [\cite{Faradaybatt}].
    \item Improved thermal stability, making them suitable for challenging environments.
\end{itemize}

\noindent However, limitations remain, such as:
\begin{itemize}
    \item Manufacturing complexity and cost at scale.
    \item Interfacial resistance and dendrite formation under certain conditions [\cite{alia2020regenerative}].
    \item Current technology readiness is lower compared to commercial lithium-ion batteries [\cite{Faradaybatt}].
\end{itemize}

For space elevator climbers, SSBs offer an attractive alternative due to safety and energy density improvements. Yet, manufacturability and long-term reliability must be proven before widespread adoption.

At a specific energy of 700 Wh/kg, storing a terajoule of energy requires a battery that masses 397 tonnes.  While less than the masses of SMES and supercapacitors, still outside the range of viability for 20 tonne climbers.

\subsubsection{Fuel cells}

Fuel cells convert chemical energy from a fuel (typically hydrogen) into electricity through electrochemical reactions. Regenerative fuel cell systems (RFCS) incorporate electrolyzers and fuel cells to store and generate energy by cycling water into hydrogen and oxygen [\cite{burke2002high}]. The electrical energy output of an RFCS over a discharge time $t_d$ is:

$$E = n \cdot V_d \cdot I_a \cdot t_d$$

\noindent where n is the number of cells, $V_d$ the discharge voltage, and $I_a$ the average discharge current [\cite{burke2002high}].
\bigskip \\
\noindent Key advantages include:
\begin{itemize}
    \item Very high theoretical energy density, experimentally reaching up to 790 Wh/kg [\cite{burke2002high}].
    \item Capability for indefinite operation with proper reactant handling.
    \item Suitability for long-duration storage with relatively low self-discharge.
\end{itemize}

\noindent Challenges and disadvantages:
\begin{itemize}
    \item System complexity due to reactant storage (hydrogen and oxygen tanks) and management.
    \item Lower round-trip efficiency compared to batteries due to electrochemical losses (approximately 53\% reported) [\cite{burke2002high}].
    \item Weight and volume penalties from auxiliary components limit mass advantage.
    \item Technology readiness remains low for many regenerative fuel cell systems [\cite{burke2002high}].
\end{itemize}

At a specific energy of 790 Wh/kg, the capacity of fuel cells is not that much higher than that of solid-state batteries.  The mass of a fuel cell to store a terajoule of energy is 352 tonnes, orders of magnitude too large for the 20 tonne climber.

\subsubsection{Conclusion on onboard storage technologies}

Some of the technologies shown in this section could potentially be used for the 50 kWh (180 MJ) storage discussed in Section \ref{sec:onboardbatt}, but they are all inadequate to store the terajoule of energy that either an ascending climber needs to climb to GEO, or a descending climber needs to store in coming from GEO.  These limitations motivate the development of other power transfer technologies for the climbers.\label{sec:EnergyStorage}

\subsection{Beaming power off the climber by microwave}
\label{sec:mwavebeamoff}

One possibility to get the power generated off the climber is to install a microwave transmitter on the descending climber which will allow radiating this energy off the climber within the limits of efficiency.

Likely efficiencies include roughly 90\% for regenerative braking and roughly 90\% for the microwave transmitter, thus total efficiency for conversion of mechanical braking to microwave would likely be roughly 80\%.  This leaves roughly 20\% to be dissipated by other means.

In order to permit operating the descending climber transmitter in LEO, the power density must be compatible with the ionosphere, which is 230 \si{W/m^2}.  At this power density, the diameter required to transmit and receive 4 MW is 150 m.  An estimate for the power density of a microwave WPT transmitter may be calculated from [\cite{mankins2014case} Table 5-1 pp.233] as
follows:

The baseline GEO space solar power satellite provides 230 \si{W/m^2} over a 10 km diameter rectenna from a 1 km transmit aperture.  Because of the Goubau relationship [\cite{jaffe2024power} pp.167], the power density scales by the square of the ratio of the diameters, $10^2$ = 100.  Thus, the power density at the 1 km SSP transmitter aperture will be 23 \si{kW/m^2}.

Summarising [\cite{mankins2014case} Table 5-1 pp.233], a WPT module weighs 281 kg and is 4 m in diameter, for 12.57 \si{m^2} area, thus 22.36 \si{kg/m^2}.  This module transmits 23 \si{kW/m^2}.  Dividing these into each other results in 1.03 kW/kg = 1.03 MW/tonne.  Again, a 4 MW transmitter would weigh over 4 tonnes, or 20\% of the weight of the whole climber.  This is probably too large for the mass budget of the climber and will require development to reduce.

\subsubsection{Transmission of power from descenders to ascenders}

The radiated microwaves from descenders may be exported to nearby ascending climbers if the ascenders are equipped with receiving arrays to help power them.  Climbers as currently conceived do not have the capability of passing one another on the same tether, so descenders would have to be on a dedicated descending tether and ascenders on an adjacent ascending tether.  

Applying the Goubau relationship for 150 m apertures for both receiving and transmitting at 10.0 GHz, the link efficiency would be better than 90\% for distances less than 380 km.

The amount of time that ascenders will be in the range of 380 km to descenders will be very small during the initial commissioning years of the space elevator.  See Section \ref{sec:passingproblem} for a more detailed discussion of the issues of ascenders and descenders on the same tether.

\subsubsection{Beaming microwave power to the ground for Earth Anchor use}

Beaming microwave power from the descending climber to the ground enables harvesting regenerative braking energy by radiating it as microwaves to a big rectenna on the ground.  At lower altitudes, this should be fairly efficient in recovering the gravitational energy generated by a descending climber.
For example, using the Goubau relationship, a descending climber transmitting with a 150 m aperture would see 90\% or more link efficiency with a 1900 m diameter rectenna on the ground for distances less than
5,000 km.

\subsection{Dissipating the power in resistors} 

If it is not technically feasible to beam the power off the descending climber, the power must be dissipated as heat.  The most straightforward way to do this is to design electrical resistors that can take the kiloamps of current coming from the motor/generators.  These resistors dissipate the energy of the current by heating up and radiating the energy away as thermal radiation.  Reference [\cite{dhanush}] shows the work of another ISEC intern, Dhanush Salunke, who designed a graphene sheet resistor that can dissipate up to 6 MW of power by heating up to 2,309K ($2,036 \degree \si{C}$).  The melting point of Single Crystal Graphene is between 5,000K and 6,000K [\cite{scgMP}], so the resistor temperature at full power is well within the capability of SCG.  At the temperature of 2,309K, the graphene resistor will glow with a warm white or yellowish white light, similar to a standard incandescent electric light bulb.  It remains to be seen in the final design how the resistor stands up to corrosion in the atmosphere at that temperature.  Of course, convective cooling in the atmosphere would lower the temperature of the graphene for the same power dissipation.

The electrical resistivity of the graphene sheet in [\cite{dhanush}] was for monolayer SCG.  As we see in Chapter \ref{chapt:ACDC}, stacking monolayer graphene to make multilayer graphene sheets reduces the electrical conductivity of the material by orders of magnitude.  The work needs to be updated to reflect the new understanding of the electrical conductivity of multilayer graphene; however, the basic design should remain viable.

\begin{figure}
\begin{centering}
\includegraphics[width=\textwidth]{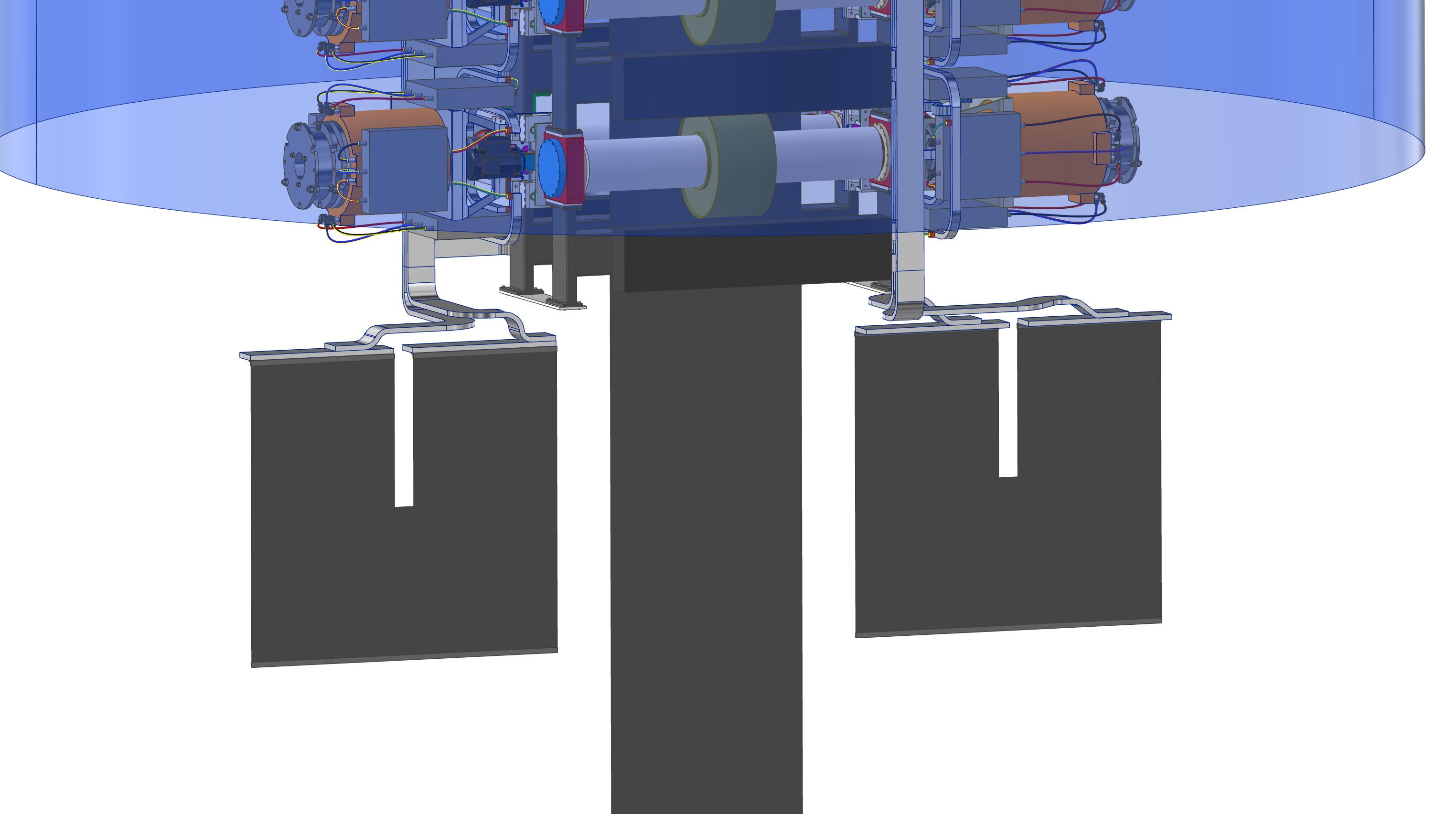}
\caption{\label{fig:resistor} Conceptual design of a pair of graphene resistors that can dissipate 6 MW of power each by heating up to 2,309K.  Additional electrically insulating support structure has not been designed yet.}
\end{centering}
\end{figure}

Figure \ref{fig:resistor} shows the conceptual design of the resistors attached to the bus bars on each half of the climber.  At 6 MW, the bus bars are carrying 7.5 kA of current at 800 V.  The basic design concept is to size the graphene sheet in terms of length, thickness, and cross-sectional area to achieve the required resistance, and to size it in planar area to provide surfaces at the desired temperature to radiate the heat away.  The hot graphene sheets are connected to the much lower melting point aluminum bus bars by tungsten transition bars.  The tungsten bars provide a thermally resistant attachment with a high melting point, ensuring that the graphene sheet neither melts the tungsten nor the aluminum busbar.  

The mass of both resistor assemblies totals 119 kg, which is readily accommodated in the mass budget of the climber.

A future, more detailed study of the resistors will examine their performance along the whole descent from GEO to make sure that the amount they heat up dissipates the correct amount of heat at lower power levels than the maximum.  The thermal resistive length of the tungsten attachment bars also needs to be studied in a thermal finite element analysis to make sure the surfaces touching the aluminum bus bar do not exceed the melting point of aluminum.  A subtlety of designing thermal and electrical contacts is that they have different properties in a vacuum from the properties they exhibit in the atmosphere.

\begin{figure}[hbt]
\begin{centering}
\includegraphics[width=\textwidth]{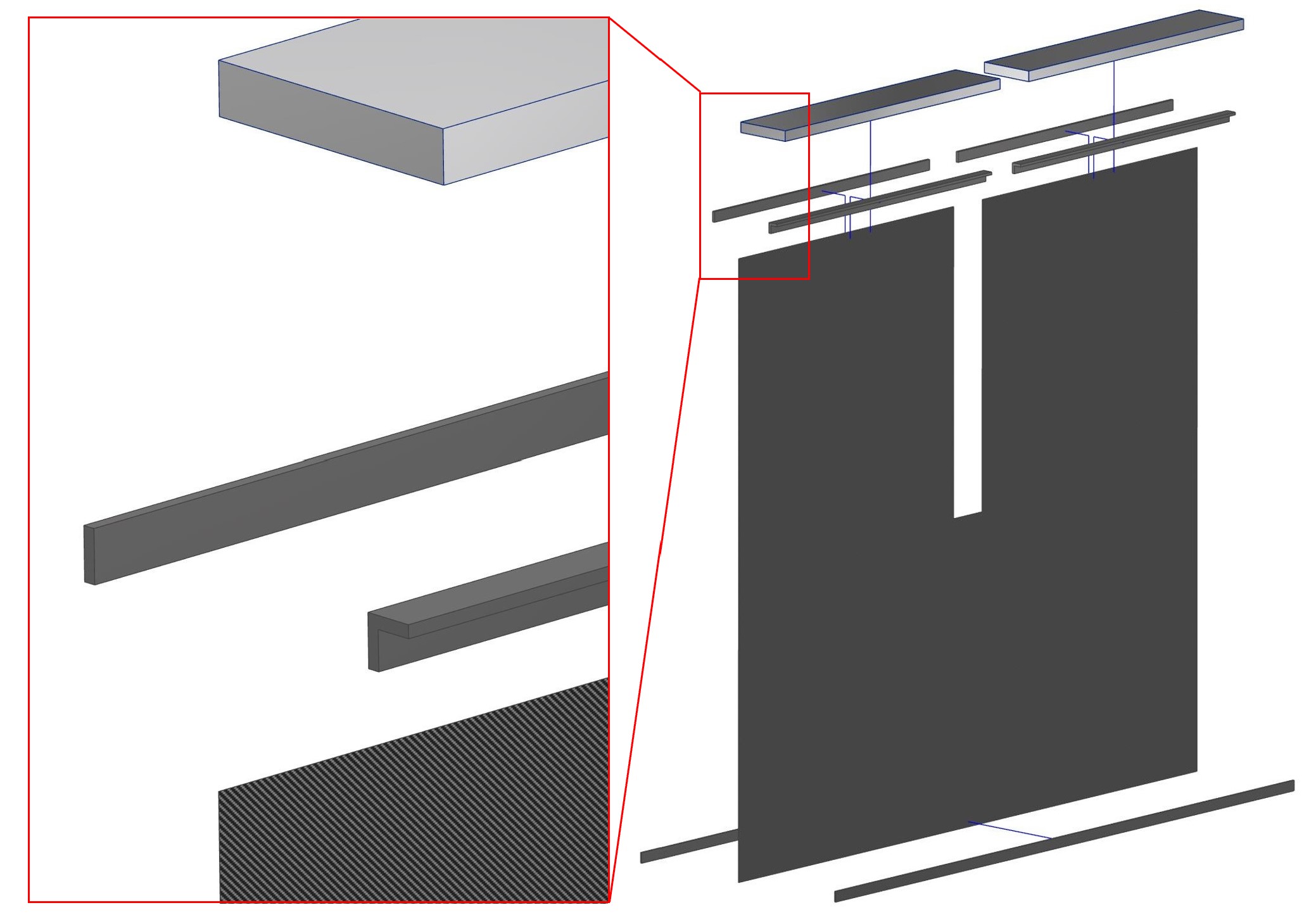}
\caption{\label{fig:resistexp} Exploded view of a graphene sheet resistor and the tungsten parts connecting it to the aluminum busbar.  The view on the left is a detail view showing the shapes of the tungsten bar and angle that connects the graphene sheet to the aluminum busbar.  The two bars shown at the bottom are to keep the graphene sheet held taut so the thermal radiation surface area is maintained.  No bolts are shown, but the assembly is designed to be bolted together.}
\end{centering}
\end{figure}

An additional detail is that the electrical resistance of the tungsten brackets is actually in series with the electrical resistance of the graphene, and this series connection was not accounted for in the conceptual design.  Figure \ref{fig:resistexp} shows a close-up of the attachment tungsten parts.  The assembly is intended to be bolted together, but none of the bolts and bolt holes are shown. 

\subsection{Sending power into tether vibrations}

As we shall see in Section~\ref{sec:vibrations-descending-climbers} a
descending climber can also dissipate substantial power into tether
vibrations, which in turn can be absorbed by an ascending climber, at the
base or apex anchor, or in the atmosphere.

\newpage
\section{Conclusions about managing the energy released by a descender}
\label{sec:descentcon}

\begin{itemize}
    \item There is currently no technology capable of storing a terajoule of energy onboard the climber at a mass that works within the climber mass budget.
    \item If the specific mass of microwave transmitters on the climber is 1.03 MW/tonne, to be able to dissipate up to 12 MW would require transmitters that mass 11.65 tonnes.  This is too large a fraction of the mass of a 20 tonne climber.  There isn't room for payload and traction drive as well.
    \item 150 m diameter rectennas on the climber are very problematic to support mechanically at one gee of gravity near the Earth.  See Figure \ref{fig:Fig14rectennasizes} for a comparison of different diameter rectennas and the 20 tonne climber.  Given the power to be dissipated will be on the order of 12 MW near the surface of the Earth, 150 m diameter rectennas may not be large enough to transmit that amount of power and keep the power density at or below 230 \si{W/m^2}.  This means that even larger rectennas could be required.
    \item Beaming microwave power from descenders to ascenders is only efficient at small distances between climbers (380 km).  The ascending and descending tethers would have to be populated with far more climbers than the IOC tether can handle.  It is also not currently known how far apart ascending tethers and descending tethers would have to be located from each other to avoid entanglement at high altitudes.  Since ascending climbers are forced West by Coriolis forces, and descenders are forced East, the ascending tether should be located to the West of the descending tether.
    \item Transmitting power to the ground by microwave beaming from descenders would be a useful feature to have if transmitters can be built light enough.  The combination of heavy transmitters and very large rectennas makes this scenario unlikely in the near future.
    \item The technology that appears easiest to deploy and fits within the mass budget of the climber is to dissipate the electrical energy of the regenerative braking system in graphene resistors that heat up to radiate the energy away as thermal radiation.  More work is needed to finalize the design of the graphene resistors.
\end{itemize}

One question that could be asked here is why the discussion did not include optical power beaming by mounting a laser on the climber.  Chapter \ref{chapt:PowerBeaming} should provide insight that the transmitting and receiving ends of laser power beaming are very different and that the inefficiency of lasers would produce a very difficult heat dissipation problem for the descender.

It would be fair to say that there is no power dissipation technology today that offers energy recovery from descenders.  The resistive option just wastes the released energy as heat into space.

\chapter{Other Means of Sending Power Along the Tether}\label{chapt:OtherMethods}

This chapter discusses three different methods of moving a climber along the tether in which power is transferred along the tether itself.  The first method, reel-out launch, may become the default way to get the climber out of the atmosphere for almost any power system, since we have shown that wind drag can be detrimental to both the tether and the climber.

The other two methods, using mechanical vibrations on the tether and microwaves externally guided by the tether, are included to illustrate the diversity of approaches to powering the climbers on the space elevator.  These methods are at an earlier stage of conceptual design than other methods previously discussed and will require much work to bring them to the same level of definition.

\section{Reel-out launch}
\label{sec:reelout}
\label{sec:pull-down}

The first tens of kilometers are particularly challenging for the climber. Wind loads on the climber’s rectenna or photovoltaic arrays pose structural difficulties and increase the lateral forces on the elevator ribbon. Likewise, in order for the climber to climb the ribbon, it needs to have sufficient width to be climbable, which increases wind susceptibility. If power beaming is used, this phase of travel also drives requirements on how close the power beaming station must be to the elevator’s anchor station or how inclined the beam may be.

\begin{figure}
\begin{centering}
\includegraphics[width=3cm]{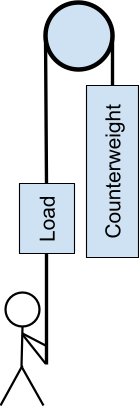}    
\caption{\label{fig:pulley-lift} An analogy to the reel-out launch method. If the stick figure lets out rope, the counterweight descends causing the load to raise. The same principle works with a space elevator except that the Apex Anchor is pulled up by centrifugal force, making the pulley unnecessary.}
\end{centering}
\end{figure}

In this section, we look at the reel-out method for powering the climber out of the atmosphere. In this method, which was first published in [\cite{knapman2014design}], the climber is launched from the ground simply by reeling out the ribbon at the anchor from a large spool. The upward force from the Apex Anchor keeps the ribbon in tension. This is a close analog to the pulley system shown in Figure~\ref{fig:pulley-lift}. In this system, a load can be raised by feeding out rope beneath it, allowing the counterweight to lower, pulling up the lighter load. As it descends, the counterweight loses potential energy which is transferred in part to the load, and in part to the person paying out the rope. Once the load is removed, the person resets the system to its initial condition by pulling down on the rope, providing work. The counterweight goes up, reacquiring the potential energy it had lost. In this system, the pulley is needed to convert the downward force on the counterweight into an upward force on the load. For the space elevator no such pulley is needed. Instead, we rely on the fact that the force caused by gravity and centrifugal force is downward near the ground but upward beyond GEO.

\begin{figure}
\begin{centering}
\includegraphics[width=\textwidth]{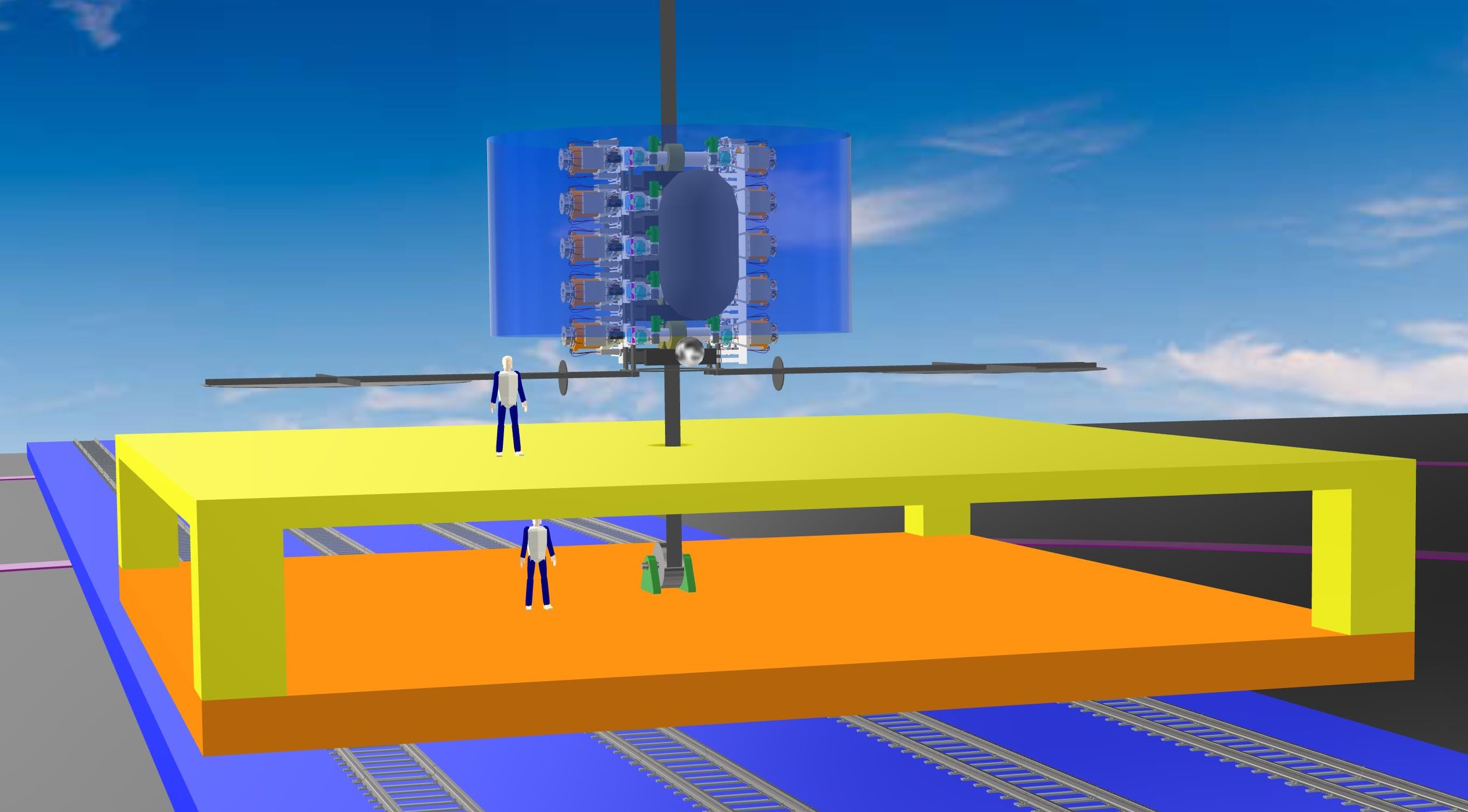}    
\caption{\label{fig:turntable} Concept model of the reel on a lower level of a moving platform. This moving platform concept applies to tether base motion of +/-100 m to avoid orbital debris and satellites.}
\end{centering}
\end{figure}

\begin{figure}
\begin{centering}
\includegraphics[width=\textwidth]{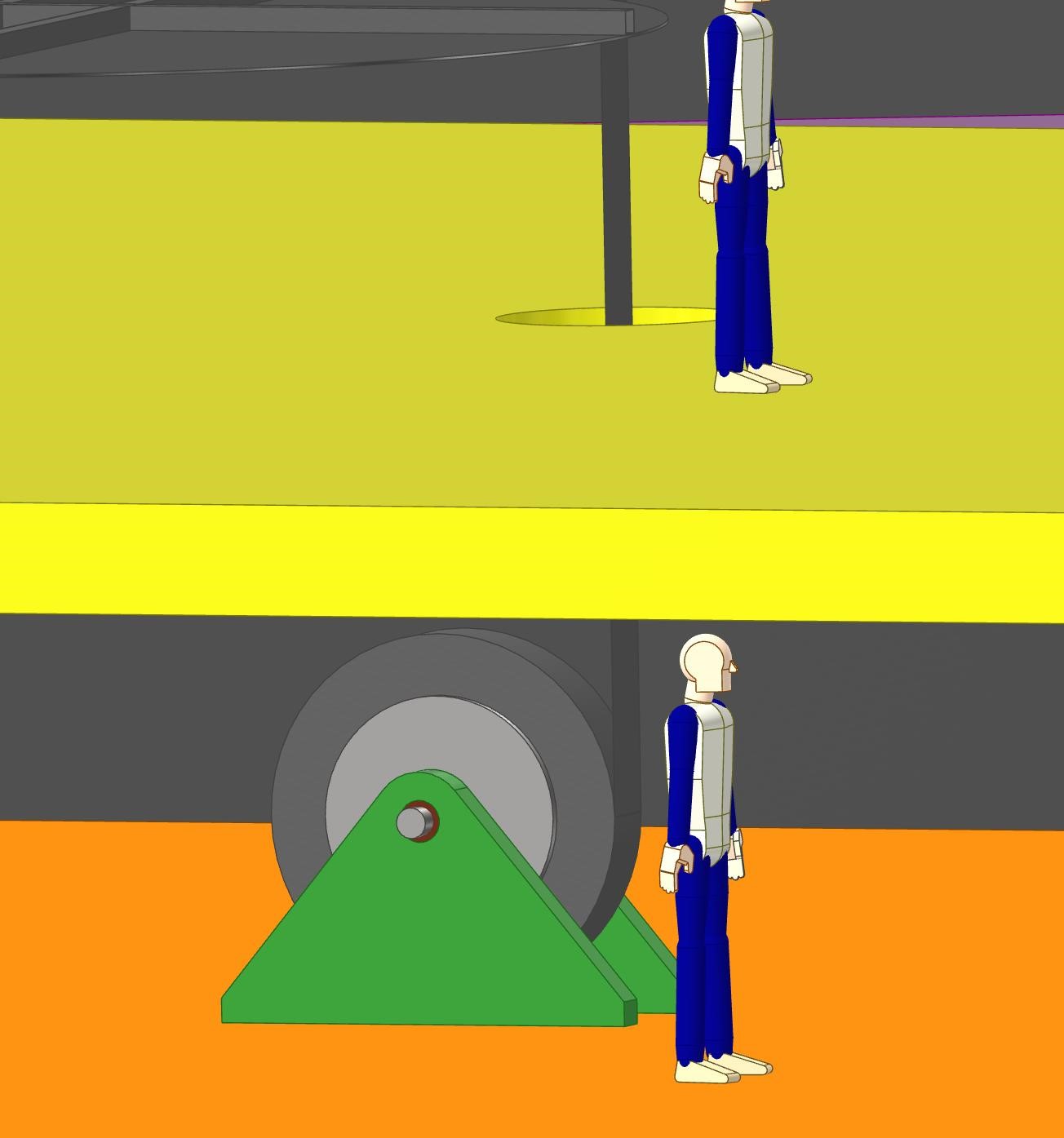}    
\caption{\label{fig:reelclose} Close-up view of a concept reel holding about 100 km of tether.  No actuating mechanism is shown in the model.  The tether is shown as a ribbon in these pictures, ignoring the effect of wind drag on the tether.  The tether cross-section may be circular, like a rope, to reduce drag.  Changes would be needed on the climber wheels for this.  The topic remains for a future study to determine.}
\end{centering}
\end{figure}

Figure \ref{fig:turntable} shows a concept model of a reel holding about 100 km of tether material that can be reeled in or out.  This figure and Figure \ref{fig:reelclose} were made as part of a study of the Earth Anchor platform assuming that the base of the tether was only required to move back and forth by +/- 100 m to allow climbers to clear satellites and orbital debris.

In previous literature, this launch method has been called the spring-forward method, because of the finite speed of propagation of waves along the elevator tether. When ribbon is fed out from the Earth Anchor, it takes hours before that information makes it up to the Apex Anchor through wave propagation. Yet, the climber does immediately get lifted, maintaining the tether below it in tension. In the short term, the energy to lift the climber is coming from the elastic energy in the tether. Once the wave reaches the Apex Anchor, the Apex Anchor will begin to move upward, transferring its potential energy back into tether elastic energy. The fact that the energy initially came from the elastic energy in the ribbon gave the spring-forward method its name, and while accurate, the name distracts from the fundamental simplicity of the scheme. By calling this the reel-out method, we allow it to be understood as being essentially Figure~\ref{fig:pulley-lift}, which anybody can understand. In practice, we will need to understand those dynamics if we wish to minimize the energy dumped into tether waves, but that understanding is not necessary to understand the basic concept.

\subsection{Benefits of reel-out launch}

One immediate benefit of the reel-out method is that it allows rectenna or photovoltaic arrays to be stowed while the climber is in the atmosphere, minimizing their wind drag. This should allow flimsier structures to be used, keeping weight low. Another wind-driven concern is that excessive lateral forces on the elevator can cause it to be pulled down, particularly when the tension in the ribbon is low as is the case right after a climber has launched [\cite{gassend-wind}]. Keeping the climber’s cross-section minimized in the atmosphere can help with this concern. Moreover, the ribbon being reeled out below the climber does not need to be climbable, and it is not exposed to a space debris environment. It can therefore be as narrow as possible to minimize its wind cross-section, reducing elevator pull-down susceptibility.

The second benefit of the reel-out method is the added flexibility it gives to power beaming installations. Beaming power to the climber as it first leaves the ground adds constraints to the power beaming system: either it has to be located directly beside the tether’s anchor, or the power beaming has to be operable with a near-horizontal beam. Near-horizontal operation would force receive arrays to be swivelable, adding weight and complexity to the climber, and would add large elevation range requirements to the beaming station used for the first part of the ascent. Being able to locate the power beaming facility at a distance from the anchor also conveniently decouples the power beaming facility from the motions of the anchor that will be needed for orbital debris avoidance.

The reel-out launch likely has other benefits. For instance, in an electrically powered tether, it could limit the range of ambient atmospheric pressures in which the climber has to be powered, which could remove constraints from the design caused by corona discharge and arcing.

\subsection{Spooling}

One practical consideration with reel-out launch is the spooling of the tether. The spooled tether has a narrow cross-section for wind loading, and hence it will have to be wound on a spool in a way that avoids tangling. This type of difficulty is already present with winches and cranes and can presumably be solved with careful design.

The total volume of the spool can be computed from the climber mass. The cross-section of the ribbon is given by $A=m_cgf/\sigma$ where $m_c$ is the climber mass, $g$ is the acceleration of gravity, f is the safety factor on stress in the tether, and $\sigma$ is the maximum working stress of the ribbon. Multiplying by $L$, the length of tether we wish to reel out, the volume of the spool is $V=mgfL/\sigma$. Plugging in reasonable values, we get 
$$V=\frac{20\times10^3\ \mathrm{kg} \times 9.81\ \mathrm{m/s^2} \times 1.5 \times 100\ \mathrm{km}}{130\ \mathrm{GPa}}=0.23\ \mathrm{m^3}$$
\
which is a manageable volume.

\subsection{Coriolis effect}

With reel out launch, we are moving the whole space elevator rather than
just a climber

\subsection{How far can reel-out launch go?}

The reel-out launch concept is so simple that one wonders whether it could be used to power a climber all the way to GEO. Unfortunately, this would not be straightforward.

The first problem that comes to mind is that we want a narrow tether in the atmosphere and a wide one within a couple of hundred km where orbital debris starts picking up. This limits the reel-out launch to a few hundred km unless some scheme can be found to have a tether that can be converted from narrow to wide as it moves from the atmosphere into the debris belts.

\begin{figure}
\begin{centering}
\includegraphics[width=\columnwidth]{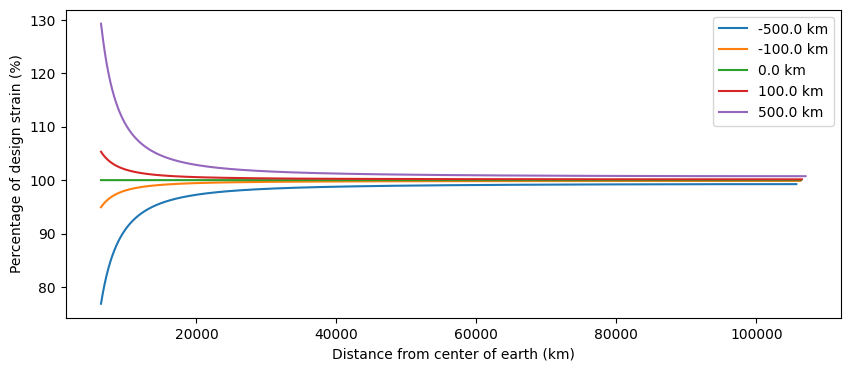}    
  \caption{\label{fig:reel-out-strain} Strain in the elevator relative to
  the design strain for an elevator at equilibrium with
  various amounts of tether reeled in or out. A
  100~km displacement will barely be noticed. 500~km down will erode the
  tension to the point where it may no longer be possible to support a
  climber, and 500~km up erodes a large fraction of the factor of safety.
  There is roughly a 5.2\% change in relative strain per 100km of reeling,
  and the equilibrium counterweight location moves 31\% more than the
  amount of reeling due to tether elasticity.}
\end{centering}
\end{figure}

Ignoring these tether width considerations, we encounter additional difficulties. As more tether is reeled out, the Apex Anchor gets farther from Earth and experiences an increasing centrifugal force, which causes an increase in tension in the whole ribbon, quickly exceeding structural safety factors. Figure~\ref{fig:reel-out-strain} shows the tension at the Earth anchor and Apex Anchor of the elevator if it is reeled in or out. The reel-out approach will work well as long as that tension does not deviate much from the nominal tension.  As can be seen, reel-out can easily work for a few hundred kilometers, but starts to notably erode factors of safety past that point. This is sufficient to get out of the atmosphere, and allows the beaming facilities used at the lowest altitudes to be at least a few kilometers from the anchor site. With this much reeling, we can have a narrow ribbon in the atmosphere for the wind, a wide ribbon in LEO for debris impact robustness, and a region between the two where both a wide and a narrow tether are acceptable, and will occur as the ribbon is reeled in and out.

The problem of excess tension at the anchor and Apex Anchor might be mitigable by increasing the factor of safety on the tether strength, at a high material and deployment cost. Alternatively, the Apex Anchor could be equipped with a reel-in capability, which some are advocating for anyhow [\cite{Robinson2023}], to maintain itself at a constant distance from Earth. Reeling in at the Apex Anchor would require power, however, which would have to be provided 100,000 km from Earth. This could be done using solar panels, but maintenance needs would have to be addressed. When the Earth Anchor reels the tether back in, the Apex Anchor would now have to reel out, generating power which would have to be dissipated.

Additionally, this mode of operation would make inefficient use of the elevator, as the elevator would have to frequently alternate between upward and downward reeling. In a first phase, climbers would attach to the tether and be reeled out until they get to GEO. At some point, this would have to stop, and the tether would have to be reeled back in so the cycle can repeat itself. The downward period could be used for descending payloads, but each change of direction would still create windows of time where payloads put onto the elevator would only go part-way to GEO before the elevator’s direction is reversed. While these part-way payloads could possibly be put to use, they would add operational constraints and complexity to space elevator operations.

\subsection{Other reeling approaches}

Reeling is a very effective way of transferring energy to a climber and moving it vertically. Energy is easy to provide at the Earth Port and is transferred with minimal losses to a climber anywhere along the tether. The climber is very simple: it just has to hold on. And the reeling at the base could be done with 19th-century technology. This is in stark contrast to the challenges of beaming power using microwaves or light. So even though the simple reel-out method can't get us past LEO, it is worth exploring whether there are other ways of using reeling to power climbers.

First, the need to alternate between upward and downward motion could be avoided by having a loop elevator. This type of concept has been explored [\cite{cablelift}] and [\cite{tapertether}], and may be practical in shallower gravity wells than Earth’s as long as difficulties such as tangling between the upward and downward tether can be avoided. The need for tapering, however, makes these schemes costly in tether material or impossible for an Earth elevator, depending on the actual tether strength-to-weight ratio. One could consider proposals in which there are multiple loops with different cross-sections, each covering a different height range, and transferring motion and climbers from one to the next. While this class of concept is appealing, there have not yet been any concrete proposals.

Instead of a loop, one could imagine two tethers alternating between upward and downward motion at opposite times, with a climber transferring its weight from one to the other as necessary to make continuous progress. This approach avoids the problems associated with keeping the ribbon’s taper where it belongs, and with the Apex Anchor’s height varying excessively, but does not avoid concerns about the two tethers tangling. More importantly, the ribbon’s elasticity would make it difficult to efficiently transfer between the two ribbons. If done naively, each transfer would result in the climber losing height as the tether it was on sprung up, while the one it transferred to sagged as it took on the climber’s load. If there is a viable solution here, it has yet to be uncovered and fleshed out.

Another approach to alternating motion is to use a single tether but increase the frequency at which alternation happens, allowing vibrations of the ribbon to transfer power to the climber, which the climber then uses for climbing. This will be the topic of section \ref{sec:mechanical-vibrations}.

\clearpage

\section {Mechanical vibrations}
\label{sec:mechanical-vibrations}
Like any tether under tension, the space elevator can carry mechanical waves. These waves can be used to carry energy, which could potentially be used to power a climber. In this section we shall introduce the reader to this approach and review prior work.

\subsection{Waves in a tether}

\begin{figure}[h]
\begin{centering}
\includegraphics[width=.9\textwidth]{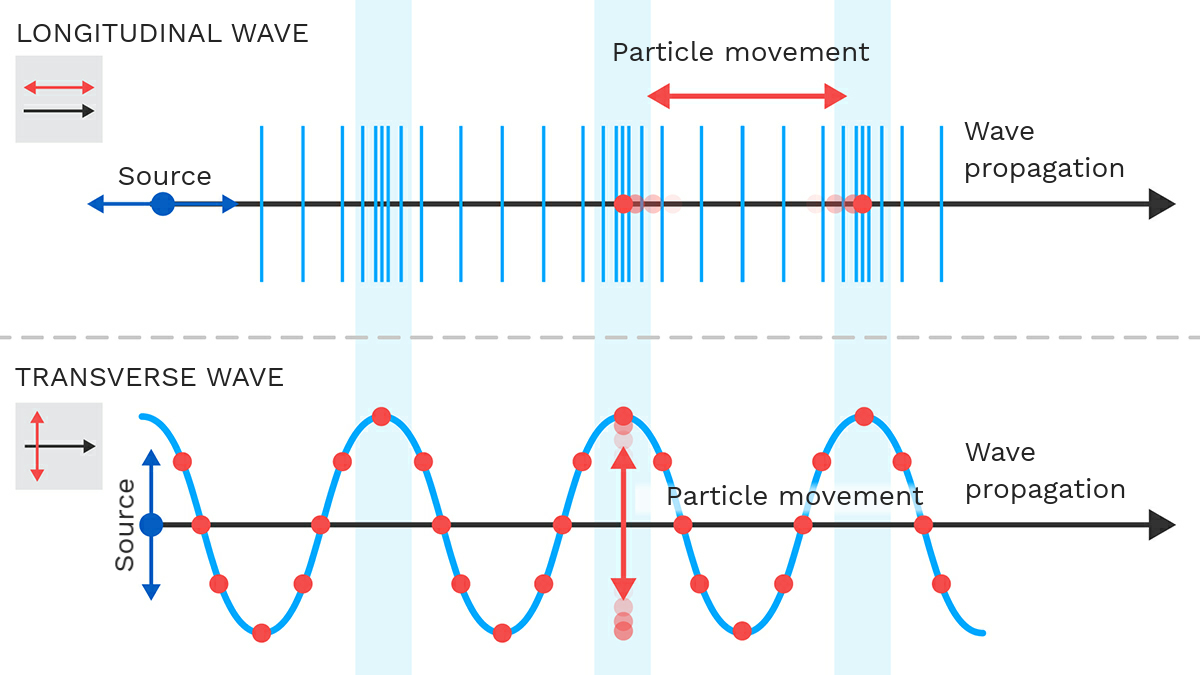}    
\caption{\label{fig:trans-long} The top picture shows longitudinal waves propagating along the tether as regions of higher and lower elastic strain.  (The elastic strain is higher where the blue bars are further apart.) The lower picture shows transverse waves in the tether.  Particle motion in longitudinal waves is parallel to the direction of propagation, and it transverse waves it is perpendicular. From [\cite{Labster}].}
\end{centering}
\end{figure}

There are two types of waves that propagate along a tether under tension: longitudinal waves in which the tether stretches and compresses, and transverse waves in which the tether undergoes side-to-side motion. See Figure \ref{fig:trans-long} for a diagram of how particles in the tether material move in these two oscillations. We shall use $\perp$ when referring to quantities that relate to transverse
waves, and $\parallel$ for longitudinal waves.

These two types of waves propagate at respective velocities:

\begin{align}
  c_\perp & = \sqrt{\frac{\sigma}{\rho}} \approx 6.3\ \mathrm{km/s} \\
  c_\parallel & = \sqrt{\frac{E}{\rho}} \approx 21\ \mathrm{km/s},
\end{align}
where $\sigma$ is the nominal stress at which the elevator is being operated, and assuming $\sigma = 90\ \mathrm{GPa}$, $\rho = 2300\ \mathrm{kg/m^3}$ and $E = 1\ \mathrm{TPa}$. 

When one moves the end of a tether in tension, one experiences a drag force $F_\mathrm{Z}$.\footnote{One also experiences other forces caused by the existing motion of the tether, and if one wiggles one side of a tether and waits for the wave to propagate to the other end of the tether and back, one will experience forces due to those reflections too.} It is a linear drag force, with coefficient $Z$, which we will call the mechanical impedance of the tether.

\begin{equation}
F_\mathrm{Z} = Z v
\end{equation}

If you start to move the base of the tether at a steady speed $v$ a wave will propagate up the tether. At locations not yet reached by the wave, the speed is zero. At locations at which the wave has arrived, the speed of the tether is $v$. As the wave passes a point, the tether suddenly accelerates. The drag force experienced at the end of the tether is exactly the force needed to cause that acceleration, which is given by:

\begin{equation}
  F = c \rho A v,
\end{equation}
where $A$ is the tether cross-sectional area and $\rho$ is its density.

This yields the impedance:

\begin{align}
  Z & = c \rho A \\
  \frac{Z_\perp}{A} & = \sqrt{\sigma \rho} \approx 14\ \mathrm{N/(m/s)/mm^2} \\
  \frac{Z_\parallel}{A} & = \sqrt{E \rho} \approx 48\ \mathrm{N/(m/s)/mm^2} \\
\end{align}

The presence of this drag force also indicates that power is being injected into the tether as it is wiggled. That power is given by:

\begin{align}
  \label{eq:vibration-power-limit-speed}
  \mathcal{P}  = F_\mathrm{Z} v & = Z v^2  \\
  \label{eq:vibration-power-limit-force}
              & = \frac{F_\mathrm{Z}^2}{Z}, 
\end{align}
where one can observe that the unit for $Z$ can also be written $\mathrm{W/(m/s)^2}$.

Thus, we see that the power that can be sent into the tether is limited by the speed $v$ at which the tether can be moved, as well as by the drag force $F_\mathrm{Z}$ the tether can withstand. Speed will be limited by aerodynamic heating and the usual material-strength problems faced by high-speed rotating machines such as wheels or flywheels. 

In the case of longitudinal waves, $F_\mathrm{Z}$ is a load that has to be carried by the tether. Hence, additional tether cross-sectional area will have to be provided to carry this load. For a tether whose sole purpose is to transfer power using waves, and with maximum working strength $\sigma_\mathrm{max}$, you could set the nominal tension in the tether to $\frac{\sigma_\mathrm{max}}{2}$ and transfer power by alternating between pulling at $\sigma_\mathrm{max}$, and pushing at near-zero stress. In either direction the power would be:

\begin{equation}
  \frac{\mathcal{P}_\parallel}{A}
   = \frac{\left(\frac{\sigma_\mathrm{max}}{2} A \right)^2}{Z_\parallel A}
   = \frac{\sigma_\mathrm{max}^2}{4 \sqrt{E \rho}}
   \approx 42\ \mathrm{MW/mm^2}
\end{equation}

But this power would be achieved at the substantial speed of $\frac{\sigma_\mathrm{max}A}{2 Z_\parallel} \approx 940~\mathrm{m/s}$.

For transverse waves, to first order, the waves do not cause an increase in tether stress, so one would have to look at second-order effects to understand the maximum power that can be transferred. For transverse waves, the drag is sufficiently high that speed is the practical limit.

Table~\ref{tab:max_vib_power} shows the power per unit-cross-section that
can be put in a graphene tether based on velocity and stress limits.

\begin{table}[h]
\begin{centering}
\begin{tabular}{l r@{\,} l r@{\,} l}
\hline
& \multicolumn{2}{c}{Transverse}
& \multicolumn{2}{c}{Longitudinal} \\
\hline
$Z/A$
  & 14  & $\mathrm{N/(m/s)/mm^2}$
  & 48  & $\mathrm{N/(m/s)/mm^2}$ \\

$\mathcal{P}/A$ @ 45\,GPa
  & \multicolumn{2}{c}{Ill-defined}
  & 42  & $\mathrm{MW/mm^2}$ \\

$\mathcal{P}/A$ @ 200\,km/h
  & 45  & $\mathrm{kW/mm^2}$
  & 150 & $\mathrm{kW/mm^2}$ \\

$\mathcal{P}/A$ @ 1000\,km/h
  & 1.1 & $\mathrm{MW/mm^2}$
  & 3.7 & $\mathrm{MW/mm^2}$ \\
\hline
\end{tabular}  
\caption{\label{tab:max_vib_power}
The impedance and stress/velocity limited power per unit area for a graphene tether.
 }
\end{centering}
\end{table}

\subsection{Prior work}

A few published works exist on using mechanical waves to power a space elevator climber.

[\cite{lofstrom2017acoustic}] proposed using longitudinal waves between 100 and 1000 Hz (wavelength between 25 and 250~m) with 100~m/s oscillation velocity to transfer 2-6~MW. The corresponding displacement is 1.2 to 18 cm. The paper contains numerous insights. He leaves the designs of wave transmitters and receivers open, but sketches an anchor where the tether could be attached to a rotating wheel used to inject the vibrations, and a climber in which the ribbon follows a back and forth path suggesting that it could be desirable for the length of tether contained within the climber to be variable. He mentions that descending climbers could inject vibrations into a tether to dissipate power. He considers the effect of tether taper on the waves, but incorrectly assumes that the taper will cause large amounts of power to be back-reflected. 

\begin{figure}[h]
\begin{centering}
\includegraphics[width=.8\textwidth]{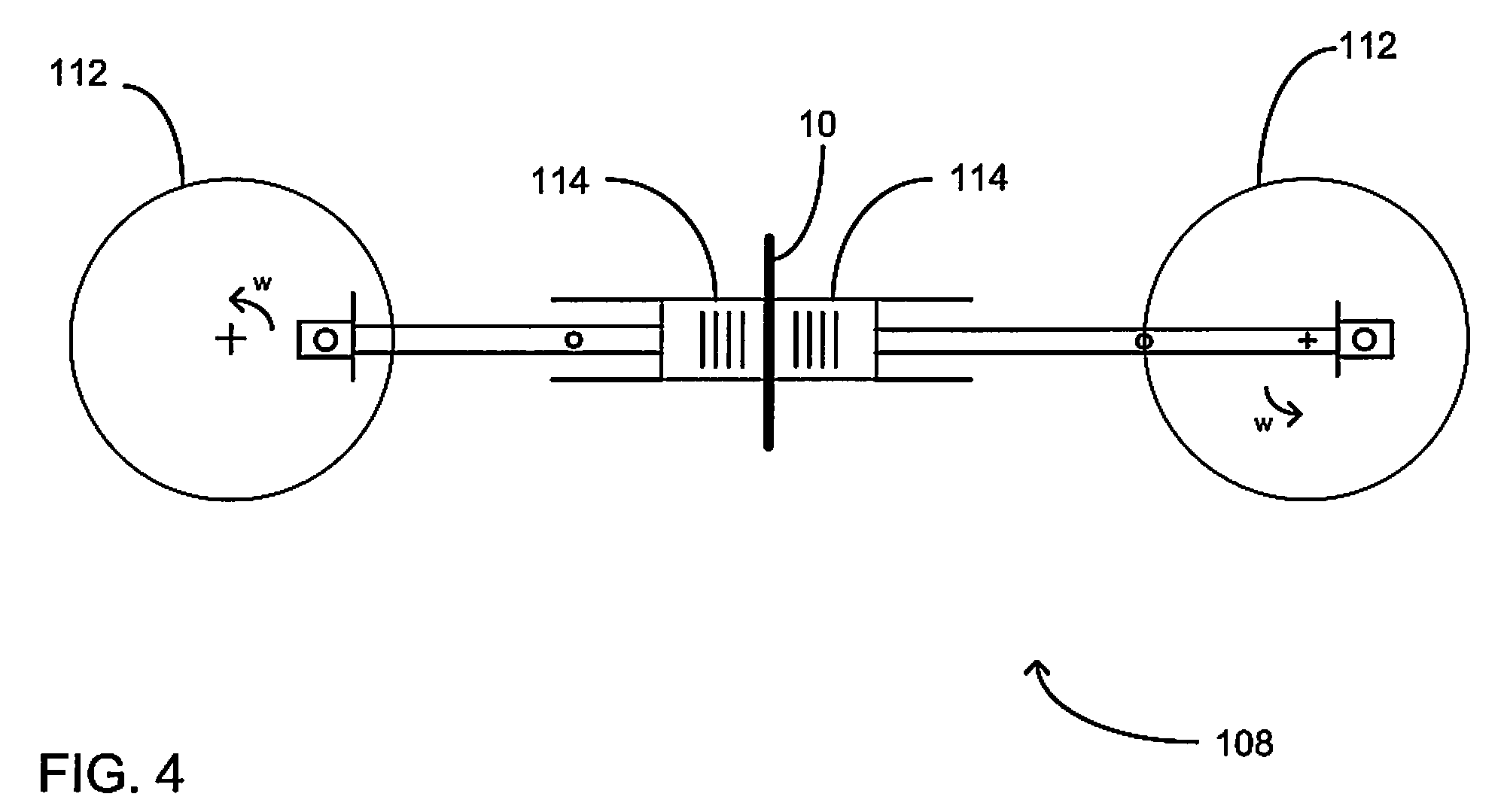}    
\caption{\label{fig:pistonpatent} Figure 4 from Mark Wessels' US Patent 8,196,867, \textit{Space elevator propulsion system using mechanical waves}. This design has a slider crank on each side of the tether and pistons push the tether back and forth, causing transverse waves. }
\end{centering}
\end{figure}

\begin{figure}
\begin{centering}
\includegraphics[width=.8\textwidth]{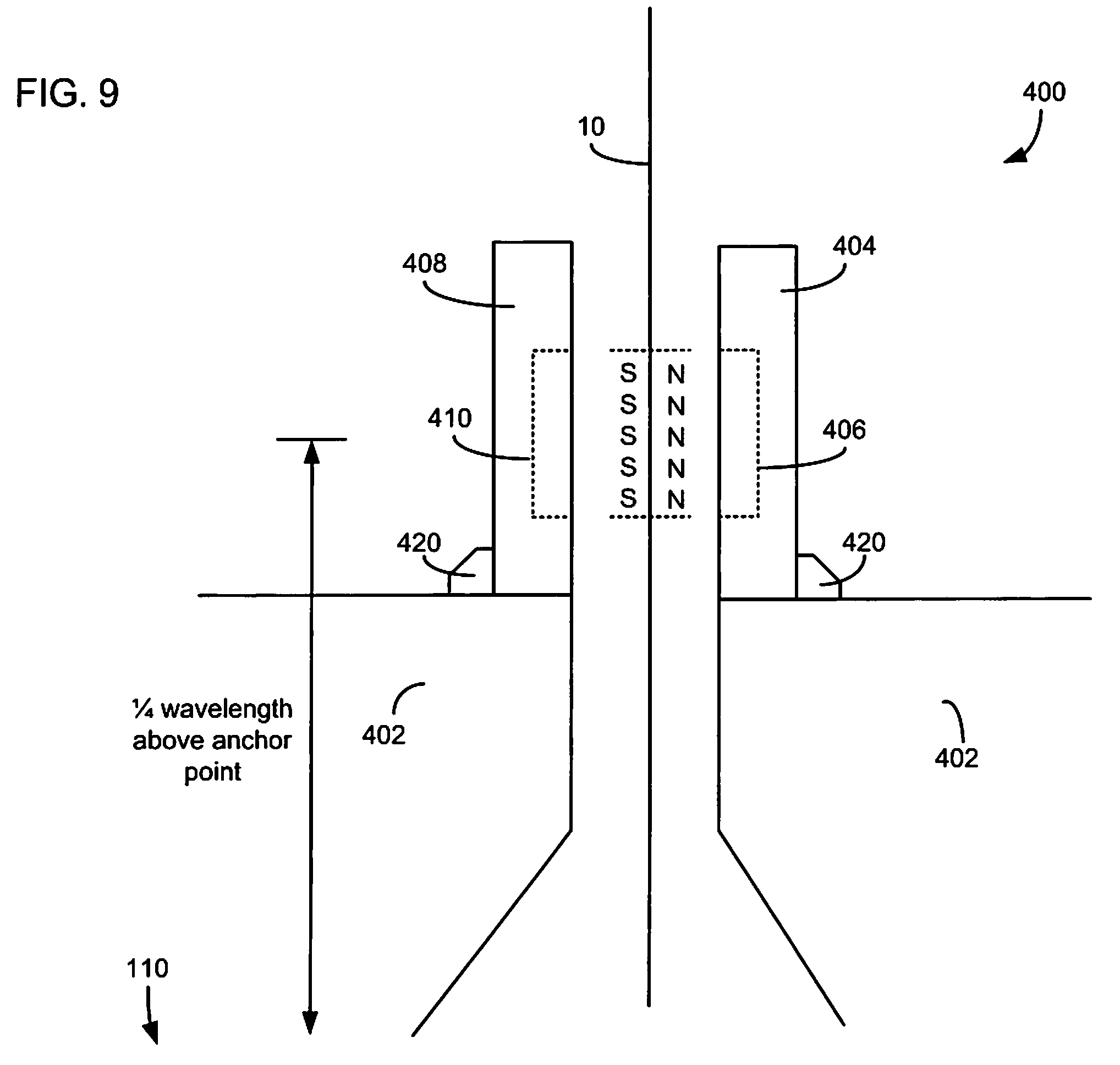}    
\caption{\label{fig:electpatent} Figure 9 from Mark Wessels' US Patent 8,196,867, \textit{Space elevator propulsion system using mechanical waves}. This design uses electromagnets on opposite sides of the tether to produce transverse waves.}
\end{centering}
\end{figure}

[\cite{wessels2018mechanical}] independently published a 3.23 MW design using transverse waves at 92 Hz with 40 m wavelength and 60 cm displacement, which should correspond to an oscillation velocity around 350 m/s. He suggests having a fixed attachment point for the tether and driving it with hydraulics at a quarter-wavelength above the attachment point, so that reflections from the attachment point would interfere constructively. He also suggests a rotation-driven design to drive the vibrations (see Figure \ref{fig:pistonpatent}) and proposes an electromagnetically driven one (see Figure \ref{fig:electpatent}). On the climber side, he focuses on the need for a design that does not reflect waves and proposes a design where a series of extractors absorbs part of the incoming wave energy. In a good design, the reflections caused by each of the extractors should interfere destructively, avoiding reflected power. (This would be the case if the extractors are spread over a whole number of wavelengths.) The extractors are implemented with pistons in contact with the ribbon. As they move, they cause a magnet to move within a coil, transducing the power into an electrical form that can drive motors.

[\cite{gassend2025mechanicalwaves}] took a more general approach, which the present section is based on, considering both transverse and longitudinal waves at a variety of wavelengths, and focusing on power transfer per square millimeter of tether. He studied the effect of atmospheric drag, finding that it is a showstopper for transverse waves, but only a minor penalty for longitudinal waves. He also recognized the high speeds needed to drive sufficient power into the tether as a major challenge, and proposed techniques which we shall describe, to reduce them.

\subsection{Mechanical waves in a space elevator}

Mechanical wave propagation in a space elevator is substantially the same as it is in any other tether. If one carefully writes out the equations for waves in a space elevator, one will find terms relating to gradients in the gravity/centrifugal field that the elevator is in. While these terms are important for understanding the lowest multi-hour frequency modes of the elevator, once one gets down to periods of a few minutes or less, they are completely negligible and need not be considered.

The space elevator is also a tapered structure. This taper will influence wave propagation along its length. If there are sudden changes in cross-section, an impedance mismatch will form and reflections will result. On the other hand, when there are variations in cross-section over length scales much greater than the wavelength, we get an adiabatic transition where the amplitude of the wave slowly varies to keep the power flux constant without causing reflections. This is the case for the large-scale tapering of the space elevator. It will cause the amplitude to be proportional to $\frac{1}{\sqrt{A}}$, but will not cause any reflections, contrary to what is stated in~[\cite{lofstrom2017acoustic}].

Localized defects in the elevator could cause reflections. These defects could be caused by damage or by repairs/reinforcements. In the case of punctures there could be some missing mass as well as reduced stiffness. Repairs will likely show up as sections of tether to which a patch has been applied. These defects will cause an amount of reflection related to the relative change in mass and stiffness of one wavelength of tether between the pristine state and the damaged state. To minimize reflections, the wavelength should be kept large relative to the expected size of defects. When patching, patches should either be very short compared with the wavelength, be sized such that the reflections from the start of the patch interfere destructively with the reflections from the end of the patch, or include a taper at each end that is long compared with the wavelength.

During normal operation, the tension in the space elevator tether can undergo large variations, particularly near the base where the tension below a freshly launched climber may be many times less than the tension just above that climber. For longitudinal waves, the impedance and wave speed remain unchanged when the tension varies, but if the tension gets too low, the power transmission capability may be limited because the tension can't go negative, limiting the upwards speed at which the base of the tether can be moved. For transverse waves, variations in tension have an even greater effect since they will cause the speed and impedance to change. This means that the amplitude with which we drive at the base has to be large enough to get sufficient power transfer with a freshly launched climber. It also means that the wavelength is variable, which could make the design of a climber that receives the waves more challenging, or could limit/complicate the ability to use impedance steps to engineer wave reflectors or other structures along the tether.

Sending waves along the space elevator tether is not free. These waves
induce stresses in the tether, which will have to be designed for the sum
of the climber loads and the vibration loads. The elevator as a whole will
get heavier. But if we can transfer a few $\mathrm{MW/mm^2}$ as suggested
by~\cite{gassend2025mechanicalwaves}, then the amount of extra tether
material needed to support power transmission is a may be acceptable.

For vibrational power transmission to work, it will be important to consider the effects of fatigue. This may be a reason to prefer lower frequency vibrations.

Another vital consideration is to understand loss mechanisms for waves propagating in the tether. We shall look at atmospheric drag in detail, but there may also be losses at the material level. [\cite{lofstrom2017acoustic}] assumes they are sufficiently low, but this will have to be proven. Here again, lower frequencies are likely to be preferable.

\subsection{Atmospheric Drag}

\begin{figure}
\begin{centering}
\includegraphics[width=.7\textwidth]{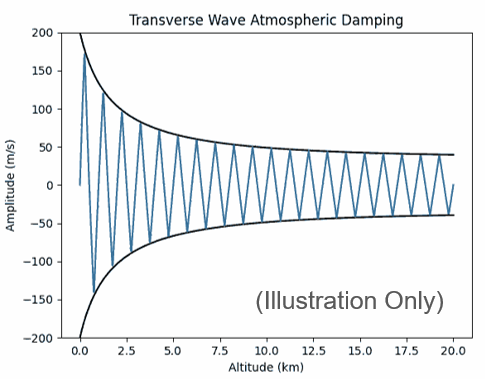}    
\caption{\label{fig:atmos-damping-illustration} Qualitative illustration of the amplitude of transverse vibrations decreasing as they propagate through the atmosphere. [\cite{gassend2025mechanicalwaves}]}
\end{centering}
\end{figure}

In the atmosphere, the space elevator tether undergoes drag that is
proportional to the square of its velocity. Since mechanical vibrations
imply motions of the ribbon, they will cause drag, which will cause power
loss. Figure~\ref{fig:atmos-damping-illustration} illustrates how
atmospheric drag will dampen the propagation of the wave as it progresses
through the atmosphere.

[\cite{gassend2025mechanicalwaves}] presents a formula that relates the speed $v_1$ of the vibration at the top of the atmosphere relative to the speed $v_0$ at the bottom of the atmosphere for a cylindrical cross-section tether and oscillations whose wavelength is short compared with the thickness of the atmosphere:

\begin{align}
  v_1 & = \frac{v_0}{1+v_0/v_\mathrm{max}}\mathrm{, where} \\
  v_\mathrm{max} & = \frac{2Z}{C_d\rho_a h_a}\sqrt{\frac{\pi}{A}},
\end{align}
where $C_d$ is the drag coefficient, $\rho_a$ is the density of the atmosphere at the surface of the earth, and $h_a$ is the characteristic height of the atmosphere.

\begin{figure}[h]
\begin{centering}
\includegraphics[width=.7\textwidth]{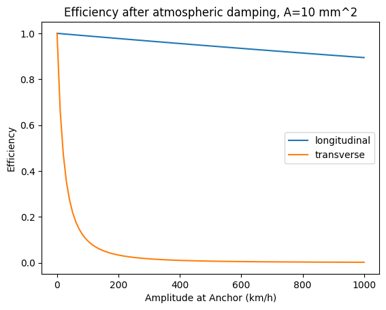}    
\caption{\label{fig:atmos-damping-efficiency} The efficiency of transferring vibrational energy through the Earth's atmosphere as a function of vibration amplitude. Transverse waves see excessive damping before substantial power can be transferred, whereas longitudinal waves see only minimal damping even at speeds where a launcher will be challenging to implement. [\cite{gassend2025mechanicalwaves}]}
\end{centering}
\end{figure}

For small amplitudes, the quadratic drag has little effect, but as $v_0$ approaches $v_\mathrm{max}$, the drag becomes substantial to the point that $v_1$ asymptotically approaches $v_\mathrm{max}$. No matter how hard the base of the elevator is wiggled, the amplitude at the top of the atmosphere will never exceed $v_\mathrm{max}$. Approaching that limit will lead to poor efficiency. For a $10~\mathrm{mm^2}$ cross-section tether, the value of $v_\mathrm{max}$ is about 13~m/s for transverse waves and 2400~m/s for longitudinal waves. This result completely rules out transverse waves as a viable option for powering climbers, initially. Payloads would need to be increased by several orders of magnitude for transverse waves to be viable. On the other hand, longitudinal waves are almost unaffected by drag.\footnote{This result assumes a cylindrical tether with $C_d=0.02$ in the longitudinal direction. A flat 30~cm wide tether would have more drag, but would also not be able to withstand wind loads [\cite{gassend-wind}]}. Figure~\ref{fig:atmos-damping-efficiency} shows the efficiency hit for transverse and longitudinal waves traversing the atmosphere.

\subsection{Reducing tether speed near the elevator base}

One of the major challenges with using vibrations to power the climber is the relatively low impedance of the tether, which forces it to move at velocities on the order of 300~m/s to be effective. These speeds, comparable to the speed of sound, imply significant structural difficulties for the machinery that creates them, and if there is any transverse motion of that machinery, substantial drag will result. For these reasons [\cite{gassend2025mechanicalwaves}] considered ways to reduce the drag.

\subsubsection{Thicker tether at injection point}
\label{sec:thicker-tether}

A first approach considered was to increase the tether cross-section near the Earth. Increasing the cross-section for the first 50 to 100~km increases the impedance and hence decreases the speed that the tether needs to be driven at. This is a relatively weak lever as $v=\sqrt{\frac{\mathcal{P}}{\rho A}}$ so there is only a square root dependence of tether speed on tether cross-section.



Increasing the tether cross-section only in the lower portion of the ribbon could cause reflections when the cross-section is reduced higher up. These reflections could be prevented by using slow tapering. An anti-reflection layer, in which the cross-section is reduced in two steps separated by a quarter wavelength, a technique common in the microwave and optical fields, can also be used.

\subsubsection{Using resonance}

\begin{figure}
\begin{centering}
\includegraphics[width=.7\textwidth]{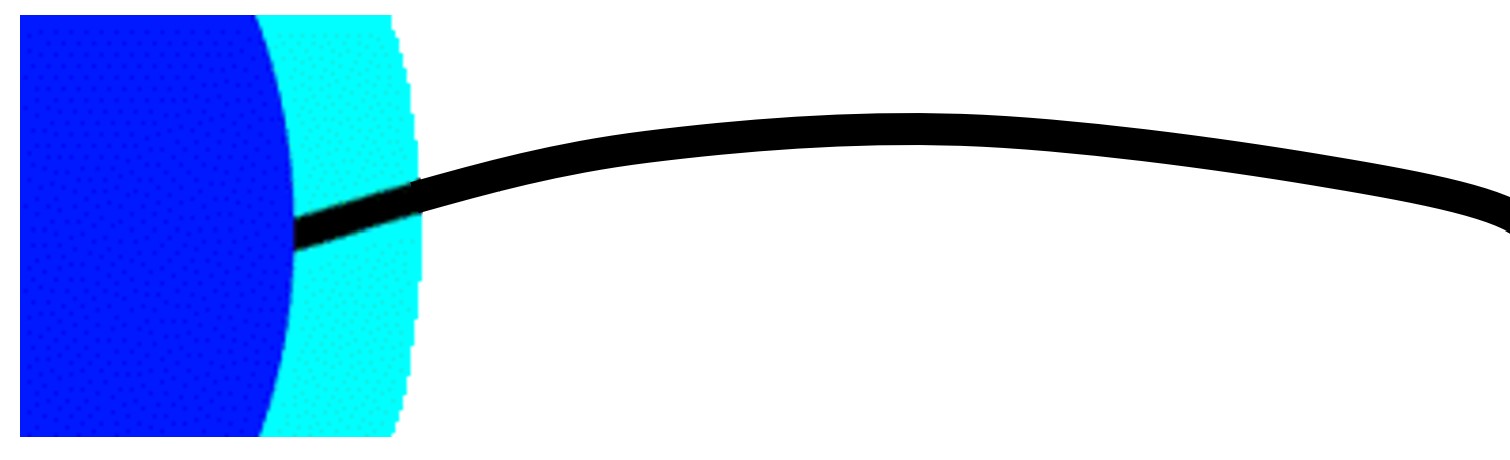}    
\caption{\label{fig:standwave} The earth, atmosphere, and one
  half-wavelength of tether showing the space between two nodes. The
  displacement needed from the machinery at the base is reduced, and if the
  half-wavelength is long enough, damping in the atmosphere is reduced.
  [\cite{gassend2025mechanicalwaves}]}
\end{centering}
\end{figure}

Another promising technique is shown in Figure~\ref{fig:standwave}. It uses
resonance by creating a standing wave on the ribbon. The drive frequency
would be set such that there is a node at the climber and at the ground. At
nodes the velocity is low, but the force is high, corresponding to a high
effective impedance. However, at anti-nodes, the velocity is high, so
ideally the frequency would be selected to be low enough that the first
anti-node is well above the atmosphere. For transverse waves this means
periods in the high tens of seconds, or low tens of seconds for
longitudinal waves; though in the case of longitudinal waves the lower drag
coefficient may make higher frequencies possible.

Because a whole number of half-periods is needed between the ground and the
climber, there is a minimum altitude at which resonance can work for a
given frequency. This means that while the climber is low this technique
breaks down. Thus, even if transverse waves can see reduced drag once the
climber is several times higher than the atmosphere, at low altitude it
won't help, and the climber will have to climb more slowly. Pull-down
approaches (see Section~\ref{sec:pull-down}) can't get the climber high
enough for resonance to take over since they can't get the climber up to
LEO altitudes. So
this technique is of limited value in making transverse waves practical for
power transmission.

On the other hand, keeping the speed at the ground low means that the
machinery needed to launch waves and receive them at the climber doesn't
have to move as fast, which should make that machinery easier to implement.
Increasing the tether cross-section near the ground had a similar effect on
the ground side, but not on the climber side except when the climber is
very low. Ideally, the increased cross-section would be long enough that
vibration velocities are always kept low: by an enlarged cross-section near
the ground, and by resonance higher up.

As always, in the resonant configuration, the power transferred into the ribbon at the base is given by the product of the force and the velocity. As before the force is limited by $\frac{\sigma_\mathrm{max}}{2}$, and the velocity is limited to $v_\mathrm{max}$ by practical considerations related to the machinery. Thus power is limited by:

\begin{equation}
  \mathcal{P} < 
    \frac{\sigma_\mathrm{max}}{2} A v_\mathrm{max} =
    \sqrt{\mathcal{P}_\mathrm{\sigma\_lim}
    \mathcal{P}_\mathrm{v\_lim}},
\end{equation}
where $\mathcal{P}_\mathrm{\sigma\_lim}$ and $\mathcal{P}_\mathrm{v\_lim}$
are the maximum powers from Equation~\ref{eq:vibration-power-limit-speed}
and~\ref{eq:vibration-power-limit-force} for $F_\mathrm{Z} =
\frac{\sigma_\mathrm{max}}{2} A$ and $v_\mathrm{max}$, respectively.
So by using resonance, it is possible to go beyond the lower of $\mathcal{P}_\mathrm{\sigma\_lim}$ and \\
$\mathcal{P}_\mathrm{v\_lim}$, up to the geometric mean of these two powers.

Resonance has the challenge that the drive frequency has to be continuously adjusted as the climber climbs to remain in resonance. Another way to see this is that the wave reflected off the climber undergoes a Doppler shift, so to maintain the standing wave, the frequency of the wave launched from the ground also has to decrease. To avoid the frequency going too low, it will have to be stepped up periodically. Making the switch will require several round trips to build up a resonance at the new frequency while absorbing energy at the previous frequency. During this time, the amount of power that can be transferred may be reduced.

When there are multiple climbers on the tether, this problem becomes even
more challenging because the resonance frequencies for the length of tether
between the ground and the first climber will, in general, not be the same
as the resonance frequencies for the length of tether between the first and
second climber. Indeed, the lengths, rates of change of length, and Doppler
shifts are different. One way to address this problem is to only use
resonance between the ground and the first climber. This option is not ideal
because when one climber launches, the previous climber is still operating
at 4~MW, so oscillation velocity reduction remains important. The third
climber may be high enough that this approach may be acceptable. 

Another multi-climber option is to try to have the two climbers go at different speeds such that both tether segments can be in resonance as the climb progresses. Assuming that the higher climber is going at $v_h$, the lower climber's speed would be constrained to $v_l = v_h \frac{h_l}{h_h}$, which will make for a very slow start, even using the pull-down method.

Other approaches may be possible, such as operating two modes at once, one for the first climber and one for the second, or putting subtle impedance steps on the tether to produce suitable adjustments to the resonance frequency. Whether any of these options are viable remains to be studied.

\subsection{Injecting waves at the ground}

Compared with other forms of power beaming, injecting power into tether
vibrations is conceptually quite straightforward, particularly if we are
able to keep the velocity amplitude of the vibrations small. A few
strategies that have been considered:

\begin{itemize}
  \item \textbf{Rotation-driven injector}: In these designs, a wheel rotates
    and an off-center feature is connected to the tether causing it to move
    in a near-sinusoidal manner. In the simplest incarnation, the tether
    would simply be connected to the off-center feature, and two dimensions
    would be driven: either transverse and longitudinal if the disk is
    vertical, or a circularly polarized transverse wave if the disk is
    horizontal. With some extra refinements, a pure transverse or
    longitudinal wave is possible as in Figures~\ref{fig:longvib} and \ref{fig:transvibe}. 
    The simplicity of these approaches, which could be built with
    century-old technology, is tantalizing, and a stark contrast to, say,
    laser-beaming using a free-electron laser.

\begin{figure}[h]
    \begin{minipage}[t]{.45\textwidth}
        \centering
        \includegraphics[width=\textwidth]{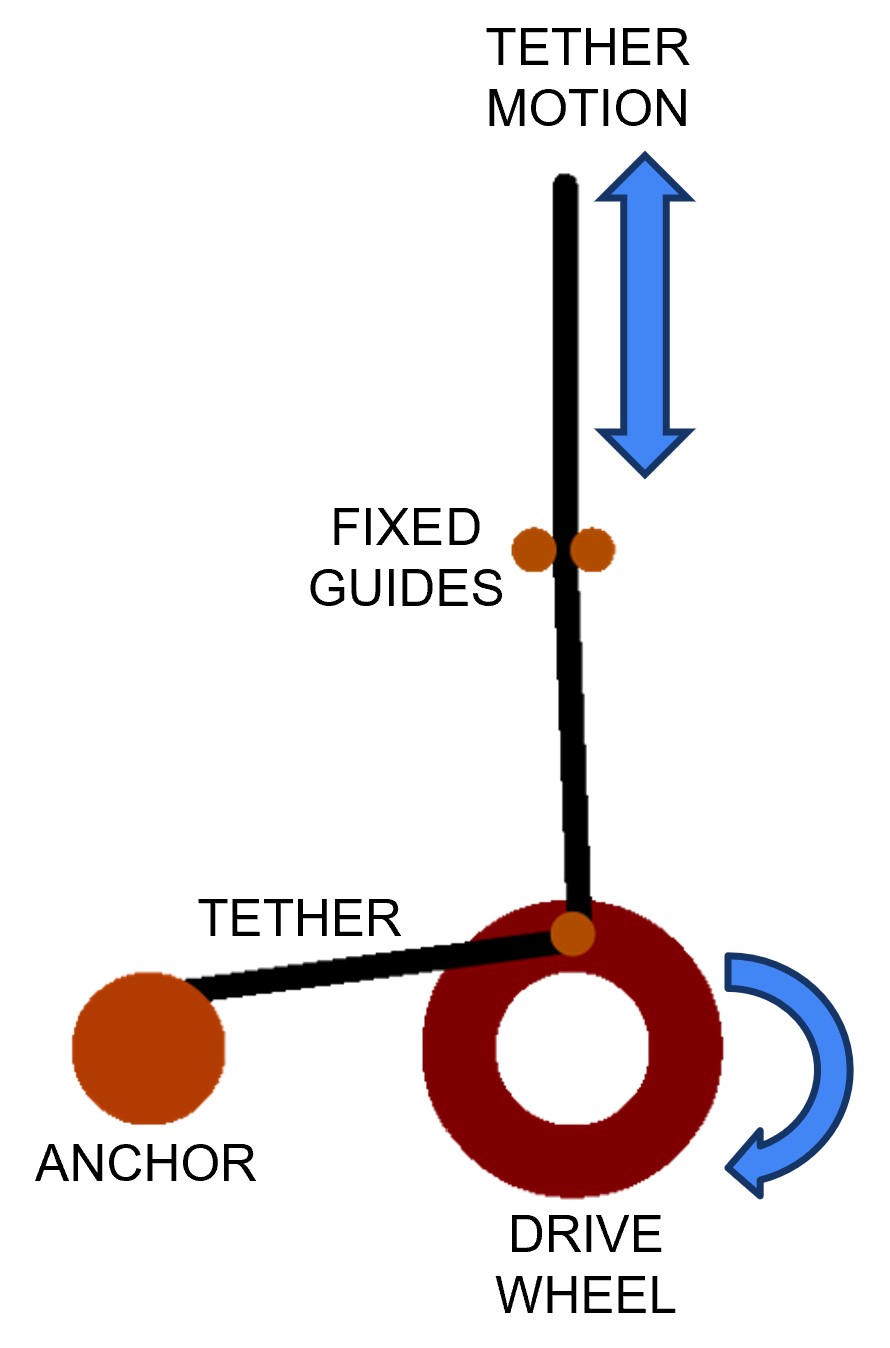}   
        \captionsetup{justification=centering}
        \caption{Concept sketch for a launcher for longitudinal vibrations on the tether}
        \label{fig:longvib}
    \end{minipage}
    \hfill
    \begin{minipage}[t]{.5\textwidth}
        \centering
        \includegraphics[width=\textwidth]{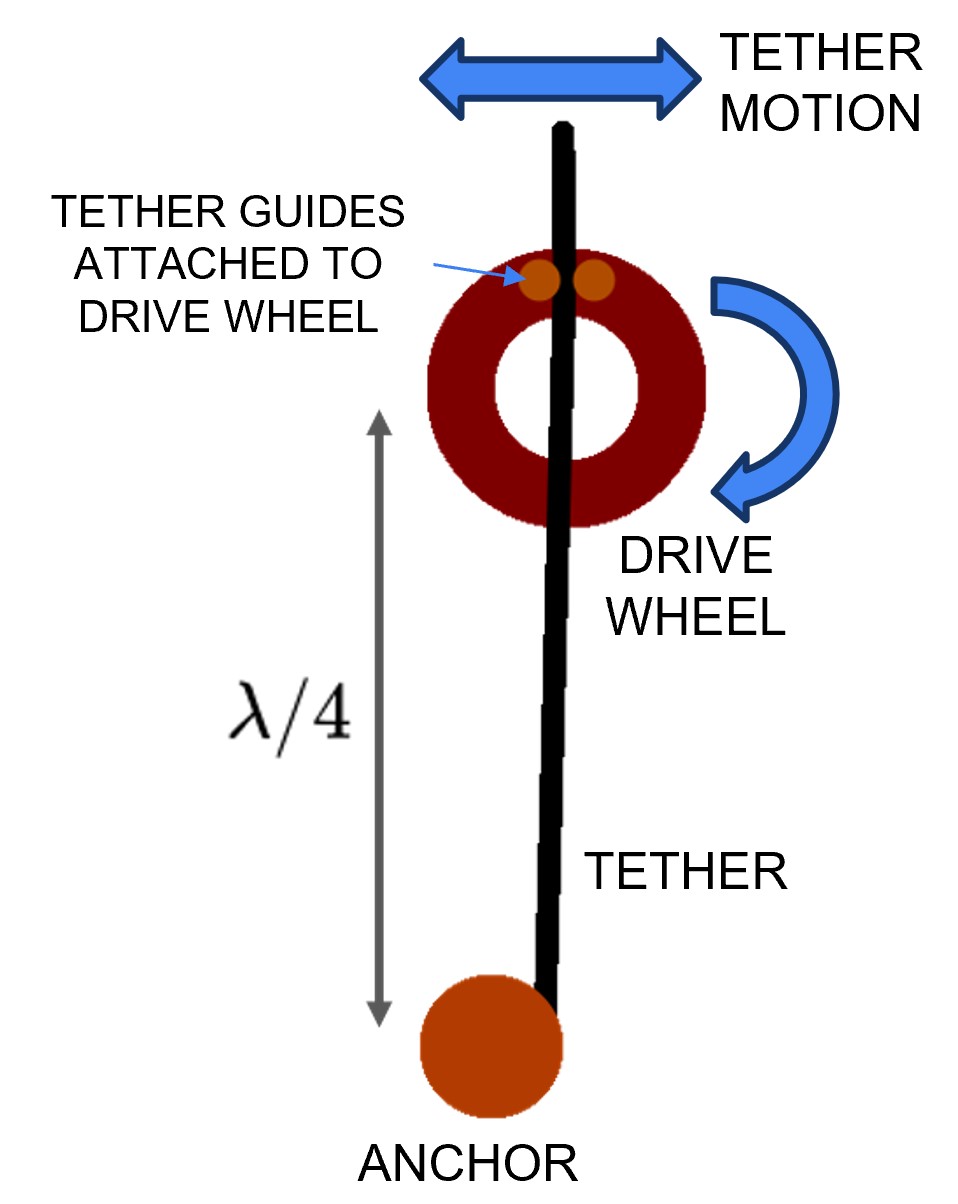}   
        \captionsetup{justification=centering}
        \caption{Concept sketch for a launcher for transverse waves on the tether}
        \label{fig:transvibe}
    \end{minipage}
\end{figure}

  \item \textbf{Linear actuators}: Various linear actuators are possible,
    such as magnetic~[\cite{wessels2018mechanical}] or hydraulic as in a jack
    hammer.  See Figures \ref{fig:pistonpatent} and \ref{fig:electpatent}.

  \item \textbf{Reel based}: If low frequencies are desired then the
    position amplitude of the vibrations may be too large for the methods
    above. For longitudinal waves, the ribbon could instead be attached to
    a reel driven by a motor. By successively reeling in and out, waves can
    be injected. The rotational inertia of the reel will be a challenge for
    higher frequencies.

\end{itemize}

Among all the approaches for sending power to the climber, these are
conceptually some of the simplest, which is one of the major reasons why sending power to the
climber via vibrations deserves further study.

\subsection{Absorbing energy at the climber}

\subsubsection{Ideal properties}

The problem of absorbing energy at the climber is still a wide-open
problem. While in principle it is a very similar problem to wave injection,
the climber does have some additional challenges due to the fact that it is
continuously moving relative to the tether: the tether can't be tapered
specifically near the attachment point (see
Section~\ref{sec:thicker-tether}) and the climber has to be able to move
relative to the tether at the same time as it is transducing vibrational
power off the tether.  There are a number of desirable properties to look
out for:

\begin{itemize}

  \item Controllable transmission and reflection coefficients: Ideally, the
    climber should be able to adjust the transmission and reflection
    coefficients experienced by an incoming wave. This capability is needed
    to control the power going to the next climber up and to set up
    resonances.  See Figure \ref{fig:incident}.

\begin{figure}[h]
\begin{centering}
\includegraphics[width=.7\textwidth]{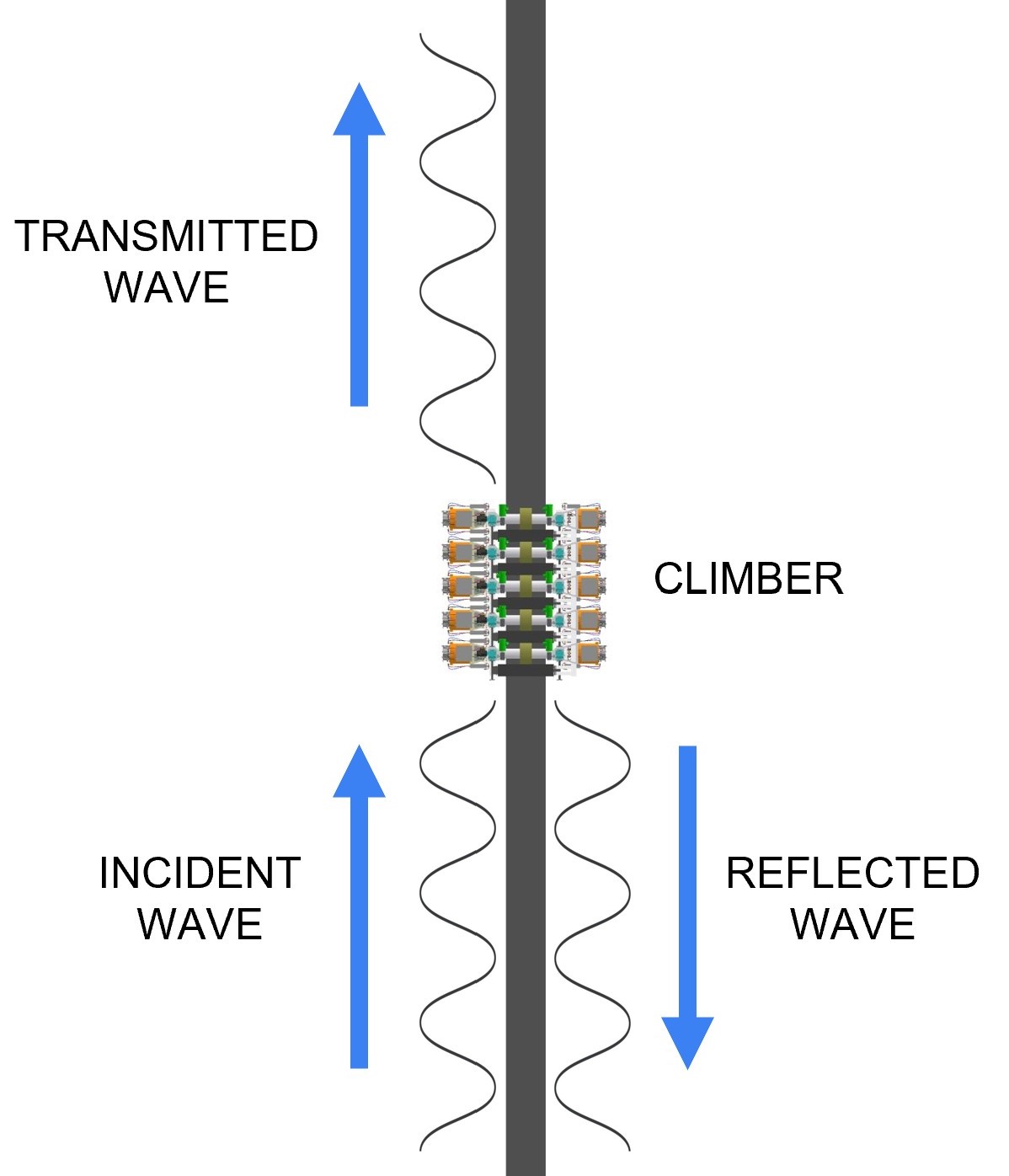}    
\caption{\label{fig:incident} Picture showing a climber receiving an incident wave, having some wave energy reflect back down the tether, and transmitting some wave energy up to the next climber higher on the tether. Some of the energy may also be absorbed, and in more complex scenarios there may be energy incoming from above.}
\end{centering}
\end{figure}

 \item Constant speed for the climber: The vibrations of the tether can
   cause a cyclic force to be applied to the climber, which would cause its
    speed to vary in time with the tether vibrations. For cargo these
    vibrations may be acceptable as long as the cargo is designed to
    support them, but for humans they could be unacceptable. For high
    frequencies, vibrating or rotating masses may be able to compensate for
    vibrations induced from the tether, or mechanical filters can isolate
    parts of the climber. For low frequencies once the mass of a wavelength
    of tether stops being negligible relative to the mass of the climber,
    these approaches will become impractical.

 \item High efficiency: The mechanism has to provide good efficiency.

\end{itemize}

\subsubsection{Basic climber forms}

The climber can take on several different forms:


\begin{itemize}
  \item \textbf{Short Climber with Constant Tether Path}: This is a
    climber that is short compared to the wavelength of the vibrations and
    that has a constant length of tether within it. This climber has tight
    constraints on how much energy it can absorb. All it can do is wiggle
    the tether at the point at which it is located. This wiggling causes a
    wave to propagate both up and down the tether. The result is a tether
    with three waves propagating along it: the incoming wave unperturbed by
    the climber, a wave going up from the climber and a wave
    going down from the climber. Both waves emanating from the climber have
    the same amplitude since they are caused by the same motion.

    If the upward wave is in the opposite phase from the incoming wave,
    then the overall wave above the climber has a lower amplitude than the
    incoming wave: some energy is absorbed or reflected. If the upward wave
    has the same amplitude as the incoming wave, but opposite phase, then
    there is no wave above the climber, but the amplitude of the downwards
    wave equals the amplitude of the incoming wave: total reflection. In
    order to maximize energy transfer to the climber, a balance between
    reducing the upward wave and not producing too much downward wave has
    to be struck. The optimum is when the upward and
    downward waves have to have half the amplitude of the incoming wave,
    and hence a quarter of the power each, leaving the remaining half of 
    the power to be absorbed by the climber.

    The inability to absorb all the power is not as bad as it sounds. If we
    are using resonance, we will need a large fraction of power to be
    reflected, and likewise we will need power to be transmitted if we are
    powering multiple climbers. Even for
    the top climber, transmitted power will be reflected from the
    counterweight (or from a Bragg mirror created using subtle impedance
    variations) and can be recovered after it is reflected down.

    In this approach, the climber is applying a net oscillating force to
    the tether, and its center of gravity will therefore have an
    oscillating speed, which it may be possible to isolate the payload
    from, as long as the frequency is high enough that the
    mass of a wavelength worth of tether is well below the climber mass.

  \item \textbf{Long Climber}: If the climber is long relative to the
    wavelength, ideally a quarter wavelength long, then it has more freedom
    in controlling the transmitted and reflected waves. Indeed, if forces
    are applied at two points a quarter wavelength apart that are out of
    phase by a quarter cycle, these waves will interfere constructively in
    one direction, and destructively in the other. So now the climber can transmit
    a wave in only one direction. If the dephasing is reversed, it can
    transmit in only the other direction. The long climber can therefore
    completely control the energy reflection, absorption, and transmission.
    This capability comes at a cost, however: for a longitudinal wave propagating
    at 21~km/s, a climber $l_c = 10~\mathrm{m}$ long is only a quarter wave
    long for frequencies above $\frac{c_\parallel}{4 l_c} \approx 525~Hz$. So
    to be long, we need to be operating at high frequencies, or the climber
    has to be quite long.

  \item \textbf{Climber with Variable Tether Path}: Another way the climber can
    gain more control is by varying the length of tether within it. The
    variable path means that the climber can now independently wiggle the
    tether above and below it since those two pieces are no longer rigidly
    bound to each other. Once again, complete control over reflection,
    transmission and absorption becomes possible, but without needing a
    long climber. However, the complexity of the climber goes up, and
    depending on the amplitude and wavelength, the amount of variability of
    the tether path may be up to twice the amplitude of the vibration. 

\end{itemize}
    
\begin{figure}
\begin{centering}
\includegraphics[width=\textwidth]{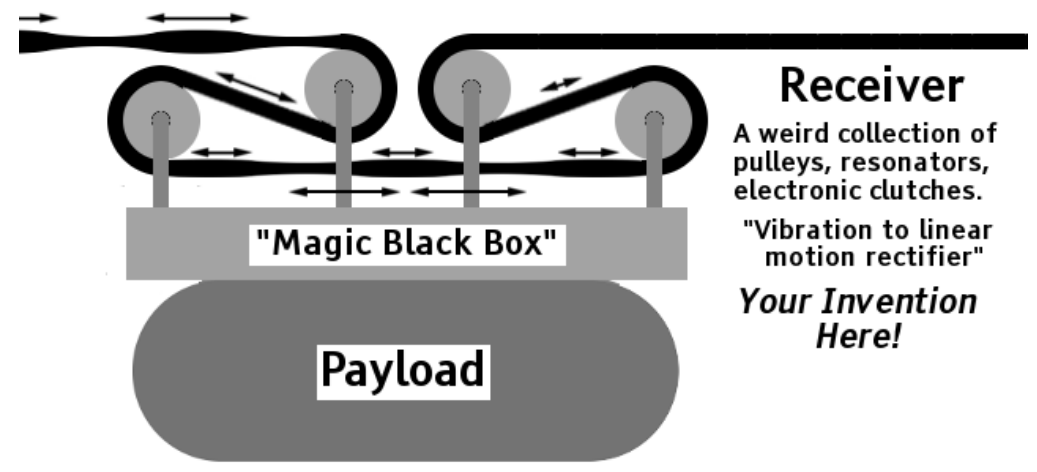}    
\caption{\label{fig:lofstrom-climber}
  Climber envisioned by \cite{lofstrom2017acoustic} showing what could be a
  long or variable tether path. Republished with permission.}
\end{centering}
\end{figure}

[\cite{lofstrom2017acoustic}] seems to have
envisioned the desirability of a variable tether path and/or a long climber in the climber
concept he sketched reproduced in Figure~\ref{fig:lofstrom-climber}. In his
concept, instead of the climber being long, it contains a long length of
tether. A climber containing a long length of tether without getting too
many reflections from rollers may be a viable alternative to a long climber. 

Overall, tether design pushes towards higher frequency operation. As the
frequency goes down, the change in climber velocity during a vibration
cycle and the variability of the tether path become unwieldy.

\subsubsection{Energy storage on climber}

The climber will have to have some energy storage on it. Indeed, incoming
power will be time-varying as the incoming wave alternates between upward
and downward velocity. Assuming a waveform that is symmetric between upward
and downward motion, the amount of energy that needs to be stored is
between zero for a triangular wave, and one half cycle of energy for
impulses, with neither extreme being practical.

This energy storage can take many forms. It could just
be kinetic energy of the climber, if some oscillatory acceleration of the
climber is permissible. In a mechanical implementation, it may be contained
in a spring or in the kinetic energy of some component of the climber. With
an electrical transmission, super-capacitors would be a good candidate
because of the extremely high charge/discharge rate that is needed.

If transverse waves were practical, circularly polarized waves would be
possible where both the North-South and East-West waves are used. This
would allow the power to the climber to be constant in time, and storage
would not be needed. To produce these transverse waves, one could imagine 
that a conversion device could be attached to the tether above the
atmosphere that is able to convert longitudinal waves into transverse ones.
Launch would be done by pulling the climber down and attaching it above
this device. It is unclear whether the conversion device could be made
light enough to make the removal of the on-climber energy storage
worthwhile.

\subsubsection{Some climber concepts}

Various climber concepts can be envisioned. The following non-exhaustive list contemplates
a few options, each needing fleshing out.


\begin{itemize}
  \item \textbf{The ratchet wheel climber:} A ratchet mechanism connected to the
    climber's wheels could allow the climber to be accelerated upward when the tether
    rises, and to keep coasting as the tether is pulled back down. An
    animation of this concept can be found in~[\cite{gassend2025mechanicalwaves}]. The 
    frequency has to be high enough that the climber doesn't lose too much
    momentum while it is coasting up. Some unpublished simulations suggest
    that this approach can be made to work, but a lot of power gets
    transmitted through the climber, and resonance is needed to get
    sufficient tension below the climber to actually get it to start
    coasting up.

    The ratcheted wheels can be replaced with wheels that can be either
    free or clamped. This simple variation may allow more reliable or
    efficient implementations than a ratchet. As long as the clamping
    occurs when the wheels are not rotating, as would be the case with the
    ratchet, there will be no energy dissipated in the clamping mechanism.

  \item \textbf{The clamping climber:} Instead of clamping wheels, the
    wheels could be eliminated and the clamping could be directly on the tether
    such that when the climber is coasting it is not touching
    the tether at all, and when it is being accelerated it is simply
    clamped to the tether. While it is clamped, the climber would have to
    apply suitable lateral forces to transfer
    Coriolis forces and avoid contact during the coast phase. The benefit
    of this approach is that the climber can reach very high velocities
    relative to the tether in the coast phase without needing any rolling
    mechanisms and the corresponding high tip-speed structural problems.
    However, the moment at which clamping happens is very critical. The
    transition from no-contact to fully loaded has to occur when there is low relative speed
    between the tether and climber, and be fast to
    avoid dissipating significant energy into wear-causing slippage between
    the tether and the clamps.

  \item \textbf{The electric climber:} The baseline design climber could
    in principle use its motors as generators. By varying the speed of the
    motors in a manner that oscillates at the same frequency as the tether
    vibrations, energy could be absorbed from the vibrations by applying a
    force that counters the vibrational motion of the tether, and the
    absorbed energy could be put into climber motion by having an average
    torque on the motor equivalent to what is needed to lift the climber.
    The main challenge with this method is likely to be that the moment of inertia
    of the wheels and motors could dominate the drag force we
    are trying to create to absorb energy from the tether. The motors could
    in principle be used to counter the inertia of the wheels, but
    the large torque involved in continually accelerating the wheels will
    likely cause unacceptable power dissipation in the motors. 

    Another way to address this wheel acceleration problem
    could be to have wheels that have a rotational
    spring constant tuned such that there is a resonance at the frequency
    of the tether vibrations. The resonance would then take care of
    accelerating the wheels to match the tether motion, and the motors
    would only have to apply the drag force. This approach is equivalent to
    the electrical approach, but it may be possible to achieve a higher
    quality factor using mechanical means.

  \item \textbf{Eccentric wheels/loads:} If a climber was progressing along the
    tether with eccentric wheels, it would induce vibrations into the
    tether. It should be possible to reverse this effect by phasing the
    eccentricity of the wheels such that the vibration they produce cancels
    the vibrations traveling up the tether.
    
    Of course, with roller pairs, if the surface of the wheel is eccentric,
    it will be non-trivial to avoid slippage and maintain constant
    pressure. Instead, the eccentricity could be limited to the mass
    distribution of the wheel. Or the eccentricity could be moved away from
    the wheels altogether into an eccentric flywheel, which would have
    the same effect. Once it is at a speed that matches the tether vibrations, 
    the flywheel's phase could be adjusted such that tether vibrational
    energy is absorbed. The motor driving the flywheel would then be used
    as a generator used to power the climber. Unfortunately, the climber would
    now have to be equipped with a set of generators able to handle the full
    power needed to climb, which will be similar in mass to the climbing
    motors.
                                                    
    Eccentric rotating parts will couple both to transverse and
    longitudinal waves. To couple only with one type of wave, simultaneous
    rotations in opposite directions of off-center masses could be used, as 
    could a linear oscillatory motion.

    More investigation is needed to determine how large and eccentric
    these masses need to be. The speeds involved should be expected to be
    larger than the already large speeds involved in the tether vibrations,
    so these approaches may simply be impractical.
                                                      
  \item \textbf{Absorbing pistons:} For transverse waves the approach from 
    [\cite{wessels2018mechanical}] with several energy absorbing pistons driven
    by contact with the tether could be used. This method is possible for
    transverse waves because the direction from which energy is being
    harvested is perpendicular to the direction of motion. For longitudinal
    waves the tip of the piston in contact with the tether would have to
    pull itself along the tether at the climb speed which would be
    impractical.

\end{itemize}

We have only scratched the surface of what is possible here. Whether a
clever design exists that does this energy transfer without needing
impractical speeds or masses remains to be seen and is one of the biggest
challenges of vibrational power transfer to the climber.

When multiple climbers are considered, it becomes important to be able to
transfer power through one climber to the ones above it. With some schemes
such as the electric climber, this comes in a very natural and controlled
way. With others such as the ratchet, some power will get through, but
as yet unpublished simulations show that it is hard to control the amount
of power that makes it through and to reach a steady state of operation.

\subsection{Descending climbers}
\label{sec:vibrations-descending-climbers}

If sufficiently high wheel speeds are allowed, a descending climber can
easily dissipate its power into the tether. Some of the same approaches we have
described to convert vibrational power into climbing power can be used in
reverse. The eccentric wheels/loads climber would be particularly simple to
implement, as long as we are able to allow a sufficiently off-balance load
to spin fast enough. Likewise, the electric climber is directly reversible.
The ratchet wheel climber, on the other hand, is not reversible, as the
ratchet only allows motion in one direction. The clamped wheel approach
seems challenging because in the absence of an incoming wave that allows
the climber and tether to match speeds, clamping will have to happen with a
large relative speed between the climber and the tether. Indeed, when the
climber releases the tether, the tether will snap up at thousands of km/s
under the full climber weight of tension difference. Beyond the extreme
tip-speed related problems on the wheels, the potentially large rotational
energy of the rollers when clamping happens would be dissipated in the
clamping mechanism. The clamping climber could conceivably be viable if the
clamping can be done fast enough that little energy is dissipated in
friction.

If there was a way for climbers to pass each other, energy could be
transferred using vibrations from a descending climber to an ascending one.
Or energy could be recovered at the elevator's base or counterweight.

With multiple descending climbers, it becomes necessary
for climbers to let vibrational power pass through them, which may be more
or less challenging depending on the scheme used to couple the climber to the tether. 
The clamping climber approach, for instance could still be used as long as the clamping is
synchronized between the various climbers in a way that favors transmission
of energy to the desired destination, rather than reflection which will
cause standing waves and potentially excessive relative velocities between
the tether and the climber.

If energy recovery was not desired, transverse waves could be used; they
would largely be dissipated in the atmosphere.

\subsection{Climber assembly onto tether}

Assembly of the climber onto the tether will pose some challenges. It is
clearly impractical to attempt to mount a climber onto a tether that
is vibrating with amplitudes of hundreds of km/h. 

The simplest approach would be to immobilize the anchor point of the
tether while the climber is mounted. For wavelengths that are very long
compared to the climber this immobilization will be sufficient, even if
there are waves descending along the tether. These descending waves will be
reflected by the fixed anchor, placing the climber near a node of the
oscillations, where the motion due to the vibrations will be small.

\begin{figure}[h]
\begin{centering}
\includegraphics[width=.8\textwidth]{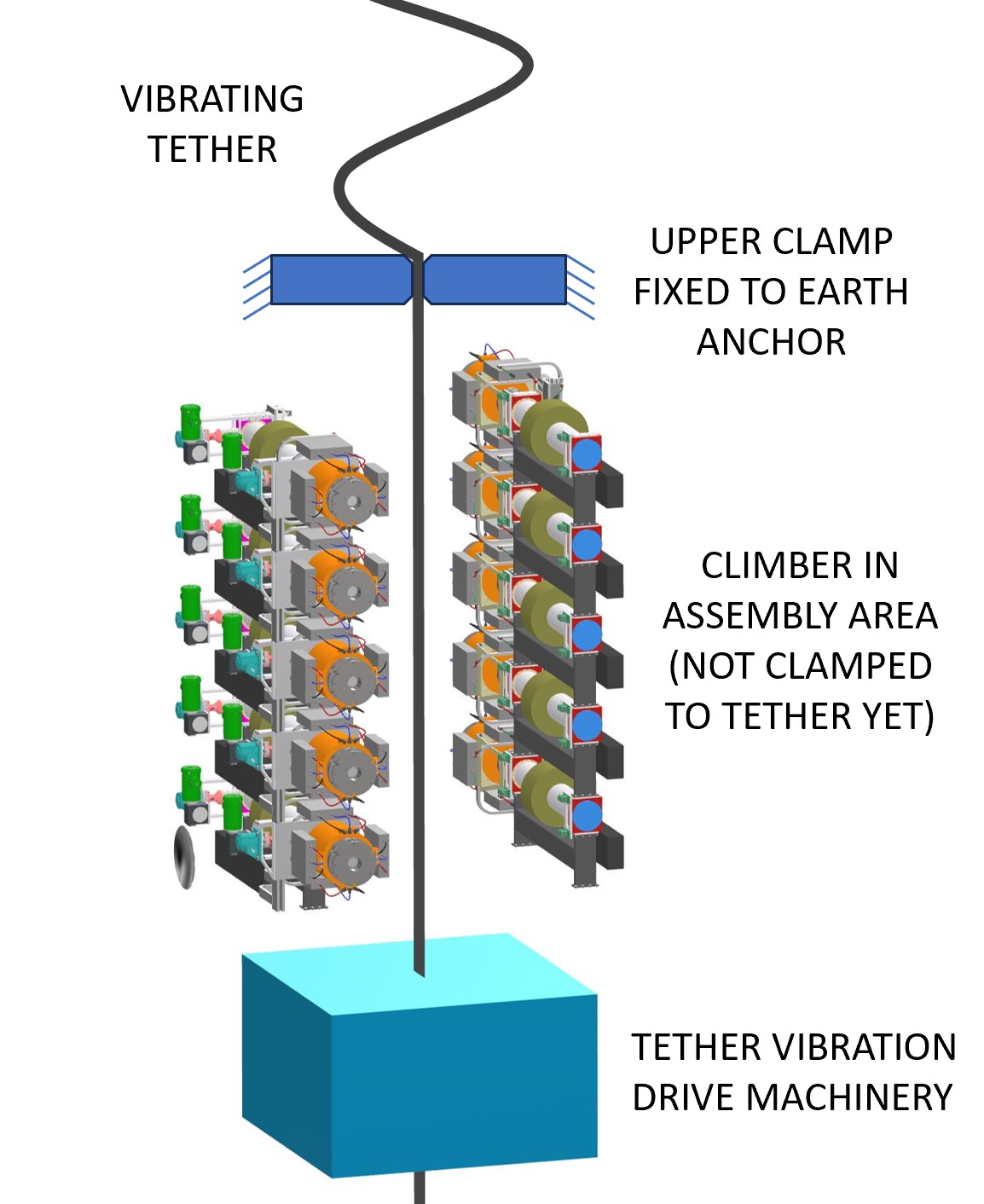}    
\caption{\label{fig:tethvibeclamp} If immobilizing the tether at the base of the elevator is insufficient to provide conditions suitable for climber assembly onto the tether, a clamp can be used to provide a length of tether that is truly immobilized. Transverse waves were used only for ease of visualization; in practice longitudinal waves would be more practical.}
\end{centering}
\end{figure}

For higher frequencies, there may be excessive motions of the tether due to
down-propagating waves, even with the anchor point immobilized. In this
case, it may be desirable to provide an additional clamp point above the
climber such that the tether at the climber truly is at rest.  See Figure \ref{fig:tethvibeclamp} for a concept sketch of this idea.

Naturally, the climber will have to be optimized to minimize assembly time,
as during that time power is not being sent up to the
climbers that are underway, reducing the utilization of the space elevator. If
the immobilization of the tether can be kept to a fraction of an hour, the
utilization reduction should be manageable.

\FloatBarrier

\subsection{Concluding remarks}

Considering mechanical waves for powering the climber, we have concluded
that transverse waves are impractical because of atmospheric
drag\footnote{While resonance can help, it requires such long wavelengths
that it is likely to make other aspects of the system too challenging.}, but
longitudinal waves look promising. Using techniques that adjust the impedance
seen by the hardware driving the oscillations in a way that keeps the
vibrational speeds low, the desired power can be injected into the tether
with simple century-old mechanical technology. Collecting that power at the
climber and transducing it into climbing power is more challenging.

It is not clear what the right frequency to operate at will be. Periods
beyond a few tens of seconds seem unnecessary as the mass of a wavelength of tether
becomes greater than the mass of the climber, making it increasingly
challenging for the climber to absorb the incoming energy except by reeling
large amounts of tether or moving extremely fast along the tether.
Frequencies above a few kHz are likewise unlikely to provide additional
benefit, as by that point the wavelength is a few meters and we have the
benefits of a long climber. Between those two extremes, the various schemes
considered here will have to be evaluated to find a combination that works. 

At the high frequency end, material properties like fatigue and attenuation
are likely to be the limiting factors, as well as backscattering from tether
damage. The highest frequency that meets those needs may be a good place to
operate, particularly if the wavelength remains less than ten or so meters
such that rotation-driven injectors remain possible. Otherwise, lower
frequencies where reeling is practical may be preferable. Straw man designs
and analysis will be needed to pin down what constitutes too high/low a 
frequency for the various schemes considered.

Mechanical vibrations are clearly a promising approach, with a lot of work
still needed to determine if there is a viable solution among all the
possibilities.

\clearpage

\section{Tether-guided microwaves}
\label{sec:external-microwaves}
\begin{figure}[h]
\begin{centering}
\includegraphics[width=\textwidth]{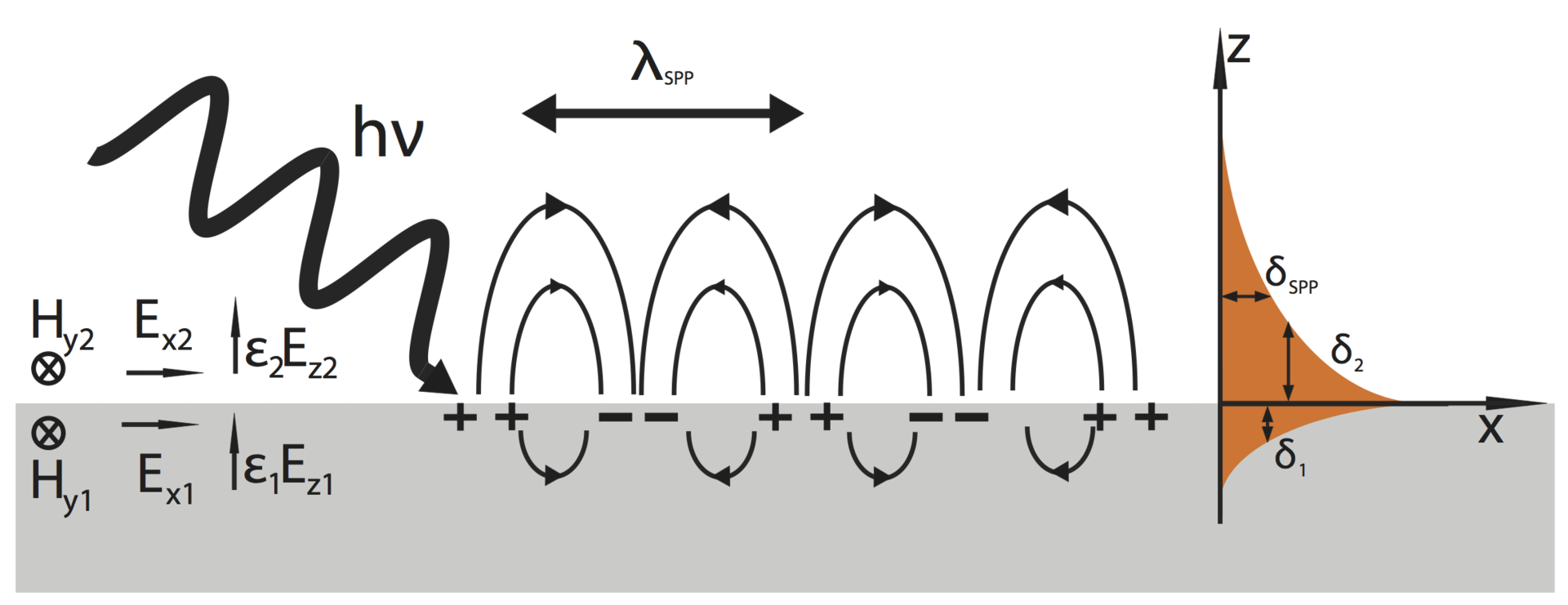}
\caption{\label{fig:surfwave} A surface wave, in this case, a surface plasmon polariton, propagating at a metal-dielectric interface. [\cite{surfacewave}]}
\end{centering}
\end{figure}

Another potential way of transmitting power to the climber is to use
electromagnetic waves guided by the tether. Surface waves are a type of
wave that propagates at the interface between two media, as in
Figure~\ref{fig:surfwave}. Unlike waves in free space, these waves have
a mode shape perpendicular to the direction of propagation that does not
vary or expand with propagation distance. The only change with distance is that there
is some attenuation. As a result, these modes do not suffer from
diffraction.

Several of these types of waves have been studied in the case of a cylindrical medium in free space, constituting a so-called single wire transmission line. \cite{sommerfeld1899ueber} (translated by \cite{warnock2019propagation}) was the first to consider them. He showed that an axially symmetric TM wave could propagate at the surface of a conductor with finite conductivity in a mode whose only variation with axial distance is an exponential decrease in overall amplitude, and which also decreases exponentially at a different rate with distance from the conductor. These waves tend to be quite large. Goubau discovered that by coating the conductor with a dielectric, the mode size could be reduced [\cite{goubau1950surface}]. Goubau also found that surface texture could have a similar effect. Finally, guided modes can be produced along a dielectric rod [\cite{hondros1910elektromagnetische}], [\cite{elsasser1949attenuation}], and [\cite{chandler1949investigation}].  These are the same modes that are found in an optical fiber. As the dielectric rod becomes small relative to the wavelength, it becomes single-mode, and the energy of the field is increasingly in the medium around the rod rather than in the rod itself.

For a possible space elevator application, having most of the energy in the mode on the outside of the medium is crucial. In the region where the field is interacting with either a non-ideal conductor or a dielectric, loss occurs. With a larger mode, the fraction of the energy that is lost per unit length is lower and the attenuation rate goes down. Typical RF/microwave guided waves have an attenuation quoted in dB/m, while optical fibers use dB/km. To be of use for a space elevator we need attenuations well below the dB/Mm range. This will only be possible by letting the bulk of the mode be in free space.

Finding modes whose shape does not change with distance along a medium with translational symmetry involves solving the 2-D Helmholtz equation [\cite{collin1990field}]. For the cylindrically symmetric cases we have considered, solutions are sought in cylindrical coordinates and can be expressed using Bessel functions, typically with complex arguments because of the attenuation with distance. Some transcendental equation solving is typically needed to determine the coefficients used in the Bessel functions.

For a ribbon shape like the space elevator, the natural coordinate system to use is elliptical coordinates, since the cross-section of the ribbon is essentially a line, which can be seen as the limit of an ellipse as eccentricity approaches unity. In elliptical coordinates, solutions will naturally be expressed using Mathieu functions. However, unlike the cylindrical case, the boundary conditions may not be satisfied by a single Mathieu function term, and solutions will have to be approximated using a series of Mathieu functions [\cite{yeh1962thesis}]. Several other papers have considered elliptical configurations such as [\cite{karbowiak1954elliptic}] and [\cite{heydari2019analysis}].

\begin{figure}[h]
\begin{centering}
\includegraphics[width=.8\textwidth]{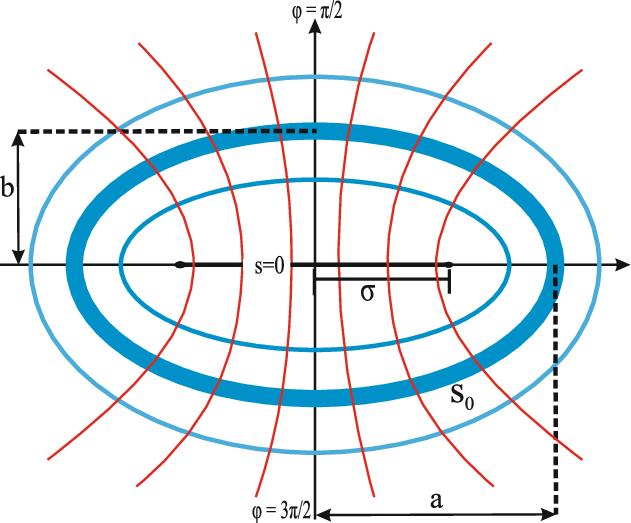}
\caption{\label{fig:elliptical} A sketch of elliptical coordinates in the xy plane. The elliptical coordinates are (s, $\phi$). The curves of s = constant are the confocal ellipses and curves of $\phi$ = constant are the orthogonal hyperbolae. The thick confocal ellipse represents the boundary of the tube, $s = S_0$. $\sigma$ is the distance between the center of the ellipse and the focal points. From [\cite{ellipticalcoords}].}
\end{centering}
\end{figure}

To determine whether these modes are practical for a space elevator, we need to know whether modes exist that are small enough that they can be received by a reasonably sized climber while having sufficiently small attenuation. To be practical, the mode would ideally be at most tens of meters in size, and would decay on a distance scale of tens of thousands of kilometers.

\begin{figure}[h]
\begin{centering}
\includegraphics[width=.95\textwidth]{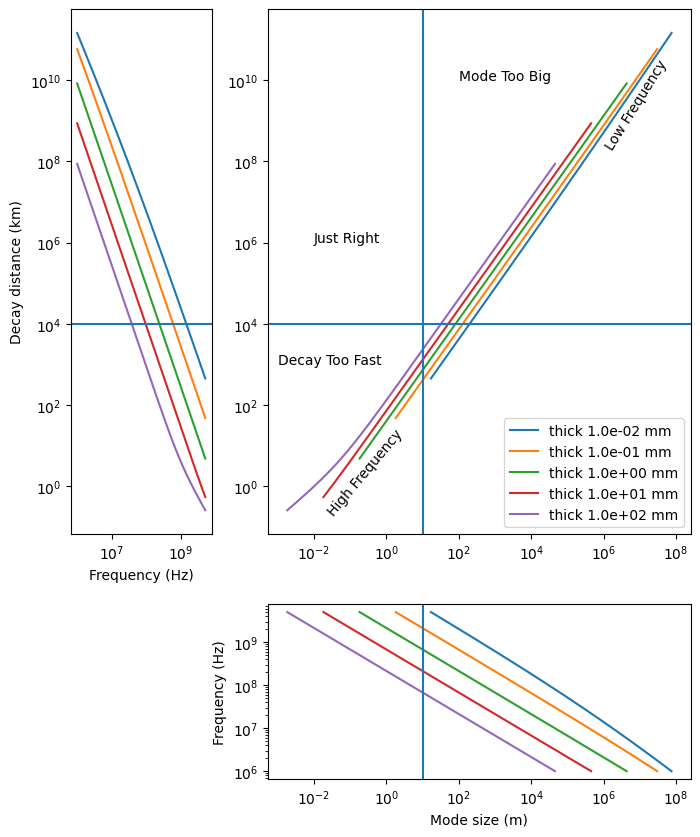}
\caption{\label{fig:single-wire-waveguide-attenuation} 
  Parametric plot of decay distance and mode size for a conducting plane
  coated with various thicknesses of polystyrene, as the frequency is
  varied. The frequency can be read off the left or bottom plots.
  The plotted configuration is merely a toy configuration with no direct
  relevance to the space elevator. The plot shows how for a given guide
  structure, frequency determines the parameters that are of interest to
  the space elevator designer: decay distance and mode size. Similar plots would
  need to be produced with various material configurations and with realistic
  tether geometries to determine if a viable externally guided mode solution exists.}
\end{centering}
\end{figure}

To get a sense of the mode sizes and attenuations, while avoiding the analytical challenges of Bessel and Mathieu functions, we have considered the case of waves propagating at the surface of an infinite planar structure. This case may be pessimistic if the field's decay away from the interface is at a similar rate regardless of geometry, with the area of the interface, where loss happens will be larger in the plane geometry. Figure~\ref{fig:single-wire-waveguide-attenuation} shows how mode size and attenuation are related as frequency is varied for a toy example. This particular planar configuration does not appear to be viable, but it is close enough that a ribbon configuration may work. Whether that is the case would be an interesting future investigation. As with free-space microwave transmission, the mode will have to be large enough to avoid excessive nonlinear interaction with the ionosphere as discussed in Section~\ref{sec:microwave-assumptions}.

We have mentioned various configurations that can support surface waves: non-ideal conductor, dielectric, non-ideal conductor coated with dielectric. Any of these configurations could potentially be viable depending on the ribbon material: graphene, hexagonal boron nitride or graphene with some surface layers of hexagonal boron nitride. It is possible that in the ionosphere, the mode will be perturbed by index changes due to the ambient plasma. The different plasma properties in the Debye sheath around the ribbon cold turn out to be sufficient to produce similar effects to a dielectric coating, allowing a plain graphene ribbon to behave like a Goubau line.

\begin{figure}
\begin{centering}
\includegraphics[width=0.75\textwidth]{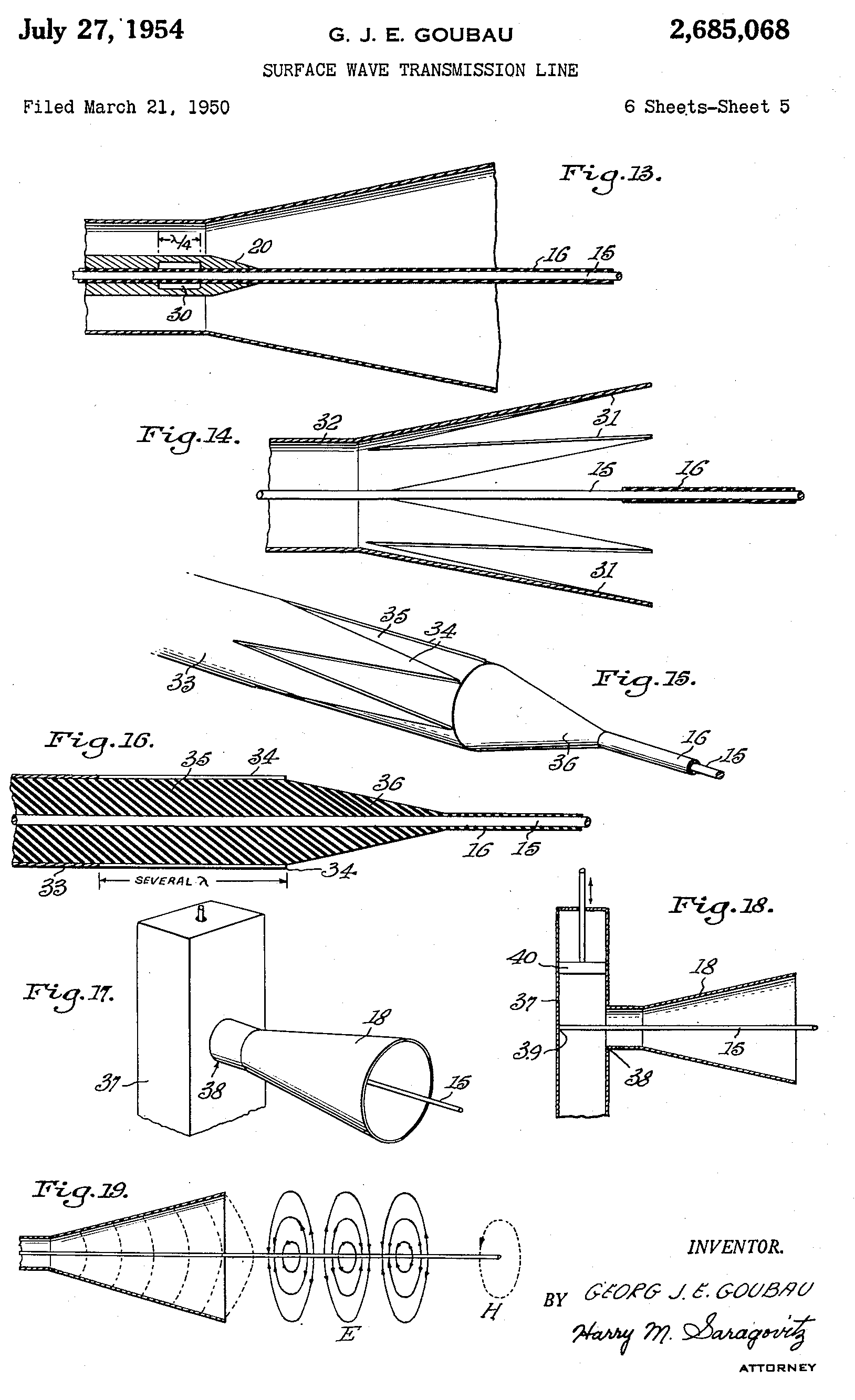}
  \caption{\label{fig:horn} Various horns and other devices envisioned to launch a surface wave on a wire by~[\cite{goubau_horn_patent}].}
\end{centering}
\end{figure}

Launching or receiving waves could be done using horns as described in [\cite{goubau1950surface}] and illustrated in Figure~\ref{fig:horn}.   These horns could convert from a moderately sized feed into the large dimensions of the surface wave mode. At the climber, it might also be possible to just sample the mode using a rectenna array designed to let a suitable proportion of the field through to supply climbers that are farther along. Alternatively, receive and transmit horns may be used to capture the beam, sample a suitable fraction of its power, and retransmit what is left. 

Bends in the ribbon will perturb the mode and cause losses. If the bend radius is large compared with the size of the mode, then these losses should be minimized. At the climber, the need to overcome Coriolis forces will always make a sudden bend which will likely need to be compensated for. This might be achieved by using a prism to deflect the beam, or if a pair of horns is used, these can be angled relative to each other appropriately.

Similarly, changes in ribbon width between the atmosphere and the ionosphere will likely cause impedance changes which will have to be done over large enough distances to prevent reflections.

At this point, it is uncertain whether there is a viable solution using tether-guided waves. The clear next steps in evaluating this option would be to determine how mode size and attenuation are related in the case of an actual ribbon geometry. If a suitable mode exists, it would then be time to flesh out the other practical considerations we have discussed.

\chapter*{Acknowledgements\vspace{-1cm}}
\addcontentsline{toc}{chapter}{\text{Acknowledgements}}
The authors wish to thank the following people for their preliminary review of the manuscript and helpful suggestions.\par
\bigskip
\noindent From Larry Bartoszek:

I can't thank all of the members of the PoweringTheSE group enough for all of the time and effort everyone put in over the last couple of years.  The expertise, professionalism, and patience of everyone who attended meetings every other Saturday morning from 2024 to 2026 and produced this report are appreciated more than you know.  I learned something every time we got together.  The fact that we all did it for no pay shows that this report was a labor of love.

Peter Robinson deserves special thanks for all the versions he proofread and improved with his suggestions.  

Martin Lades taught me more about optics and LaTeX than I've ever known.  I appreciate all the time he spent with me working through the errors and warnings in the documents and setting up the structure of the files.

I would also like to acknowledge my wife, Laura, who makes all of my work possible.  I'm sorry for all the time spent away from you writing and working on my "hobby/obsession".  You have been my rock for the last 47 years.
\par
\bigskip
\noindent From Avery Davis:

Section \ref{sec:microwave} is the result of email and Zoom discussions of the ISEC group, PoweringTheSE, starting in mid-2024. This author would especially like to thank the group leader, Larry Bartoszek, plus Blaise Gassend, Peter Robinson, John Knapman, Dennis Wright, Martin Lades, Adrian Nixon, and others for their helpful review comments. This author would also like to gratefully acknowledge the helpful review comments from James McSpadden of RTX, and Darel Preble of Georgia Tech. 

Peter Robinson provided very helpful tips in formatting the graphs in the microwave power beaming section as well.\par

\par
\bigskip
\noindent From Blaise Gassend:

A big thank you to the members of the PoweringTheSE group. Having a group
of people you can bounce research ideas off of as they mature, and having
deadlines to meet, makes the whole process more enjoyable and productive.
Without having that team around to report back to, none of the
presentations and writing I have done in the past 18 months would have
happened. Special thanks to Larry Bartoszek who helped wherever he could
with figures, proof reading and bibliography work on my sections of this
report. He really helped get them landed in time!
Thanks also to my wife Valerie for putting up with all the time I have sunk
into this effort.

\titleformat{\chapter}[display]{\normalfont\sffamily\LARGE\bfseries\color{docblue}}{Appendix~\Alph{chapter}}{11pt}{\Large}
\begin{appendices}
\renewcommand{\thechapter}{\Alph{chapter}}

\chapter{International Space Elevator Consortium}\label{chapt:whoweare}

\centerline{\textbf{Who We Are}} 
The International Space Elevator Consortium (ISEC) is composed of individuals and organizations
from around the world who share a vision of humanity in space.
\\

\centerline{\textbf{Our Vision}}
Space Elevators are the Green Road to Space, enabling humanity's most important missions by
moving massive tonnage to GEO and beyond. They will accomplish this safely, routinely, inexpensively
and daily.  They are environmentally neutral.
\\

\centerline{\textbf{Our Mission}}
The ISEC promotes the development, construction and operation of a space elevator infrastructure
as a revolutionary and efficient way to space for all humanity.
\\

\centerline{\textbf{What We Do}}
\begin{itemize}
  \item Provide technical leadership promoting development, construction, and operation of space
           elevator infrastructures.

  \item Become the “go to” organization for all things space elevator.

  \item Energize and stimulate the public and the space community to support a space elevator
           for low cost access to space.

   \item Stimulate science, technology, engineering and mathematics educational activities while
            supporting educational gatherings, meetings, workshops, classes and other similar events
            to carry out this mission.
\end{itemize}

\centerline{\textbf{A Brief History of ISEC}}
The idea for an organization like ISEC had been discussed for years, but it wasn’t until the Space
Elevator Conference in Redmond, Washington, in July of 2008, that things became serious. Interest
and enthusiasm for a space elevator had reached an all-time peak and, with Space Elevator
conferences upcoming in both Europe and Japan, it was felt that this was the time to formalize an
international organization.  An initial set of directors and officers were elected and they immediately
began the difficult task of unifying the disparate efforts of space elevator supporters worldwide.

ISEC's first Strategic Plan was adopted in January of 2010 and it is now the driving force behind
ISEC's efforts. This Strategic Plan calls for adopting a yearly theme to focus ISEC activities. Because
of our common goals and hopes for the future of mankind off-planet, ISEC became an Affiliate of the
National Space Society in August of 2013. In addition, ISEC works closely with the Japanese Space
Elevator Association.
\\
\\

\centerline{\textbf{Our Approach}}
ISEC’s activities are pushing the concept of space elevators forward. These cross all
disciplines and encourage people from around the world to participate. The following
activities are being accomplished in parallel:

\begin{itemize}
  \item Yearly conference: International Space Elevator Conferences were initiated by
            Dr. Brad Edwards in the Seattle area in 2002. Follow-on conferences were in
            Santa Fe (2003), Washington DC (2004), Albuquerque (2005/6 –smaller sessions),
            Seattle (2008 to 2019) and online meetings (2020-2022). Each of these
            conferences had multiple discussions across the whole arena of space elevators
            with remarkable concepts and presentations.

   \item Year-long technical studies: ISEC sponsors research into a focused topic each
            year to ensure progress in a discipline within the space elevator project. The first
            such study was conducted in 2010 to evaluate the threat of space debris. The
            products of these studies are reports that are published to document progress in
            the development of space elevators. They can be downloaded at www.isec.org.

   \item International cooperation: ISEC supports many activities around the globe to
            ensure that space elevators keep progressing towards a developmental program.
            International activities include coordinating with other major societies focusing
            on space, including the National Space Society (NSS), the British Interplanetary
            Soceity and the Japanese Space Elevator Association. In addition, ISEC supports
            symposia and presentations at the International Space Development Conference
            (NSS) and the International Academy of Astronautics and the International
            Astronautical Federation Congress each year.

   \item Publications: ISEC publishes a monthly e-Newsletter and yearly study reports.

   \item Reference material: ISEC is building a Space Elevator Library, including a reference database
            of space-elevator-related papers and publications. (See www.isec.org .)

   \item Outreach: People need to be made aware of the idea of a space elevator. Our outreach
            activity is responsible for providing the blueprint to reach societal, governmental, educational
            and media institutions and expose them to the benefits of space elevators. ISEC members are
            readily available to speak at conferences and other public events in support of the space
            elevator. In addition to our monthly e--Newsletter, we are also on Facebook, Instagram,
            LinkedIn, Twitter and YouTube.

   \item Legal: The space elevator is going to break new legal ground. Existing space treaties may
            need to be amended, new treaties may be needed and international cooperation must be sought.
            Insurability will be a requirement. Legal activities encompass the legal environment of a space
            elevator including international maritime, air and space law. Also, there will be interest within
            intellectual property, liability and commerce law. Starting work on the legal foundation well in
            advance will result in a more rational product.

   \item History committee: ISEC supports a small group of volunteers to document the history of space
            elevators. The committee’s purpose is to provide insight into the progress being achieved
            currently and over the last century.

   \item Research committee: ISEC is gathering the insight of researchers from around the world with
            respect to the future of space elevators. As scientific papers, reports and books are published,
            the research committee is pulling together this relative progress to assist academia and industry
            to progress towards an operational space elevator infrastructure.
\end{itemize}

ISEC is a traditional U.S. not-for-profit 501 (c) (3) organization with a board of directors
and four officers: President, Vice President, Treasurer, and Secretary.
inbox@isec.org / www.isec.org


\chapter{ISEC and IAA Studies}\label{chapt:studieslist}
\section{ISEC Study Reports}
All past and current ISEC reports are listed here. They are available for sale in hardcopy
or at no cost in pdf format at www.isec.org.
\par
\bigskip
\noindent \textbf{2010 – Space Elevator Survivability, Space Debris Mitigation} evaluated the
threat of space debris and proposed ways to reduce it.
\par
\bigskip
\noindent \textbf{2012 – Space Elevator Concept of Operations} focused on how space elevators
would be organized and operated.
\par
\bigskip
\noindent \textbf{2013 – Design Considerations for the Tether Climber} studied the physical,
operational and economic aspects which constrain space elevator climber design.
\par
\bigskip
\noindent \textbf{2014 – Space Elevator Architecture and Roadmaps} delineated the development
of the space elevator as a transportation infrastructure.
\par
\bigskip
\noindent \textbf{2015 – Design Considerations for the Earth Port} examined the interface of the
space elevator and the Earth’s surface in terms of a harbor and transportation hub.
\par
\bigskip
\noindent \textbf{2016 – Design Considerations for the Space Elevator GEO Node and Apex Anchor}
 considered the functional aspects of the GEO and Apex Anchor regions in terms of the
 transportation infrastructure and space enterprise.
\par
\bigskip
\noindent \textbf{2017 – Design Considerations for a Software Space Elevator Simulation} studied
the requirements for a software space elevator simulator and provides an outline for its
development.
\par
\bigskip
\noindent \textbf{2018 – Design Considerations for a Multi-Stage Space Elevator} examined the
use of a rotating-bolt space tower at lower altitudes connecting to a tether at higher
altitudes.
\par
\bigskip
\noindent \textbf{2019 – Today's  Space Elevator} looked at the development of a space elevator 
into a global and interplanetary transportation infrastructure.
\par
\bigskip
\noindent \textbf{2020 \#1 – Today's Space Elevator Assured Survivability Approach for Space Debris}
re-examined and updated the space debris situation for 2020 and its effect on the space
elevator.
\par
\bigskip
\noindent \textbf{2020 \#2 – Space Elevators Are the Transportation Story of the 21st Century}
considered new mission capabilities supported by the space elevator infrastructure.
\par
\bigskip
\noindent \textbf{2021 - Space Elevators Are the Green Road to Space} examined the environmental 
impact of space elevators as compared to rockets.
\par
\bigskip
\noindent \textbf{2023 \#1 - The Climber-Tether Interface} The conditions at the interface between the space elevator tether and climber determine whether or not climbing is possible.  The construction of a tether material and a device to climb it seem possible in the near future. Study was also reported in Acta Astronautica.
\par
\bigskip
\noindent \textbf{2023 \#2 - Leverage Dual Space Access Architecture  Advanced Rockets and Space Elevators} This Dual Space Access Strategy will ensure the future we dream of becomes a reality. Space Elevators are needed to do the “heavy lifting” for movement off planet. Their transformational characteristics will enable so many of humanity’s dreams by moving huge mass to far away destinations. Printed paperback version available for purchase from Lulu.com.
\par
\bigskip
\noindent \textbf{2024 - Apex Anchor: Full-Service Logistics Transportation Node at the Top of the Gravity Well} The Apex Anchor will become a Full-Service Transportation Node focused around the “Logistics Center at the top of the Gravity Well.” The inherent rapid release velocity (7.76 km/sec) at an Apex Anchor – and the ability to release every day of the year towards any destination, such as the Moon, Mars and even Pluto – opens up the universe to humanity. Printed paperback version available for purchase from Lulu.com.
\par
\bigskip
\section{IAA Studies on Space Elevators}
The following studies were sponsored by the International Academy of Astronautics (IAA)
(www.iaaweb.org) and published by the Virginia Edition Publishing Company, Heinlein Prize Trust.
They are available at https://www.heinleinbooks.com/book-store.
\par
\bigskip
\noindent \textbf{2014 – Space Elevators: An Assessment of the Technological Feasibility and the Way Forward}
\par
\bigskip
\noindent \textbf{2019 – The Road to the Space Elevator Era}

\end{appendices} 
\nocite{*}
\printbibliography

{\section*{}}

 \pagecolor{teal}
 \thispagestyle{empty}

 \vspace{-3.0 cm}

 \begin{minipage}{15em}
 \textcolor{yellow}{\normalsize Front cover: CAD rendering of a 20 metric ton climber gripping the space elevator tether and receiving beamed laser power from four 5 MW free electron lasers.}
\end{minipage}

 \vspace{2.0 cm}

\begin{minipage}{.9\textwidth}
\includegraphics[width=\linewidth]{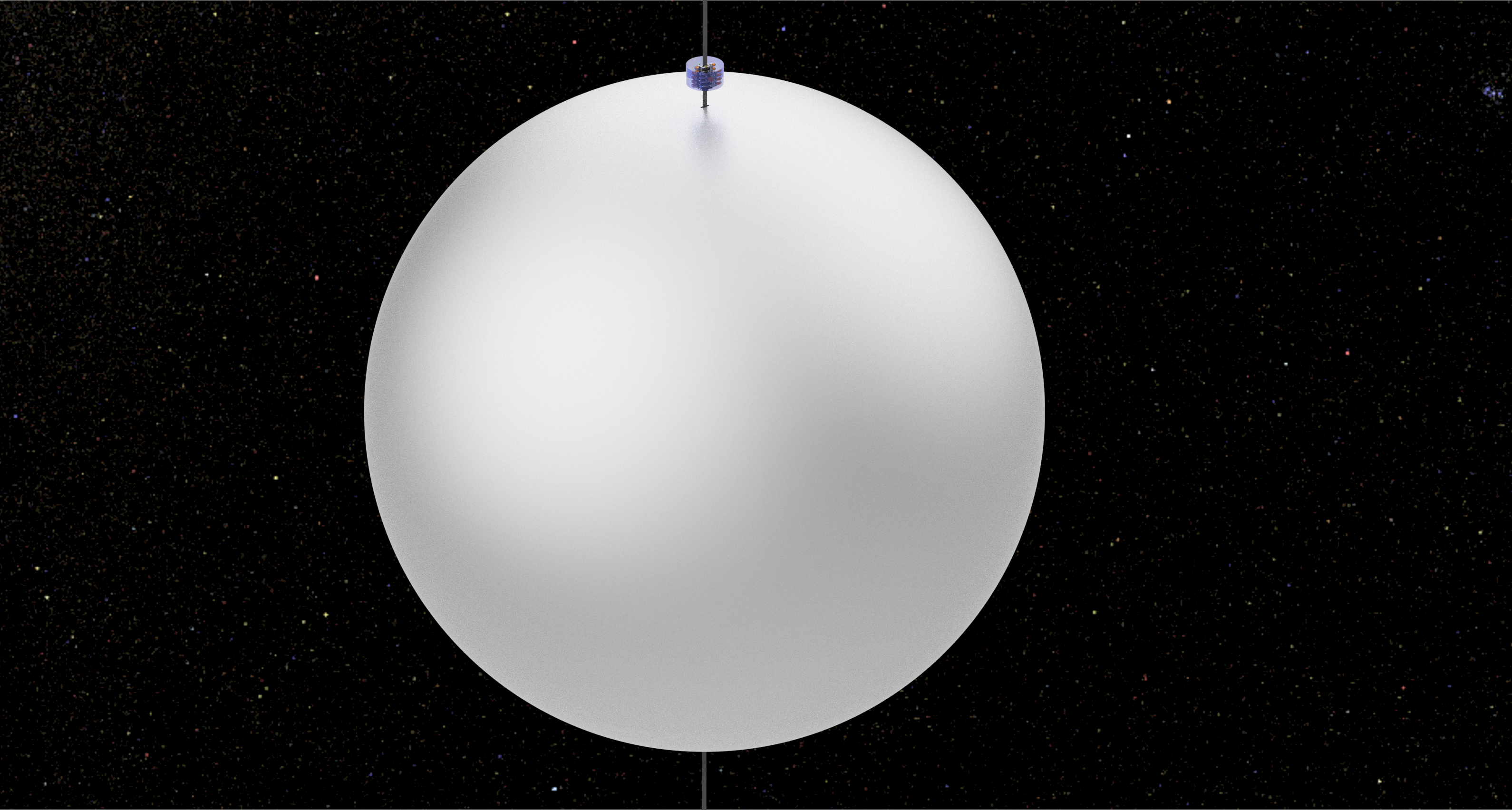}
\textcolor{yellow}{\normalsize Conceptual design for a solar-powered climber.  The sphere is an array of photocells that send power up to the climber.}
\end{minipage}

 \vspace{2.0cm}
 \begin{center}
 \textbf{\textcolor{yellow}{\Large }}
 \vspace{1.0cm}
                                       
 \textbf{\textcolor{yellow}{\Large A major step in developing the architecture of the Space Elevator will be to choose the power system for the climbers.  Each power system has advantages and disadvantages that must be weighed in the context of the total economics of the space elevator.}}
 \end{center}

\end{document}